\RequirePackage[l2tabu,orthodox]{nag}

\documentclass[headsepline,footsepline,footinclude=false,oneside,fontsize=11pt,paper=a4,listof=totoc,bibliography=totoc,DIV=12]{scrbook} 

\PassOptionsToPackage{table,svgnames,dvipsnames}{xcolor}

\usepackage[utf8]{inputenc}
\usepackage[T1]{fontenc}
\usepackage[english]{babel}
\usepackage[autostyle]{csquotes}
\usepackage[%
  backend=bibtex,
  url=false,
  style=numeric,
  maxnames=4,
  minnames=3,
  maxbibnames=99,
  giveninits,
  uniquename=init]{biblatex} 
\usepackage{graphicx}
\usepackage{scrhack} 
\usepackage{listings}
\usepackage{lstautogobble}
\usepackage{tikz}
\usepackage{pgfplots}
\usepackage{pgfplotstable}
\usepackage{booktabs}
\usepackage[final]{microtype}
\usepackage{caption}
\usepackage[hidelinks]{hyperref} 
\usepackage{amsmath,amssymb,amscd,amsthm,mathrsfs}
\usepackage{dsfont}
\usepackage{bbm}

\setkomafont{disposition}{\normalfont\bfseries} 
\bibliography{../../../../../control21vJan20}

\BeforeTOCHead[toc]{{\cleardoublepage\pdfbookmark[0]{\contentsname}{toc}}}

\pgfplotsset{compat=newest}
\pgfplotsset{
  cycle list={TUMBlue\\TUMAccentOrange\\TUMAccentGreen\\TUMSecondaryBlue2\\TUMDarkGray\\},
}

\theoremstyle{plain}
\newtheorem{definition}{Definition}[section]
\newtheorem{lemma}[definition]{Lemma}
\newtheorem{corollary}[definition]{Corollary}
\newtheorem{theorem}[definition]{Theorem}
\newtheorem{proposition}[definition]{Proposition}

\theoremstyle{definition}
\newtheorem{remark}[definition]{Remark}

\newtheorem{example}[definition]{Example}
\newcommand{\tops}{\tau_{\mathrm{s}}}
\newcommand{\topw}{\tau_{\mathrm{w}}}
\newcommand{\topn}{\tau_{\mathrm{n}}}
\newcommand{\topuw}{\tau_{\mathrm{uw}}}
\newcommand{\basiss}{\mathcal B_{\mathrm{s}}}
\newcommand{\basisw}{\mathcal B_{\mathrm{w}}}
\newcommand{\unitvector}{\mathbbm{e}}

\usepackage{longtable}
\usepackage{imakeidx}
\makeindex[options=-s mkidx]
\makeindex[name=symbol, title=Index of Notation, options=-s mkidx_symb]

\counterwithout{footnote}{chapter} 

\newcommand*{\getUniversity}{Technische Universit\"at M\"unchen}
\newcommand*{\getFaculty}{Fakult{\"a}t f{\"u}r Chemie}
\newcommand*{\getTitle}{Reachability in Controlled Markovian Quantum Systems}
\newcommand*{\getSubtitle}{An Operator-Theoretic Approach}
\newcommand*{\getAuthor}{Frederik vom Ende}
\newcommand*{\getSupervisor}{Prof.~Dr.~Bernd Reif}
\newcommand*{\getAdvisor}{Prof.~Dr.~Steffen J.~Glaser}
\newcommand*{\getExaminerOne}{Prof.~Dr.~Robert K\"onig}
\newcommand*{\getExaminerTwo}{Prof.~Dr.~Dariusz Chru{\'s}ci{\'n}ski (Universit{\"a}t Toru{\'n})}
\newcommand*{\getSubmissionDate}{28.09.2020}
\newcommand*{\getAcceptanceDate}{28.10.2020}
\newcommand*{\getSubmissionLocation}{Garching}

\begin{document}
\pagenumbering{alph}

\frontmatter{}

\begin{titlepage}
  \centering
\vspace*{-23mm}
	\includegraphics[width=0.24\textwidth , bb=0 0 740 740]{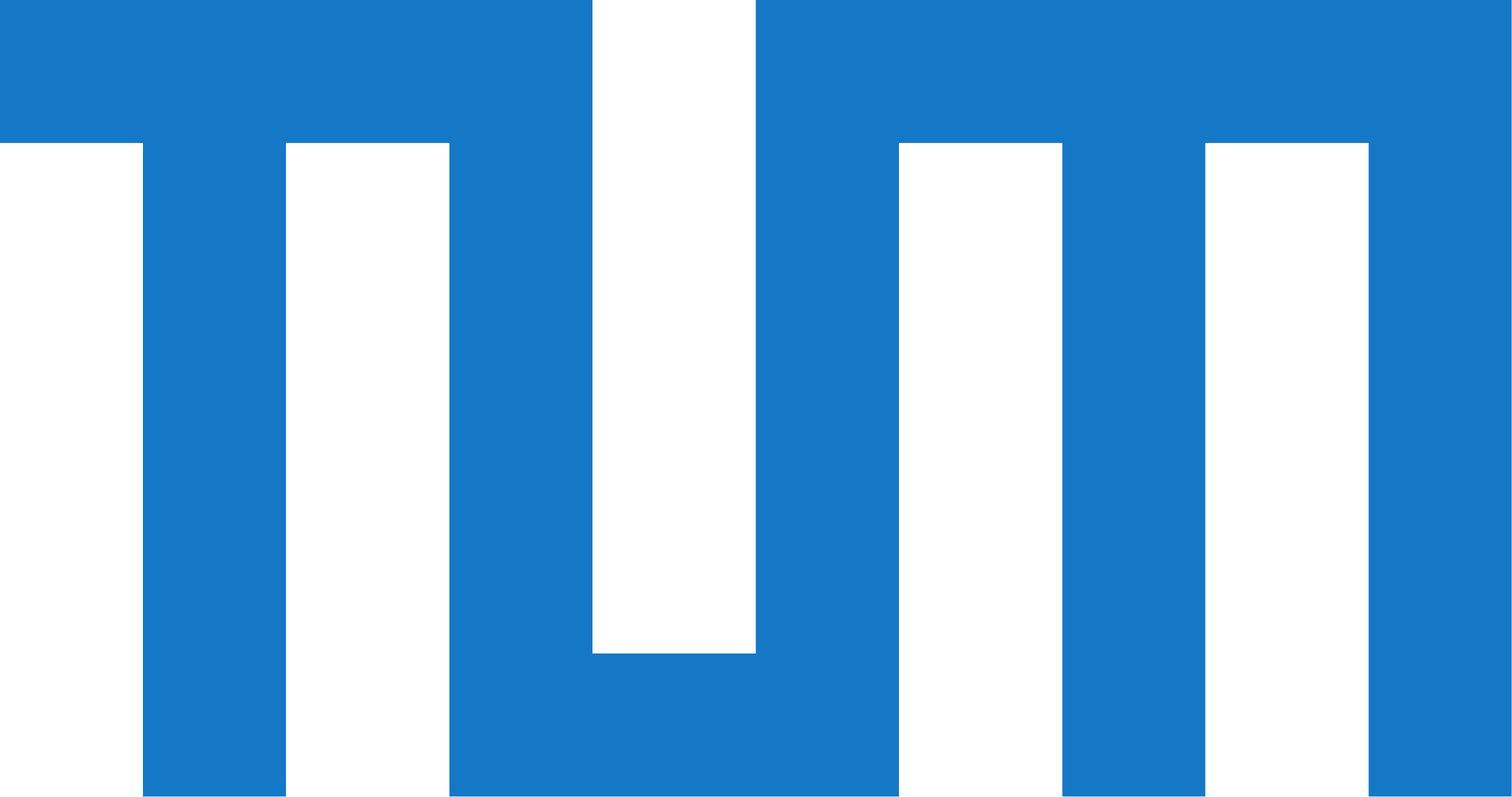}

  \vspace{5mm}
  {\huge\MakeUppercase{\getUniversity{}}}\\

  \vspace{5mm}
  {\large\MakeUppercase{\getFaculty{}}}\\

  \vspace{10mm}
 \hrule  
  \vspace{10mm}
  {\huge\bfseries \getTitle{}}
  \vspace{3mm}
  
  {\LARGE\bfseries \getSubtitle{}}
  \vspace{10mm}
  
 \hrule  
 
  \vspace{15mm}

  {\Large 
  \getAuthor{}}


  \vspace{15mm}
  
\begin{center}
  Vollst{\"a}ndiger Abdruck der von der Fakult{\"a}t f{\"u}r Chemie der Technischen\\
Universit{\"a}t M{\"u}nchen zur Erlangung des akademischen Grades eines\medskip

\textit{Doktors der Naturwissenschaften}\medskip

genehmigten Dissertation.

  \end{center}  
  \vspace{12mm}
  \begin{tabular}{l l}
    Vorsitzender:      &\hphantom{1. }\getSupervisor{} \medskip\\
    Pr{\"u}fer der Dissertation:         & 1. \getAdvisor{} \\
           & 2. \getExaminerOne{} \\
    &3. \getExaminerTwo{}
  \end{tabular}
  
\vspace{15mm}  
  
  Die Dissertation wurde am \getSubmissionDate\ bei der Technischen Universit{\"a}t
  M{\"u}nchen\\eingereicht und durch die Fakult{\"a}t f{\"u}r Chemie am \getAcceptanceDate\ angenommen.

\end{titlepage}

\newpage\null\thispagestyle{empty}\newpage
 
\thispagestyle{empty}
\section*{Declaration}\vspace{10mm}
\noindent
I hereby declare that the content of my thesis is original work and is based on the following
publications, which have already been submitted to or planned to be submitted to scientific
journals:\vspace{5mm}
\begin{itemize}
\item F.~vom Ende, G.~Dirr: \textit{The $d$-Majorization Polytope}. (2020) \href{https://arxiv.org/abs/1911.01061}{arXiv:1911.01061}\vspace{4pt}
\item F.~vom Ende: \textit{Strict Positivity and $D$-Majorization}. Accepted to Lin. Multilin. Alg. (2020) \href{https://arxiv.org/abs/2004.05613}{arXiv:2004.05613}\vspace{4pt}
\item G.~Dirr, F.~vom Ende: \textit{Von Neumann Type of Trace Inequalities for Schatten-Class Operators}. J. Oper. Theory 84 (2020), pp.~323--338. DOI: \href{https://doi.org/10.7900/jot.2019jun03.2241}{10.7900/jot.2019jun03.2241}\vspace{4pt}
\item G.~Dirr, F.~vom Ende, T.~Schulte-Herbr{\"u}ggen. \textit{Reachable Sets from Toy Models to Controlled Markovian Quantum Systems}. Proc. IEEE Conf. Decision Control (IEEE-CDC) 58 (2019), p.~2322. DOI: \href{https://doi.org/10.1109/CDC40024.2019.9029452}{10.1109/CDC40024.2019.9029452}\vspace{4pt}
\item F.~vom Ende, G.~Dirr, M.~Keyl, T.~Schulte-Herbr{\"u}ggen: \textit{Reachability in Infinite-Dimensional Unital Open Quantum Systems with Switchable GKS-Lindblad Generators.} Open Syst. Inf. Dyn. 26 (2019), p.~122702. DOI: \href{https://doi.org/10.1142/S1230161219500148}{10.1142/S1230161219500148}\vspace{4pt}
\item F.~vom Ende, G.~Dirr: \textit{Unitary Dilations of Discrete-Time Quantum-Dynamical Semigroups}. J. Math. Phys. 60 (2019), p.~122702. DOI: \href{https://doi.org/10.1063/1.5095868}{10.1063/1.5095868}\vspace{4pt}
\item G.~Dirr, F.~vom Ende: \textit{Author's Addendum to ``The $C$-Numerical Range in Infinite Dimensions''}. Lin. Multilin. Alg. 68.4 (2019) pp.~867--868. DOI: \href{https://doi.org/10.1080/03081087.2019.1604624}{10.1080/03081087.2019.1604624}\vspace{4pt}
\item G.~Dirr, F.~vom Ende: \textit{The $C$-Numerical Range in Infinite Dimensions}. Lin. Multilin. Alg. 68.4 (2018) pp.~652--678. DOI: \href{https://doi.org/10.1080/03081087.2018.1515884}{10.1080/03081087.2018.1515884}
\end{itemize}

\vspace{15mm}
\noindent
\getSubmissionLocation{}, \getSubmissionDate{} \hspace{50mm} \getAuthor{}

\cleardoublepage{}
\newpage
\section*{List of Publications}\vspace{10mm}
\begin{itemize}
\item F.~vom Ende, G.~Dirr: \textit{The $d$-Majorization Polytope}. (2020) \href{https://arxiv.org/abs/1911.01061}{arXiv:1911.01061}\vspace{4pt}
\item F.~vom Ende: \textit{Strict Positivity and $D$-Majorization}. Accepted to Lin. Multilin. Alg. (2020) \href{https://arxiv.org/abs/2004.05613}{arXiv:2004.05613}\vspace{4pt}
\item G.~Dirr, F.~vom Ende: \textit{Von Neumann Type of Trace Inequalities for Schatten-Class Operators}. J. Oper. Theory 84 (2020), pp.~323--338. DOI: \href{https://doi.org/10.7900/jot.2019jun03.2241}{10.7900/jot.2019jun03.2241}\vspace{4pt}
\item S.~Chakraborty, D.~Chru{\'s}ci{\'n}ski, G.~Sarbicki, F.~vom Ende: \textit{On the Alberti-Uhlmann Condition for Unital Channels}. Quantum 4 (2020), p.~360. DOI: \href{https://doi.org/10.22331/q-2020-11-08-360}{10.22331/q-2020-11-08-360}
\item G.~Dirr, F.~vom Ende, T.~Schulte-Herbr{\"u}ggen. \textit{Reachable Sets from Toy Models to Controlled Markovian Quantum Systems}. Proc. IEEE Conf. Decision Control (IEEE-CDC) 58 (2019), p.~2322. DOI: \href{https://doi.org/10.1109/CDC40024.2019.9029452}{10.1109/CDC40024.2019.9029452}\vspace{4pt}
\item F.~vom Ende, G.~Dirr, M.~Keyl, T.~Schulte-Herbr{\"u}ggen: \textit{Reachability in Infinite-Dimensional Unital Open Quantum Systems with Switchable GKS-Lindblad Generators.} Open Syst. Inf. Dyn. 26 (2019), p.~122702. DOI: \href{https://doi.org/10.1142/S1230161219500148}{10.1142/S1230161219500148}\vspace{4pt}
\item B.~Koczor, F.~vom Ende, M.~de Gosson, S.~Glaser, R.~Zeier: \textit{Phase Spaces, Parity Operators, and the Born-Jordan Distribution}. Submitted to Comm. Math. Phys. (2018) \href{https://arxiv.org/abs/1811.05872}{arXiv:1811.05872}\vspace{4pt}
\item F.~vom Ende, G.~Dirr: \textit{Unitary Dilations of Discrete-Time Quantum-Dynamical Semigroups}. J. Math. Phys. 60 (2019), p.~122702. DOI: \href{https://doi.org/10.1063/1.5095868}{10.1063/1.5095868}\vspace{4pt}
\item G.~Dirr, F.~vom Ende: \textit{Author's Addendum to ``The $C$-Numerical Range in Infinite Dimensions''}. Lin. Multilin. Alg. 68.4 (2019) pp.~867--868. DOI: \href{https://doi.org/10.1080/03081087.2019.1604624}{10.1080/03081087.2019.1604624}\vspace{4pt}
\item G.~Dirr, F.~vom Ende: \textit{The $C$-Numerical Range in Infinite Dimensions}. Lin. Multilin. Alg. 68.4 (2018) pp.~652--678. DOI: \href{https://doi.org/10.1080/03081087.2018.1515884}{10.1080/03081087.2018.1515884}
\end{itemize}

\newpage

\chapter*{Abstract}

In quantum systems theory one of the fundamental problems boils down to: Given an initial state, which final states can be reached by the dynamic system in question? Formulated in the framework of bilinear control systems, the evolution shall be governed by an inevitable Hamiltonian drift term, finitely many control Hamiltonians allowing for (at least) piecewise constant control amplitudes, plus a bang-bang switchable noise term in Kossakowski-Lindblad form. In order to obtain constructive results for such such systems we first present new results
\begin{itemize}
\item on majorization: The set of all quantum states majorized by any initial state is trace norm-closed, in particular in infinite dimensions.
\item on $d$-majorization: The set of all vectors $d$-majorized by any initial vector from $\mathbb R^n$ forms a non-empty convex polytope which has at most $n!$ extreme points. If the initial state is non-negative, then one of these extreme points, which is unique up to permutation, classically majorizes everything from said polytope.
\item on strictly positive maps: The collection of all linear maps sending positive definite matrices to positive definite matrices forms a convex semigroup, which is open with respect to the set of positive maps.
\end{itemize}

Now assuming switchable coupling of finite-dimensional systems to a thermal bath of arbitrary temperature, the core problem boils down to studying points in the standard simplex amenable to two types of controls that can be used interleaved: Permutations within the simplex, and contractions by a dissipative one-parameter semigroup. We illustrate how the solutions of the core problem pertain to the reachable set of the original controlled Markovian quantum system. This allows us to show that for global as well as local switchable coupling to a temperature-zero bath one can (approximately) generate every quantum state from every initial state. Moreover we present an inclusion for non-zero temperatures as a consequence of our results on $d$-majorization.\medskip

Then we consider infinite-dimensional open quantum-dynamical systems following a unital Kossakowski-Lindblad master equation extended by controls. Here the drift Hamiltonian can be arbitrary, the finitely many control Hamiltonians are bounded, and the switchable noise term is generated by a single compact normal operator. Via the new majorization results mentioned above, we show that such bilinear quantum control systems allow to approximately reach any target state majorized by the initial one, as up to now only has been known in finite-dimensional analogues.

\newpage\null\thispagestyle{empty}\newpage
\chapter*{Zusammenfassung}

{\Large Titel: Erreichbarkeit in Kontrollierten Markovschen Quantensystemen: Ein Operatortheoretischer Ansatz}\vspace*{15pt}

{\small
\noindent Eines der fundamentalen Probleme der Quantensystemstheorie lautet: f{\"u}r einen gegebenen Anfangszustand, welche Endzust{\"a}nde k{\"o}nnen innerhalb eines dynamischen Systems erreicht werden? Formuliert im Rahmen bilinearer Kontrolltheorie wird die Zeitentwicklung des Systems durch einen unvermeidbaren Hamiltonschen Drift, endlich viele Kontroll-Hamiltonians mit (mindestens) st{\"u}ckweise konstanten Kontrollen, sowie ``Bang-Bang'' schaltbarer Kopplung an die Systemsumgebung in Kossakowski-Lindblad-Form beschrieben. Um konstruktive Ergebnisse zu erhalten, pr{\"a}sentieren wir zuerst neue Ergebnisse
\begin{itemize}
\item f{\"u}r Majorisierung: Die Menge aller von einem beliebigen Ausgangszustand majorisierten Quantenzust{\"a}nde ist abgeschlossen in der Spurnorm, insbesondere in unendlichen Dimensionen.
\item f{\"u}r $d$-Majorisierung: Die Sammlung aller von einem beliebigen Anfangsvektor (aus $\mathbb R^n$) $d$-majorisierten Vektoren ist ein nicht-leeres, konvexes Polytop mit maximal $n!$ Extrempunkten. Ist der Ausgangsvektor nicht-negativ, so majorisiert einer dieser Extrempunkte alles aus besagtem Polytop klassisch, und er ist bis auf Permutationen eindeutig bestimmt.
\item f{\"u}r strikt positive Abbildungen: Die Menge aller linearen Abbildungen, die aus positiv definiten Matrizen wieder positiv definite Matrizen machen, bildet eine konvexe Halbgruppe, welche offen ist bez{\"u}glich der Menge aller positiven Abbildungen.
\end{itemize}

F{\"u}r schaltbare Kopplung beliebiger endlichdimensionaler Systeme an ein thermales Bad endlicher Temperatur l{\"a}uft das Kernproblem auf die Betrachung von Punkten im Standard-Simplex heraus, welche den folgenden zwei abwechselnd einsetzbaren Kontrollen ausgesetzt sind: Permutationen im Simplex, sowie Kontraktionen durch eine dissipative Ein-Parameter Halbgruppe. Wir zeigen, wie sich L{\"o}sungen des Kernproblems auf die Erreichbarkeitsmenge des urspr{\"u}nglichen kontrollierten Markovschen Quantensystems {\"u}bertragen. Daraus folgern wir, dass man f{\"u}r globale, sowie lokale schaltbare Kopplung an ein Bad der Temperatur Null jeden Zustand von jedem Anfangszustand aus (approximativ) erreichen kann. Au{\ss}erdem pr{\"a}sentieren wir eine Obermenge f{\"u}r Temperatur ungleich Null als Konsequenz unserer neuen Ergebnisse bez{\"u}glich $d$-Majorisierung.\medskip

Weiterhin untersuchen wir unendlichdimensionale offene quanten-dynamische Systeme welche einer unitalen Kossakowski-Lindblad Mastergleichung, erweitert durch Kontrollen, folgen. Der Hamiltonsche Drift kann beliebig sein, die endlich vielen Kontrollhamiltonians sind beschr{\"a}nkt, und die schaltbare Kopplung an die Umgebung wird von einem einzigen, kompakten, normalen Operator erzeugt. Mit Hilfe der obigen neuen Majorisierungs-Resultate zeigen wir, dass innerhalb solcher bilinearen Quantenkontrollsysteme jeder Zustand, welcher vom Anfangszustand majorisiert wird, approximativ erreicht werden kann -- ein Ergebnis, welches bisher nur in endlichen Dimensionen bekannt war.
}

\newpage\null\thispagestyle{empty}\newpage
\thispagestyle{empty}

\vspace*{20mm}

\begin{center}
{\usekomafont{section} Acknowledgments}
\end{center}
First and foremost, I would like to express my deepest gratitude to Dr.~Gunther Dirr and Dr.~Thomas Schulte-Herbrüggen for their scientific guidance, supervision, and endless support throughout the last three years. I owe them a considerable portion of my scientific and personal growth as well as the beautiful experience that is mathematical control theory. \smallskip

Of course I wish to thank Prof.~Dr.~Steffen Glaser and the whole Glaser group for the good working environment they provided---in particular my introduction to NMR applications of quantum control in the weekly seminar, as well as the scientific discussions during the daily coffee breaks (despite me always drinking hot chocolate instead of espresso).\smallskip

I am grateful to Prof.~Dr.~Michael Keyl, a collaborator and close friend of ours, who supported me a lot when trying to understand dynamical systems and control theory in infinite dimensions. When becoming a PhD student I also attended his course on quantum field theory with great interest.\smallskip

Particular thanks go out to Prof.~Dr.~Robert König, Prof.~Dr.~Michael Wolf, and the whole chair M5 for their kindness and the opportunity to support them with their teaching duties. The tutoring I did for their lectures ``Analysis 3'', ``Representations of compact groups'', and ``Functional Analysis'' were a true pleasure and a great opportunity to deepen my own understanding of these subjects.\smallskip

Finally, I wish to thank Prof.~Dr.~Dariusz Chru{\'s}ci{\'n}ski for his welcoming attitude from when we first met all the way to my pleasant short stay in Toru{\'n}. My work on generalized majorization greatly benefited from my time there, in particular from illuminating discussions with him as well as Sagnik and Ujan.\smallskip

This work was supported by the Bavarian excellence network \textsc{enb} via the \mbox{International} PhD Programme of Excellence \textit{Exploring Quantum Matter} (\textsc{exqm}).

\cleardoublepage{}

\newpage\null\thispagestyle{empty}\newpage
\thispagestyle{empty}

\vspace*{50mm}

\begin{center}
{
\usekomafont{section}
\LARGE \textit{F{\"u}r meine Eltern}\\\vspace*{15pt}

\textit{Birgit und Werner}}
\end{center}
\newpage\null\thispagestyle{empty}\newpage
\microtypesetup{protrusion=false}
\tableofcontents{}
\microtypesetup{protrusion=true}

\mainmatter{}
\chapter{Introduction}\label{chapter:introduction}

Quantum systems theory and control engineering is a corner stone to unlock the potential of many
quantum devices in view of emerging technologies \cite{DowMil03,Roadmap2015}. Indeed, together with quantum information theory, this forms the foundation of the research field of quantum technologies which comprises quantum communication, quantum computation, quantum simulation, and quantum sensing \cite{Roadmap2018}. 
From a control perspective, the ability to generate certain states or even unitary gates, e.g., for communication protocols or general quantum computation (as part of the ``DiVincenzo criteria'' \cite{VincCriteria}), is of fundamental importance here. 
This is complemented by a number of optimal control tasks, such as error resistant single-qubit gates with trapped ions \cite{TGW08} for computation purposes, loading of ultra-cold atomic gas into an optical lattice \cite{Rosi13} which serves as one of the platforms of quantum simulation, or optimizing pulses against noise and other experimental imperfections \cite{KG04b,Braun14}, e.g., for quantum sensing, to name just a few\footnote{
For a more complete overview on the applications of quantum control in view of quantum technologies we refer to the European roadmaps \cite{Roadmap2018,Roadmap2015}.
}.

The great interest particularly in quantum computing within the last decades is due to the expectation that quantum computers significantly outperform classical computers. While this is based on well-founded conjectures in computational complexity theory, this advantage has been proven rigorously only recently for a certain class of problems which cannot be solved using classical constant-depth circuits \cite{Koenig18}. This advantage is a consequence of quantum non-locality, and remains under the restriction of geometrically local gates and corruption by noise \cite{Koenig20}.
Indeed the quantum circuit proposed by Bravyi et al.~in said articles is also a candidate for experimentally realizing quantum algorithms in the near future.

Among the list of platforms for quantum computation as well as simulation one finds ultra-cold atoms \cite{GMEHB02,Rosi13} and trapped ions \cite{BW08,Home09}, semiconductor nanostructures \cite{Watson18,Kim14}, and superconducting circuits \cite{Devoret04,Riste12,Campagne13}---for more detailed review articles cf.~\cite{Georgescu14,Devoret13,Gross17,Roadmap2018}.
While superconducting qubits are among the most promising for achieving fault-tolerant quantum computation \cite{Devoret13,Neill18}---also because they perform well when it comes to error correction \cite{Barends14,Reed12}---recently this field also opened up new perspectives in control engineering:
While quantum control usually is concerned with the systematic manipulation of the dynamics of nanosystems by external controls such as laser pulses or electro-magnetic fields, there are recent works on using dissipation for quantum state engineering \cite{VWC09,KMP11}, as well as fast tunable couplers for superconducting qubits \cite{Mart09,Mart13,Mart14}.
We will come back to this development later, after revisiting the mathematical foundation of quantum control theory. 
To ensure well-posedness of a large class of control tasks, e.g., in view of optimal control, it is advisable
to check first whether the desired target state is within the reachable set of the
dynamic system:\medskip

If a quantum system is closed, that is, the system is isolated from its environment, and its state space is of finite dimension, then such questions of controllability (i.e.~the possibility to generate every final state from every initial state in finite time) are well-understood in a rigorous manner: The system's evolution, originally described by the Schr\"odinger equation $\dot\psi(t)=-iH_0\psi(t)$, becomes 
\begin{equation}\label{eq:1_schr_eq_control}
\dot\psi(t)=-i\Big(H_0+\sum\nolimits_{j=1}^m u_j(t)H_j\Big)\psi(t)\quad\text{ with }\quad\psi(0)=\psi_0\in\{\psi\in\mathbb C^n\,|\,\langle \psi,\psi\rangle=1\}\,.
\end{equation}
Here the dynamics given by the inevitable drift Hamiltonian $H_0$ can be influenced by means of control Hamiltonians $H_1,\ldots,H_m$ modelling, e.g., electro-magnetic fields, and control amplitudes $u_1(t),\ldots,u_m(t)$. Now one may lift the problem from state vectors to the unitary group to obtain a differential equation for unitary propagators starting at the identity:
\begin{equation}\label{eq:unitary_prop_control_intr}
\dot U(t)=-i\Big(H_0+\sum\nolimits_{j=1}^m u_j(t)H_j\Big)U(t)\quad\text{ with }\quad U(0)=\mathbbm{1}\,.
\end{equation}
Controllability on the special unitary group turns out to be equivalent to controllability of the Liouville-von Neumann equation
$$
\dot\rho(t)=-i\Big[H_0+\sum\nolimits_{j=1}^m u_j(t)H_j,\rho(t)\Big]\quad\text{with}\quad\rho(0)=\rho_0\in\{\rho\in\mathbb C^{n\times n}\,|\,\rho\geq 0\text{ and }\operatorname{tr}(\rho)=1\}
$$
on the unitary orbit of each initial state. The idea of lifting the control problem to the group is not only strictly stronger than controllability of the Schr\"odinger equation \cite{AA03} but also gives access to strong tools from the fields of bilinear control systems \cite{Elliott09} and Lie group theory: Controllability of \eqref{eq:unitary_prop_control_intr} is fully settled by the simple Lie algebra rank condition\footnote{
One only has to check that the number of linearly independent elements within $H_0,H_1,\ldots,H_m$ together with all iterated commutators $[H_0,H_j]$, $[H_i,H_j ]$, $[H_0, [H_i,H_j ]]$, $\ldots$ is equal to the dimension of the (special) unitary algebra, cf.~Ch.~\ref{ch:control_fin_dim} for more details.
}
as described in the groundbreaking works of Jurdjevic, Sussmann \cite{SJ72,JS72}, and Brockett \cite{Bro72,Bro73}.\medskip


From this result there are three different paths one can pursue: 1. After proving the existence of a control sequence which steers an initial state to a target state, as a next step one usually asks about optimizing this sequence, e.g., to find a control scheme with high fidelity or minimized time or energy costs, or a scheme which is robust against environmental noise. While these questions undoubtedly are important in view of emerging technologies and industrial applications---as mentioned in the beginning---we, in this thesis, will instead stay at a more fundamental level and focus on reachability in two different scenarios:\medskip

2. With the finite-dimensional (closed) case being fully settled, moving to infinite dimensions 
makes things much more challenging. This step is natural due to quantum mechanics requiring 
infinite-dimensional Hilbert spaces and unbounded operators (cf., Ch.~\ref{sec_unbounded_op}, and footnote \ref{footnote_inf_dim_qc} on page 
\pageref{footnote_inf_dim_qc}). Well-studied examples of infinite-dimensional quantum control 
systems include, but are not limited to, atom-cavity systems as used in quantum optics 
\cite{walther2006cavity}. While the control aspect of such systems is understood to some 
degree\footnote{
Works in this field are restricted to Hamiltonians at most quadratic in position and momentum, 
which from an application point of view is rather restrictive.
} \cite{brockett2003controllability,rangan2004control,KZSH,boscain2015approximate,HoKe17,HeKe18}, 
``virtually all studies on infinite-dimensional quantum systems treat the controllability problem 
within the wave function picture'' \eqref{eq:1_schr_eq_control} (cf.~\cite{keyl18InfLie}, also for an overview on the methods used in this field). This is a 
serious limitation as it does not allow for a generalization to open systems and, as seen above, 
is not equivalent to controllability on the level of density operators. The more reasonable 
alternative is to study controllability of the operator lift \eqref{eq:unitary_prop_control_intr} 
which establishes a promising link to operator and representation theory, and the already rich 
infinite-dimensional Lie theory.

It turns out that exact controllability---that is, the reachable set of 
\eqref{eq:unitary_prop_control_intr} being equal to the full unitary group---in infinite dimensions 
is impossible (cf.~\cite{ball1982controllability} and Ch.~\ref{ch_control_infdim}) so one has to 
resort to an approximate version of controllability in a suitable topology. This was first 
studied in a recent paper by Keyl \cite{keyl18InfLie} whose remarkable main result was to 
prove \textit{strong approximate} controllability of \eqref{eq:unitary_prop_control_intr} for unbounded, 
self-adjoint $H_0$ with pure point spectrum, and bounded, self-adjoint $H_1,\ldots,H_m$ under 
certain assumptions on the eigenvalues of $H_0$ as well as the control Hamiltonians (Thm.~5.2 
in said paper). In this setting one finds an approximate version of the already mentioned Lie 
algebra rank condition; note that this condition for general infinite-dimensional systems is not 
sufficient anymore. Although this result does not cover unbounded operators with continuous 
spectrum or unbounded control operators---more on this in the conclusions, Ch.~\ref{ch:concl_outlook}---this is a proof of concept and a promising first 
step towards a better understanding of infinite-dimensional quantum control theory (for closed 
systems).\medskip

3. Last but not least, moving to open quantum systems (i.e.~systems which interact with its 
environment in a dissipative way) is most desirable in terms of applications. After all, the 
assumption of a system being closed is too inaccurate for a lot of experiments as shielding the 
system from its environment is often infeasible. While closed systems are rather well-studied, 
``questions of quantum state reachability in dissipative systems remain largely unresolved'' 
\cite[Ch.~4.3]{Roadmap2015}---even in the simplest case of the interaction being of Markovian 
nature. Mathematically the latter means that the (uncontrolled) evolution of a quantum system
$(T_t)_{t\geq 0}$ is a semigroup of quantum channels of some continuity type in the
time-parameter $t$, which by the pioneering results of Gorini, Kossakowski, Sudarshan
\cite{GKS76}, and Lindblad \cite{Lindblad76} are necessarily of the exponential
form $T_t=e^{tL}$ for all $t\geq 0$ with time-independent generator
$$
L(\rho)=-i[H,\rho]-\sum_{j\in I}\Big(\frac12 (V_j^*V_j\rho +\rho V_j^*V_j)-V_j\rho V_j^*\Big)\,.
$$
We recap these results in more detail in Ch.~\ref{ch:qu_dyn_sys}. For a discussion of when the evolution of a quantum system can be described by such a Markovian master equation, cf.~\cite[Ch.~3.2.1]{BreuPetr02} and Rem.~\ref{rem_control_born_markov}. Either way this simple form of the generator allows for the following adjustment to controlled Markovian quantum systems:
\begin{equation}\label{eq:controlled_markov}
\dot\rho(t)=-i\Big[H_0+\sum\nolimits_{j=1}^m u_j(t) H_j,\rho(t)\Big]-\sum_{j\in I}\Big(\frac12 (V_j^*V_j\rho +\rho V_j^*V_j)-V_j\rho V_j^*\Big)\,.
\end{equation}
Specifying reachable sets for such dissipative systems is rather challenging, to say the least---even in finite dimensions---and 
the reachable set takes the form of a (Lie) semigroup orbit \cite{DHKS08}. While for general Markovian control systems such tools from Lie semigroup theory are of great use \cite{ODS11}, there are some special cases where the reachable set can be specified more explicitly:
\begin{itemize}
\item If the system is unital, that is, the maximally mixed state $\frac{\mathbbm{1}}{n}$ is left invariant at all times, then \textit{majorization} (Ch.~\ref{ch_maj_cnr}) makes for a natural upper bound (cf.~Start of Ch.~\ref{sec:results}). However, this characterization becomes increasingly inaccurate the larger the system.
\item More recently, Bergholm et al.~\cite{BSH16} studied the case of switchable noise, meaning the dissipative part of \eqref{eq:controlled_markov} is controlled by means of a (bang-bang) control function $\gamma(t)$. This is motivated by recent experimental progress on superconducting qubits
\cite{Mart09,Mart13,Mart14,McDermott_TunDissip_2019}; thus this scenario is of physical interest and also allows for rigorous mathematical results.
\end{itemize}
Formulated as a bilinear control system this second scenario reads as follows:
\begin{equation}\label{eq:contr_mark_switch}
\dot\rho(t)=-i\Big[H_0+\sum\nolimits_{j=1}^m u_j(t) H_j,\rho(t)\Big]-\gamma(t)\Gamma(\rho(t))
\end{equation}
with $\Gamma(\rho)=\sum_{j\in I}(\frac12 (V_j^*V_j\rho +\rho V_j^*V_j)-V_j\rho V_j^*)$. 
It was shown in \cite{BSH16} for a system of one or more qubits, i.e.~the underlying Hilbert space is $\mathbb C^{2^n}$, that if the dissipation takes the form of local amplitude damping and if the closed system ($\gamma(t)=0$) allows to apply every unitary channel, then every quantum state can be approximately generated from every initial state\footnote{
Later we will formulate this as: The closure of the reachable set of \eqref{eq:contr_mark_switch} is equal to the set of all states $\mathbb D(\mathbb C^{2^n})$.
}.
Approximate controllability is the best result obtainable as Markovian control systems are never \textit{exactly} controllable (\cite[Thm.~3.10]{DiHeGAMM08} and Rem.~\ref{rem_control_sp_lamb}). 

Moreover Bergholm et al.~proved that, in the same scenario, if the dissipation takes the form of 
local bit-flip noise (instead of local amplitude damping) then one can approximately reach 
every quantum state majorized by the initial state. Thus this is also a special case of the unital 
systems described above where the upper bound of majorization can approximately be 
saturated. On top of this they numerically investigated feasibility of their results and proposed an 
implementation of these results via GMon \cite{Mart09,Mart13,Mart14}.

\section{A Guide on How to Read this Thesis}

In this thesis we will build upon these promising interdisciplinary results by considering systems \eqref{eq:contr_mark_switch} of one or more $d$-level systems (so the Hilbert space is $\mathbb C^{d^n}$ with $d\geq 2$) coupled to a thermal bath in a switchable manner, as well as infinite-dimensional control systems of this type with unital noise of a special form. More precisely a roadmap for this thesis looks as follows:\medskip

In Chapter \ref{chapter:prelim} we set the stage by recapping fundamental results about 
operator theory and operator topologies on normed (Ch.~\ref{ch_2_1_normedop}) and inner product spaces (Ch.~\ref{ch_2_3}), followed by a quick introduction to positive maps, 
quantum channels, and quantum-dynamical systems (Ch.~\ref{sec:quantum_channels}). This 
is complemented with important constructions from functional analysis in Appendix 
\ref{sec_refresh_func_ana}.
Following-up we apply these topological considerations to bounded
(Ch.~\ref{ch_3_1_top_con}) and unitary operators (Ch.~\ref{ch_3_2_unitary_gr}) on 
separable Hilbert spaces, and finally we give an introduction to bilinear and quantum control theory in 
finite (Ch.~\ref{ch:control_fin_dim}) and infinite dimensions (Ch.~\ref{ch_control_infdim}). 

We already saw that majorization is an important notion for the control problems we want to 
study in this thesis. Hence this concept, among related ones, is explored in Chapter 
\ref{ch_maj_cnr} where we develop a general toolbox necessary for said control problems. 
First we explore classical ($\prec$) and general $d$-majorization ($\prec_d$) on vectors and 
the associated convex polytope (Ch.~\ref{sec:maj_d_vec}). After this we lift these concepts to 
matrices while also coming across the notion of strict positivity (Ch.~\ref{sec:maj_d_mat}). 
Closely related to majorization is the $C$-numerical range of bounded and general Schatten 
class operators (Ch.~\ref{sec:c_num_range}) which will allow us to explore infinite-dimensional 
majorization and its properties (Ch.~\ref{maj:trace_class}). 
While this chapter is entirely new and fully based on our publications and preprints 
\cite{vomEnde19polytope,vomEnde20Dmaj,DvE18,DvE18_Schatten,vE_dirr_vonNeumann} the 
new results \textit{relevant to this thesis' controllability results} read as follows: 
\begin{itemize}
\item Given $y\in\mathbb R^n$, and a vector $d\in\mathbb R^n$ with positive entries, the set $\{x\in\mathbb R^n\,|\,x\prec_d y\}$ of vectors $d$-majorized by $y$ is a non-empty convex polytope which has at most $n!$ extreme points (Coro.~\ref{thm_convex_poly}).
\item One of these extreme points classically majorizes everything from said polytope and is unique up to permutation. In particular this extreme point $z$ satisfies 
$\frac{z_{{\pi}(1)}}{d_{{\pi}(1)}}\geq \ldots\geq \frac{z_{{\pi}(n)}}{d_{{\pi}(n)}}$ if $\pi$ is a permutation which orders $d$ decreasingly, i.e.~$d_{{\pi}(1)}\geq\ldots\geq d_{{\pi}(n)}$ (Thm.~\ref{theorem_max_corner_maj}).
\item The collection of all strictly positive maps, that is, all $T:\mathbb C^{n\times n}\to\mathbb C^{k\times k}$ linear which map positive definite matrices to positive definite matrices again, is a convex semigroup which is open with respect to the set of positive maps (Lemma \ref{lemma_app_1}).
\item Given an infinite-dimensional, separable, complex Hilbert space $\mathcal H$ one finds that the set \mbox{$\lbrace \rho\in\mathbb D(\mathcal H)\,|\,\rho\prec\rho_0\rbrace$} is trace norm-closed for all $\rho_0\in\mathbb D(\mathcal H)$ (Thm.~\ref{lemma_maj_closed}). 
\end{itemize}
Finally we come to our main results in Chapter \ref{sec:results} which is based on our publications \cite{CDC19,OSID19}. Here we show how reachability problems of (finite-dimensional) Markovian open
quantum systems can be reduced to studying hybrid control systems on the standard
simplex of $\mathbb R^n$ (Ch.~\ref{sec_reach_fin_dim}), and how the result of \cite{BSH16} about normal generators generalizes to infinite dimensions (Ch.~\ref{sec_reach_inf_dim}). 
More precisely, we show that
\begin{itemize}
\item For a Markovian control system \eqref{eq:contr_mark_switch} of one or more qudits (i.e.~arbitrary $d$-level systems), if one of the qudits is coupled to a bath of temperature zero and if the closed system allows to apply all unitary channels, then every quantum state can be approximately generated from every initial state (Coro.~\ref{cor:toy_quantum}).
\item Given a single qudit with equidistant energy levels which is coupled to a bath of arbitrary finite temperature, the reachable set for the associated toy model (on the standard simplex $\Delta^{n-1}$) is upper bounded by $\lbrace x\in\Delta^{n-1}\,|\, x\prec z \rbrace$ for arbitrary initial states $x_0$. Here $z=z(x_0,d)$ can be chosen as the special extremal point from above (Thm.~\ref{thm_3}).
\item Consider the Markovian control system \eqref{eq:contr_mark_switch} where the drift $H_0$ is self-adjoint and the controls $H_1,\ldots,H_m$ are self-adjoint and bounded. If the closed part ($\gamma(t)=0$)  is strongly approximately controllable and the dissipative term is generated by a single non-zero operator $V$ which is compact and normal, then one can approximately reach 
every quantum state majorized by the initial state (Thm.~\ref{thm_normal_V}).
\end{itemize}
Conclusions and an outlook are presented in Ch.~\ref{ch:concl_outlook}. \bigskip

\noindent \textbf{Remark.} Now if one is solely interested in the reachability results in \textit{finite dimensions} it suffices to read only the chapters \ref{sec:quantum_channels}, \ref{ch:control_fin_dim}, \ref{sec:maj_d_vec} \& \ref{sec:maj_d_mat}, and of course \ref{sec_reach_fin_dim}. \bigskip

As a final note before diving in: Although the majority of lemmata and theorems explicitly indicate the assumptions regarding the underlying Hilbert space, there are a few sections which feature a \textit{global} assumption at their beginning. To increase transparency let us state them here, as well:
\begin{itemize}
\item Starting from Ch.~\ref{sec:quantum_channels} \textbf{until the end of this thesis} all Hilbert spaces are assumed to be complex. This is the main thing to keep in mind.
\item On top of that the Hilbert spaces in Ch.~\ref{ch_schr_heis_pic} \& \ref{ch:qu_dyn_sys} will all be separable.
\item Also all Hilbert spaces in Ch.~\ref{ch_bounded12938} \& \ref{sect_C_spectrum}, \ref{maj:trace_class}, and \ref{sec_reach_inf_dim} are assumed to be infinite-dimensional, separable, and complex.
\end{itemize}
\chapter{Preliminaries}\label{chapter:prelim}

Working with controlled quantum systems of course requires understanding the mathematical description of uncontrolled quantum-dynamical systems as well as general physical operations on quantum states. Because some of our main results deal with infinite-dimensional quantum systems we have to lay the focus on topological aspects, different classes of operators (i.e.~bounded, unbounded, compact, trace class), and their relations.

Thus we will start with recapping linear and bounded operators between normed spaces (all quantum channels will fall into this class), dual spaces (duality between Schr\"odinger and Heisenberg picture), and the most common topologies on such spaces. Section \ref{ch_2_3} deals with the special case of linear operators on inner product spaces, general unbounded operators, functional calculus (how to make sense of $e^{itH}$ if $H$ is a self-adjoint, but unbounded operator), and Schatten class operators (how to define the trace in infinite dimensions without running into convergence problems). All of this paves the way for Section \ref{sec:quantum_channels} where after recapping complete positivity we explore quantum channels on Hilbert spaces of arbitrary dimension, their properties, and some of their representations. In particular this leads us to closed and open Markovian quantum-dynamical systems and the structure of their generators.

For a refresher on functional analysis---in particular topology and special classes of vector spaces, ranging from metric to Banach to Hilbert spaces---we refer to Appendix \ref{sec_refresh_func_ana}.

\section{Linear Operators between Normed Spaces}\label{ch_2_1_normedop}

Let us start with normed spaces $(X,\|\cdot\|_X)$, $(Y,\|\cdot\|_Y)$ and the collection of all linear maps $T:X\to Y$ denoted by $\mathcal L(X,Y)$\label{def_linear_op}.
Here and henceforth, we require such spaces $X$ and $Y$ to have the same base field. Moreover the image $\{Tx\,|\,x\in X\}\subseteq Y$ of a linear map $T\in\mathcal L(X,Y)$ will be denoted by $\operatorname{im}(T)$.\label{symb_image}

Of course to introduce linear maps between vector spaces there is no need for the latter to be normed. However, doing so results in a very useful characterization of continuity of linear maps \cite[Prop.~5.4]{MeiseVogt97en}.

\begin{lemma}\label{lemma_cont_norm}
Let normed spaces $(X,\|\cdot\|_X)$, $(Y,\|\cdot\|_Y)$ and $T\in\mathcal L(X,Y)$ be given. The following statements are equivalent.
\begin{itemize}
\item[(i)] $T$ is continuous.\index{operator!continuous}
\item[(ii)] $T$ is continuous at $0$.
\item[(iii)] $T$ is bounded, that is, there exists $C>0$ such that $\|Tx\|_Y\leq C\|x\|_X$ for all $x\in X$.\index{operator!bounded}
\end{itemize}
\end{lemma}

\subsection{Bounded Operators}

Lemma \ref{lemma_cont_norm} justifies the following definition.

\begin{definition}\label{def_bounded_op}
Let normed spaces $(X,\|\cdot\|_X)$, $(Y,\|\cdot\|_Y)$ be given and define $\mathcal B(X,Y)$ as the collection of all continuous linear maps between $X$ and $Y$. Then the
operator norm\index{operator norm}\label{symb_op_norm} of $T\in\mathcal B(X,Y)$ is defined as
$$
\|T\|_\mathrm{op}:=\inf\{C>0\,|\,\|Tx\|_Y\leq C\|x\|_X\text{ for all }x\in X\}
$$
For convenience we define $\mathcal B(X):=\mathcal B(X,X)$.
\end{definition}

Like this $(\mathcal B(X,Y),\|\cdot\|_\mathrm{op})$ becomes a normed space with the following properties, see \cite[Lemma 5.5~ff.]{MeiseVogt97en}.

\begin{lemma}\label{lemma_normed_space_properties}
Let normed spaces $(X,\|\cdot\|_X)$, $(Y,\|\cdot\|_Y)$, $(Z,\|\cdot\|_Z)$ be given. The following statements hold.
\begin{itemize}
\item[(i)] For all $T\in\mathcal B(X,Y)$
\begin{align*}
\|T\|_\mathrm{op}=\sup_{x\in X\setminus\{0\}}\frac{\|Tx\|_Y}{\|x\|_X}=\sup_{\|x\|_X= 1,x\in X}\|Tx\|_Y\,,
\end{align*}
and $\|Tx\|_Y\leq\|T\|_\mathrm{op}\|x\|_X$ for all $x\in X$.
\item[(ii)] For all $T\in\mathcal B(X,Y)$, $S\in\mathcal B(Y,Z)$ one has $S\circ T=:ST\in\mathcal B(X,Z)$ is bounded again with $\|ST\|_\mathrm{op}\leq\|S\|_\mathrm{op}\|T\|_\mathrm{op}$.
\item[(iii)] If $Y$ is a Banach space then $\mathcal B(X,Y)$ is a Banach space. 
\end{itemize}
\end{lemma}

Having access to a norm on domain and codomain of a linear operator also enables a strong notion of ``structurally identifying'' normed spaces with each other.
\begin{definition}\label{def_isom_isom}
Let normed spaces $(X,\|\cdot\|_X)$, $(Y,\|\cdot\|_Y)$ be given. A map $T\in\mathcal L(X,Y)$ which is an isometry\index{isometry} (i.e.~$\|Tx\|_Y=\|x\|_X$ for all $x\in X$) and also surjective is called an isometric isomorphism\index{isometric isomorphism}. If for a pair of normed spaces such a map exists then we say that $X$ and $Y$ are isometrically isomorphic\label{symb_iso_iso}, denoted by $X\simeq Y$. 
\end{definition}
\noindent This definition is backed by the fact that 
\begin{itemize}
\item every linear isometry between normed spaces is injective: If $Tx=0$ then $\|Tx\|_Y=\|x\|_X=0$ so $x=0$.
\item the inverse of a bijective isometry is again an isometry: $\|T^{-1}y\|_X=\|T(T^{-1}y)\|_Y=\|y\|_Y$.
\end{itemize}

To simplify things we henceforth drop the index of a norm wherever doing so does not result in ambiguity.

\subsection{Dual Spaces of Normed Spaces}\label{section_dual_space}

Duality is a concept familiar from quantum physics: The Schr\"odinger picture and its dual description---the Heisenberg picture---are known to be equivalent. In order to see what this means in a rigorous manner we have to introduce dual spaces as well as dual operators. For the former we follow Rudin \cite[Ch.~3 \& 4]{Rudin91}.
\begin{definition}\label{def_dual_space_tvs}
Let $X$ be a topological vector space over $\mathbb F$. 
\begin{itemize}
\item[(i)] The (topological) dual space\index{space!dual|see{space, topological dual}}\index{space!topological dual} of $X$ is the vector space whose elements are the continuous linear functionals on $X$. The dual space of $X$ is usually denoted by $X^*\subset\mathcal L(X,\mathbb F)$.
\item[(ii)] If $X$ is normed then its (topological) dual space is given by $X^*=\mathcal B(X,\mathbb F)$.
\end{itemize}
\end{definition}
\noindent Part (ii) of this definition is a direct consequence of Lemma \ref{lemma_cont_norm}. On the other hand by Lemma \ref{lemma_normed_space_properties} (iii)---because we are considering normed spaces over a \textit{complete} field---the corresponding dual space is always a Banach space. At first glance this might come as a surprise given the original normed space need not be complete for this.\medskip

Either way the question arises in which way the dual space is useful, and what information $X^*$ contains about the original space $X$. ``As a matter of fact, so far [...] we have not even ruled out the utter indignity that $X^*=\{0\}$ while $X$ is [...] infinite-dimensional'' \cite[p.~45]{Bollobas99}. An answer to this is given by the Hahn-Banach theorem\index{theorem!Hahn-Banach} as well as its spiritual descendants, one of them reading as follows.

\begin{lemma}\label{lemma_hahn_banach_coro}
Let a normed space $X$, a linear subspace $M\subseteq X$, and $x_0\in X\setminus \overline{M}$ be given. Then there exists $f\in X^*$ such that $f(x_0)=1$ but $f(x)=0$ for all $x\in M$. In particular one has $\operatorname{dim}(X)=\operatorname{dim}(X^*)$.
\end{lemma}
\begin{proof}
Every normed space is a locally convex space under the norm topology (cf.~Remark \ref{remark_normed_locally_convex}) so the existence of such a functional follows from \cite[Thm.~3.5]{Rudin91}.

For the second statement let $\{x_1,\ldots,x_n\}\subset X\setminus\{0\}$, $n\in\mathbb N$ be an arbitrary collection of vectors. Then $M_n:=\operatorname{span}\{x_1,\ldots,x_n\}$ is a finite-dimensional subspace of $X$ and thus closed \cite[Thm.~1.21]{Rudin91}. Assuming we have $x_{n+1}\in X\setminus M_n$ (i.e.~the latter is not empty) then we can find $f_n\in X^*$ with $f_n(x_{n+1})=1$ and $f_n(x)=0$ for all $x\in M_n$. 

The inductive construction then goes as follows: Starting from $x_1\in X\setminus\{0\}$ and we get a corresponding non-zero $f_1$. Then for $x_2\in X\setminus M_1$ the above procedure yields $f_2$. Now $f_1,f_2$ are linearly independent: If $\lambda_1f_1+\lambda_2f_2=0$ then $0=\lambda_1f_1(x_2)+\lambda_2f_2(x_2)=\lambda_1$ and thus also $\lambda_2=0$. Indeed if $\operatorname{dim}(X)<\infty$ one easily sees that $\{f_1,\ldots,f_{\operatorname{dim}(X)}\}$ can be turned into a basis of $X^*$ (e.g., the dual basis corresponding to the basis $\{x_1,\ldots,x_{\operatorname{dim}(X)}\}$ of $X$). If $\operatorname{dim}(X)=\infty$ then this procedure of generating linearly independent functionals never stops---because the set $X\setminus M_n$ is non-empty for all $n\in\mathbb N$---so $\operatorname{dim}(X^*)=\infty$, as well. 
\end{proof}
\noindent Thus, aside from taking the same dimension as the original space, the dual space can separate linear subspaces from points outside (the closure of) said subspace. In fact $X^*$ can separate points\footnote{The dual space can do way more such as separating arbitrary disjoint convex sets where one set is closed and the other one is compact, cf.~\cite[Thm.~3.4]{Rudin91}. However this would go beyond our applicational needs.} on $X$:

\begin{lemma}\label{lemma_point_separation_dual_space}
Let a normed space $X$ and $x_1,x_2\in X$ be given. Then $x_1=x_2$ if and only if $f(x_1)=f(x_2)$ for all $f\in X^*$.
\end{lemma}
\begin{proof}
``$\Rightarrow$'': Obvious. ``$\Leftarrow$'': If $f(x_1-x_2)=0$ for all $f\in X^*$ then by \cite[Thm.~4.3]{Rudin91} we get
$$
\|x_1-x_2\|=\sup_{f\in X^*,\|f\|\leq 1}|f(x_1-x_2)|= 0
$$
which shows $x_1=x_2$ as claimed.
\end{proof}

Finally the null space of a continuous linear functional on an infinite-dimensional normed space admits further interesting structure.

\begin{lemma}\label{lemma_ker_functionals}
Let $X$ be a normed space with $\operatorname{dim}(X)=\infty$ and let any $f_1,\ldots,f_n\in X^*$ be given. Then $\bigcap_{j=1}^n \operatorname{ker}(f_j)$ is non-trivial, i.e.~$\operatorname{dim}\big(\bigcap_{j=1}^n \operatorname{ker}(f_j)\big)\geq 1$.
\end{lemma}
\begin{proof}
Assume to the contrary that there exists some $n\in\mathbb N$ and functionals $f_1,\ldots,f_n\in X^*$ such that $\bigcap_{j=1}^n \operatorname{ker}(f_j)=\{0\}$. Now given any functional $f\in X^*$ we trivially have $f(\bigcap_{j=1}^n \operatorname{ker}(f_j))=f(0)=0$ which by \cite[Lemma 3.9]{Rudin91} forces $f\in\operatorname{span}\{f_1,\ldots,f_n\}$. But $f$ was arbitrary so $\operatorname{span}\{f_1,\ldots,f_n\}=X^*$ and thus $\operatorname{dim}(X^*)\leq n<\infty$. On the other hand $\operatorname{dim}(X^*)=\operatorname{dim}(X)=\infty$ by Lemma \ref{lemma_hahn_banach_coro}, a contradiction.
\end{proof}
\begin{remark}
Although Lemma \ref{lemma_ker_functionals} will be sufficient for our applicational needs we could obtain the stronger statement that for any finite collection of continuous linear functionals $\{f_1,\ldots,f_n\}$ one has $\operatorname{dim}\big(\bigcap_{j=1}^n \operatorname{ker}(f_j)\big)=\infty$ as a corollary: If this were not the case then $M:=\bigcap_{j=1}^n \operatorname{ker}(f_j)$ is a finite-dimensional subspace of $X$ so 
we can find a basis $\{x_1,\ldots,x_m\}$ of $M$. This by Lemma \ref{lemma_hahn_banach_coro} yields $\Lambda_1,\ldots,\Lambda_m\in X^*$ such that $\Lambda_j(x_k)=\delta_{jk}$ for all $j,k=1,\ldots,m$ so $\bigcap_{j=1}^n \operatorname{ker}(f_j)\cap \bigcap_{j=1}^m \operatorname{ker}(\Lambda_j)=\{0\}$ but this by Lemma \ref{lemma_ker_functionals} is not possible.
\end{remark}

Given a normed space $X$ one may not only be interested in its dual $X^*$ but also in its bidual\index{space!bidual} $X^{**}:=(X^*)^*$ which is well-defined because $X^*$ is a normed space. Following \cite[Ch.~7]{MeiseVogt97en} the fact that the map $\iota:X\to X^{**}$, $\iota(x)(y):=y(x)$\label{symb_can_emb_1}
for all $x\in X$, $y\in X^*$ is a linear isometry motivates the following definition.
\begin{definition}
Let $X$ be a Banach space. We say $X$ is reflexive\index{space!reflexive} if the canonical embedding\index{canonical embedding} $\iota:X\to X^{**}$ is surjective, that is, $X\simeq X^{**}$ by means of $\iota$.
\end{definition}
\noindent As so often this is a purely infinite-dimensional concept as every finite-dimensional vector space is (normable and) reflexive. Note that if a normed space $X$ were not complete then $\iota(X)\subsetneq X^{**}$ as the latter is always complete (because the underlying field $\mathbb F$ is assumed to be complete).

Following Example \ref{ex_ell_p_space} the dual spaces of sequence spaces are well-explored. As some of these results will be useful later on let us quickly summarize them, cf.~\cite[Prop.~7.9 ff.]{MeiseVogt97en} \& \cite[Ch.~IV.8 ff.]{Dunford58} (for general $L^p$-spaces).

\begin{example}\label{ex_ell_p_space_2}
Let $p,q\in(1,\infty)$ be conjugate\index{conjugate}, i.e.~$\frac{1}{p}+\frac{1}{q}=1$. Then $(\ell^p(\mathbb N))^*\simeq \ell^q(\mathbb N)$\index{space!lp@$\ell^p(\mathbb N)$} by means of the isometric isomorphism
\begin{align*}
\Psi:\ell^q(\mathbb N)&\to (\ell^p(\mathbb N))^*\\
y&\mapsto\Psi_y\qquad\qquad \text{ which acts like }\Psi_y(z):=\sum\nolimits_{n=1}^\infty y_nz_n\text{ for all }z\in\ell^p(\mathbb N)\,.
\end{align*}
Moreover $(c_0(\mathbb N))^*\simeq\ell^1(\mathbb N)$\index{space!c0@$c_0(\mathbb N)$} and $(\ell^1(\mathbb N))^*\simeq\ell^\infty(\mathbb N)$ by means of the same map (with adjusted domain and codomain, of course). Based on this $\ell^p(\mathbb N)$ is reflexive for all $p\in (1,\infty)$ whereas $c_0(\mathbb N),\ell^1(\mathbb N),\ell^\infty(\mathbb N)$ are not reflexive.\index{space!l1@$\ell^1(\mathbb N)$}\index{space!linf@$\ell^\infty(\mathbb N)$}
\end{example}

Having learned about the dual space we can now introduce \textit{dual operators}\index{operator!dual} (sometimes also referred to as ``adjoint operators''\index{operator!adjoint} although this term will have a different meaning as soon we get to Hilbert space operators), see \cite[Ch.~9]{MeiseVogt97en} or \cite[Ch.~11]{Bollobas99}.
\begin{definition}\label{def_dual_op}
Let normed spaces $X,Y$ and $T\in\mathcal B(X,Y)$ be given. The map $T':Y^*\to X^*$ defined via $T'(f)=f\circ T$ for all $f\in Y^*$ is called the dual operator of $T$.
\end{definition}

This duality admits the following important properties.
\begin{lemma}\label{lemma_dual_map_properties}
Let normed spaces $X,Y,Z$ be given. 
\begin{itemize}
\item[(i)] The map ${}':\mathcal B(X,Y)\to \mathcal B(Y^*,X^*)$, $T\mapsto T'$ is well-defined (i.e.~$T'$ is bounded), linear, and an isometry.
\item[(ii)] For all $S\in\mathcal B(Y,Z)$, $T\in\mathcal B(X,Y)$ one has $(S\circ T)'=T'\circ S'$. 
\item[(iii)] If $T\in\mathcal B(X,Y)$ is an isomorphism, i.e.~$T$ is bijective and $T^{-1}$ is continuous, then so is $T'$. In this case $(T^{-1})'=(T')^{-1}$.
\item[(iv)] If $X$ is a Banach space, then $T\in\mathcal B(X,Y)$ is invertible if and only if $T'$ is.
\end{itemize}
\end{lemma}

\subsection{Topologies on Normed Spaces and their Dual}\label{sec_topo_norm_dual}

Before we get to topologies on general operator spaces let us quickly focus on the special case $\mathcal B(X,\mathbb F)$, that is, on the dual space $X^*$ of some normed space $X$. While there is of course the usual norm topology\index{topology!norm} (more precisely the topology on $X^*$ induced by the operator norm) one can define the weak topology\index{topology!weak} as $\sigma(X,X^*)$,\label{symb_weak_top} which is the weakest topology such that all $f\in X^*$ are still continuous. Note that this definition and the following properties hold for general topological vector spaces but we will only need these results for normed spaces.\smallskip

Defining a topology in such a way immediately yields the following:

\begin{corollary}\label{coro_net_weak}
Let $X$ be a normed space. A net $(x_i)_{i\in I}$ on $X$ converges weakly to $x\in X$, i.e.~$x_i\to x$ in $\sigma(X,X^*)$, if and only if $f(x_i)\to f(x)$ for all $f\in X^*$. Moreover $\sigma(X,X^*)\subseteq \tau$ with $\tau$ being the norm topology on $X$.
\end{corollary}
\begin{proof}
The first part is a direct application of Lemma \ref{lemma_initial_top}. For the second part let $(x_i)_{i\in I}$ be a net in $X$ which converges to $x\in X$ in norm. Then for all $f\in X^*$ we get $|f(x_i)-f(x)|\leq\|f\|\|x_i-x\|\to 0$ so the net converges weakly. Hence $\sigma(X,X^*)\subseteq \tau$ by Prop.~\ref{prop_compare_topo}.
\end{proof}
\noindent It is easy to see that the weak and the norm topology on a normed space co{\"i}ncide if and only if\,\footnote{For infinite-dimensional normed spaces one can show that the closure of $\{x\in X\,|\,\|x\|=1\}$ in $\sigma(X.X^*)$ is $\{x\in X\,|\,\|x\|\leq 1\}$ whereas the former is obviously closed in norm, cf.~\cite[p.~128]{Conway90}.} $\operatorname{dim}(X)<\infty$.\smallskip

As the dual space of every normed space is again a normed space under the operator norm it can also be equipped with a weak topology $\sigma(X^*,X^{**})$, same for the bidual and so forth. But there is also ``a way back'': The weak*-topology\index{topology!weak*}
on the dual $X^*$ of a normed space is defined to be $\sigma(X^*,\iota(X))$ (usually denoted by $\sigma(X^*,X)$).\label{symb_weak_star_top}
In other words we do not want \textit{all} elements of the bidual to be continuous in this topology, but only the ones of the form $\iota(x)$ for some $x\in X$. Just as before one obtains the following characterization of convergence in the weak*-topology.
\begin{corollary}\label{coro_net_weak_star}
A net $(f_i)_{i\in I}$ on the dual $X^*$ of a normed space weak*-converges to $f\in X^*$, i.e.~$f_i\to f$ in $\sigma(X^*,X)$, if and only if $f_i(x)\to f(x)$ for all $x\in X$. Moreover $\sigma(X^*,X)\subseteq \sigma(X^*,X^{**})$.
\end{corollary}
Unsurprisingly, a Banach space $X$ is reflexive if and only if the weak and the weak*-topology on $X^*$ co{\"i}ncide, that is $\sigma(X^*,X)=\sigma(X^*,X^{**})$ \cite[Ch.~V, Thm.~4.2]{Conway90}. Moreover, separability carries over using the weak*-topology:
\begin{lemma}\label{lemma_dual_weakstar_sep}
Let $X$ be a Banach space. If $X$ is separable then $(X^*,\sigma(X^*,X))$ is separable.
\end{lemma}
\begin{proof}
Be aware that the closed unit ball $\overline{B_1}(0)=\{x\in X^*\,|\,\|x\|\leq 1\}$ of the dual space $X^*$ is weak*-compact by the Banach-Alaoglu theorem \cite[Ch.~V, Thm.~3.1]{Conway90}\index{theorem!Banach-Alaoglu}, and, because $X$ is separable, $\overline{B_1}(0)$ is weak*-metrizable \cite[Ch.~V, Thm.~5.1]{Conway90}. But by Lemma \ref{lemma_comp_met_sep} every compact metric space is separable so $(\overline{B_1}(0),\sigma(X^*,X))$---and thus $(X^*,\sigma(X^*,X))$ by linearity---is separable as claimed.
\end{proof}

\begin{example}
To get a better feeling for the weak- and weak*-topology let us again think about some sequence spaces, given their dual spaces are rather well-structured. Recall that the closure of $c_{00}(\mathbb N)$\index{space!c00@$c_{00}(\mathbb N)$} in $(\ell^\infty(\mathbb N),\|\cdot\|_\infty)$ is $c_0(\mathbb N)$\index{space!c0@$c_0(\mathbb N)$} (Ex.~\ref{ex_ell_p_space}) so the former is not norm-dense in $\ell^\infty(\mathbb N)$\index{space!linf@$\ell^\infty(\mathbb N)$} (this would make $\ell^\infty(\mathbb N)$ separable in norm which is not the case). But is $c_{00}(\mathbb N)$ dense in the bounded sequences when considering a weaker topology?

To answer this let us consider $x\in\ell^\infty(\mathbb N)$ and define $x^{(n)}:=(x_1,\ldots,x_n,0,0,\ldots)\in c_{00}(\mathbb N)$ for all $n\in\mathbb N$. Then for every $y\in\ell^1(\mathbb N)$---using the isometric isomorphism $\Psi:\ell^\infty(\mathbb N)\to (\ell^1(\mathbb N))^*$ from Ex.~\ref{ex_ell_p_space_2}---one gets
$$
|\Psi_x(y)-\Psi_{x^{(n)}}(y)|=\Big|\sum\nolimits_{j=n+1}^\infty x_jy_j\Big|\leq \|x\|_\infty \sum\nolimits_{j=n+1}^\infty |y_j|\overset{n\to\infty}\to 0\,.
$$
This shows that every bounded sequence can be weak*-approximated\footnote{Identifying $(\ell^1(\mathbb R))^*\simeq \ell^\infty(\mathbb R)$ the weak*-topology on this space is given by $\sigma(\ell^\infty(\mathbb R),\ell^1(\mathbb R))$, i.e.~a net $(y_i)_{i\in I}$ in $\ell^\infty(\mathbb R)$ weak*-converges to $y\in\ell^\infty(\mathbb R)$ if and only if
$
\Psi_{y_i}(x)=\iota(x)(\Psi_{y_i})\to \iota(x)(\Psi_y)=\Psi_y(x)$
for all $x\in\ell^1(\mathbb R)$ (Coro.~\ref{coro_net_weak_star}).} by eventually-zero sequences, hence $c_{00}(\mathbb N)$ is dense in $\ell^\infty(\mathbb N)$ in the weak*-topology.
\end{example}
Either way these constructions should give us an idea how to find weaker topologies (compared to the operator norm topology) for operators between arbitrary normed spaces.

\subsection{Topologies on $\mathcal B(X,Y)$}\label{sec_top_B_X_Y}

For a lot of applications the norm topology on $\mathcal B(X,Y)$ is too strong in the sense that some sequences we would like to converge or some continuity properties we would like to hold do not apply. Prominent examples---among many others---are 
\begin{itemize}
\item projectors onto subspaces induced by an orthonormal basis which do not converge to the identity operator: Given an orthonormal basis $(e_n)_{n\in\mathbb N}$ of a separable Hilbert space, we expect the maps $\Pi_k\in\mathcal B(\mathcal H)$ defined via $\Pi_k(x):=\sum_{i=1}^k \langle e_i,x\rangle e_i$\label{symb_proj_k_el_1} to converge to the identity operator based on the Fourier expansion (Prop.~\ref{prop_hilbert_space_basis} (ii)). However $\|\Pi_k-\Pi_j\|_{\textrm{op}}=1$ for all $j\neq k$ so this is not a Cauchy sequence (and thus not convergent in $(\mathcal B(\mathcal H),\|\cdot\|_{\textrm{op}})$).
\item one-parameter groups $(e^{-itH})_{t\in\mathbb R}$ induced by an unbounded self-adjoint operator $H$ (we will learn in Ch.~\ref{sec_unbounded_op} what this means) on some Hilbert space $\mathcal H$. Taking this as a mapping $:\mathbb R\to(\mathcal B(\mathcal H),\|\cdot\|_{\textrm{op}})$, $t\mapsto e^{-itH}$ it lacks continuity.
\end{itemize}
\noindent In order to fix those issues later on we have to introduce further (weaker) topologies on the space of bounded operators. For this we refer to Dunford \& Schwartz \cite[Ch.~VI.1]{Dunford58}.\smallskip

While everything in this (and the next) chapter in principle is known---certainly to operator theorists---the results are largely scattered across the literature, if they are to be found in the main books on functional analysis and operator theory at all (beyond merely being listed as a fact)\footnote{The reference which to my knowledge comes closest to being complete in this regard is \cite{ChoiKim08}.}.
Thus all the operator topologies, their characterizations, and their properties we are interested in for the purpose of this thesis will be listed \textit{and proven}.

\begin{lemma}\label{lemma_topsw_basis}
Let normed spaces $X,Y$ be given. Then both collections\footnote{Of course saying $A$ ($B$) is a finite subset of $X$ ($Y^*$) implicitly assumes that $A,B\neq\emptyset$.}
\begin{align*}
\basiss&:= \{N(T,A,\varepsilon)\,|\,T\in\mathcal B(X,Y),A\subset X\text{ finite, }\varepsilon>0\} \\
\basisw&:=\{N(T,A,B,\varepsilon)\,|\,T\in\mathcal B(X,Y),A\subset X\text{ and }B\subset Y^*\text{ both finite, }\varepsilon>0\} 
\end{align*}
form a basis where
\begin{align*}
N(T,A,\varepsilon)&:= \{S\in\mathcal B(X,Y)\,|\,\|Tx-Sx\|<\varepsilon\text{ for all }x\in A\} \\
N(T,A,B,\varepsilon)&:= \{S\in\mathcal B(X,Y)\,|\,|y(Tx)-y(Sx)|<\varepsilon\text{ for all }x\in A,y\in B\}
\end{align*}
for all $T\in\mathcal B(X,Y)$, $A\subset X$ and $B\subset Y^*$ both finite, $\varepsilon>0$. 
\end{lemma}
\begin{proof}
Obviously $T\in N(T,A,\varepsilon)$ and $T\in N(T,A,B,\varepsilon)$ for all $T\in\mathcal B(X,Y)$, $A\subset X$ and $B\subset Y^*$ both finite, and $\varepsilon>0$. Thus all we have to show is that for non-empty intersections of any two basis elements there is a third basis element contained in said intersection.

Indeed let $T,T_1,T_2\in\mathcal B(X,Y)$, $A_1,A_2\subset X$ finite, and $\varepsilon_1,\varepsilon_2>0$ be given such that $T\in N(T_1,A_1,\varepsilon_1)\cap N(T_2,A_2,\varepsilon_2)$. Following the idea of Lemma \ref{lemma_metric_topo} define $A:=A_1\cup A_2\subset X$ (finite!) and
$$
\varepsilon:=\min_{j=1,2}\big(\varepsilon_j-\max_{x\in A_j}\|Tx-T_jx\|\big)>0\,.
$$
Then for all $S\in N(T,A,\varepsilon)$, $x\in A_j$, and $j=1,2$ we get
\begin{align*}
\|Sx-T_jx\|&\leq \|Sx-Tx\|+\|Tx-T_jx\|<\varepsilon+\|Tx-T_jx\|\\
&\leq \varepsilon_j-\big(\max_{x'\in A_j}\|Tx'-T_jx'\|\big)+\|Tx-T_jx\|\leq\varepsilon_j
\end{align*}
so $T\in N(T,A,\varepsilon)\subseteq N(T_1,A_1,\varepsilon_1)\cap N(T_2,A_2,\varepsilon_2)$. This proves that $\basiss$ has the basis property.

For $\basisw$ choose $A:=A_1\cup A_2$, $B:=B_1\cup B_2$ and
$$
\varepsilon:=\min_{j=1,2}\big(\varepsilon_j-\max_{x\in A_j}\max_{y\in B_j}|y(Tx)-y(T_jx)|\big)>0\,.
$$
The rest is analogous.
\end{proof}

This motivates the following definition.
\begin{definition}\label{def_strong_weak_op_top}
Let normed spaces $X,Y$ be given. The topology $\tops$ generated by $\basiss$ is called the strong operator topology\index{topology!strong operator}, and the topology $\topw$ generated by $\basisw$ is called the weak operator topology\index{topology!weak operator} on $\mathcal B(X,Y)$.
\end{definition}
These are by no means the only interesting topologies $\mathcal B(X,Y)$ can be equipped with, as is elaborated on in \cite[Ch.~VI.1]{Dunford58}. Now let us list some important properties of $\tops$ and $\topw$ the lengthy proof of which is outsourced to Appendix \ref{sec:app_proof_op_topologies}.

\begin{proposition}\label{prop_strong_weak_op_top}
Let non-trivial\footnote{This means $X,Y\neq\{0\}$ which is reasonable because else $\mathcal B(X,Y)=\{0\}$ so there would not be any interesting structure to discover.} normed spaces $X,Y$, a net $(T_i)_{i\in I}$ in $\mathcal B(X,Y)$, and $T\in\mathcal B(X,Y)$ be given, and let $\topn$ denote the operator norm topology on $\mathcal B(X,Y)$. Then the following statements hold.
\begin{itemize}
\item[(i)] The collection
\begin{itemize}
\item[(a)] $\{N(T,A,\varepsilon)\,|\,A\subset X\text{ finite, }\varepsilon>0\}$ forms a neighborhood basis of $\tops$ at $T\in\mathcal B(X,Y)$.
\item[(b)] $\{N(T,A,B,\varepsilon)\,|\,A\subset X\text{ and }B\subset Y^*\text{ both finite, }\varepsilon>0\} $ forms a neighborhood basis of $\topw$ at $T\in\mathcal B(X,Y)$.
\end{itemize}
\item[(ii)] One has $T_i\to T$ in $\tops$ if and only if $T_ix\to Tx$ for all $x\in X$ and, moreover, $T_i\to T$ in $\topw$ if and only if $y(T_ix)\to y(Tx)$ for all $x\in X$, $y\in Y^*$.
\item[(iii)] Both $\tops$ and $\topw$ are Hausdorff.
\item[(iv)] The following statements hold:
\begin{itemize}
\item[(a)] $\topw\subseteq\tops\subseteq\topn$.
\item[(b)] $\tops=\topn$ if and only if $\operatorname{dim}(X)<\infty$. 
\item[(c)] $\topw=\tops$ if and only if $\operatorname{dim}(Y)<\infty$.
\item[(d)] $\topw=\tops=\topn$ if and only if $\operatorname{dim}(X),\operatorname{dim}(Y)<\infty$.
\end{itemize}
\item[(v)] \begin{itemize}
\item[(a)] $\tops$ is the topology induced by the seminorms $\{T\mapsto \|Tx\|\}_{x\in X}$. Equivalently it is the weakest topology such that all evaluation maps $\{T\mapsto Tx\}_{x\in X}$ are continuous. Moreover $(\mathcal B(X,Y),\tops)$ is a locally convex space.
\item[(b)] $\topw$ is the topology induced by the seminorms $\{T\mapsto |y(Tx)|\}_{x\in X,y\in Y^*}$. Equivalently it is the weakest topology such that all evaluation maps $\{T\mapsto y(Tx)\}_{x\in X,y\in Y^*}$ are continuous. Moreover $(\mathcal B(X,Y),\topw)$ is a locally convex space.
\end{itemize}
\end{itemize}
\end{proposition}

There is a lot of information to digest here. First off $\tops$ contains information regarding ``pointwise convergence of operators'' ($T_ix\to Tx$ for all $x\in X$) while $\topw$ is about convergence ``on matrix elements'' ($y(T_ix)\to y(Tx)$ for all $x\in X$, $y\in Y^*$). Thus there is no point in considering either of these topologies when the involved spaces are finite-dimensional. However as soon as the domain or the codomain (or both) are infinite-dimensional one gets access to topologies on $\mathcal B(X,Y)$ strictly weaker than the norm topology.

Secondly there is an important distinction to make: While $\tops$ is the topology induced by the seminorms $\{T\mapsto \|Tx\|\}_{x\in X}$ it in general is \textbf{not} the \textbf{initial} topology with respect to the family $\{T\mapsto \|Tx\|\}_{x\in X}$---one has to be similarly cautious regarding $\topw$. 

Lastly while $(\mathcal B(X,Y),\tops)$, $(\mathcal B(X,Y),\topw)$ are locally convex spaces they are not metrizable if the underlying spaces are infinite-dimensional\footnote{If $(\mathcal B(X,Y),\tops)$ were metrizable then sequential completeness of the former together with the open mapping theorem would imply $\tops=\topn$ which is not the case if $\operatorname{dim}(X)=\infty$, cf.~\cite{Wengenroth18}.\label{footnote_tops_not_metrizable}}. However, metrizability can be restored when restricting oneself to a bounded subset of $\mathcal B(X,Y)$ as we will see now. This can be surprisingly advantageous because nets then become superfluous and everything is handled solely by sequences, see also Remark \ref{rem_metrizable_topo}.

\begin{proposition}\label{prop_op_top_metrizable}
Let $X,Y$ be normed spaces and let $S\subseteq B(X,Y)$ be bounded (i.e.~there exists $C>0$ such that $\|T\|\leq C$ for all $T\in S$). The following statements hold.
\begin{itemize}
\item[(i)] If $X$ is separable then $(S,\tops)$ is metrizable. 
\item[(ii)] If $X$ and $Y^*$ are both separable then $(S,\topw)$ is metrizable.
\item[(iii)] If $X^*$ is separable and $S\subseteq\mathcal B(X)$ is bounded then $(S,\topw)$ is metrizable.
\end{itemize}

\end{proposition}
\begin{proof}
When we talk about $(S,\tops)$ or $(S,\topw)$ really we mean $S$ equipped with the subspace topology induced by $(\mathcal B(X,Y),\tops)$ or $(\mathcal B(X,Y),\topw)$ (cf.~Def.~\ref{defi_subspace_top}), e.g., $(S,\tops(S))$ where $\tops(S)=\{S\cap U\,|\,U\in\tops\}$. However for simplicity we will write $\tops$ instead of $\tops(S)$.

(i): Let $X$ be separable so we can find a subset $S_0:=\{x_n\}_{n\in\mathbb N}$ of the closed unit ball $\overline{B_1}(0)=\{x\in X\,|\,\|x\|\leq 1\}$ which is (norm-)dense, i.e.~$\overline{S_0}=\overline{B_1}(0)$. With this define $d:\mathcal B(X,Y)\times \mathcal B(X,Y)\to\mathbb R$ via $d(T_1,T_2):=\sum_{n=1}^\infty\frac{\|T_1x_n-T_2x_n\|_Y}{2^n}$ which is a metric on $\mathcal B(X,Y)$ as is readily verified\footnote{The only non-trivial step is definiteness of the metric: If $d(T_1,T_2)=0$ then $T_1x_n=T_2x_n$ for all $n\in\mathbb N$. But $T_1,T_2$ are continuous so $T_1x=\lim_{j\to\infty}T_1x_{n_j}=\lim_{j\to\infty}T_2x_{n_j}=T_2x$ for all $x\in\overline{B_1}(0)$ which by linearity shows $T_1=T_2$.}. We want to show that the topology induced by the metric $d$ on $S$ co{\"i}ncides with $\tops$. Keep in mind that boundedness of $S$ guarantees the existence of some $C>0$ such that $\|T\|\leq C$ for all $T\in S$. 

``$\Rightarrow$'': Let $(T_i)_{i\in I}$ be a net in $S$ which converges to $T\in S$ in $\tops$. Now given $\varepsilon>0$ one finds $N\in\mathbb N$ such that $\sum_{n=N+1}^\infty \frac{1}{2^n}<\frac{\varepsilon}{4C}$. Moreover because $T_i\overset{\tops}\to T$ one finds $i_0\in I$ such that
$$
T_i\in N(T,\{x_1,\ldots,x_N\},\varepsilon/2)\cap S\qquad\text{ (i.e.~} \|Tx_j-T_ix_j\|<\varepsilon/2\text{ for all }j=1,\ldots,N )
$$
for all $i\succeq i_0$. Thus
\begin{align*}
d(T,T_i)&=\sum\nolimits_{n=1}^N \frac{\|Tx_n-T_ix_n\|}{2^n}+\sum\nolimits_{n=N+1}^\infty \frac{\|Tx_n-T_ix_n\|}{2^n}\\
&< \frac{\varepsilon}{2}\underbrace{ \sum\nolimits_{n=1}^N \frac{1}{2^n} }_{\leq 1}+\underbrace{(\|T\|+\|T_i\|)}_{\leq 2C}\sum_{n=N+1}^\infty \frac{1}{2^n}\underbrace{\|x_n\|}_{\leq 1}<\frac{\varepsilon}{2}+\frac{\varepsilon}{2}=\varepsilon
\end{align*}
for all $i\succeq i_0$ so $T_i\overset{d}\to T$.

``$\Leftarrow$'': Assume $T_i\overset{d}\to T$ so by Prop.~\ref{prop_strong_weak_op_top} (ii) we have to show that $(T_ix)_{i\in I}$ converges to $Tx$ for all $x\in\overline{B_1}(0)$ (and thus for all $x\in X$ by linearity). Thus let $x\in X$, $\|x\|\leq 1$ as well as $\varepsilon>0$ be given. By density of $S_0$ we find $m\in\mathbb N$ such that $\|x_m-x\|<\frac{\varepsilon}{3C}$. Also by assumption one finds $i_0\in I$ such that $d(T_i,T)<\frac{\varepsilon}{3\cdot 2^m}$ for all $i\succeq i_0$. Altogether
\begin{align*}
\|T_ix-Tx\|&\leq \|T_i(x-x_m)\|+\|T_ix_m-Tx_m\|+\|T(x_m-x)\|\\
&\leq \|T_i\| \|x-x_m\|+2^m\frac{\|T_ix_m-Tx_m\|}{2^m}+ \|T\| \|x-x_m\|\\
&< \frac{\varepsilon}{3}+2^m d(T_i,T)+\frac{\varepsilon}{3}<\frac{2\varepsilon}{3}+2^m\cdot \frac{\varepsilon}{3\cdot 2^m}=\varepsilon
\end{align*}
for all $i\succeq i_0$ as desired.\smallskip

(ii): Given dense subsets $\{x_m\}_{m\in\mathbb N}$, $\{y_n\}_{n\in\mathbb N}$ of the respective closed unit ball in $X$, $Y^*$ define the metric $d(T_1,T_2):=\sum_{m,n\in\mathbb N}\frac{|y_n(T_1x_m)-y_n(T_2x_m)|}{2^{m+n}}$ for all $T_1,T_2\in\mathcal B(X,Y)$. As before one sees that $(S,\topw)=(S,\tau_d)$ so the former is metrizable as claimed.\smallskip

(iii): If $X^*$ is separable then so is $X$ \cite[Ch.~II.3, Lemma 16]{Dunford58} so this follows from (ii).
\end{proof}

\subsection{Topologies on $\mathcal B(Y^*,X^*)$}

Another issue which arises here is concerned with topologies on the conjugate operator space $\mathcal B(Y^*,X^*)$ for some normed spaces $X,Y$. As before this operator space can be equipped with the norm, the strong operator, and the weak operator topology.
The latter is of particular interest because a net $T_i\in\mathcal B(Y^*,X^*)$ converges to $T$ in $\topw$ if and only if $\tilde x(T_iy)\to \tilde x(Ty)$ for all $\tilde x\in X^{**}$, $y\in Y^*$ (Prop.~\ref{prop_strong_weak_op_top}). In the case of the domain being reflexive (i.e.~$X^{**}\simeq X$ by means of the canonical embedding $\iota$) one has
$$
\iota(x)(T'y)=(T'y)(x)=(y\circ T)(x)=y(Tx)
$$
for all $x\in X$, $y\in Y^*$, and $T\in\mathcal B(X,Y)$, meaning $\topw$ on $\mathcal B(Y^*,X^*)$ here leads us back to the weak operator topology on $\mathcal B(X,Y)$. Unfortunately, a lot of normed spaces one deals with in infinite-dimensional quantum theory are not reflexive as their second dual is ``too large'' ($\iota(X)\subsetneq X^{**}$). This becomes a problem if one wants to formulate some concepts involving operators on a normed space \textit{equivalently} on its dual space, and we will fix this as follows:

\begin{definition}\label{defi_weak_star_top}
Let $X,Y$ be normed spaces. Define $\topw^*$ as the weakest topology on $\mathcal B(Y^*,X^*)$ such that all evaluation maps $\{T\mapsto (Ty)(x)\}_{y\in Y^*,x\in X}$ are continuous.
\end{definition}
\begin{remark}\label{rem_properties_weak_star_op_top}
Just as before one can work out that a basis of $\topw^*$ is given by
\begin{align*}
\basisw^*&:=\{N^*(T,A,B,\varepsilon)\,|\,T\in\mathcal B(Y^*,X^*),A\subset X\text{ and }B\subset Y^*\text{ both finite, }\varepsilon>0\} \\
N^*(T,A,B,\varepsilon)&:= \{S\in\mathcal B(Y^*,X^*)\,|\,|(Ty)(x)-(Sy)(x)|<\varepsilon\text{ for all }x\in A,y\in B\}
\end{align*}
and a neighborhood basis of $\topw^*$ at $T\in\mathcal B(Y^*,X^*)$ is given by $\{N^*(T,A,B,\varepsilon)\,|\,A\subset X\text{ and }B\subset Y^*\text{ both finite, }\varepsilon>0\}$. A net $(T_i)_{i\in I}$ in $\mathcal B(Y^*,X^*)$ converges to $T\in\mathcal B(Y^*,X^*)$ in $\topw^*$ if and only if $(T_iy)(x)\to (Ty)(x)$ for all $x\in X$, $y\in Y^*$. 
With this it is easy to see that $\topw^*$, equivalently, is the topology induced by the complete family of seminorms $\{T\mapsto |(Ty)(x)|\}_{x\in X,y\in Y^*}$, hence $\topw^*$ is Hausdorff and $(\mathcal B(Y^*,X^*),\topw^*)$ is a locally convex space.
Finally---given a bounded subset $S\subset\mathcal B(Y^*,X^*)$---if $X$ and $Y^*$ are separable then $(S,\topw^*)$ is metrizable.
\end{remark}

The topology $\topw^*$ from Def.~\ref{defi_weak_star_top} is called the \textit{weak*-operator topology}\index{topology!weak*-operator} (or \textit{$\sigma$-weak topology})\index{topology!sigma-weak@$\sigma$-weak}. These names are obviously motivated by the weak*-topology from Section \ref{sec_topo_norm_dual}. 
This fits our needs from the beginning of this section as the latter---by definition---focusses on $\iota(X)$ instead of the whole second dual $X^{**}$.

\begin{definition}
Let $X,Y$ be normed spaces. An operator $T\in\mathcal B(Y^*,X^*)$ is said to be weak*-continuous\index{continuous!weak*} if it is continuous as a map $T:(Y^*,\sigma(Y^*,Y))\to(X^*,\sigma(X^*,X))$, that is, if for every net $(\tilde y_i)_{i\in I}$ on $Y^*$ and $\tilde y\in Y^*$ one has
$$
\tilde y_i(z)\to \tilde y(z)\quad\text{ for all }z\in Y \qquad\Longrightarrow\qquad(T\tilde y_i)(x)\to (T\tilde y)(x)\quad\text{ for all }x\in X\,.
$$
\end{definition}
\noindent Thus a functional $f\in X^{**}=\mathcal B(X^*,\mathbb F)$ is weak*-continuous if $f(\tilde x_i)\to f(\tilde x)$ for every net $(\tilde x_i)_{i\in I}$ on $X^*$ which weak*-converges to $x\in X^*$. 
An important feature of this construction is that every $f\in X^{**}$ which is weak*-continuous 
can be written as $f=\iota(x)$ for some $x\in X$ (see also \cite[Ch.~3.14]{Rudin91}).

\begin{proposition}\label{prop_weak_star_weak_comp}
Let $X,Y$ be non-trivial normed spaces. The following statements hold.
\begin{itemize}
\item[(i)] $\topw^*\subseteq\topw$ on $\mathcal B(Y^*,X^*)$.
\item[(ii)] Let $X$ be a Banach space. Then $\topw^*=\topw$ if and only if $X$ is reflexive. 
\end{itemize}
\end{proposition}
\begin{proof}
(i): Let $(T_i)_{i\in I}$ be a net in $\mathcal B(Y^*,X^*)$ which converges to $T\in\mathcal B(Y^*,X^*)$ with respect to $\topw$. This means $\tilde x(T_iy)\to\tilde x(Ty)$ for all $y\in Y^*$, $\tilde x\in X^{**}$. Choosing $\tilde x=\iota(x)\in X^{**}$ for arbitrary $x\in X$ yields
$$
(T_iy)(x)=\iota(x)(T_iy)\to\iota(x)(Ty)=(Ty)(x)
$$
for all $y\in Y^*$. But this by Remark \ref{rem_properties_weak_star_op_top} means $T_i\to T$ in $\topw^*$, hence $\topw^*\subseteq\topw$ by Prop.~\ref{prop_compare_topo}.\smallskip

(ii): ``$\Leftarrow$'': Let $X$ be reflexive, and let $(T_i)_{i\in I}$ be a net in $\mathcal B(Y^*,X^*)$ which converges to $T$ in $\topw^*$. Then for every $\tilde x\in X^{**}$ there exists $x\in X$ such that $\tilde x=\iota(x)$. Thus for all $y\in Y^*$ we obtain
$$
\tilde x(T_iy)=\iota(x)(T_iy)=(T_iy)(x)\to (Ty)(x)=\iota(x)(Ty)=\tilde x(Ty)\,,
$$
meaning $T_i\to T$ in $\topw$. 

``$\Rightarrow$'': Let $X$ be a non-reflexive Banach space so one finds $\tilde x\in X^{**}\setminus\iota(X)$. Also because $Y$ is non-trivial we by Lemma \ref{lemma_hahn_banach_coro} find $y_0\in Y^*$, $\|y_0\|=1$. Just as in the proof of Prop.~\ref{prop_strong_weak_op_top} (iv),(b) \& (c), our goal is to show $N^*(0,A,B,\varepsilon)\not\subset N(0,\{y_0\},\{\tilde x\},\frac{1}{2})$ for all $A\subset X$, $B\subset Y^*$ both finite and all $\varepsilon>0$.

Our main concern for now is to ``distinguish'' $\tilde x$ from $\iota(X)$ by means of a linear functional, ideally acting on $X$ itself. First off Lemma \ref{lemma_image_closed_isometry} shows that $\iota(X)$ is norm-closed in $X^{**}$ because $\iota$ is an isometry and $X$ is complete by assumption. Thus $\tilde x\in X^{**}\setminus\iota(X)=X^{**}\setminus\overline{\iota(X)}$ so Lemma \ref{lemma_hahn_banach_coro} yields $\psi\in X^{***}$ such that $\psi(\tilde x)=1$ and $\psi(\iota(X))=0$. Although there is no reason for $\psi$ to be in $\iota(X^*)$, the latter is weak*-dense in $X^{***}$ \cite[Ch.~V, Prop.~4.1]{Conway90} hence one can find $\phi\in X^*$ such that\footnote{To add a little more detail for those interested: The weak*-density guarantees the existence of a net $(f_j)_{j\in J}$ in $X^*$ such that $\iota(f_j)\to\psi$ in $\sigma(X^{***},\iota(X^{**}))$ (i.e.~in the weak*-topology on $X^{***}$). This by Lemma \ref{lemma_initial_top} means $x'(f_j)=\iota(x')(\iota(f_j))\to\iota(x')(\psi)=\psi(x')$ for all $x'\in X^{**}$. We have to turn this pointwise convergence into an approximation on finitely many elements of $X^{**}$.

On one hand for every $x\in A$ we find $j_x\in J$ such that $|\iota(x)(f_j)-\psi(\iota(x))|=|f_j(x)|<\frac{\varepsilon}{C}$ for all $j\succeq j_x$ (here we used $\psi(\iota(X))=0$). Because $A$ is finite there exists $j_A\in J$ such that $j_A\succeq j_x$ for all $x\in A$ by the directed set property of $J$. On the other hand one finds $\tilde j\in J$ such that $|\tilde x(f_j)-\psi(\tilde x)|<\frac12$ for all $j\succeq\tilde j$. Again this yields $j_0\in J$ with $j_0\succeq j_A$ and $j_0\succeq \tilde j$. But now $\phi:=f_{j_0}\in X^*$ satisfies \eqref{eq:psi_approx} as desired.}
\begin{equation}\label{eq:psi_approx}
|\phi(x)|<\frac{\varepsilon}{C}\quad\text{ for all }x\in A\qquad\text{ as well as }\qquad|\tilde x(\phi)-\psi(\tilde x)|<\frac{1}{2}
\end{equation}
where $C:=\max_{y\in B}\|y\|+1>0$ (here we use that $B$ is finite). 

This will allow us to define an operator $T\in N^*(0,A,B,\varepsilon)\setminus N(0,\{y_0\},\{\tilde x\},\frac{1}{2})$ as follows: First one finds $f\in Y^{**},\|f\|= 1$ such that $f(y_0)=\|y_0\|=1$ \cite[p.~59]{Rudin91} which allows us to define $T:Y^*\to X^*$ via $ y\mapsto f(y)\phi$. Obviously $T$ is linear and bounded ($\|T\|\leq\|f\|\|\phi\|<\infty$). Then $T\in N^*(0,A,B,\varepsilon)$ because for all $x\in A$, $y\in B$
$$
|(Ty)(x)|=|f(y)|\cdot|\phi(x)|<\underbrace{\|f\|}_{=1}\underbrace{\|y\|}_{<C}\cdot\frac{\varepsilon}{C}<\varepsilon
$$
but $T\not\in N(0,\{y_0\},\{\tilde x\},\frac{1}{2})$ by the reverse triangle inequality:
\begin{equation}
|\tilde x(Ty_0)|=\underbrace{|f(y_0)|}_{=1}\cdot |\tilde x(\phi)|\geq \underbrace{|\psi(\tilde x)|}_{=1}-|\psi(\tilde x)-\tilde x(\phi)|>\frac{1}{2}\,.\tag*{\qedhere}
\end{equation}
\end{proof}

Finally the concept of weak*-continuity gives us a one-to-one correspondence between bounded operators and their dual operators. This should not come as a surprise because this in some sense is baked into the definition of the weak* (operator) topology. 

\begin{proposition}\label{prop_dual_weak_star_cont}
Let normed spaces $X,Y$ as well as $T\in\mathcal B(Y^*,X^*)$ be given. The following statements are equivalent.
\begin{itemize}
\item[(i)] $T$ is weak*-continuous.
\item[(ii)] There exists unique $\tilde T\in\mathcal B(X,Y)$ such that $\tilde T'\equiv T$. 
\end{itemize}
Thus $\mathcal B(X,Y)\simeq \{T\in\mathcal B(Y^*,X^*)\,|\,T\text{ weak*-continuous}\}$ by means of the map ${}'$ from Lemma \ref{lemma_dual_map_properties}. 
\end{proposition}
\begin{proof}
``(ii) $\Rightarrow$ (i)'': Let $(\tilde y_i)_{i\in I}$ be a net on $Y^*$ which weak*-converges to some $\tilde y\in Y^*$. Then $(T'\tilde y_i)(x)=\tilde y_i(Tx)\to \tilde y(Tx)=(T'y)(x)$ for all $x\in X$ as claimed.

``(i) $\Rightarrow$ (ii)'': Assume $T$ is weak*-continuous and let any $x\in X$ be given. Then the map $f_x:Y^*\to\mathbb F$, $\tilde y\mapsto (T\tilde y)(x)$ has the following properties:
\begin{itemize}
\item[$\bullet$] $f_x\in Y^{**}$: Linearity transfers from $T$ to $f_x$. (Norm-)continuity follows from
$$
|f_x(y)|=|(Ty)(x)|\leq \|Ty\|\|x\|\leq\|T\|\|x\|\|y\|
$$
for all $y\in Y^*$ (so $\|f_x\|\leq\|T\|\|x\|<\infty $). 
\item[$\bullet$] $f_x$ is weak*-continuous: Consider a net $(\tilde y_i)_{i\in I}$ as well as $\tilde y\in Y^*$ such that $\tilde y_i\to\tilde y$ in $\sigma(Y^*,Y)$. Then $T\tilde y_i\to T\tilde y$ in $\sigma(X^*,X)$ as $T$ is weak*-continuous so for all $x\in X$
$$
f_x(\tilde y_i)=(T\tilde y_i)(x)\to (T\tilde y)(x)=f_x(\tilde y)\,.
$$
\end{itemize}
Hence by definition of the weak*-topology---as elaborated on before---there exists (unique) $y=y_x\in Y$ such that $f_x=\iota(y_x)$. Because $x$ was chosen arbitrarily lets us define a map $\tilde T:X\to Y$, $\tilde Tx:=y_x$ which is linear and bounded\footnote{Linearity of $\tilde T$ is a direct consequence of linearity of $T$: For all $y\in Y^*$, $\lambda_1,\lambda_2\in\mathbb F$, $x_1,x_2\in X$
\begin{align*}
y\big(\tilde T(\lambda_1x_1+\lambda_2x_2)\big)&=y(y_{\lambda_1x_1+\lambda_2x_2})=\iota(y_{\lambda_1x_1+\lambda_2x_2})(y)=f_{\lambda_1x_1+\lambda_2x_2}(y)\\
&=(Ty)(\lambda_1x_1+\lambda_2x_2)=\lambda_1( Ty)(x_1)+\lambda_2( Ty)(x_2)=\ldots=y\big( \lambda_1\tilde Tx_1+\lambda_2\tilde Tx_2 \big)
\end{align*}
so $\tilde T(\lambda_1x_1+\lambda_2x_2)= \lambda_1\tilde Tx_1+\lambda_2\tilde Tx_2$ by Lemma \ref{lemma_point_separation_dual_space}. Similarly one shows $\|\tilde T\|\leq\|T\|<\infty$ (i.e.~boundedness) using \cite[Thm.~4.3]{Rudin91}: For all $x\in X$
\begin{align*}
\|\tilde Tx\|=\sup_{y\in Y^*,\|y\|\leq 1}|y(\tilde Tx)|= \sup_{y\in Y^*,\|y\|\leq 1}| (Ty)(x) |\leq \|T\|\|x\| \sup_{y\in Y^*,\|y\|\leq 1}\|y\|=\|T\|\|x\|\,.
\end{align*}} 
and satisfies
$$
(\tilde T'y)(x)=y(\tilde Tx)=y(y_x)=\iota(y_x)(y)=f_x(y)=(Ty)(x)
$$
for all $x\in X$, $\tilde y\in Y^*$ so $\tilde T'=T$. Uniqueness is easy: Given $\tilde T_1,\tilde T_2\in\mathcal B(X,Y)$ with $\tilde T_1'=T=\tilde T_2'$ one computes
$$
0=\|T-T\|=\|\tilde T_1'-\tilde T_2'\|=\|(\tilde T_1-\tilde T_2)'\|=\|\tilde T_1-\tilde T_2\|
$$
by Lemma \ref{lemma_dual_map_properties} (i) so $\tilde T_1=\tilde T_2$.\smallskip

For the additional statement we already know by Lemma \ref{lemma_dual_map_properties} that $'$ is a linear isometry (regardless of the codomain). On top of that, the map
$${}':\mathcal B(X,Y)\to \{T\in\mathcal B(Y^*,X^*)\,|\,T\text{ weak*-continuous}\}
$$
is well-defined (``(ii) $\Rightarrow$ (i)'') and surjective (``(i) $\Rightarrow$ (ii)'') which concludes the proof.
\end{proof}

Finally one can show that, roughly speaking, the weak*-operator topology on the dual operators is the ``dual topology'' of the weak operator topology on the original operator space. More precisely, we get the following result.

\begin{proposition}\label{prop_connection_weak_weakstar_op}
Let $X,Y$ be normed spaces and $S\subset {}'(\mathcal B(X,Y))$ be given (i.e.~for all $T\in S$ there exists unique $\tilde T\in\mathcal B(X,Y)$ such that $\tilde T'=T$). Moreover let $\tilde S\subset\mathcal B(X,Y)$ denote the set of all these pre-dual operators $\tilde T$. Then the following statements are equivalent.
\begin{itemize}
\item[(i)] $S$ is $\topw^*$-closed in ${}'(\mathcal B(X,Y))$.
\item[(ii)] $\tilde S$ is $\topw$-closed.
\end{itemize}
\end{proposition}
\begin{proof}
``(i) $\Rightarrow$ (ii)'': Let $(\tilde T_i)_{i\in I}$ be a net in $\tilde S$ which converges to $\tilde T\in\mathcal B(X,Y)$ in $\topw$. If we can show that $\tilde T\in S$ then $S$ is closed (in $\topw$) by Lemma \ref{lemma_topol_connect} (ii). Indeed for all $x\in X$, $y\in Y^*$
$$
(\tilde T_i'y)(x)=y(\tilde T_ix)\to y(\tilde Tx)=(\tilde T'y)(x)
$$
because $\tilde T_i\to\tilde T$ in $\topw$ (Prop.~\ref{prop_strong_weak_op_top} (ii)) so $\tilde T_i'\to\tilde T'$ in $\topw^*$ (Remark \ref{rem_properties_weak_star_op_top}). But closedness of $S$ implies $\tilde T'\in S$ so $\tilde T\in\tilde S$ as desired.

``(ii) $\Rightarrow$ (i)'': Again let $(T_i)_{i\in I}\subseteq S$ converge to\footnote{We assume that the limit of the net is in ${}'(\mathcal B(X,Y))$ (and not in the whole operator space $\mathcal B(Y^*,X^*)$) because we want to show that $S$ is closed \textit{within the subspace} ${}'(\mathcal B(X,Y))\subseteq(\mathcal B(Y^*,X^*),\topw^*)$ with the corresponding subspace topology (i.e.~$\topw^*\cap{}'(\mathcal B(X,Y))$, cf.~also Section \ref{section_product_subsp_top}).} $ T\in{}'(\mathcal B(X,Y))$ in $\topw^*$. Moreover let $(\tilde T_i)_{i\in I}\subseteq\mathcal B(X,Y)$ be the corresponding net of pre-dual operators, and let $\tilde T\in\mathcal B(X,Y)$ denote the pre-dual of $T$ (i.e.~$\tilde T_i'=T_i$ for all $i\in I$ as well as $\tilde T'=T$). Just as before for all $x\in X$, $y\in Y^*$ one gets 
$
y(\tilde T_ix)=(\tilde T_i'y)(x)=(T_iy)(x)\to (Ty)(x)= (\tilde T'y)(x)=y(\tilde Tx)
$
because $T_i'\to T$ in $\topw^*$ so $\tilde T_i\to\tilde T$ in $\topw$. Again closedness of $\tilde S$ implies $ \tilde T\in \tilde S$ so $T=\tilde T'\in S$.
\end{proof}

\begin{remark}\label{rem_closed_in_prime_nec}
The restriction of $S$ being $\topw^*$-closed in ${}'(\mathcal B(X,Y))$ (as opposed to the whole space $(\mathcal B(Y^*,X^*),\topw^*)$) in Prop.~\ref{prop_connection_weak_weakstar_op} is necessary: One can find normed spaces $X,Y$ and a $\topw^*$-convergent net of operators in ${}'(\mathcal B(X,Y))$ such that their limit is in $\mathcal B(Y^*,X^*)\setminus {}'(\mathcal B(X,Y))$. In other words---using Prop.~\ref{prop_dual_weak_star_cont}---one can find a $\topw^*$-convergent net of weak*-continuous operators the limit of which is not weak*-continuous anymore. For more details on this counterexample we refer to \cite{Glueck20} (and for the special case $X=Y$ to \cite{Furber20}).
\end{remark}

\subsection{The Mean Ergodic Theorem}\label{ch:mean_erg}\index{theorem!mean ergodic}

Let us take a small detour here before coming to operator theory on Hilbert spaces. While ergodic theory in general is concerned with dynamical systems, their underlying statistics, and the concept of equilibrium---the most prominent result from the perspective of quantum theory probably being von Neumann's mean ergodic theorem \cite{vonNeumann32} (cf.~also \cite[Ch.~VIII.4 \& VIII.5]{Dunford58})---there also are operator theoretic approaches to this theory. As we are concerned with quantum-\textit{dynamical} (control) systems later on, familiarizing ourselves with some of the base concepts will not do any harm. We orient ourselves towards Eisner et al.~\cite[Ch.~8]{Eisner15}.\smallskip

Given a bounded operator $T$ on a Banach space $X$ one can ask about the time mean $\frac{1}{n}\sum\nolimits_{k=0}^{n-1} T^k(x)$ of\footnote{Writing $T^k(x)$ is a short-hand notation for applying $T$ $k$-times to $x$, i.e.~$(T\circ\ldots\circ T)(x)$.} some $x\in X$ under $T$ and its behaviour as $n\to\infty$. From a more applied perspective this could mean to ask about the long-term behaviour of a discrete quantum-dynamical semigroup $(T_n)_{n\in\mathbb N}$ (evaluated on some state $\rho$ of the underlying quantum system).

\begin{definition}[Def.~8.4, \cite{Eisner15}]\label{def_mean_erg}
Let $X$ be a Banach space and $T\in\mathcal B(X)$. Then the operator $P_T$ defined by
$$
P_Tx:=\lim_{n\to\infty}\frac{1}{n}\sum\nolimits_{k=0}^{n-1} T^kx
$$
on the space $Z$ of all $x\in X$ where this limit exists is called the mean ergodic projection\index{mean ergodic projection} associated with $T$. The operator $T$ is called mean ergodic\index{operator!mean ergodic} if $Z= X$, that is, if the above limit exists for every $x\in X$.
\end{definition}

Now $Z$ is a $T$-invariant linear subspace of $X$ which contains all fixed points of $T$ (where $\operatorname{fix}(T):=\operatorname{ker}(\mathbbm{1}_X-T)$)\label{symb_fixed_points}
and $P_T$ is a projection onto said fixed point space which satisfies $P_TT=TP_T=T$ on $Z$. Further results read as follows, cf.~\cite[Lemma 8.3, Thms.~8.5 \& 8.22]{Eisner15}.

\begin{proposition}\label{prop_mean_ergodic}
Let a Banach space $X$ as well as $T\in\mathcal B(X)$ be given and define $P_T,Z$ as above.\begin{itemize}
\item[(i)] Suppose that $\sup_{n\in\mathbb N}\|\frac{1}{n}\sum_{k=0}^{n-1}T^k\|<\infty$ and $T^nx\to 0$ as $n\to\infty$ for all $x\in X$. Then $Z$ is closed and decomposes into $Z=\operatorname{ker}(\mathbbm{1}_X-T)\oplus\overline{\operatorname{im}(\mathbbm{1}_X-T)}$. Moreover the operator $T|_Z\in\mathcal B(Z)$ is mean ergodic.
\item[(ii)] If $T$ is a power-bounded operator (i.e.~$\|T^n\|\leq c$ for some $c\geq 1$ and all $n\in\mathbb N$), and $X$ is reflexive then $Z= X$, that is, $T$ is mean ergodic with $\|P_T\|\leq c$.
\end{itemize}
\end{proposition}

This result will have immediate implications for certain Hilbert space operators and will spark some intriguing questions regarding mean ergodicity of quantum channels in Chapter \ref{sec:quantum_channels}.

\section{Linear Operators between Hilbert Spaces}\label{ch_2_3}

As soon as one works with vector spaces which have a (reasonable) notion of a basis, linear maps can be defined via their action on any such basis. While Banach spaces may fail to have this feature (called a ``Schauder basis'') as proven by Enflo \cite{Enflo73} in the 70s, as soon as we turn to Hilbert spaces such pathologies cannot occur anymore. Indeed because every Hilbert space has an orthonormal basis 
it suffices to define an operator on that, which---assuming boundedness on the (finite) linear span of said basis---yields a unique operator on the full Hilbert space of same norm which extends the original operator. A special case is treated in Lemma \ref{lemma_unitary_ONB}.

Luckily, the literature on Hilbert space operators is much more rich and well-documented than for (operators on) general normed spaces which allows as to take a lighter path through this chapter with less proofs, more concepts, and more references. To name a few one may consider Berberian \cite[Ch.~VI]{Berberian76} and Dunford \& Schwartz \cite[Ch.~X]{Dunford63}, as well as \cite{MeiseVogt97en,Conway90,Kadison83,Pedersen89,Ringrose71,Rudin91}. 
\subsection{Bounded Operators}

Hilbert spaces are particularly nice spaces because, among other reasons, their dual space is (isometrically isomorphic to) the original space, cf.~\cite[Thm.~11.9]{MeiseVogt97en}.

\begin{lemma}[Riesz-Fr{\'e}chet]\label{lemma_riesz_frechet}\index{theorem!Riesz-Fr{\'e}chet}
Let $\mathcal H$ be a Hilbert space. For every $f\in\mathcal H^*$ there exists unique $x\in\mathcal H$ such that $f(y)=\langle x,y\rangle$ for all $y\in\mathcal H$ where $\|f\|=\|x\|$. In other words the map $\Phi:\mathcal H\to\mathcal H^*$, $x\mapsto \Phi_x$ (acting via $\Phi_x(y)=\langle x,y\rangle$ for $x,y\in\mathcal H$) is a real-linear bijective isometry.
\end{lemma}
\begin{remark}
\begin{itemize}
\item[(i)] Note that---although $\Phi$ is always real-linear---if the underlying field of $\mathcal H$ is $\mathbb C$ then $\Phi$ is not (complex-)linear but conjugate-linear (sometimes called antilinear):
$$
\Phi_{x+\lambda y}(z)=\langle x+\lambda y,z\rangle=\langle x,z\rangle+\overline{\lambda}\langle y,z\rangle=(\Phi_x+\overline{\lambda}\Phi_y)(z)\qquad\text{ for all }z\in\mathcal H\,.
$$
\item[(ii)] The Riesz-Fr{\'e}chet theorem motivates us to write $\langle x,\cdot\rangle$ or $\langle x|$ (bra-ket\label{symb_bra_ket}
notation)\index{bra-ket notation} for some $x\in\mathcal H$ by which we mean the associated dual space element $\Phi_x\in\mathcal H^*$. Moreover, given any $x\in\mathcal H$, $y\in\mathcal G$, this lets us define $T_{x,y}:\mathcal H\to\mathcal G$ via $T_{x,y}(z):=\langle x,z\rangle y$ for all $z\in\mathcal H$ which is obviously linear and bounded. We will usually write $|y\rangle\langle x|$ 
(instead of $T_{x,y}$).
\end{itemize}
\end{remark}

The fact that the dual space of a Hilbert space can be structured nicely has three immediate consequences. 

\begin{corollary}\label{coro_hilbert_reflexive}
Every Hilbert space is reflexive.\index{space!reflexive}
\end{corollary}
\begin{proof}[Proof idea]
The key here is that the map $\Phi$ ``transfers'' the inner product of $\mathcal H$ onto $\mathcal H^*$ via $\langle f,g\rangle_{\mathcal H^*}:=\langle \Phi^{-1}(g),\Phi^{-1}(f)\rangle_{\mathcal H}$ which turns $\mathcal H^*$ into a Hilbert space. Then, using Fr{\'e}chet-Riesz, every element of $\mathcal H^{**}$ can be traced back first to $\mathcal H^*$ and then to $\mathcal H$ (via the embedding $\iota:\mathcal H\to\mathcal H^{**}$). The details are carried out in \cite[Coro.~11.10]{MeiseVogt97en}.
\end{proof}

The second result is concerned with operators of rank one.

\begin{lemma}\label{lemma_rank_one}
Let Hilbert spaces $\mathcal H,\mathcal G$ and $T\in\mathcal B(\mathcal H,\mathcal G)$ be given. If $\operatorname{dim}(\operatorname{im}(T))=1$ then there exist $x\in\mathcal H$, $y\in\mathcal G$ such that $T=|y\rangle\langle x|$. Moreover $\|T\|=\|x\|\|y\|$.
\end{lemma}
\begin{proof}
By assumption there exists non-zero $y\in\mathcal G$ such that $\operatorname{span}\{y\}=\operatorname{im}(T)$ and $\|y\|=1$ (which is always possible by appropriate scaling). This lets us define $f:\mathcal H\to\mathbb F$, $z\mapsto \langle y,Tz\rangle$ which is obviously linear and bounded so $f\in\mathcal H^*$. Thus by Lemma \ref{lemma_riesz_frechet} one finds unique $x\in\mathcal H$ such that $\langle x,z\rangle=f(z)=\langle y,Tz\rangle$ for all $z\in\mathcal H$. Now if we can show that $\langle v,Tw\rangle=\langle v,(|y\rangle\langle x|)w\rangle=\langle x,w\rangle\langle v,y\rangle$ for all $v,w\in\mathcal H$ then $T=|y\rangle\langle x|$ by Lemma \ref{lemma_point_separation_dual_space} \& \ref{lemma_riesz_frechet}.

Indeed for all $w\in\mathcal H$ by assumption there exists $\lambda_w\in\mathbb F$ such that $Tw=\lambda_w y$ which implies
\begin{align*}
\langle v,Tw\rangle&=\big\langle \langle v,y\rangle y,Tw\big\rangle+\big\langle (v-\langle v,y\rangle y),Tw\big\rangle\\
&=\langle y,v\rangle \langle y,Tw\rangle +\lambda_w \underbrace{\big\langle (v-\langle v,y\rangle y),y\big\rangle}_{\langle v,y\rangle-\langle v,y\rangle\|y\|^2=0}=\langle y,v\rangle\langle x,w\rangle\,.
\end{align*}
Finally
$
\|T\|=\sup_{z\in\mathcal H,\|z\|=1}|\langle x,z\rangle|\|y\|=\|\Phi_x\|\|y\|=\|x\|\|y\|
$ which concludes the proof.
\end{proof}

The third result refines the concept of dual operators.\index{operator!dual}

\begin{proposition}\label{prop_adjoint_op_hilbert}
Let Hilbert spaces $\mathcal H,\mathcal G,\mathcal K$ as well as $T\in\mathcal B(\mathcal H,\mathcal G)$ be given. Then there exists unique $T^*\in\mathcal B(\mathcal G,\mathcal H)$ such that
$$
\langle y,Tx\rangle=\langle T^*y,x\rangle\qquad\text{ for all }x\in\mathcal H,y\in\mathcal G\,.
$$
Moreover the following statements hold for all $T\in\mathcal B(\mathcal H,\mathcal G)$, $S\in\mathcal B(\mathcal G,\mathcal K)$.
\begin{itemize}
\item[(i)] $T^{**}=T$ as well as $(ST)^*=T^*S^*$.
\item[(ii)] $\|T\|=\|T^*\|$ as well as $\|T^*T\|=\|TT^*\|=\|T\|^2$.
\item[(iii)] The map ${}^*:\mathcal B(\mathcal H,\mathcal G)\to\mathcal B(\mathcal G,\mathcal H)$ is a conjugate-linear bijective isometry.
\item[(iv)] If $T$ is invertible then $T^*$ is invertible with $(T^{-1})^*=(T^*)^{-1}$. 
\end{itemize}
\end{proposition}
\begin{proof}
The idea is to refine the notion of a dual operator using the fact that $\mathcal H^*\simeq\mathcal H$. Indeed $T^*:= \Phi^{-1}_{\mathcal H}\circ T'\circ\Phi_{\mathcal G}$ explicitly constructs the adjoint operator (using the dual operator $T'$ from Chapter \ref{section_dual_space}). The details are carried out in \cite[Prop.~11.11]{MeiseVogt97en}. Finally (iv) follows from Lemma \ref{lemma_dual_map_properties} (iii) together with the fact that the inverse of a continuous linear map between Banach spaces is automatically continuous (``bounded inverse theorem'', cf.~\cite[Thm.~8.6]{MeiseVogt97en}).\index{theorem!bounded inverse}
\end{proof}
\begin{definition}
Given Hilbert spaces $\mathcal H,\mathcal G$ and $T\in\mathcal B(\mathcal H,\mathcal G)$ the operator $T^*$ from Prop.~\ref{prop_adjoint_op_hilbert} is called the adjoint operator of $T$.\index{operator!adjoint}
\end{definition}

Now for two Hilbert spaces $\mathcal H,\mathcal G$ (as usual over the same field) an operator $U\in\mathcal B(\mathcal H,\mathcal G)$ is an isometric isomorphism (i.e.~a surjective linear isometry) if and only if $U$ is invertible and $U^{-1}=U^*$ (cf.~\cite[Ch.~II, Prop.~2.5]{Conway90}). As we know isometric isomorphisms are a fundamental tool to identify different spaces with each other, thus we may give such operators an explicit name; more generally the following classes of operators are of importance:

\begin{definition}
Let $\mathcal H$ be a Hilbert space. Then an operator $T\in\mathcal B(\mathcal H)$ is called
\begin{itemize}
\item[(i)] finite-rank operator\index{operator!finite-rank} if $\operatorname{dim}(\operatorname{im}(T))<\infty$. The set of all such operators is denoted by $\mathcal F(\mathcal H)$\label{symb_finite_rank}.
\item[(ii)] normal\index{operator!normal} if $T^*T=TT^*$.
\item[(iii)] self-adjoint if $T^*=T$.\index{operator!self-adjoint}
\item[(iv)] positive semi-definite\label{symb_positive_semi_def}\index{operator!positive semi-definite} if $T$ is self-adjoint and $\langle x,Tx\rangle\geq 0$ for all $x\in\mathcal H$. If $T$ is positive semi-definite we write $T\geq 0$. The set of all positive semi-definite operators on $\mathcal H$ shall be denoted by $\mathfrak{pos}(\mathcal H)$. 
\item[(v)] positive definite\label{symb_positive_def}\index{operator!positive definite}, denoted by $T> 0$, if $T$ is self-adjoint and $\langle x,Tx\rangle>0$ for all $x\in\mathcal H\setminus\{0\}$. 
\item[(vi)] unitary\index{operator!unitary} if $T$ is bijective and $T^{-1}=T^*$. The collection of all unitary operators on $\mathcal H$ is denoted by $\mathcal U(\mathcal H)$. (If the underlying field is $\mathbb R$ then such an operator is also called ``orthogonal''). \label{symb_unitary}
\item[(vii)] projection\index{operator!projection} if $T^2=T$, and orthogonal projection\index{operator!orthogonal projection} if $T$ is a self-adjoint projection.
\item[(viii)] partial isometry\index{operator!partial isometry} if $T^*T$ is an orthogonal projection (i.e.~if $T^*TT^*T=T^*T$).
\end{itemize}
\end{definition}
\begin{remark}\label{rem_op_between_diff_spaces}
Obviously,
\begin{itemize}
\item[(i)] the notion of finite rank operators makes sense for operators between different Hilbert spaces or even normed spaces, and $\mathcal F(\mathcal H,\mathcal G)$ is a linear subspace of $\mathcal B(\mathcal H,\mathcal G)$.
\item[(ii)] the positive semi-definite operators form a convex cone, meaning for all $\lambda,\mu\geq 0$ and all $S,T\in\mathfrak{pos}(\mathcal H)$ one has $\lambda S+\mu T\geq 0$. 
\item[(iii)] the notion of unitary operators makes sense for operators between different Hilbert spaces over the same field. Such maps are sometimes called \textit{unitary transformations} (as opposed to ``unitary operators'', similarly for orthogonal operators).
\end{itemize}
\end{remark}


Until now we allowed all normed spaces, so in particular all Hilbert spaces, to have base field $\mathbb R$ or $\mathbb C$. However dealing with \textit{complex} Hilbert spaces is beneficial as it simplifies a few things:

\begin{lemma}\label{lemma_complex_H_space}
Let $\mathcal H$ be a complex Hilbert space and $T\in\mathcal B(\mathcal H)$. The following statements hold.
\begin{itemize}
\item[(i)] $T$ is self-adjoint if and only if $\langle x,Tx\rangle\in\mathbb R$ for all $x\in\mathcal H$.
\item[(ii)] If $\langle x,Tx\rangle\geq 0$ for all $x\in\mathcal H$ then $T$ is positive semi-definite. Analogously if $\langle x,Tx\rangle>0$ for all $x\in\mathcal H\setminus\{0\}$ then $T$ is positive definite.
\item[(iii)] If $\langle x,Tx\rangle=0$ for all $x\in\mathcal H$ then $T=0$.
\item[(iv)] $T$ is an isometry, i.e.~$\|Tx\|=\|x\|$ for all $x\in\mathcal H$, if and only if $T^*T=\mathbbm{1}_{\mathcal H}$. 
\item[(v)] $T$ is normal if and only if $\|Tx\|=\|T^*x\|$ for all $x\in\mathcal H$.
\item[(vi)] $T$ is unitary if and only if $\|Tx\|=\|x\|=\|T^*x\|$ for all $x\in\mathcal H$.
\end{itemize}
\end{lemma}
\begin{proof}
(i): See for example \cite[Prop.~2.12]{Conway90}. (ii): Direct consequence of (i). (iii): \cite[Coro.~2.14]{Conway90}. (iv): $\|Tx\|^2=\|x\|^2$ for all $x\in\mathcal H$ $\Leftrightarrow$ $\langle x,T^*Tx\rangle=\langle x,x\rangle$ for all $x\in\mathcal H$ $\Leftrightarrow$ $\langle x,(T^*T-\mathbbm{1}_{\mathcal H})x\rangle=0$ for all $x\in\mathcal H$ $\Leftrightarrow$ (by (iii)) $T^*T=\mathbbm{1}_{\mathcal H}$. (v): Analogous to the proof of (iv). (vi): Obviously $T$ is unitary if and only if $T^*T=TT^*=\mathbbm{1}_{\mathcal H}$ if and only if $T$ is a normal isometry so this follows from (iv) \& (v).
\end{proof}
\begin{remark}\label{rem_real_complex_hs}
\begin{itemize}
\item[(i)] The assumption of $\mathcal H$ being a complex Hilbert space in Lemma \ref{lemma_complex_H_space} is necessary. The most prominent counterexample to (iii) is a simple $\frac{\pi}{2}$ rotation on $\mathbb R^2$ (equipped with the standard inner product), i.e.~$T(x_1,x_2):= (x_2,-x_1)$. Evidently $\langle x,Tx\rangle=0$ for all $x\in\mathbb R^2$ but $T\neq 0$ and $T$ is not self-adjoint (indeed $T={\scriptsize\begin{pmatrix} 0&1\\-1&0 \end{pmatrix}}$ with respect to the standard basis).
\item[(ii)] Working on complex Hilbert spaces, in contrast to real ones, is not only advantageous from the perspective of operator theory but is also necessary from the point of quantum mechanics. Indeed Stueckelberg has shown in the early 60s \cite{Stueckelberg60} that in order to have an uncertainty principle over real Hilbert spaces one has to introduce an operator $J$ which satisfies $J^2=-\mathbbm{1}$ and which commutes with all observables.

On the other hand one can ask whether it is beneficial, mathematically or physically, to go \textit{beyond} complex Hilbert spaces and consider (left-)quaternionic Hilbert spaces\footnote{
While the quaternions $\mathbb H$ are the only further associative division algebra over the reals (aside from $\mathbb R$ and $\mathbb C$) they do not constitute a field anymore because multiplication in $\mathbb H$ is not commutative. This is also why one has to specify whether the Hilbert space is left- or right-quaternionic, i.e.~whether one considers scalar multiplication from the left or from the right.

The only further division algebra over the reals, the octernions $\mathbb O$, is non-associative which suffices to make quantum mechanics non-extendable to $\mathbb O$, cf.~\cite[Ch.~2.7]{Adler95}.}.
While there is a quaternionic formulation of quantum mechanics\index{quantum mechanics!quaternionic}, ``all presently known physical phenomena appear to be very well described by complex quantum mechanics'' \cite[p.~497]{Adler95} which is why in this thesis we will stick to complex Hilbert spaces. For more details on quaternionic quantum mechanics we, unsurprisingly, refer to the book of Adler \cite{Adler95}.
\end{itemize}
\end{remark}

Unitary transformations have a simple but special connection to orthonormal bases of Hilbert spaces.
\begin{lemma}\label{lemma_unitary_ONB}
Let $\mathcal H,\mathcal G$ be Hilbert spaces. The following statements hold.
\begin{itemize}
\item[(i)] Let any orthonormal basis\index{orthonormal basis} $(e_i)_{i\in I}$ of $\mathcal H$ and a family of pairwise orthogonal vectors $(y_i)_{i\in I}\subseteq\mathcal G$ with $\sup_{i\in I}\|y_i\|<\infty$ be given. Then there exists unique $T\in\mathcal B(\mathcal H,\mathcal G)$ with $Te_i=y_i$ for all $i\in I$. In this case $\|T\|=\sup_{i\in I}\|y_i\|$.
\item[(ii)] Given $T\in\mathcal B(\mathcal H,\mathcal G)$ the following are equivalent.
\begin{itemize}
\item[(a)] $T$ is unitary.
\item[(b)] For every orthonormal basis $(e_i)_{i\in I}$ of $\mathcal H$, $(Te_i)_{i\in I}$ is an orthonormal basis of $\mathcal G$.
\item[(c)] There exists an orthonormal basis $(e_i)_{i\in I}$ of $\mathcal H$ such that $(Te_i)_{i\in I}$ is an orthonormal basis of $\mathcal G$.
\end{itemize}
\item[(iii)] Let $(e_i)_{i\in I}$, $(f_i)_{i\in I}$ be an arbitrary orthonormal basis of $\mathcal H$, $\mathcal G$, respectively. Then there exists unique $T\in\mathcal B(\mathcal H,\mathcal G)$ with $Te_i=f_i$ for all $i\in I$. In fact this $T$ is unitary. 
\end{itemize}
\end{lemma}
\begin{proof}
(i): We only have to prove existence of such an operator as uniqueness is evident: Assume there are $T_1,T_2\in\mathcal B(\mathcal H,\mathcal G)$ which satisfy $T_1e_i=y_i=T_2e_i$ for all $i\in I$. In particular $(T_1-T_2)e_i=0$ for all $i\in I$ so $T_1=T_2$ on $\operatorname{span}\{e_i\,|\,i\in I\}$. By continuity (and because $\overline{\operatorname{span}\{e_i\,|\,i\in I\}}=\mathcal H$) the two operators co{\"i}ncide.

For existence define $\mathcal H_0:=\operatorname{span}\{e_i\,|\,i\in I\}$ and $T_0\in\mathcal L(\mathcal H_0,\mathcal G)$ via $T_0e_i:=y_i$ for all $i\in I$ as well as its linear extension onto all of $\mathcal H_0$. Now for every $x\in\mathcal H_0$ there exist $i_1,\ldots,i_m\in I$ such that $x=\sum_{j=1}^m \langle e_{i_j},x\rangle e_{i_j}$ (so $\|x\|^2=\sum_{j=1}^m |\langle e_{i_j},x\rangle|^2$ by Lemma \ref{lemma_pyth_thm}) and thus
\begin{align*}
\|T_0x\|^2&=\Big\|\sum\nolimits_{j=1}^m \langle e_{i_j},x\rangle y_{i_j}\Big\|^2\overset{\text{Lemma }\ref{lemma_pyth_thm}}=\sum\nolimits_{j=1}^m |\langle e_{i_j},x\rangle|^2\|y_{i_j}\|^2\\
&\leq \big(\sup_{i\in I}\|y_i\|^2\big)\sum\nolimits_{j=1}^m |\langle e_{i_j},x\rangle|^2= \big(\sup_{i\in I}\|y_i\|^2\big)\|x\|^2\,.
\end{align*}
This shows $\|T_0\|\leq \sup_{i\in I}\|y_i\|<\infty$ so $T_0\in\mathcal B(\mathcal H_0,\mathcal G)$; actually one readily verifies $\|T_0\|\geq\sup_{i\in I}\|T_0e_i\|=\sup_{i\in I}\|y_i\|$ so the latter is equal to $\|T_0\|$. Then \cite[Prop.~2.1.11]{Pedersen89} yields (unique) $T\in\mathcal B(\mathcal H,\mathcal G)$ with $T|_{\mathcal H_0}=T_0$ (i.e.~$Te_i=T_0e_i=y_i$ for all $i\in I$) and $\|T\|=\|T_0\|=\sup_{i\in I}\|y_i\|$ as desired.

(ii): For the proof we orient ourselves towards \cite[Prop.~1.49]{Heinosaari12} ``(a) $\Rightarrow$ (b)': Let $(e_i)_{i\in I}$ be any orthonormal basis of $\mathcal H$. Then $(Te_i)_{i\in I}$ is obviously an orthonormal system in $\mathcal G$ because $\langle Te_i,Te_j\rangle=\langle e_i,T^*Te_j\rangle=\langle e_i,e_j\rangle=\delta_{ij}$ for all $i,j\in I$. Now for all $y\in\mathcal G$ using Parseval's equation (Prop.~\ref{prop_hilbert_space_basis} (ii)) we find
\begin{align*}
\|y\|^2=\langle y,TT^*y\rangle=\langle T^*y,T^*y\rangle=\|T^*y\|^2=\sum\nolimits_{i\in I}|\langle e_i,T^*y\rangle|^2=\sum\nolimits_{i\in I}|\langle Te_i,y\rangle|^2
\end{align*}
so $(Te_i)_{i\in I}$ is an orthonormal basis of $\mathcal G$ (again by Prop.~\ref{prop_hilbert_space_basis} (ii)).

``(b) $\Rightarrow$ (c)'': Trivial because every Hilbert space has an orthonormal basis (Prop.~\ref{prop_hilbert_space_basis} (iii)).

``(c) $\Rightarrow$ (a)'': By assumption---using (i)---there exists unique $S\in\mathcal B(\mathcal G,\mathcal H)$ such that $S(Te_i)=e_i$ for all $i\in I$. But $S\circ T$ and $\mathbbm{1}_{\mathcal H}$ act the same on $\operatorname{span}\{e_i\,|\,i\in I\}$ so because this is an orthonormal basis, by continuity $S\circ T=\mathbbm{1}_{\mathcal H}$. Now because $(Te_i)_{i\in I}$ is an orthonormal basis of $\mathcal G$ we for every $y\in\mathcal G$ find $y=\sum_{i\in I}\langle Te_i,y\rangle Te_i$ (Prop.~\ref{prop_hilbert_space_basis}) which yields
$$
Sy=\sum\nolimits_{i\in I}\langle Te_i,y\rangle S(Te_i)=\sum\nolimits_{i\in I}\langle Te_i,y\rangle e_i=\sum\nolimits_{i\in I}\langle e_i,T^*y\rangle e_i=T^*y\,.
$$
Hence $S=T^*$ and thus $T^*T=\mathbbm{1}_{\mathcal H}$ which shows that ($T$ is injective and) $T^*$ is surjective. On the other hand for all $y\in\mathcal G$
$$
\|T^*y\|^2=\sum\nolimits_{i\in I}|\langle e_i,T^*y\rangle|^2=\sum\nolimits_{i\in I} |\langle Te_i,y\rangle|^2=\|y\|^2
$$
so $T^*$ is a surjective linear isometry which shows that $T^*$ is bijective with $(T^*)^{-1}=(T^*)^*$ \cite[Ch.~II, Prop.~2.5]{Conway90}. But Prop.~\ref{prop_adjoint_op_hilbert} (iv) implies that $T=(T^*)^*$ is bijective with $((T^*)^{-1})^*=((T^{-1})^*)^*=T^{-1}$ so $T^{-1}=((T^*)^{-1})^*=T^{***}=T^*$ which lets us conclude that $T$ is unitary.\smallskip

(iii): By (i) such $T$ exists and is unique. By (ii) $T$ is unitary.
\end{proof}

The notion of positive (semi-definite) operators enables us to carry over square roots as well as absolute values to Hilbert space operators. The following statement is proven for example in \cite[Prop.~3.2.11 \& Thm.~3.2.17]{Pedersen89}.

\begin{lemma}
Let $\mathcal H$ be a Hilbert space and $T\in\mathcal B(\mathcal H)$. The following statements hold.
\begin{itemize}
\item[(i)] If $T\geq 0$ then there exists unique $\sqrt{T}\in\mathfrak{pos}(\mathcal H)$\label{symb_sqrt}
such that $(\sqrt{T})^2=T$. Moreover if an operator commutes with $T$ then it commutes with $\sqrt{T}$.
\item[(ii)] There exists unique $|T|\in\mathfrak{pos}(\mathcal H)$ such that $\|Tx\|=\|\,|T|x\|$ for all $x\in\mathcal H$ and one has $|T|=\sqrt{T^*T}$\label{symb_abs}. Moreover there exists a unique partial isometry $U\in\mathcal B(\mathcal H)$ such that $T=U|T|$, $\operatorname{ker}(U)=\operatorname{ker}(T)$. In particular $U^*U|T|=|T|$, $U^*T=|T|$, and $U^*UT=T$. 
\end{itemize}
\end{lemma} 

Given $T\in\mathfrak{pos}(\mathcal H)$ the operator $\sqrt{T}$ is termed \textit{square root}\index{operator!square root} of $T$. Writing general $T\in\mathcal B(\mathcal H)$ as $T=U|T|$ in the above sense is called the \textit{polar decomposition} of $T$.\index{operator!polar decomposition}

\subsection{Unbounded Operators}\label{sec_unbounded_op}

As Reed \& Simon nicely put it in the first volume of their renowned series \textit{Methods of Mathematical Physics}: ``it is a fact of life that many of the most important operators which occur in mathematical physics are not bounded'' \cite[p.~249]{ReedSimonI}. After all, the \textit{canonical commutation relations}\index{canonical commutation relations} $PQ-QP=i\mathbbm{1}_{\mathcal H}$ for linear operators $P,Q\in\mathcal L(\mathcal H)$ on some Hilbert space $\mathcal H$, which are fundamental in quantum physics, require that either $P$ or $Q$ has to be unbounded\footnote{
The following proof is taken from \cite[Thm.~13.6]{Rudin91}: If any two bounded operators $P,Q\in\mathcal B(\mathcal H)$ would satisfy $PQ-QP=\lambda\mathbbm{1}_{\mathcal H}$ for some $\lambda\in\mathbb C\setminus\{0\}$ then
\begin{align*}
PQ^n-Q^nP&=\sum\nolimits_{j=0}^{n-1}\big( Q^jPQ^{n-j}-Q^{j+1}PQ^{n-j-1} \big)\\
&=\sum\nolimits_{j=0}^{n-1}Q^j(PQ-QP)Q^{n-j-1}=\lambda\sum\nolimits_{j=0}^{n-1}Q^j\mathbbm{1}_{\mathcal H}Q^{n-j-1}=\lambda nQ^{n-1}
\end{align*}
for all $n\in\mathbb N$ which would imply $|\lambda |n\|Q^{n-1}\|\leq 2\|P\|\|Q^n\|\leq 2\|P\|\|Q\|\|Q^{n-1}\|$. Now if $Q^n\neq 0$ for all $n\in\mathbb N$ then we may divide out its norm to obtain $|\lambda |n\leq 2\|P\|\|Q\|$ for \textit{all} $n\in\mathbb N$, contradicting boundedness of $P,Q$ as $\lambda \neq 0$. Thus there has to exist some $N\in\mathbb N_0$ with $Q^N\neq 0$ but $Q^{N+1}=0$ 
which ends in the contradiction
$$
0=0-0=PQ^{N+1}-Q^{N+1}P=\lambda NQ^N\neq 0\,.
$$
}. This in turn means that $\mathcal H$ has to be of infinite dimension because in finite dimensions every linear operator is automatically bounded.

In the usual formulation of quantum theory, observables are described by self-adjoint Hilbert space operators, that is, operators which satisfy $\langle x,Ay\rangle=\langle Ax,y\rangle$ where $x,y$ are chosen appropriately. By the Hellinger-Toeplitz theorem\index{theorem!Hellinger-Toeplitz} \cite[p.~84]{ReedSimonI} such operators can only be unbounded if their domain is a strict subset of the underlying Hilbert space. This is a consequence of the closed graph theorem \cite[Thm.~III.12]{ReedSimonI} which states that for a linear map $T\in\mathcal L(X,Y)$ between Banach spaces $X,Y$, boundedness of $T$ is equivalent to closedness of the graph of $T$, i.e.~$\{(x,Tx)\,|\,x\in X\}\subseteq X\times Y$ being closed (in the product topology). Therefore we have to be careful about the domain of unbounded operators. For the following definition we orient ourselves towards \cite[Ch.~VIII.1]{ReedSimonI}.

\begin{definition}
Let $X$ be an arbitrary Banach space. An operator\index{operator} $T$ on $X$
\begin{itemize}
\item[(i)] is a linear map from its domain\index{operator!domain of}, a linear subspace of $ X$ denoted by $D(T)$, into $ X$. If $\overline{D(T)}= X$ then we say $T$ is densely defined\index{operator!densely defined}. 
\item[(ii)] is called closed\index{operator!closed} if its graph\index{graph} $\mathrm{gr}(T):=\{(x,Tx)\,|\,x\in D(T)\}$\label{symb_graph}
is a closed subset of $ X\times X$ (in the product topology).
\item[(iii)] is an extension\index{operator!extension of} of an operator $S$ if $D(S)\subseteq D(T)$ and $Sx=Tx$ for all $x\in D(S)$. This is equivalent to $\mathrm{gr}(S)\subseteq\mathrm{gr}(T)$.
\item[(iv)] is closable if it has a closed extension. Every closable operator\index{operator!closable} has a smallest closed extension, called its closure\index{operator!closure of} (denoted by $\overline{T}$).\label{symb_closure_op}
\end{itemize}
\end{definition}
\begin{remark}\label{rem_densely_def_bounded}
To emphasize the necessity of these domain considerations for unbounded operators be aware that if $T:D(T)\to X$ is densely defined and bounded, then there exists a unique extension $\tilde T\in\mathcal B( X)$ of $T$ to the whole space \cite[Thm.~I.7]{ReedSimonI}; thus in the bounded case there is no point in specifying a (dense) domain. Moreover, this extension---just like \textit{every} bounded linear Banach space operator---is closed by the closed graph theorem \cite[Thm.~III.12]{ReedSimonI}.
\end{remark}

In the spirit of the adjoint operator from the bounded case (Prop.~\ref{prop_adjoint_op_hilbert}) we want to extend this notion to general (unbounded) Hilbert space operators. Because of the previous domain discussion we get three related notions of ``self-adjointness'' (or similar) which for bounded operators all co{\"i}ncide.
\begin{definition}\label{def_adjoint_unbounded}
Let $\mathcal H$ be a Hilbert space and $T$ be a densely defined operator on $\mathcal H$.
\begin{itemize}
\item[(i)] Let $D(T^*)$ denote the set of all $x\in\mathcal H$ for which there exists $y\in\mathcal H$ such that
$$
\langle x,Tz\rangle=\langle y,z\rangle\qquad\text{ for all }z\in D(T)\,.
$$
This defines the adjoint map\index{operator!adjoint} $T^*:D(T^*)\to\mathcal H$ of $T$ via $T^*x:=y$.
\end{itemize}
With this $T$ is called
\begin{itemize}
\item[(ii)] symmetric\index{operator!symmetric} if $\langle y,Tx\rangle=\langle Ty,x\rangle$ for all $x,y\in D(T)$ which is equivalent to $D(T)\subseteq D(T^*)$ together with $Tx=T^*x$ for all $x\in D(T)$.
\item[(iii)] self-adjoint\index{operator!self-adjoint} if $T=T^*$, i.e.~$T$ is symmetric and $D(T)=D(T^*)$.
\item[(iv)] essentially self-adjoint\index{operator!essentially self-adjoint} if $T$ is symmetric and $\overline{T}$ is self-adjoint.
\end{itemize}
\end{definition}
\noindent Often one deals with symmetric operators which are not closed (but closable by considering a larger domain) so essential self-adjointness guarantees the existence of a unique self-adjoint extension.
\begin{remark}\label{rem_adjoint_unbounded}
\begin{itemize}
\item[(i)] By Riesz-Fr{\'e}chet the domain of the adjoint can be written as
$$
D(T^*)=\{x\in\mathcal H\,|\, \exists_{C>0}\ \forall_{z\in D(T)}\ |\langle x,Tz\rangle|\leq C\|z\|\}\,.
$$
This also explains the requirement of $T$ being densely defined in Def.~\ref{def_adjoint_unbounded} (i), otherwise one could not apply Lemma \ref{lemma_riesz_frechet} and uniqueness of $y$ could not be guaranteed.
 \item[(ii)] It might happen that $D(T^*)$ is not dense in $\mathcal H$ although $D(T)$ is. For an example we refer to \cite[Ch.~VIII.1, Ex.~4]{ReedSimonI}. 
\end{itemize}
\end{remark}

Now for some basic connections between the introduced notions.
\begin{lemma}\label{lemma_symmetric_closable}
Let $\mathcal H$ be a Hilbert space and $T$ be a densely defined operator on $\mathcal H$. The following statements hold.
\begin{itemize}
\item[(i)] The adjoint $T^*$ is closed and, moreover, $T$ is closable if and only if $T^*$ is densely defined in which case $\overline{T}=T^{**}$
\item[(ii)] If $T$ is self-adjoint then $T$ is closed.
\end{itemize}
Now if $T$ a densely defined, symmetric operator on $\mathcal H$
\begin{itemize}
\item[(iii)] then $T$ is closable with $\overline{T}=T^{**}$. 
\item[(iv)] and $\overline{\operatorname{im}(T)}=\mathcal H$ then $T$ is injective.
\item[(v)] and $T$ is surjective, then $T$ is bijective, self-adjoint, and has bounded self-adjoint inverse.
\end{itemize}
\end{lemma}
\begin{proof}
(i): \cite[Thm.~VIII.1]{ReedSimonI}. (ii): By (i) the adjoint $T^*$ is closed but $T=T^*$ by definition of self-adjointness. (iii): If $T$ is symmetric then $\mathcal H=\overline{D(T)}\subseteq\overline{D(T^*)}\subseteq\mathcal H$ so $T^*$ is densely defined which by (i) concludes the proof. (iv) \& (v): \cite[Thm.~13.11]{Rudin91}. For self-adjointness of $T^{-1}$ note that $T^{-1}x,T^{-1}y\in D(T)$ for all $x,y\in\mathcal H$, so by symmetry of $T$
\begin{equation}
\langle x,T^{-1}y\rangle=\langle T(T^{-1}x),T^{-1}y\rangle=\langle T^{-1}x,T(T^{-1}y)\rangle=\langle T^{-1}x,y\rangle\,.\tag*{\qedhere}
\end{equation}
\end{proof}

After this flood of definitions and concepts, presenting an example is in order (and hopefully illuminating). For this let us consider one of the few quantum systems which can be solved analytically:

\begin{example}\label{ex_quantum_harm_osc}
The quantum harmonic oscillator\index{quantum harmonic oscillator} (in one dimension) can be written as a model on the square-summable sequences\footnote{
To be more precise the quantum harmonic oscillator is formulated on the Hilbert space of complex-valued square-integrable functions $\mathcal H=L^2(\mathbb R)$ and the Hamiltonian describing the particle is of the form $$H|\psi\rangle(x)=-\frac{1}{2m}\frac{d^2}{dx^2}|\psi(x)\rangle+\frac{m\omega^2}{2}x^2|\psi(x)\rangle$$ for $|\psi\rangle\in D(H)\subset L^2(\mathbb R)$ from a suitable domain (cf.~\cite[Ch.~11]{Hall13}). One can show that $H$ has discrete spectrum of the form $\{\frac{\hbar\omega}{2}(2n-1)\,|\,n\in\mathbb N\}$ and the corresponding eigenvectors (countably many weighted hermite polynomials) form an orthonormal basis of $L^2(\mathbb R)$. In particular $\mathcal H$ is separable so one finds a unitary transformation from $L^2(\mathbb R)$ to $\ell^2(\mathbb N)$ (Rem.~\ref{rem_ell2_sep_HS}) which transforms $H$ into ``$\frac{\hbar\omega}{2}\operatorname{diag}(1,3,5,7,\ldots)$'' in a suitable basis.\label{footnote_qhosc}},
i.e.~on the Hilbert space $\mathcal H=\ell^2(\mathbb N)$ from Ex.~\ref{ex_ell2_Hilbert}. Up to positive
constants the Hamiltonian of the system is of the form $H=\operatorname{diag}(1,3,5,7,\ldots)$:
\begin{equation}\label{eq:def_Ham}
H:D(H)\to\ell^2(\mathbb N)\qquad x=(x_1,x_2,x_3,\ldots)\mapsto (x_1,3x_2,5x_3,\ldots)\,.
\end{equation}
Already the imprecise form $H=\operatorname{diag}(1,3,5,7,\ldots)$ strongly suggests that $H$ is unbounded so we have to think of a reasonable dense domain for $H$; after all $H$ cannot be defined everywhere because $x=(\frac{1}{2n-1})_{n\in\mathbb N}\in\ell^2(\mathbb N)$ but $Hx=(1,1,1,\ldots)\not\in\ell^2(\mathbb N)$ so the choice $D(H)=\ell^2(\mathbb N)$ would violate the codomain of $H$. 

A first na\"ive approach to find a domain for $H$ is to set $D(H)=c_{00}(\mathbb N)$ because
\begin{itemize}
\item[$\bullet$] every scaling of an eventually-zero sequence stays in $c_{00}(\mathbb N)\subset\ell^2(\mathbb N)$ so $H$ is well-defined.
\item[$\bullet$] $c_{00}(\mathbb N)$ is a dense linear subspace of $\ell^2(\mathbb N)$ (Ex.~\ref{ex_ell_p_space}). 
\end{itemize}
Therefore this choice turns $H$ into a densely defined linear operator. Unsurprisingly $H$ is symmetric because $\langle x,Hy\rangle=\langle Hx,y\rangle$ for all $x,y\in c_{00}(\mathbb N)$ as is readily verified.
Unfortunately, however, $H$ (with the current domain) is not closed so it cannot be self-adjoint as a consequence of Lemma \ref{lemma_symmetric_closable} (ii). To see this define $x:=(\frac{1}{(2n-1)^2})_{n\in\mathbb N}$ as well as the truncated sequences $x^{(n)}:=(x_1,\ldots,x_n,0,0,\ldots)$ for all $n\in\mathbb N$. Then the sequence $(x^{(n)},Hx^{(n)})_{n\in\mathbb N}$ in $\mathrm{gr}(H)\subset\ell^2(\mathbb N)\times \ell^2(\mathbb N)$ converges to $(x,(\frac{1}{2n-1})_{n\in\mathbb N})$ in the product topology, but $x\not\in c_{00}(\mathbb N)$ so the limit point lives outside of the graph $\mathrm{gr}(H)$ which shows that the latter is not closed.

This whole dilemma is a consequence of $D(H)\subsetneq D(H^*)$ and is resolved by adjusting the domain of $H$. Motivated by Lemma \ref{lemma_symmetric_closable}---which tells us that the densely defined, symmetric operator $T$ is closable---we consider the linear map $H$ from \eqref{eq:def_Ham} now with maximal domain
\begin{align*}
D(H):=&\{x\in\ell^2(\mathbb N)\,|\,Hx\in\ell^2(\mathbb N)\}\\=
&\Big\{x\in\ell^2(\mathbb N)\,\Big|\,\sum\nolimits_{n=1}^\infty (2n-1)^2|x_{n}|^2<\infty\Big\}= \Big\{\Big(\frac{y_n}{2n-1}\Big)_{n\in\mathbb N}\,\Big|\,y\in\ell^2(\mathbb N)\Big\} \,.
\end{align*}
With this $H$ is still densely defined ($c_{00}(\mathbb N)\subset D(H)$), symmetric, and, moreover, for any $y\in\ell^2(\mathbb N)$ the $\ell^2$-sequence $\tilde y:=(\frac{y_n}{2n-1})_{n\in\mathbb N}$ satisfies $H\tilde y=y$ so $H$ is surjective. Thus Lemma \ref{lemma_symmetric_closable} (v) implies $H$ is bijective, self-adjoint, and has bounded inverse
\begin{equation*}
H^{-1}:\ell^2(\mathbb N)\to D(H)\subset\ell^2(\mathbb N)\qquad x=(x_1,x_2,x_3,\ldots)\mapsto (x_1,\tfrac{x_2}{3},\tfrac{x_3}{5},\ldots)\,.
\end{equation*}
\end{example}

\begin{remark}\label{rem_I_plus_a_a_star}
The above example merely is an incarnation of a much broader class of operators: it turns out that the Hamiltonian for the quantum harmonic oscillator from footnote \ref{footnote_qhosc} can be written as $H=\hbar\omega(\frac{1}{2}\mathbbm{1}+a^*a)$ where
$$
a:D(a)\to\ell^2(\mathbb N)\qquad x=(x_1,x_2,x_3,\ldots)\mapsto ( x_2,\sqrt{2}x_3,\sqrt{3}x_4 ,\ldots)
$$
with $D(a)=\{x\in\ell^2(\mathbb N)\,|\,\sum_{n=1}^\infty n|x_{n+1}|^2<\infty\}$ is the ``lowering operator''\index{operator!lowering} or ``annihilation operator''\index{operator!annihilation}, cf.~\cite[Ch.~11.2]{Hall13} \& \cite[Ch.~7.2]{Schmuedgen12}. Because $a$ is densely defined ($c_{00}(\mathbb N)\subset D(a)$) and closed (Appendix \ref{appendix_lowering_op_proof}) the Hamiltonian $H:D(a^*a)\to\ell^2(\mathbb N)$ has to be bijective and self-adjoint with bounded, self-adjoint inverse \cite[Prop.~3.18]{Schmuedgen12}.
\end{remark}

%
%
%
%
%
%

\subsection{Spectral Theorem and Functional Calculus}\label{ch:func_calc}

Given some bounded self-adjoint operator $H\in\mathcal B(\mathcal H)$ on a complex Hilbert space $\mathcal H$ the solution to the ordinary differential equation\footnote{Physicists will immediately recognize this as the \textit{time-dependent Schr{\"o}dinger equation}\index{Schr{\"o}dinger equation} which describes the evolution of a closed quantum system, cf.~also Chapter \ref{ch:qu_dyn_sys}.}
$$
\frac{dx(t)}{dt}=-iHx(t)
$$
for all $t>0$ with initial condition $x(0)=x_0\in\mathcal H$ is obviously given by $x(t)=e^{-itH}x_0$. Because $H$ is bounded the exponential $e^{-itH}$ can be defined via $\sum_{n=0}^\infty (-it)^n\frac{H^n}{n!}$ where the sum converges in the operator norm.

In Chapter \ref{sec_unbounded_op} we learned that such evolutions are often described by unbounded (self-adjoint) operators. While we still desire a solution of the form $x(t)=e^{-itH}x_0$, the exponential series $\sum_{n=0}^\infty (-it)^n\frac{H^n}{n!}$ is at best defined on a common domain $\bigcap_{n=1}^\infty D(H^n)$ which might not be dense anymore (or, even worse, just $\{0\}$), not to mention possible convergence problems. This motivates finding formalisms which turn $e^{-itH}$ into a well-defined object.

One way of achieving this is via \textit{functional calculus}, the idea of which simply goes as follows: Decompose an operator in terms of its spectrum and define the action of ``sufficiently nice'' functions (e.g., $z\mapsto e^{itz}$) on the operator by applying it to said spectrum. Our main references for this chapter are \cite[Ch.~III.6]{Kato80}, \cite[Ch.~VII.9]{Dunford58} \& \cite[Ch.~4 \& 5]{Schmuedgen12}.

\begin{definition}
Let $X$ be a complex Banach space and $T$ be a densely defined operator on $X$. Then the resolvent\index{resolvent} of $T$ is defined to be\label{symb_resolvent}
$$
\mathbbm{r}(T):=\{\lambda\in\mathbb C\,|\,T-\lambda\mathbbm{1}\,\text{is bijective with bounded inverse}\,\}
$$
and the spectrum\index{spectrum} $\sigma(T):=\mathbb C\setminus\mathbbm{r}(T)$ is the complement of the resolvent. Moreover\label{symb_spectrum}
\begin{align*}
\sigma_{\mathrm{p}}(T)&:=\{\lambda\in\mathbb C\,|\, \operatorname{ker}(T-\lambda\mathbbm{1})\neq\{0\}\,\}\\
\sigma_{\mathrm{c}}(T)&:=\{\lambda\in\mathbb C\,|\,\operatorname{ker}(T-\lambda\mathbbm{1})=\{0\}\text{ and }\operatorname{im}(T-\lambda\mathbbm{1})\neq X\text{ is dense}\, \} \\
\sigma_{\mathrm{r}}(T)&:= \{\lambda\in\mathbb C\,|\,\operatorname{ker}(T-\lambda\mathbbm{1})=\{0\}\text{ and }\operatorname{im}(T-\lambda\mathbbm{1})\neq X\text{ is not dense}\, \}
\end{align*}
where $\sigma_{\mathrm{p}}$ is called the point spectrum\index{spectrum!point} (and its elements are called eigenvalues\index{eigenvalue}), $\sigma_{\mathrm{c}}$ is the continuous spectrum\index{spectrum!continuous}, and $\sigma_{\mathrm{r}}$ is the residual spectrum\index{spectrum!residual} of $T$. 
\end{definition}
Here $\mathbbm{1}$ is short for $\mathbbm{1}_{D(T)}$ to make sense of $T-\lambda\mathbbm{1}:D(T)\to X$. As for some further remarks:
\begin{remark}
\begin{itemize}
\item[(i)] If $T$ is closed then it follows from the closed graph theorem\index{theorem!closed graph} \cite[Thm.~5.20]{Kato80} that
$$
\mathbbm{r}(T)=\{\lambda\in\mathbb C\,|\,T-\lambda\mathbbm{1}_X\,\text{is bijective}\,\}
$$
and, moreover, $\mathbbm{r}(T)$ is open so $\sigma(T)$ is closed \cite[Prop.~19.11]{MeiseVogt97en}. In this case the spectrum obviously decomposes as $\sigma(T)=\sigma_{\mathrm{p}}(T)\cup\sigma_{\mathrm{c}}(T)\cup\sigma_{\mathrm{r}}(T)$, cf.~also \cite[Ch.~XII.1, Lemma 3]{Dunford63}. In particular this is true for every bounded operator (Remark \ref{rem_densely_def_bounded}).
\item[(ii)] The spectrum behaves nicely under taking the dual or the adjoint \cite[Ch.~12, Thm.~11]{Bollobas99}: If $T\in\mathcal B(X)$ where $X$ is a Banach space then $\sigma(T')=\sigma(T)$. If $T\in\mathcal B(\mathcal H)$ where $\mathcal H$ is a Hilbert space then $\sigma(T^*)=(\sigma(T))^*$. 
\item[(iii)] For bounded operators $T$ on a complex Banach space the spectrum is non-empty and bounded by $\|T\|$ \cite[Ch.~12, Thm.~6]{Bollobas99}. For unbounded operators, however, the spectrum may be bounded, unbounded, empty or the whole complex plane, refer to \cite[Ch.~VII, Ex.~10.1]{Dunford63}.
\item[(iv)] If $T$ is a self-adjoint (possibly unbounded) operator on a complex Hilbert space $\mathcal H$ then every isolated point 
of $\sigma(T)$ is an eigenvalue of $T$, i.e.~is in $\sigma_{\mathrm{p}}(T)$ \cite[Coro.~5.11]{Schmuedgen12}. This is usually proven via functional calculus (which we are yet to develop) but for the statement itself one only needs to know what the spectrum is (and what self-adjoint operators are). This, in our eyes, justifies presenting it here already.
\item[(v)] If $T$ is a densely defined, symmetric operator---this of course includes the self-adjoint operators---then $T$ has non-empty spectrum\footnote{
To see this consider the following standard argument: Assume to the contrary that $\sigma(T)=\emptyset$ so in particular $0\in\mathbbm{r}(T)$ meaning $T$ is surjective. This by Lemma \ref{lemma_symmetric_closable} (v) means that $T$ is bijective and self-adjoint with bounded self-adjoint inverse $T^{-1}$. If we can show that $\sigma(T^{-1})=\{0\}$ then $\|T^{-1}\|=0$ by self-adjointness \cite[Ch.~12, Thm.~11 (c)]{Bollobas99} which would imply the obvious contradiction $T^{-1}=0$. Indeed for all $\lambda\in\mathbb C\setminus\{0\}$ and all $y\in D(T)$ one has
$
-Ty=(\frac{1}{\lambda}\mathbbm{1}-T)y-\frac{1}{\lambda}y
$
so
$$
-T\Big(\frac{1}{\lambda}\mathbbm{1}-T\Big)^{-1}x=\Big(\mathbbm{1}-\frac{1}{\lambda}\Big(\frac{1}{\lambda}\mathbbm{1}-T\Big)^{-1}\Big)x
$$
for all $x\in\mathcal H$; here we used $\frac{1}{\lambda}\in\mathbbm{r}(T)$, that is, bijectivity of $(\frac{1}{\lambda}\mathbbm{1}-T):D(T)\to\mathcal H$. With this one readily verifies that $\frac{1}{\lambda}T(\frac{1}{\lambda}\mathbbm{1}-T)^{-1}$ is a bounded inverse of $T^{-1}-\lambda\mathbbm{1}$, hence $\lambda\in\mathbbm{r}(T^{-1})$. Now $T^{-1}$ is bounded so the spectrum of $T^{-1}$ is non-empty by (iii); but the only point which could possibly lie within the spectrum now is $0$, hence $\sigma(T^{-1})=\{0\}$ and we are done.
}.

%
%
\end{itemize}
\end{remark}

Interestingly enough the spectrum is a footprint of some special classes of Hilbert space operators, and under further assumptions it even characterizes them.

\begin{lemma}\label{lemma_spectrum}
Let $\mathcal H$ be a complex Hilbert space and $T\in\mathcal B(\mathcal H)$ be given. The following statements hold.
\begin{itemize}
\item[(i)] If $T$ is self-adjoint then $\sigma(T)\subseteq \mathbb R$.
\item[(ii)] If $T\geq 0$ then $\sigma(T)\subseteq [0,\infty)$.
\item[(iii)] If $T$ is a projection then $\sigma(T)\subseteq \{0,1\}$.
\item[(iv)] If $T$ is unitary then $\sigma(T)\subseteq \{z\in\mathbb C\,|\,|z|=1\}$.
\end{itemize}
If $T$ is normal then the converses to (i)-(iv) are valid.
\end{lemma}
\begin{proof}
\cite[Thm.~3.2.14 \& 4.4.5]{Kadison83}
\end{proof}

Not only does this lemma show that the spectrum characterizes normal operators, but the connection between the two is even deeper: Given a normal matrix $A\in\mathbb C^{n\times n}$ there exists an orthonormal basis $(g_\lambda)_{\lambda\in\sigma(A)}$ of $\mathbb C^n$ such that $A=\sum_{\lambda\in\sigma(A)}\lambda |g_\lambda\rangle\langle g_\lambda|$ \cite[Thm.~2.5.4]{HJ1}. Then any continuous function $f:\mathbb C\to\mathbb C$ acts on such $A$ via $f(A)=\sum_{\lambda\in\sigma(A)}f(\lambda) |g_\lambda\rangle\langle g_\lambda|$. In infinite dimensions things become a bit more delicate as $\sigma(T)$ in general is different from $\sigma_{\mathrm{p}}(T)$ or $\overline{\sigma_{\mathrm{p}}(T)}$ in which case obtaining a sum of the form $T=\sum_{\lambda\in\sigma(T)}\lambda |g_\lambda\rangle\langle g_\lambda|$ is not possible. However if one passes over from (discrete) sums to (continuous) integrals one finds a similar result. Readers unfamiliar with spectral measures and spectral integrals we first relegate to Appendix \ref{sec:spectral_integrals}.

\begin{proposition}\label{prop_func_calc}
Let a complex Hilbert space $\mathcal H$ and a self-adjoint operator $T$ on $\mathcal H$ be given. Then there exists a unique spectral measure $E$ on the Borel-$\sigma$-algebra $\mathbb B(\mathbb R)$ such that
$$
T=\int_{-\infty}^\infty t\,dE(t)\,,
$$
i.e.~$\langle x,Ty\rangle=\int_{-\infty}^\infty t\,d\langle x,E(t)y\rangle$ for all $x\in \mathcal H$, $y\in D(T)$. Moreover, $E$ is concentrated on $\sigma(T)\subset\mathbb R$ in the sense that $E(\sigma(T))=\mathbbm{1}_{\mathcal H}$.
\end{proposition}
\begin{proof}
\cite[Thm.~5.7]{Schmuedgen12}
\end{proof}
This result (called ``spectral theorem''\index{theorem!spectral} or ``spectral decomposition''\index{spectral decomposition})---which by the way holds analogously for normal operators---enables functional calculus\index{functional calculus}, meaning we can make sense of expressions $f(A)$ (e.g., $e^{itA}$ or $e^{-tA}$) under certain assumptions. 

\begin{proposition}\label{prop_spectral_mapping}
Let $\mathcal H$ be a complex Hilbert space, $T$ be a self-adjoint operator on $\mathcal H$, and $f:\mathbb R\to\mathbb C$ be continuous. The following statements hold.
\begin{itemize}
\item[(i)] If $f|_{\sigma(T)}$ is bounded then $f(T)=\int_{-\infty}^\infty f(t)\,dE(t)$ defines a bounded linear operator on $\mathcal H$ with $\|f(T)\|=\sup_{t\in\sigma(T)} |f(t)|<\infty$.
\item[(ii)] $\sigma(f(T))=\overline{f(\sigma(T))}$. If $\sigma(T)$ is compact or $f$ has bounded support then one even has $\sigma(f(T))=f(\sigma(T))$.
\end{itemize}
\end{proposition}
\begin{proof}
(ii) is shown in \cite[Prop.~5.25]{Schmuedgen12} so we only prove (i). Because $f$ is continuous 
it is Borel measurable\footnote{
The argument is straightforward: Let topological spaces $(X,\tau_X),(Y,\tau_Y)$ and $f:X\to Y$ continuous be given. Define $Z:=\{S\in Y\,|\,f^{-1}(S)\in\mathbb B(X)\}$ where as usual $\mathbb B(X)$ is the smallest $\sigma$-algebra which contains $\tau_X$. By continuity $\tau_Y\subseteq Z$ and one readily verifies that $Z$ is a $\sigma$-algebra itself. As $\mathbb B(Y)$ is the smallest $\sigma$-algebra which contains $\tau_Y$ this implies $\mathbb B(Y)\subseteq Z$. Thus every pre-image of some $S\in\mathbb B(Y)$ under $f$ is in $\mathbb B(X)$, that is, $f$ is Borel measurable.
}.
By \cite[Thm.~5.9]{Schmuedgen12} this means $f(T)$ is bounded if and only if $f\in L^\infty(\mathbb R,E)$, i.e.~\label{symb_L_infty}
$$
\|f\|_\infty=\inf_{\substack{N\in\mathbb B(\mathbb R)\\ E(N)=0}}\sup_{t\in\mathbb R\setminus N}|f(t)|<\infty
$$
in which case $\|f(T)\|=\|f\|_\infty$. Now by \cite[Prop.~5.10]{Schmuedgen12} the support\footnote{
The support\index{spectral measure!support} of a spectral measure $E$ on some Borel-$\sigma$-algebra is the complement of the union of all open sets $N$ such that $E(N)=0$ \cite[Def.~4.3]{Schmuedgen12}.
}
of the spectral measure $E$ is equal to $\sigma(T)$. Thus $E(\mathbb R\setminus\sigma(T))=0$, and using continuity of $f$ we find
\begin{equation}
\|f(T)\|=\|f\|_\infty=\inf_{\substack{N\in\mathbb B(\mathbb R)\\ E(N)=0}}\sup_{t\in\mathbb R\setminus N}|f(t)|=\sup_{t\in\mathbb R\setminus (\mathbb R\setminus\sigma(T))}|f(t)|=\sup_{t\in\sigma(T)}|f(t)|<\infty\,.\tag*{\qedhere}
\end{equation}
\end{proof}
\noindent Statement (ii) of this proposition is usually known as the spectral mapping theorem\index{theorem!spectral mapping}.\smallskip

Let us consider an easy example which demonstrates the power of functional calculus: Given some self-adjoint operator $H$ we can simply plug it into the function $f(t):=e^{it}$. Because $|f(t)|=1$ for all $t\in\mathbb R$, $e^{iH}$ is bounded and, moreover, because $e^{iH}$ is normal (Prop.~\ref{prop_properties_func_calc}) Lemma \ref{lemma_spectrum} tells us that $e^{iH}$ even is a unitary operator. Interestingly enough this form characterizes unitary operators (cf.~proof of \cite[Thm.~12.37]{Rudin91}) and it is essential for describing the dynamics of closed quantum systems (Section \ref{ch:qu_dyn_sys}, Lemma \ref{lemma_stone}).

\subsection{Compact Operators and the Schatten Classes}\label{compact_wot_trace_class}

In the usual formulation of quantum mechanics\index{quantum mechanics} the state of an isolated system\index{isolated system}---characterized by some Hilbert space $\mathcal H$---is described by a pure state\footnote{
Many introductory books to quantum mechanics argue that the state of the system is described by a \textit{vector} $\psi$ which is problematic mainly for two reasons (among a few others): 
\begin{itemize}
\item[1.] Global phases would make a difference, that is, nominally $\psi$ and $e^{i\phi}\psi$ for any $\phi\in(0,2\pi)$ are different states; but such phases can never be detected because they vanish in the expectation value and thus in the measurement.
\item[2.] Such an approach denies us the possibility of describing states of non-isolated systems: The requirement $\langle\psi,\psi\rangle=1$ ensures the probabilistic interpretation of quantum mechanics so this leaves nothing physically reasonable in $\mathcal H$ which is more general than states of isolated systems.
\end{itemize}
Thus we prefer to work in the projective representation\index{projective representation} from the start.
} $|\psi\rangle\langle\psi|\in\mathcal B(\mathcal H)$ for some vector $\psi\in\mathcal H$ with $\langle\psi,\psi\rangle=1$. Now the expectation value of an observable $A\in\mathcal B(\mathcal H)$ in this state is given by $\langle A\rangle=\langle \psi,A\psi\rangle$.

But strictly speaking no physical system, with exception of the \textit{whole universe}, is isolated as there are always correlations between the system and its environment. Following this idea, that is, defining a world vector\index{world vector} and ``getting rid of'' the environmental degrees of freedom, one finds that a (for now: finite-dimensional) system is described by a matrix $\rho\in\mathbb C^{n\times n}$ which is positive semi-definite and has trace one\footnote{
This approach is carried out in detail in the Primas lectures \cite[Ch.~2]{AM11} (in german).
}. This matrix contains all the information necessary to compute expectation values for every possible observable and thus is indispensable when studying quantum systems. 

There of course is no guarantee that a system of interest satisfies $\operatorname{dim}(\mathcal H)<\infty$ which raises the question: Can the trace be generalized to $\mathcal B(\mathcal H)$ or, at least, a non-trivial subset? The path we will take in order to answer this, for which we refer to \cite[Ch.~15 \& 16]{MeiseVogt97en}, first leads us to the notion of a compact operator.

\begin{definition}\label{def_compact_op}
Let $X,Y$ be normed spaces and $T\in\mathcal L(X,Y)$ be given. We say $T$ is compact\index{operator!compact} if $T(\overline{B_1}(0))$ is relatively compact in $Y$, where $\overline{B_1}(0)=\{x\in X\,|\,\|x\|\leq 1\}$ denotes the closed unit ball as usual. The collection of all compact operators from $X$ to $Y$ will be denoted by $\mathcal K(X,Y)$, as well as $\mathcal K(X):=\mathcal K(X,X)$. 
\end{definition}

It is easy to see that the image of every bounded subset under a compact operator is relatively compact. If $Y$ even is a Banach space then some $T\in\mathcal L(X,Y)$ is compact if and only if for every bounded sequence $(x_n)_{n\in\mathbb N}$ in $X$ the image sequence $(Tx_n)_{n\in\mathbb N}$ has a convergent subsequence \cite[Coro.~4.10]{MeiseVogt97en}. Let us summarize a few of the key properties of compact operators now:

\begin{lemma}
Let $W,X,Y,Z$ be normed spaces. The following statements hold.
\begin{itemize}
\item[(i)] $\mathcal K(X,Y)$ is a closed linear subspace of $\mathcal B(X,Y)$.
\item[(ii)] Given $A\in\mathcal B(W,X)$, $T\in\mathcal K(X,Y)$, and $B\in\mathcal B(Y,Z)$ one has $A\circ T\circ B\in\mathcal K(W,Z)$.
\item[(iii)] $\mathcal F(X,Y)\subseteq\mathcal K(X,Y)$
\item[(iv)] Let $X$ be a Banach space. Given $T\in\mathcal K(X)$ one finds 
\begin{itemize}
\item[(a)] that $\operatorname{ker}(\mathbbm{1}-T)$ is finite-dimensional.
\item[(b)] that $\operatorname{im}	(\mathbbm{1}-T)$ is closed.
\item[(c)] that $\operatorname{im}	(\mathbbm{1}-T)$ has finite codimension, i.e.~$X\setminus (\operatorname{im}(\mathbbm{1}-T))$ is finite-dimensional.
\item[(d)] an $n\in\mathbb N$ such that $\operatorname{ker}\big((\mathbbm{1}-T)^n\big)=\operatorname{ker}\big((\mathbbm{1}-T)^{n+1}\big)$.
\end{itemize}
\end{itemize}
\end{lemma}
\begin{proof}
(i) \& (ii): \cite[Prop.~15.1]{MeiseVogt97en}. The proof given there does not use that any of the normed spaces are complete. (iii): Given $T\in\mathcal F(X,Y)$ we find that $\overline{T(\overline{B_1}(0))}$ is a closed and bounded subset of the finite-dimensional normed space $\operatorname{im}(T)\subseteq Y$, hence compact \cite[Ch.~4, Coro.~6]{Bollobas99}. (iv): \cite[Prop.~15.8 \& Lemma 15.9]{MeiseVogt97en}.
\end{proof}
Diving into the ideas in more detail would take up a substantial amount of time (and space, hence space-time). It suffices to know that these results pave the way to fully structure the spectrum of compact operators. The following is shown in \cite[Lemma 15.11 \& Prop.~15.12]{MeiseVogt97en}. 
\begin{proposition}\label{prop_compact_spectrum}
Let $X$ be an infinite-dimensional Banach space and $T\in\mathcal K(X)$ be given. The following statements hold.
\begin{itemize}
\item[(i)] Every $\lambda\in\sigma(T)\setminus\{0\}$ is an eigenvalue of $T$ with finite geometric and algebraic multiplicity, that is, 
$$
0<\operatorname{dim}\big(\operatorname{ker}(\lambda\mathbbm{1}-T)\big)\leq\operatorname{dim}\Big(\bigcup_{k\in\mathbb N}\operatorname{ker}\big((\lambda\mathbbm{1}-T)^k\big)\Big)<\infty\,.
$$
\item[(ii)] There exists a null sequence $(\lambda_n)_{n\in\mathbb N}$ such that $\sigma(T)=\{0\}\cup\{\lambda_n\,|\,n\in\mathbb N\}$.
\end{itemize}
Therefore the eigenvalue sequence\index{eigenvalue sequence} of $T$, obtained by arranging the (necessarily countably many) non-zero eigenvalues in decreasing order with respect to their absolute values
and each eigenvalue is repeated as many times as its algebraic multiplicity $\operatorname{dim}(\bigcup_{k\in\mathbb N}\operatorname{ker}((\lambda\mathbbm{1}-T)^k))$, is well-defined. If $\sigma(T)$ is finite then the sequence is filled up with zeros.
\end{proposition}
Often it is said that compact operators are the generalization of finite-dimensional operators due to the related spectral behaviour: in both cases the spectrum---aside from $\{0\}$---is fully discrete and consists only of eigenvalues with finite multiplicities. Even better, in the case of Hilbert spaces one gets an analogue of the singular value decomposition of a matrix:
\begin{proposition}[Schmidt representation]\label{prop_compact_SVD}\index{Schmidt representation}
Let $\mathcal H,\mathcal G$ be infinite-dimensional Hilbert spaces (as usual over the same field $\mathbb R$ or $\mathbb C$). For $T\in\mathcal K(\mathcal H,\mathcal G)$ there exists a unique decreasing null sequence $(s_j)_{j\in\mathbb N}$ in $[0,\infty)$, and orthonormal systems $(e_j)_{j\in\mathbb N}$ in $\mathcal H$ and $(f_j)_{j\in\mathbb N}$ in $\mathcal G$ such that
\begin{align}\label{eq:compact_SVD}
T=\sum\nolimits_{j=1}^\infty s_j|f_j\rangle\langle e_j|
\end{align}
where the series converges in operator norm.
\end{proposition}
\begin{proof}[Proof idea]
First one considers the compact, positive semi-definite operator $T^*T$ and shows via Prop.~\ref{prop_compact_spectrum} that it can be written as $\sum_{j=1}^\infty s_j^2|e_j\rangle\langle e_j|$ for some decreasing null sequence $(s_j)_{j\in\mathbb N}$ in $[0,\infty)$ and some orthonormal system $(e_j)_{j\in\mathbb N}$ in $\mathcal H$. Then one defines $f_j:=\frac{Te_j}{s_j}$ whenever $s_j>0$ which yields the orthonormal system in $\mathcal G$ we were looking for. This shows \eqref{eq:compact_SVD}. Finally for arbitrary $x\in\mathcal H$, $n\in\mathbb N$ one by the Pythagorean theorem as well as Bessel's inequality (Lemma \ref{lemma_pyth_thm} \& Prop.~\ref{prop_hilbert_space_basis}) finds
\begin{align*}
\Big\|Tx-\sum\nolimits_{j=1}^ns_j\langle e_j,x\rangle f_j\Big\|^2=\sum\nolimits_{j=n+1}^\infty s_j^2|\langle e_j,x\rangle|^2\leq \big(\|x\|\sup_{j>n}s_j\big)^2
\end{align*}
which by taking the supremum over all $x$ with $\|x\|=1$ concludes the proof.
\end{proof}
This of course is a characterization: a linear operator is compact if and only if it is of form \eqref{eq:compact_SVD}. However, in contrast to the finite-dimensional case \cite[Thm.~7.3.5]{HJ1} there is no diagonalization result for compact operators in infinite dimensions:
\begin{example}
Consider the weighted left shift $T=\sum_{n=1}^\infty \frac{1}{n}|e_n\rangle\langle e_{n+1}|$ with $(e_n)_{n\in\mathbb N}$ being the usual standard basis of $\ell^2(\mathbb N)$ so
$$
T=\begin{pmatrix} 0&1&0&0&\cdots\\ 0&0&\frac12&0&\cdots\\0&0&0&\frac13&\cdots\\\vdots&\vdots&\vdots&\vdots&\ddots \end{pmatrix}\,.
$$
Indeed $T$ is of Schmidt form meaning $T$ is compact. Now assume that there exist $U,V\in\mathcal U(\ell^2(\mathbb N))$ which diagonalize $T$, that is, $\langle e_j,UTVe_k\rangle=\frac{\delta_{jk}}{\sqrt{jk}}$ for all $j,k\in\mathbb N$. Then $V$ maps $(e_n)_{n\in\mathbb N}$ (orthonormal basis) to $(e_{n+1})_{n\in\mathbb N}$ (no orthonormal basis), so $V$ is not unitary by Lemma \ref{lemma_unitary_ONB}, a contradiction.
\end{example}
One can, however, unitarily diagonalize a compact operator if and only if said operator is normal:
\begin{theorem}\label{thm_compact_normal_unit_diag}
Let $\mathcal H$ be a complex Hilbert space and $T\in\mathcal K(\mathcal H)$ with corresponding eigenvalue sequence $(\lambda_n)_{n\in\mathbb N}$ be given. The following statements are equivalent.
\begin{itemize}
\item[(i)] $T$ is normal, that is, $T^*T=TT^*$.\index{operator!compact normal}
\item[(ii)] There exists an orthonormal sequence of eigenvectors $(e_n)_{n\in\mathbb N}$ associated to the eigenvalue sequence of $T$ such that $T=\sum_{n\in\mathbb N}\lambda_n|e_n\rangle\langle e_n|$.
\end{itemize}
\end{theorem}
\begin{proof}
``(i) $\Rightarrow$ (ii)'': \cite[Thm.~1.9.2]{Ringrose71}. ``(ii) $\Rightarrow$ (i)'': Direct computation.
\end{proof}

Imposing further structure on compact operators now can be done via the sequence of singular values $(s_j)_{j\in\mathbb N}\in c_{0}(\mathbb N)$. For example said sequence is in $c_{00}(\mathbb N)$\index{space!c00@$c_{00}(\mathbb N)$} if and only if the operator in question is finite-rank; thus density of $c_{00}(\mathbb N)$ in $(c_0(\mathbb N),\|\cdot\|_\infty)$\index{space!c0@$c_0(\mathbb N)$} transfers to the operator case meaning $\mathcal F(\mathcal H,\mathcal G)$ is dense in $(\mathcal K(\mathcal H,\mathcal G),\|\cdot\|)$ \cite[Coro.~16.4]{MeiseVogt97en}. Other sequence spaces from Example \ref{ex_ell_p_space} come into play as follows:
\begin{definition}\label{def_schatten_class}
Given infinite-dimensional Hilbert spaces $\mathcal H,\mathcal G$ as well as $p\in[1,\infty)$ one defines the Schatten-$p$ class\index{Schatten-p class@Schatten-$p$ class} to be
$$
\mathcal B^p(\mathcal H,\mathcal G):=\big\{A\in\mathcal K(\mathcal H,\mathcal G)\,\big|\,(s_j(A))_{j\in\mathbb N}\in\ell^p(\mathbb N)\big\}
$$
as well as $\mathcal B^p(\mathcal H):=\mathcal B^p(\mathcal H,\mathcal H)$. Then the Schatten-$p$ norm\index{Schatten-p norm@Schatten-$p$ norm}
$$
\|T\|_p:=\Big(\sum\nolimits_{j=1}^\infty s_j(T)^p\Big)^{\frac1p}
$$
is a norm on $\mathcal B^p(\mathcal H,\mathcal G)$. Moreover $\mathcal B^\infty(\mathcal H,\mathcal G):=\mathcal K(\mathcal H,\mathcal G)$ with $\|\cdot\|_\infty:=s_1(\cdot)=\|\cdot\|$ being the usual operator norm\footnote{
It is shown in \cite[Lemma 16.6]{MeiseVogt97en} that the largest singular value of any compact operator $T$ satisfies $s_1(T)=\|T\|$.
}. 
\end{definition}
\noindent Obviously one could define the Schatten classes\index{operator!Schatten class} for finite-dimensional Hilbert spaces but then $\mathcal B^p(\mathcal H,\mathcal G)=\mathcal B(\mathcal H,\mathcal G)=\mathcal L(\mathcal H,\mathcal G)$. Therefore we will drop the prefix ``infinite-dimensional'' and simply write ``Hilbert space'' for the remainder of this section.

It turns out that density if $c_{00}(\mathbb N)$ in $(\ell^p(\mathbb N),\|\cdot\|_p)$ as well as the ideal property of the compact operators transfer onto the Schatten classes. 
\begin{proposition}\label{prop_Schatten_p_properties}
Let Hilbert spaces $\mathcal H,\mathcal G,\mathcal F,\mathcal E$ and $1\leq p\leq q\leq \infty$ be given. The following statements hold.
\begin{itemize}
\item[(i)] The finite-rank operators $\mathcal F(\mathcal H,\mathcal G)$ are dense in the Banach space $(\mathcal B^p(\mathcal H,\mathcal G),\|\cdot\|_p)$. 
\item[(ii)] For all $A\in\mathcal B(\mathcal G,\mathcal H)$, $B\in\mathcal B(\mathcal E,\mathcal F)$, $T\in\mathcal B^p(\mathcal F,\mathcal G)$ one has $ATB\in\mathcal B^p(\mathcal E,\mathcal H)$ as well as $ \|ATB\|_p\leq\|A\|\|T\|_p\|B\| $.
\item[(iii)] One has $\mathcal B^p(\mathcal H,\mathcal G)\subseteq\mathcal B^q(\mathcal H,\mathcal G)$ and $\|T\|_p\geq\|T\|_q$ for all $T\in\mathcal B^p(\mathcal H,\mathcal G)$.
\end{itemize}
\end{proposition}
\begin{proof}
(i): If $T\in\mathcal B^p(\mathcal H,\mathcal G)$ then \eqref{eq:compact_SVD} obviously converges in the $p$-norm. The fact that the Schatten classes are Banach spaces is shown for example in \cite[Coro.~16.34]{MeiseVogt97en}. (ii): \cite[Lemma 16.6 \& 16.7]{MeiseVogt97en}. (iii): Follows from the corresponding statement for the $\ell^p$-spaces which is readily verified.
\end{proof}

If $\mathcal H=\mathcal G$ the Schatten-$p$ class has an interesting characterization which will lead us to the trace of infinite-dimensional operators.

\begin{lemma}\label{lemma_schatten_trace_p}
Let $\mathcal H$ be a complex Hilbert space and $T\in\mathcal B(\mathcal H)$ as well as $p\in[1,\infty)$ be given. The following are equivalent.
\begin{itemize}
\item[(i)] $T\in\mathcal B^p(\mathcal H)$
\item[(ii)] For all orthonormal systems $(f_i)_{i\in I}$ in $\mathcal H$ one has $\sum_{i\in I}|\langle f_i,Tf_i\rangle|^p<\infty$.
\item[(iii)] For all orthonormal bases $(f_i)_{i\in I}$ of $\mathcal H$ one has $\sum_{i\in I}|\langle f_i,Tf_i\rangle|^p<\infty$.
\end{itemize}
\end{lemma}
\begin{proof}
(i) $\Leftrightarrow$ (ii): Let $\mathscr C^p(\mathcal H)$ denote the set of all bounded operators on $\mathcal H$ which satisfy (ii)---this notation then matches \cite[Def.~2.1]{Ringrose71}. Indeed Ringrose shows that $(\mathscr C^p(\mathcal H),\|\cdot\|_p)$ is a Banach space and $\overline{\mathcal F(\mathcal H)}^{\|\cdot\|_p}=\mathscr C^p(\mathcal H)$ \cite[Thm.~2.3.8]{Ringrose71}. But by Prop.~\ref{prop_Schatten_p_properties} (i) this means
$
\mathcal B^p(\mathcal H)=\overline{\mathcal F(\mathcal H)}^{\|\cdot\|_p}=\mathscr C^p(\mathcal H)\,.
$
(ii) $\Rightarrow$ (iii): Obvious as every orthonormal basis is an orthonormal system. (iii) $\Rightarrow$ (ii): Given an orthonormal system $(f_i)_{i\in I}$ we by Prop.~\ref{prop_hilbert_space_basis} (iii) can extend it to an orthonormal basis $(f_j)_{j\in J}$, $I\subseteq J$ of $\mathcal H$ so $\sum_{i\in I}|\langle f_i,Tf_i\rangle|^p\leq \sum_{j\in J}|\langle f_j,Tf_j\rangle|^p<\infty$.
\end{proof}
\noindent If $\mathcal H$ is a real Hilbert space then this Lemma is not valid anymore: A counterexample---the idea of which is very similar to Remark \ref{rem_real_complex_hs} (i)---is given in \cite[Ex.~16.19]{MeiseVogt97en}.\smallskip

We have to be careful to not let this turn into a circular argument: For the proof of (i) $\Leftrightarrow$ (ii) Ringrose defined and used the trace. Therefore we, as said before, will use this result \textit{only} as an inspiration. Indeed for $p=1$ the expression $\sum_{i\in I}|\langle f_i,Tf_i\rangle|<\infty$ from Lemma \ref{lemma_schatten_trace_p} looks like the trace, or rather a version of it where absolute convergence is checked. 

\begin{lemma}\label{lemma_schatten_1_summable}
For all $T\in\mathcal B^1(\mathcal H)$ and every orthonormal basis $(e_i)_{i\in I}$ of $\mathcal H$, $(\langle e_i,Te_i\rangle)_{i\in I}$ is summable with
\begin{equation}\label{eq:lemma_schatten_1_summable_1}
\Big|\sum\nolimits_{i\in I}\langle e_i,Te_i\rangle\Big|\leq\sum\nolimits_{i\in I}|\langle e_i,Te_i\rangle|\leq\|T\|_1<\infty\,.
\end{equation}
\end{lemma}
\begin{proof}
For this proof we follow \cite[Prop.~16.16]{MeiseVogt97en}. By Proposition \ref{prop_compact_SVD} we can write $T=\sum_{j=1}^\infty s_j|f_j\rangle\langle g_j|$ so for an arbitrary orthonormal basis $(e_i)_{i\in I}$ of $\mathcal H$ we by the triangle inequality, the Cauchy-Schwarz inequality for the Hilbert space $\ell^2(I)$ \cite[Ex.~12.11]{MeiseVogt97en}, and Parseval's equation get
\begin{align*}
\sum_{i\in I}|\langle e_i,Te_i\rangle|&\leq \sum_{j=1}^\infty s_j \sum_{i\in I} |\langle e_i,f_j\rangle\langle g_j,e_i\rangle|\\
&\leq \sum_{j=1}^\infty s_j \underbrace{\Big(\sum_{i\in I} |\langle e_i,f_j\rangle|^2\Big)^{1/2}}_{=\|f_j\|=1}\underbrace{\Big(\sum_{i\in I}\langle g_j,e_i\rangle|^2\Big)^{1/2}}_{=\|g_j\|=1}=\sum_{j=1}^\infty s_j=\|T\|_1<\infty\,.
\end{align*}
Therefore $(\langle e_i,Te_i\rangle)_{i\in I}$ is summable and \eqref{eq:lemma_schatten_1_summable_1} holds \cite[Lemma 1.2.5]{Ringrose71}.
\end{proof}

Similar to the proof presented just now, one sees that the expression $\sum_{i\in I}\langle e_i,Te_i\rangle$ for any $T\in\mathcal B^1(\mathcal H)$ does not depend on the chosen orthonormal basis $(e_i)_{i\in I}$ of $\mathcal H$ \cite[Prop.~16.16 (2)]{MeiseVogt97en}. Thus the following definition is meaningful:
\begin{definition}\label{def_trace}
For $T\in\mathcal B^1(\mathcal H)$ define the trace of $T$ via $\operatorname{tr}(T):=\sum_{i\in I}\langle e_i,Te_i\rangle$ where $(e_i)_{i\in I}$ is an arbitrary orthonormal basis of $\mathcal H$.\index{trace}
\end{definition}
Actually $\mathcal B^1(\mathcal H)$ is the largest subset of the bounded operators which allows for a reasonable definition of the trace \cite[Rem.~3.4.6]{Pedersen89}. This is why the Schatten-$1$-class $\mathcal B^1(\mathcal H,\mathcal G)$ is usually called the \textit{trace class}\index{trace class} and the respective norm $\|\cdot\|_1$ is called \textit{trace norm}\index{trace norm}. 
\begin{remark}
There is an even deeper meaning to this terminology when looking at it topologically. Using \eqref{eq:lemma_schatten_1_summable_1} one immediately sees that $\operatorname{tr}:(\mathcal B^1(\mathcal H),\|\cdot\|_1)\to\mathbb C$ is a continuous linear map. But the trace class is a subset of the bounded operators so one could also use the usual topologies from $\mathcal B(\mathcal H)$. However the trace $\operatorname{tr}:(\mathcal B^1(\mathcal H),\tau)\to\mathbb C$ becomes discontinuous when choosing $\tau=\topn,\tops,\topw$, that is, the operator norm, the strong operator topology, or the weak operator topology.

To see this define $T_n:\ell^2(\mathbb N)\to \ell^2(\mathbb N)$ for any $n\in\mathbb N$ via $T_n(e_k):=\frac1ne_k$ for all $k=1,\ldots,n$, and $T_n(e_k)=0$ for all $k>n$ as well as its linear extension onto all of $\ell^2(\mathbb N)$. One readily verifies that $T_n\in\mathcal F(\ell^2(\mathbb N))\subset\mathcal B^1(\ell^2(\mathbb N))$ has operator norm $\frac1n$ and trace $\operatorname{tr}(T_n)=\sum_{k=1}^n\frac1n=1$. Therefore $\|T_n\|=\frac1n\to0$ so $T_n$ converges to $0$ in operator norm, but $\lim_{n\to\infty}\operatorname{tr}(T_n)=1\neq 0=\operatorname{tr}(0)$. Because $\topw\subseteq\tops\subseteq\topn$ (Prop.~\ref{prop_strong_weak_op_top} (iv)) this example is valid for the weaker topologies, as well.
\end{remark}
If $\mathcal H$ is a separable, complex Hilbert space then $ \mathcal B^1(\mathcal H)=\{B\in\mathcal B(\mathcal H)\,|\,\operatorname{tr}(\sqrt{B^*B})<\infty\}$, cf.~\cite[Thm.~VI.21]{ReedSimonI}. Further elementary properties of trace, trace norm, and the Schatten norms in general are summarized in the following lemma:
\begin{lemma}\label{lemma_schatten_prop_pq}
The following statements hold. 
\begin{itemize}
\item[(i)] $\operatorname{tr}(T^*)=\overline{\operatorname{tr}(T)}$ for all $T\in\mathcal B^1(\mathcal H)$
\item[(ii)] If $A\in\mathcal K(\mathcal H)$ and $B\in\mathcal B(\mathcal H)$ satisfy $AB,BA\in\mathcal B^1(\mathcal H)$ then $\operatorname{tr}(AB)=\operatorname{tr}(BA)$. 
\item[(iii)] Given $T\in\mathcal B^p(\mathcal H)$, $S\in\mathcal B^	q(\mathcal H)$ with $p,q\in[1,\infty]$ conjugate, i.e.~$\frac{1}{p}+\frac1q=1$, one has $TS,ST\in\mathcal B^1(\mathcal H)$ and $|\operatorname{tr}(TS)|\leq \|TS\|_1\leq\|T\|_p\|S\|_q$. This result remains valid if $\mathcal B^\infty(\mathcal H)$ is replaced by $\mathcal B(\mathcal H)$.
\end{itemize}
\end{lemma}
\begin{proof}
(i) \& (ii): \cite[Lemma 16.20]{MeiseVogt97en}. (iii): \cite[Ch.~XI.9, Lemma 14]{Dunford63} \& Lemma \ref{lemma_schatten_1_summable}. The additional statement was already shown in Prop.~\ref{prop_Schatten_p_properties}.
\end{proof}

After motivating things from a physics perspective, we finally made it all the way from compact operators to the trace on infinite-dimensional Hilbert spaces. To wrap up this chapter let us bring together the Schatten classes and some topological notions such as separability, dual spaces, and the like. In fact the Schatten classes nicely enhance the strong operator topology:
\begin{lemma}\label{lemma_schatten_p_approx}
Let $p\in[1,\infty]$, $T \in \mathcal B^p(\mathcal H)$, and $(B_n)_{n\in\mathbb N}$ be a sequence in $\mathcal B(\mathcal H)$ which converges strongly to $B\in\mathcal B(\mathcal H)$. Then one has
$B_n T \to BT$, $TB_n^* \to TB^*$, and $B_nTB_n^* \to BTB^*$ in the $p$-norm for $n \to \infty$
with respect to the norm $\|\cdot\|_p$.
\end{lemma}
\begin{proof}
\cite[Prop.~2.1]{Widom76}
\end{proof}

Thus the following result is immediate.
\begin{corollary}\label{coro_schatten_sep}
Let $p\in[1,\infty]$ be given. Then $(\mathcal B^p(\mathcal H,\mathcal G),\|\cdot\|_p)$ is separable if and only if $\mathcal H,\mathcal G$ are separable. 
\end{corollary}
\begin{proof}
``$\Rightarrow$'': Copy (\textit{Step 3} of) the proof of Prop.~\ref{prop_bounded_op_norm_sep}.
``$\Leftarrow$'': W.l.o.g.~let $\mathcal H,\mathcal G$ be infinite-dimensional separable Hilbert spaces so one finds countable orthonormal bases $(e_n)_{n\in\mathbb N}$, $(g_n)_{n\in\mathbb N}$ of $\mathcal H,\mathcal G$, respectively (Prop.~\ref{prop_hilbert_space_basis} (iv)). By Lemma \ref{lemma_approx_strong_top} we know that the corresponding projections $\Pi_n^{\mathcal H},\Pi_n^{\mathcal G}$ converge to $\mathbbm{1}_{\mathcal H},\mathbbm{1}_{\mathcal G}$ in the strong operator topology, respectively. Thus, given some $T\in\mathcal B^p(\mathcal H,\mathcal G)$, Lemma \ref{lemma_schatten_p_approx} (i) yields $\|\Pi_n^{\mathcal H}T\Pi_n^{\mathcal G}-T\|_p\to 0$ as $n\to\infty$. Because $\Pi_n^{\mathcal H}T\Pi_n^{\mathcal G}\in\operatorname{span}\{|g_j\rangle\langle e_i|\,|\,i,j\in\mathbb N\}$ for all $n\in\mathbb N$ and all operators $T$, separability follows from Lemma \ref{lemma_separable_linear_span}.
\end{proof}

The final parallel we draw between sequence spaces and their ``big brothers'', the Schatten classes, pursues the dual space considerations from Example \ref{ex_ell_p_space_2}. For example the fact that $(c_0(\mathbb N))^*$ is isometrically isomorphic to $\ell^1(\mathbb N)$ should in light of the singular value sequence become: ``The dual space of the compact operators can be identified with the trace class''. This idea actually works out and leads to the next proposition.

\begin{proposition}\label{prop_schatten_dual}
Let Hilbert spaces $\mathcal G,\mathcal H$ be given. The following statements hold.
\begin{itemize}
\item[(i)] The dual space $(\mathcal K(\mathcal H,\mathcal G))^*$ is isometrically isomorphic to $\mathcal B^1(\mathcal G,\mathcal H)$ by means of the map $\varphi:\mathcal B^1(\mathcal G,\mathcal H)\to (\mathcal K(\mathcal H,\mathcal G))^*$, $T\mapsto\varphi_T$ where
$$
\varphi_T(A)=\operatorname{tr}(TA)\qquad\text{ for all }A\in \mathcal K(\mathcal H,\mathcal G)\,.
$$
\item[(ii)] The dual space $(\mathcal B^1(\mathcal G,\mathcal H))^*$ is isometrically isomorphic to $\mathcal B(\mathcal H,\mathcal G)$ by means of the map $\psi:\mathcal B(\mathcal H,\mathcal G)\to (\mathcal B^1(\mathcal G,\mathcal H))^*$, $B\mapsto\psi_B$ where
$$
\psi_B(T)=\operatorname{tr}(BT)\qquad\text{ for all }T\in \mathcal B^1(\mathcal G,\mathcal H)\,.
$$
\item[(iii)] Let $p,q\in(1,\infty)$ be conjugate, that is, $\frac{1}{p}+\frac1q=1$. Then $(\mathcal B^p(\mathcal H,\mathcal G))^*$ is isometrically isomorphic to $\mathcal B^q(\mathcal G,\mathcal H)$ by means of the map $\phi:\mathcal B^q(\mathcal G,\mathcal H)\to (\mathcal B^p(\mathcal H,\mathcal G))^*$, $T\mapsto\phi_T$ where
$$
\phi_T(S)=\operatorname{tr}(TS)\qquad\text{ for all }S\in \mathcal B^p(\mathcal H,\mathcal G)\,.
$$
\end{itemize}
In particular $\mathcal B^p(\mathcal H,\mathcal G)$ is reflexive for all $p\in(1,\infty)$ whereas---if $\mathcal G,\mathcal H$ are infinite-dimensional---then $\mathcal K(\mathcal H,\mathcal G)$, $\mathcal B^1(\mathcal G,\mathcal H)$ and $\mathcal B(\mathcal H,\mathcal G)$ are not reflexive.
\end{proposition}
\begin{proof}
(i) \& (ii): \cite[Prop.~16.24 \& 16.26]{MeiseVogt97en}. (iii): \cite[Ch.~V, Thm.~15]{Schatten50}. Then the additional statement follows at once from (i)--(iii).
\end{proof}

The most important duality for quantum physics is the one between the trace class (which contains the quantum states) and the bounded operators (which contains the observables). As we shall see in Chapter \ref{sec:quantum_channels} this will establish the duality between the Schr{\"o}dinger and the Heisenberg picture.

To do this we consider the bounded operators as a normed space---and not as an operator space as done in Chapter \ref{sec_top_B_X_Y}---so because $\mathcal B(\mathcal H,\mathcal G)$ is the dual space of the trace class up to isometric isomorphism we can equip it with a weak*-topology:

\begin{definition}\label{def_uw_topology}
Given Hilbert spaces $\mathcal G,\mathcal H$ the ultraweak topology\index{topology!ultraweak} $\topuw$ on $\mathcal B(\mathcal H,\mathcal G)$ is the weak*-topology of $(\mathcal B^1(\mathcal G,\mathcal H))^*$ under the map $\psi$ from Prop.~\ref{prop_schatten_dual} (ii).
\end{definition}

Thus by Coro.~\ref{coro_net_weak_star} a net $(B_i)_{i\in I}$ in $\mathcal B(\mathcal H,\mathcal G)$ converges to $B\in\mathcal B(\mathcal H,\mathcal G)$ in $\topuw$ if and only if $\operatorname{tr}(B_iT)\to\operatorname{tr}(BT)$ for all $T\in\mathcal B^1(\mathcal G,\mathcal H)$. This immediately shows $\topw\subseteq\topuw$ because choosing $T_{xy}:=|y\rangle\langle x|\in\mathcal B^1(\mathcal G,\mathcal H)$ for any $x\in\mathcal G$, $y\in\mathcal H$ by continuity of the inner product yields
\begin{align*}
\operatorname{tr}(B_iT_{xy})=\sum_{j\in I}\langle e_j,B_iy\rangle\langle x,e_j\rangle=\Big\langle\sum_{j\in I}\langle e_j,x\rangle e_j,B_iy\Big\rangle=\langle x,B_iy\rangle\to\langle x,By\rangle=\operatorname{tr}(BT_{xy})\,.
\end{align*}
Without much effort we can now adjust Coro.~\ref{coro_schatten_sep} to the ultraweak topology and obtain another separability result:

\begin{corollary}
Let $\mathcal G,\mathcal H$ be separable Hilbert spaces. Then $\mathcal B(\mathcal H,\mathcal G)$ with the ultraweak topology is separable.
\end{corollary}
\begin{proof}
Because $\mathcal G,\mathcal H$ are separable $(\mathcal B^1(\mathcal G,\mathcal H),\|\cdot\|_1)$ is separable (Coro.~\ref{coro_schatten_sep}) so $(\mathcal B^1(\mathcal G,\mathcal H))^*$ is separable in the weak*-topology by Lemma \ref{lemma_dual_weakstar_sep}. Therefore one finds a countable weak*-dense subset $(f_i)_{i\in\mathbb N}\subseteq(\mathcal B^1(\mathcal G,\mathcal H))^*$ so for all $B\in\mathcal B(\mathcal H,\mathcal G)$ there exists a subsequence $(f_{i_n})_{n\in\mathbb N}$ such that
$
\operatorname{tr}(\psi^{-1}(f_{i_n})T)=f_{i_n}(T)\to \psi_B(T)=\operatorname{tr}(BT)
$.
Thus by the above characterization of the ultraweak topology the countable subset $(\psi^{-1}(f_i))_{i\in\mathbb N}\subseteq\mathcal B(\mathcal H,\mathcal G)$ is $\topuw$-dense.
\end{proof}
Let us conclude with an example 
which brings together functional calculus, compact operators, and the Schatten classes. 
\begin{example}\label{ex_func_calc_schatten}
Given a monotonically increasing sequence $(E_n)_{n\in\mathbb N}\subset\mathbb R$ which tends to infinity as $n\to\infty$, as well as an orthonormal basis $(g_n)_{n\in\mathbb N}$ of $\ell^2(\mathbb N)$ consider the linear operator $H=\sum_{n=1}^\infty E_n|g_n\rangle\langle g_n|$ defined on $D(H)=\{x\in\ell^2(\mathbb N)\,|\,\sum_{n=1}^\infty E_k^2|\langle g_k,x\rangle|^2<\infty\}\,$\footnote{
Note that operators of this form are sometimes called ``discrete''\index{operator!discrete}, cf., e.g., \cite{Shirokov19}.
}.
Picking up the idea from Rem.~\ref{rem_I_plus_a_a_star} one finds $H=-E_0+a_E^*a_E$ with a weighted lowering operator $a_E((x_j)_{j\in\mathbb N})=(\sqrt{E_{j+1}-E_1}\,x_{j+1})_{j\in\mathbb N}$. Thus---as $a_E$ is densely defined and closed (as is seen similarly to Appendix \ref{appendix_lowering_op_proof})---$H$, unsurprisingly, is a self-adjoint operator with spectrum $\sigma(H)=\{E_n\,|\,n\in\mathbb N\}$. 

Now introduce a real parameter $T>0$ so the continuous function $t\mapsto e^{-t/T}$ is bounded on $\sigma(H)$ due to $\sup_{n\in\mathbb N}e^{-E_n/T}=e^{-E_1/T}$. This enables using functional calculus, that is, the normal operator $e^{-H/T}\in\mathcal B(\mathcal H)$ has spectrum $\{e^{-E_n/T}\,|\,n\in\mathbb N\}$ (Prop.~\ref{prop_spectral_mapping}) and is even positive semi-definite (Lemma \ref{lemma_spectrum}). Actually because $H$ is so nicely structured we find the explicit expression $e^{-H/T}=\sum_{n=1}^\infty e^{-E_n/T} |g_n\rangle\langle g_n|$ (Example \ref{ex_sep_spectr_integral}); hence $e^{-H/T}$ is even compact as we found a Schmidt representation. By definition of the Schatten classes, $e^{-H/T}\in\mathcal B^p(\mathcal H)$ for some $p\in[1,\infty)$ holds if and only if $(e^{-E_n/T})_{n\in\mathbb N}\in\ell^p(\mathbb N)$. 
\end{example}

\section{Quantum Channels}\label{sec:quantum_channels}

As explained at the start of Section \ref{compact_wot_trace_class} knowing the state of a quantum system enables computing the expectation value of the system under any observable, as well as probabilities of certain measurement outcomes. This state can be described by a positive semi-definite trace class operator with unit trace, denoted by \label{symb_q_states}\index{quantum state}\index{density operator|see{quantum state}}
\begin{equation*}
\mathbb D(\mathcal H):=\{\rho\in\mathcal B^1(\mathcal H)\,|\,\rho\geq 0\,\text{ and }\operatorname{tr}(\rho)=1\}
\end{equation*}
with $\mathcal H$ being the Hilbert space which describes the system. As we are diving into the foundations of quantum physics now we will---motivated by Remark \ref{rem_real_complex_hs}---assume \textbf{here and henceforth} that all Hilbert spaces are complex.

\begin{remark}[Gibbs state]\label{rem_gibbs}\index{Gibbs state}
The temperature $T > 0$ given as a macroscopic parameter of a bath relates to the
equilibrium state $\rho_\textsf{Gibbs}$ (henceforth called \emph{Gibbs state})\label{symb_gibbs_state}
of an \mbox{$n$-level} quantum system with Hamiltonian $H\in\mathbb C^{n\times n}$ (i.e.~$H$ is a Hermitian matrix) once the system is `\/opened\/' by coupling it to the bath and letting it
equilibrate.
In equilibrium, the quantum system is assumed to adopt the bath temperature in the sense
that\index{quantum system!equilibrium} $\rho_\textsf{Gibbs}$ exhibits the same eigenbasis as $H$ and its corresponding eigenvalues can be
interpreted as populations of the energy levels of $H$ following the Boltzmann distribution: 
$$ 
\frac{\lambda(\rho_\textsf{Gibbs})_{k}} {\lambda(\rho_\textsf{Gibbs})_{k'}} = \frac{e^{-E_k/T} } {e^{-E_{k'}/T} }
$$
for $k,k'=1,2,\dots, n$, and $T>0$. This obviously
leads to 
$$
\rho_\textsf{Gibbs}=\frac{e^{-H/T}}{\operatorname{tr}(e^{-H/T})}
$$
(see, e.g., \cite{AlickiLendi07}). Note that different $T > 0$ and different $H$ may lead to the same Gibbs state.
To port this to infinite dimensions let an orthonormal basis $(g_n)_{n\in\mathbb N}$ of a separable Hilbert space $\mathcal H$ as well as a monotonically increasing real sequence $(E_n)_{n\in\mathbb N}$ be given such that $(e^{-E_n/T})_{n\in\mathbb N}\in\ell^1(\mathbb N)$ for all $T>0$. Then the operator $H=\sum_{n=1}^\infty E_n|g_n\rangle\langle g_n|$ gives rise to the positive semi-definite trace-class operator $e^{-H/T}=\sum_{n=1}^\infty e^{-E_n/T} |g_n\rangle\langle g_n|$ (Ex.~\ref{ex_func_calc_schatten}) so
$$
\rho_\textsf{Gibbs}=\frac{e^{-H/T}}{\operatorname{tr}(e^{-H/T})}=\sum_{n=1}^\infty \Big(\frac{e^{-E_n/T}}{\sum_{n=1}^\infty e^{-E_n/T}} \Big)|g_n\rangle\langle g_n|\in\mathbb D(\mathcal H)
$$
is well-defined for all $T>0$. 
\end{remark}
\begin{example}
To illustrate this let us compute the Gibbs state of the quantum harmonic oscillator\index{quantum harmonic oscillator} (Ex.~\ref{ex_quantum_harm_osc}) so $E_n=2n-1$ in Rem.~\ref{rem_gibbs}, up to a positive global constant which we can neglect. An easy computation yields $\rho_\textsf{Gibbs}=2\sinh(\frac1T)\operatorname{diag}((e^{-(2n-1)/T})_{n\in\mathbb N})$ for any $T>0$. Now if one cools the system to $T=0$ then the Gibbs state becomes the ground state:
\begin{align*}
\lim_{T\to 0^+}\|e^{-H/T}-|g_1\rangle\langle g_1|\,\|_1&=\lim_{T\to 0^+}\Big(\Big|2\sinh(\tfrac1T)e^{-1/T}-1\Big|+2\sinh(\tfrac1T)\sum\nolimits_{n=2}^\infty e^{-(2n-1)/T}\Big)\\
&= 2\lim_{T\to 0^+} \big(1-2\sinh(\tfrac1T)e^{-1/T}\big) = 2\lim_{T\to 0^+}e^{-2/T}=0\,.
\end{align*}
\end{example}
Let us start with some fundamental properties of the set of all quantum states. 
\begin{lemma}\label{lemma_pure_extreme_general}
The set of states $\mathbb D(\mathcal H)$ is convex, closed and bounded. Moreover
\begin{equation}\label{eq:D_conv_pure}
\mathbb D(\mathcal H)=\overline{ \operatorname{conv}\big\{ |\psi\rangle\langle\psi| \,\big|\, \psi\in\mathcal H, \langle\psi,\psi\rangle=1 \big\} }
\end{equation}
with the closure being taken w.r.t.~the trace norm, and the rank-one projections (called pure states\index{pure state}) are precisely the extreme points of $\mathbb D(\mathcal H)$. 
\end{lemma}
\begin{proof}
Convexity is evident. Boundedness holds because $\|\rho\|_1=\operatorname{tr}(\rho)=1$ using $\rho\geq 0$. For closedness consider a sequence in $\mathbb D(\mathcal H)$ which converges in trace norm to some $A\in\mathcal B^1(\mathcal H)$. Then in particular it converges in trace ($|\operatorname{tr}(\cdot)|\leq\|\cdot\|_1$) and weakly ($|\langle x,(\cdot)x\rangle|\leq \|\cdot\|\|x\|^2\leq \|\cdot\|_1\|x\|^2$ for all $x\in\mathcal H$) hence $A\in\mathbb D(\mathcal H)$. 

To prove \eqref{eq:D_conv_pure} note that obviously $\operatorname{conv}\{ |\psi\rangle\langle\psi| \,|\, \psi\in\mathcal H, \langle\psi,\psi\rangle=1 \}\subseteq\mathbb D(\mathcal H)$ so this still holds after taking the closure. On the other hand because any $\rho\in\mathbb D(\mathcal H)$ is positive semi-definite---so in particular self-adjoint---it can be written as $\rho=\sum_{n\in\mathbb N}\lambda_n|e_n\rangle\langle e_n|$ for some null sequence $(\lambda_n)_{n\in\mathbb N}$ in $[0,1]$ which sums up to one and an associated sequence of eigenvectors $(e_n)_{n\in\mathbb N}$ (Thm.~\ref{thm_compact_normal_unit_diag}). Because $\rho$ is trace class, meaning $(\lambda_n)_{n\in\mathbb N}\in\ell^1(\mathbb N)$, the sum $\sum_{n\in\mathbb N}\lambda_n|e_n\rangle\langle e_n|$ converges in trace norm so $\overline{ \operatorname{conv}\big\{ |\psi\rangle\langle\psi| \,\big|\, \psi\in\mathcal H, \langle\psi,\psi\rangle=1 \big\} }\supseteq\mathbb D(\mathcal H)$.
 
Finally the statement regarding the extreme points is shown, e.g., in \cite[Thm.~2.3]{Holevo12} \footnote{Although said theorem considers finite-dimensional Hilbert spaces the proof carries over to infinite dimensions without further change when allowing for $d=\infty$.}.
\end{proof}

Taking the closure obviously is only necessary if $\mathcal H$ is of infinite dimension so, also using the Heine-Borel theorem \cite[Thm.~2.41]{Rudin76}\index{theorem!Heine-Borel}, we get the following refinement for finite-dimensional systems:
\begin{corollary}\label{coro_states_comp}
For all $n\in\mathbb N$ the set $\mathbb D(\mathbb C^n)=\operatorname{conv}\big\{ |\psi\rangle\langle\psi| \,\big|\, \psi\in\mathbb C^n, \langle\psi,\psi\rangle=1 \big\}$ is convex and compact, and has precisely the pure states as extreme points.
\end{corollary}

\begin{remark}
The fact that every state has a canonical decomposition into pure states allows for a statistical interpretation: Consider an arbitrary state $\rho=\sum_i\lambda_i|e_i\rangle\langle e_i|$ and a random vector $\psi$ which is equal to each of the $e_i$'s with probability $\lambda_i$. It turns out that measuring any observable with the state $\rho$ or with the random state $|\psi\rangle\langle\psi|$ gives the same outcome with the same probabilities. A proof can be found in \cite[Ch.~5.3.4]{Attal5}. Be aware that this ``convex decomposition of a mixed state into pure states is highly non-unique'' \cite[p.~56 ff.]{Heinosaari12}.
\end{remark}
Of course every quantum system undergoes changes, let it be natural or man-made, so we need a formalism to describe such changes for which we orient ourselves towards \cite[Ch.~4]{Heinosaari12}.

\subsection{Positive and Completely Positive Maps}
When assuming a model which splits an experiment into preparation (creation of a quantum state) and measurement (converting a state into a measurement outcome) one might want to place an operation in between these two which modifies the state of the system. This immediately leads to the following requirements on such operations:
\begin{itemize}
\item \textit{Linearity:} This is forced by consistency with the statistical interpretation of quantum mechanics. Indeed, every non-pure quantum state has uncountably many different convex combinations into pure states \cite[p.~56 ff.]{Heinosaari12} all of which are statistically indistinguishable. Therefore an operation on states has to preserve convex combinations which due to $\operatorname{span}(\mathbb D(\mathcal H))=\mathcal B^1(\mathcal H)$ translates into linearity on the trace class.
\item \textit{Positivity- \& trace-preservation:} As positive semi-definiteness and unit trace characterize quantum states these properties have to be preserved.\index{trace-preserving}
\end{itemize}
Indeed a linear map $T:\mathcal B^1(\mathcal H)\to\mathcal B^1(\mathcal G)$ with $\mathcal H,\mathcal G$ arbitrary complex Hilbert spaces is said to be positivity-preserving\index{positivity-preserving} (for short: \textit{positive}\index{positive|see{positivity-preserving}})\footnote{
Clearly this definition makes sense for maps between bounded operators instead of the trace class, as well.
} if $T(A)\geq 0$ for all $A\geq 0$. Although we will see in a bit that physically valid operations on states have to satisfy an even stronger condition it is still advisable to investigate the notion of positivity first. 

First and foremost, positivity automatically implies boundedness and thus continuity. While in finite dimensions this is trivial as every linear map is automatically continuous, for arbitrary Hilbert spaces this is a remarkable first result:
\begin{lemma}\label{lemma_pos_cont}
Let $\mathcal H,\mathcal G$ be complex Hilbert spaces and $T:\mathcal B^1(\mathcal H)\to\mathcal B^1(\mathcal G)$ be linear and positive. Then $T$ is continuous.
\end{lemma}
\begin{proof}
Every $A\in\mathcal B^1(\mathcal H)$ can be written as $A=A_1-A_2+iA_3-iA_4$ for some positive semi-definite $A_1,A_2,A_3,A_4\in\mathcal B^1(\mathcal H)$ where $\|A_j\|_1\leq\|A\|_1$ for $j=1,2,3,4$ \cite[Coro.~4.2.4]{Kadison83}. Thus it suffices to prove that $T$ is bounded when restricted to $\mathfrak{pos}(\mathcal H)\,$\footnote{
Indeed if $\|T(P)\|_1\leq c\|P\|_1$ for all $P\in\mathcal B^1(\mathcal H)\cap\mathfrak{pos}(\mathcal H)$ then $
\|T(A)\|_1\leq \sum_{j=1}^4\|T(A_j)\|_1\leq c\sum_{j=1}^4\|A_j\|_1\leq 4c\|A\|_1
$
for all $A\in\mathcal B^1(\mathcal H)$
.}. For the rest of the argument we follow \cite[Ch.~2, Lemma 2.1]{Davies76}: Assume $T$ is not bounded on $\mathcal B^1(\mathcal H)\cap\mathfrak{pos}(\mathcal H)$, that is,
$$
\sup_{A\in\mathbb D(\mathcal H)}\|T(A)\|_1=\sup_{A\in\mathbb D(\mathcal H)}\operatorname{tr}(T(A))=\infty
$$
where we used that trace norm and trace co{\"i}ncide precisely on the positive semi-definite elements. Now for every $n\in\mathbb N$ one finds $\rho_n\in\mathbb D(\mathcal H)$ such that $\operatorname{tr}(T(\rho_n))\geq 4^n$. This lets us define $\rho:=\sum_{n=1}^\infty 2^{-n}\rho_n\in\mathbb D(\mathcal H)$ which satisfies $2^{-n}\rho_n\leq\rho$ and thus $ 2^{-n}T(\rho_n)\leq T(\rho)$ because $T$ is positive and linear. But this would mean $2^n\leq 2^{-n}\operatorname{tr}(T(\rho_n))\leq \operatorname{tr}(T(\rho))$ for all $n\in\mathbb N$, an obvious contradiction; hence $T$ has to be bounded and thus continuous (Lemma \ref{lemma_cont_norm}).
\end{proof}

This result is the very foundation of the equivalence of different descriptions of Schr{\"odinger} and Heisenberg picture, more on that in the next section. For now we note that continuity of positive linear maps enables looking at the associated dual operator\index{operator!dual}\index{positivity-preserving!dual of} from Def.~\ref{def_dual_op}:
\begin{corollary}\label{coro_pos_dual_equal}
Let $\mathcal H,\mathcal G$ be complex Hilbert spaces. The following statements hold.
\begin{itemize}
\item[(i)] Given $T:\mathcal B^1(\mathcal H)\to\mathcal B^1(\mathcal G)$ linear and positive there exists unique $T^*:\mathcal B(\mathcal G)\to\mathcal B(\mathcal H)$ linear, positive, and ultraweakly continuous\footnote{
So $T^*$ is continuous as a map $(\mathcal B(\mathcal G),\topuw)\to(\mathcal B(\mathcal H),\topuw)$ with $\topuw$ being the ultraweak topology from Def.~\ref{def_uw_topology}.
}
such that
\begin{equation}\label{eq:duality_trace_map}
\operatorname{tr}(T(A)B)=\operatorname{tr}(AT^*(B))\quad\text{ for all }A\in\mathcal B^1(\mathcal H), B\in\mathcal B(\mathcal G)\,.
\end{equation}
\item[(ii)] Given some $S:\mathcal B(\mathcal G)\to\mathcal B(\mathcal H)$ linear, positive, and ultraweakly continuous there exists unique $T:\mathcal B^1(\mathcal H)\to\mathcal B^1(\mathcal G)$ linear and positive such that $\operatorname{tr}(T(A)B)=\operatorname{tr}(AS(B))$ for all $A\in\mathcal B^1(\mathcal H)$, $B\in\mathcal B(\mathcal G)$.
\end{itemize}
\end{corollary}
\begin{proof}
This follows directly from Prop.~\ref{prop_dual_weak_star_cont}, together with the following two facts:
\begin{itemize}
\item From the weak*-continuous dual map $T':(\mathcal B^1(\mathcal G))^*\to(\mathcal B^1(\mathcal H))^*$ one gets to the ultraweakly continuous map $T^*:=\psi_{\mathcal H}^{-1}\circ T'\circ \psi_{\mathcal G}:\mathcal B(\mathcal G)\to\mathcal B(\mathcal H)$ (and back) using the isometric isomorphism $\psi$ from Prop.~\ref{prop_schatten_dual}.
\item Positivity transfers via \eqref{eq:duality_trace_map} together with the fact that a linear map $T:\mathcal B^1(\mathcal H)\to\mathcal B^1(\mathcal G)$ is positive if and only if $\operatorname{tr}(T(A)B)\geq 0$ for all $A\in\mathcal B^1(\mathcal H)$, $B\in\mathcal B(\mathcal G)$ both positive semi-definite \cite[Lemma 3]{vE_dirr_semigroups}, and similarly for a map between bounded operators.\qedhere
\end{itemize}
\end{proof}
\begin{remark}
Ultraweak continuity not only allows for the equivalence in the previous corollary but also connects our approach of defining states via trace-class operators to the algebraic approach usually taken for quantum field theory. There, a state\index{quantum state} is a positive linear functional $\varphi$ on a $C^*$--algebra which satisfies $\varphi(\mathbbm{1})=1$, and if one even deals with $W^*$-algebras then ultraweak continuity of $\varphi$ is equivalent to the existence of $\rho\in\mathbb D(\mathcal H)$ such that $\varphi(B)=\operatorname{tr}(\rho B)$ for all $B\in\mathcal B(\mathcal H)$ \cite[Thm.~2.4.21]{Bratteli87}. While states which are not ultraweakly continuous (usually called: ``non-normal'', ``singular'' or in some situations ``disjoint'') are not of further importance in usual quantum mechanics they do turn up in quantum field theory \cite{Bratteli97} in relation to inequivalent representations.
\end{remark}

Following \cite[Ch.~9.2]{Davies76} let us do a quick thought experiment: Imagine a (spatially bounded) physical system described by a Hilbert space $\mathcal H$, and a particle with $n\in\mathbb N$ degrees of freedom so far away that there is no interaction between the two. Sometimes taken as an axiom of quantum mechanics, the Hilbert space of the full system then is given by the tensor product\index{tensor product}\footnote{
Readers unfamiliar with the tensor product of Hilbert spaces may check Appendix \ref{app_tensor_hilbert} 
}
$\mathcal H\otimes\mathbb C^n$, and an operation which only acts on the original system is of the form $T\otimes\mathbbm{1}_n$. While $T$ has to be linear and positive as argued above, that does \textit{not} guarantee positivity of $T\otimes\mathbbm{1}_n$. This motivates the following definition:

\begin{definition}
A linear map $T:\mathcal B^1(\mathcal H)\to\mathcal B^1(\mathcal G)$ with $\mathcal H,\mathcal G$ arbitrary complex Hilbert spaces is said to be completely positive if $T\otimes\mathbbm{1}_n$ is positive for every $n\in\mathbb N$.\index{completely positive}
\end{definition}
Again this definition makes sense for maps between bounded operators instead of the trace class, as well. Either way our thought experiment lets us conclude that the reasonable physical transformations have to form a subset of the completely positive maps. This is also of interest from a mathematical viewpoint as complete positivity is characterized by the Kraus representation:

\begin{proposition}\label{prop_kraus_op_general}
Let $\mathcal H,\mathcal G$ be complex Hilbert spaces and $S:\mathcal B(\mathcal G)\to\mathcal B(\mathcal H)$ be a linear and ultraweakly continuous map. The following statements are equivalent.
\begin{itemize}
\item[(i)] $S$ is completely positive.
\item[(ii)] There exist $(K_i)_{i\in I}\subset\mathcal B(\mathcal H,\mathcal G)$---called Kraus operators---such that \index{operator!Kraus}
\begin{equation}\label{eq:kraus_form}
S(B)=\sum\nolimits_{i\in I}K_i^*BK_i
\end{equation}
for all $B\in\mathcal B(\mathcal G)$ where the sum converges in the strong operator topology. 
\end{itemize}
If $\mathcal H,\mathcal G$ both are separable then one can choose the index set $I$ to be countable.
\end{proposition}
\begin{proof}
For $\mathcal H=\mathcal G$ this is shown in \cite[Ch.~9, Thm.~2.3]{Davies76} or \cite[Thm.~1]{Kraus83}. Based on this, one proves the general case, refer to Appendix \ref{app_kraus_gen}.
\end{proof}
\begin{remark}
\begin{itemize}
\item[(i)] Using relation \eqref{eq:duality_trace_map} \& Coro.~\ref{coro_pos_dual_equal} this readily implies that a linear map $T:\mathcal B^1(\mathcal H)\to\mathcal B^1(\mathcal G)$ is completely positive if and only if there exist $(K_i)_{i\in I}\subset\mathcal B(\mathcal H,\mathcal G)$ such that $T(A)=\sum\nolimits_{i\in I}K_iBK_i^*$ for all $A\in\mathcal B^1(\mathcal H)$ where the sum converges in trace norm (again cf., e.g., \cite[Thm.~1]{Kraus83}). Note that the convergence behaviour of this sum as well as \eqref{eq:kraus_form} is well-known, see also \cite[Prop.~6.3 \& 6.10]{Attal6}.
\item[(ii)] Taking only finite-dimensional auxiliary systems in the definition of complete positivity is sufficient because using the Kraus representation one finds that the tensor product of any two completely positive (normal) linear maps is completely positive again.
\end{itemize}
\end{remark}

\subsection{Channels in the Schr{\"o}dinger and the Heisenberg Picture}\label{ch_schr_heis_pic}

As renowned mathematical physicist Barry Simon once wrote: ``Throughout, all our Hilbert spaces will be complex and separable (are there any others?)'' \cite[p.~1]{Simon05}, a piece of advice we shall follow for the remainder of this chapter, as well.

After going over the building blocks of channels let us actually define them: A \textit{(Schr\"odinger) quantum channel}\index{quantum channel!Schr{\"o}dinger}\index{cptp@\textsc{cptp}|see{quantum channel, Schr{\"o}dinger}}---sometimes \textsc{cptp}\label{not_cptp} map---is a linear, completely positive, and trace-preserving map
$T:\mathcal B^1(\mathcal H)\to\mathcal B^1(\mathcal G)$. Define\label{symb_schrodinger_ch}
\begin{align*}
Q_S (\mathcal H,\mathcal G):=
\lbrace T:\mathcal B^1(\mathcal H)\to\mathcal B^1(\mathcal G)\, |\, T\text{ is Schr\"odinger quantum channel}\rbrace
\end{align*}
and $Q_S (\mathcal H):= Q_S (\mathcal H,\mathcal H)$. For its dual concept be aware that complete positivity transfers back and forth just like positivity \cite[p. 35]{Kraus83} so Coro.~\ref{coro_pos_dual_equal} can be readily adjusted to that. Trace-preservation on the other hand behaves like
\begin{align*}
\operatorname{tr}(A)=\operatorname{tr}(T(A))=\operatorname{tr}(T(A)\mathbbm{1})=\operatorname{tr}(AT^*(\mathbbm{1}))
\end{align*}
for all $A$ of trace class. This means the dual channel\index{dual channel} has to preserve the identity---sometimes called \textit{unital}\index{unital}---so $T^*(\mathbbm{1})=\mathbbm{1}$. Therefore a \textit{Heisenberg quantum channel}\index{quantum channel!Heisenberg} is a linear, ultraweakly continuous, completely 
positive, and unital map $S:\mathcal B(\mathcal G)\to\mathcal B(\mathcal H)$. Furthermore, we define\label{symb_heisenberg_ch}
\begin{align*}
Q_H(\mathcal G,\mathcal H):=\lbrace S:\mathcal B(\mathcal G)\to\mathcal B(\mathcal H)\, |\, S\text{ is Heisenberg quantum channel}\rbrace
\end{align*}
and $Q_H (\mathcal H):= Q_H (\mathcal H,\mathcal H)$. With both these concepts introduced it is evident that the map 
$*:Q_S(\mathcal H,\mathcal G)\to Q_H(\mathcal G,\mathcal H)$
from Coro.~\ref{coro_pos_dual_equal}---which to any quantum channel assigns its dual channel---is well-defined. Interestingly enough this map is even bijective:

\begin{corollary}\label{thm_dual}
For every $S\in Q_H(\mathcal G,\mathcal H)$ there exists unique $T\in Q_S(\mathcal H,\mathcal G)$ with $T^*=S$.
\end{corollary}
\begin{proof}
By Coro.~\ref{coro_pos_dual_equal} one finds a positive linear map $T:\mathcal B^1(\mathcal H)\to\mathcal B^1(\mathcal G)$ such that $T^*=S$. As argued above complete positivity transfers and $S$ being unital becomes $T$ being trace-preserving.
\end{proof}

In other words this corollary tells us that the Schr{\"o}dinger and the Heisenberg picture\index{Schr{\"o}dinger picture}\index{Heisenberg picture} are equivalent as $Q_S(\mathcal H,\mathcal G)\simeq Q_H(\mathcal G,\mathcal H)$ by means of the isometric isomorphism\footnote{
The map $*$ is an isometry as $'$ is an isometry (Lemma \ref{lemma_dual_map_properties}) and the map $\psi$ ``transforming $'$ into $*$'' is an isometric isomorphism, as well (Prop.~\ref{prop_schatten_dual}).
}
$*$. Thus the structure of the set of all physical transformations does not change when switching the picture. Further algebraic and topological properties of the set of channels read as follows:

\begin{proposition}\label{thm_monoid}
The following statements hold.
\begin{itemize}
\item[(i)] The set $Q_S (\mathcal H)$ is a convex subsemigroup of $\mathcal B(\mathcal B^1(\mathcal H))$ with unity
element $\mathbbm{1}_{\mathcal B^1(\mathcal H)}$. Moreover, $Q_S(\mathcal H,\mathcal G)$ is closed in
$\mathcal B(\mathcal B^1(\mathcal H),\mathcal B^1(\mathcal G))$ with respect to the weak operator, strong operator, and operator norm topology\index{topology!operator norm|see{topology, norm}}. However, as soon as $\mathcal G$ is of infinite dimension, then $Q_S(\mathcal H,\mathcal G)$ is not compact in either of these topologies. 
\item[(ii)] The set $Q_H (\mathcal H)$ is a convex subsemigroup of $\mathcal B(\mathcal B(\mathcal H))$ with unity
element $\mathbbm{1}_{\mathcal B(\mathcal H)}$. Moreover, $Q_H(\mathcal G,\mathcal H)$ is closed in
${}^*(\mathcal B(\mathcal B^1(\mathcal G),\mathcal B^1(\mathcal H)))$ with respect to the ultraweak operator, weak operator, strong operator, and operator norm topology. However, as soon as $\mathcal G$ is of infinite dimension, then $Q_H(\mathcal G,\mathcal H)$ is not compact in either of these topologies.
\end{itemize}
\end{proposition}
\begin{proof}
(i): Convexity, semigroup property, and closedness are due to \cite[Thm.~1]{vE_dirr_semigroups}. The 
generalization from $Q_S(\mathcal H)$ to $Q_S(\mathcal H,\mathcal G)$ is briefly discussed 
in \cite[Remark 15]{vE_dirr_semigroups}. As for lack of compactness: We will construct a 
sequence of channels which has no w.o.t.-convergent subsequence so the set of channels 
cannot be w.o.t.-compact; this implies lack of compactness in any stronger topology (Lemma 
\ref{lemma_separable_topology_comp} (iii)) such as s.o.t. and the operator norm topology. W.l.o.g.~let 
$\mathcal G$ be infinite-dimensional but separable so one finds an orthonormal basis 
$(g_n)_{n\in\mathbb N}$ which lets us define $T_n:\mathcal B^1(\mathcal H)\to\mathcal B^1(\mathcal G)$ 
via $A\mapsto\operatorname{tr}(A)|g_n\rangle\langle g_n|$. Indeed $T_n\in Q_S(\mathcal H,\mathcal G)$ 
for all $n\in\mathbb N$ \cite[Ex.~4.10]{Heinosaari12} as $(|e_n\rangle\langle e_m|)_{m\in\mathbb N}$ 
is a set of Kraus operators for $T_n$. Now assume one finds a subsequence $(T_{n_k})_{k\in\mathbb N}$ 
which converges in w.o.t.~to some $S\in Q_S(\mathcal H,\mathcal G)$. Then for all $i\in\mathbb N$, $A\in\mathcal B^1(\mathcal H)$
\begin{align*}
0\underset{\text{b.a.}}{\overset{k\to\infty}{\longleftarrow}}&\big|\operatorname{tr}\big(|g_i\rangle\langle g_i|T_{n_k}(A)\big)-\operatorname{tr}\big(|g_i\rangle\langle g_i|S(A)\big)\big|\\
=&\big|\operatorname{tr}(A)\underbrace{|\langle g_i,g_{n_k}\rangle|^2}_{\to 0}-\langle g_i,S(A)g_i\rangle\big|\to|\langle g_i,S(A)g_i\rangle|
\end{align*}
so $\langle g_i,S(A)g_i\rangle=0$. This implies $\operatorname{tr}(S(A))=\sum_{i=1}^\infty \langle g_i,S(A)g_i\rangle=0$ for all $A\in\mathcal B^1(\mathcal H)$ so $S$ is not trace-preserving, a contradiction.

 (ii): First note that a net $(T_i)_{i\in I}\subseteq {}'(\mathcal B(\mathcal B^1(\mathcal H),\mathcal B^1(\mathcal G))$ converges to $T\in {}'(\mathcal B(\mathcal B^1(\mathcal H),\mathcal B^1(\mathcal G))$ in the weak*-operator topology if and only if $\psi_{\mathcal H}^{-1}\circ T_i \circ\psi_{\mathcal G}$ converges to $\psi_{\mathcal H}^{-1}\circ T \circ\psi_{\mathcal G}$ in the ultraweak operator topology. Now applying Prop.~\ref{prop_connection_weak_weakstar_op} to (i) we get that $'Q_S(\mathcal H,\mathcal G)$ is weak*-closed in $'(\mathcal B(\mathcal B^1(\mathcal H),\mathcal B^1(\mathcal G))$ which by the previous fact translates to ultraweak closedness of ${}^*Q_S(\mathcal H,\mathcal G)=Q_H(\mathcal G,\mathcal H)$ in ${}^*(\mathcal B(\mathcal B^1(\mathcal G),\mathcal B^1(\mathcal H)))$. Similarly, one sees that $Q_S(\mathcal H,\mathcal G)$ is $\topw$-compact if and only if $Q_H(\mathcal G,\mathcal H)$ is $\topuw$-compact so the counterexample from (i) carries over. The fact that the ultraweak operator topology is weaker than the weak, strong, and operator norm topology concludes the proof.
\end{proof}
\noindent It remains unknown whether $Q_H(\mathcal G,\mathcal H)$ is closed as a subset of $\mathcal B(\mathcal B(\mathcal G),\mathcal B(\mathcal H))$. If true it cannot be a mere corollary of Prop.~\ref{thm_monoid} (ii) as one can find general examples where the weak*-operator topology does not transfer weak*-continuity to the limit (Remark \ref{rem_closed_in_prime_nec}).\smallskip

\begin{remark}
The example which shows that $Q_S(\mathcal H,\mathcal G)$ is not w.o.t.-compact is, after slight modification, the same which shows that $\mathbb D(\mathcal H)$ is not weakly compact in infinite dimensions.
\end{remark}

Perhaps the most important property of quantum channels for the purpose of analyzing reachable sets is the fact that every channel is a contraction:

\begin{proposition}\label{thm_q_norm_1}
Let $T\in Q_S(\mathcal H,\mathcal G)$ and $S\in Q_H(\mathcal G,\mathcal H)$. Then $\|T\|=1$ and $\|S\|=1$.
\end{proposition}
\begin{proof}
As each $S\in Q_H(\mathcal G,\mathcal H) $ is linear, positive, and unital it has operator norm 
$\|S\|=1$ as a consequence of the Russo-Dye Theorem\index{theorem!Russo-Dye} \cite[Cor.~1]{Russo66}. This directly implies $\|T\|=\|T^*\|=1$.
\end{proof}
\noindent Note that this result holds even without complete positivity as the proof only needs that the linear map is positive and trace-preserving (resp.: positive and unital).\smallskip

For the remainder of this section there are two things we want to do. First off we will characterize invertibility of channels\index{quantum channel!invertibility of}; we do this for Schr{\"o}dinger channels, the Heisenberg case reads analogously:
\begin{proposition}\label{ch_3_Theorem_14}
Let $T\in Q_S(\mathcal H)$ be bijective. Then the following statements are equivalent.
\begin{itemize}
\item[(i)] $T^{-1}$ is positive.
\item[(ii)] There exists unitary $U\in\mathcal B(\mathcal H)$ such that $T(A)=U AU^*$ for all
$A\in\mathcal B^1(\mathcal H)$.
\end{itemize}
\end{proposition}
\begin{proof}
\cite[Prop.~1]{vE_dirr_semigroups}.
\end{proof}
\noindent Of course there are still channels which are invertible in the sense of a linear operators: Channels of the form $e^L$ for an appropriate generator $L$ (more on that in Section \ref{ch:qu_dyn_sys}) are of course bijective but there is no guarantee that the inverse $e^{-L}$ is a channel again, as it may fail to be positive.\smallskip

As for the second thing, following Ch.~\ref{ch:mean_erg} let us prove that not all \textsc{cptp} maps are mean ergodic. While every finite-dimensional normed space is reflexive so in finite dimensions all channels are mean ergodic (Prop.~\ref{prop_mean_ergodic})---this is investigated further in \cite{Burgarth13}---not much is known for the general case. First off let us consider the following beautiful and well-known representation result for Schr\"odinger quantum channels which can be traced back to Kraus:

\begin{theorem}\label{thm_stinespring_schr}
For every $T\in Q_S(\mathcal H)$ there exists a separable Hilbert space $\mathcal K$, a pure state $\omega\in\mathbb D(\mathcal K)$, and unitary $U\in\mathcal B(\mathcal H\otimes\mathcal K)$ such that
\begin{align*}
T(A)=\operatorname{tr}_{\mathcal K}(U(A\otimes \omega)U^*)
\end{align*}
for all $A\in\mathcal B^1(\mathcal H)$. Moreover if $T\in Q_S(\mathcal H,\mathcal G)$ then there exists a separable Hilbert space $\mathcal K$, pure $\omega_G\in\mathbb D(\mathcal G)$, $\omega_K\in\mathbb D(\mathcal K)$, and unitary 
$U\in\mathcal B(\mathcal H\otimes\mathcal G\otimes\mathcal K)$ such that
\begin{align*}
T(A)=(\operatorname{tr}_{\mathcal H}\circ\operatorname{tr}_{\mathcal K})(U(A\otimes \omega_G\otimes\omega_K)U^*)
\end{align*}
for all $A\in\mathcal B^1(\mathcal H)$.
\end{theorem}
\begin{proof}
\cite[second part of Thm.~2]{Kraus83} \& \cite[Thm.~2 \& Coro.~1]{vE_dirr_semigroups}. Be aware that the separable auxiliary space $\mathcal K$ can be chosen independently of $T$ (e.g., $\mathcal K := \ell_2(\mathbb N)$). Moreover, once $\mathcal K$ is fixed $\omega\in\mathbb D(\mathcal K)$ can be chosen as any orthogonal rank-$1$ projection. Thus $\omega$ is pure and independent of $T$, too.
\end{proof}

\noindent
Here $\operatorname{tr}_{\mathcal K}:\mathcal B^1(\mathcal H\otimes\mathcal K)\to\mathcal B^1(\mathcal H)$
is the partial trace \textit{with respect to} $\mathcal K$\index{partial trace} which is defined via\label{eq:partial_trace_1}
\begin{align*}
\operatorname{tr}(B\operatorname{tr}_{\mathcal K}(A))=\operatorname{tr}((B\otimes\mathbbm{1}_{\mathcal K})A)
\end{align*}
for all $B\in\mathcal B(\mathcal H)$ and all $A\in\mathcal B^1(\mathcal H\otimes\mathcal K)$. In other words $\operatorname{tr}_{\mathcal K}$ is the (unique) pre-dual of the extension channel $B\mapsto B\otimes\mathbbm{1}$.
\begin{remark}
Like in Section \ref{app_kraus_gen} one can define the partial trace analogously \textit{with respect to a state} $\omega\in\mathbb D(\mathcal K)$ via 
$
\operatorname{tr}(\operatorname{tr}_{\omega}(B)A)=\operatorname{tr}(B(A\otimes\omega))
$\label{eq:partial_trace_2}
for all $B\in\mathcal B(\mathcal H\otimes\mathcal K)$, $A\in\mathcal B^1(\mathcal H)$, 
cf.~\cite[Ch.~9, Lemma~1.1]{Davies76}; so it is the dual of the extension channel $A\mapsto A\otimes\omega$. With this one can carry over Thm.~\ref{thm_stinespring_schr} to the Heisenberg picture: Given $S\in Q_H(\mathcal G,\mathcal H)$ there exists a separable Hilbert space $\mathcal K$, 
pure states $\omega_G\in\mathbb D(\mathcal G)$, $\omega_K\in\mathbb D(\mathcal K)$, and a unitary 
$U\in\mathcal B(\mathcal H\otimes\mathcal G\otimes\mathcal K)$ such that
\begin{align*}
S(B)=(\operatorname{tr}_{\omega_G}\circ\operatorname{tr}_{\omega_K})(U^*(\mathbbm{1}_{\mathcal H}\otimes B\otimes \mathbbm{1}_{\mathcal K})U)
\end{align*}
for all $B\in\mathcal B(\mathcal G)$. For $\mathcal G=\mathcal H$ this reduces to
\begin{align}\label{eq:stinespring_q_h}
S(B)=\operatorname{tr}_{\omega_K}(U^*(B\otimes \mathbbm{1}_{\mathcal K})U)
\end{align}
for all $B\in\mathcal B(\mathcal H)$ where the unitary operator $U$ now acts on $\mathcal H\otimes\mathcal K$. This is a more structured version of Stinespring's theorem\index{theorem!Stinespring's} \cite{Stinespring55}
for Heisenberg quantum channels due to the following: Given $S\in Q_H(\mathcal H)$ (the same argument works
for $S\in Q_H(\mathcal G,\mathcal H)$), let $\omega_K\in\mathbb D(\mathcal K)$ be the state from \eqref{eq:stinespring_q_h} of rank one, 
i.e.~$\omega_K=|y\rangle\langle y|$ for some $y\in\mathcal K$ with $\|y\|=1$. As the isometric embedding $V_y:\mathcal H\to\mathcal H\otimes\mathcal K$, $x\mapsto x\otimes y$, satisfies $\operatorname{tr}_{\omega_K}(B)=V_y^* BV_y$ for all $B\in\mathcal B(\mathcal H\otimes\mathcal K)$ (Lemma \ref{lemma_V_tensor}), \eqref{eq:stinespring_q_h} becomes
$
S(\cdot)=V^*\pi(\cdot) V
$
with the auxiliary Hilbert space $\mathcal H\otimes\mathcal K$ being of tensor form, the 
Stinespring isometry $V = UV_y$, and the unital $*$-homomorphism 
$\pi:\mathcal B(\mathcal H)\to\mathcal B(\mathcal H\otimes\mathcal K) $ being 
$\pi(B):= B\otimes \mathbbm{1}_{\mathcal K}$.
To the best of our knowledge, the above representation \eqref{eq:stinespring_q_h} so far only 
appeared in an unpublished (as of now) book by Attal \cite[Thm.~6.15]{Attal6}.
\end{remark}
%
%
%
%

With this let us come to the promised counterexample, the idea of which is due to G.~Dirr (private communication). 
\begin{example}
Let $U:\ell^2(\mathbb Z)\to \ell^2(\mathbb Z)$, $e_n\mapsto e_{n-1}$ for all $n\in\mathbb Z$ be the bilateral shift on the Hilbert space $\ell^2(\mathbb Z)$. Then $U$ is unitary, has no eigenvalues, and $\sigma(U)=\{z\in\mathbb C\,|\,|z|=1\}$ \cite[Solution 84]{Halmos82}. We will show that $T_U:\mathcal B^1(\ell^2(\mathbb Z))\to \mathcal B^1(\ell^2(\mathbb Z))$, $A\mapsto UAU^*$ is not mean ergodic\index{quantum channel!not mean ergodic} by means of the following computation: Let $m,n\in\mathbb N$ with w.l.o.g.~$m\geq n$ be given. Then
\begin{align*}
\Big\|\frac1n\sum\nolimits_{k=0}^{n-1}T_U^k(|e_0\rangle\langle e_0|)-\frac1m\sum\nolimits_{k=0}^{m-1}T_U^k(|e_0\rangle\langle e_0|)\Big\|_1=n\cdot\Big(\frac1n-\frac1m\Big)+(m-n)\cdot\frac1m=2-\frac{2n}{m}\,.
\end{align*}
This shows that $(\frac1n\sum\nolimits_{k=0}^{n-1}T_U^k(|e_0\rangle\langle e_0|))_{n\in\mathbb N}$ cannot be a Cauchy sequence: Given any $N\in\mathbb N$ choose $n=N$ and $m=2N$ as then
\begin{align*}
\Big\|\frac1{N}\sum\nolimits_{k=0}^{{N}-1}T_U^k(|e_0\rangle\langle e_0|)-\frac1{2N}\sum\nolimits_{k=0}^{{2N}-1}T_U^k(|e_0\rangle\langle e_0|)\Big\|_1=2-\frac{2N}{2N}=1\geq\frac12\,.
\end{align*}
Therefore the operator $P_T$ associated to $T_U$ from Def.~\ref{def_mean_erg} does not converge on all of $\mathcal B^1(\ell^2(\mathbb Z))$ meaning $T_U$ is not mean ergodic.
\end{example}
Roughly speaking, the problem here is the mentioned spectral behaviour of $U$: Hypothetically if $U$ could be diagonalized as a countable sum (so $\sigma(U)=\overline{\sigma_{\mathrm{p}}(U)}$, i.e.~$U$ has ``almost only'' eigenvalues) then the corresponding channel would be mean ergodic as can be directly computed. Moreover if such spectral behaviour could be guaranteed for all ``physical'' unitaries---whatever that would mean---then the result would carry over to all quantum channels via the Stinespring dilation (Thm.~\ref{thm_stinespring_schr}), using a corresponding discrete-time dilation result \cite[Thm.~4]{vE_dirr_semigroups}.\smallskip

This example shows that mean ergodicity is not a general feature of quantum channels as soon as those act on infinite-dimensional Hilbert spaces. This result is non-trivial because on such spaces the trace class is not reflexive \cite[Coro.~16.27]{MeiseVogt97en} which by Prop.~\ref{prop_mean_ergodic} \textit{would} be enough to guarantee mean ergodicity.
\subsection{Special Case: Finite Dimensions}

All our considerations in this chapter so far were concerned with arbitrary---or ``at worst'' separable---Hilbert spaces, meaning these results in particular include the finite-dimensional case. However, of course, in finite dimensions a lot of things simplify: For example all of the topological considerations are obsolete as ultraweak, weak, and strong operator topology co{\"i}ncide with the norm topology (Prop.~\ref{prop_strong_weak_op_top} \& \ref{prop_weak_star_weak_comp}). But even beyond this more structure unfolds, such as the following characterization of complete positivity\index{completely positive} \cite{Choi75}:
\begin{lemma}\label{lemma_choi_matrix}
Let linear $T:\mathbb C^{n\times n}\to\mathbb C^{k\times k}$ be given. The following are equivalent.
\begin{itemize}
\item[(i)] $T$ is completely positive.
\item[(ii)] The Choi matrix\index{Choi matrix} $
C(T):=\big(T(|e_j\rangle\langle e_k|)\big)_{j,k=1}^n$ of $T$ is positive semi-definite.
\item[(iii)] There exist Kraus operators\index{operator!Kraus} $\{K_i\}_{i\in I}\subset\mathbb C^{n\times k}$ with $|I|\leq nk$ such that $T(A)=\sum_{i\in I}K_i^* AK_i$ for all $A\in\mathbb C^{n\times n}$.
\end{itemize}
\end{lemma}
To simplify notation let us write $Q(n,k):=Q_S(\mathbb C^n,\mathbb C^k)$ for the collection of all \textsc{cptp} maps from $\mathbb C^{n\times n}$ to $\mathbb C^{k\times k}$, as well as $Q(n):=Q(n,n)$.\label{symb_Q_n}
For some applications it is advantageous to define the set of channels with a common fixed point $X\in\mathbb D(\mathbb C^n)$ which will be denoted by\label{symb_Q_X_n}
$
Q_X(n):=\lbrace T\in Q(n)\,|\,T(X)=X\rbrace
$. One finds\footnote{
The positive trace-preserving maps are precisely those linear maps $T$ which satisfy $T(\mathbb D(\mathbb C^n))\subseteq\mathbb D(\mathbb C^n)$, hence this also holds for every quantum channel. As the states form a convex and compact set (Coro.~\ref{coro_states_comp}), by the Brouwer fixed-point theorem\index{theorem!Brouwer fixed-point} \cite{Brouwer11} every such $T$ has a fixed point in $\mathbb D(\mathbb C^n)$. This fails in infinite dimensions---even if one uses Schauder's fixed-point theorem\index{theorem!Schauder fixed-point} \cite{Schauder30} for topological vector spaces---for lack of compactness of generic $T(\mathbb D(\mathcal H))$, or any superset for that matter.
}
$Q(n)=\bigcup_{X\in\mathbb D(\mathbb C^n)}Q_X(n)$. 
%
Now the following is a simple consequence of Prop.~\ref{thm_monoid} due to the Heine-Borel theorem\index{theorem!Heine-Borel}; one readily verifies that the corresponding result still holds when replacing $Q(n)$ by $Q_X(n)$:

\begin{corollary}\label{lemma_convex_subsemigroup}
The set $Q(n)$, as well as $Q_X(n)$ for arbitrary $X\in\mathbb D(\mathbb C^{n})$, forms a convex and compact semigroup with identity element $\mathbbm{1}_{n\times n}$.
\end{corollary}
Finally one can characterize when a transition from one Hermitian matrix to another via a quantum channel is possible.

\begin{proposition}\label{prop_exist_channel}
Let $A,B\in\mathbb C^{n\times n}$ be Hermitian. Then the following are equivalent.
\begin{itemize}
\item[(i)] $\operatorname{tr}(A)=\operatorname{tr}(B)$ and $\|A\|_1\leq\|B\|_1$. 
\item[(ii)] There exists $T\in Q(n)$ such that $T(B)=A$. 
\item[(iii)] There exists $T:\mathbb C^{n\times n}\to \mathbb C^{n\times n}$ linear, positive, and trace-preserving such that $T(B)=A$. 
\end{itemize}
Moreover, if (i) holds and $0$ is an eigenvalue of $B$, then there exists $\psi\in\mathbb C^n$ with $\langle\psi,\psi\rangle=1$ such that $T(|\psi\rangle\langle\psi|)$ can be chosen arbitrarily from $\mathbb D(\mathbb C^n)$. 
\end{proposition}
\noindent Note that if $A\in\mathbb D(\mathbb C^n)$ then one can simply choose $T(B):=\operatorname{tr}(B)A$, but the general case is not as elementary. In order to prove this we need the following notation from Ch.~\ref{ch_maj_cnr}:
\begin{itemize}
\item Let $\unitvector=(1,\ldots,1)^T$\label{symb_unit_vector} be the column-vector of all ones.
\item A matrix $A\in\mathbb R^{n\times n}$ is called column-stochastic if it is non-negative---that is, $a_{jk}\geq 0$ for all $j,k=1,\ldots,n$---and satisfies $\unitvector^TA=\unitvector^T$ so all the columns sum up to one.
\end{itemize} 
Obviously if a vector $x\in\mathbb R^n$ has non-negative entries, then so does $Ax$ for any $A$ column-stochastic. Also such matrices cannot change the sum of entries of $x$ as $\unitvector^T(Ax)=(\unitvector^TA)x=\unitvector^Tx$. Therefore column-stochastic matrices are the ``classical'' analogue of positive and trace-preserving maps which is why we first want to verify Prop.~\ref{prop_exist_channel} for the vector case:
\begin{lemma}\label{lemma_stoch_transf}
For $x,y\in\mathbb R^n$ the following statements are equivalent.
\begin{itemize}
\item[(i)] $\unitvector^Tx=\unitvector^Ty$ and $\|x\|_1\leq\|y\|_1$. Here $\|\cdot\|_1=\sum_{i=1}^n|(\cdot)_i|$ is the usual vector-$1$-norm.
\item[(ii)] There exists a column-stochastic matrix $A\in\mathbb R^{n\times n}$ such that $Ay=x$.
\end{itemize}
\end{lemma}
One can prove this by explicitly constructing $A$ as is done in \cite[Thm.~3.3]{Li11}.
Now all that is left is to lift this result to the matrix case:
\begin{proof}[Proof of Prop.~\ref{prop_exist_channel}]
``(ii) $\Rightarrow$ (iii)'': Obvious. ``(iii) $\Rightarrow$ (i)'': Every positive trace-preserving map is trace-norm contractive, see Prop.~\ref{thm_q_norm_1} or for an explicit proof in finite dimensions \cite[Thm.~2.1]{Wolf06}.

``(i) $\Rightarrow$ (ii)'': Using \cite[Thm.~4.1.5]{HJ1} one finds unitaries $U,V\in\mathbb C^{n\times n}$ and vectors $x,y\in\mathbb R^n$ such that $A=U\operatorname{diag}(x) U^*$, $B=V\operatorname{diag}(y) V^*$. By assumption $\unitvector^Tx=\operatorname{tr}(A)=\operatorname{tr}(B)=\unitvector^Ty$ and $\|x\|_1=\|A\|_1\leq\|B\|_1=\|y\|_1$. Hence Lemma \ref{lemma_stoch_transf} yields a column-stochastic matrix $M\in\mathbb R^{n\times n}$ with $My=x$. Define a map $\tilde{T}:\mathbb C^{n\times n}\to \mathbb C^{n\times n}$ via
\begin{align*}
|e_i\rangle\langle e_j|\mapsto\begin{cases}
0&\text{ if }i\neq j\\
\sum\nolimits_{k=1}^n M_{ki} |e_k\rangle\langle e_k|&\text{ if }i= j
\end{cases}
\end{align*}
and its linear extension onto all of $\mathbb C^{n\times n}$. The Choi matrix of $\tilde{T}$ is diagonal with non-negative entries because $M_{jk}\geq 0$ for all $j,k$ so $C(\tilde{T})\geq 0$ and $\tilde{T}$ is completely positive by Lemma \ref{lemma_choi_matrix}. Moreover $\tilde{T}$ is trace preserving because
$$
\operatorname{tr}\big(\tilde{T}( |e_j\rangle\langle e_j|)\big)=\sum\nolimits_{k=1}^n M_{kj}\operatorname{tr}( |e_k\rangle\langle e_k|)=(\unitvector^TM)_j=1=\operatorname{tr}( |e_j\rangle\langle e_j|)
$$
for all $j=1,\ldots,n$. This shows $\tilde{T}\in Q(n)$. Also 
\begin{align*}
\tilde{T}(\operatorname{diag}(y))=\sum\nolimits_{j=1}^n y_j \tilde{T}( |e_j\rangle\langle e_j|)&=\sum\nolimits_{i=1}^n\Big(\sum\nolimits_{j=1}^nM_{ij}y_{j}\Big) |e_i\rangle\langle e_i|\\
&=\sum\nolimits_{i=1}^n(My)_i |e_i\rangle\langle e_i|=\sum\nolimits_{i=1}^nx_i |e_i\rangle\langle e_i|=\operatorname{diag}(x)
\end{align*}
so $T(\cdot):=U \tilde T(V^* (\cdot)V)U^*$ ($\in Q(n)$ as a composition of quantum channels, Coro.~\ref{lemma_convex_subsemigroup}) satisfies $T(B)=A$. 
Now if one of the $y_j$ (eigenvalues of $B$) is $0$ then the action of $\tilde T( |e_j\rangle\langle e_j|)=:\omega$ can obviously be chosen freely without affecting \mbox{$\tilde T(\operatorname{diag} y)=\operatorname{diag} x$}, that is, $T(B)=A$. If $\omega\in\mathbb D(\mathbb C^n)$ then $\tilde T,T$ remain in $Q(n)$ by the above argument, so defining $\psi:=Ve_j$ concludes the proof. 
\end{proof}

Of course this section merely scratches the surface of finite-dimensional quantum channels and their special properties, but with this we covered all we will need later on.
\subsection{Quantum-Dynamical Systems}\label{ch:qu_dyn_sys}

Up until now our considerations were of static nature which, of course, cannot be the end of the story. After all, control theory and differential equations in general---among a plethora of other fields in mathematics and physics---involve an additional real parameter modelling time and subsequent changes of physical systems.

For quantum dynamics\index{quantum dynamics}, arguably, the simplest dynamics a system with Hilbert space $\mathcal H$ might undergo are of semigroup structure: If one works in the Schr\"odinger picture this is described by a mapping\footnote{
Of course one may consider a smaller domain $I=[a,b]\subseteq\mathbb R$ of $T$ where $-\infty\leq a\leq 0<b\leq\infty$. However by the semigroup property $T(t+s)=T(t)T(s)$, there exists a unique extension of $T$ to a one-parameter semigroup on $\mathbb R_+$ (or even $\mathbb R$ if $a<0$) as is readily verified; thus w.l.o.g.~assume $T$ is already defined on the latter.
} 
$T:\mathbb R_+\to Q_S(\mathcal H)$ which satisfies $T(0)=\mathbbm{1}_{\mathcal B^1(\mathcal H)}$ and $T(t+s)=T(t)T(s)$ for all $t,s\geq 0$; similarly this idea can be adjusted to the Heisenberg picture. Clearly, this description requires that the evolution does not depend on the past of the system or, from a physical perspective, that one may neglect such memory effects due to ``short environmental correlation times'' (cf.~\cite[Ch.~3.2.1]{BreuPetr02} or Rem.~\ref{rem_control_born_markov}). No matter the physical motivation this allows for a mathematically much more structured description; we will come to this in a bit. 

To start off let us make things as simple as possible by assuming that the system $\mathcal H$ is isolated from its environment. This means that, as ``almost all known laws of physics are invariant under time reversal and time translation'' \cite[Ch.~9.1]{Davies76}, the one-parameter semigroup $T:\mathbb R_+\to Q_S(\mathcal H)$ can be extended to domain $\mathbb R$ so in particular
$$
T(t)T(-t)=T(0)=\mathbbm{1}_{\mathcal B^1(\mathcal H)}=T(-t)T(t)\qquad\text{ for all }t\in \mathbb R_+\,.
$$
Then the channel $T(t)$ has positive (even \textsc{cptp}) inverse $T(-t)$ meaning $T(t)=U(t)(\cdot) U(t)^*$ has to be a unitary channel by Prop.~\ref{ch_3_Theorem_14}.
This motivates this section's first definition:
\begin{definition}
Let $\mathcal H$ be a complex Hilbert space and consider a mapping $U:\mathbb R\to\mathcal U(\mathcal H)$ such that $U(0)=\mathbbm{1}_\mathcal H$ and $U(s+t)=U(s)U(t)$ for all $s,t\in\mathbb R$. Then $\{U(t)\}_{t\in\mathbb R}$ is called a one-parameter unitary group. It is called\index{one-parameter!unitary group}
\begin{itemize}
\item[(i)] norm continuous if, additionally, $U:\mathbb R\to(\mathcal U(\mathcal H),\topn)$ is continuous in $0$.
\item[(ii)] strongly continuous if, additionally, $U:\mathbb R\to(\mathcal U(\mathcal H),\tops)$ is continuous in $0$.
\end{itemize}
\end{definition}

Under sufficiently strong continuity assumptions such one-parameter unitary groups can be expressed via the exponential map, making use of functional calculus:

\begin{lemma}\label{lemma_stone}
Let a complex Hilbert space $\mathcal H$ as well as a mapping $U:\mathbb R\to\mathcal U(\mathcal H)$ be given. The following statements are equivalent.
\begin{itemize}
\item[(i)] $\{U(t)\}_{t\in\mathbb R}$ is a strongly continuous one-parameter unitary group.
\item[(ii)] There exists a self-adjoint operator $A$ on $\mathcal H$ such that $U(t)=e^{-itA}$ for all $t\in\mathbb R$.
\end{itemize}
In this case the differential equation
$$
\frac{d}{dt}U(t)\psi =-iAU(t)\psi
$$
holds for all $\psi\in D(A)$. The generator $A$ can be chosen to be bounded if and only if the one-parameter group is norm continuous.
\end{lemma}
\begin{proof}
\cite[Thm.~VIII.7 \& VIII.8]{ReedSimonI}. The additional statement is shown, e.g., in \cite[Thm.~13.35 \& 13.36]{Rudin91}. Note that the statement (i) $\Rightarrow$ (ii) is also known as ``Stone's theorem''.\index{theorem!Stone's} 
\end{proof}
This suggests that to any closed system one can associate a self-adjoint operator $H$ (called ``Hamiltonian'')\index{Hamiltonian} such that every state $\rho\in\mathbb D(\mathcal H)$ of the system evolves in time via $\rho(t)=e^{-itH}\rho e^{itH}$ for all $t\geq 0$. Actually if one leaves the projective representation for a moment and goes back to state vectors $\psi\in\mathcal H$ this recovers Schr\"odinger's equation\index{Schr{\"o}dinger equation} \cite[Ch.~3.1]{BreuPetr02}
$$
\frac{d}{dt}\psi(t)=-iH(t)\psi(t)\qquad \text{ with }\psi(0)=\psi_0\in D(H)
$$
for the special case $H(t)\equiv H$, that is, if the system's Hamiltonian does not change over time.

Back to states being described by trace-class operators: given any self-adjoint operator $H$ one wants to find a similar exponential generator of the associated one-parameter (semi)group of channels. For this consider the next definition:

\begin{definition}
Let $X$ be a Banach space and consider a mapping $T:[0,\infty)\to\mathcal B(X)$ such that $T(0)=\mathbbm{1}_X$ and $T(s+t)=T(s)T(t)$ for all $s,t\geq 0$. Then $\{T(t)\}_{t\geq 0}$ is called a one-parameter semigroup\index{one-parameter!semigroup}. It is called
\begin{itemize}
\item[(i)] norm continuous if, additionally, $T:[0,\infty)\to\mathcal (\mathcal B(X),\topn)$ is continuous in $0$.
\item[(ii)] strongly continuous if, additionally, $T:[0,\infty)\to\mathcal (\mathcal B(X),\tops)$ is continuous in $0$.
\end{itemize}
\end{definition}
\noindent Analogously one defines weakly continuous semigroups which, however, turn out to co{\"i}ncide with the strongly continuous semigroups \cite[Ch.~I, Thm.~5.8]{EngelNagel00}. In this case one finds constants $M\geq 1$ and $w\in\mathbb R$ such that $\|T(t)\|\leq Me^{wt}$ for all $t\geq 0$ \cite[Ch.~I, Prop.~5.5]{EngelNagel00}.\medskip

Let us first assess the norm-continuous case. As expected this guarantees the existence of a bounded generator $A\in\mathcal B(X)$ such that $T(t)=e^{tA}$ \cite[Ch.~I, Thm.~3.7]{EngelNagel00} invoking the usual exponential map. Therefore if the Hamiltonian $H$ is bounded---meaning $t\mapsto e^{-itH}$ is norm continuous---then 
\begin{align*}
\|e^{-itH}(\cdot)e^{itH}-(\cdot)\|_{\textrm{op}}&\leq \|e^{-itH}(\cdot)e^{itH}-e^{-itH}(\cdot)\|_{\textrm{op}}+\|e^{-itH}(\cdot)-(\cdot)\|_{\textrm{op}}\\
&\leq \|e^{itH}-\mathbbm{1}_{\mathcal H}\|+\|e^{itH}-\mathbbm{1}_{\mathcal H}\|\overset{t\to 0}\to 0
\end{align*}
so the semigroup of channels $\{e^{-itH}(\cdot)e^{itH}\}_{t\geq 0}$ is norm continuous. Now by simple differentiation one finds that its generator\index{generator of one-parameter semigroup} is given by the (bounded) map\label{symb_ad_rep}
\begin{align*}
-i\operatorname{ad}_H:\mathcal B^1(\mathcal H)&\to\mathcal B^1(\mathcal H)\\
A&\mapsto -i[H,A]=-i(HA-AH)\,,
\end{align*}
cf.~\cite[p.~21]{EngelNagel00} which reproduces the well-known Liouville-von Neumann equation\index{Liouville-von Neumann equation} \cite[Eq.~3.10]{BreuPetr02} $ \frac{d}{dt}\rho(t)=-i[H(t),\rho(t)] $ (again for the special case $H(t)\equiv H$). 

The assumption this analysis is based on---namely, that the system is closed---is
too inaccurate for a lot of experiments as shielding the system from its environment is often unfeasible. Luckily under the assumption of norm continuity one can fully characterize the generators of quantum-dynamical semigroups\index{quantum-dynamical semigroup} (\textsc{qds}),\label{not_qds}
i.e.~one-parameter semigroups $T:\mathbb R_+\to Q_S(\mathcal H)$:
\begin{theorem}\label{thm_gksl}
Let a complex Hilbert space $\mathcal H$ and a map $T:\mathbb R_+\to Q_S(\mathcal H)$ be given. The following statements are equivalent.
\begin{itemize}
\item[(i)] $\{T(t)\}_{t\geq 0}$ is a norm continuous quantum-dynamical semigroup.
\item[(ii)] There exists self-adjoint $H\in\mathcal B(\mathcal H)$ as well as a family $\{V_j\}_{j\in I}$ of bounded operators such that $\sum_{j\in I}V_j^*V_j\in\mathcal B(\mathcal H)$, which satisfy $T(t)=e^{tL}$ for all $t\geq 0$ where
\begin{equation}\label{eq:gksl_gen}
L(\rho)=-i[H,\rho]-\sum_{j\in I}\Big(\frac12 (V_j^*V_j\rho +\rho V_j^*V_j)-V_j\rho V_j^*\Big)\qquad\text{ for all }\rho\in\mathcal B^1(\mathcal H)\,.
\end{equation}
\end{itemize}
If the index set $I$ is infinite then $\sum_{j\in I}V_j^*V_j$ is taken in the weak operator topology so the sum in \eqref{eq:gksl_gen} converges in trace norm.
\end{theorem}
\begin{proof}
\cite[Coro.~1 \& Thm.~2]{Lindblad76}
\end{proof}
This result is due to Gorini, Kossakowski, and Sudarshan \cite{GKS76} for finite-dimensional systems and Lindblad \cite{Lindblad76} for arbitrary separable spaces; both were published in 1976 \footnote{
For a more detailed recap of these historical events we refer to \cite{ChruPas17}.
}.
This is why equation \eqref{eq:gksl_gen} is usually referred to as \textsc{gksl}-equation\label{not_gksl}\index{gksl-equation@\textsc{gksl}-equation}, as
well as \textsc{gksl}-form or standard form of the generator (of a norm-continuous \textsc{qds}),
and the operators $V_j$ in \eqref{eq:gksl_gen} are sometimes called \textit{Lindblad-$V$}.\label{symb_lindblad_V}\index{Lindblad-V@Lindblad-$V$}
 While $H$ in general is not the same as the Hamiltonian of the shielded-off system \cite[p.~822, Remark]{GKS76} the standard form still bears resemblance to the closed case analyzed before. Thus $-i\operatorname{ad}_H$ is called the \textit{Hamiltonian part} of the generator, and the rest (i.e.~$L+i\operatorname{ad}_H$) is called the \textit{dissipative part}\index{dissipative} which models the interaction of the system with the environment. 
\begin{remark}[Markovianity and Notions of Divisibility]\label{rem_markov_div}
Processes which are described by one-parameter semigroups are also called \textit{(time-independent) Markovian}\index{Markovian!time-independent}, and so is every channel $T\in Q_S(\mathcal H)$ which can be written like $T=e^{L}$ where $L$ is the generator of a strongly continuous \textsc{qds}. On one hand it is NP-hard to decide whether a given channel is Markovian \cite{Cubitt12}, and on the other there are enough applications where memory effects do occur (see \cite{Breuer16} for an overview). Although not directly relevant to the setting in this work, some notions which generalize Markovianity are infinitesimal divisibility\index{infinitesimally divisible} of channels \cite{Wolf08a} as well as P-divisibility\index{P-divisible} of dynamical processes \cite{Chru18}. In finite dimensions, notable related results are that the Markovian channels are precisely those which are bijective and infinitely divisible\index{infinitely divisible}\footnote{
A channel $T\in Q(n)$ is called infinitely divisible if for all $n\in\mathbb N$ there exists $T_n\in Q(n)$ such that $T=T_n^n$. It is known that such a channel can be written as $T=T_0 e^L$ for some \textsc{gksl}-generator $L$ and some $T_0\in Q(n)$ which satisfies $T_0^2=T_0$ and $T_0L=T_0LT_0$ \cite{Wolf08a}. Thus if an infinitely divisible $T$ is bijective then so is the product $T_0=Te^{-L}$. Hence idempotence reduces to $T_0=\mathbbm{1}$ meaning $T=e^L$ is Markovian.
},
and that the closure of the set of all time-dependent Markovian\index{Markovian!time-dependent} channels co{\"i}ncides with the closure of the set of all infinitesimal divisible channels \cite[Ch.~V]{Wolf08a}.
\end{remark}
While Thm.~\ref{thm_gksl} fully settles the norm-continuous case in terms of \textsc{qds}-generators, if we relax the continuity type then things become much more difficult as then, in general, one has to deal with unbounded operators. Indeed given a strongly continuous one-parameter semigroup $\{T(t)\}_{t\geq 0}$ on a Banach space $X$ its generator $A:D(A)\subseteq X\to X$ is given by
$$
Ax=\lim_{h\to 0^+}\frac{1}{h}(T(h)x-x)
$$
for all $x\in D(A)=\{x\in X\,|\,\lim_{h\to 0^+}\frac{1}{h}(T(h)x-x)\text{ exists}\}$. The generator is a closed, densely defined linear operator which determines the semigroup completely \cite[Ch.~II, Thm.~1.4]{EngelNagel00}. This justifies writing ``$(A,D(A))$ generates a strongly continuous one-parameter semigroup'' instead of ``there exists a (unique) strongly continuous one-parameter semigroup $\{T(t)\}_{t\geq 0}$ with generator $(A,D(A))$''. 

As we are in the fortunate situation $T(t)\in Q_S(\mathcal H)$---so $\|T(t)\|=1$ for all $t\geq 0$ (Prop.~\ref{thm_q_norm_1})---one can characterize their generators as follows:
\begin{lemma}[Hille-Yosida for Contractions]\label{lemma_hille_yosida}\index{theorem!Hille-Yosida}
For a linear operator $(A,D(A))$ on a Banach space $X$ the following are equivalent.
\begin{itemize}
\item[(i)] $(A,D(A))$ generates a strongly continuous contraction\footnote{
This means $\|T(t)\|\leq 1$ for all $t\geq 0$.
} semigroup\index{contraction semigroup} $\{T(t)\}_{t\geq 0}$.
\item[(ii)] $(A,D(A))$ is closed, densely defined, $(0,\infty)\subseteq\mathbbm{r}(A)$ (i.e.~every $\lambda>0$ is in the resolvent of $A$), and $\|\lambda (\lambda-A)^{-1}\|\leq 1$ \footnote{
Here and henceforth, expressions like $(\lambda-A)^{-1}$ where scalars and operators get mixed are short for $(\lambda\mathbbm{1}-A)^{-1}$, i.e.~$\lambda$ represents a scalar multiple of the identity.
}.
\item[(iii)] $(A,D(A))$ is closed, densely defined, dissipative\index{operator!dissipative} (i.e.~$\|(\lambda-A)x\|\geq\lambda\|x\|$ for all $\lambda>0$, $x\in D(A)$), and $\lambda-A$ is surjective for some $\lambda>0$.
\item[(iv)] $(A,D(A))$ is closed, densely defined, and every $\lambda\in\mathbb C$, $\operatorname{Re}\lambda>0$ is in the resolvent $\mathbbm{r}(A)$ of $A$ with $\| (\lambda-A)^{-1}\|\leq \frac{1}{\operatorname{Re}\lambda}$.
\end{itemize}
\end{lemma}
\begin{proof}
``(i) $\Leftrightarrow$ (ii) $\Leftrightarrow$ (iv)'': \cite[Ch.~II, Thm.~3.5]{EngelNagel00}. For the original works refer to \cite{Hille48,Yosida48}. ``(ii) $\Rightarrow$ (iii)'': Obvious. ``(iii) $\Rightarrow$ (ii)'': Because $A$ is dissipative, surjectivity of $\lambda-A$ for any $\lambda>0$ implies $(0,\infty)\subseteq\mathbbm{r}(A)$ \cite[Ch.~II, Prop.~3.14]{EngelNagel00}. This together with dissipativity readily implies the norm condition.
\end{proof}
There are more general characterizations for semigroups which are not contractive but ``only'' exponentially bounded, i.e.~$\|T(t)\|\leq e^{wt}$ for some $w\in\mathbb R$; however, as explained before this is beyond our needs. Now for a general strongly continuous semigroup one for all $x\in D(A)$ and all $t\geq 0$ finds $T(t)x\in D(A)$ with
$$
\frac{d}{dt}(T(t)x)=A(T(t)x)\,,
$$
cf.~\cite[Ch.~II, Lemma 1.3]{EngelNagel00}. This suggests an exponential relation between the semigroup and its generator:\index{generator of one-parameter semigroup}
\begin{lemma}[Post-Widder Inversion Formula]\label{lemma_post_widder}\index{Post-Widder inversion formula}
For every strongly continuous one-parameter semigroup $\{T(t)\}_{t\geq 0}$ on a Banach space $X$ with generator $(A,D(A))$ one has
$$
T(t)x=\lim_{n\to\infty} \Big(1-\frac{t}{n}A\Big)^{-n}x
$$
for all $x\in X$. The convergence is uniform in $t$ on compact intervals.
\end{lemma}
\begin{proof}
\cite[Ch.~III, Coro.~5.5]{EngelNagel00}
\end{proof}
\noindent So while the usual exponential series $e^t=\sum_{n=0}^\infty \frac{t^n}{n!}$ might not be sufficiently defined anymore when plugging in the generator $A$, the basic formula $e^t=\lim_{n\to\infty}(1-\frac{t}{n})^{-n}$ does the job---this justifies the \textit{formal} expression $T(t)=e^{tA}$ we will use occasionally. 
\begin{remark}
Combining this lemma with the Hille-Yosida theorem we now have a complete characterization of the generators---although without explicit form---of strongly continuous contraction semigroups, and we even have a way to recover the semigroup just from the generator.
\end{remark}
To apply this to \textsc{qds} let us for the moment revert to the easier case of closed systems\index{closed quantum system}. 
For any self-adjoint operator $H$, following Lemma \ref{lemma_schatten_p_approx} \& \ref{lemma_stone} one sees that strong continuity of the unitary (semi)group $\{e^{-itH}\}_{t\in\mathbb R}$ implies strong continuity of the one-parameter semigroup $\{T(t)\}_{t\geq 0}$, $T(t)=e^{-itH}(\cdot)e^{itH}$. Thus we may ask if---and if so, how---the form of the generator $-i\operatorname{ad}_H$ from the bounded case carries over:
\begin{lemma}\label{lemma_ad_H_unbounded}
Given a self-adjoint operator $(H,D(H))$ on a complex Hilbert space $\mathcal H$ define
\begin{align*}
\operatorname{ad}_H:D(\operatorname{ad}_H)\subseteq \mathcal B^1(\mathcal H)&\to\mathcal B^1(\mathcal H)\\
\rho&\mapsto H\rho-\rho H
\end{align*}
with domain
\begin{align*}
D(\operatorname{ad}_H)=\{\rho\in\mathcal B^1(\mathcal H)\,|\,&\rho(D(H))\subseteq D( H)\text{ and }H\rho-\rho H\text{ is norm bounded }\\
&\text{with an extension to a trace-class operator on }\mathcal H\}\,.
\end{align*}
Then the generator of the strongly continuous semigroup $\{T(t)\}_{t\geq 0}$, $T(t)=e^{-itH}(\cdot)e^{itH}$ is given by $-i\operatorname{ad}_H$.
\end{lemma}
\begin{proof}
\cite[Ch.~5, Lemma 5.1]{Davies76}
\end{proof}
Therefore the Liouville-von Neumann equation from before still describes the dynamics of an isolated quantum system\index{isolated system}\index{closed quantum system}. While the differential equation in the unbounded case only holds on a dense subspace of $\mathcal B^1(\mathcal H)$ the corresponding semigroup is defined on the whole trace class. 

For the general case of open quantum systems\index{open quantum system} there is at least an abstract result which characterizes the generators of strongly continuous \textsc{qds} as is hinted at in \cite[Sec.~II.C]{OSID_Werner_17}. 
For convenience we switch from $A$ to $L$ for the generator, in spirit of the Lindbladian\index{Lindbladian} ``$L$'' from the \textsc{gksl}-theorem.
\begin{corollary}
Let $\mathcal H$ be a complex Hilbert space and $(L,D(L))$ be a linear operator on $\mathcal B^1(\mathcal H)$. The following statements are equivalent.
\begin{itemize}
\item[(i)] $(L,D(L))$ generates a strongly continuous quantum-dynamical semigroup.
\item[(ii)] $(L,D(L))$ is closed, densely defined, $(0,\infty)\subseteq\mathbbm{r}(L)$, and $\lambda(\lambda-L)^{-1}$ is \textsc{cptp} for all $\lambda>0$.
\item[(iii)] $(L,D(L))$ is closed, densely defined, trace-annihilating (i.e.~$\operatorname{tr}(L(A))=0$ for all $A\in D(L)$), $(0,\infty)\subseteq\mathbbm{r}(L)$, and $(\lambda-L)^{-1}$ is completely positive for all $\lambda>0$.
\end{itemize}
In this case $D(L)=\operatorname{im}((\lambda-L)^{-1})$ and $L(\lambda-L)^{-1}\rho=\lambda (\lambda-L)^{-1}\rho-\rho$ for any $\lambda>0$ and all $\rho\in\mathcal B^1(\mathcal H)$.
\end{corollary}
\begin{proof}
``(i) $\Rightarrow$ (ii)'': By Lemma \ref{lemma_hille_yosida} $(L,D(L))$ is closed, densely defined, and $(0,\infty)\subseteq\mathbbm{r}(L)$ holds. To show $\lambda(\lambda-L)^{-1}\in Q_S(\mathcal H)$ for all $\lambda>0$ we use the integral representation of the resolvent: given any $\lambda\in\mathbbm{r}(L)$, $\operatorname{Re}\lambda>0$ one has\footnote{
More precisely, this integral is to be read as an improper Riemann integral, i.e.~$(\lambda-L)^{-1}\rho=\int_0^\infty e^{-\lambda s}T(s)\rho\,ds$ for all $\rho\in\mathcal B^1(\mathcal H)$.
}
$(\lambda-L)^{-1}=\int_0^\infty e^{-\lambda s}T(s)\,ds$ \cite[Ch.~II, Thm.~1.10]{EngelNagel00}. Now for $\lambda>0$ we compute
\begin{align*}
\operatorname{tr}\big(\lambda(\lambda-L)^{-1}(\rho)\big)=\int_0^\infty\lambda e^{-\lambda s}\underbrace{\operatorname{tr}( e^{tL}(\rho))}_{=\operatorname{tr}(\rho)}\,ds=\operatorname{tr}(\rho)\underbrace{\int_0^\infty\lambda e^{-\lambda s}\,ds}_{=1}=\operatorname{tr}(\rho)
\end{align*}
for all $\rho\in\mathcal B^1(\mathcal H)$. Analogously one shows complete positivity of $\lambda(\lambda-L)^{-1}$.

``(ii) $\Rightarrow$ (i)'': Because $\lambda(\lambda-L)^{-1}\in Q_S(\mathcal H)$ for all $\lambda>0$ we use Prop.~\ref{thm_q_norm_1} to deduce $\|\lambda(\lambda-L)^{-1}\|=1$. This together with the fact that $(L,D(L))$ is closed, densely defined, and $(0,\infty)\subseteq\mathbbm{r}(L)$ by Lemma \ref{lemma_hille_yosida} shows that $(L,D(L))$ generates a strongly continuous semigroup. Now all we have to show is that this semigroup is \textsc{cptp} at all times. Let $t>0$. By assumption $\frac{n}{t}(\frac{n}{t}-L)^{-1}=(1-\frac{t}{n}L)^{-1}\in Q_S(\mathcal H)$ for all $n\in\mathbb N$, but the quantum channels form a semigroup (Prop.~\ref{thm_monoid} (i)) so $(1-\frac{t}{n}L)^{-n}\in Q_S(\mathcal H)$ for all $n\in\mathbb N$. Indeed this sequence $((1-\frac{t}{n}L)^{-n})_{n\in\mathbb N}$ converges to the generated semigroup $e^{tL}$ in the strong operator topology (Lemma \ref{lemma_post_widder}) so closedness of $(Q_S(\mathcal H),\tops)$ (again Prop.~\ref{thm_monoid} (i)) lets us conclude $e^{tL}\in Q_S(\mathcal H)$ where $t>0$ was chosen arbitrarily. 

``(iii) $\Leftrightarrow$ (ii)'': Let $\lambda>0$. Because $\lambda$ is in the resolvent, that is, $\lambda-L:D(L)\to\mathcal B^1(\mathcal H)$ is bijective with bounded inverse, given any $A\in\mathcal B^1(\mathcal H)$ one finds (unique) $B\in D(L)$ such that $A=(\lambda-L)(B)$. By assumption $\operatorname{tr}((\lambda-L)(B))=\lambda\operatorname{tr}(B)$ so
\begin{align*}
\operatorname{tr}(\lambda(\lambda-L)^{-1}(A))=\lambda\operatorname{tr}(B)=\operatorname{tr}((\lambda-L)(B))=\operatorname{tr}(A)\,.
\end{align*}
The converse is shown analogously. 

Now the additional statement is obvious from $(0,\infty)\subseteq\mathbbm{r}(L)$ meaning the inverse $(\lambda-L)^{-1}:\mathcal B^1(\mathcal H)\to D(L)$ is surjective.
\end{proof}
\begin{remark}\label{rem_unbounded_lindblad_genV}
This corollary tells us that---assuming sufficient care regarding domain considerations---the standard form can still be used to describe open system dynamics. An example of this is given by the quantum harmonic oscillator\index{quantum harmonic oscillator} with multi-photon drive and damping without drift \cite{Azouit16} among other more standard examples like the quantum birth process\index{quantum birth process} \cite[Ch.~3.2]{OSID_Werner_17}. Sufficient ``simple-to-verify'' conditions on the generators of a \textsc{qds} are given, e.g., in \cite{Chebotarev98}.
Unfortunately, however, there is no hope obtaining a result like Thm.~\ref{thm_gksl} for the strongly continuous case: there exist generators of quantum-dynamical semigroups which are not of standard form \cite[Ch.~4]{OSID_Werner_17}. For a broader overview on unbounded \textsc{qds}-generators we refer to \cite[p.~110]{AlickiLendi07}.
\end{remark}

A common technique, e.g., in quantum mechanics is to add a bounded term to an unbounded generator of a semigroup. The following result can (in a more general form) be found in \cite[Ch.~5, Thm.~5.2]{Davies76}.
\begin{proposition}\label{prop_bounded_pert}
Let self-adjoint $(H,D(H))$ on a complex Hilbert space $\mathcal H$ as well as $(V_j)_{j\in I}\subset\mathcal B(\mathcal H)$ be given. If $I$ is infinite then assume that $\sum_{j\in I}V_j^*V_j$ converges to a bounded operator in the weak operator topology. Recalling the domain $D(\operatorname{ad}_H)$ from Lemma \ref{lemma_ad_H_unbounded},
\begin{align*}
L:D(\operatorname{ad}_H)&\to\mathcal B^1(\mathcal H)\\
\rho&\mapsto-i[H,\rho]-\sum_{j\in I}\Big(\frac12 (V_j^*V_j\rho +\rho V_j^*V_j)-V_j\rho V_j^*\Big)
\end{align*}
is the generator of a strongly continuous quantum-dynamical semigroup.
\end{proposition}
\begin{proof}
Because $(-i\operatorname{ad}_H,D(\operatorname{ad}_H))$ generates a strongly continuous contraction semigroup (Lemma \ref{lemma_ad_H_unbounded}) and because $-\Gamma(\rho):=-\sum_{j\in I}(\frac12 (V_j^*V_j\rho +\rho V_j^*V_j)-V_j\rho V_j^*)$\label{symb_gamma_qds}
is bounded\footnote{This follows from the assumed boundedness of $\sum_{j\in I}V_j^*V_j$ as well as (complete) positivity of the well-defined map $\rho\mapsto\sum_{j\in I}V_j\rho V_j^*$ (\cite[Prop.~6.3]{Attal6} \& Lemma \ref{lemma_pos_cont}).
}
and dissipative (Thm.~\ref{thm_gksl} \& Lemma \ref{lemma_hille_yosida} (iii)) their sum $L=-i\operatorname{ad}_H-\Gamma$ with domain $D(\operatorname{ad}_H)$ generates a strongly continuous contraction semigroup $\{e^{tL}\}_{t\geq 0}$ \cite[Ch.~III, Thm.~2.7]{EngelNagel00}. Thus we may apply the Trotter product formula (Lemma \ref{lemma_trotter_banach}) to find $\lim_{n\to\infty}\| e^{tL}(\rho)-(e^{-it\operatorname{ad}_H/n}e^{-t\Gamma/n})^n(\rho) \|_1=0$ for all $t\geq 0$, $\rho\in\mathcal B^1(\mathcal H)$. All that is left to show is that $e^{tL}$ is \textsc{cptp} at all times.

Let $t>0$. By Lemma \ref{lemma_ad_H_unbounded} $-i\operatorname{ad}_H$ generates unitary channels, and $e^{-t\Gamma}$ is \textsc{cptp} at all times by the \textsc{gksl}-theorem. Thus $(e^{-it\operatorname{ad}_H/n}e^{-t\Gamma/n})^n\in Q_S(\mathcal H)$ for all $n\in\mathbb N$ as the quantum channels form a semigroup (Prop.~\ref{thm_monoid} (i)). But $Q_S(\mathcal H)$ is closed in the strong operator topology so the (s.o.t.-)limit of $((e^{-it\operatorname{ad}_H/n}e^{-t\Gamma/n})^n)_{n\in\mathbb N}$---that is, $e^{tL}$---is in $ Q_S(\mathcal H)$, as well.
\end{proof}

While this approach (called ``bounded perturbation''\index{bounded perturbation}) certainly is useful on a handful of occasions---and will be of great use for us later---one hits the brick wall as soon as one wants to model a system with an unbounded Hamiltonian $H$ \textit{as well as} an unbounded Lindblad-$V$ (e.g., a ladder operator\index{operator!ladder}).

\chapter{Quantum Control Systems}

Being able to describe the dynamics of a quantum system is not where the story ends; indeed ``it is control that turns scientific knowledge into useful technology'' \cite{Roadmap2015}. Thus a fundamental question for applications is how to influence a quantum system to one's liking, and how to describe these manipulations in a mathematically rigorous manner. 
Breaking down this question we find the following three fundamental aspects of it:
\begin{itemize}
\item \textit{(State) Controllability:}\index{controllable} given an initial state of a system and a set of controls to choose from at any point in time, can one steer to a given target state?
\item \textit{Approximation:} If generating a target state cannot be achieved exactly can one at least ensure that such a transition can be done up to some (arbitrary) precision?
\item \textit{Control Design:} once it is ensured that a target state can be reached (approximately) from some initial state how can this transition be done (e.g., time-)optimally or in a robust manner (e.g., robust against noise)?
\end{itemize}
The second point is what adjusts the question of state controllability to an engineering perspective as generating states with arbitrary precision is perfectly reasonable for applications.
Moreover, in particular for infinite-dimensional systems and subsequent numerical considerations, one has to consider truncations to finite-dimensional subspaces. While this does of course not turn the \textit{whole} problem into a finite-dimensional one it emphasizes the necessity of factoring in approximability when asking about generating quantum states or synthesizing unitary gates.

\section{Topological Considerations on $\mathcal B(\mathcal H,\mathcal G)$}\label{ch_3_1_top_con}

From a mathematical point of view such approximations mean that one does not consider the set of states (or gates) one can reach, but rather their closure. This leads us back to topology and related notions such as continuity and separability. Let us investigate which topologies are suited for approximations via finite-dimensional projections in order to rigorously incorporate these aspects into our control theoretic considerations later on. Recall that
\begin{itemize}
\item[$\bullet$] convergence in the strong operator topology is pointwise convergence (i.e.~$T_i\to T$ in $(\mathcal B(\mathcal H,\mathcal G),\tops)$ if and only if $T_ix\to Tx$ for all $x\in\mathcal H$)
\item[$\bullet$] convergence in the weak operator topology is convergence of matrix elements (i.e.~$T_i\to T$ in $(\mathcal B(\mathcal H,\mathcal G),\topw)$ if and only if $\langle y,T_ix\rangle\to \langle y,Tx\rangle$ for all $x\in\mathcal H,y\in\mathcal G$) 
\item[$\bullet$] $\topw$ is weaker than $\tops$ on $\mathcal B(\mathcal H,\mathcal G)$ with equality if and only if $\mathcal G$ is finite-dimensional
\item[$\bullet$] $\tops$ is weaker than the (operator) norm topology on $\mathcal B(\mathcal H,\mathcal G)$ with equality if and only if $\mathcal H$ is finite-dimensional
\end{itemize}
by Prop.~\ref{prop_strong_weak_op_top} (together with Lemma \ref{lemma_riesz_frechet}). To shorten the proofs a bit we will repeatedly use those facts in this as well as the next chapter without further reference. Also recall that $\topn$ is short for the operator norm topology.

\begin{lemma}\label{lemma_continuity_adjoint}
Let $\mathcal H,\mathcal G$ arbitrary Hilbert spaces. The following statements hold.
\begin{itemize}
\item[(i)] ${}^*:(\mathcal B(\mathcal H,\mathcal G),\topn)\to(\mathcal B(\mathcal G,\mathcal H),\topn)$, $T\mapsto T^*$ is continuous.
\item[(ii)] ${}^*:(\mathcal B(\mathcal H,\mathcal G),\tops)\to(\mathcal B(\mathcal G,\mathcal H),\tops)$, $T\mapsto T^*$ is continuous if and only if $\operatorname{dim}(\mathcal H)<\infty$.
\item[(iii)] ${}^*:(\mathcal B(\mathcal H,\mathcal G),\topw)\to(\mathcal B(\mathcal G,\mathcal H),\topw)$, $T\mapsto T^*$ is continuous.
\end{itemize}
\end{lemma}
\begin{proof}
The key idea here will be Lemma \ref{lemma_topol_connect} (i), that is, to look at the image of generalized sequences (nets) of operators under ${}^*$.\smallskip

(i): Consider a sequence $(T_n)_{n\in\mathbb N}$ in $\mathcal B(\mathcal H,\mathcal G)$ which norm-converges to $T$. Then using the properties of ${}^*$, obviously $\|T_n^*-T^*\|=\|(T_n-T)^*\|=\|T_n-T\|\to 0$ as $n\to\infty$ so ${}^*$ is norm-continuous.\smallskip

(ii): ``$\Leftarrow$'': Let $\mathcal H$ be finite-dimensional so $(\mathcal B(\mathcal H,\mathcal G),\tops)=(\mathcal B(\mathcal H,\mathcal G),\topn)$ and $(\mathcal B(\mathcal G,\mathcal H),\tops)=(\mathcal B(\mathcal G,\mathcal H),\topw)$. Thus we have to show that ${}^*:(\mathcal B(\mathcal H,\mathcal G),\topn)\to(\mathcal B(\mathcal G,\mathcal H),\topw)$ is continuous. But $\topw$ is weaker than the norm topology so $\mathbbm{1}_{\topn\to\topw}:(B(\mathcal G,\mathcal H),\topn)\to(B(\mathcal G,\mathcal H),\topw)$ is continuous (Prop.~\ref{prop_compare_topo}) and ${}^*$ is a composition of  $\mathbbm{1}_{\topn\to\topw}$ and ${}^*_{\topn\to\topn}$, i.e.~a composition of two continuous maps (by (i)), hence continuous itself.

``$\Rightarrow$'': Let $\mathcal H$ be infinite-dimensional so we have to show that ${}^*$ is not continuous w.r.t.~$\tops$. By Prop.~\ref{prop_hilbert_space_basis} we can find an orthonormal basis $(e_i)_{i\in I}$ of $\mathcal H$---where $I$ by assumption is infinite---from which we can select a countable orthonormal system $(e_{i_n})_{n\in\mathbb N}$. Also one finds $y\in\mathcal G$, $\|y\|=1$ so for all $n\in\mathbb N$ we may define $T_n\in\mathcal B(\mathcal H,\mathcal G)$ via $T_n(x):=\langle e_{i_n},x\rangle y$ for all $x\in\mathcal H$. Again by Prop.~\ref{prop_hilbert_space_basis} we know $\|x\|^2=\sum_{i\in I}|\langle e_i,x\rangle|^2$ so Lemma \ref{lemma_unordered_summation} (iv) in particular shows $\lim_{n\to\infty}|\langle e_{i_n},x\rangle|=0$ for all $x\in\mathcal H$. In other words $\|T_n(x)\|=|\langle e_{i_n},x\rangle|\|y\|\to 0$ as $n\to\infty$ which shows $T_n\to 0$ in $\tops$. Now if ${}^*$ were $\tops$-$\tops$-continuous then $T_n^*\to 0$ in $\tops$ as well. However, $\|T_n^*y\|=\| \langle y,y\rangle e_{i_n} \|=\|y\|^2\|e_{i_n}\|=1$ for all $n\in\mathbb N$.\smallskip

(iii): Let $(T_i)_{i\in I}$ be a net in $\mathcal B(\mathcal H,\mathcal G)$ which converges to $T\in\mathcal B(\mathcal H,\mathcal G)$ in $\topw$, that is, \mbox{$|\langle y,T_ix\rangle-\langle y,Tx\rangle|\to 0$} for all $x\in\mathcal H$, $y\in\mathcal G$. But $|\langle x,T_i^*y\rangle-\langle x,T^*y\rangle|=|\langle y,T_ix\rangle-\langle y,Tx\rangle|\to 0$ so $T_i^*\to T^*$ in $\topw$, showing continuity of $\topw$. 
\end{proof}
\noindent This might seem bizarre at first because $\topw\subseteq\tops\subseteq\topn$ but continuity of ${}^*$ breaks down midway (and then recovers). However for a general map $T:X\to Y$ between topological spaces one can only transfer continuity statements if the topology on $X$ is made stronger or on $Y$ is made weaker, but not if \textit{both} topologies are made stronger (or weaker) at the same time.\smallskip

An important tool for operator approximation results are projections onto finite-dimensional subspaces:

\begin{lemma}\label{lemma_approx_strong_top}
Let $\mathcal H,\mathcal G$ be separable Hilbert spaces of infinite dimension and $(e_n)_{n\in\mathbb N}$, $(g_n)_{n\in\mathbb N}$ an arbitrary orthonormal basis of $\mathcal H,\mathcal G$, respectively. Define the operators $\Pi_n^{\mathcal H}:=\sum_{j=1}^n |e_j\rangle\langle e_j|$ and $\Pi_n^{\mathcal G}:=\sum_{j=1}^n |g_j\rangle\langle g_j|$ for all $n\in\mathbb N$. The following statements hold.
\begin{itemize}
\item[(i)] $\|\Pi_n^{\mathcal H}\|=1$ for all $n\in\mathbb N$.
\item[(ii)] $\Pi_n^{\mathcal H}\to\mathbbm{1}_{\mathcal H}$ in $(\mathcal B(\mathcal H),\tops)$. 
\item[(iii)] For all $T\in\mathcal B(\mathcal H,\mathcal G)$ one has $\Pi_n^{\mathcal H}T\Pi_n^{\mathcal G}\to T$ in $(\mathcal B(\mathcal H,\mathcal G),\tops)$ as $n\to\infty$. In other words \mbox{$T=\sum_{j,k=1}^\infty \langle g_k,Te_j\rangle |g_k\rangle\langle e_j|$} where the sum converges in $\tops$.
\end{itemize}
\end{lemma}
\begin{proof}
(i): Consider arbitrary $x\in\mathcal H$. By Lemma \ref{lemma_pyth_thm} and Prop.~\ref{prop_hilbert_space_basis} (i) we immediately get
$$
\|\Pi_n^{\mathcal H}x\|^2=\Big\|\sum\nolimits_{j=1}^n \langle e_j,x\rangle e_j\Big\|^2=\sum\nolimits_{j=1}^n |\langle e_j,x\rangle|^2\leq \|x\|^2\,.
$$
On the other hand $\|\Pi_n^{\mathcal H}\|\geq \|\Pi_n^{\mathcal H} e_1\|=1$ so indeed $\|\Pi_n^{\mathcal H}\|=1$ (where $n\in\mathbb N$ was chosen arbitrarily).\smallskip

(ii): For all $x\in\mathcal H$ one has $\Pi_n^{\mathcal H}x=\sum_{j=1}^n \langle e_j,x\rangle e_j\overset{n\to\infty}\to\sum_{j=1}^\infty \langle e_n,x\rangle e_n=x$ due to the Fourier expansion (Prop.~\ref{prop_hilbert_space_basis} (ii)) so indeed $\Pi_n\to\mathbbm{1}_{\mathcal H}$ in $\tops$.\smallskip

(iii): Again for all $x\in\mathcal H$, using (ii) we get
\begin{align*}
\|\Pi_n^{\mathcal G}T\Pi_n^{\mathcal H}x-Tx\|&\leq\|\Pi_n^{\mathcal G}T\Pi_n^{\mathcal H}x-\Pi_n^{\mathcal G}Tx\|+\|\Pi_n^{\mathcal G}Tx-Tx\|\\
&\leq\underbrace{\|\Pi_n^{\mathcal H}\|}_{=1\text{ for all }n\in\mathbb N}\underbrace{\|T\|}_{<\infty}\|\Pi_n^{\mathcal H}x-x\|+\|\Pi_n^{\mathcal G}(Tx)-Tx\|\overset{n\to\infty}\to 0\,.\qedhere
\end{align*}
\end{proof}
\noindent Recalling the start of Chapter \ref{sec_top_B_X_Y} this is one thing we wanted to fix because the sequence of projections $(\Pi_n)_{n\in\mathbb N}$ does not converge in norm (as it is not a Cauchy sequence: For all $m>n$ one has $\|\Pi_m-\Pi_n\|\geq\|\Pi_m(e_m)-\Pi_n(e_m)\|=1$).\smallskip

Either way this lets us answer the question of whether separability of the underlying Hilbert spaces transfers to the operator space; first for the norm topology:

\begin{proposition}\label{prop_bounded_op_norm_sep}
Let $\mathcal G,\mathcal H$ be non-trivial Hilbert spaces. Then $\mathcal B(\mathcal H,\mathcal G)$ is norm-separable if and only if $\mathcal H$ is finite dimensional and $\mathcal G$ is separable or vice versa. Thus $\mathcal B(\mathcal H)$ is norm-separable if and only if $\mathcal H$ is finite-dimensional.
\end{proposition}
\begin{proof}
A normed space can either be (a) finite-dimensional, (b) infinite-dimensional and separable or (c) non-separable, so as this applies to domain and codomain there are nine classes $\mathcal B(\mathcal H,\mathcal G)$ can fall into. This allows us to subdivide the proof into the three following steps.\smallskip

\textit{Step 1:} If $\mathcal H$ is separable and $\mathcal G$ is finite-dimensional or vice versa, then $\mathcal B(\mathcal H,\mathcal G)$ is separable.\smallskip

\noindent If $\mathcal H,\mathcal G$ are both of finite dimension $n\in\mathbb N$ then $\mathcal B(\mathcal H,\mathcal G)=\mathcal L(\mathcal H,\mathcal G)\simeq\mathbb C^{n\times n}$ (or $\mathbb R^{n\times n}$) is finite-dimensional, hence norm-separable (Lemma \ref{lemma_separable_linear_span}). Now let $\mathcal G$ be infinite-dimensional and separable (with orthonormal basis $(g_n)_{n\in\mathbb N}$) and $\mathcal H$ be finite-dimensional (with orthonormal basis $(e_n)_{n=1}^m$). Then $(\mathcal B(\mathcal H,\mathcal G),\tops)=(\mathcal B(\mathcal H,\mathcal G),\topn)$ so for any $T\in\mathcal B(\mathcal H,\mathcal G)$ we by Lemma \ref{lemma_approx_strong_top} know that $\Pi_n^{\mathcal G}\circ T\overset{n\to\infty}\to T$ in norm. By definition of $\Pi_n^{\mathcal G}$ as well as the Fourier expansion on $\mathcal H$
$$
\Pi_n^{\mathcal G}\circ T=\Pi_n^{\mathcal G}\circ T\circ\mathbbm{1}_{\mathcal H}=\sum\nolimits_{j=1}^n\sum\nolimits_{k=1}^m \langle g_j,Te_k\rangle |g_j\rangle\langle e_k|\in \operatorname{span}\{|g_j\rangle\langle e_k|\,|\,k=1,\ldots,m\,,\,j\in\mathbb N\}
$$
which converges in norm to $T$ as shown before. Because $T$ was chosen arbitrarily we find that $\operatorname{span}\{|g_j\rangle\langle e_k|\,|\,k=1,\ldots,m\,,\,j\in\mathbb N\}$ is indeed norm-dense in $\mathcal B(\mathcal H,\mathcal G)$. But this again by Lemma \ref{lemma_separable_linear_span} proves separability. Finally if the roles of $\mathcal H$ and $\mathcal G$ are reversed, i.e.~$\mathcal H$ is infinite-dimensional and separable $\mathcal G$ is finite-dimensional, then norm-separability transfers from $\mathcal B(\mathcal G,\mathcal H)$ to $\mathcal B(\mathcal H,\mathcal G)$ via the conjugate-linear bijective isometry ${}^*$ as can be easily seen.\smallskip

\textit{Step 2:} If $\mathcal G,\mathcal H$ are infinite-dimensional and separable then $\mathcal B(\mathcal H,\mathcal G)$ is not separable.\smallskip

\noindent Again we find an orthonormal basis $(e_n)_{n\in\mathbb N}$ of $\mathcal H$ and $(g_n)_{n\in\mathbb N}$ of $\mathcal G$. Given any $z\in\ell^\infty(\mathbb N)$ consider $(z_ng_n)_{n\in\mathbb N}$ which is a family of pairwise orthogonal vectors with $\sup_{n\in\mathbb N}\|z_ng_n\|=\|z\|_\infty<\infty$ so by Lemma \ref{lemma_unitary_ONB} (i) one finds unique $T_z\in\mathcal B(\mathcal H,\mathcal G)$ which maps $(e_n)_{n\in\mathbb N}$ to $(z_ng_n)_{n\in\mathbb N}$ with $\|T_z\|=\|z\|_\infty$. Indeed the map $z\mapsto T_z$ is linear as is readily verified. Thus the proof that $\ell^\infty(\mathbb N)$ is non-separable (Ex.~\ref{ex_ell_p_space}, footnote \ref{footnote_sequence_space_separable}) transfers: The set $\{T_z\}_{z\in\{0,1\}^{\mathbb N}}$\footnote{$\{0,1\}^{\mathbb N}\subset\ell^\infty(\mathbb N)$ is the subset of all sequences with values in $\{0,1\}$.\label{footnote_0_1_N}} is uncountable \cite[Thm.~2.14]{Rudin76} and for $z_1,z_2\in\{0,1\}^{\mathbb N}$ with $z_1\neq z_2$ one has $\|T_{z_1}-T_{z_2}\|=\|T_{z_1-z_2}\|=\|z_1-z_2\|=1$. Therefore $\{B_{1/2}(T_z)\}_{z\in\{0,1\}^{\mathbb N}}$ is an uncountable set of disjoint open balls, showing non-separability of $\mathcal B(\mathcal H,\mathcal G)$ in this case (Lemma \ref{lemma_non_sep}).\smallskip

\textit{Step 3:} If $\mathcal H$ or $\mathcal G$ is non-separable then $\mathcal B(\mathcal H,\mathcal G)$ is non-separable.\smallskip

\noindent W.l.o.g.~$\mathcal H$ is non-separable \& $\mathcal G$ arbitrary, but non-trivial. Thus we find $y\in \mathcal G$, $\|y\|=1$ as well as an orthonormal basis $\{e_i\}_{i\in I}$ of $\mathcal H$ where $I$ is uncountable (Prop.~\ref{prop_hilbert_space_basis} (iv)). This lets us define $T_i:=|y\rangle\langle e_i|\in \mathcal B(\mathcal H,\mathcal G)$ so for all $i\neq j$, using Lemma \ref{lemma_pyth_thm} we get
$$
\|T_i-T_j\|^2=\||y\rangle\langle e_i-e_j|\|^2=\|e_i-e_j\|^2 \|y\|^2=\|e_i\|^2+\|e_j\|^2=2>1\,,
$$
i.e.~$B_{1/2}(T_i)\cap B_{1/2}(T_j)=\emptyset$ for all $i\neq j$. Thus we found an uncountable family $\{B_{1/2}(T_i)\}_{i\in I}$ of disjoint open sets which shows that $\mathcal B(\mathcal H,\mathcal G)$ is not separable (again by Lemma \ref{lemma_non_sep}). 
\end{proof}
\begin{remark}
When considering separability of bounded operators between general Banach spaces things get messier: As an example consider the sequence spaces $c_0(\mathbb N),\ell^1(\mathbb N)$ from Ex.~\ref{ex_ell_p_space} and arbitrary $T\in\mathcal B(c_0(\mathbb N),\ell^1(\mathbb N))$. Then $T$ is compact by Pitt's theorem\index{theorem!Pitt's} \cite{Pitt79,Delpech09}. In particular---because $c_0(\mathbb N),\ell^1(\mathbb N)$ are Banach spaces with a Schauder basis---$T$ can be norm-approximated by operators of finite-rank so $\mathcal B(c_0(\mathbb N),\ell^1(\mathbb N))$ is norm-separable. Also be aware that this statement is non-trivial as $\operatorname{dim}(\mathcal B(c_0(\mathbb N),\ell^1(\mathbb N)))=\infty$ because every $(x,y)\in\ell^1(\mathbb N)\times \ell^1(\mathbb N)$ gives rise to a bounded operator $T_{x,y}$ via $T_{x,y}(z):=\Psi_x(z)y$ where $\Psi:\ell^1(\mathbb N)\to(c_0(\mathbb N))^*$ is the isometric isomorphism from Ex.~\ref{ex_ell_p_space_2}.
\end{remark}

The most important difference in Prop.~\ref{prop_bounded_op_norm_sep} between the norm and the strong operator topology concerns the case of both spaces being infinite-dimensional but separable.

\begin{corollary}\label{coro_bounded_op_strong_sep}
Let $\mathcal G,\mathcal H$ be separable Hilbert spaces
. Then $(\mathcal B(\mathcal H,\mathcal G),\tops)$ and $(\mathcal B(\mathcal H,\mathcal G),\topw)$ are separable. In particular $(\mathcal B(\mathcal H),\tops),(\mathcal B(\mathcal H),\topw)$ is separable for all separable Hilbert spaces $\mathcal H$.
\end{corollary}
\begin{proof}
If either of the Hilbert spaces is finite-dimensional (and the other one is still separable) then we know that $\mathcal B(\mathcal H,\mathcal G)$ is norm-separable by Prop.~\ref{prop_bounded_op_norm_sep} so we immediately have separability in the weaker topologies $\tops$ and $\topw$ (Lemma \ref{lemma_separable_topology_comp} (ii)). 

Now let $\mathcal G,\mathcal H$ both be infinite-dimensional and separable with respective orthonormal basis $(g_n)_{n\in\mathbb N}$, $(e_n)_{n\in\mathbb N}$. One has $\Pi_n^{\mathcal G}T\Pi_n^{\mathcal H}\in\operatorname{span}\{|g_k\rangle\langle e_j|\}_{j,k\in\mathbb N}$ for all $n\in\mathbb N$ with $\Pi_n^{\mathcal G}T\Pi_n^{\mathcal H}\to T$ in $\tops$ (Lemma \ref{lemma_approx_strong_top} (iii)). Then separability follows from Lemma \ref{lemma_separable_linear_span} which we are allowed to apply because $(\mathcal B(\mathcal H,\mathcal G),\tops)$ is a locally convex space (Prop.~\ref{prop_strong_weak_op_top} (v)) so in particular it is a topological vector space, cf.~Lemma \ref{lemma_locally_convex} ff.
\end{proof}

\section{The Unitary Group}\label{ch_3_2_unitary_gr}
Unitary operators are of fundamental importance as they describe the evolution of isolated quantum systems (Section \ref{ch:qu_dyn_sys}) and they model how external electro-magnetic fields influence, e.g., an atom or a molecule; more on that later. This warrants studying such operators further, also in light of the topological and approximation questions posed previously.
Indeed we learned that for approximation problems on $\mathcal B(\mathcal H)$ the strong operator topology is a better choice than the operator norm, and it turns out that on a separable (complex) Hilbert space one can approximate every unitary operator via a sequence of unitary matrices embedded into $\mathcal B(\mathcal H)$:
\begin{lemma}\label{lemma_unitary_approx_matrices}
Let an infinite-dimensional separable complex Hilbert space $\mathcal H$, and any orthonormal basis $(e_n)_{n\in\mathbb N}$ of $\mathcal H$ be given. Consider the map
$$
J_n:\mathbb C^{n\times n}\to\mathcal B(\mathcal H)\quad\text{ defined via }\quad J_n(A):=\sum\nolimits_{j,k=1}^n a_{jk}|e_j\rangle\langle e_k|
$$
for all $A=(a_{jk})_{j,k=1}^n\in\mathbb C^{n\times n}$. The following statements hold.
\begin{itemize}
\item[(i)] $J_n$ is well-defined, linear, and satisfies $\|J_n\|\leq 1$ for all $n\in\mathbb N$, when $\mathbb C^{n\times n}
$ is equipped with the usual operator norm (under the identification $\mathbb C^{n\times n}\simeq\mathcal B(\mathbb C^n)$ on the Hilbert space $\mathbb C^n$).
\item[(ii)] For every $U\in\mathcal U(\mathcal H)$ there exists a sequence $( U_n)_{n\in\mathbb N}$ with $ U_n\in\mathbb C^{2n\times 2n}$ unitary such that $J_{2n}( U_n)\to U$ in $\tops$.
\end{itemize}
\end{lemma}
\begin{proof}[Proof idea]
(i): Note that $J_n$ is obviously well-defined and linear
. To see that $J_n$ is a contraction we use that $J_n(\cdot)=\Gamma_n(\cdot)\Gamma_n^*$ where $\Gamma_n\in\mathcal L(\mathbb C^n,\mathcal H)$ maps the standard basis $(\hat e_i)_{i=1}^n$ of $\mathbb C^n$ to $(e_i)_{i=1}^n\subset\mathcal H$ (cf.~Ch.~\ref{sec:c_num_range}) so in particular $\Gamma_nx=\sum_{j=1}^n x_je_j$ for all $x\in\mathbb C^n$. Using the usual euclidean norm (which turns $\mathbb C^n$ into a Hilbert space as desired) we by Lemma \ref{lemma_pyth_thm} get
\begin{align*}
\|\Gamma_n\|^2=\sup_{x\in\mathbb C^n,\|x\|^2=1}\|\Gamma_nx\|^2 &=\sup_{x\in\mathbb C^n,\sum_{j=1}^n|x_j|^2=1}\Big\|\sum_{j=1}^n x_je_j\Big\|^2=\sup_{x\in\mathbb C^n,\sum_{j=1}^n|x_j|^2=1}\sum_{j=1}^n |x_j|^2=1\,.
\end{align*}
This for all $A\in\mathbb C^{n\times n}$ shows
$$
\|J_n(A)\|=\|\Gamma_nA\Gamma_n^*\|\leq\|\Gamma_n\|\|A\|\|\Gamma_n^*\|=\|\Gamma_n\|^2\|A\|=\|A\|\,.
$$ so $\|J_n\|\leq 1$ as claimed.

(ii): The idea is to ``cut out'' the upper left $n\times n$ corner of (the matrix representation with respect to $(e_n)_{n\in\mathbb N}$) of $U$. This matrix can be ``completed'' to a larger unitary matrix $ U_n\in\mathbb C^{2n\times 2n}$ and embedded into $\mathcal B(\mathcal H)$ via $J_{2n}$. Convergence in $\tops$ then follows from direct computation. For a full proof see Lemma \ref{U-approximation}.
\end{proof}

Sometimes it is more desirable to get such an approximation but with unitary operators instead of ``just'' embedded unitary matrices (the mere embedding of which is of course not unitary). With relatively little effort we obtain the following result.

\begin{corollary}\label{coro_unitary_approx_block_unitary}
Let $\mathcal H$ be an infinite-dimensional separable complex Hilbert space and $(e_n)_{n\in\mathbb N}$ any orthonormal basis of $\mathcal H$. Consider the map
$$
J'_n:\mathbb C^{n\times n}\to\mathcal B(\mathcal H)\qquad A\mapsto J_n(A)+\mathbbm{1}_{\mathcal H}-\Pi_n\,.
$$
Here $\Pi_n$ is the projection onto the first $n$ elements of $(e_n)_{n\in\mathbb N}$ from Lemma \ref{lemma_approx_strong_top} and $J_n$ is the map from Lemma \ref{lemma_unitary_approx_matrices}. The following statements hold.
\begin{itemize}
\item[(i)] For all $n\in\mathbb N$ if $U\in\mathbb C^{n\times n}$ is unitary, then $J_n'(U)\in\mathcal U(\mathcal H)$.
\item[(ii)] Let any $U\in\mathcal U(\mathcal H)$ be given. Then there exists a sequence $( U_n)_{n\in\mathbb N}$ with $ U_n\in\mathbb C^{2n\times 2n}$ unitary such that $J'_{2n}( U_n)\to U$ in $\tops$.
\end{itemize}
In other words every $U\in\mathcal U(\mathcal H)$ on a separable Hilbert space can be strongly approximated by a sequence of unitary operators the ``main information'' of which reduces to a $2n\times 2n$ unitary matrix (in the ``top left corner''). 
\end{corollary}
\begin{proof}
(i): Obviously $J'_n$ is well-defined. Given $U=(u_{jk})_{j,k\in\mathbb N}\in\mathbb C^{n\times n}$ unitary one readily verifies $J_n'(U)e_j=\sum_{k=1}^n u_{kj} e_k $ for all $j=1,\ldots,n$ as well as $J_n'(U)e_j=e_j$ for all $j>n$. If we can show that
\begin{align}\label{eq:image_orthonormal_set}
\Big\{\sum\nolimits_{k=1}^n u_{k1}e_k\,,\,\ldots\,,\,\sum\nolimits_{k=1}^n u_{kn}e_k\,,\,e_{n+1}\,,\,e_{n+2}\,,\,\ldots\Big\}
\end{align}
is an orthonormal basis of $\mathcal H$ then $J_n'(U)$ is unitary by Lemma \ref{lemma_unitary_ONB} (ii) and we are done. For orthonormality of \eqref{eq:image_orthonormal_set} the only non-trivial thing is $\langle J_n'(e_j),J_n'(e_k)\rangle=\delta_{jk}$ for all $j,k=1,\ldots,n$. But this is obvious from
\begin{align*}
\langle J_n'(e_j),J_n'(e_k)\rangle=\sum\nolimits_{\alpha,\beta=1}^n \overline{u_{\alpha j}} u_{\beta k}\underbrace{\langle e_\alpha,e_\beta\rangle}_{=\delta_{\alpha\beta}}=\sum\nolimits_{\alpha=1}^n (U^*)_{j\alpha} U_{\alpha k}=(U^*U)_{jk}=\delta_{jk}\,.
\end{align*}
Moreover, obviously
$
\operatorname{span}\{J_n'(U)e_j\,|\,j\in\mathbb N\}=\operatorname{span}\{ e_j\,|\,j\in\mathbb N\}
$, 
so as the latter is dense in $\mathcal H$, $\{J_n'(U)e_j\}_{j\in\mathbb N}$ is an orthonormal basis of $\mathcal H$ by Prop.~\ref{prop_hilbert_space_basis} (ii) as desired.\smallskip

(ii): Let $U\in\mathcal U(\mathcal H)$ be given and let $( U_n)_{n\in\mathbb N}$ be the sequence of unitary $2n\times 2n$ matrices from Lemma \ref{lemma_unitary_approx_matrices}. For all $x\in\mathcal H$
\begin{align*}
\|Ux-J_{2n}'( U_n)x\|&=\|Ux-J_{2n}( U_n)x\|+\|J_{2n}( U_n)x-J_{2n}'( U_n)x\|\\
&=\underbrace{\|Ux-J_{2n}( U_n)x\|}_{\to 0\text{ (Lemma \ref{lemma_unitary_approx_matrices})}}+\underbrace{\|x-\Pi_{2n}(x)\|}_{\to 0 \text{ (Lemma \ref{lemma_approx_strong_top} (iii))} } \to 0\,.
\end{align*}
Thus $( U_n)_{n\in\mathbb N}$ is the sequence of unitary matrices we were looking for.
\end{proof}

This paves the way for a detailed exploration of unitary operators, which turn out to form a group living on the unit sphere. 

\begin{theorem}\label{theorem_unitary_group_properties}
Let $\mathcal H$ be a complex Hilbert space. Then $\mathcal U(\mathcal H)$
\begin{itemize}
\item[(i)] is a subset of the unit sphere $S_1(0)=\{T\in\mathcal B(\mathcal H)\,|\,\|T\|=1\}$.
\item[(ii)] forms a group (under the usual composition of operators).\index{unitary group}
\item[(iii)] is closed in norm.
\item[(iv)] is path-connected in $\topw,\tops$, and norm.\index{path-connected}
\item[(v)] satisfies $\overline{\mathcal U(\mathcal H)}^{\,\tops}\subseteq\{T\in\mathcal B(\mathcal H)\,|\,T\text{ isometry}\}\subseteq S_1(0)$ and $\overline{\mathcal U(\mathcal H)}^{\,\topw}\subseteq \overline{B_1}(0)$.
\item[(vi)] is closed in $\tops$ if and only if $\operatorname{dim}(\mathcal H)<\infty$. The same holds for $\topw$.
\item[(vii)] is a topological group\index{topological group}\footnote{A topological group $(G,\tau)$ is a group $G$ together with a topology $\tau$ (on $G$) such that the group operations $\cdot:G\times G\to G$, $(x,y)\mapsto x\cdot y$ as well as $\iota:G\to G$, $x\mapsto x^{-1}$ are continuous (the former with respect to the product topology on $G\times G$).} when equipped with either the norm topology or with $\tops$ or with $\topw$.
\item[(viii)] is norm-separable if and only if $\operatorname{dim}(\mathcal H)<\infty$.
\item[(ix)] is separable in $\tops$ (and $\topw$) if $\mathcal H$ is separable.
\item[(x)] is metrizable (when equipped with $\tops$ and $\topw$) if $\mathcal H$ is separable.
\item[(xi)] is compact (in norm, $\tops$ or $\topw$) if and only if $\operatorname{dim}(\mathcal H)<\infty$.
\item[(xii)] consists of mean ergodic operators (i.e.~every unitary operator is mean ergodic).\index{operator!mean ergodic}
\end{itemize}
\end{theorem}
The proof of these results is rather lengthy which is why we outsourced it to Appendix \ref{subsec:unitary_group}.
\begin{remark}
\begin{itemize}
\item[(i)] Thm.~\ref{theorem_unitary_group_properties} (as well as its proof) shows that the limit of a sequence of unitaries which converges in $\tops$ may only be an isometry. This is a consequence of ${}^*$ not being continuous in $\tops$ as soon as $\mathcal H$ is infinite-dimensional (Lemma \ref{lemma_continuity_adjoint}). Similarly given an infinite-dimensional Hilbert space one can find a sequence of unitary operators which converges to $0$ in $\topw$ so the inclusions in Thm.~\ref{theorem_unitary_group_properties} (iv) cannot be strengthened.
\item[(ii)] While $(\mathcal U(\mathcal H),\tops)$ is a separable and metrizable topological group (if $\mathcal H$ is separable), the corresponding metric cannot be complete for $\operatorname{dim}(\mathcal H)=\infty$ as then $(\mathcal U(\mathcal H),\tops)$ is not closed. However, although \textit{this} metric is not complete, there still \textit{exists} a complete metric on $\mathcal U(\mathcal H)$ which generates $\tops$; this is also referred to as ``$(\mathcal U(\mathcal H),\tops)$ is a completely metrizable topological group'' as shown in \cite[Prop.~II.1]{Neeb97}. Indeed the above separability considerations then imply that $(\mathcal U(\mathcal H),\tops)$ is a Polish group\index{Polish group}\footnote{A Polish group is a topological group which is separable and its topology is completely metrizable, cf.~\cite[Ch.~IV]{Fabec00}.}.
\item[(iii)] One can repair the fact that $\mathcal U(\mathcal H)$ is not closed in $(\mathcal B(\mathcal H),\tops)$ by considering the strong*-topology $\tops^*$ on $\mathcal B(\mathcal H)$ which is the weakest topology such that all the evaluation maps $\{x\mapsto Tx\,,\,x\mapsto T^*x \}_{x\in\mathcal H}$ are continuous or, equivalently, the locally convex vector space topology induced by the seminorms $\{T\mapsto \|Tx\|\,,\,T\mapsto\|T^*x\|\}_{x\in\mathcal H}$.
Evidently $\mathcal U(\mathcal H)$ is closed in $\tops^*$ is closed because for a net of unitaries which converges to $T$ in $\tops^*$ (i.e.~$U_i\to T$ and $U_i^*\to T^*$ in $\tops$) one knows that $T$ as well as $T^*$ are isometries (Thm.~\ref{theorem_unitary_group_properties} (v)) hence $T^*T=TT^*=\mathbbm{1}_{\mathcal H}$ so the limit is unitary again. Because $\topw$, $\tops$, and $\tops^*$ co{\"i}ncide on $\mathcal U(\mathcal H)$ (i.e.~the induced subspace topologies are the
same \cite[Lemma 1.5]{Espinoza14}) either of these turn the unitary group (over a separable Hilbert space) into a Polish group. An explicit proof that $(\mathcal U(\mathcal H),\tops^*)$ is a Polish group can be found in \cite[Thm.~IV.1]{Fabec00}.
\end{itemize} 
\end{remark}

Again there is a lot of information to digest here but the points to focus on are the following: Assuming separability of the Hilbert space, the unitary group equipped with the strong operator topology is not closed, but is separable and metrizable. Indeed separability---which comes from the fact that every unitary can be strongly approximated by unitaries on finite-dimensional subspaces---will be a key feature for infinite-dimensional controllability.

\section{Bilinear Control Systems}\label{ch:bilinear_control}

Influencing atoms or molecules via external forces is usually done via electro-magnetic fields as this is the only classical long-range\footnote{
This means the force decreases with distance $r$ not quicker than $r^{-(d-1)}$ with $d$ being the spatial dimension.
} force which can be manipulated sufficiently well for experiments. This leads to an adjustment of the uncontrolled system---described by a time-independent self-adjoint operator $H_0$---to a time-dependent Hamiltonian\index{Hamiltonian!time-dependent} $H(t)=H_0+\sum_{j=1}^mu_j(t) H_j$. Here $\{H_j\}_{j=1}^m$ are the (time-independent) control Hamiltonians\index{control Hamiltonian}, and $\{u_j(t)\}_{j=1}^m$ are the input functions, also called ``control functions''\index{control function} or ``control amplitudes'', taken from some suitable class of functions. Following Section \ref{ch:qu_dyn_sys} this means that the dynamics of the controlled system are, at least formally, described by the differential equation
$$
\dot\rho(t)= -i\operatorname{ad}_{H_0}(\rho(t))+\sum\nolimits_{j=1}^m u_j(t) \big(\!-i\operatorname{ad}_{H_j}\!\big)(\rho(t)) =-i\Big[H_0+\sum\nolimits_{j=1}^m u_j(t) H_j,\rho(t)\Big]
$$
with initial value $\rho(0)=\rho_0\in\mathbb D(\mathcal H)$. Thus turning the dynamics into a control problem grants access to a family of possible trajectories instead of just a single one. For more on this we refer to \cite{DiHeGAMM08,dAll08}, \cite[Ch.~2.10 \& 6.5]{Elliott09} or \cite[Ch.~3.2]{AM11}. 

Mathematically speaking this puts us in the realm of bilinear control systems\index{bilinear control system}
\cite{Jurdjevic97,Sontag,Elliott09}: 
\begin{equation}\label{eq:bilinear_1}
\dot y(t) = \Big(A + \sum\nolimits_{j=1}^{m} u_j(t) B_j\Big) y(t)\quad\text{with}\quad y(0)=y_0\in X\,,
\end{equation}
with (for now) bounded linear operators $A,B_1,\ldots,B_m\in\mathcal B(X)$ acting on a Banach space $X$, and corresponding group lift\index{group lift}
\begin{equation}\label{eq:bilinear_2}
\dot Y(t) = \Big(A + \sum\nolimits_{j=1}^{m} u_j(t) B_j\Big) Y(t)\quad\text{with}\quad Y(0)=Y_0\in\operatorname{GL}(X)\,.
\end{equation}
Here $\operatorname{GL}(X)\subset\mathcal B(X)$\label{symb_GL}
denotes the general linear group\index{general linear group} on $X$, that is, the collection of all bounded linear operators on $X$ which are bijective\footnote{
Recall that if $Y\in\mathcal B(X)$ is bijective then its inverse $Y^{-1}$ is automatically bounded as a consequence of the open mapping theorem \cite[Coro.~2.12]{Rudin91}. Hence $\operatorname{GL}(X)$, as the name suggests, is indeed a group. 
}. Also the operator $A$ is called drift\index{operator!drift}\index{drift|see{operator, drift}} and the $B_j$ are called control operators\index{control operator}.

\begin{remark}
Systems of the above form are called bilinear because they involve terms of the form $u_jB_jy$ which are linear in $y$ for fixed $u$, and vice versa\footnote{In contrast to this, linear control systems do not feature such cross terms, i.e.~they are of the form $\dot y=A(t)y+B(t)u$.}. The system \eqref{eq:bilinear_2} is also known as right-invariant (control-affine) system\index{right-invariant control system} on a Lie group.

\end{remark}

Following \cite[Ch.~1.3]{Elliott09} the function space the controls $u(t):=(u_1(t),\ldots,u_m(t)):\mathbb R_+\to\mathbb R^m$ live in---which must be invariant under time-shifts and concatenations for technical reasons \cite[Ch.~1.5.2]{Elliott09}---is commonly chosen from the following:
\begin{itemize}
\item the locally integrable functions $\mathcal{LI}$\label{symb_LI}\index{control function!locally integrable}, which is the largest class one commonly needs: This is the collection of all $\mathbb R^m$-valued functions $u$ for which the Lebesgue integral $\int_s^t\|u(\tau)\|\,d\tau$ exists for all $0\leq s\leq t<\infty$.
\item the piecewise continuous functions $\mathcal{PC}$:\label{symb_PC}\index{control function!piecewise continuous}
this is the shift-invariant subspace of $\mathcal{LI}$ which contains all the $\mathbb R^m$-valued functions $u$ such that, given any interval $[0, T]$, there exists a finite partition of $[0,T]$ such that $u$ is continuous on the corresponding open intervals (cf.~also footnote \ref{footnote_partition} in Appendix \ref{sec:spectral_integrals}). It is assumed that the limits at the endpoints of the pieces exist and are finite.
\item the piecewise constant functions $\mathcal{PK}$:\label{symb_PK}\index{control function!piecewise constant}
analogous to $\mathcal{PC}$ these are all functions such that, again given any interval $[0, T]$, there exists a finite partition of $[0,T]$ such that $u$ takes constant values on the corresponding open intervals. In other words the image $u([0,T])$ is a finite set, and the pre-image $u^{-1}(\{y\})$ for all $y\in\mathbb R$ is either empty or a union of finitely many intervals.
\end{itemize}
Obviously, $\mathcal{PK}\subset\mathcal{PC}\subset\mathcal{LI}$. While the class $\mathcal{PK}$, unsurprisingly, is the easiest to handle for reachability questions as it allows for explicit solutions of the control problem, we will see that it is as powerful as using the broader classes $\mathcal{PC}$ and $\mathcal{LI}$.
\begin{remark}\label{rem_solution}
Allowing for controls to be discontinuous means we run into the problem of (formal) non-differentiable solutions of \eqref{eq:bilinear_1}, \eqref{eq:bilinear_2}. Therefore we have to clarify what we mean by a ``solution'' of an initial value problem
\begin{equation}\label{eq:ivp}
\dot x=f(t,x(t))\qquad\text{with}\qquad x(t_0)=x_0\ ,\ x(t)\in \Omega\subseteq\mathbb R^n\,.
\end{equation}
Following Sontag \cite[Appendix C.2]{Sontag} the simplest approach is to define a solution \eqref{eq:ivp} on an interval $I$ to be an absolutely continuous function $x:I\to\Omega$ such that the corresponding integral equation
\begin{equation}\label{eq:integ}
x(t)=x_0+\int_{t_0}^t f(\tau,x(\tau))\,d\tau
\end{equation}
holds for all $t\in I$. In finite dimensions, it is well-known that a function $f:[a,b]\to\mathbb R^n$ is absolutely continuous if and only if it is differentiable almost everywhere and $f$ can be written as $f(x)=f(a)+\int_a^x g(\tau)\,d\tau$ on $[a,b]$ for some $L^1$-function $g$ \cite[Thm.~7.20]{Rudin86}. Thus solutions of \eqref{eq:bilinear_1}, \eqref{eq:bilinear_2} \textit{in this sense} are the ``almost-everywhere classical solutions'', that is, functions which solve equation \eqref{eq:bilinear_1} or \eqref{eq:bilinear_2} for almost all times $t$.
\end{remark}
Therefore whenever we say ``solution'' in the following, we mean it in the sense of Rem.~\ref{rem_solution}.
\begin{lemma}\label{lemma_pk_control_solution}
Let $A,B_1,\ldots,B_m\in\mathcal B(X)$, $T>0$, and $u\in\mathcal{PK}$ be given. Because $u$ is piecewise constant, there exists $N\in\mathbb N$, times $\tau_1,\ldots,\tau_{N-1}\in(0,T)$ and values $u_0,\ldots,u_{N-1}\in\mathbb R^m$ such that
$$
u(s)=\begin{cases}
u_0&0=\tau_0< s<\tau_1\\
u_{i-1}&\tau_{i-1}< s<\tau_i\text{ for }i=2,\ldots,N-1\\
u_{N-1}&\tau_{N-1}< s<\tau_{N}=T
\end{cases}\,.
$$
Then for all $0\leq k\leq N-1$ the descending order product
\begin{align*}
Y(t,u):=\exp\Big( (t-\tau_k)\Big(A+\sum_{j=1}^mu_j(\tau_k)B_j \Big) \Big)\prod_{i=k}^1\exp\Big( (\tau_i-\tau_{i-1})\Big(A+\sum_{j=1}^mu_j(\tau_{i-1})B_j \Big) \Big)
\end{align*}
is the unique solution of \eqref{eq:bilinear_2} on $t\in[\tau_k,\tau_{k+1})$ for $Y_0=\mathbbm{1}_X$.
\end{lemma}
\begin{proof}
For $\operatorname{dim}(X)<\infty$ this is shown in \cite[Ch.~1.5.1]{Elliott09} or \cite[Prop.~3.3]{Lawson99}. The proof remains the same for bounded operators on arbitrary Banach spaces.
\end{proof}

This paves the way for constructing solutions of \eqref{eq:bilinear_2} if $u$ is not piecewise constant but only locally integrable, which is more relevant for applications: given $u\in\mathcal{LI}$ we can approximate it via a sequence $(u^{(n)})_{n=1}^\infty$ from $\mathcal{PK}$ in the $L^1$-sense, meaning $\lim_{n\to\infty}\int_0^T\|u(t)-u^{(n)}(t)\|\,dt=0$. In this case the sequence of operators $(Y(t,u^{(n)}))_{n=1}^\infty$ converges uniformly on $[0,T]$ to a limit $Y(t,u)$ called product integral. Indeed $Y(t,u)$ then is well-defined, absolutely continuous, and the unique solution of \eqref{eq:bilinear_2} \cite[Ch.~1.8 \& 3.4 ff.]{DF84}. This approximation result justifies the following assumption which will be valid for the remainder of this thesis:\bigskip

\noindent\textbf{Assumption PK:}\label{not_ass_pk}
The control function $u:\mathbb R_+\to\Omega$, possibly restricted to a subset $\Omega$ of $\mathbb R^m$, is piecewise constant.\bigskip

Introducing constraints on the controls can be motivated either by experimental limitations (e.g.~$u(\mathbb R_+)\subseteq B_C(0)$ so the control amplitude is upper bounded by some $C>0$), or by considering special types of controls (such as ``bang-bang''-controls\index{bang-bang control} $u_j:\mathbb R_+\to\{0,1\}$). Either way we are now ready to define controllability and accessibility of bilinear control systems, for which we refer to \cite[Def.~1.6 \& Ch.~3.3]{Elliott09}:

\begin{definition}\label{def_contr_bounded}
Let a Banach space $X$, operators $A,B_1,\ldots,B_m\in\mathcal B(X)$, $\Omega\subseteq\mathbb R^m$ be given, and let assumption PK be true.
\begin{itemize}
\item[(i)] Let $S_\Omega$ denote the smallest semigroup in $\mathcal B(X)$ which contains the set
$$
\Big\{\exp\Big(\tau\Big(A+\sum\nolimits_{j=1}^mu_jB_j\Big)\Big)\,\Big|\,\tau\geq 0,u\in\Omega\Big\}\,.
$$
Usually $S_\Omega$ is called the system semigroup\index{system semigroup} associated with \eqref{eq:bilinear_2}.
\item[(ii)] Given $T>0$, $y_0\in X$, and $Y_0\in\operatorname{GL}(X)$ define
\begin{align*}
\mathfrak{reach}_{[0,T]}(y_0)&:=\bigcup\nolimits_{0\leq t\leq T}\{Y(t,u)y_0\,|\,u(\cdot)\in\Omega\}\\
\mathfrak{reach}(y_0)&:=S_\Omega\;\! y_0
\end{align*}
and analogously $\mathfrak{reach}_{[0,T]}(Y_0)$, $\mathfrak{reach}(Y_0)$ for the group-lifted problem. The elements of $\mathfrak{reach}_{[0,T]}(\cdot)$ are called reachable from $(\cdot)$ in time $T$ and $\mathfrak{reach}(\cdot)$ is called the reachable set\index{reachable set}.
\item[(iii)] Given connected non-empty sets $U\subseteq X$, $V\subseteq\operatorname{GL}(X)$ let, here and henceforth, $\overline{(\cdot)}$ denote the closure in $U$ with respect to the norm, $\overline{(\cdot)}^{\,\mathrm{u}}$\label{symb_subspace_top_cl} denote the closure in $V$ with respect to the operator norm, and $\overline{(\cdot)}^{\,\mathrm{s}}$ denote the closure in $V$ with respect to the strong operator topology. Then
\begin{itemize}
\item[(a)] system \eqref{eq:bilinear_1} is called accessible\index{accessible} (approximately accessible)\index{accessible!approximately} on $U$ if $\mathfrak{reach}(y_0)$ ( $\overline{\mathfrak{reach}(y_0)}$) $\subseteq U$ has an interior point with respect to $U$ for all $y_0\in U$.
\item[(b)] system \eqref{eq:bilinear_2} is called accessible on $V$ if $\mathfrak{reach}(Y_0)\subseteq V$ has an interior point with respect to $V$ for all $Y_0\in V$. It is called uniformly approximately (strongly approximately) accessible\index{accessible!strongly approximately}\index{accessible!uniformly approximately} on $V$ if $\overline{\mathfrak{reach}(Y_0)}^{\,\mathrm{u}}$ ( $\overline{\mathfrak{reach}(Y_0)}^{\,\mathrm{s}}$) $\subseteq V$ has an interior point with respect to $(V,\|\cdot\|_\mathrm{op})$ for all $Y_0\in V$.
\end{itemize} 
\item[(iv)] Given connected non-empty sets $U\subseteq X$, $V\subseteq\operatorname{GL}(X)$
\begin{itemize}
\item[(a)] system \eqref{eq:bilinear_1} is called controllable\index{controllable} (approximately controllable)\index{controllable!approximately} on $U$ if $\mathfrak{reach}(y_0)$ ( $\overline{\mathfrak{reach}(y_0)}$) is equal to $U$ for all $y_0\in U$.
\item[(b)] system \eqref{eq:bilinear_2} is called controllable on $V$ if $\mathfrak{reach}(Y_0)=V$ for all $Y_0\in V$. Moreover, it is called uniformly approximately\index{controllable!strongly approximately}\index{controllable!uniformly approximately} (strongly approximately) controllable on $V$ if $\overline{\mathfrak{reach}(Y_0)}^{\,\mathrm{u}}$ ( $\overline{\mathfrak{reach}(Y_0)}^{\,\mathrm{s}}$) is equal to $V$ for all $Y_0\in V$.
\end{itemize} 
\end{itemize}
\end{definition}

The following standard result tells us that accessibility and controllability analysis of the lifted control problem simplifies considerably, assuming the problem is formulated on a group:
\begin{lemma}\label{lemma_con_acc_id}
Let $X$ be a Banach space and $G\subseteq\operatorname{GL}(X)$ be a subgroup.
\begin{itemize}
\item[(i)] System \eqref{eq:bilinear_2} is accessible on $G$ if and only if $\mathfrak{reach}(\mathbbm{1}_X)$ has an interior point w.r.t.~$G$.
\item[(ii)] System \eqref{eq:bilinear_2} is controllable on $G$ if and only if $\mathfrak{reach}(\mathbbm{1}_X)=G$.
\end{itemize} 
These statements stay valid if accessibility (controllability) gets replaced by approximate accessibility (approximate controllability), and $\mathfrak{reach}(\mathbbm{1}_X)$ gets replaced by its closure; both in the respective topology.
\end{lemma}
\begin{proof}
The simple but fundamental observation here is $\mathfrak{reach}(Y_0)=\mathfrak{reach}(\mathbbm{1}_X)Y_0$ which holds for all $Y_0\in\operatorname{GL}(X)$. 
(i): A straightforward calculation shows
$$
B_{\varepsilon/\|Y_0^{-1}\|}(AY_0)\subseteq B_{\varepsilon}(A)Y_0\subseteq B_{\varepsilon\|Y_0\|}(AY_0)
$$
for all $\varepsilon>0$ and all $A\in\mathcal B(X),Y_0\in\operatorname{GL}(X)$. 
Now assume $\mathfrak{reach}(\mathbbm{1}_X)$ has an interior point with respect to $G$, that is, there exists an element $Z\in \mathfrak{reach}(\mathbbm{1}_X)$ as well as $\varepsilon>0$ such that $B_\varepsilon(Z)\cap G\subseteq \mathfrak{reach}(\mathbbm{1}_X)$. Then for all $Y_0\in G$ we find
\begin{align*}
B_{\varepsilon/\|Y_0^{-1}\|}(ZY_0)\cap G\subseteq B_\varepsilon(Z)Y_0\cap G&=B_\varepsilon(Z)Y_0\cap GY_0\\
&=(B_\varepsilon(Z)\cap G)Y_0\subseteq \mathfrak{reach}(\mathbbm{1}_X)Y_0=\mathfrak{reach}(Y_0)\,.
\end{align*}
In the second step we used the group property $GY_0=G$ for all $Y_0\in G$. Thus we found $\tilde Z:=ZY_0\in\mathfrak{reach}(\mathbbm{1}_X)Y_0=\mathfrak{reach}(Y_0)$ and $\tilde\varepsilon:=\varepsilon/\|Y_0^{-1}\|>0$ such that $B_{\tilde\varepsilon}(\tilde Z)\cap G\subseteq\mathfrak{reach}(Y_0)$. Because $Y_0$ was chosen arbitrarily from $G$, system \eqref{eq:bilinear_2} is accessible.

(ii): Let $Y_0\in G$ be arbitrary. If $\mathfrak{reach}(\mathbbm{1}_X)=G$ then 
$
\mathfrak{reach}(Y_0)=\mathfrak{reach}(\mathbbm{1}_X)Y_0=GY_0=G
$
where in the last step we again used that $G$ is a group. 

The additional statements are shown the same way which concludes the proof.
\end{proof}
\noindent Usually the above is shown via the fact that left- and right-multiplication on Lie groups (topological groups) are diffeomorphisms (homeomorphisms) so interior points are mapped to interior points. Yet we presented a more explicit proof to convey the idea from an operator-theoretic perspective, and to highlight where the assumption of $G$ being a group comes into play.\medskip

For general systems, controllability obviously implies approximate controllability as well as (approximate) accessibility; now for the group lift the converse holds, as well: 
\begin{lemma}\label{rem_approx_contr_acc}
Let $X$ be a Banach space and $G\subseteq\operatorname{GL}(X)$ be a subgroup. The following statements are equivalent.
\begin{itemize}
\item[(i)] System \eqref{eq:bilinear_2} is controllable.
\item[(ii)] System \eqref{eq:bilinear_2} is uniformly approximately controllable and accessible.
\end{itemize}
\end{lemma}
\begin{proof}
``(i) $\Rightarrow$ (ii)'': Trivial. ``(ii) $\Rightarrow$ (i)'': Consider an accessible control system on $G$ and any final point $z\in G$. Due to the group nature the backwards control system (i.e.~for negative times) is accessible, as well: because $\mathfrak{reach}(\mathbbm 1)$ has an interior point, so does the reachable set of the left-invariant backwards system $\dot Y(t)=Y(t)(-A-\sum_{j=1}^m u_j(t)B_j)$, $Y_0=\mathbbm 1$ due to norm-continuity of inverting bounded operators. But our initial point is the identity, meaning the reachable set of the left-invariant and the right-invariant system co{\"i}ncide.

Thus one finds $z_0\in G$ and $\varepsilon>0$ such that $z'\in\mathfrak{reach}(z)$ for all $z'\in B_\varepsilon(z_0)\cap G$. But by approximate controllability for every initial point $y_0\in G$ there exists $y'\in B_\varepsilon(z_0)\cap\mathfrak{reach}(y_0)$ so using the (exact) control sequence $y_0\to y'\to z$ one finds $z\in\mathfrak{reach}(y_0)$. Because $y_0,z$ were chosen arbitrarily from $G$ we can conclude that the system is controllable. 
\end{proof}

While controllability is the stronger of the two notions---because, trivially, $\operatorname{int}(U)=U$ for all topological spaces $(U,\tau)$---for systems where controllability cannot be achieved knowing whether ``all directions can be generated'' (i.e.~accessibility) may be of interest. For example open quantum systems are never controllable\index{open quantum system!never controllable}, regardless of whether one considers the group lift or the state problem on $\mathbb D(\mathcal H)$ \cite[Thm.~3.10]{DiHeGAMM08}. This is due to the dissipative part of the \textsc{gksl}-generator. Of course this neither rules out approximate controllability nor accessibility (cf.~Ch.~\ref{sec_reach_fin_dim}), because Lemma \ref{rem_approx_contr_acc} holds for groups, but not for control problems on arbitrary homogeneous spaces.

\subsection{Finite Dimensions}\label{ch:control_fin_dim}

Given a bilinear control system the question now is how to characterize or easily decide whether (approximate) controllability or at least (approximate) accessibility holds. It turns out that for the group-lifted control problem in finite dimensions---although the used techniques (differential geometry and Lie group theory) are more involved---things simplify a lot; thus Lemma \ref{lemma_con_acc_id} \& \ref{rem_approx_contr_acc} are not the only reasons why the group lift is of interest.

For example if $\operatorname{dim}(X)<\infty$ and if $G$ is a closed\footnote{
By this we mean closed with respect to the subspace topology on $\operatorname{GL}(X)$. In other words if a sequence $(A_n)_{n\in\mathbb N}$ in $G$ converges to some $A\in\mathcal B(X)$ then either $A\in G$ or $A\not\in\operatorname{GL}(X)$ (cf.~Appendix \ref{section_product_subsp_top} and \cite[Def.~1.4]{Hall15}).
}
Lie subgroup of $\operatorname{GL}(X)$, then $G$ is a Lie group\index{Lie group}, that is, a smooth manifold equipped with a group structure such that group multiplication and inversion are smooth \cite[Coro.~3.45]{Hall15}. This grants us access to the Lie algebra\index{Lie algebra}\footnote{\label{footnote_lie_algebra}
An abstract Lie algebra is a vector space $\mathfrak g$ together with a bilinear and skew-symmetric map $[\cdot,\cdot]:\mathfrak g\to\mathfrak g$ which satisfies the Jacobi identity $[A,[B,C]]+[B,[C,A]]+[C,[A,B]]=0$ for all $A,B,C\in\mathfrak g$. It turns out that for matrix Lie groups the bracket of the corresponding Lie algebra is given by the commutator $[A,B]=AB-BA$ for all $A,B\in\mathfrak g$ \cite[Thm.~3.20]{Hall15}.
} of $G$,
i.e.~the tangent space at the identity which, remarkably, turns out to co{\"i}ncide with the set of all matrices $A$ such that the whole one-parameter subgroup $\{e^{tA}\}_{t\in\mathbb R}$ lies in $G$ \cite[Coro.~3.46]{Hall15}. Thus the Lie algebra can be viewed as the generator of a Lie group via the exponential map, which due to its linear nature is a lot easier to handle than the (non-linear) differential geometric object $G$.

Passing from Lie group to Lie algebra is also the key when characterizing accessibility and controllability. We will focus on the underlying concepts and try to highlight how they are interconnected, thus omitting most of the proofs.
\begin{proposition}\label{prop_bilinear_fin_dim}
Let $X$ be a finite-dimensional vector space and let assumption PK hold.
\begin{itemize}
\item[(i)] Given a closed subgroup $G$ of $\operatorname{GL}(X)$ and assuming $\Omega=\mathbb R^m$ the following are equivalent.
\begin{itemize}
\item[(a)] System \eqref{eq:bilinear_2} is accessible on $G$.
\item[(b)] The system Lie algebra\label{symb_Lie_closure}
$
\langle A,B_1,\ldots,B_m\rangle_\mathrm{Lie}
$, 
that is, the smallest linear subspace of $\mathfrak g$ which contains $A,B_1,\ldots,B_m$ together with all iterated Lie brackets $[A,B_j]$, $[B_i,B_j ]$, $[A, [B_i,B_j ]]$, $\ldots$ is equal to $\mathfrak g$.
\end{itemize}
\item[(ii)] Given a compact and connected subgroup $G$ of $\operatorname{GL}(X)$ and assuming $\Omega=\mathbb R^m$ the following are equivalent.
\begin{itemize}
\item[(a)] System \eqref{eq:bilinear_2} is controllable on $G$.
\item[(b)] $\langle A,B_1,\ldots,B_m\rangle_\mathrm{Lie}=\mathfrak g$.
\end{itemize}
\item[(iii)] Given a closed, connected, and simple\footnote{
A finite-dimensional Lie algebra is called simple if it is not abelian (i.e.~there exist $X,Y\in\mathfrak g$ such that $[X,Y]\neq 0$) and if it contains no non-trivial ideals (i.e.~the only subspaces $\mathfrak h\subseteq\mathfrak g$ which satisfy $[\mathfrak h,\mathfrak h]\subseteq\mathfrak h $ and $ [\mathfrak g,\mathfrak h]\subseteq\mathfrak h $ are $\{0\}$ and $\mathfrak g$) \cite{Hall15}.
Then a simple Lie group is a connected Lie group whose Lie algebra is simple.
}
subgroup $G$ of $\operatorname{GL}(X)$ the following hold.
\begin{itemize}
\item[(a)] System \eqref{eq:bilinear_2} is controllable on $G$ if and only if it is approximately controllable
on $G$.
\item[(b)] System \eqref{eq:bilinear_2} is accessible on $G$ if and only if it is approximately\footnote{
There is no need here to distinguish between operator norm and strong operator topology as they co{\"i}ncide in finite dimensions (Prop.~\ref{prop_strong_weak_op_top} (iv)).
}
accessible.
\end{itemize}
\end{itemize}
\end{proposition}
\begin{proof}
(i):\footnote{
Note that while the respective theorems are sometimes formulated for broader classes of control functions the proofs rely on piecewise constant controls or even just ``bang-bang''-controls (so the result naturally extends to said broader classes).
}
\cite[Coro.~4.6 \& Ex.~5.2]{SJ72}. (ii): \cite[Thm.~7.1]{JS72}.
(iii),(a): \cite[Thm.~17]{boscain2015approximate}. Note that the proof solely relies on simplicity of $G$ \cite[Thm.~19]{boscain2015approximate} which excludes problematic situations such as dense windings on a torus. (iii),(b): Approximate accessibility implies that the closure of the group generated by $S_\Omega$ is equal to $G$ (due to the group nature, simply shift the interior point to the identity). Hence $\langle A,B_1,\ldots,B_m,-A,-B_1,\ldots,-B_m\rangle_\mathrm{Lie}=\mathfrak g$ by (iii),(i) together with (ii). But $\langle A,B_1,\ldots,B_m,-A,-B_1,\ldots,-B_m\rangle_\mathrm{Lie}=\langle A,B_1,\ldots,B_m\rangle_\mathrm{Lie}$ so the system is already accessible by (i).
\end{proof}
\noindent Condition (i),(b) (resp.~(ii),(b)) is usually referred to as the Lie algebra rank condition (\textsc{larc}).\index{Lie algebra rank condition}\index{larc@\textsc{larc}|see{Lie algebra rank condition}}\medskip

To apply this to closed quantum systems\index{closed quantum system!controlled} quickly recall the corresponding control problems
\begin{align}
\dot\psi(t)&=-i\Big(H_0+\sum\nolimits_{j=1}^m u_j(t)H_j\Big)\psi(t)\qquad\psi(0)=\psi_0\in S_1(\mathcal H)=\{\psi\in\mathcal H\,|\,\langle \psi,\psi\rangle=1\}\label{eq:controlled_schr} \\
\dot\rho(t)&=-i\Big[H_0+\sum\nolimits_{j=1}^m u_j(t)H_j,\rho(t)\Big]\qquad\rho(0)=\rho_0\in\mathbb D(\mathcal H)\label{eq:controlled_LvN}\\
\dot U(t)&=-i\Big(H_0+\sum\nolimits_{j=1}^m u_j(t)H_j\Big)U(t)\qquad U(0)=\mathbbm{1}\in\mathcal U(\mathcal H)\,,\label{eq:controlled_propagators}
\end{align}
that is, the controlled Schr{\"o}dinger equation\index{Schr{\"o}dinger equation!controlled} \eqref{eq:controlled_schr}, the controlled Liouville-von Neumann equation\index{Liouville-von Neumann equation!controlled} \eqref{eq:controlled_LvN}, and the control problem lifted to the unitary propagators \eqref{eq:controlled_propagators}. Note that the reachable set of \eqref{eq:controlled_propagators} readily transfers to \eqref{eq:controlled_LvN}, that is, 
$\mathfrak{reach}(\rho_0)=\{U\rho_0 U^*\,|\,U\in \mathfrak{reach}(\mathbbm{1})\} $ for all $\rho_0\in\mathbb D(\mathcal H)$ as a direct consequence of Lemma \ref{lemma_ad_H_unbounded}. It turns out that in finite dimensions the converse holds, as well:
\begin{lemma}[\cite{AA03}]\label{lemma_unit_state_cont_equiv}
Let $\mathcal H$ be a finite-dimensional complex Hilbert space. Then system \eqref{eq:controlled_propagators} is controllable on the special unitary group\index{special unitary group} $SU(\mathcal H)=\{U\in\mathcal U(\mathcal H)\,|\,\operatorname{det}(U)=1\}$ if and only if $\mathfrak{reach}(\rho_0)=\{U\rho U^*\,|\,U\in\mathcal U(\mathcal H)\}$, that is, system \eqref{eq:controlled_LvN} is controllable on the unitary orbit of each initial state $\rho_0\in\mathbb D(\mathcal H)$. If either of these is true, then \eqref{eq:controlled_schr} is controllable on $ S_1(\mathcal H)$ which is equivalent to $\mathfrak{reach}(|\psi\rangle\langle\psi|)=\{U|\psi\rangle\langle\psi|U^*\,|\,U\in\mathcal U(\mathcal H)\}$ for all $\psi\in S_1(\mathcal H)$.
\end{lemma}
\noindent This motivates us to primarily consider the problem on density matrices \eqref{eq:controlled_LvN} instead of state vectors (or equivalently on rank-1 projectors) \eqref{eq:controlled_schr}. For more on these connections we refer to the book of D'Alessandro \cite[Ch.~3.6 \& Fig.~3.2]{dAll08}.	\medskip

Next let us apply the above characterizations of controllability to our system of unitary propagators: 

\begin{corollary}\label{coro_contr_su}
Let $\mathcal H$ be a finite-dimensional complex Hilbert space and let assumption PK hold. Given $H_0,H_1,\ldots,H_m\in\mathcal B(\mathcal H)$ Hermitian and traceless, and assuming $\operatorname{dim}(\mathcal H)>1$, $\Omega=\mathbb R^m$ the following statements are equivalent.
\begin{itemize}
\item[(i)] System \eqref{eq:controlled_propagators} is controllable on $ \operatorname{SU}(\mathcal H)$.\label{symb_SU}
\item[(ii)] System \eqref{eq:controlled_propagators} is approximately controllable on $ \operatorname{SU}(\mathcal H)$.
\item[(iii)] $\langle iH_0, iH_j\,|\,j=1,\dots,m\rangle_\textsf{Lie}$ equals $\mathfrak{su}(\mathcal H)=\{A\in\mathcal B(\mathcal H)\,|\,A^*=-A\text{ and }\operatorname{tr}(A)=0\}$.\label{symb_su}
\end{itemize}
\end{corollary}
\begin{proof}
Because $\operatorname{dim}(\mathcal H)>1$, by a standard result $\operatorname{SU}(\mathcal H)$ is a compact, connected, and simple Lie group \cite[Ch.~2.5]{Arv03}, \cite{BroeDie85} so this equivalence follows from Prop.~\ref{prop_bilinear_fin_dim} (ii) \& (iii).
\end{proof}
Although $\mathcal U(\mathcal H)$ is not simple---because $\mathfrak{su}(\mathcal H)$ is a non-trivial ideal of $\mathfrak u(\mathcal H)$---the previous result still extends to the general unitary case:
\begin{corollary}\label{coro_contr_u}
Let $\mathcal H$ be a finite-dimensional complex Hilbert space and let assumption PK hold. Given $H_0,H_1,\ldots,H_m\in\mathcal B(\mathcal H)$ Hermitian and assuming $\Omega=\mathbb R^m$ the following statements are equivalent.
\begin{itemize}
\item[(i)] System \eqref{eq:controlled_propagators} is controllable on $\mathcal U(\mathcal H)$.
\item[(ii)] System \eqref{eq:controlled_propagators} is approximately controllable on $\mathcal U(\mathcal H)$.
\item[(iii)] $\langle iH_0, iH_j\,|\,j=1,\dots,m\rangle_\textsf{Lie}$ equals $\mathfrak u(\mathcal H)=\{A\in\mathcal B(\mathcal H)\,|\,A^*=-A\}$.\label{symb_un_alg}
\end{itemize}
\end{corollary}
\begin{proof}
``(i) $\Leftrightarrow$ (iii)'': In finite dimensions $\mathcal U(\mathcal H)$ is a compact and (path-)connected subgroup of $\operatorname{GL}(\mathcal H)$ by Thm.~\ref{theorem_unitary_group_properties} so this is a direct consequence of Prop.~\ref{prop_bilinear_fin_dim} (ii). ``(i) $\Leftrightarrow$ (ii)'': \cite[Thm.~17]{boscain2015approximate}. Note that this part of the proof is independent of the choice of $\Omega$.
\end{proof}


The control problem \eqref{eq:controlled_LvN} does not care whether the group lift is controllable on $\mathcal U(\mathcal H)$ or $\operatorname{SU}(\mathcal H)$ for the following reason: Given $H_0,H_1,\ldots,H_m\in\mathcal B(\mathcal H)$ Hermitian let $\tilde H_j:=H_j-\operatorname{tr}(H_j)\frac{\mathbbm{1}}{n}$ denote their traceless part. Then
\begin{align*}
e^{-it(H_0+\sum_{j=1}^mu_jH_j)}(\cdot)e^{it(H_0+\sum_{j=1}^mu_jH_j)}
=e^{-it(\tilde H_0+\sum_{j=1}^mu_j\tilde H_j)}(\cdot)e^{it(\tilde H_0+\sum_{j=1}^mu_j\tilde H_j)}
\end{align*}
for all $t\in\mathbb R$, $u\in\mathbb R^m$ as is readily verified, meaning the unitary and the special unitary similarity orbit of any initial state co{\"i}ncide. Indeed $\operatorname{tr}(H_j)\in\mathbb R$ implies that the additional term is just a phase factor which vanishes under conjugation. Hence the full unitary orbit gets generated either way, which is why both scenarios (Coro.~\ref{coro_contr_su} \& \ref{coro_contr_u}) will be referred to as \textit{unitary controllability}\index{unitary controllability}.

\begin{remark}
\begin{itemize}
\item[(i)] For the special case of qubit systems, that is, $\mathcal H=\mathbb C^{2^n}$ with $n$ being the number of qubits, there is a criterion more powerful than the Lie algebra rank condition: Given $H_0,H_1,\ldots,H_m$ Hermitian and traceless, controllability of \eqref{eq:controlled_propagators} on the special unitary group is equivalent to\footnote{
Two elements $A,B$ of a Lie algebra $\mathfrak g$ are said to commute if $[A,B]=0$. Then the commutant\index{commutant} to any subset $S\subset\mathfrak g$ is defined via $S':=\{A\in\mathfrak g\,|\,[A,B]=0\text{ for all }B\in S\}$.
}
$$
\operatorname{dim}\big(\{\mathbbm{1}\otimes H_j+H_j\otimes\mathbbm{1}\,,\,\mathbbm{1}\otimes H_j-H_j^T\otimes \mathbbm{1}\,|\,j=0,1,\ldots,m\}'\big)=2\,.
$$
Because said commutant is usually referred to as quadratic symmetries this criterion for controllability is also called symmetry criterion\index{symmetry criterion}, cf.~\cite[Thm.~4]{OSID17} and \cite{ZS11,ZZ15}.
\item[(ii)] The above results fully settle when generating all unitary propagators (resp.~all unitary channels) is possible which is all we need for our main results. Follow-up questions would be how to find \textit{explicit} control schemes which generate some target unitary \cite{dAle09a} or how to find such a scheme, e.g., with minimal control time (``time optimal torus theorem'' \cite[Thm.~2.13]{DiHeGAMM08}) or under further constraints $\Omega\subsetneq\mathbb R^m$, see the roadmap \cite{Roadmap2015} for an overview.
\end{itemize}
\end{remark}

For (Markovian) open systems\index{open quantum system!controlled} things become more difficult, even in finite dimensions. Recalling the \textsc{gksl}-form of a continuous quantum-dynamical semigroup (Thm.~\ref{thm_gksl}) the corresponding control system usually looks like
\begin{align*}
\dot\rho(t)&=-i\Big[H_0+\sum\nolimits_{j=1}^m u_j(t) H_j,\rho(t)\Big]-\sum_{k\in I}\Big(\frac12 (V_k^*V_k\rho (t)+\rho (t)V_k^*V_k)-V_k\rho(t) V_k^*\Big)\\
&= \Big(\big(-i\operatorname{ad}_{H_0}- \Gamma\big)+\sum\nolimits_{j=1}^m u_j(t) \big(\!-i\operatorname{ad}_{H_j}\!\big)\Big)\rho(t)
\end{align*}
with initial value $\rho(0)=\rho_0\in\mathbb D(\mathbb C^n)$. Here $ \Gamma=\sum\nolimits_{k\in I} \Gamma_k$ is short for the dissipative part where $ \Gamma_k:\rho\mapsto\frac12 (V_k^*V_k\rho +\rho V_k^*V_k)-V_k\rho V_k^*\in\mathcal L(\mathbb C^{n\times n})$ for all $k\in I$. In the finite-dimensional case, an unambiguous 
separation of the dissipative part and the coherent part results from choosing the
$V_k$ traceless \cite{GKS76}. Equivalently one can also vectorize\index{vectorization}\footnote{
Vectorization is the linear map $\operatorname{vec}:\mathbb C^{m\times n}\to\mathbb C^{mn}$ which turns a matrix into a column vector by stacking its columns one underneath the other \cite[Ch.~2.4]{MN07}. One finds $\operatorname{vec}(ABC)=(C^T\otimes A)\operatorname{vec}(B)$ where $\otimes$ denotes the usual Kronecker product\index{Kronecker product|see{tensor product, Kronecker}}\index{tensor product!Kronecker} \cite[Ch.~2.4, Thm.~2]{MN07}.\label{footnote_vec}
}
the system yielding a standard bilinear control system on $\mathbb C^{n^2}$:
\begin{align*}
\dot y(t)= \Big(\big(-i\hat H_0-\hat \Gamma\big)+\sum\nolimits_{j=1}^m u_j(t) \big(-i\hat H_j\big)\Big)y(t)
\end{align*}
with $y(0)=\operatorname{vec}(\rho_0)$ for some $\rho_0\in\mathbb D(\mathbb C^n)$. One readily verifies $\hat H_j=\mathbbm{1}\otimes H_j-H_j^T\otimes\mathbbm{1}\in\mathbb C^{n^2\times n^2}$ for all $j=0,\ldots,m$ as well as $\hat \Gamma=\sum_{k\in I}\hat \Gamma_k$ where $\hat \Gamma_k=\frac12(\mathbbm{1}\otimes V_k^*V_k+V_k^T\overline{V_k}\otimes \mathbbm{1})-\overline{V_k}\otimes V_k$, cf.~also \cite[Ch.~2]{OSID17}. This covers a broad class of quantum control problems including 
coherent and incoherent feedback\index{coherent feedback}\index{incoherent feedback}
\cite{mirrahimi2004controllability,DongPetersen2010,Haroche11,Haroche13}.

Both formulations allow for an operator lift $\dot F(t)=((-i\operatorname{ad}_{H_0}- \Gamma)+\sum_{j=1}^m u_j(t) (-i\operatorname{ad}_{H_j}))F(t)$ or $\dot F(t)=((-i\hat H_0-\hat \Gamma)+\sum_{j=1}^m u_j(t) (-i\hat H_j))F(t)$ with $F(0)=\mathbbm{1}$ (in the respective space) the solutions of which (by definition, cf.~also Rem.~\ref{rem_markov_div}) are time-dependent Markovian channels. Indeed one can show that every time-dependent Markovian\index{Markovian!time-dependent} channel is infinitesimal divisible into products of exponentials of \textsc{gksl}-generators \cite{Wolf08a} hence leading to Lie semigroup\index{Lie semigroup} (and not Lie group) structure \cite{DHKS08}. 

This motivates defining the \textit{Kossakowski-Lindblad algebra}\index{Kossakowski-Lindblad algebra} as the Lie algebra comprising all \textsc{gksl}-generators $\mathfrak g^\textrm{KL}:=\langle i\hat H,\hat \Gamma\,|\,H\in\mathfrak{su}(n)\rangle_\textrm{Lie}$ \footnote{
While the Jacobi identity yields $[-i\operatorname{ad}_{H_1},-i\operatorname{ad}_{H_2}]=-i\operatorname{ad}_{i[H_1,H_2]}$, commutators of the form $[-i\operatorname{ad}_{H}, \Gamma_k]$ are in general not of \textsc{gksl}-form anymore. This is why one has to consider the Lie algebra \textit{generated} by all \textsc{gksl}-generators.
}
which yields the dynamic system Lie group $G^\textrm{KL}$ generated by $\mathfrak g^\textrm{KL}$. 
Now accessibility on $G^\textrm{KL}$ boils down to accessibility at the identity (Lemma \ref{lemma_con_acc_id}) which is equivalent to the system algebra $\langle (i\hat H_0+\hat \Gamma),i\hat H_j\,|\,j=1,\ldots,m\rangle_\textrm{Lie}$ being all of $\mathfrak g^\textrm{KL}$ (Prop.~\ref{prop_bilinear_fin_dim}). Moreover one finds symmetry conditions related to the system algebra which are necessary for accessibility, for more detail on this we refer to \cite[Ch.~6.2]{OSID17}.

\subsection{Infinite Dimensions}\label{ch_control_infdim}

It comes as no surprise that establishing unitary controllability in infinite dimensions is considerably more intricate. The most fundamental problem we run into is guaranteeing ``reasonable'' solutions of \eqref{eq:bilinear_1}; else the definition of accessibility and controllability may go down the drain. For this let us quickly recap some terminology regarding infinite-dimensional initial value problems \cite[Ch.~II.6]{EngelNagel00}: Given a linear operator $A:D(A)\subset X\to X$ on a Banach space $X$ the initial value problem
\begin{equation}\label{eq:acp}
\dot y(t)=Ay(t)\quad\text{ with }\quad y(0)=y_0\in X
\end{equation}
is called the abstract Cauchy problem\index{Cauchy problem} associated to $(A,D(A))$. Now if $A$ is unbounded then this problem is not defined everywhere but only on a dense domain. This leads to two different notions of a solution $y:[0,\infty)\to X$ of \eqref{eq:acp}:
\begin{itemize}
\item If $y$ is continuously differentiable, $y(t)\in D(A)$ for all $t\geq 0$ and \eqref{eq:acp} holds then $y$ is called a \textit{classical solution}\index{classical solution} of \eqref{eq:acp}.
\item If $y$ is continuous, $\int_0^t y(s)\,ds\in D(A)$, and $y(t)=y_0+A\int_0^t y(s)\,ds$ holds\footnote{
While for a continuous function $f:J\subseteq\mathbb R\to X$ into a Banach space $X$ the integral $\int_J f(s)\,ds$ can---as in the scalar case---be defined as the limit of Riemann sums, this is often too restrictive. The more general notion then is Bochner integration which carries a lot of the properties known from Lebesgue integration, and for which we refer to \cite[Appendix C]{EngelNagel00}. 
}
for all $t\geq 0$ then $y$ is called a \textit{mild solution}\index{mild solution} of \eqref{eq:acp}.
\end{itemize}
Note that this only makes a difference in the unbounded case: If $A\in\mathcal B(X)$ and $y$ is a mild solution then one readily verifies $\frac{d}{dt}y(t)=Ay(t)$ so $y$ is a classical solution, as well. Recalling the notion of strongly continuous semigroups from Ch.~\ref{ch:qu_dyn_sys} one gets the following first result:
\begin{lemma}\label{lemma_solution_const_gen}
Let $(A,D(A))$ be the generator of a strongly continuous semigroup $(T(t))_{t\geq 0}$. The following statements hold.
\begin{itemize}
\item[(i)] For every $y_0\in D(A)$ the unique classical solution of \eqref{eq:acp} is given by $y(t)=T(t)y_0$.
\item[(ii)] For every $y_0\in X$ the unique mild solution of \eqref{eq:acp} is given by $y(t)=T(t)y_0$.
\end{itemize}
Moreover, for every sequence $(y_n)_{n\in\mathbb N}\subseteq D(A)$ with $\lim_{n\to\infty} y_n=0$ one has $\lim_{n\to\infty}T(t)y_n=0$ uniformly on compact intervals.
\end{lemma}
\begin{proof}
\cite[Ch.~II, Prop.~6.2, 6.4 \& Thm.~6.7]{EngelNagel00}
\end{proof}
In our notation this means that $e^{tA}y_0$ is always a mild solution, and even becomes a classical solution once $y_0\in D(A)$. Thus for all self-adjoint operators $H$ on a complex Hilbert space---because $-i\operatorname{ad}_H$ is the generator of a strongly continuous semigroup (Lemma \ref{lemma_ad_H_unbounded})---the Liouville-von Neumann equation $\frac{d}{dt}\rho(t)=-i[H,\rho(t)]$ has unique mild solution $\rho(t)=e^{-itH}\rho_0 e^{itH}$ for all $\rho_0\in\mathbb D(\mathcal H)$ which is also the unique classical solution if $\rho_0\in D(\operatorname{ad}_H)\cap\mathbb D(\mathcal H)$. Of course one can make a similar statement regarding the Schr{\"o}dinger equation.

The bilinear control problems we are interested in fall into the class of semilinear equations\index{semilinear equation}
\begin{equation}\label{eq:semilinear_de}
\dot y(t)=Ay(t)+f(t,y(t))\quad\text{ with }\quad y(0)=y_0\in X
\end{equation}
with $(A,D(A))$ the generator of a strongly continuous semigroup and $f:[0,\infty)\times X\to X$ a (for now arbitrary) function. Based on the homogeneous case ($f\equiv 0$) a function $y:[0,T]\to X$ is a 
\begin{itemize}
\item classical solution of \eqref{eq:semilinear_de} if $y$ is continuous, is continuously differentiable on $(0,T]$, and satisfies \eqref{eq:semilinear_de},
\item mild solution of \eqref{eq:semilinear_de} if $y$ is continuous, and satisfies the corresponding integral equation\footnote{
To see that every classical solution is also a mild solution keep in mind that
$$
\frac{d}{dt}\int_0^t g(t,s)\,ds=\frac{\partial}{\partial r}\Big(\int_0^r g(u,s)\,ds\Big)(t,t)+\frac{\partial}{\partial u}\Big(\int_0^r g(u,s)\,ds\Big)(t,t)=g(t,t)+\int_0^t \frac{\partial}{\partial u}g(u,s)\Big|_{u=t}\,ds
$$
via the chain rule (assuming $g$ is a ``sufficiently nice'' function).
}
$y(t)=e^{tA}y_0+\int_0^t e^{(t-s)A}f(s,y(s))\,ds$ (``Duhamel's formula'' \cite[Ch.~1, Thm.~5.1]{DF84}),
\end{itemize}
cf.~\cite[Ch.~2.2]{Goldstein17}. Given a control operator $B\in\mathcal B(X)$ and a control function $u:[0,T]\to\mathbb R$ choosing $f(t,y(t))=u(t)By(t)$ leads to a bilinear control system as desired. Therefore the notion of a mild solution is in spirit of the finite-dimensional case (Rem.~\ref{rem_solution}) where one also relaxes the notion of a solution to allow for discontinuous control functions; although now, additionally, we had to account for unbounded operators. Under certain constraints on $u$ one can guarantee existence and uniqueness of mild solutions:
\begin{proposition}\label{prop_control_eq}
Let $(A,D(A))$ be the generator of a strongly continuous semigroup on a Banach space $X$, and let $B_1,\ldots,B_m\in\mathcal B(X)$ as well as $T>0$ be given. The following hold:
\begin{itemize}
\item[(i)] If $u\in L^1([0,T],\mathbb R^m)$ (i.e.~$u$ is Lebesgue measurable and satisfies $\int_0^T\|u(s)\|\,ds<\infty$) then
\begin{equation}\label{eq:control_eq}
\dot y(t)=\Big(A+\sum\nolimits_{j=1}^mu_j(t)B_j\Big) y(t)\quad\text{ with }\quad y(0)=y_0
\end{equation}
for all $y_0\in X$ has a unique mild solution on $[0,T]$, denoted by $y(t,u,y_0)$. 
\item[(ii)] If $u$ is piecewise constant then $y(t,u,y_0)=Y(t,u)y_0$ (with $Y(t,u)$ from Lemma \ref{lemma_pk_control_solution}) is the unique mild solution of \eqref{eq:control_eq}. 
\item[(iii)] If a sequence $(u^{(n)})_{n\in\mathbb N}\subseteq L^1([0,T],\mathbb R^m)$ converges to $u\in L^1([0,T],\mathbb R^m)$ in the weak topology then $\lim_{n\to\infty}y(t,u^{(n)},y_0)=y(t,u,y_0)$ uniformly on $[0,T]$.
\item[(iv)] If $y_0\in D(A)$ and the control $u$ is continuously differentiable then $y(t,u,y_0)$ is even a classical solution of \eqref{eq:control_eq}.
\end{itemize}
\end{proposition}
\begin{proof}
(i) \& (iii): The case $m=1$ is shown in \cite[Thm.~2.5 \& 3.6]{ball1982controllability} but the proof remains valid when replacing $|p(s)|$ by $\|p(s)\|$ (for us: $\|u(s)\|$). (iv): \cite[Rem.~2.7]{ball1982controllability}. (ii): For all $u\in\mathbb R^m$ by the bounded perturbation theorem \cite[Ch.~III, Thm.~1.3]{EngelNagel00} $(A+\sum_{j=1}^mu_jB_j,D(A))$ is the generator of a strongly continuous semigroup. Therefore $Y(\cdot,u):[0,T]\to\mathcal B(X)$ is well-defined and $\tops$-continuous so $y:[0,T]\to X$, $t\mapsto Y(t,u)y_0$ is continuous, as well. If we can show that this is a mild solution of \eqref{eq:control_eq} then uniqueness follows from (i). Indeed on each interval $(\tau_{i-1},\tau_i)$, $i=1,\ldots,N$ where $u$ is constant, $Y(t,u)y_0$ is a mild solution (by Lemma \ref{lemma_solution_const_gen} as the generator is time-independent on said interval). Then piecing together the solutions yields that $Y(t,u)y_0$ is a mild solution on $[0,T]$ because the missing times are of Lebesgue measure zero.
\end{proof}
With this the definitions for bilinear control problems carry over: Given the generator of a strongly continuous semigroup $(A,D(A))$, bounded operators $B_1,\ldots,B_m\in\mathcal B(X)$, and a control region $\Omega\subseteq\mathbb R^m$ we again define the system semigroup\index{system semigroup} $S_\Omega$ as the smallest semigroup in $\mathcal B(X)$ which contains the set
\begin{equation}\label{eq:system_semigroup_unbounded}
\Big\{\exp\Big(\tau\Big(A+\sum\nolimits_{j=1}^mu_jB_j\Big)\Big)\,\Big|\,\tau\geq 0,u\in\Omega\Big\}\,.
\end{equation}
Then given $T>0$ and $y_0\in X$---under the assumption PK---define the reachable sets\index{reachable set}
\begin{align*}
\mathfrak{reach}_{[0,T]}(y_0)&:=\bigcup\nolimits_{0\leq t\leq T}\{y(t,u,y_0)\,|\,u(\cdot)\in\Omega\}=\bigcup\nolimits_{0\leq t\leq T}\{Y(t,u)y_0\,|\,u(\cdot)\in\Omega\}\\
\mathfrak{reach}(y_0)&:=\bigcup\nolimits_{T>0}\mathfrak{reach}_{[0,T]}(y_0)=S_\Omega\;\! y_0\,.
\end{align*}
which is the collection of all mild solutions to \eqref{eq:control_eq}. With this \eqref{eq:system_semigroup_unbounded} is the reachable set for the group lift
$$
\dot Y(t)=\Big(A+\sum\nolimits_{j=1}^mu_j(t)B_j\Big) Y(t)\quad\text{ with }\quad Y(0)=\mathbbm{1}_X
$$
which now has to be taken in the strong sense\footnote{
Note that given $U\subseteq\mathbb R$ open, $t\in U$, and $f:U\to(\mathcal B(X),\tau)$ such that $(\mathcal B(X),\tau)$ is a topological vector space, one readily verifies that $f$ is differentiable in $t$ only if $f$ is continuous in $t$. We learned in Ch.~\ref{ch:qu_dyn_sys} if the generator $(A,D(A))$ of a semigroup $(T(t))_{t\geq 0}$ is unbounded then $t\mapsto T(t)$ fails to be norm continuous meaning one cannot have differentiability in norm. This justifies differentiating ``in the strong sense'', i.e.~considering $\lim_{h\to 0}\frac{Y(t+h)x-Y(t)x}{h}$ as a limit in $X$ for all $x\in X$, which by Prop.~\ref{prop_strong_weak_op_top} is nothing but differentiating $Y:U\to(\mathcal B(X),\tops)$.
}.
As before one finds (in slight abuse of notation) $\mathfrak{reach}(y_0)=\mathfrak{reach}(\operatorname{id_X})y_0$. Thus the messages to take home from this chapter so far are the following:
\begin{itemize}
\item After relaxing the notion of a ``solution'' we can guarantee existence and uniqueness of such solutions for our quantum control problems\footnote{
This at least is true if the control operators are bounded. For unbounded $H_j$ or even unbounded \textsc{gksl}-operators $V_j$ things, just like in Ch.~\ref{ch:qu_dyn_sys}, become even more difficult.
}
which for piecewise constant controls are of the same form as in finite dimensions.
\item As the piecewise continuous functions are dense in $L^1([0,T],\mathbb R^m)$ \cite[Ch.~III.3, Coro.~8]{Dunford58} point (iii) of Prop.~\ref{prop_control_eq} tells us $\mathfrak{reach}(y_0)|_{u \textrm{ PK}}\subseteq\mathfrak{reach}(y_0)|_{u\in L^1}\subseteq\overline{\mathfrak{reach}(y_0)|_{u \textrm{ PK}}}$ for all $y_0\in X$. Therefore $\overline{\mathfrak{reach}(y_0)|_{u \textrm{ PK}}}=\overline{\mathfrak{reach}(y_0)|_{u \in L^1}}$ which is why considering piecewise constant controls\index{control function!piecewise constant} is sufficient, as carries over from the bounded case.
\item Given $y_0\in D(A)$ and piecewise constant $u$, approximating $u$ via continuously differentiable controls $u^{(n)}$ leads to a uniform approximation of the mild solution $y(t,u,y_0)$ via classical solutions $y(t,u^{(n)},y_0)$.
\end{itemize}

After adjusting the formulation of bilinear control problems and establishing their well-posedness we can finally come back to the case of closed quantum systems:\index{closed quantum system!controlled}

\begin{definition}\label{def_inf_dim_unit_contr}
Let self-adjoint $H_0,H_1,\ldots,H_m$ acting on a complex Hilbert space $\mathcal H$, and a control region $\Omega\subseteq\mathbb R^m$ be given. Assume that the bilinear system (taken in the strong sense)
\begin{equation}\label{eq:bilinear_inf_dim_closed}
\dot U(t)=-i\Big(H_0+\sum\nolimits_{j=1}^mu_j(t)H_j\Big) U(t)\quad\text{ with }\quad U(0)=\mathbbm{1}
\end{equation}
is well-posed, that is, $H_0,\ldots,H_m$ admit a joint dense domain on which $H_0+\sum_{j=1}^mu_jH_j$ is essentially self-adjoint for all $u\in\mathbb R^m$. Then the system semigroup $S_\Omega$ generated by
$$
\Big\{\exp\Big(-i\tau\Big(H_0+\sum\nolimits_{j=1}^mu_jH_j\Big)\Big)\,\Big|\,\tau\geq 0,u\in\Omega\Big\}
$$
is well-defined, and given a subgroup $U_0\subseteq\mathcal U(\mathcal H)$ we call
\begin{itemize}
\item[(i)] system \eqref{eq:bilinear_inf_dim_closed} accessible\index{accessible} on $U_0$ if $S_\Omega$ has an interior point with respect to $U_0$. It is called uniformly approximately (strongly approximately) accessible\index{accessible!strongly approximately}\index{accessible!uniformly approximately} on $U_0$ if $\overline{S_\Omega}^{\,\mathrm{u}}$ ( $\overline{S_\Omega}^{\,\mathrm{s}}$) has\footnote{
As before $\overline{(\cdot)}^{\,\mathrm{u}}$ denotes the closure in $U_0$ with respect to the operator norm and $\overline{(\cdot)}^{\,\mathrm{s}}$ denotes the closure in $U_0$ with respect to the strong operator topology.
}
an interior point with respect to $(U_0,\|\cdot\|_\mathrm{op})$.
\item[(ii)] system \eqref{eq:bilinear_inf_dim_closed} is called controllable\index{controllable} on $U_0$ if $S_\Omega=U_0$. Moreover, it is called uniformly approximately (strongly approximately) controllable\index{controllable!strongly approximately}\index{controllable!uniformly approximately} on $U_0$ if $\overline{S_\Omega}^{\,\mathrm{u}}$ ( $\overline{S_\Omega}^{\,\mathrm{s}}$) is equal to $U_0$.
\end{itemize} 
\end{definition}
We already ``applied'' Lemma \ref{lemma_con_acc_id} to the original definition \ref{def_contr_bounded} by making controllability and accessibility only depend on the respective property at the identity. Thus this definition is as general as possible, although we note that Prop.~\ref{prop_control_eq} (i.e.~existence of the solution for arbitrary $L^1$ controls, continuity of $y(t,u,y_0)$ in $u$, connection to the Liouville-von Neumann equation) does not apply if the control operators are unbounded.

The notion of approximate state controllability in the more subtle case of unbounded control operators has already been studied, to name just a few examples,
\begin{itemize}
\item under the restriction of a finite-dimensional state space (despite $\operatorname{dim}(\mathcal H)=\infty$) \cite{Huang83,bloch2010finite}.
\item more generally via Galerkin approximations (cf.~\cite{boscain2012weak,boscain2015approximate,caponigro2018exact} and the references therein).
\item using finite-dimensional Lie algebraic techniques in connection with invariant subspaces \cite{KZSH,HoKe17}.
\end{itemize}
The path we will take, however, will be to study the group-lifted system meaning we get access to powerful operator- as well as Lie-theoretic methods \cite{keyl18InfLie}. This is backed up by the following important result
: 
\begin{lemma}[State Approximation Lemma]\label{lemma_state_approx}\index{state approximation lemma}
Consider a non-empty subset $U_0\subseteq\mathcal U(\mathcal H)$ and $R\subseteq U_0$ such that $\overline{ R }^{\,\mathrm{s}}=U_0$ (i.e.~$R$ is dense in the $U_0$ with respect to the strong operator topology on $\mathcal U(\mathcal H)$). Given $T\in\mathcal B^1(\mathcal H)$, $U\in U_0$, and $\varepsilon>0$ one finds $\tilde U\in R$ such that $\|UT U^*-\tilde UT\tilde U^*\|_1<\varepsilon$.
\end{lemma}
\begin{proof}
The case $U_0=\mathcal U(\mathcal H)$ is shown in \cite[Lemma 6]{OSID19} and the proof remains valid for arbitrary (non-empty) subsets $U_0$. 
\end{proof}
\noindent In other words strong approximate controllability of the propagators on a subgroup $U_0\subseteq \mathcal U(\mathcal H)$ becomes approximate controllability of the Liouville-von Neumann equation on $\{U\rho_0 U^*\,|\,U\in U_0\}$ for all $\rho_0\in\mathbb D(\mathcal H)\subseteq(\mathcal B^1(\mathcal H),\|\cdot\|_1)$.\medskip

Remarkably, Ball, Marsden, and Slemrod showed that given $H_0$ self-adjoint and $H_1,\ldots,H_m\in\mathcal B(\mathcal H)$ the controlled Schr\"odinger equation can never be (exactly) controllable if $\operatorname{dim}(\mathcal H)=\infty$, no matter the initial state $|\psi\rangle\in \mathcal H$ (slightly adjust \cite[Thm.~3.6]{ball1982controllability} to the state space $ S_1(\mathcal H)$). Therefore the lifted system cannot be (exactly) controllable on $\mathcal U(\mathcal H)$ meaning we have to opt for an approximate notion of controllability.
However, even in the bounded case the topology induced by the operator norm is still too strong: $\mathfrak{reach}(\mathbbm{1})$ is norm-separable for all $H_0,H_1,\ldots,H_m\in\mathcal B(\mathcal H)$ by \cite[Thm.~3.6]{ball1982controllability}\footnote{
If $\{T_t\}_{t\geq 0}$ is a norm-continuous semigroup on a Banach space $X$ then the induced multiplication operators $\hat T_t(B):=T_t\cdot B$ form a norm-continuous semigroup on $\mathcal B(X)$---and if $A\in\mathcal B(X)$ is the generator of $T_t$ then the map $B\mapsto A\cdot B$ is the generator of $\hat T_t$---as is readily verified. Apply said theorem with $X=\mathcal B(\mathcal H)$ and use Lemma \ref{lemma_separable_countable_union} \& \ref{lemma_comp_met_sep} to see that the countable union of compact sets in a metric space is separable.
}.
But then $\overline{S}^{\,\mathrm{u}}=\mathcal U(\mathcal H)$ would imply that the unitary group is norm-separable, a contradiction (Thm.~\ref{theorem_unitary_group_properties} (viii)). Therefore, as is also hinted at by Lemma \ref{lemma_state_approx}, it is most reasonable to aim for strong approximate controllability of closed, infinite-dimensional quantum systems.\medskip

In finite dimensions the phenomenon of recurrence\index{recurrence} automatically turns the system semigroup $S_\Omega$ of a closed system into a subgroup of the unitaries: Given $H\in\mathbb C^{n\times n}$ Hermitian one finds
$
\overline{\{e^{-i\tau H}\,|\,\tau>0\}}=\overline{\{e^{-i\tau H}\,|\,\tau\in\mathbb R\}}
$ due to compactness of the unitary group in finite dimensions. This readily implies that the closure of the system semigroup $\overline{S_\Omega}=:\mathcal G$ is a closed subgroup of the unitaries called the \textit{dynamical group}\index{dynamical group}. Thus we can study the reachable set via its Lie algebra $\mathfrak{g}$, called the \textit{dynamical Lie algebra}\index{dynamical Lie algebra}. 
In infinite dimensions we lose access to compactness (Prop.~\ref{theorem_unitary_group_properties} (xi)) so this construction in general does not work anymore. Rare exceptions are special cases such as self-adjoint operators which have only eigenvalues in its spectrum \cite[Prop.~3.1]{keyl18InfLie}. Thus checking (approximate) controllability in infinite dimensions can be done via the following general strategy \cite[Ch.~2]{keyl18InfLie}:
\begin{itemize}
\item[1.] Check that the solutions of \eqref{eq:bilinear_inf_dim_closed} are well-defined and generate the solutions of the controlled Schr{\"o}dinger / Liouville-von Neumann equation (assuming piecewise constant controls).
\item[2.] Prove that $\overline{S_\Omega}^{\mathrm{s}}$ is a group which co{\"i}ncides with the generated dynamical group $\mathcal G$. If it fails to be a group then one already has a no-go result.
\item[3.] Show for all bounded control operators $H_j$ that $iH_j$ is in the dynamical Lie algebra.
\item[4.] If some of the $H_j$ are unbounded find a set of generators (in $\mathfrak g$) which replace it.
\item[5.] Calculate repeated commutators of all (replaced) generators and show that any element of $\mathfrak u(\mathcal H)$ can be strongly approximated be them.
\end{itemize}

One of the most general results currently known 
is the following~\cite{keyl18InfLie}: Let $H_0, . . . ,H_m$ be self-adjoint operators on a separable Hilbert space $\mathcal H$.
Further assume that
\begin{itemize}
\item[(i)] $H_0$ is bounded or unbounded, but has only pure point spectrum. The eigenvalues
$x_k, k \in \mathbb N$ are non-degenerate and rationally independent.
\item[(ii)] The operators $H_1, . . . ,H_m$ are bounded and the set $\{H_1, . . . ,H_m\}$ 
is connected\footnote{This means that the associated graph (which roughly
speaking indicates whether a transition from energy level $k$ to $l$ is possible) has to be
connected, cf.~\cite{keyl18InfLie}.} with respect to a complete set of eigenvectors $\phi_k \in \mathcal H$,
$k \in \mathbb N$ of $H_0$.
\end{itemize}
Then \eqref{eq:bilinear_inf_dim_closed} is strongly approximately controllable on $\mathcal U(\mathcal H)$. The result can be generalized to eigenvalues $x_k$, $k \in \mathbb{N}$ with finite multiplicities, but 
this requires more technical conditions on the control Hamiltonians: One has to ensure that trace-free 
finite-rank operators commuting with all eigenprojections of $H_0$ are contained in the strong closure of
the Lie algebra generated by the $H_j$, $j=1,\dots,m$. More challenging are drift
Hamiltonians with rationally dependent eigenvalues which, however, can be studied in terms of certain
non-Abelian von Neumann algebras.\medskip

To conclude this chapter let us address the ``elephant in the room'': Does the powerful Lie algebra rank condition\index{Lie algebra rank condition} from finite dimensions carry over to the general case? Interestingly enough one can find simple counterexamples which show that $\overline{ \langle A,B_1,\ldots,B_m\rangle_\mathrm{Lie} }$ in general does not imply that the corresponding group lifted problem is (approximately) accessible (due to G.~Dirr, private communication). While there are some candidates for a suitable replacement this problem is still subject of current research.


\chapter{Majorization and the $C$-Numerical Range}\label{ch_maj_cnr}

After setting the stage for studying (Markovian) quantum control problems the final thing we need are appropriate tools to characterize the corresponding reachable sets: While there will be situations where we can show $\overline{\mathfrak{reach}(\rho_0)}=\mathbb D(\mathcal H)$ for arbitrary initial states $\rho_0$ this is the exception rather than the rule. Because we will be particularly interested in bath couplings as the dissipative action, the notion of majorization will be of utmost importance (and later on we will see why).

The concept of majorization as first introduced by Muirhead \cite{Muirhead02} and more widely spread 
by Hardy, Littlewood, and P{\'o}lya \cite{Hardy52}, roughly speaking describes if a vector with real entries is ``less 
or more nearly equal'' than another which found numerous applications in various fields of science, e.g., 
\cite{Lorenz05,Dalton20,Parker80,BSH16,OSID17}. More precisely, one says that a vector $x\in\mathbb R^n$ majorizes \index{majorization!on vectors}
$y\in\mathbb R^n$, denoted by $x\prec y$\label{symb_maj_1}, if $\sum\nolimits_{i=1}^n x_i=\sum\nolimits_{i=1}^n y_i$ and 
$\sum\nolimits_{i=1}^k x_i^\downarrow\leq \sum\nolimits_{i=1}^k y_i^\downarrow$ for all $k=1,\ldots,n$, where $x_i^\downarrow,y_i^\downarrow$ 
are the components of $x,y$ in decreasing order\label{symb_downarrow_vec}. 
 A comprehensive survey on classical majorization as well as its applications can be found in Marshall \& Olkin \cite{MarshallOlkin}.
 
As for the route we will take: In Ch.~\ref{sec:maj_d_vec}---after a quick recap of convex polytopes and their different descriptions---majorization will be generalized from the maximally mixed distribution $(1,\ldots,1)^T$ as reference, to arbitrary distributions with strictly positive entries, called \textit{$d$-majorization}. By looking at majorization from the viewpoint of convex polytopes we will learn about the underlying ``geometry'' which will be essential to upper bound some reachable sets in Ch.~\ref{sec_reach_fin_dim}. Following up we will learn how to generalize majorization from vectors to (Hermitian) matrices for which the notion of strict positivity will be useful (Ch.~\ref{sec:maj_d_mat}). While the chapter up until then dealt with finite-dimensional systems, Ch.~\ref{sec:c_num_range} features the $C$-numerical range in infinite dimensions which does not only find application in optimal control problems \cite{DiHeGAMM08,SDHG08} but is also deeply connected to majorization on matrices. Therefore we can apply these results when generalizing majorization to infinite dimensions, that is, to trace-class operators in Ch.~\ref{maj:trace_class}.

\section{Majorization on Vectors}\label{sec:maj_d_vec}
This section is entirely based on one of our preprints \cite{vomEnde19polytope}. 
%
%
Over the last few years, sparked by Brand\~ao, Horodecki, Oppenheim \cite{Brandao15,Horodecki13}, and others \cite{Faist17,Gour15,Lostaglio18,Sagawa19,Mazurek19}, thermomajorization has been a widely discussed and researched 
topic in quantum physics and in particular quantum thermodynamics\index{quantum thermodynamics}. Mathematically 
speaking, this is about majorization relative to an entrywise positive vector $d\in\mathbb R^n$ as introduced by Veinott \cite{Veinott71} 
and (in the quantum regime) Ruch, Schranner, and Seligman \cite{Ruch78}. For such positive $d$, some vector 
$x$ is said to $d$-majorize $y$, denoted by $x\prec_d y$, if there exists a column-stochastic
matrix $A$ with 
$Ad=d$ and $x=Ay$. Such $A$ is called a $d$-stochastic matrix, cf. Def.~\ref{defi_d_stochastic_matrix}. 
A variety of characterizations of $\prec_d$ and $d$-stochastic matrices can be found in the work of Joe \cite{Joe90}, or Prop.~\ref{lemma_char_d_vec} below.
%
%
%
For this purpose be aware of the following notions and notations:
\begin{itemize}
\item In accordance with Marshall and Olkin \cite{MarshallOlkin}, $\mathbb R_+^n$ ($\mathbb R_{++}^n$)\label{symb_R_plus}
denotes the set of all real vectors with non-negative (strictly positive) entries. Whenever it is clear that $x$ is a real vector of length $n$ we occasionally write $x>0$ to express strict positivity of its entries, i.e.~$x\in\mathbb R_{++}^n$.
\item $\unitvector$ shall denote the column vector of ones, i.e.~$\unitvector=(1,\ldots,1)^T$.
\item $S_n$ is the symmetric group, that is, the group of all permutations of order $n$.\label{symb_sym_group}\index{symmetric group}
\item The standard simplex $\Delta^{n-1}\subseteq\mathbb R^n$\label{symb_simplex}\index{standard simplex}
is given by the convex hull of all standard basis vectors $e_1,\ldots,e_n$ and precisely contains all probability vectors, i.e.~all vectors $x\in\mathbb R_+^n$ with $\unitvector^Tx=1$.
\item For simplicity we use the convention that $\min$ and $\max$ operates entrywise on vectors, meaning $\min\{b,b'\}=(\min\{b_j,b_j'\})_{j=1}^m$ for all $b,b'\in\mathbb R^m$.
\end{itemize}
\subsection{Convex Polytopes and Majorization}\label{sec_prelim_vector}

Convex polytopes\index{convex polytope} usually are introduced as subsets of $\mathbb R^n$ which can be written as the convex hull of finitely many vectors from $\mathbb R^n$, cf.~\cite[Ch.~7.2]{Schrijver86}, \cite[Ch.~3]{Gruenbaum03}. Now it is well-known that such polytopes can be characterized via finitely many affine half-spaces; more precisely a set $P\subset \mathbb R^n$ is a convex polytope if and only if $P$ is bounded and there exist $m\in\mathbb N$, $A\in\mathbb R^{m\times n}$, and $b\in\mathbb R^m$ such that $P=\{x\in\mathbb R^n\,|\,Ax\leq b\}$ \cite[Coro.~7.1c]{Schrijver86}. These characterizations of convex polytopes are also known as $\mathscr V$- and $\mathscr H$-description\index{V-description@$\mathscr V$-description}\index{H-description@$\mathscr H$-description}, respectively \cite[Ch.~3.6]{Gruenbaum03}. 
\begin{remark}\label{rem:halfspace_actions}
Let any $A\in\mathbb R^{m\times n}$, $b,b'\in\mathbb R^m$, and $p\in\mathbb R^n$ be given. The following observations are readily verified.
\begin{align*}
\{x\in\mathbb R^n\,|\,Ax\leq b\}\cap \{x\in\mathbb R^n\,|\,Ax\leq b'\}&=\{x\in\mathbb R^n\,|\,Ax\leq\min\{b,b'\}\}\\
\{x\in\mathbb R^n\,|\,Ax\leq b\}+p&=\{x\in\mathbb R^n\,|\,Ax\leq b+Ap\}\\
\{x\in\mathbb R^n\,|\,Ax\leq b\}&\subseteq \{x\in\mathbb R^n\,|\,Ax\leq b'\}\quad\text{ if and only if }b\leq b'\,.
\end{align*}
This is not too surprising as the matrix $A$ in some sense describes the geometry of the polytope which intuitively should not change under the above operations.
\end{remark}
Over the course of this chapter we want to explore sets $\{x\in\mathbb R^n\,|\,Mx\leq b\}$ where\footnote{Note that $\sum_{j=1}^{n-1}\binom{n}{j}+2=\sum_{j=0}^n\binom{n}{j}=2^n$ by the binomial theorem\index{theorem!binomial}, which shows $M\in\mathbb R^{2^n\times n}$.}
\begin{equation}\label{eq:M_maj}
M:={\footnotesize\begin{pmatrix} M_1\\M_2\\\vdots\\M_{n-1}\\\unitvector^T\\-\unitvector^T \end{pmatrix}}\in\mathbb R^{2^n\times n}
\end{equation}
and the rows of $M_j\in\mathbb R^{\binom{n}{j}\times n}$ are made up of all elements of 
$$
\Big\{(x_1,\ldots,x_n)\,\Big|\,x_1,\ldots,x_n\in\{0,1\}\text{ and }\sum\nolimits_{i=1}^n x_i=j\Big\}
$$
in an arbitrary, but fixed order. 
In particular we are interested in the case where $b\in\mathbb R^{2^n}$ includes a trace equality condition, meaning $b_{2^n-1}+b_{2^n}=0$ or, equivalently,
$$
b={\footnotesize \begin{pmatrix} b_1\\\vdots\\b_{n-1}\\b_n\\-b_n\end{pmatrix} }\in\mathbb R^{2^n}\,.
$$
with $b_j\in\mathbb R^{\binom{n}{j}}$ for all $j=1,\ldots,n$. 

\begin{lemma}\label{lemma_Mx_b_convex_polytope}
Let $M$ be the matrix \eqref{eq:M_maj} and $b\in\mathbb R^{2^n}$ with $b_{2^n-1}+b_{2^n}=0$ be given. If $\{x\in\mathbb R^n\,|\,Mx\leq b\}$ is non-empty then it is a convex polytope of at most $n-1$ dimensions.
\end{lemma}
\begin{proof}
Because $b_{2^n-1}+b_{2^n}=0$ by assumption, all solutions to $Mx\leq b$ have to satisfy $\unitvector^Tx=b_{2^n-1}$ which reduces the dimension of $\{x\in\mathbb R^n\,|\,Mx\leq b\}$ by $1$. Now if we can show that $\{x\in\mathbb R^n\,|\,Mx\leq 0\}=\{0\}$ then the set in question is bounded (cf.~\cite[Ch.~8.2]{Schrijver86}) so by the previous characterization of convex polytopes we would be done.
Indeed if $x\in\mathbb R^n$ satisfies $Mx\leq 0$ then $M_1x=x\leq 0$ and---because $M_{n-1}$ is of the form $\unitvector\unitvector^T-\mathbbm{1}_n$ (up to row permutation)---one has $M_{n-1}x\leq 0$ which together with $\unitvector^Tx=b_{2^n-1}=0$ yields $0\geq x\geq \unitvector(\unitvector^Tx) =0$, so $x=0$ as desired. 
\end{proof}

Now an immediate question is the one concerning the extreme points\index{convex polytope!extreme point} (``vertices''\index{vertex}) of said convex polytope. This will be the topic of the remaining part of this section.
For this we need a characterization of the extreme points of a convex polytope given in $\mathscr H$-description.
\begin{definition}\label{def_mathfrak_b}
Let $b\in\mathbb R^{2^n}$ with $b_{2^n-1}+b_{2^n}=0$ as well as $p\in\{0,1\}^n$, $p\neq 0$ be given. Then the row vector $p$ corresponds to a unique row of $M\in\mathbb R^{2^n\times n}$ as well as a corresponding entry $\mathfrak b(p)$ in $b$ (when considering the vector inequality $Mx\leq b$). Setting $\mathfrak b(0):=0$ this defines a map $\mathfrak b:\{0,1\}^n\to\mathbb R$, $p\mapsto\mathfrak b(p)$ which naturally generalizes to arbitrary $A\in\{0,1\}^{m\times n}$, $b\in\mathbb R^m$, $m\in\mathbb N$ via\footnote{
For this map to be well-defined we need the assumption that no row $a^T$ of $A$ appears twice. But this is rather natural because one of the two inequalities $a^Tx\leq b$, $a^Tx\leq b'$ is redundant and can be disregarded.
}
$$
\mathfrak b(A)=\mathfrak b\Big( \begin{pmatrix} a_1\\a_2\\\vdots\\a_m \end{pmatrix} \Big)=\begin{pmatrix} \mathfrak b(a_1)\\\mathfrak b(a_2)\\\vdots\\\mathfrak b(a_m) \end{pmatrix}\,.
$$
\end{definition}
\begin{lemma}\label{lemma_extreme_points_h_descr}
Let $b\in\mathbb R^{2^n}$ with $b_{2^n-1}+b_{2^n}=0$ as well as $p\in\mathbb R^n$ be given such that $Mp\leq b$. The following statements are equivalent.
\begin{itemize}
\item[(i)] $p$ is an extreme point of $\{x\in\mathbb R^n\,|\,Mx\leq b\}$.
\item[(ii)] There exists a submatrix $M'\in\mathbb R^{n\times n}$ of $M$---one row of $M'$ being equal to $\unitvector^T$---such that $M'p=\mathfrak b(M')=:b'$ and $\operatorname{rank}M'=n$.
\end{itemize}
\end{lemma}
\begin{proof}
%
``(ii) $\Rightarrow$ (i)'': Assume there exist $x_1,x_2\in\{x\in\mathbb R^n\,|\,Mx\leq b\}$ and $\lambda\in (0,1)$ such that $p=\lambda x_1+(1-\lambda)x_2$. Then $M'x_1\leq b'$, $M'x_2\leq b'$ by assumption and thus
$$
b'=M'p=\lambda M'x_1+(1-\lambda)M'x_2\leq \lambda b'+(1-\lambda)b'=b'\quad\Rightarrow\quad M'x_1=M'x_2=b'\,.
$$
But $M'\in\mathbb R^{n\times n}$ is of full rank so the system of linear equations $M'y=b$ has a unique solution in $\mathbb R^n$. This implies $x_1=x_2=p$ so $p$ is in fact an extreme point of $\{x\in\mathbb R^n\,|\,Mx\leq b\}$.

``(i) $\Rightarrow$ (ii)'': 
Each extreme point $p$ of $\{x\in\mathbb R^n\,|\,Mx\leq b\}$ is determined by $n$ linearly independent equations from $Mx=b$ so there exists a submatrix $\hat M\in\mathbb R^{n\times n}$ of $M$ of full rank such that $\hat Mp=\mathfrak b(\hat M)=:\hat b$, cf.~\cite[Thm.~8.4 ff.]{Schrijver86}. If one row of $\hat M$ equals $\unitvector^T$ then we are done. Otherwise define
$$
\tilde{M}:={\begin{pmatrix} \hat M\\\unitvector^T \end{pmatrix}}\in\mathbb R^{(n+1)\times n}\quad\text{ and }\quad\tilde{b}:=\mathfrak b(\tilde M)={\begin{pmatrix} \hat b\\b_{2^n-1} \end{pmatrix}}\in\mathbb R^{n+1}
$$
so $\tilde{M}p=\tilde{b}$ because $p$ satisfies the trace condition. But this system of linear equations is now overdetermined so there exists a row of $\tilde{M}$---aside from $\unitvector^T$---which is redundant and can be removed. The resulting matrix $M'\in\mathbb R^{n\times n}$ is of full rank, contains $\unitvector^T$, and 
satisfies $M'p=\mathfrak b(M')$.
\end{proof}

This enables---in some special cases---an explicit description of the extreme points of the polytope induced by $M$ and $b$.

\begin{definition}\label{def_Ebsigma}
Let $b\in\mathbb R^{2^n}$ with $b_{2^n-1}+b_{2^n}=0$ and arbitrary ${\pi}\in S_n$ be given. Denote by $\underline{{\pi}}$ the permutation matrix\footnote{
Given some permutation ${\pi}\in S_n$ the corresponding permutation matrix is 
given by $\sum_{i=1}^n e_i e_{{\pi}(i)}^T$. In particular the identities $(\underline{{\pi}}x)_j=x_{{\pi}(j)}$ and $\underline{{\pi}\circ\tau}=\underline{\tau}\cdot\underline{{\pi}}$ hold for all $\pi,\tau\in S_n$, $x\in\mathbb C^n$, $j\in\{1,\ldots,n\}$.\label{footnote_permutation_matrix}
}
induced by ${\pi}$. Then the unique solution to
\begin{equation}\label{eq:E_sigma_unique_sol}
{\footnotesize\begin{pmatrix}
1&0&\cdots&0\\
\vdots&\ddots&\ddots&\vdots\\
\vdots&&\ddots&0\\
1&\cdots&\cdots&1
\end{pmatrix}}\underline{{\pi}}x=\mathfrak b\Big( {\footnotesize\begin{pmatrix}
1&0&\cdots&0\\
\vdots&\ddots&\ddots&\vdots\\
\vdots&&\ddots&0\\
1&\cdots&\cdots&1
\end{pmatrix}}\underline{{\pi}} \Big)=:b_{\pi}
\end{equation}
shall be denoted by $x=E_b({\pi})$.
\end{definition} 

Now $E_b({\pi})$ is of the following simple form.
\begin{lemma}\label{lemma_E_b_sigma_properties}
Let $b\in\mathbb R^{2^n}$ with $b_{2^n-1}+b_{2^n}=0$, arbitrary ${\pi}\in S_n$, as well as $p\in\mathbb R^n$ be given. 
Then for all $j=1,\ldots,n$
\begin{equation}\label{eq:E_sigma_property}
(E_b({\pi}))_{{\pi}(j)}=(b_{\pi})_{j}-(b_{\pi})_{j-1}=\mathfrak b\big(\sum\nolimits_{i=1}^j e_{{\pi}(i)}^T\big)-\mathfrak b\big(\sum\nolimits_{i=1}^{j-1} e_{{\pi}(i)}^T\big)
\end{equation}
and $E_{b+Mp}({\pi})=E_b({\pi})+p$ for all ${\pi}\in S_n$.
\end{lemma}
\begin{proof}
The $j$-th row of \eqref{eq:E_sigma_unique_sol} for $x=E_{b}({\pi})$, $j=1,\ldots,n$ reads
\begin{align*}
\mathfrak b\big(\big(\sum\nolimits_{i=1}^j e_i^T\big)\underline{{\pi}}\big)=\mathfrak b\big(\sum\nolimits_{i=1}^j e_{{\pi}(i)}^T\big)=\big(\sum\nolimits_{i=1}^j e_{{\pi}(i)}^T\big) E_{b}({\pi})=\sum\nolimits_{i=1}^j (E_{b}({\pi}))_{{\pi}(i)}
\end{align*}
which implies \eqref{eq:E_sigma_property}. Also one readily verifies
$$
(b+Mp)_{\pi}=b_{\pi}+\underbrace{\big(\sum\nolimits_{i=1}^j p_{{\pi}(i)}\big)_{j=1}^n}_{\text{corresponding entry in }Mp}=b_{\pi}+{\footnotesize\begin{pmatrix}
1&0&\cdots&0\\
\vdots&\ddots&\ddots&\vdots\\
\vdots&&\ddots&0\\
1&\cdots&\cdots&1
\end{pmatrix}}\underline{{\pi}}p
$$
for any ${\pi}\in S_n$ so $E_{b+Mp}({\pi})=E_b({\pi})+p$ by uniqueness of the solution of \eqref{eq:E_sigma_unique_sol}. 
\end{proof}

Clearly, if $E_b({\pi})\in\{x\in\mathbb R^n\,|\,Mx\leq b\}$ for some ${\pi}\in S_n$ then it is an extreme point by Lemma \ref{lemma_extreme_points_h_descr} although, in general, not every $E_b({\pi})$ needs to be in $\{x\in\mathbb R^n\,|\,Mx\leq b\}$ for arbitrary $b\in\mathbb R^{2^n}$ with $b_{2^n-1}+b_{2^n}=0$ (cf.~Example \ref{ex_proof_ext_point_fail}). However for the well-structured polytopes we will deal with later on all of these $E_b({\pi})$ lie within the polytope, in which case these are the only extreme points:
\begin{theorem}\label{thm_Ebsigma_extreme}
Let $b\in\mathbb R^{2^n}$ with $b_{2^n-1}+b_{2^n}=0$ be given such that $\{E_b({\pi})\,|\,{\pi}\in S_n\}\subset \{x\in\mathbb R^n\,|\,Mx\leq b\}$. Then every extreme point of $\{x\in\mathbb R^n\,|\,Mx\leq b\}$ is of the form $E_b({\pi})$ for some ${\pi}\in S_n$ and therefore $\{x\in\mathbb R^n\,|\,Mx\leq b\}=\operatorname{conv}\{E_b({\pi})\,|\,{\pi}\in S_n\}$.
\end{theorem}
For the rather technical proof see Appendix \ref{app_thm_minmax_matrix}. Following Remark \ref{rem_extend_lemma_minmax} one can even improve upon this result: If (in the proof) the matrix $M'$ corresponding to $p$ contains two rows $m_1^T,m_2^T$ such that $m_1^T\not\geq m_2^T\not\geq m_1^T$, i.e.~the rows are incomparable and do not belong to some equation \eqref{eq:E_sigma_unique_sol}, then there exist at least two permutations ${\pi}_1,{\pi}_2\in S_n$ such that $E_b({\pi}_1)=p=E_b({\pi}_2)$. Notably in such a situation the map ${\pi}\mapsto E_b({\pi})$ is not injective.\medskip


In order to generalize majorization to arbitrary weight vectors it, unsurprisingly, is advisable to first recap and explore classical vector majorization. 
As described before, the common definition of vector majorization goes as follows: given $x,y\in\mathbb R^n$ one says $x$ majorizes 
$y$, denoted by $x\prec y$, if $\sum\nolimits_{i=1}^n x_i=\sum\nolimits_{i=1}^n y_i$ and 
\begin{equation}\label{eq:maj_cond}
\sum\nolimits_{i=1}^j x_i^\downarrow\leq \sum\nolimits_{i=1}^j y_i^\downarrow\quad\text{ for all }j=1,\ldots,n-1
\end{equation}
where $x_i^\downarrow,y_i^\downarrow$ 
are the components of $x,y$ in decreasing order, respectively. Given how well-explored this concept is there are a handful of 
characterizations for $\prec$, cf.~\cite[Ch.~1, Point A.3]{MarshallOlkin}. The most notable one for our purposes is 
the following: $y$ majorizes $x$ if and only if there exists a doubly stochastic matrix, i.e.~a matrix $A\in\mathbb R_+^{n\times n}$ which satisfies $\unitvector^TA=\unitvector^T$ and $A\unitvector=\unitvector$, such that $x=Ay$. Related to this, Birkhoff's theorem 
\cite[Ch.~2, Thm.~A.2]{MarshallOlkin} states that the set of doubly stochastic matrices equals the convex hull 
of all permutation matrices (cf.~footnote \ref{footnote_permutation_matrix}) and that these permutation matrices are precisely the extreme points of said set. 

Thus Birkhoff's theorem directly implies that for vectors $x,y\in\mathbb R^n$ one has $x\prec y$ if and only if $x$ lies in the convex hull of the $n!$ permutations of $y$---as also shown in \cite{Rado52}---so the set $\{x\in\mathbb R^n\,|\,x\prec y\}$ is a convex polytope with at most $n!$ corners. This motivates finding its half-space description.

\begin{proposition}\label{prop_maj_halfspace}
Let $y\in\mathbb R^n$. Then
$$
\{x\in\mathbb R^n\,|\,x\prec y\}=\{x\in\mathbb R^n\,|\,Mx\leq b_y\}
$$
where $M$ is the matrix from \eqref{eq:M_maj} and $b_y\in\mathbb R^{2^n}$ is of the following form:
the first $\binom{n}{1}$ entries equal $y_1^\downarrow$, the next $\binom{n}{2}$ entries equal $y_1^\downarrow+y_2^\downarrow$, and so forth until $\binom{n}{n-1}$ entries equaling $\sum_{i=1}^{n-1} y_i^\downarrow$. The last two entries are $\unitvector^Ty$ and $-\unitvector^Ty$, respectively. 
\end{proposition}
\begin{proof}
First be aware that by construction $Mx\leq b_y$ translates to $M_jx\leq ( \sum_{i=1}^j y_i^\downarrow)\unitvector$ for all $j=1,\ldots,n-1$ as well as $\unitvector^Tx\leq \unitvector^Ty$, $ -\unitvector^Tx\leq -\unitvector^Ty$, the latter obviously being equivalent to $\unitvector^Tx=\unitvector^Ty$. This equality of trace together with 
$
\sum\nolimits_{i=1}^j x_i^\downarrow\leq \sum\nolimits_{i=1}^j y_i^\downarrow
$
for all $j=1,\ldots,n-1$ by definition would show $x\prec y$. 

``$\subseteq$'': Let any $j=1,\ldots,n-1$. Given how we constructed $M_j$ every entry of $M_jx$ is of the form $\sum_{i=1}^j x_{{\pi}(i)}$ for some permutation ${\pi}$, but $\sum_{i=1}^j x_{{\pi}(i)}\leq \sum_{i=1}^j x_i^\downarrow$ which in turn is upper bounded by $\sum_{i=1}^j y_i^\downarrow$ by assumption. Thus $Mx\leq b_y$.

``$\supseteq$'': If $M_jx\leq ( \sum_{i=1}^j y_j^\downarrow)\unitvector$ then $\sum_{i=1}^j x_{{\pi}(i)}\leq \sum_{i=1}^j y_i^\downarrow$ for all permutations ${\pi}$, so in particular for the permutation which picks the $j$ largest values of $x$; thus \eqref{eq:maj_cond} holds.
\end{proof}
The $\mathscr H$-description of majorization\index{majorization!H-description@$\mathscr H$-description} enables a (to our knowledge) new proof to a well-known result already stated before.
\begin{corollary}\label{coro_classical_maj_extreme}
Let $y\in\mathbb R^n$. Then $\{x\in\mathbb R^n\,|\,x\prec y\}=\operatorname{conv}\{\underline{{\pi}}y\,|\,{\pi}\in S_n\}$ where every $\underline{{\pi}}y$ is an extreme point, so in particular this set has at most $n!$ extreme points. 
\end{corollary}
\begin{proof}
Given $y$ there exists a permutation $\tau\in S_n$ such that $\underline{\tau}\,y=(y_1^\downarrow,y_2^\downarrow,\ldots,y_n^\downarrow)^T$. Then for all ${\pi}\in S_n$ the unique solution of
$$
{\footnotesize\begin{pmatrix}
1&0&\cdots&0\\
\vdots&\ddots&\ddots&\vdots\\
\vdots&&\ddots&0\\
1&\cdots&\cdots&1
\end{pmatrix}}\,\underline{{\pi}^{-1}\circ \tau}\,p={\footnotesize\begin{pmatrix}
1&0&\cdots&0\\
\vdots&\ddots&\ddots&\vdots\\
\vdots&&\ddots&0\\
1&\cdots&\cdots&1
\end{pmatrix}}\,\underline{\tau}\,\underline{{\pi}^{-1}}\,p=\big(\sum\nolimits_{i=1}^j y_i^\downarrow\big)_{j=1}^n=b_{{\pi}^{-1}\circ \tau}
$$
is obviously given by $p=\underline{{\pi}}y\in \{x\in\mathbb R^n\,|\,x\prec y\}$ which by Lemma \ref{lemma_extreme_points_h_descr} is an extreme point. On the other hand $p=E_b({\pi}^{-1}\circ \tau)$ (Definition \ref{def_Ebsigma}) so
$$
\{\underline{{\pi}}y\,|\,{\pi}\in S_n\}= \{E_b({\pi}^{-1}\circ \tau)\,|\,{\pi}\in S_n\}=\{E_b(\tilde{\pi})\,|\,\tilde{\pi}\in S_n\}\subset \{x\in\mathbb R^n\,|\,Mx\leq b\}\,.
$$
But this by Thm.~\ref{thm_Ebsigma_extreme} concludes the proof.
\end{proof}

\subsection{Characterizations and Order Properties of $\prec_d$}\label{sec_order_d_maj}
Having developed tools surrounding classical vector majorization as well as general convex polytopes we are finally prepared to investigate the non-symmetric case, that is, the case where the fixed point $\unitvector$ of the doubly stochastic\index{matrix!doubly stochastic} matrices becomes an arbitrary but fixed element from $\mathbb R_{++}^n$. For the definition we mostly follow \cite[p.~585]{MarshallOlkin}. 
\begin{definition}\label{defi_d_stochastic_matrix}
Let $d\in\mathbb R_{++}^n$ and $x,y\in\mathbb R^n$ be given.
\begin{itemize}
\item[(i)] A quadratic matrix $A\in\mathbb R^{n\times n}$ is said to be \textit{column-stochastic}\index{matrix!column-stochastic} if
\begin{itemize}
\item[(a)] $a_{ij}\geq 0$ for all $i,j$ (i.e.~$A\in\mathbb R_+^{n\times n}$),
\item[(b)] $\unitvector^TA=\unitvector^T$.
\end{itemize}
\noindent If, additionally, $Ad=d$ then $A$ is said to be \textit{$d$-stochastic}\index{matrix!d-stochastic@$d$-stochastic}. The set of all $d$-stochastic $n\times n$ matrices is denoted by $ s_d(n)$.\label{symb_d_stoc_mat}
\item[(ii)] Furthermore $x$ is said to be $d$-majorized by $y$, denoted by $x\prec_d y$, if there exists $A\in s_d(n)$ such that $x=Ay$.\index{d-majorization@$d$-majorization}
\end{itemize}
\end{definition}
\noindent In particular, $x\prec_d y$ implies
$
\unitvector^Tx=\unitvector^TAy=\unitvector^Ty\,.
$
Note that this definition of $\prec_d$ naturally extends to complex vectors, cf.~also \cite{Goldberg77}.
\begin{remark}\label{rem_maj_vector}
\begin{itemize}
\item[(i)] In the mathematical literature $d$-stochastic matrices are defined via $dA=d$ and $A\unitvector^T=\unitvector^T$ which is equivalent to the definition above as it only differs by transposing once. This is because we consider $d,x,y$ to be usual column vectors whereas \cite{Joe90,MarshallOlkin} consider row vectors.
\item[(ii)] For any $d\in\mathbb R_{++}^n$, $ s_d(n)$ constitutes a convex and compact subsemigroup of $\mathbb C^{n\times n}$ with identity element $\mathbbm{1}_n$. In particular it acts contractively in the $1$-norm: for all $z\in\mathbb R^n$ and every $A\in\mathbb R_+^{n\times n}$ which satisfies $\unitvector^TA=\unitvector^T$ one finds the estimate
\begin{equation}\label{eq:doubly_stoch_trace_norm}
\|Az\|_1=\sum_{i=1}^n\Big|\sum_{j=1}^n a_{ij}z_j\Big|\leq \sum_{i,j=1}^n \underbrace{a_{ij}}_{\geq 0}|z_j|=\sum_{j=1}^n \Big(\hspace*{-10pt} \underbrace{\sum_{i=1}^n a_{ij}}_{=(\unitvector^TA)_j=(\unitvector^T)_j=1}\hspace*{-10pt} \Big) |z_j|=\|z\|_1\,.
\end{equation}
\item[(iii)] By Minkowski's theorem\index{theorem!Minkowski} \cite[Thm.~5.10]{Brondsted83} the previous point (ii) implies that $s_d(n)$ can be written as the convex hull of its extreme points. However---unless $d=\unitvector$---this does not prove to be all too helpful as stating said extreme points (for $n>2$) becomes quite delicate\footnote{The number of extreme points of $s_d(n)$ is lower bounded by $n!$ and upper bounded by $\binom{n^2}{2n-1}$, cf.~\cite[Rem.~4.5]{Joe90}.}. To substantiate this the extreme points for $n=3$ and non-degenerate $d\in\mathbb R_{++}^3$ can be found in Lemma \ref{lemma_extreme_points} (Appendix \ref{sec:ex}).
\item[(iv)] If some entries of the $d$-vector co{\"i}ncide, then $\prec_d$ is known to be a preordering but not a partial ordering\index{partial order}. Contrary to what is written in \cite[Rem.~4.2]{Joe90} this does not change if all entries of $d$ are pairwise distinct. To see this, consider $d=(3,2,1)^T$, $x=(1,0,0)^T$, $y=(0,\tfrac23,\tfrac13)^T$, and
$$
A=\begin{pmatrix} 0&1&1\\ \frac23&0&0\\\frac13&0&0 \end{pmatrix}\in s_d(3)\,.
$$
Then $Ax=y$ and $Ay=x$ so $x\prec_d y\prec_d x$ but obviously $x\neq y$. This counterexample can be easily modified to any $d\in\mathbb R^3_{++}$ with $d_1=d_2+d_3$.
\end{itemize}
\end{remark}


Now let us summarize the known characterizations of $\prec_d$: The equivalence of (i) through (v) in the following proposition is due to Joe \cite{Joe90}, and (vi) will be a new result of ours. Moreover, (vii) is related to the definition most prominent among the physics literature, called ``thermo-majorization curves'' \cite{Horodecki13}---indeed the criterion (vii) we present here is a more explicit version of \cite[Thm.~4]{Alhambra16}---more on this later.

\begin{proposition}\label{lemma_char_d_vec}
Let $d\in\mathbb R_{++}^n$ and $x,y\in\mathbb R^n$ be given. The following are equivalent.
\begin{itemize}
\item[(i)] $x\prec_dy$
\item[(ii)] $\sum_{j=1}^n d_j \psi(\frac{x_j}{d_j})\leq \sum_{j=1}^n d_j \psi(\frac{y_j}{d_j})$ for all continuous convex functions $\psi:D(\psi)\subseteq\mathbb R\to\mathbb R$ such that $\{\frac{x_j}{d_j}\,|\,j=1,\ldots,n\},\{\frac{y_j}{d_j}\,|\,j=1,\ldots,n\}\subseteq D(\psi)$.
\item[(iii)] $\sum_{j=1}^n (x_j-td_j)_+\leq\sum_{j=1}^n (y_j-td_j)_+$ for all $t\in\mathbb R$ where $(\cdot)_+:=\max\{\cdot,0\}$.
\item[(iv)] $\sum_{j=1}^n (x_j-td_j)_+\leq\sum_{j=1}^n (y_j-td_j)_+$ for all $t\in\{ \frac{x_i}{d_i},\frac{y_i}{d_i}\,|\,i=1,\ldots,n \}$.
\item[(v)] $\|x-td\|_1\leq\|y-td\|_1$ (i.e.~$\sum_{j=1}^n |x_j-td_j|\leq\sum_{j=1}^n |y_j-td_j|$) for all $t\in\mathbb R$.
\item[(vi)] $\unitvector^Tx=\unitvector^Ty$, and $\|x-\frac{y_i}{d_i}d\|_1\leq\|y-\frac{y_i}{d_i}d\|_1$ for all $i=1,\ldots,n $.
\item[(vii)] $\unitvector^Tx=\unitvector^Ty$, and for all $j=1,\ldots,n-1 $
$$
\sum\nolimits_{i=1}^jx_{\sigma(i)}\leq \min_{i=1,\ldots,n} \unitvector^T\Big(y-\frac{y_i}{d_i}d\Big)_++\frac{y_i}{d_i}\Big(  \sum\nolimits_{k=1}^jd_{\sigma(k)}\Big)
$$
where $\sigma\in S_n$ is a permutation such that $\frac{x_{\sigma(1)}}{d_{\sigma(1)}}\geq\ldots\geq \frac{x_{\sigma(n)}}{d_{\sigma(n)}}$. 
\end{itemize}
\end{proposition}
\begin{proof}
(v) $\Rightarrow$ (vi): For $t$ large enough all entries of $x-td,y-td$ are non-positive so
$$
-\unitvector^T(x-td)=\|x-td\|_1\leq \|y-td\|_1=-\unitvector^T(y-td)
$$
and thus $\unitvector^Tx\geq \unitvector^Ty$. Doing the same for $-t$ large enough gives $\unitvector^Tx\leq \unitvector^Ty$ so together $\unitvector^Tx= \unitvector^Ty$.

(vi) $\Rightarrow$ (v): Define $P:=\{ \frac{x_i}{d_i},\frac{y_i}{d_i}\,|\,i=1,\ldots,n \}$. As argued before, $\unitvector^Tx=\unitvector^Ty$ implies $\|x-td\|_1=\|y-td\|_1$ on $t\in(-\infty,\min P]\cup[\max P,\infty)$. Now define
\begin{align*}
g_x:[\min P,\max P]&\to\mathbb R_+\\
t&\mapsto \|x-td\|_1=\sum\nolimits_{i=1}^n d_i \Big|\frac{x_i}{d_i}-t\Big|
\end{align*}
and $g_y$ analogously.  Evidently, $g_x$ is convex, we have $g_x(\min P)=g_y(\min P)$, $g_x(\max P)=g_y(\max P)$, and $g_y$ is continuous piecewise linear with change in slope only if $t=\frac{y_i}{d_i}$ for some $i=1,\ldots,n$. But at those changes in slope we by assumption have $g_x(\frac{y_i}{d_i})\leq g_y(\frac{y_i}{d_i})$ so Lemma \ref{lemma_convex_cpl_compare} (ii) implies $g_x(t)\leq g_y(t)$ for all $t\in[\min P,\max P]$, and thus $\|x-td\|_1\leq\|y-td\|_1$ for all $t\in\mathbb R$.\medskip

As stated before the equivalence of (i) through (v) is due to \cite[Thm.~2.2]{Joe90}. However for the sake of this work being self-contained (and possibly filling some gaps in the literature) let us show a proof, or at least sketch the ideas. First of all (ii) $\Rightarrow$ (iii) $\Rightarrow$ (iv) is obvious.

(i) $\Rightarrow$ (v): There exists $A\in s_d(n)$ which maps $y$ to $x$ so $A(y-td)=Ay-tAd=x-td$ for all $t\in\mathbb R$ so this is a direct consequence of \eqref{eq:doubly_stoch_trace_norm}.

(v) $\Rightarrow$ (iv): Because $\unitvector^Tx=\unitvector^Ty$, just as in the proof of Lemma \ref{lemma_trace_norm_ball_maj} trace equality and trace norm inequality imply the inequality for the positive part of the vectors.

(iv) $\Rightarrow$ (ii): Let continuous convex $\psi:D(\psi)\subseteq\mathbb R\to\mathbb R$ such that $P:=\{ \frac{x_j}{d_j},\frac{y_j}{d_j}\,|\,j=1,\ldots,n \}\subseteq D(\psi)$ be given. In particular one can construct a continuous function $\tilde\psi:\mathbb R\to\mathbb R$ such that
\begin{itemize}
\item[$\bullet$] $\tilde\psi(\frac{x_j}{d_j})=\psi(\frac{x_j}{d_j})$ and $\tilde\psi(\frac{y_j}{d_j})=\psi(\frac{y_j}{d_j})$ for all $j=1,\ldots,n$
\item[$\bullet$] $\tilde\psi$ is piecewise linear with change in slope only at the elements of $P$.
\item[$\bullet$] $\tilde\psi$ is convex (evident because $\psi$ is convex).
\end{itemize}
In other words $\tilde\psi$ is the ``piecewise linearization'' of $\psi$ (with respect to $P$). Thus it suffices to prove (ii) for all such $\tilde\psi$ because then
$$
\sum\nolimits_{j=1}^n d_j \psi\big(\frac{x_j}{d_j}\big)=\sum\nolimits_{j=1}^n d_j \tilde\psi\big(\frac{x_j}{d_j}\big)\leq \sum\nolimits_{j=1}^n d_j \tilde\psi\big(\frac{y_j}{d_j}\big)= \sum\nolimits_{j=1}^n d_j \psi\big(\frac{y_j}{d_j}\big)\,.
$$

Now let $\phi:\mathbb R\to\mathbb R$ continuous, convex, and piecewise linear (with respect to $P$) be given. Then $\phi$ can be written as a (non-negative) linear combination of the maps\footnote{
This is true up to an affine linear map which due to $\unitvector^Tx=\unitvector^Ty$ yields equality in (ii), thus can be disregarded.
} $\{\phi_p\,|\,p\in P\}$ where $\phi_p(t):=(p-t)_+$. But all $\phi_p$ satisfy (ii) by assumption, hence $\phi$ does as well.

(ii) $\Rightarrow$ (i): The idea here is much in the spirit of Kemperman \cite[Thm.~2]{Kemperman75}. Finding $A\in\mathbb R_+^{n\times n}$ with $\unitvector^TA=\unitvector^T$, $Ad=d$, and $Ay=x$ is equivalent (by vectorization, cf.~footnote \ref{footnote_vec}) to finding a solution $z\in\mathbb R_+^{n^2}$ to
$$
\begin{pmatrix} y^T\otimes \mathbbm{1}_n\\d^T\otimes\mathbbm{1}_n\\\mathbbm{1}_n\otimes \unitvector^T \end{pmatrix}z=\begin{pmatrix} x\\d\\\unitvector \end{pmatrix}
$$
where $\otimes$ is the usual Kronecker product \cite[Ch.~2.2]{MN07} and $z=\operatorname{vec} A$. By Farkas' lemma\footnote{Farkas' lemma states that for $m,n\in\mathbb N$, $A\in\mathbb R^{m\times n}$, $b\in\mathbb R^m$, the system of linear equations $Ax=b$ has a solution in $\mathbb R_+^n$ if and only if for all $y\in\mathbb R^m$ which satisfy $A^Ty\leq 0$ one has $b^Ty\leq 0$, refer to \cite[Coro.~7.1.d]{Schrijver86} (when replacing $A,b$ by $-A,-b$).} such a solution exists if (and only if) for all $w\in\mathbb R^{3n}$ which satisfy
\begin{align}
\Big(\begin{pmatrix} y^T\otimes \mathbbm{1}_n\\d^T\otimes\mathbbm{1}_n\\\mathbbm{1}_n\otimes \unitvector^T \end{pmatrix}^T\begin{pmatrix} \vec{w_1}\\\vec{w_2}\\\vec{w_3} \end{pmatrix}\Big)_{n(j-1)+k}&=\Big(\begin{pmatrix} (y\otimes \mathbbm{1}_n)\vec{w_1}&(d\otimes \mathbbm{1}_n)\vec{w_2}&(\mathbbm{1}_n\otimes \unitvector)\vec{w_3}\end{pmatrix}\Big)_{n(j-1)+k}\notag\\
&=y_jw_k+d_jw_{n+k}+w_{2n+j}\leq 0\label{eq:w_y_ineq}
\end{align}
for all $j,k=1,\ldots,n$, one has
$$
\sum\nolimits_{j=1}^n (x_jw_j+d_jw_{n+j}+w_{2n+j})\leq 0\,.
$$
Consider the convex (because affine linear) functions $\psi_j:\mathbb R\to\mathbb R$, $t\mapsto w_jt+w_{n+j}$ for all $j=1,\ldots,n$. Then
$$
\psi:\mathbb R\to\mathbb R\qquad t\mapsto\max_{j=1,\ldots,n}\psi_j(t)
$$
is convex and continuous as well so by assumption and because $d>0$
\begin{align*}
\sum\nolimits_{j=1}^n (x_jw_j+d_jw_{n+j}+w_{2n+j})&=\sum\nolimits_{j=1}^n d_j\psi_j(\tfrac{x_j}{d_j})+w_{2n+j}\\
&\leq \sum\nolimits_{j=1}^n d_j\psi(\tfrac{x_j}{d_j})+w_{2n+j}\\
&\leq \sum\nolimits_{j=1}^n d_j\psi(\tfrac{y_j}{d_j})+w_{2n+j}\,.
\end{align*}
But now for every $j=1,\ldots,n$ exists $k=k(j)$ such that $\psi(\tfrac{y_j}{d_j})=\psi_{k(j)}(\tfrac{y_j}{d_j})$ by definition of $\psi$ (the maximum has to be attained by at least one of the $\psi_k$). Hence
\begin{align*}
\sum\nolimits_{j=1}^n (x_jw_j+d_jw_{n+j}+w_{2n+j})&\leq \sum\nolimits_{j=1}^n d_j\psi(\tfrac{y_j}{d_j})+w_{2n+j}\\
&= \sum\nolimits_{j=1}^n d_j\psi_{k(j)}(\tfrac{y_j}{d_j})+w_{2n+j}\\
&=\sum\nolimits_{j=1}^n \underbrace{y_jw_{k(j)}+d_jw_{n+k(j)}+w_{2n+j}}_{\leq 0\text{ by }\eqref{eq:w_y_ineq}}\leq 0
\end{align*}
so we are done. Note that we needed access not to all but only to the piecewise linear convex functions; this is the same effect as in the proof of (iv) $\Rightarrow$ (ii).

(i) $\Leftrightarrow$ (vii): This will be a direct consequence of our considerations in Section \ref{sec_d_maj_poly} so we will postpone this part of the proof to after Rem.~\ref{rem_thermomaj_phys_lit}. Note that we will not use this result anywhere in this thesis, meaning we will not run into a circular argument.
\end{proof}

Recently Alhambra et al.~\cite{Alhambra19} were able to find conditions under which classical majorization implies $d$-majorization for $d$ from some parameter range
. As their result was obtained in the context of dephasing thermalization let us reformulate it by casting it into our notation:
\begin{proposition}
The following statements hold.
\begin{itemize}
\item[(i)] Let $x,y\in\mathbb R^n$ and $d\in\mathbb R_{++}^n$ be given. If $x,y$ are similarly $d$-ordered, i.e.~there exists a permutation $\sigma\in S_n$ such that $\frac{x_{\sigma(1)}}{d_{\sigma(1)}}\geq\ldots\geq\frac{x_{\sigma(n)}}{d_{\sigma(n)}}$ and $\frac{y_{\sigma(1)}}{d_{\sigma(1)}}\geq\ldots\geq \frac{y_{\sigma(n)}}{d_{\sigma(n)}}$, then $x\prec_d y$ is equivalent to $x\prec y$.\smallskip
\item[(ii)] Let $y\in\mathbb R_+^n$ and $d\in\mathbb R_{++}^n$. If $y_1\leq\ldots\leq y_n$ and $d_1\geq\ldots\geq d_n$, then $\underline{\sigma}y\prec_d y$ for all $\sigma\in S_n$.\smallskip
\item[(iii)] Let $y\in\mathbb R_+^n$ with $y_1\leq\ldots\leq y_n$ be given. Then for all $x\in\mathbb R^n$ one has $x\prec y$ if and only if $x\prec_d y$ for all $d\in\mathbb R_{++}^n$ with $d_1\geq\ldots\geq d_n$.
\end{itemize}
\end{proposition}
\begin{proof}
(i): \cite[Coro.~2.5]{Joe90}. Note that $x\prec_d y$ if and only if $\underline{\sigma}x\prec_{\underline{\sigma}d}\underline{\sigma}y$ for all $\sigma\in S_n$ as is readily verified so it suffices to have $x,y$ similarly $d$-ordered. (ii): One can explicitly write down generalized T-transforms which first shift $y_1$ to $y_{\sigma(1)}$, then $y_2$ to $y_{\sigma(2)}$, and so on. The details are carried out in \cite[p.~13 \& 14]{Alhambra19}. (iii), $\Leftarrow$: Obvious. (iii), $\Rightarrow$: Let $x\in\mathbb R^n$ with $x\prec y$ be given and let $\tau\in S_n$ be a permutation such that $\underline{\tau}x_1\leq\ldots\leq \underline{\tau}x_n$. Then (ii) implies $\underline{\sigma}\,\underline{\tau}x\prec_d\underline{\tau}x$ for all $\sigma\in S_n$ and all $d\in\mathbb R_{++}^n$ with $d_1\geq\ldots\geq d_n$, so choosing $\sigma=\tau^{-1}$ yields $x\prec_d\underline{\tau}x$. On the other hand $\underline{\tau}x$ and $y$ are similarly $d$-ordered for all such $d$---because $\frac{(\underline{\tau}x)_1}{d_1}\leq\ldots\leq\frac{(\underline{\tau}x)_n}{d_n}$ and $\frac{y_1}{d_1}\leq\ldots\leq \frac{y_n}{d_n}$---so we have $\underline{\tau}x\prec_d y$ by (i). Using that $\prec_d$ is a preorder this yields $x\prec_d\underline{\tau}x\prec_dy$, that is, $x\prec_d y$ as claimed.
\end{proof}

To conclude this section we make some statements about minimal and maximal elements of the preorder $\prec_d$.
\begin{theorem}\label{prop_1}
Let $d\in\mathbb R_{++}^n$ be given. The following statements hold.
\begin{itemize}
\item[(i)] $d$ is the unique minimal element within $\lbrace x\in\mathbb R^n\,|\,\unitvector^Tx=\unitvector^Td\rbrace$ (i.e.~the trace hyperplane ``spanned'' by $d$) with respect to $\prec_d\,$.
\item[(ii)] $(\unitvector^Td) e_k$ is maximal within $(\unitvector^Td)\Delta^{n-1}=\lbrace x\in\mathbb R_+^n\,|\,\unitvector^Tx=\unitvector^Td\rbrace$ with respect to $\prec_d$ where $k$ is chosen such that $d_k$ is minimal in $d$. It is the unique maximal element in $(\unitvector^Td)\Delta^{n-1}$ with respect to $\prec_d$ if and only if $d_k$ is the unique minimal element of $d$.
\end{itemize}
\end{theorem}
\begin{proof}
(i) Consider $d\unitvector^T/(\unitvector^Td)\in s_d(n)$ which maps any $x\in\mathbb R^n$ with $\unitvector^Tx=\unitvector^Td$ to $d$ so $d\prec_d x$. Uniqueness is obvious because $d$ is a fixed point of every $d$-stochastic matrix.

(ii): W.l.o.g.~$\unitvector^Td=1$ (else we can rescale the problem accordingly), so $\lbrace x\in\mathbb R_+^n\,|\,\unitvector^Tx=\unitvector^Td\rbrace$ is equal to the standard simplex $\Delta^{n-1}$. In light of convexity of $\prec_d$ it suffices to show that $e_k$ $d$-majorizes all extreme points of $\Delta^{n-1}$ which by definition are given by the standard basis vectors $e_1,\ldots,e_n$. Let $j\in\lbrace1,\ldots,n\rbrace$, $j\neq k$. Choose $A$ as the identity matrix aside from
\begin{align*}
\begin{pmatrix} A_{jj}&A_{jk}\\A_{kj}&A_{kk} \end{pmatrix}=\begin{pmatrix} 1-\frac{d_k}{d_j}&1\\\frac{d_k}{d_j}&0 \end{pmatrix}\text{ if }j<k\quad\text{ or }\quad
\begin{pmatrix} A_{kk}&A_{kj}\\A_{jk}&A_{jj} \end{pmatrix}=\begin{pmatrix} 0&\frac{d_k}{d_j}\\1&1-\frac{d_k}{d_j} \end{pmatrix}\text{ if }k<j\,.
\end{align*}
One readily verifies that $A$ is $d$-stochastic (because $d_k\leq d_j$ by assumption) with $Ae_k=e_j$, so $e_j\prec_d e_k$ which together with convexity of $\prec_d$ shows $x\prec_d e_k$ for all $x\in \Delta^{n-1}$.

For uniqueness first assume that $d_k$ is the unique minimal element in $d$ and further that there exists $M'\in\Delta^{n-1}$ such that $x\prec_d M'$ for all $x\in\Delta^{n-1}$. In particular, $e_k\prec_d M'\prec_d e_k$ so Prop.~\ref{lemma_char_d_vec} (iv) yields
\begin{align*}
\sum\nolimits_{i=1}^n d_i\Big(\frac{M_i'}{d_i}-t\Big)_+=1-d_kt
\end{align*}
for all $t\in \lbrace \frac{M_1'}{d_1},\ldots,\frac{M_n'}{d_n},\frac{1}{d_k},0\rbrace$. Choose any $j\in\lbrace 1,\ldots,n\rbrace$, $j\neq k$. Then considering the index set $J=\big\{i\in\{1,\ldots,n\}\,\big|\, \frac{M_i'}{d_i}\geq \frac{M_j'}{d_j} \big\}$ one finds
\begin{align*}
1-d_k\frac{M_j'}{d_j}=\sum_{i=1}^n d_i \Big( \frac{M_i'}{d_i}-\frac{M_j'}{d_j}\Big)_+ =\sum_{i\in J} d_i \Big( \frac{M_i'}{d_i}-\frac{M_j'}{d_j}\Big)=\sum_{i\in J}M_i' -\frac{M_j'}{d_j}\sum_{i\in J}d_i\,.
\end{align*}
Note that $J\neq\emptyset$ as $j\in J$. Because of $\unitvector^TM'=1$, the above equation yields
\begin{align*}
\sum\nolimits_{i\in\lbrace1,\ldots,n\rbrace\setminus J} M_i'=1-\sum\nolimits_{i\in J}M_i'=\frac{M_j'}{d_j}\Big( d_k-d_j-\sum\nolimits_{i\in J\setminus\lbrace j\rbrace}d_i \Big)\,.
\end{align*}
As the l.h.s.~is non-negative, the same has to hold for the r.h.s.~so in particular $M_j'=0$ due to $d_k-d_j<0$. As $j\neq k$ was chosen arbitrarily, this implies $M'=e_k$.

On the other hand, assume there exist $k,k'\in\lbrace1,\ldots,n\rbrace$ with $k\neq k'$ such that $d_k=d_{k'}$ is minimal in $d$. Then $e_k$ and $e_{k'}$ are both maximal with respect to $\prec_d$ by the same argument as above, hence uniqueness does not hold which concludes the proof.
\end{proof}
\begin{remark}\label{rem_pure_state_transform}
The fact that every $e_1,\ldots,e_n$ is maximal in the standard simplex $\Delta^{n-1}$ for $d=\unitvector$ is lost in the general setting (consider the example from Rem.~\ref{rem_maj_vector} (iii)).

However, for strictly positive vectors $z\in\mathbb R_{++}^n$ one still has $(\unitvector^Tz )e_k\not\prec_d z$ for all $k=1,\ldots,n$. More generally, if $y\prec_d z$ then $y$ has to be strictly positive as well; otherwise the corresponding transformation matrix (non-negative entries) would contain a row of zeros which due to $d>0$ contradicts $d$ being a fixed point.
\end{remark}

\subsection{Characterizing the $\prec_d$-Polytope}\label{sec_d_maj_poly}

To explore the ``geometry'' of $d$-majorization, we shall consider the set of all vectors which are $d$-majorized by some $y\in\mathbb R^n$. For this, we introduce the map
\begin{align*}
M_d:\mathcal P(\mathbb R^n)&\to\mathcal P(\mathbb R^n)\\
S&\mapsto \bigcup\nolimits_{y\in S}\lbrace x\in\mathbb R^n\,|\,x\prec_d y\rbrace
\end{align*}
where $\mathcal P$ denotes the power set. 
For convenience $M_d(y):=M_d(\{y\})$ for any $y\in\mathbb R^n$ which then equals the set of all vectors which are $d$-majorized by $y$. Note that the idea here is close to---but should not be confused with---the ($d$-)majorization polytope\index{d-majorization polytope@$d$-majorization polytope} of two vectors, which is the set of all ($d$-)stochastic matrices which map one vector to the other as studied, e.g., in \cite{Dahl99a,Dahl99b}. 
\begin{lemma}\label{lemma_closure_op}
Let $d\in\mathbb R_{++}^n$. Then $M_d$ is a closure operator\footnote{
Recall that an operator $J$ on the power set $\mathcal P(S)$ of a set $S$ is called \textit{closure operator} or \textit{hull operator} if it is extensive ($X\subseteq J(X)$), increasing ($X\subseteq Y\,\Rightarrow\,J(X)\subseteq J(Y)$) and idempotent ($J(J(X))=J(X)$) for all $X,Y\in\mathcal P(S)$, cf., e.g., \cite[p.~42]{Cohn81}.\label{footnote_closure_operator}
}. In particular, for any $x,y\in\mathbb R^n$ one has $x\prec_dy$ if and only if $M_d(x)\subseteq M_d(y)$.
\end{lemma}
\begin{proof}
The first statement is a simple consequence of the $d$-stochastic matrices $s_d(n)$ forming a semigroup with identity. For the second statement note that $x\prec_dy$, that is, $x\in M_d(y)$ implies $M_d(x)\subseteq M_d(M_d(y))=M_d(y)$.
\end{proof}
Now Prop.~\ref{lemma_char_d_vec} (vi) directly implies
\begin{align}
M_d(y)&=\bigcap\nolimits_{i=1}^n\Big\lbrace x\in\mathbb R^n\,\Big|\, \unitvector^Tx=\unitvector^Ty\ \wedge\ \Big\|x-\frac{y_i}{d_i}d\Big\|_1\leq\Big\|y-\frac{y_i}{d_i}d\Big\|_1\Big\rbrace\notag\\
&=\bigcap\nolimits_{i=1}^n\Big\lbrace x\in\mathbb R^n\,\Big|\, \unitvector^T\Big(x-\frac{y_i}{d_i}d\Big)=\unitvector^T\Big(y-\frac{y_i}{d_i}d\Big)\ \wedge\ \Big\|x-\frac{y_i}{d_i}d\Big\|_1\leq\Big\|y-\frac{y_i}{d_i}d\Big\|_1\Big\rbrace\notag\\
&=\bigcap\nolimits_{i=1}^n\Big(\Big\lbrace \tilde x\in\mathbb R^n\,\Big|\, \unitvector^T\tilde x=\unitvector^T\Big(y-\frac{y_i}{d_i}d\Big)\ \wedge\ \|\tilde x\|_1\leq\Big\|y-\frac{y_i}{d_i}d\Big\|_1\Big\rbrace+\frac{y_i}{d_i}d\Big)\label{eq:decomp_m_d}
\end{align}
for all $y\in\mathbb R^n$, $d\in\mathbb R_{++}^n$. 

\begin{lemma}\label{lemma_trace_norm_ball_maj}
Let $z\in\mathbb R^n$
. Then
\begin{align*}
\{x\in\mathbb R^n\,|\,\unitvector^Tx=\unitvector^Tz\ \wedge\ \|x\|_1\leq\|z\|_1\rbrace=\Big\lbrace x\in\mathbb R^n\,\Big|\,x\prec( \unitvector^Tz_+,-\unitvector^Tz_-,0,\ldots,0)^T\Big\}
\end{align*}
where $z=z_+-z_-$ is the unique decomposition of $z$ into positive and negative part, i.e.~$z_+=(\max\{z_j,0\})_{j=1}^n,z_-=(-z)_+=(\max\{-z_j,0\})_{j=1}^n)\in\mathbb R_+^n$.
\end{lemma}
\begin{proof}
For what follows let $ \hat z:=(\unitvector^Tz_+,-\unitvector^Tz_-,0,\ldots,0)$.
`` $\supseteq$ '': Majorization by definition forces the two vectors to be in the same hyperplane: $\unitvector^Tx=\unitvector^Tz_+-\unitvector^Tz_-=\unitvector^T(z_+-z_-)=\unitvector^Tz$. Also if $x\prec \hat z$ then there exists a doubly stochastic matrix $A$ which maps $\hat z$ to $x$ so
$$
\|x\|_1=\|A\hat z\|_1\overset{\eqref{eq:doubly_stoch_trace_norm}}\leq \|\hat z\|_1= \unitvector^Tz_++\unitvector^Tz_-=\sum\nolimits_{j=1}^n |z_j|=\|z\|_1\,.
$$

`` $\subseteq$ '': Decompose $x=x_+-x_-$ with $x_+,x_-\in\mathbb R_+^n$ as above. By assumption
\begin{align*}
\unitvector^Tx=\unitvector^Tx_+-\unitvector^Tx_-&=\unitvector^Tz_+-\unitvector^Tz_-=\unitvector^Tz\\
\|x\|_1=\unitvector^Tx_++\unitvector^Tx_-&\leq \unitvector^Tz_++\unitvector^Tz_-=\|z\|_1
\end{align*}
so taking the sum of these two gives $\unitvector^Tx_+\leq \unitvector^Tz_+$. Thus for all $k=1,\ldots,n-1$
$$
\sum\nolimits_{i=1}^k x_i^\downarrow \leq \sum\nolimits_{i=1}^k (x_i^\downarrow)_+ \leq \unitvector^Tx_+\leq \unitvector^Tz_+= \unitvector^Tz_++\underbrace{0+\ldots+0}_{k-1\text{ zeros}} =\sum\nolimits_{i=1}^k \hat z_i^\downarrow 
$$
which---together with $\unitvector^Tx=\unitvector^Tz$---shows $x\prec\hat z$. 
\end{proof}
\begin{theorem}\label{thm_maj_halfspace}
Let $y\in\mathbb R^n$, $d\in\mathbb R_{++}^n$. Then\index{d-majorization polytope@$d$-majorization polytope!H-description@$\mathscr H$-description}
$
M_d(y)=\{x\in\mathbb R^n\,|\,Mx\leq b\}
$
with $M$ being the matrix \eqref{eq:M_maj} and
\begin{equation}\label{eq:d_majorization_b}
b= \min_{i=1,\ldots,n} {\footnotesize\begin{pmatrix} 
\unitvector^T(y-\frac{y_i}{d_i}d)_+\unitvector+\frac{y_i}{d_i}M_1d\\
\vdots\\
 \unitvector^T(y-\frac{y_i}{d_i}d)_+\unitvector+\frac{y_i}{d_i}M_{n-1}d\\
\unitvector^Ty\\
-\unitvector^Ty
\end{pmatrix}} \in\mathbb R^{2^n}
\,.
\end{equation}
\end{theorem}
\begin{proof}
By \eqref{eq:decomp_m_d}, Lemma \ref{lemma_trace_norm_ball_maj}, Prop.~\ref{prop_maj_halfspace}, and Remark \ref{rem:halfspace_actions}
\begin{align*}
M_d(y)&=\bigcap\nolimits_{i=1}^n\Big(\Big\lbrace \tilde x\in\mathbb R^n\,\Big|\, \unitvector^T\tilde x=\unitvector^T\Big(y-\frac{y_i}{d_i}d\Big)\ \wedge\ \|\tilde x\|_1\leq\Big\|y-\frac{y_i}{d_i}d\Big\|_1\Big\rbrace+\frac{y_i}{d_i}d\Big)\\
&=\bigcap\nolimits_{i=1}^n\Big(\Big\lbrace x\in\mathbb R^n\,\Big|\,x\prec\big( \unitvector^T(y-\tfrac{y_i}{d_i}d)_+,-\unitvector^T(y-\tfrac{y_i}{d_i}d)_-,0,\ldots,0\big)^T\Big\}+\frac{y_i}{d_i}d\Big)\\
&=\bigcap\nolimits_{i=1}^n\Big(\Big\lbrace x\in\mathbb R^n\,\Big|\,Mx\leq {\footnotesize\begin{pmatrix} \unitvector^T(y-\frac{y_i}{d_i}d)_+\\\vdots\\ \unitvector^T(y-\frac{y_i}{d_i}d)_+\\\unitvector^T(y-\frac{y_i}{d_i}d)\\-\unitvector^T(y-\frac{y_i}{d_i}d)\end{pmatrix}} \Big\}+\frac{y_i}{d_i}d\Big)\\
&=\Big\lbrace x\in\mathbb R^n\,\Big|\,Mx\leq \min_{i=1,\ldots,n}{\footnotesize\begin{pmatrix} 
\unitvector^T(y-\frac{y_i}{d_i}d)_+\unitvector+\frac{y_i}{d_i}M_1d\\
\vdots\\
 \unitvector^T(y-\frac{y_i}{d_i}d)_+\unitvector+\frac{y_i}{d_i}M_{n-1}d\\
\unitvector^Ty\\
-\unitvector^Ty
\end{pmatrix}} \Big\}\,.\qedhere
\end{align*}
\end{proof}
\noindent Setting $d=\unitvector$ in Thm.~\ref{thm_maj_halfspace} together with Lemma \ref{lemma_maj_sum_recovery} (iii) recovers the $\mathscr H$-description of classical majorization\index{majorization!H-description@$\mathscr H$-description} (Prop.~\ref{prop_maj_halfspace}) as was to be expected.

The previous theorem shows that, roughly speaking, $\prec_d$ and $\prec$ share the same geometry, meaning the faces of $M_d(y)$ for arbitrary $y\in\mathbb R^n$, $d\in\mathbb R_{++}^n$ are all parallel to some face of a classical majorization polytope; but the precise location of the halfspaces (respectively faces) may differ. 

\begin{remark}
While the description of $d$-majorization via halfspaces is conceptionally interesting---as seen above---it also enables an algorithmic computation of the extreme points of $M_d(y)$. This conversion (from $\mathscr H$- to $\mathscr V$-description) is known as the vertex enumeration problem\index{vertex enumeration problem} which is a well-studied problem in the field of convex polytopes and computational geometry, see \cite{Avis97} for an overview. 
However the polytopes we are concerned with are of such convenient structure that one can even do this analytically
.
\end{remark}

If $M_d$ acts on a set consisting of more than one vector we can state further geometric and topological results. This will be of use when treating continuity questions of the map $(d,P)\mapsto M_d(P)$ afterwards.
\begin{theorem}\label{coro_0_1}
Let $d\in\mathbb R_{++}^n$ and an arbitrary subsets $P,P'\subseteq \mathbb R^n$ be given. Then the following statements hold.
\begin{itemize}
\item[(i)] If $P$ lies within a trace hyperplane, i.e.~there exists $ c\in\mathbb R$ such that $\unitvector^Tx= c$ for all $x\in P$, then $M_d(P)$ is star-shaped with respect to $\frac{ c}{\unitvector^Td}d$.
\item[(ii)] If $P$ is convex, then $M_d(P)$ is path-connected.
\item[(iii)] If $P$ is compact, then $ M_d(P)$ is compact.
\end{itemize}
\end{theorem}
\begin{proof}
(i): Every $x\in P$ is directly connected to $\frac{\unitvector^Tx}{\unitvector^Td}d$ within $M_d(P)$ (cf.~Thm.~\ref{prop_1} together with convexity of $\prec_d$). (ii): Let $y,z\in P$, $\lambda\in[0,1]$ be arbitrary. Then $\lambda y+(1-\lambda)z\in P$ and
$$
\lambda \frac{\unitvector^Ty}{\unitvector^Td}d+(1-\lambda)\frac{\unitvector^Tz}{\unitvector^Td}d=\frac{\unitvector^T(\lambda y+(1-\lambda)z)}{\unitvector^Td}d\in M_d( \lambda y+(1-\lambda)z)\subseteq M_d(P)
$$
where in the last step we used that $M_d$ is increasing. Thus $\frac{\unitvector^Ty}{\unitvector^Td}d,\frac{\unitvector^Tz}{\unitvector^Td}d$ are path-connected in $M_d(P)$, which together with (i) shows (ii).

(iii): 
As $P$ is bounded by assumption, and $s_d(n)$ is bounded (cf.~Remark \ref{rem_maj_vector} (ii)) this readily implies that $M_d(P)$ is bounded. For closedness, consider a sequence $( x_n)_{n\in\mathbb N}$ in $ M_d(P)$ which converges to some $ x\in\mathbb R^n$. By definition there exists a sequence $(y_n)_{n\in\mathbb N}$ in $P$ and a sequence $(A_n)_{n\in\mathbb N}$ in $s_d(n)$ such that $A_ny_n=x_n$. Because $P$ is compact there exists a subsequence $(y_{n_j})_{j\in\mathbb N}$ of $(y_n)_{n\in\mathbb N}$ which converges to some $y\in P$. On the other hand compactness of $s_d(n)$ yields a subsequence $(A_{m_j})_{j\in\mathbb N}$ of $(A_{n_j})_{j\in\mathbb N}$ which converges to some $A\in s_d(n)$. Combining these two yields 
\begin{align*}
\|Ay-A_{m_j}y_{m_j}\|&\leq \|Ay-A_{m_j}y\|+\|A_{m_j}y-A_{m_j}y_{m_j}\|\\
&\leq \|A-A_{m_j}\|\|y\|+\underbrace{\|A_{m_j}\|}_{\substack{\leq c\text{ for some }c\in\mathbb R_+\\\text{ (boundedness of }s_d(n)\text{)}}}\|y-y_{m_j}\|_1\to 0\quad\text{ as }j\to\infty\,.
\end{align*}
Therefore 
$
 x=\lim_{j\to\infty} x_{m_j}=\lim_{l\to\infty}A_{m_j}y_{m_j}=Ay\,,
$
so $ x\in M_d(P)$ because $y\in P$ which concludes the proof. 
\end{proof}
\noindent The previous theorem still holds when extending $\prec_d$ to complex vectors. 
Also one might hope that Thm.~\ref{coro_0_1}.(ii) is not optimal in the sense that convexity of general $P$ implies convexity of $M_d(P)$. Example \ref{example_1}
, however, gives a negative answer.

The description of $M_d(y)$ as a convex polytope is powerful enough to answer continuity questions regarding the map $M_d$.
\begin{theorem}\label{thm_cont_Md}
Let $\mathcal P_c(\mathbb R^n)$ denote the collection of all compact subsets of $(\mathbb R^n,\|\cdot\|_1)$ and let $\Delta$ be the Hausdorff metric (cf.~Appendix \ref{app_hausdorff})
on $\mathcal P_c(\mathbb R^n)$. Then the following statements hold.
\begin{itemize}
\item[(i)] For all $d\in\mathbb R_{++}^n$ and all $P,P'\in\mathcal P_c(\mathbb R^n)$ one has $\Delta(M_d(P),M_d(P'))\leq\Delta(P,P')$, i.e.~$M_d$ is non-expansive under $\Delta$.
\item[(ii)] The following map is continuous:
\begin{align*}
M:\mathbb R_{++}^n\times(\mathcal P_c(\mathbb R^n),\Delta)&\to (\mathcal P_c(\mathbb R^n),\Delta)\\
(d,P)&\mapsto M_d(P)
\end{align*}
\end{itemize}
\end{theorem}
\begin{proof}
Note that the image of a compact set under $M_d$ remains compact by Thm.~\ref{coro_0_1} (iii) so the statements to be proven are well-defined.

(i): This is a direct consequence of \eqref{eq:doubly_stoch_trace_norm}:
\begin{align*}
\max_{z\in M_d(P)}\min_{w\in M_d(P')}\|z-w\|_1&=\max_{\substack{A\in s_d(n)\\z_1\in P}}\min_{\substack{B\in s_d(n)\\z_2\in P'}}\|Az_1-Bz_2\|_1\\
&\leq \max_{\substack{A\in s_d(n)\\z_1\in P}}\min_{z_2\in P'}\|A(z_1-z_2)\|_1\\
&\leq \max_{\substack{A\in s_d(n)\\z_1\in P}}\min_{z_2\in P'}\|z_1-z_2\|_1=\max_{z\in P}\min_{w\in P'}\|z-w\|_1\,,
\end{align*}
and analogously when interchanging $P$ and $P'$.

(ii): Our proof can be divided into the following five steps.

\noindent\textit{Step 1:} For all $y\in\mathbb R^n$ the vector $b$ from \eqref{eq:d_majorization_b} continuously depends on $d\in\mathbb R_{++}^n$.

This is due to the following facts:
\begin{itemize}
\item The map $x\mapsto x_+=\frac{x+|x|}{2}$ is continuous on $\mathbb R$.
\item The map $x\mapsto \frac{1}{x}$ is continuous on $\mathbb R_{++}$.
\item The map $(x_1,\ldots,x_n)\mapsto x_j$ for all $j=1,\ldots,n$ is continuous on $\mathbb R^n$.
\end{itemize}
Thus every component $b_i:\mathbb R_{++}^n\to\mathbb R$, $d\mapsto b_i(d)$ (the latter still corresponding to the vector $b$ from \eqref{eq:d_majorization_b}) is continuous as a composition and a finite sum of continuous functions, using that the minimum over finitely many continuous functions remains continuous.

\noindent\textit{Step 2:} If a sequence $(b^{(m)})_{m\in\mathbb N}\subset\mathbb R^{2^n}$ with $b^{(m)}_{2^n-1}+b^{(m)}_{2^n}=0$ for all $m\in\mathbb N$ converges to $b\in\mathbb R^{2^n}$ in norm and all the induced convex polytopes $\{x\in\mathbb R^n\,|\,Mx\leq b^{(m)}\}$, $\{x\in\mathbb R^n\,|\,Mx\leq b\}$ are non-empty then
$$
\lim_{m\to\infty}\Delta( \{x\in\mathbb R^n\,|\,Mx\leq b^{(m)}\},\{x\in\mathbb R^n\,|\,Mx\leq b\} )=0\,.
$$


This follows directly from Lemma \ref{lemma_cont_in_b} which yields a constant $c_M>0$ (which only depends on $M$) such that
$$
\Delta( \{x\in\mathbb R^n\,|\,Mx\leq b^{(m)}\},\{x\in\mathbb R^n\,|\,Mx\leq b\} )\leq c_M\|b^{(m)}-b\|_1\overset{m\to\infty}\to 0\,.
$$

\noindent \textit{Step 3:} $d\mapsto M_d(y)$ is continuous on $\mathbb R_{++}^n$ for all $y\in\mathbb R^n$.

Let $d^{(m)}\subset\mathbb R_{++}^n$ be a sequence which converges to $d\in\mathbb R_{++}^n$. As shown in Step 1 this implies that $b^{(m)}=b(d^{(m)})\subset\mathbb R_{++}^{2^n}$ converges to $b(d)\in\mathbb R_{++}^{2^n}$ so Step 2 together with Thm.~\ref{thm_maj_halfspace} yields
$$
\lim_{m\to\infty} M_{d^{(m)}}(y)=\lim_{m\to\infty} \{x\in\mathbb R^n\,|\,Mx\leq b^{(m)}\}=\{x\in\mathbb R^n\,|\,Mx\leq b\}=M_d(y)\,.
$$
This proves continuity because the subspace topology (inherited from $(\mathbb R^n,\|\cdot\|_1$) on $\mathbb R_{++}^n$ is induced by the restricted norm $\|\cdot\|_1:\mathbb R_{++}^n\to \mathbb R_+$; thus $\mathbb R_{++}^n$ is a metric space and continuity is the same as sequential continuity.

\noindent \textit{Step 4:} $d\mapsto M_d(P)$ is continuous on $\mathbb R_{++}^n$ for all $P\in\mathcal P_c(\mathbb R^n)$.

As before let $d^{(m)}\subset\mathbb R_{++}^n$ be a sequence which converges to $d\in\mathbb R_{++}^n$ and let $\varepsilon>0$ be given. Because $P$ is compact one finds $y_1,\ldots,y_k\in P$, $k\in\mathbb N$ with $P\subseteq\bigcup_{i=1}^k B_{\varepsilon/3}(y_i)$. On the other hand (by Step 3) for every $i=1,\ldots,k$ one finds $N_i\in\mathbb N$ such that $ \Delta(M_{d^{(m)}}(y_i),M_d(y_i))<\frac{\varepsilon}{3} $ for all $m\geq N_i$. We want to show $\Delta(M_{d^{(m)}}(P),M_d(P))<\varepsilon$ for all $m\geq N:=\max\{N_1,\ldots,N_k\}$ which would imply the claim. 

Let any $m\geq N$ and $x\in M_{d^{(m)}}(P)$ so one finds $A\in s_{d^{(m)}}(n)$ and $y\in P$ such that $x=Ay$. First, compactness of $P$ yields $i=1,\ldots,k$ such that $\|y-y_i\|<\frac{\varepsilon}{3}$. Then $Ay_i$ is in $ M_{d^{(m)}}(y_i)$ which lets us pick $B\in s_d(n)$ with $\|Ay_i-By_i\|_1<\frac{\varepsilon}{3}$ (because $\Delta(M_{d^{(m)}}(y_i),M_d(y_i))<\frac{\varepsilon}{3}$). Finally $\tilde x:=By\in M_d(P)$ satisfies
\begin{align*}
\|x-\tilde x\|_1=\|Ay-By\|_1&\leq \|Ay-Ay_i\|_1+\|Ay_i-By_i\|_1+\|By_i-By\|_1\\
&\overset{\eqref{eq:doubly_stoch_trace_norm}}\leq \|y-y_i\|_1+\|Ay_i-By_i\|_1+\|y_i-y\|_1<\frac{\varepsilon}{3}+\frac{\varepsilon}{3}+\frac{\varepsilon}{3}=\varepsilon\,.
\end{align*}
Analogously for every $\tilde x\in M_{d}(P)$ one finds $x\in M_{d^{(m)}}(P)$ such that $\|x-\tilde x\|_1<\varepsilon$ which implies $\Delta(M_{d^{(m)}}(P),M_d(P))<\varepsilon$ for all $m\geq N$ (cf.~Appendix \ref{app_hausdorff}).

\noindent \textit{Step 5:} $M$ is continuous (in the product topology).

As both spaces which make up the domain of $M$ are metric spaces the product topology is metrizable and continuity, again, can be decided by sequences. Let $(d^{(m)},P_m)_{m\in\mathbb N}\subset \mathbb R_{++}^n\times \mathcal P_c(\mathbb R^n)$ be an arbitrary sequence which converges to $(d,P)\in \mathbb R_{++}^n\times \mathcal P_c(\mathbb R^n)$ in the product topology; this is equivalent to $\lim_{m\to\infty}\|d^{(m)}-d\|_1=0$ and $\lim_{m\to\infty}\Delta(P^{(m)},P)=0$. The former (by Step 4) gives $\lim_{m\to\infty}\Delta(M_{d^{(m)}}(P),M_d(P))=0$ so altogether
\begin{align*}
\Delta(M_{d^{(m)}}(P^{(m)}),M_d(P))&\leq \Delta(M_{d^{(m)}}(P^{(m)}),M_{d^{(m)}}(P))+\Delta(M_{d^{(m)}}(P),M_d(P))\\
&\overset{\text{(i)}}\leq \Delta(P^{(m)},P)+\Delta(M_{d^{(m)}}(P),M_d(P))\overset{m\to\infty}\to 0\,.\qedhere
\end{align*}
\end{proof}
\begin{remark}
\begin{itemize}
\item[(i)] Continuity of the map $M$ is supported by the fact that the half-spaces limiting $M_d(y)$ are independent of $d,y$. An example of a discontinuous relation between $A\in\mathbb R^{m\times n}$ and the induced polytope $\{x\in\mathbb R^n\,|\,Ax\leq b\}$ can be found in Example \ref{example_not_cont_A}.
\item[(ii)] While $M_d$ is defined for arbitrary $d\in\mathbb R^n$, the continuity statement from Thm.~\ref{theorem_max_corner_maj} (ii) becomes wrong if the domain is extended to $\mathbb R_+^n\times P_c(\mathbb R^n)$. A counterexample is given in Example \ref{ex_discont_IR_plus}.
\end{itemize}
\end{remark}


So far we learned that the majorization polytope $M_d(y)$ induced by a single vector $y\in\mathbb R^n$ with respect to some $d\in\mathbb R_{++}^n$ differs from the classical majorization polytope not in the orientation of the faces but only their precise location. By Thm.~\ref{thm_maj_halfspace} this difference is fully captured by the following map:
\begin{equation}\label{eq:f_b_vec}
f:[0,\unitvector^Td]\to\mathbb R\qquad c\mapsto \min_{i=1,\ldots,n} \unitvector^T\Big(y-\frac{y_i}{d_i}d\Big)_++\frac{y_i}{d_i}c
\end{equation}
Thus if we want to learn more about the $d$-majorization polytope we are well-advised to study the properties of \eqref{eq:f_b_vec}. 
\begin{lemma}\label{lemma_properties_of_f_min}
Let $y\in\mathbb R^n$, $d\in\mathbb R_{++}^n$ be given and ${\pi}\in S_n$ be a permutation which orders $(\frac{y_i}{d_i})_{i=1}^n$ decreasingly, i.e.~$\frac{y_{{\pi}(1)}}{d_{{\pi}(1)}}\geq\ldots\geq\frac{y_{{\pi}(n)}}{d_{{\pi}(n)}}$. Then the map $f$ from \eqref{eq:f_b_vec} has the following properties.
\begin{itemize}
\item[(i)] $f$ is continuous, piecewise linear, and concave.
\item[(ii)] For arbitrary $j=1,\ldots,n$ and $c\in(\sum_{i=1}^{j-1}d_{{\pi}(i)},\sum_{i=1}^{j}d_{{\pi}(i)})$
\begin{equation*}
f(c)=\sum\nolimits_{i=1}^{j-1} y_{{\pi}(i)}+\frac{y_{{\pi}(j)}}{d_{{\pi}(j)}}\Big(c-\sum\nolimits_{i=1}^{j-1}d_{{\pi}(i)}\Big)
\end{equation*}
as well as $f'(c)=\frac{y_{{\pi}(j)}}{d_{{\pi}(j)}}$ so the (weak) derivative of $f$ is monotonically decreasing.
\item[(iii)] For all $j=0,\ldots,n$ one has $f( \sum_{i=1}^j d_{{\pi}(i)} )= \sum_{i=1}^j y_{{\pi}(i)} $ so in particular $f(0)=0$ and $f(\unitvector^Td)=\unitvector^Ty$.
\item[(iv)] Let $k=1,\ldots,n-1$, pairwise different $\alpha_1,\ldots,\alpha_k\in\{1,\ldots,n\}$, and $\tau\in S_n$ be given. Then
$$
\sum\nolimits_{j=1}^k f\Big(\sum\nolimits_{i=1}^{\alpha_j-1}d_{\tau(i)}\Big)+f\Big(\sum\nolimits_{i=1}^k d_{\tau(\alpha_i)}\Big)\geq\sum\nolimits_{j=1}^k f\Big(\sum\nolimits_{i=1}^{\alpha_j}d_{\tau(i)}\Big)\,.
$$
\item[(v)] For all $j=1,\ldots,n-1$
$$
\frac{f\big(\sum\nolimits_{i=1}^j d_i^\downarrow\big)-f\big(\sum\nolimits_{i=1}^{j-1} d_i^\downarrow\big)}{d_j^\downarrow}\geq \frac{f\big(\sum\nolimits_{i=1}^{j+1} d_i^\downarrow\big)-f\big(\sum\nolimits_{i=1}^j d_i^\downarrow\big)}{d_{j+1}^\downarrow}\,.
$$
\end{itemize}
\end{lemma}
\begin{proof}
(i): The minimum over finitely many affine linear functions (in particular these functions are continuous \& concave) is piecewise linear, continuous, and concave. (ii): Direct consequence of Lemma \ref{lemma_maj_sum_recovery}. (iii): Follows from (ii) together with continuity of $f$. (iv): Because $f$ is continuous \& concave ($-f$ is continuous \& convex) this is a direct consequence of Prop.~\ref{lemma_char_d_vec} (for $d=\unitvector$) together with Lemma \ref{lemma_d_ordering_maj} and $f(0)=0$ from (iii). (v): Define $\tilde d:=(d_j^\downarrow,d_{j+1}^\downarrow)^T\in\mathbb R_{++}^2$. Evidently
$$
\Big(\sum\nolimits_{i=1}^j d_i^\downarrow\Big)\tilde d=\begin{pmatrix}
 d_j^\downarrow\sum\nolimits_{i=1}^j d_i^\downarrow\\
 d_{j+1}^\downarrow\sum\nolimits_{i=1}^j d_i^\downarrow
\end{pmatrix}\prec_{\tilde d} \begin{pmatrix} d_j^\downarrow\sum\nolimits_{i=1}^{j+1} d_i^\downarrow \\
d_{j+1}^\downarrow\sum\nolimits_{i=1}^{j-1} d_i^\downarrow
\end{pmatrix}
$$
due to minimality of $\tilde d$ w.r.t.~$\prec_{\tilde d}$ and because the two vectors are of same trace\footnote{Direct computation:
\begin{align*}
d_j^\downarrow\sum\nolimits_{i=1}^j d_i^\downarrow+ d_{j+1}^\downarrow\sum\nolimits_{i=1}^j d_i^\downarrow=d_j^\downarrow\sum\nolimits_{i=1}^j d_i^\downarrow+ d_{j+1}^\downarrow d_{j}^\downarrow+ d_{j+1}^\downarrow\sum\nolimits_{i=1}^{j-1} d_i^\downarrow= d_j^\downarrow\sum\nolimits_{i=1}^{j+1}d_i^\downarrow+ d_{j+1}^\downarrow\sum\nolimits_{i=1}^{j-1} d_i^\downarrow
\end{align*}
}
(Thm.~\ref{prop_1} (i)). Again, using convexity of $-f$, Prop.~\ref{lemma_char_d_vec} (for $d=\tilde d$) yields
\begin{align*}
d_j^\downarrow f\Big(\sum\nolimits_{i=1}^j d_i^\downarrow\Big)+d_{j+1}^\downarrow f\Big(\sum\nolimits_{i=1}^j d_i^\downarrow\Big)&=
d_j^\downarrow f\Big(\frac{ d_j^\downarrow\sum\nolimits_{i=1}^j d_i^\downarrow}{d_j^\downarrow}\Big)+d_{j+1}^\downarrow f\Big( \frac{d_{j+1}^\downarrow\sum\nolimits_{i=1}^j d_i^\downarrow}{d_{j+1}^\downarrow}\Big)\\
&\geq d_j^\downarrow f\Big(\frac{ d_j^\downarrow\sum\nolimits_{i=1}^{j+1} d_i^\downarrow}{d_j^\downarrow}\Big)+d_{j+1}^\downarrow f\Big( \frac{d_{j+1}^\downarrow\sum\nolimits_{i=1}^{j-1} d_i^\downarrow}{d_{j+1}^\downarrow}\Big)\\
&= d_j^\downarrow f\Big(\sum\nolimits_{i=1}^{j+1} d_i^\downarrow\Big)+d_{j+1}^\downarrow f\Big(\sum\nolimits_{i=1}^{j-1} d_i^\downarrow\Big)
\end{align*}
which readily implies (v).
\end{proof}
For all rows $m^T\in\{0,1\}^n$ of $M$, the $b$-vector of the $d$-majorization polytope satisfies $\mathfrak b(m^T)=f(m^Td)$ so, in slight abuse of notation, we may write $M_d(y)=\{x\in\mathbb R^n\,|\,Mx\leq f(Md)\}$. 
\begin{remark}\label{rem_thermomaj_phys_lit}
Recall that in the physics literature, thermo-majorization is usually defined via curves of the following form: Given a vector $x\in\mathbb R^n$ and $d\in\mathbb R_{++}^n$ consider the piecewise linear, continuous curve fully characterized by the elbow points $\{ \big(\sum_{i=1}^j d_{\sigma(i)},\sum_{i=1}^j x_{\sigma(j)}\big) \}_{j=0}^n$, where $\sigma$ is a permutation such that $\frac{x_{\sigma(1)}}{d_{\sigma(1)}}\geq\ldots\geq\frac{x_{\sigma(n)}}{d_{\sigma(n)}}$. Then a vector $y$ is said to thermomajorize $x$ if $\unitvector^T x=\unitvector^T y$ and if the curve induced by $y$ is never below the curve induced by $x$ \cite{Horodecki13}.
But by the previous lemma this thermo-majorization curve is precisely the function $f$ which characterizes the polytope, so $x\prec_dy$ is equivalent to $f_x(c)\leq f_y(c)$ for all $c\in[0,\unitvector^Td]$ (more on this in a second).
\end{remark}
While this confirms the (well-known) equivalence of $d$-majorization and thermo-majorization, we can reduce the comparison of the two curves to just the ``elbow points'' of the lower curve---as already observed in \cite[Thm.~4]{Alhambra16}---by means of the following elegant proof:
\begin{proof}[Proof of Prop.~\ref{lemma_char_d_vec} (i) $\Leftrightarrow$ (vii)]
By Lemma \ref{lemma_closure_op}, $x\prec_dy$ is equivalent to $M_d(x)\subseteq M_d(y)$ which by Thm.~\ref{thm_maj_halfspace} and Remark \ref{rem:halfspace_actions} holds if and only if $\unitvector^Tx=\unitvector^Ty$ and $f_x((Md)_i)\leq f_y((Md)_i)$ for all $i=1,\ldots,2^n-2$. Now we may apply Lemma \ref{lemma_properties_of_f_min} \& Lemma \ref{lemma_convex_cpl_compare} to arrive at the equivalent condition: $\unitvector^Tx=\unitvector^Ty$ and $\sum_{i=1}^j x_{\sigma(i)}=f_x(\sum_{i=1}^jd_{\sigma(i)})\leq f_y(\sum_{i=1}^jd_{\sigma(i)})$ for all $j=1,\ldots,n-1$, which concludes the proof.
\end{proof}

Another advantage of introducing and studying the function $f$ is that its properties transfer to $M_d(y)$ which suffices to fully characterize the extreme points of the $d$-majorization polytope, thus generalizing Corollary \ref{coro_classical_maj_extreme}:

\begin{theorem}\label{thm_Eb_sigma}
Let $y\in\mathbb R^n$, $d\in\mathbb R_{++}^n$. Then the extreme points of $M_d(y)$ are precisely the $E_b({\pi})$, ${\pi}\in S_n$ from Definition \ref{def_Ebsigma}. In particular $M_d(y)=\operatorname{conv}\{E_b({\pi})\,|\,{\pi}\in S_n\}$.
\end{theorem}
\begin{proof}
By Thm.~\ref{thm_Ebsigma_extreme} (and Thm.~\ref{thm_maj_halfspace}) all we have to show is that $\{E_b({\pi})\,|\,{\pi}\in S_n\}\subseteq M_d(y)$. Let arbitrary ${\pi}\in S_n$ be given. Showing $E_b({\pi})\in M_d(y)$ by Thm.~\ref{thm_maj_halfspace} is equivalent to showing $ME_b({\pi})\leq b$, i.e.
$$
\sum\nolimits_{j=1}^k (E_b({\pi}))_{{\pi}(\alpha_j)}=\Big(\sum\nolimits_{i=1}^k e_{{\pi}(\alpha_i)}\Big)^T E_b({\pi})\leq \mathfrak b\Big(\sum\nolimits_{i=1}^k e_{{\pi}(\alpha_i)}^T\Big)=f\Big(\sum\nolimits_{i=1}^k d_{{\pi}(\alpha_i)}\Big)
$$
for all $k=1,\ldots,n-1$ and all pairwise different $\alpha_1,\ldots,\alpha_k\in\{1,\ldots,n\}$ where $f$ is the map from \eqref{eq:f_b_vec}. Be aware that adding ${\pi}$ in the above indices yields an equivalent problem as the $\alpha_i$ can be chosen arbitrarily anyway. This will be advantageous because the expression $(E_b({\pi}))_{{\pi}(\alpha_i)}$ is easier to handle than $(E_b({\pi}))_{\alpha_i}$; indeed by Lemma \ref{lemma_E_b_sigma_properties}
\begin{align*}
(E_b({\pi}))_{{\pi}(\alpha_j)}=\mathfrak b\Big(\sum\nolimits_{i=1}^{\alpha_j} e_{{\pi}(i)}^T\Big)-\mathfrak b\Big(\sum\nolimits_{i=1}^{\alpha_j-1} e_{{\pi}(i)}^T\Big)=f\Big(\sum\nolimits_{i=1}^{\alpha_j} d_{{\pi}(i)}\Big)-f\Big(\sum\nolimits_{i=1}^{\alpha_j-1} d_{{\pi}(i)}\Big)
\end{align*}
for all $j=1,\ldots,k$. Hence showing $E_b({\pi})\in M_d(y)$ is equivalent to
\begin{align*}
\sum\nolimits_{j=1}^k \Big(f\Big(\sum\nolimits_{i=1}^{\alpha_j} d_{{\pi}(i)}\Big)&-f\Big(\sum\nolimits_{i=1}^{\alpha_j-1} d_{{\pi}(i)}\Big)\Big)\leq f\Big(\sum\nolimits_{i=1}^k d_{{\pi}(\alpha_i)}\Big)\\
\Leftrightarrow\quad \sum\nolimits_{j=1}^k f\Big(\sum\nolimits_{i=1}^{\alpha_j} d_{{\pi}(i)}\Big)&\leq \sum\nolimits_{j=1}^kf\Big(\sum\nolimits_{i=1}^{\alpha_j-1} d_{{\pi}(i)}\Big)+f\Big(\sum\nolimits_{i=1}^k d_{{\pi}(\alpha_i)}\Big)
\end{align*}
but this holds due to Lemma \ref{lemma_properties_of_f_min} (iv).
\end{proof}
%
%
%

We immediately obtain the following result.

\begin{corollary}\label{thm_convex_poly}
Let $y\in\mathbb R^n$, $d\in\mathbb R_{++}^n$. Then $M_d(y)$ is a non-empty convex polytope\index{d-majorization polytope@$d$-majorization polytope} of at most $n-1$ dimensions and, moreover, has at most $n!$ extreme points. 
\end{corollary}
\begin{proof}
For non-emptiness note that the identity matrix is $d$-stochastic so $y\in M_d(y)$. 
By Thm.~\ref{thm_maj_halfspace} there exists $b\in\mathbb R^{2^n}$ such that $M_d(y)=\{x\in\mathbb R^n\,|\,Mx\leq b\}$ so $M_d(y)$ is a convex polytope of at most $n-1$ dimensions (Lemma \ref{lemma_Mx_b_convex_polytope}). Finally the extreme points of $M_d(y)$ are contained given by $\{E_b({\pi})\,|\,{\pi}\in S_n\}$ (Thm.~\ref{thm_Eb_sigma}) which due to $|S_n|=n!$ concludes the proof.
\end{proof}

\begin{remark}
These results (Thm.~\ref{thm_Eb_sigma} \& Coro.~\ref{thm_convex_poly}) recently appeared in the physics literature for the special case $y\geq 0$ \cite[Sec.~2.2]{Alhambra19} but with an entirely different proof strategy: Alhambra et al.~explicitly constructed a family $\{P^{(\pi,\alpha)}\}_{\alpha}$ of $d$-stochastic matrices called ``$\beta$-permutations'' with the property that $\{P^{(\pi,\alpha)}y\}_{\alpha}$ contains all extreme points of $M_d(y)$, which---in our language---necessarily have to be of the form $E_b(\pi)$ \cite[Lemma 12]{Lostaglio18}.
\end{remark}

Now one of these extreme points has the property of classically majorizing every other vector from the $d$-majorization polytope. The result which is of particular interest, e.g., to answer reachability questions in Ch.~\ref{sec_reach_fin_dim} reads as follows.

\begin{theorem}\label{theorem_max_corner_maj}
Let $y\in\mathbb R_+^n$, $d\in\mathbb R_{++}^n$. Then there exists $z\in M_d(y)$ such that $x\prec z$ for all $x\in M_d(y)$, i.e.~$M_d(y)\subseteq M_\unitvector(z)$, and this $z$ is unique up to permutation.

More precisely if ${\pi}\in S_n$ is a permutation which orders $d$ decreasingly, i.e.~$d_{{\pi}(1)}\geq\ldots\geq d_{{\pi}(n)}$, then $M_d(y)\subseteq M_\unitvector(E_b({\pi}))$, i.e.~$z$ can be chosen to be $E_b({\pi})$ which is the solution to
\begin{equation}\label{eq:implicit_solution_z}
{\footnotesize\begin{pmatrix}
1&0&\cdots&0\\
\vdots&\ddots&\ddots&\vdots\\
\vdots&&\ddots&0\\
1&\cdots&\cdots&1
\end{pmatrix}}\underline{{\pi}}z=
\min_{i=1,\ldots,n}{\footnotesize\begin{pmatrix}
\unitvector^T(y-\frac{y_i}{d_i}d)_++\frac{y_i}{d_i}d_{{\pi}(1)}\\
\vdots\\
\unitvector^T(y-\frac{y_i}{d_i}d)_++\frac{y_i}{d_i}\sum_{j=1}^{n-1}d_{{\pi}(j)}\\
\unitvector^Ty
\end{pmatrix}} \overset{\eqref{eq:f_b_vec}}=\Big(f\big(\sum\nolimits_{i=1}^jd_{{\pi}(i)}\big)\Big)_{j=1}^n \,.
\end{equation}
Moreover this $z=E_b({\pi})$ has the following properties.
\begin{itemize}
\item[(i)] $z$ is an extreme point of $M_d(y)$.
\item[(ii)] $M_d(z)\subseteq M_\unitvector(z)$.
\item[(iii)] $\frac{z_{{\pi}(1)}}{d_{{\pi}(1)}}\geq \ldots\geq \frac{z_{{\pi}(n)}}{d_{{\pi}(n)}}$.%
\end{itemize}
\end{theorem}
\begin{proof}

Uniqueness of such $z$ (up to permutation) is the easiest to show so let us start with that: If there exist $z_1,z_2\in M_d(y)$ such that $M_d(y)\subseteq M_\unitvector(z_i)$ for $i=1,2$. In particular one has $z_{2-i}\in M_d(y)\subseteq M_\unitvector(z_i)$ so $z_1\prec z_2\prec z_1$ and one finds a permutation $\tau\in S_n$ such that $z_1=\underline{\tau}z_2$.

For existence let ${\pi}\in S_n$ be a permutation which orders $d$ decreasingly. From Thm.~\ref{thm_Eb_sigma} we know that $z:=E_b({\pi})\in M_d(y)$ and $z$ is even an extreme point (this shows (i)).
By \eqref{eq:implicit_solution_z} 
one for all $j=1,\ldots,n-1$ finds
\begin{equation*}
z_{{\pi}(j)}=f\Big(\sum\nolimits_{i=1}^{j}d_{{\pi}(i)}\Big)-f\Big(\sum\nolimits_{i=1}^{j-1}d_{{\pi}(i)}\Big)=f\Big(\sum\nolimits_{i=1}^{j}d_i^\downarrow\Big)-f\Big(\sum\nolimits_{i=1}^{j-1}d_i^\downarrow\Big)\,.
\end{equation*}
Therefore $\frac{z_{{\pi}(j)}}{d_{{\pi}(j)}}\geq \frac{z_{{\pi}(j+1)}}{d_{{\pi}(j+1)}}$ is equivalent to 
\begin{align*}
\frac{f\big(\sum\nolimits_{i=1}^j d_i^\downarrow\big)-f\big(\sum\nolimits_{i=1}^{j-1} d_i^\downarrow\big)}{d_j^\downarrow}\geq \frac{f\big(\sum\nolimits_{i=1}^{j+1} d_i^\downarrow\big)-f\big(\sum\nolimits_{i=1}^j d_i^\downarrow\big)}{d_{j+1}^\downarrow}
\end{align*}
which holds due to Lemma \ref{lemma_properties_of_f_min} (v) so $\frac{z}{d}$ and $d$ are indeed similarly ordered (this shows (iii)). More importantly because $z\in\mathbb R_+^n$ (as stochastic matrices preserve non-negativity of $y$) one even has $z_{{\pi}(j)}=z_j^\downarrow$ for all $j=1,\ldots,n$ because
$$
z_{{\pi}(j)}\geq \frac{d_{{\pi}(j)}}{d_{{\pi}(j+1)}}z_{{\pi}(j+1)}=\underbrace{\frac{d_j^\downarrow}{d_{j+1}^\downarrow}}_{\geq 1} z_{{\pi}(j+1)}\geq z_{{\pi}(j+1)}\,.
$$
Now recall that $M_\unitvector(z)=\{x\in\mathbb R^n\,|\,Mx\leq b_z\}$ (Prop.~\ref{prop_maj_halfspace}) where $b_z$ is of the following form:
the first $\binom{n}{1}$ entries equal $z_1^\downarrow=z_{{\pi}(1)}$, the next $\binom{n}{2}$ entries equal $z_1^\downarrow+z_2^\downarrow=z_{{\pi}(1)}+z_{{\pi}(2)}$ and so forth until $\binom{n}{n-1}$ entries equaling $\sum_{i=1}^{n-1} z_i^\downarrow=\sum_{i=1}^{n-1} z_{{\pi}(i)}$. Writing $M_d(y)=\{x\in\mathbb R^n\,|\,Mx\leq b\}$ (Thm.~\ref{thm_maj_halfspace}), if we can show that $b\leq b_z$ then we get $M_d(y)\subseteq M_\unitvector(z)$ (Remark \ref{rem:halfspace_actions}) as desired. 

For all $k=1,\ldots,n-1$ and all $\tau\in S_n$ by Lemma \ref{lemma_minmax} (which we may apply because $\frac{y_i}{d_i}\geq 0$ for all $i$)
\begin{align*}
\mathfrak b\Big(\sum\nolimits_{j=1}^k e_{\tau(j)}^T\Big)&=\min_{i=1,\ldots,n}\unitvector^T(y-\frac{y_i}{d_i}d)_++\frac{y_i}{d_i}\sum\nolimits_{j=1}^k d_{\tau(j)}\\
&\leq\max_{\tau\in S_n}\min_{i=1,\ldots,n}\unitvector^T(y-\frac{y_i}{d_i}d)_++\frac{y_i}{d_i}\sum\nolimits_{j=1}^k d_{\tau(j)}\\
&=\min_{i=1,\ldots,n}\unitvector^T(y-\frac{y_i}{d_i}d)_++\frac{y_i}{d_i}\Big(\max_{\tau\in S_n}\sum\nolimits_{j=1}^k d_{\tau(j)}\Big)\\
&=\min_{i=1,\ldots,n}\unitvector^T(y-\frac{y_i}{d_i}d)_++\frac{y_i}{d_i}\Big(\sum_{j=1}^kd_j^\downarrow\Big)\overset{\eqref{eq:implicit_solution_z}}=\sum_{i=1}^k z_{{\pi}(i)}=\mathfrak b_z\Big(\sum_{j=1}^k e_{\tau(j)}^T\Big)
\end{align*}
so $b\leq b_z$ as claimed. Now the only statement left to prove is (ii): 
%
$z\in M_d(y)$ by Lemma \ref{lemma_closure_op} implies $M_d(z)\subseteq (M_d\circ M_d)(y)=M_d(y)\subseteq M_\unitvector(z)$.
\end{proof}
\noindent Non-negativity of $y$ in Thm.~\ref{theorem_max_corner_maj} is actually necessary as Example \ref{ex_pos_y_necess} shows.\smallskip

Given our knowledge of this maximal point (w.r.t.~classical majorization) in the $d$-majorization polytope one can now give necessary conditions for when the initial vector $y$ itself is this maximal element.

\begin{corollary}\label{coro_y_max_corner}
Let $y\in\mathbb R_+^n$, $d\in\mathbb R_{++}^n$. If there exists a permutation ${\pi}\in S_n$ such that $d_{{\pi}(1)}\geq \ldots\geq d_{{\pi}(n)}$ and $\frac{y_{{\pi}(1)}}{d_{{\pi}(1)}}\geq \ldots\geq \frac{y_{{\pi}(n)}}{d_{{\pi}(n)}}$ then $M_d(y)\subseteq M_\unitvector(y)$.
\end{corollary}
\begin{proof}
If $d$ and $\frac{y}{d}$ are similarly ordered (by means of ${\pi}$) then Thm.~\ref{theorem_max_corner_maj} tells us $M_d(y)\subseteq M_\unitvector(E_b({\pi}))$ where
\begin{align*}
{\footnotesize\begin{pmatrix}
1&0&\cdots&0\\
\vdots&\ddots&\ddots&\vdots\\
\vdots&&\ddots&0\\
1&\cdots&\cdots&1
\end{pmatrix}}\underline{{\pi}}E_b({\pi})=
\Big(f\Big(\sum_{i=1}^jd_{{\pi}(i)}\Big)\Big)_{j=1}^n=
\Big(\sum_{i=1}^jy_{{\pi}(i)}\Big)_{j=1}^n=
{\footnotesize\begin{pmatrix}
1&0&\cdots&0\\
\vdots&\ddots&\ddots&\vdots\\
\vdots&&\ddots&0\\
1&\cdots&\cdots&1
\end{pmatrix}}\underline{{\pi}}y
\end{align*}
by Lemma \ref{lemma_properties_of_f_min} (iii), hence $E_b({\pi})=y$.
\end{proof}
To see how the $d$-majorization polytope behaves (aside from continuity) when changing only $d$ while leaving the initial vector $y$ untouched we refer to Example \ref{ex_wandering_d_vector}. This example also illustrates Coro.~\ref{coro_y_max_corner} because the whole trajectory $\{d(\lambda)\,|\,\lambda\in[0,1]\}$ taken by the $d$-vector satisfies $ \frac{y_1}{(d(\lambda))_1}\geq\ldots\geq \frac{y_n}{(d(\lambda))_n} $ so maximality of $y$ (w.r.t.~classical majorization) is preserved throughout.

\section{Majorization on Matrices}\label{sec:maj_d_mat}
%
%

A fundamental aspect of resource theories is finding conditions which characterize state-transfers via ``allowed'' operations. In quantum thermodynamics\index{quantum thermodynamics}, for example, one usually asks whether a state can be generated from an initial state via a quantum channel which preserves the Gibbs state of the system \cite{Brandao15,DallArno20,Gour15,Horodecki13}. 
As Gibbs states are of the form $e^{-\beta H}/\operatorname{tr}(e^{-\beta H})$ for some system's Hamiltonian $H$ 
and some inverse temperature $\beta>0$ (see also Rem.~\ref{rem_gibbs}) these states in particular are of full rank. Actually every full-rank state $D$ is the Gibbs state of some system (up to trace) by simply choosing $\beta=1$ and $H=\ln(D^{-1})$. Therefore gaining a better understanding of the set of all channels with a given full-rank fixed point and its geometry, properties, etc.~would greatly benefit the aforementioned state-conversion problem. Channels with a full-rank fixed point, sometimes called \textit{faithful}\index{quantum channel!faithful} \cite{Albert19}, are characterized by not containing a decaying subspace (under the asymptotic projection). Also conserved quantities of such channels commute with all Kraus operators up to a phase \cite[Prop.~1]{Albert19}. 
Topics related to faithful channels and fixed-point analysis of \textsc{cptp} maps are (mean-)ergodic channels \cite{Burgarth13,Ohya81}, irreducible channels \cite{Davies70,Sanz10}, and zero-error \cite{BlumeKohout10}, \cite[Ch.~4]{Gupta15} and relaxation properties of discrete-time \cite{Cirillo15} and continuous-time \cite{Frigerio77,Spohn76,Spohn77} Markovian systems. 
As a notable special case if the initial and the final state commute with the Hamiltonian $H$ then one is in the classical realm which is handled by vector $d$-majorization again.
Recall from the last section that the above state-conversion problem in this classical case reduces to $n$ vector-$1$-norm inequalities and the vectors $d$-majorized by some initial vector form a convex polytope with at most $n!$ (analytically computable) extreme points.

When studying channels with a full-rank fixed point one finds that these belong to the larger class of linear maps which preserve positive definiteness, sometimes called \textit{strictly positive} maps. While this is a rather large class of maps it turns out that some results regarding positive and strictly positive maps will be useful tools when generalizing $d$-majorization from vectors to matrices. 
As before this whole section is based on one of our preprints \cite{vomEnde20Dmaj}. 
\subsection{Strict Positivity}\label{sec:strictpos}

While the term ``strict positivity'' in the context of Perron-Frobenius theory refers to maps which send positive semi-definite operators to positive definite ones \cite{Farenick96,Gaubert17,Rahaman20} we want it to mean the following:

\begin{definition}[\cite{Bhatia07}, Ch.~2.2]
A linear map $T:\mathbb C^{n\times n}\to\mathbb C^{k\times k}$ is called strictly positive\index{strictly positive} (\textsc{sp}\label{not_sp}) if $T(X)>0$ whenever $X>0$. 
Moreover $T$ is called completely strictly positive\index{completely strictly positive} (\textsc{csp}\label{not_csp}) if $T\otimes\mathbbm{1}_{m\times m}$ is strictly positive for all $m\in\mathbb N$.
\end{definition}

\begin{figure}[!htb]
\centering
\includegraphics[width=0.55\textwidth]{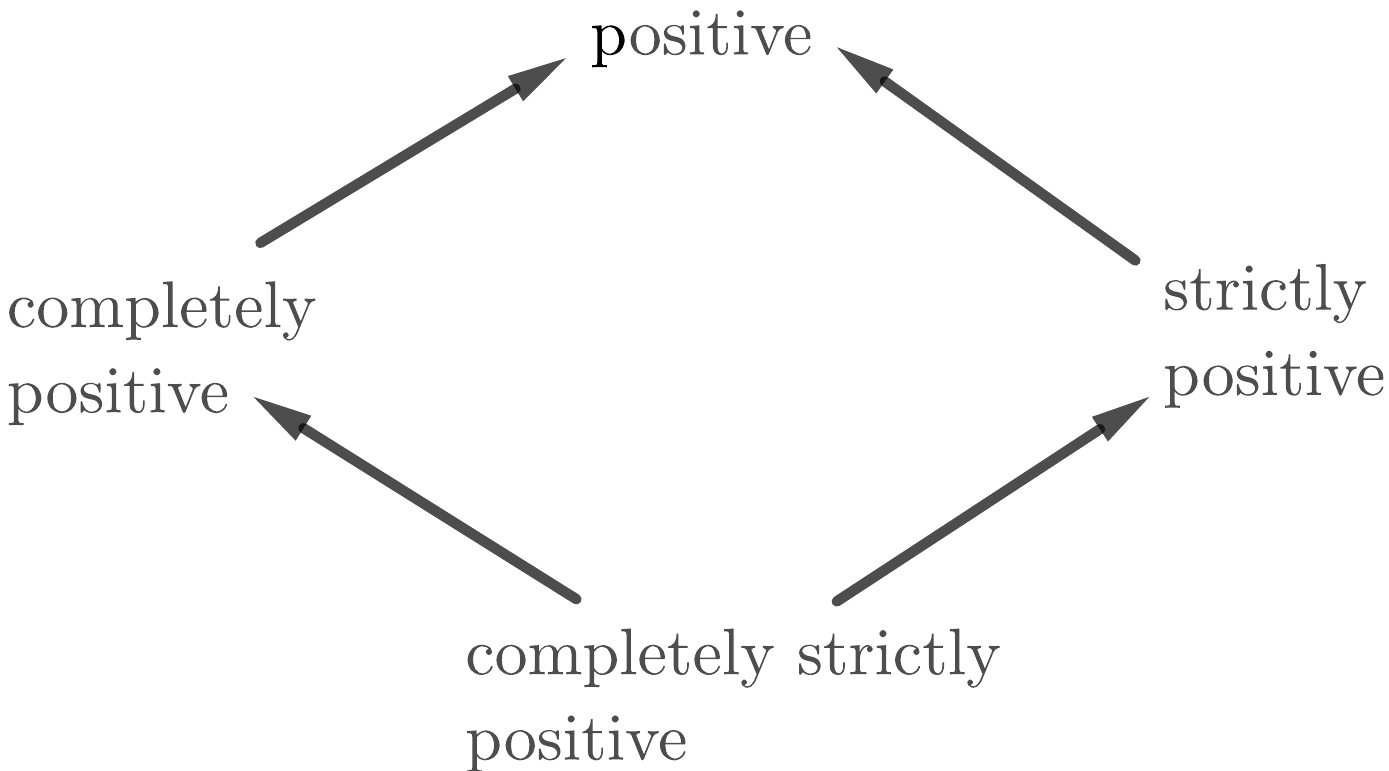}
\caption{Relation between \textsc{p}, \textsc{cp}, \textsc{sp}, and \textsc{csp}. By definition \textsc{csp} implies \textsc{sp}. A simple continuity-type argument shows that \textsc{sp} $\Rightarrow$ \textsc{p} 
and \textsc{csp} $\Rightarrow$ \textsc{cp}. Moreover \textsc{sp} and \textsc{cp} are incomparable: The transposition map is obviously \textsc{sp} but well-known to not be \textsc{cp}, and, on the other hand, the trace projection $X\mapsto \operatorname{tr}(X)|\psi\rangle\langle\psi|$ for some pure state $\psi$ is a quantum channel \cite[Ex.~5.3]{Hayashi06} (hence \textsc{cp}) but evidently not \textsc{sp}, unless the dimension equals one.}
\label{fig1}
\end{figure}

\noindent
Comparison of these concepts to usual positivity (\textsc{p}\label{not_p}) and 
complete positivity (\textsc{cp}\label{not_cp}) is shown in Fig.~\ref{fig1}.
${}$

Bhatia \cite{Bhatia07} observed that ``a positive linear map $\Phi$ is strictly positive if and only if $\Phi(\mathbbm{1})>0$'' so strict positivity can be easily checked. This, however, turns out to be a mere corollary of the following stronger result: The image of positive linear maps admit a ``universal kernel'' completely characterized by the action on any positive definite matrix.
\begin{proposition}\label{prop_positivity}
Let $T:\mathbb C^{n\times n}\to \mathbb C^{k\times k}$ linear and positive be given. For all \mbox{$X,Y,A\in\mathbb C^{n\times n}$}, $X,Y>0$ one has
\begin{equation}\label{eq:ker_T}
\operatorname{ker}(T(X))=\operatorname{ker}(T(Y))\subseteq\operatorname{ker}(T(A))\,.
\end{equation}
In particular the following are equivalent:
\begin{itemize}
\item[(i)] $T$ is strictly positive.
\item[(ii)] $T(\mathbbm{1})>0$
\item[(iii)] There exists $X>0$ such that $T(X)>0$.
\end{itemize}
\end{proposition}
\begin{proof}
Let $X\in\mathbb C^{n\times n}$, $X>0$, and $\psi\in\operatorname{ker}(T(X))$ be given. Then for all $Y\in\mathbb C^{ n \times n }$ positive semi-definite one finds $\lambda\in\mathbb R$ such that\footnote{
E.g., choose $\lambda=\frac{\max\sigma(Y)}{\min\sigma(X)}$ and note $\min\sigma(X)>0$ by assumption.
}
$\lambda X-Y\geq 0$. But then $T(\lambda X-Y)\geq 0$ by positivity of $T$ so linearity shows
$$
0\leq \langle \psi,T(\lambda X-Y)\psi\rangle=-\langle \psi,T(Y)\psi\rangle\leq 0\,.
$$
Now $\|\sqrt{T(Y)}\psi\|^2=\langle \psi,T(Y)\psi\rangle=0$ is equivalent to $T(Y)\psi=0$ which shows $\ker(T(X))\subseteq\ker(T(Y))$. If $Y>0$ then we can interchange the roles of $X,Y$ in the above argument to obtain $\ker(T(X))=\ker(T(Y))$. Finally the case of a general $A\in\mathbb C^{n\times n}$ follows from the fact that every matrix can be written as a linear combination of four positive semi-definite matrices \cite[Coro.~4.2.4]{Kadison83}, together with linearity of $T$. 

For the second statement---while (i) $\Rightarrow$ (ii) $\Rightarrow$ (iii) is obvious---for (iii) $\Rightarrow$ (i) note that if $T(X)>0$ for some $X>0$, meaning $\ker(T(X))=\{0\}$, then the same holds for all positive definite matrices by \eqref{eq:ker_T}. 
\end{proof}

Prop.~\ref{prop_positivity} shows that a fixed point of full rank guarantees a quantum channel to be strictly positive, whereas the converse does not hold (Example \ref{example_1z}).

\begin{remark}
\begin{itemize}
\item[(i)] While strict positivity tells us that one cannot leave the relative interior of all states (the invertible states), the boundary (the non-invertible states) can be mapped either onto the boundary or into the interior. The former is achieved for example by every unitary channel while the latter can be done via a trace projection onto some positive definite state.
\item[(ii)] In the generic case the inclusion in \eqref{eq:ker_T} is not an equality. Even worse there is no general statement one can make about the rank regarding the image of strictly positive maps (see Example \ref{example_2z}).
\end{itemize}
\end{remark}

So far we looked at \textsc{sp} and \textsc{cp} separately so let us study their interplay via the Kraus representation next. Unsurprisingly, the ``universal kernel" property from the previous proposition appears in this representation, as well, now in connection with the kernel of the Kraus operators.
\begin{lemma}
Let $T\in \mathbb C^{n\times n}\to \mathbb C^{k\times k}$ be linear and completely positive. Then the following are equivalent.
\begin{itemize}
\item[(i)] $T$ is strictly positive.
\item[(ii)] For all sets of Kraus operators $\{K_i\}_{i\in I}$ of $T$ one has $\bigcap_{i\in I}\operatorname{ker}(K_i)=\{0\}$.
\item[(iii)] There exist Kraus operators $\{K_i\}_{i\in I}$ of $T$ with $|I|\leq nk$ s.t.~$\bigcap_{i\in I}\operatorname{ker}(K_i)=\{0\}$.
\end{itemize}
If one, and hence all, of these conditions hold then $T$ is completely strictly positive. 
\end{lemma}
\begin{proof}
By Prop.~\ref{prop_positivity} strict positivity of a positive linear map $T$ is equivalent to $T(\mathbbm{1})>0$. Observe that, given any set of Kraus operators $\{K_i\}_{i\in I}$ of $T$, one has
$$
\langle \psi,T(\mathbbm{1})\psi\rangle=\sum\nolimits_{i\in I}\langle \psi,K_i^* K_i\psi\rangle=\sum\nolimits_{i\in I}\|K_i\psi\|^2
$$
so $\langle\psi,T(\mathbbm{1})\psi\rangle=0$ holds if and only if $\psi\in\operatorname{ker}(K_i)$ for all $i\in I$. Combining these two things readily implies the above equivalence.

For the additional statement we have to show that $T\otimes\mathbbm{1}_{m\times m}$ is strictly positive for all $m\in\mathbb N$. But again---because $T\otimes\mathbbm{1}_{m\times m}$ is positive by assumption---Prop.~\ref{prop_positivity} states that strict positivity is equivalent to $\det((T\otimes\mathbbm{1}_{m\times m})(\mathbbm{1}))\neq 0$. This holds due to
\begin{align*}
\det((T\otimes\mathbbm{1}_{m\times m})(\mathbbm{1}_n\otimes\mathbbm{1}_m))=\det(T(\mathbbm{1}))^m\cdot\det(\mathbbm{1}_m)^k=\det(T(\mathbbm{1}))^m\neq 0
\end{align*}
using that $T$ is strictly positive as well as the determinant formula for the Kronecker product \cite[Sec.~4.2]{HJ2}.
\end{proof}

Thus while \textsc{cp} and \textsc{sp} are incomparable, \textit{together} they are equivalent to \textsc{csp}. We conclude this section with some topological considerations.

\begin{lemma}\label{lemma_app_1}
Taking \textsc{p}, \textsc{cp}, and \textsc{sp} as subsets of the set of all linear maps $\mathcal L$ (with the induced subspace topology) the following statements hold. 
\begin{itemize}
\item[(i)] \textsc{p} and \textsc{cp} are closed, convex subsets of $\mathcal L$.
\item[(ii)] \textsc{sp} is convex and dense in \textsc{p}.
\item[(iii)] \textsc{sp} is open in the subspace topology induced by \textsc{p}.
\end{itemize}
\end{lemma}
\begin{proof}
The convexity statements are obvious so we only prove what remains. (i): Let $(T_m)_{m\in\mathbb N}$ be a sequence in \textsc{p} which converges to $T\in\mathcal L$. Then for all $A\geq 0$, $\psi\in\mathbb C^n$
$$
\langle \psi,T(A)\psi\rangle=\lim_{m\to\infty} \underbrace{\langle \psi,T_m(A)\psi\rangle}_{\geq 0}\geq 0
$$
so $T(A)\geq 0$, hence $T$ is in \textsc{p}. The proof for \textsc{cp} is analogous. (ii): Density can be shown constructively: if $T$ is in \textsc{p} then $((1-\frac1m)T+\frac1m\mathbbm{1})_{m\in\mathbb N}$ is a sequence in \textsc{sp} which approximates $T$.
(iii): We will show, equivalently, that its complement is closed so let $(T_m)_{m\in\mathbb N}$ be a sequence in $\textsc{p}\setminus\textsc{sp}$ which converges to some $T\in\textsc{p}$. Using Prop.~\ref{prop_positivity} we get $\det(T(\mathbbm{1}))=\lim_{m\to\infty}\det(T_m(\mathbbm{1}))=0$ by continuity of the determinant so $T\in \textsc{p}\setminus\textsc{sp}$ as claimed.
\end{proof}
\noindent 
Be aware that \textsc{sp} is not open when taken as a subset of $\mathcal L$, cf.~Example \ref{example_3new}.\smallskip
%


The fact that every positive map ``produces'' the same kernel on \textit{all} full-rank states begs the question: What if this kernel is non-zero, that is, what is the footprint of positive maps which are not strictly positive? And how does this kernel manifest? 
It turns out that such channels map into a subalgebra of $\mathbb C^{k \times k }$ the size of which is determined by their action on the identity:
\begin{theorem}\label{mainthm}
Let $T: \mathbb C^{ n \times n }\to \mathbb C^{ k \times k }$ linear and positive be given. Define $m:=\operatorname{dim}\operatorname{ker}(T(\mathbbm{1}))$. Then the following are equivalent.
\begin{itemize}
\item[(i)] $T$ is not strictly positive.
\item[(ii)] There exist pairwise orthonormal vectors $\psi_1,\ldots,\psi_m\in\mathbb C^n$ such that $T(A)\psi_j=0$ as well as $\psi_j^* T(A)=0$ for all $A\in\mathbb C^{n\times n}$, $j=1,\ldots,m$ where $m\geq 1$.
\item[(iii)] There exists unitary $U\in\mathbb C^{ k\times k }$ such that $\operatorname{im}(\operatorname{Ad}_{U^*}\circ T)\subseteq \mathbb C^{ (k-m) \times (k-m) }\oplus 0^{m\times m}$ where $m\geq 1$, that is, for all $A\in\mathbb C^{ n\times n}$
\begin{equation}\label{eq:T_block_form}
U^* T(A)U=\begin{pmatrix} *&0\\0&0_m \end{pmatrix}.
\end{equation}
\item[(iv)] There exists an orthogonal projection $\pi\in\mathbb C^{k\times k}$ of rank $k-m$ where $m\geq 1$ such that \mbox{$\pi T(A)\pi=T(A)$} for all $A\in\mathbb C^{n\times n}$.
\end{itemize}
If $T$, in addition, is trace-preserving then $m<k$. 
\end{theorem}
\begin{proof}
(i) $\Rightarrow$ (ii): Define $\mathcal K:=\operatorname{ker}(T(\mathbbm{1}))$ and be aware that $\mathcal K=\bigcap_{A\in \mathbb C^{n \times n }}{\ker}(T(A))$ by Prop.~\ref{prop_positivity}. By assumption $T$ {is} not strictly positive so $T(\mathbbm{1})$ is not invertible and thus $m=\operatorname{dim}\mathcal K\geq 1$. Now one finds an orthonormal basis $\{\psi_1,\ldots,\psi_m\}$ of $\mathcal K$, thus $T(A)\psi_j=0$ as well as
$
\psi_j^* T(A)=(T(A^* )\psi_j)^* =0^* =0
$
for all $j=1,\ldots,m$
because positivity of $T$ in particular means that $T$ preserves hermiticity.

(ii) $\Rightarrow$ (iii): Define $U$ via $e_{k+1-j}\mapsto \psi_j$ for all $j=1,\ldots,m$ and 
choose $U$ on $e_1,\ldots, e_{k-m}$ such that $U$ becomes unitary. Then for all $j=1,\ldots, m$ and all $A$ one gets
$
U^* T(A)Ue_{k+1-j}=U^* T(A)\psi_j=0$ as well as $e_{k+1-j}^T U^* T(A)U=\psi_j^* T(A)U=0
$ which shows \eqref{eq:T_block_form}.

(iii) $\Rightarrow$ (iv): Take $\pi=U(\sum_{i=1}^{k-m}|e_i\rangle\langle e_i|)U^*=\sum_{i=1}^{k-m}|Ue_i\rangle\langle Ue_i|$. 
Finally (iv) $\Rightarrow$ (i) is obvious. This establishes the equivalence of (i) through (iv). If $T$ additionally is trace-preserving then $T(\mathbbm{1})\neq 0$ so $m<k$.
\end{proof}

In other words lack of strict positivity means that the image of such maps show a ``loss of dimension''. Thus, as a direct application, if the codomain are the $2\times 2$ matrices then the action of the map is limited to one entry (or it is zero, altogether):

\begin{corollary}\label{coro1}
Let a qudit-to-qubit channel---i.e.~$T: \mathbb C^{ n \times n}\to\mathbb C^{ 2 \times 2 }$ \textsc{cptp}---be given which is not strictly positive. Then there exists $\psi\in\mathbb C^2$, $\langle\psi,\psi\rangle=1$ such that \mbox{$T(A)=\operatorname{tr}(A)|\psi\rangle\langle\psi|$} for all $A\in \mathbb C^{ n \times n}$ so $T$ is the trace projection onto a pure state.
\end{corollary}
\begin{proof}
By Thm.~\ref{mainthm} there exists $U\in \mathbb C^{ 2\times 2 }$ unitary such that 
$$
U^* T(A)U=\begin{pmatrix} *&0\\0&0 \end{pmatrix}
$$
for all $A\in\mathbb C^{ n \times n}$. Because $T$ is trace-preserving $*$ has to be of size $1$ and thus is equal to $\operatorname{tr}(A)$. Hence $T(A)=\operatorname{tr}(A)|Ue_1\rangle\langle Ue_1|$ for all $A$. Choosing $\psi:=Ue_1$ concludes the proof.
\end{proof}
Of course Coro.~\ref{coro1} holds not only for channels but all \textsc{ptp}\label{not_ptp}
maps as the proof does not exploit complete positivity. Moreover the requirement of the final system being a qubit system is essential as one can, unsurprisingly, construct \textsc{cptp} maps which are not strictly positive but are not a trace projection (Example \ref{example_5} (i)). 

\begin{remark}\label{rem_ch_heisenberg}
So far we analyzed channels in the Schr{\"o}dinger picture but how does the above phenomenon manifest in the Heisenberg picture where the channels are completely positive and satisfy $T(\mathbbm{1})=\mathbbm{1}$, hence they are \textsc{sp} by definition? Let $T$ be positive but not strictly positive so using the relation between a linear map and its dual \eqref{coro_pos_dual_equal} we get
\begin{align*}
\operatorname{tr}\big(AT^*(B)\big)=\operatorname{tr}\big(T(A)B\big)=\operatorname{tr}\big(\pi T(A)\pi B\big)=\operatorname{tr}\big(T(A) \pi B\pi\big)=\operatorname{tr}\big(AT^*(\pi B\pi)\big)
\end{align*}
with $\pi$ being the corresponding projection from Thm.~\ref{mainthm} (iv). This shows that the (pre-)dual channel $T$ of a Heisenberg channel $T^*$ is not strictly positive iff there exists an orthogonal projection such that $T^*(\pi B\pi)=T^*(B)$ for all $B$. In other words $T^*$ in some basis is fully determined by a $(k-m)^2$-dimensional subspace of the input $B$. To substantiate this we refer to Example \ref{example_5} (ii). 
\end{remark}

The final result of this section is motivated by Lemma \ref{lemma_app_1}: The identity is \textsc{sp} and the latter forms an open set (relative to \textsc{p}) so one finds $\varepsilon>0$ such that every positive linear map $\varepsilon$-close to the identity is strictly positive as well. Now somewhat surprisingly 
using Thm.~\ref{mainthm}
this $\varepsilon$ can be lower bounded by one, and in the case of trace-preserving maps be specified explicitly:

\begin{proposition}\label{lemma_opnorm_dist}
Let $T: (\mathbb C^{ n \times n },\|\cdot\|_1)\to(\mathbb C^{ n \times n },\|\cdot\|_1)$ be linear and positive but not strictly positive. Then
$\|T-\mathbbm{1}\|
\geq 1$.
If $T$ additionally is trace-preserving one has $\|T-\mathbbm{1}\|=2$ so the distance between $T$ and the identity channel is maximal.
\end{proposition}
\begin{proof}
By Thm.~\ref{mainthm} one finds $U\in\mathcal U(\mathbb C^n)$ such that $\operatorname{im}(\operatorname{Ad}_{U^*}\circ T)\subseteq \mathbb C^{ (n-m) \times (n-m) }\oplus 0^{m\times m}$ where $m=\operatorname{dim}\operatorname{ker}(T(\mathbbm{1}))\geq 1$. Because $\|\,|Ue_n\rangle\langle Ue_n|\,\|_1=1$ we by unitary equivalence of the trace norm compute
\begin{align}
\|T-\mathbbm{1}\|&\geq \big\|T\big( U|e_n\rangle\langle e_n|U^* \big)-U|e_n\rangle\langle e_n|U^*\big\|_1\notag\\
&=\big\|(\operatorname{Ad}_{U^*}\circ T)\big( U|e_n\rangle\langle e_n|U^* \big)-|e_n\rangle\langle e_n|\big\|_1\notag\\
&=\|*_{n-m}\,\oplus \,0_{m-1}\oplus (-1)\|_1= \|*_{n-m}\|_1+|-1|\notag\\
&=\big\|T\big(U|e_n\rangle\langle e_n|U^*\big)\big\|_1+1=\operatorname{tr}\big(T\big(U|e_n\rangle\langle e_n|U^*\big)\big)+1\geq 1\,.\label{eq:laststep}
\end{align}
In the last step we used that $\|A\|_1=\operatorname{tr}(A)$ for all $A\geq 0$, as well as positivity of $T$. If $T$ additionally is trace-preserving then \eqref{eq:laststep} is obviously equal to $2$. To show that this is also an upper bound recall that $\|T\|=\|\mathbbm{1}\|=1$ because every \textsc{ptp} map is trace norm-contractive (Prop.~\ref{thm_q_norm_1}) hence by the triangle inequality
$
\|T-\mathbbm{1}\|\leq\|T\|+\|\mathbbm{1}\|=2
$
which concludes the proof.
\end{proof}
Prop.~\ref{lemma_opnorm_dist} gives a necessary criterion for lack of strict positivity which, however, is not sufficient (cf.~Example \ref{example_5c}). This is linked to the fact that $\|T-\operatorname{id}\|< 1$ is well-known to be sufficient for invertibility of $T$, but fails to be necessary. Indeed an elegant alternative proof of the above proposition could use this very argument, because Theorem \ref{mainthm} shows that if $n=k$ then a positive, but not strictly positive linear map cannot be invertible.
\begin{remark}[Application to quantum dynamics]\label{rem_markov}
So far we learned that lack of strict positivity comes along with a loss of dimension which, when approaching from a more physical point of view, motivates the following question: Given a quantum-\textit{dynamical} semigroup\index{quantum-dynamical semigroup} (recall Ch.~\ref{ch:qu_dyn_sys}) can one determine the exact time when this dimension loss occurs, if at all?

Indeed the semigroup structure guarantees strict positivity at all times. 
As for a short proof: Given any $t>0$ using continuity of the semigroup in $t$ one finds $m\in\mathbb N$ such that $\|T_{t/m}-T_0\|=\|T_{t/m}-\mathbbm{1}\|<2$. This by Prop.~\ref{lemma_opnorm_dist} implies that $T_{t/m}$ is a strictly positive channel so---because \textsc{sp} forms a semigroup---$T_t=(T_{t/m})^m$ is strictly positive as well.
%
\end{remark}
The result from Remark \ref{rem_markov} also shows that (Markovian) cooling processes\index{Markovian!cooling process}---or any relaxation process of semigroup structure the steady state of which is not invertible---have to take infinitely long. This is not too surprising as the dissipation 
happens exponentially in time.
\subsection{Majorization on Matrices}\label{sec:matrix_d_maj}
Moving to the matrix case, classical majorization on the level of Hermitian matrices uses their ``eigenvalue vector'' $\vec\lambda(\cdot)$ arranged in any order with multiplicities counted.
More precisely for $A,B\in \mathbb C^{n\times n}$ Hermitian, $A$ is said to be majorized\index{majorization!on matrices} by $B$ if \mbox{$\vec\lambda(A)\prec\vec\lambda(B)$}, cf. \cite{Ando89}. The most na{\"i}ve approach to define \mbox{$D$-majorization} on matrices would be to replace $\prec$ by $\prec_d$ and leave the rest as it is. However, just as in the vector case such a definition depends on the eigenvalues' arrangement in $\vec\lambda$ which is infeasible due to the lack of permutation invariance of $d$, unless, of course, $d=\unitvector$. 
The most natural way out of this dilemma is to characterize classical majorization on matrices via quantum channels which have the identity matrix as a fixed point \cite[Thm.~7.1]{Ando89}:
\begin{lemma}\label{thm_ando_7_1}
Let $A,B\in\mathbb C^{n\times n}$ be Hermitian. The following are equivalent.
\begin{itemize}
\item[(i)] $A\prec B$, that is, $\vec\lambda(A)\prec\vec\lambda(B)$.\label{symb_maj_2}
\item[(ii)] There exists $ T\in Q(n)$ such that $T(B)=A$, $ T(\mathbbm{1})=\mathbbm{1}$.
\item[(iii)] There exists $ T:\mathbb C^{n\times n}\to \mathbb C^{n\times n}$ linear and \textsc{ptp} such that $T(B)=A$, $ T(\mathbbm{1})=\mathbbm{1}$.
\end{itemize}
\end{lemma}
Therefore it seems utmost reasonable to generalize majorization on square matrices as follows:
\begin{definition}\label{defi_matrix_D_maj}
Let $D\in\mathbb C^{n\times n}$ positive definite and $A,B\in\mathbb C^{n\times n}$ be given. Now $A$ is 
said to be \textit{$D$-majorized}\index{D-majorization@$D$-majorization} by $B$, denoted by $A\prec_D B$, if there exists $T\in Q(n)$ such that $T(B)=A$ 
and $T(D)=D$.
\end{definition}
\noindent Shortly after submission of our manuscript \cite{vomEnde20Dmaj} the concept of $D$-majorization has, independently, been introduced in the physics literature by Sagawa \cite[Ch.~6.3]{Sagawa20}. However---beyond a mere definition---said reference ``only'' contributes a R{\'e}nyi divergence-criterion for $D$-majorization, which is sufficient but by far not necessary.\medskip

Coro.~\ref{lemma_convex_subsemigroup} immediately implies that $\prec_D$ admits convex structure:
\begin{lemma}\label{lemma_conv_d_maj_matrices}
Let $A,B,C,D\in\mathbb C^{n\times n}$ with $D>0$ be given. If $A\prec_D C$ and $B\prec_D C$, then $\lambda A+(1-\lambda)B\prec_D C$ for all $\lambda\in [0,1]$.
\end{lemma}
\begin{remark}[Semigroup majorization]\label{rem_semigroup}
Comparing to \cite[Def.~4.6]{Parker96} $\prec_D$ co{\"i}ncides with the semigroup majorization\index{semigroup majorization} induced by $Q_D(n)$, a concept also treated in \cite[Ch.~14.C]{MarshallOlkin}. Moreover, as $Q_D(n)$ is convex and compact (Coro.~\ref{lemma_convex_subsemigroup})
one can consider the set $Q^E_D(n)$ of extreme points of $Q_D(n)$\label{symb_Q_X_E_n}
 which can be abstractly characterized using, e.g., \cite[Thm.~5]{Choi75} or \cite[Thm.~4]{Verstraete}. Now for any $A,B\in\mathbb C^{n\times n}$, $A\prec_D B$ holds if and only if $A$ lies in the convex hull of the set $\lbrace T(B)\,|\,T\in Q^E_D(n)\rbrace$, cf. \cite[Ch.~14, Obs.C2.(iii)]{MarshallOlkin}.
\end{remark}
Be aware that there is a global unitary degree of freedom here: Given square matrices $A,B,D$ with $D>0$ and any unitary transformation $U$ one has $A\prec_D B$ if and only if $UAU^*\prec_{UDU^*} UBU^*$, with the proof being a straightforward computation. This we use to require w.l.o.g.~that $D=\operatorname{diag}(d)$ for some strictly positive vector $d\in\mathbb R_{++}^n$. Now the relation between $\prec_D$ and $\prec_d$ reads as follows.
\begin{corollary}\label{coro_d_maj_diag}
Let $d\in\mathbb R_{++}^n$ and matrices $A,B\in\mathbb C^{n\times n}$ be given. Then the following statements hold.
\begin{itemize}
\item[(i)] If $A$ is diagonal and $(\langle e_j,Ae_j\rangle)_{j=1}^n\prec_d (\langle e_j,Be_j\rangle)_{j=1}^n$, then $A\prec_D B$.
\item[(ii)] If $B$ is diagonal and $A\prec_D B$, then $(\langle e_j,Ae_j\rangle)_{j=1}^n\prec_d (\langle e_j,Be_j\rangle)_{j=1}^n$.
\end{itemize}
Additionally if $A,B$ both are diagonal, then $A\prec_D B$ holds if and only if the diagonal of $B$ $d$-majorizes that of $A$, that is, $(\langle e_j,Ae_j\rangle)_{j=1}^n\prec_d (\langle e_j,Be_j\rangle)_{j=1}^n$.
\end{corollary}
\begin{proof}
We only prove (i) and (ii) because the additional statement directly follows. For convenience let $a_{j}:=\langle e_j,Ae_j\rangle$, $b_{j}:=\langle e_j,Be_j\rangle$ for all $j=1,\ldots,n$ as well as $a:=(a_{j})_{j=1}^n$, $b:=(b_{j})_{j=1}^n\in\mathbb C^n$.

(i): By assumption there exists $d$-stochastic $M\in\mathbb R^{n\times n}$ with $a=Mb$. Define $T$ via
\begin{align}
T:\mathbb C^{n\times n}&\to \mathbb C^{n\times n}\notag\\
|e_i\rangle\langle e_j|&\mapsto\begin{cases}
0&\text{ if }i\neq j\\
\sum\nolimits_{k=1}^n M_{ki} |e_k\rangle\langle e_k|&\text{ if }i= j
\end{cases}\label{eq:T}
\end{align}
and its linear extension onto all of $\mathbb C^{n\times n}$. For any $X\in \mathbb C^{n\times n}$ we find
\begin{align*}
T(X)=\sum_{i,j=1}^n X_{ij}T(|e_i\rangle\langle e_j|)=\sum_{i,k=1}^n X_{ii}M_{ki} |e_k\rangle\langle e_k|=\sum_{k=1}^n \Big(\sum_{i=1}^n X_{ii}M_{ki} \Big) |e_k\rangle\langle e_k|
\end{align*}
where $X_{ij}=\langle e_i,Xe_j\rangle$. Thus $M$ being column-stochastic implies that $T$ is trace-preserving.
Moreover the Choi matrix of $T$ is diagonal with non-negative entries by \eqref{eq:T} so $C(T)\geq 0$ which shows $T$ is completely positive by Lemma \ref{lemma_choi_matrix}.
All that is left to check now is $T(B)=A$ and $T(D)=D$. Indeed
\begin{align*}
T(B)=\sum\nolimits_{j=1}^n b_j T( |e_j\rangle\langle e_j|)&=\sum\nolimits_{i=1}^n\Big(\sum\nolimits_{j=1}^nM_{ij}b_j\Big) |e_i\rangle\langle e_i|\\
&=\sum\nolimits_{i=1}^n(Mb)_i |e_i\rangle\langle e_i|=\sum\nolimits_{i=1}^na_i |e_i\rangle\langle e_i|=A
\end{align*}
and
\begin{align*}
T(D)=\sum_{j=1}^n d_j T( |e_j\rangle\langle e_j|)=\sum_{i=1}^n\Big(\sum_{j=1}^nM_{ij}d_j\Big) |e_i\rangle\langle e_i|=\sum_{i=1}^n\underbrace{(Md)_i}_{=d_i} |e_i\rangle\langle e_i|=D\,.
\end{align*}

(ii): Given $T\in Q_D(n)$ with $A=T(B)$ define
$
M=\big(\langle e_i,T( |e_j\rangle\langle e_j|)e_i\rangle\big)_{i,j=1}^n\in\mathbb R^{n\times n}\,.
$
As above one verifies that $M$ is $d$-stochastic and
\begin{equation}
Mb=\sum_{i=1}^n (Mb)_i \,e_i=\sum_{i,j=1}^n \langle e_i,T( |e_j\rangle\langle e_j|)e_i\rangle b_j \,e_i = \sum_{i=1}^n \langle e_i,T(B)e_i\rangle\,e_i =a\,.\tag*{\qedhere}
\end{equation}
\end{proof}
As explained before, for the rest of this chapter \textbf{we w.l.o.g.~require} the positive definite matrix $D\in\mathbb R^{n\times n}$ to be diagonal, i.e.~$D=\operatorname{diag}(d)$ for some $d\in\mathbb R_{++}^n$.
\begin{proposition}\label{thm_char}
Let $d\in\mathbb R_{++}^2$ and $A,B\in \mathbb C^{2\times 2}$ Hermitian be given. Then the following statements are equivalent.
\begin{itemize}
\item[(i)] $A\prec_D B$, that is, there exists $T\in Q_D(2)$ such that $T(B)=A$.
\item[(ii)] There exists $T:\mathbb C^{2\times 2}\to \mathbb C^{2\times 2}$ linear, \textsc{ptp} such that $T(D)=D$ and $T(B)=A$.
\item[(iii)] $\|A-tD\|_1\leq \|B-tD\|_1$ for all $t\in\mathbb R$.
\item[(iv)] $\operatorname{tr}(A)=\operatorname{tr}(B)$ and $\|A-b_iD\|_1\leq \|B-b_iD\|_1$ for $i=1,2$, as well as for the generalized fidelity
$$
\big\|\sqrt{A-b_1D}\sqrt{b_2D-A}\big\|_1\geq \big\|\sqrt{B-b_1D}\sqrt{b_2D-B}\big\|_1
$$
where $\sigma(D^{-1/2}BD^{-1/2})=\{b_1,b_2\}$ ($b_1\leq b_2$) with $\sigma(\cdot)$ being the spectrum as usual.
\end{itemize}
\end{proposition}
\begin{proof}
``(i) $\Rightarrow$ (ii)'': Obvious. ``(ii) $\Rightarrow$ (iii)'': By assumption $T(B-tD)=A-tD$ for all $t\in\mathbb R$ so the claim follows from Prop.~\ref{thm_q_norm_1}.
``(iii) $\Rightarrow$ (i)'': Define 
\begin{align*}
a_1&:=\min\sigma(D^{-1/2}AD^{-1/2})\qquad b_1:=\min\sigma(D^{-1/2}BD^{-1/2})\\
a_2&:=\max\sigma(D^{-1/2}AD^{-1/2})\qquad b_2:=\max\sigma(D^{-1/2}BD^{-1/2})\,.
\end{align*}
Then\footnote{
The key here is the following well-known result: Let $X\in\mathbb C^{n\times n}$ be Hermitian with smallest eigenvalue $x_m$ and largest eigenvalue $x_M$. Then
$
-X+t\mathbbm{1}\geq 0$ if and only if $t\geq x_M$ and $X-t\mathbbm{1}\geq 0$ if and only if $t\leq x_m\,.
$
This is evident due to $x_m\|y\|^2\leq \langle y,Xy\rangle\leq x_M\|y\|^2$ for all $y\in\mathbb C^n$ (cf.~\cite[Thm.~4.2.2]{HJ1}).\label{footnote_herm_pos}
}
using that $Y\mapsto D^{1/2}YD^{1/2}$ is positive with positive inverse, we get
\begin{align*}
\begin{split}
\hphantom{-}A-tD\geq 0, \hphantom{-}B-tD\geq 0&\qquad\text{ for all }t\leq s:=\min\{a_1,b_1\}\\
-A+tD\geq 0,-B+tD\geq 0&\qquad\text{ for all }t\geq r:=\max\{a_2,b_2\}\,.
\end{split}
\end{align*}
Because the trace norm of a Hermitian matrix is equal to its trace if and only if it is positive semi-definite this implies
$$
\operatorname{tr}(A)-s\operatorname{tr}(D)=\|A-sD\|_1\leq\|B-sD\|_1=\operatorname{tr}(B)-s\operatorname{tr}(D)\ \Rightarrow\ \operatorname{tr}(A)\leq\operatorname{tr}(B)
$$
and $\|A-rD\|_1\leq\|B-rD\|_1$ shows $-\operatorname{tr}(A)\leq-\operatorname{tr}(B)$ so combined $\operatorname{tr}(A)=\operatorname{tr}(B)$. Thus for arbitrary $t_0<s $ we may define
\begin{equation}\label{eq:tilde_A_B}
\tilde A:=\frac{A-t_0D}{\operatorname{tr}(A)-t_0\operatorname{tr}(D)}\qquad \tilde B:=\frac{B-t_0D}{\operatorname{tr}(A)-t_0\operatorname{tr}(D)}
\end{equation}
so by our previous considerations $\tilde A,\tilde B>0$, $\tilde A,\tilde B\in\mathbb D(\mathbb C^2)$, and $\|\tilde A-t'D\|_1\leq \|\tilde B-t'D\|_1$ for all $t'\in\mathbb R$ by direct computation. Now the Alberti-Uhlmann theorem\index{theorem!Alberti-Uhlmann} \cite{AlbertiUhlmann80} guarantees the existence of a \textsc{cptp} map $T$ such that $T(\tilde B)=\tilde A$ and $T(D)=D$. Because $T$ is linear and $\operatorname{tr}(A)=\operatorname{tr}(B)$ one even has $T(B)=A$ which shows $A\prec_DB$.

``(i) $\Leftrightarrow$ (iv)'': Again we want to reduce this problem from Hermitian matrices to states, to make use of \cite[Thm.~6]{HeinosaariWolf12}. To see this---due to $\operatorname{tr}(A)=\operatorname{tr}(B)$---as before one finds $t_0<s$ such that $A-t_0D,B-t_0D>0$ so define $\tilde A,\tilde B\in\mathbb D(\mathbb C^2)$ as in \eqref{eq:tilde_A_B}. Using that for all $X\in\mathbb C^{n\times n}$ Hermitian and all $c_1,c_2\in\mathbb R$ one has $\sigma(c_1X+c_2\mathbbm{1})=c_1\sigma(X)+c_2$
$$
\sigma(D^{-1/2}\tilde BD^{-1/2})=\frac{\sigma(D^{-1/2}BD^{-1/2})-t_0}{\operatorname{tr}(A)-t_0\operatorname{tr}(D)}\,.
$$
With this it is easy to see that $\|A-tD\|_1\leq\|B-tD\|_1$ for all $t\in\sigma(D^{-1/2}BD^{-1/2})$ is equivalent to $\|\tilde A-t'D\|_1\leq\|\tilde B-t'D\|_1$ for all $t'\in\sigma(D^{-1/2}\tilde BD^{-1/2})=\{\tilde b_1,\tilde b_2\}$, $0<\tilde b_1\leq\tilde b_2$. By the same argument as in ``(iii) $\Rightarrow$ (i)''
\begin{align*}
\begin{split}
\tilde B-t'D\geq 0\quad&\text{ if and only if }\quad t'\leq \min\sigma(D^{-1/2}\tilde BD^{-1/2})=\tilde b_1\\
\tilde B-t'D\leq 0\quad&\text{ if and only if }\quad t'\geq \max\sigma(D^{-1/2}\tilde BD^{-1/2})=\tilde b_2
\end{split}
\end{align*}
which shows $\inf(\tilde B/D):=\sup\{t'\in\mathbb R\,|\,\tilde B-tD\geq 0\}=\tilde b_1$ and
\begin{align*}
\inf (D/\tilde B)&=\sup\{t'\in\mathbb R\,|\,D-t'\tilde B\geq 0\}\overset{D>0}=\sup\{t'>0\,|\,D-t'\tilde B\geq 0\}\\
&=\sup\{t'>0\,|\,B-\tfrac{1}{t'}D\leq 0\}=(\tilde b_2)^{-1}\,.
\end{align*}
With this the following statements are equivalent:
\begin{itemize}
\item $A\prec_D B$
\item $\tilde A\prec_D \tilde B$ (linearity)
\item $\tilde A-\tilde b_1D\geq 0$, $\tilde A-\tilde b_2D\leq 0$ as well as the trace norm inequality $\|\sqrt{\vphantom{\frac11}\tilde A-\tilde b_1D}\sqrt{\vphantom{\frac11}\tilde b_2D-\tilde A}\|_1\geq \|\sqrt{\vphantom{\frac11}\tilde B-\tilde b_1D}\sqrt{\vphantom{\frac11}\tilde b_2D-\tilde B}\|_1$ (due to \cite[Thm.~6]{HeinosaariWolf12} \& pulling out positive constants).
\item $\|\tilde A-t'D\|_1\leq\|\tilde B-t'D\|_1$ for all $t'\in\sigma(D^{-1/2}\tilde BD^{-1/2})$ as well as $\|\sqrt{\vphantom{\frac11}\tilde A-\tilde b_1D}\sqrt{\vphantom{\frac11} \tilde A-\tilde b_2D}\|_1\geq \|\sqrt{\vphantom{\frac11}\tilde B-\tilde b_1D}\sqrt{\vphantom{\frac11}\tilde B-\tilde b_2D}\|_1$
\end{itemize}
For the latter note that
$
\|\tilde B-\tilde b_1D\|_1=\operatorname{tr}(\tilde B-\tilde b_1D)=\operatorname{tr}(\tilde A-\tilde b_1D)
$
and similarly $\|\tilde B-\tilde b_2D\|_1=-\operatorname{tr}(\tilde A-\tilde b_2D)$; therefore the trace norm conditions are equivalent to the positivity conditions $\tilde A-\tilde b_1D,-\tilde A+\tilde b_2D\geq 0$ because a Hermitian matrix $X\in\mathbb C^{n\times n}$ is positive semi-definite if and only if $\|X\|_1=\operatorname{tr}(X)$ if and only if $\|X\|_1\leq\operatorname{tr}(X)$. Now by construction the last point from the above list is in turn equivalent to (iv) as $\tilde A-\tilde b_iD,\tilde B-\tilde b_iD$ equal $A-b_iD$, $B-b_iD$ for $i=1,2$ up to global positive constant.
\end{proof}

It may be possible to prove Prop.~\ref{thm_char} (iv) $\Rightarrow$ (i) by applying Prop.~\ref{prop_exist_channel} to $A-b_1D, B-b_1D$ and $A-b_2D,B-b_2D$, respectively, to get two trace-preserving maps $T_1,T_2$ mapping $B$ to $A$ and having $D$ as fixed point---because $B-tD$ is rank-deficient if and only if $t\in\sigma(D^{-1/2}BD^{-1/2})$---and the fidelity condition might ensure that at least one of these two is completely positive. However, even if this works then one would, most likely, end up with an argument rather close to \cite{AlbertiUhlmann80} so we save ourselves the bother. 

\begin{remark}
\begin{itemize}
\item[(i)] The characterizations from Prop.~\ref{thm_char} do not generalize to dimensions larger than $2$. To see this Heinosaari et al.~\cite{HeinosaariWolf12} gave a counterexample to the Alberti-Uhlmann theorem in higher dimensions which pertains to our case: Consider the Hermitian matrices
\begin{equation}\label{eq:counterex_heinosaari}
A=\begin{pmatrix} 2&1&0\\1&2&-i\\0&i&2 \end{pmatrix}\quad B=\begin{pmatrix} 2&1&0\\1&2&i\\0&-i&2 \end{pmatrix}\quad D=\begin{pmatrix} 2&1&0\\1&2&1\\0&1&2 \end{pmatrix}\,.
\end{equation}
Indeed $\sigma(D)=\{2,2+\sqrt{2},2-\sqrt{2}\}$ so $D>0$. Obviously $B^T=A$ and $D^T=D$ so because the transposition map is well-known to be linear, positive, and trace-preserving one has
$
\|A-tD\|_1=\|(B-tD)^T\|_1\leq \|B-tD\|_1
$
for all $t\in\mathbb R$ by Prop.~\ref{thm_q_norm_1}. But there exists no \textsc{cptp} map, i.e.~no $T\in Q(n)$, such that $T(B)=A$ and $T(D)=D$ as shown in \cite[Prop.~6]{HeinosaariWolf12}.
\item[(ii)] Usually a characterization of any generalized form of majorization via convex functions is much sought-after. A reasonable extension of Prop.~\ref{thm_char}
\textit{would} be that $A\prec_D B$ if and only if
\begin{equation}\label{eq:matrix_convex_cond}
\operatorname{tr}\big(D\psi(D^{-1/2}AD^{-1/2})\big)\leq \operatorname{tr}\big(D\psi(D^{-1/2}BD^{-1/2})\big)
\end{equation}
for all $\psi:\mathbb R\to\mathbb R$ matrix convex\footnote{
A matrix convex function (cf.~\cite{Kraus36,Bendat55,Ando79,Bhatia}) is a map $\psi:\mathbb R\to\mathbb R$ which acts on Hermitian matrices via the spectral theorem and then satisfies
$
\psi(\lambda A+(1-\lambda)B)\leq \lambda \psi(A)+(1-\lambda) \psi(B)$
for all $\lambda\in [0,1], A,B\in\mathbb C^{n\times n}$ Hermitian, and all $n\in\mathbb N$. Here $\leq$ is the partial ordering on the Hermitian matrices induced by positive semi-definiteness.
}; 
this is supported by the fact that if $A,B,D$ are all diagonal then \eqref{eq:matrix_convex_cond} reduces to the convex function-condition from the vector case (Prop.~\ref{lemma_char_d_vec} (ii)). One can even show that \eqref{eq:matrix_convex_cond} is necessary for some $T\in Q_D(n)$ to satisfy $T(B)=A$ \cite[Thm.~2.1]{Li96}. 
However, condition \eqref{eq:matrix_convex_cond} is also disproven by the matrices in \eqref{eq:counterex_heinosaari} in the same way as above for the following reason: Because $D=D^T$ one has
$$
\sigma(D^{-1/2}AD^{-1/2})=\sigma((D^{-1/2}AD^{-1/2})^T)=\sigma(D^{-1/2}A^TD^{-1/2})
$$
so no matrix convex $\psi$---as those act via functional calculus---can distinguish $D^{-1/2}AD^{-1/2}$ from $D^{-1/2}A^TD^{-1/2}$ which readily implies equality in \eqref{eq:matrix_convex_cond}.
\end{itemize}
\end{remark}
While there exist general conditions for the existence of a quantum channel which maps a finite input set of states to an output set of same cardinality \cite{Huang12} these are rather technical and not really applicable in practice. For now characterizing $\prec_D$ beyond two dimensions (via some easy-to-verify inequalities) remains an open problem.
\subsection{Order, Geometric, and Other Properties of $\prec_D$}\label{subsec:2}
There are two results we will present for which our analysis of strict positivity in Chapter \ref{sec:strictpos} was essential. The first one of these is almost immediate. Recall that, still, w.l.o.g.~$D=\operatorname{diag}(d)$ for some strictly positive vector $d\in\mathbb R_{++}^n$.
\begin{corollary}\label{thm_full_rank}
Let $d\in\mathbb R_{++}^n$ and $\rho,\omega\in\mathbb D(\mathbb C^{n})$ be given. If $\rho$ is of full rank and $\omega\prec_D\rho$, then $\omega$ is of full rank, as well.
\end{corollary}
\begin{proof}
By assumption there exists $T\in Q_D(n)$ such that $T(\rho)=\omega\in\mathbb D(\mathbb C^{n})$. Now \mbox{$T(D)=D>0$} by Prop.~\ref{prop_positivity} implies that $\rho>0$ is mapped to something positive definite (hence of full rank) again. 
\end{proof}
For the second connection we have to dive into order properties of $D$-majorization. Some simple observations: Just like in the vector case $\prec_D$ is a preorder but it is not a partial order. To see the latter---even if the eigenvalues of $D$ differ pairwise---consider the counterexample for $\prec_d$ given in \cite[Remark 2 (iv)]{vomEnde19polytope} which transfers onto $\prec_D$ via Coro.~\ref{coro_d_maj_diag}. Next let us investigate minimal and maximal elements of $\prec_D$ for which we need the following lemma.
\begin{lemma}\label{thm_channel_pure_max_state}
Let $d\in\mathbb R_{++}^n$, $\rho\in\mathbb D(\mathbb C^{n})$, and $j\in\lbrace 1,\ldots,n\rbrace$ be given. Then $\rho\prec_D |e_j\rangle\langle e_j|$ if and only if $D-d_j\rho\geq 0$.
\end{lemma}
\begin{proof}
``$\Rightarrow$'' : By definition there exists $T\in Q_D(n)$ such that $T( |e_j\rangle\langle e_j|)=\rho$. Note that $D-d_j |e_j\rangle\langle e_j|\geq 0$ as the l.h.s.~is a diagonal matrix with non-negative entries, so linearity and positivity of $T$ imply
\begin{align*}
0\leq T\big(D-d_j |e_j\rangle\langle e_j|\big)=T(D)-d_j T\big( |e_j\rangle\langle e_j|\big)=D-d_j\rho\,.
\end{align*}
\noindent ``$\Leftarrow$'' : The case $n=1$ is trivial so assume $n> 1$. As $D-d_j\rho\geq 0$ by assumption, 
$
\omega:=\frac{D-d_j\rho}{\unitvector^Td-d_j}\in\mathbb D(\mathbb C^{n})
$ satisfies $d_j\rho+(\unitvector^Td-d_j)\omega=D$. With this, define a linear map $T:\mathbb C^{n\times n}\to\mathbb C^{n\times n}$ via $T(| e_i\rangle\langle e_k|)=0$ whenever $i\neq k$ and
\begin{align*}
T( |e_i\rangle\langle e_i|)=\begin{cases} \omega &i\neq j\\ \rho &i=j \end{cases}
\end{align*}
as well as its linear extension to all of $\mathbb C^{n\times n}$. Now $T$ is trace-preserving and
\begin{align*}
T(D)=d_jT( |e_j\rangle\langle e_j|)+\sum\nolimits_{i=1,i\neq j}^n d_i T( |e_i\rangle\langle e_i|)=d_j\rho+(\unitvector^Td-d_j)\omega=D\,,
\end{align*}
as well as $T( |e_j\rangle\langle e_j|)=\rho$. For complete positivity, consider the Choi matrix
\begin{align*}
C(T)=\begin{pmatrix} T( |e_1\rangle\langle e_1|)&&0\\&\ddots&\\0&&T( |e_n\rangle\langle e_n|) \end{pmatrix}=\underbrace{\omega\oplus\ldots\oplus \omega}_{j-1\text{ times}}\oplus \rho \oplus\underbrace{\omega\oplus\ldots\oplus \omega}_{n-j\text{ times}}
\end{align*}
which is a block-diagonal matrix built from states so $C(T)\geq 0$ and thus $T$ is completely positive by Lemma \ref{lemma_choi_matrix}. Hence we constructed $T\in Q_D(n)$ with $T( |e_j\rangle\langle e_j|)=\rho$ which concludes the proof.
\end{proof}
\begin{remark}
For $d=\unitvector$ one has $D-d_j\rho=\mathbbm{1}-\rho\geq 0$ for all $\rho\in\mathbb D(\mathbb C^{n})$ so we recover the well-known result that every pure state is maximal in $\mathbb D(\mathbb C^{n})$ w.r.t.~$\prec\,$. 
This implies that if the rank of some $\rho\in\mathbb D(\mathbb C^{n})$ is larger than one then there exists no $\psi\in\mathbb C^n$ such that $ |\psi\rangle\langle\psi|\prec\rho$. For general $D$-majorization this fails: consider again the example from \cite[Remark 2 (iv)]{vomEnde19polytope} together with Coro.~\ref{coro_d_maj_diag}.
However there still is the weaker result that full rank is preserved under $\prec_D$ (Coro.~\ref{thm_full_rank}).
\end{remark}

With these tools at hand we, like in the vector case, can prove the existence of a minimal and maximal state with respect to $\prec_D$, and we can even characterize uniqueness: 
\begin{theorem}\label{d_matrix_minmax}
Let $d\in\mathbb R_{++}^n$ be given and let
\begin{align*}
\mathfrak h_d&:=\lbrace X\in\mathbb C^{n\times n}\,|\,\operatorname{tr}(X)=\unitvector^Td\rbrace\\
\mathfrak h_d^+&:=\lbrace X\in\mathbb C^{n\times n}\,|\,X\geq 0\text{ and }\operatorname{tr}(X)=\unitvector^Td\rbrace
\end{align*}
be the trace hyperplane induced by $d$ within the complex, and the positive semi-definite matrices, respectively. The following statements hold.
\begin{itemize}
\item[(i)] $D$ is the unique minimal element in $\mathfrak h_d$ with respect to $\prec_D$.
\item[(ii)] $(\unitvector^Td)|e_k\rangle\langle e_k|$ is maximal in $\mathfrak h_d^+$ with respect to $\prec_D$ where $k$ is chosen such that $d_k$ is minimal in $d$. It is the unique maximal element in $\mathfrak h_d^+$ with respect to $\prec_D$ if and only if $d_k$ is the unique minimal element of $d$.
\end{itemize}
\end{theorem}
\begin{proof}
(i): To see $D\prec_D A$ for arbitrary $A\in\mathfrak h_d$, consider $T:\mathbb C^{n\times n}\to \mathbb C^{n\times n}$, $X\mapsto D \operatorname{tr}(X)/\unitvector^Td$ which is in $Q(n)$ \cite[Ex.~5.3]{Hayashi06} and satisfies $T(D)=D=T(A)$. Uniqueness is evident as $D$ is a fixed point of every $T\in Q_D(n)$.

(ii): W.l.o.g.~$\unitvector^Td=1$ so $\mathfrak h_d^+=\mathbb D(\mathbb C^{n})$. Let arbitrary $\rho\in\mathbb D(\mathbb C^{n})$ be given. Because $d_k$ is the minimal eigenvalue of $D$ one finds $D-d_k\rho\geq 0$ due to
\begin{equation}\label{eq:D_pos_matrix}
\begin{split}
\langle x,(D-d_k\rho)x\rangle=\langle x,Dx\rangle-d_k\langle x,\rho x\rangle&\geq d_k\|x\|^2-d_k\|x\|^2\|\rho\|\\
&\geq d_k\|x\|^2(1-\|\rho\|_1)=0
\end{split}
\end{equation}
which holds for all $x\in\mathbb C^n$. Here we used $\|\rho\|_1=\operatorname{tr}(\rho)=1$ as $\rho\geq 0$. Now by Lemma \ref{thm_channel_pure_max_state} this implies $\rho\prec_D |e_k\rangle\langle e_k|$.

To prove uniqueness first assume that $d_k$ is the unique minimal element of $d$, and that $\omega\in\mathbb D(\mathbb C^{n})$ is also maximal w.r.t. $\prec_D$. Thus $ |e_k\rangle\langle e_k|\prec_D \omega$, that is, there exists $T\in Q_D(n)$ such that $T(\omega)= |e_k\rangle\langle e_k|$. We can diagonalize \mbox{$\omega=\sum_{i=1}^r w_i|g_i\rangle\langle g_i|$} with $w_1,\ldots, w_r> 0$, $\sum_{i=1}^r w_i=1$, and some orthonormal system $(g_i)_{i=1}^r$ in $\mathbb C^n$ where $r\in\lbrace 1,\ldots,n\rbrace$. Then
$$
 |e_k\rangle\langle e_k|=T(\omega)=\sum\nolimits_{i=1}^r w_i T(|g_i\rangle\langle g_i|)
$$
meaning we expressed a pure state as a convex combination of density matrices. But by Coro.~\ref{coro_states_comp} this forces $T(|g_i\rangle\langle g_i|)= |e_k\rangle\langle e_k|$ for all $i=1,\ldots,r$.
Now $D-d_k|g_i\rangle\langle g_i|\geq 0$ for all $i$ by \eqref{eq:D_pos_matrix}; actually this matrix is positive definite if and only if $g_i$ and $e_k$ are linearly independent if and only if\,\footnote{
While these equivalences are straightforward to check the main ingredients are the estimate 
$$
\langle x,Dx\rangle=\sum\nolimits_{i=1}^n d_i|\langle e_i,x\rangle|^2\geq d_k\sum\nolimits_{i=1}^n|\langle e_i,x\rangle|^2=d_k\|x\|^2
$$
for all $x\in\mathbb C^n$---with equality if and only if $x=\lambda e_k$ for some $\lambda\in\mathbb C$ because $d_k$ is the unique minimal entry of $d$---as well as the renowned fact that equality in the Cauchy-Schwarz inequality holds if and only if one vector is a multiple of the other.
}
$|g_i\rangle\langle g_i|\neq |e_k\rangle\langle e_k|$. However, $D-d_k|g_i\rangle\langle g_i|>0$ would imply $T(D-d_k|g_i\rangle\langle g_i|)>0$ by Prop.~\ref{prop_positivity}---due to $T(D)=D>0$---so
\begin{align*}
0<\langle e_k,T(D-d_k|g_i\rangle\langle g_i|)e_k\rangle&= \langle e_k,T(D)e_k\rangle-d_k\langle e_k,T(|g_i\rangle\langle g_i|)e_k\rangle\\
&= \langle e_k,De_k\rangle-d_k|\langle e_k,e_k\rangle|^2=0
\end{align*}
for all $i=1,\ldots,r$, an obvious contradiction. Hence $|g_i\rangle\langle g_i|= |e_k\rangle\langle e_k|=\omega$.

Finally, assume there exist $k,k'\in\lbrace1,\ldots,n\rbrace$ with $k\neq k'$ such that $d_k=d_{k'}$ is minimal in $d$. Then $ |e_k\rangle\langle e_k|$ and $|e_{k'}\rangle\langle e_{k'}|$ are both maximal with respect to $\prec_D$ by the same argument as above, hence no uniqueness. This concludes the proof.
\end{proof}
Note that strict positivity of $T\in Q_D(n)$ was the key in proving uniqueness of the maximal element of $\prec_D$, assuming the corresponding eigenvalue of $D$ is simple.
\begin{remark}
From a physical point of view this is precisely what one expects: from the state with the largest energy one can generate every other state (in an equilibrium-preserving manner) and there is no other state with this property.
\end{remark}

As described at the start of Chapter \ref{sec:matrix_d_maj}, just like in the vector case, considering the set of all matrices which are $D$-majorized by some $X\in\mathbb C^{n\times n}$ is of interest for analyzing reachable sets of certain quantum control problems. Therefore define\label{eq:def_MD}
\begin{equation*}
\begin{split}
M_D:\mathcal P(\mathbb C^{n\times n})&\to\mathcal P(\mathbb C^{n\times n})\\
S&\mapsto \bigcup\nolimits_{Y\in S}\lbrace X\in\mathbb C^{n\times n}\,|\,X\prec_D Y\rbrace
\end{split}
\end{equation*}
where $\mathcal P$ as usual denotes the power set. For convenience $M_D(X):=M_D(\{X\})$ for any \mbox{$X\in\mathbb C^{n\times n}$}. Then Prop.~\ref{thm_q_norm_1} as well as Remark \ref{rem_semigroup} lead to the following.
\begin{theorem}\label{lemma_R_d_closed}
Let $d\in\mathbb R_{++}^n$, $B\in\mathbb C^{n\times n}$, and a subset $P\subseteq \mathbb C^{n\times n}$ be given. The following statements hold.
\begin{itemize}
\item[(i)] $M_D(B)$ is convex for all $B\in\mathbb C^{n\times n}$.
\item[(ii)] $M_D$ as an operator on $\mathcal P(\mathbb C^{n\times n})$ 
is a closure operator (cf.~footnote \ref{footnote_closure_operator} on p.~\pageref{footnote_closure_operator}).
\item[(iii)] If $P$ is compact, then $ M_D(P)$ is compact.
\item[(iv)] If $A$ is an extreme point of $M_D(B)$ then there exists an extreme point $T$ of $Q_D(n)$ such that $T(B)=A$.
\end{itemize}
\end{theorem}
\begin{proof}
(i): Simple consequence of Lemma \ref{lemma_conv_d_maj_matrices}. (ii): Obviously, $M_D$ is extensive and increasing. For idempotence ($M_D\circ M_D=M_D$), ``$\,\subseteq\,$'' follows from $Q_D(n)$ forming a semigroup and ``$\,\supseteq\,$'' is due to $\mathbbm{1}_{n\times n}\in Q_D(n)$.
(iii): Proven just like Thm.~\ref{coro_0_1} (iii). 
(iv): Following Remark \ref{rem_semigroup} \mbox{$M_D(B)= \operatorname{conv}\lbrace T(B)\,|\, T\in Q^E_D(n) \rbrace$} so the statement in question follows from Minkowski's theorem\index{theorem!Minkowski} \cite[Thm.~5.10]{Brondsted83}, that is, the extreme points of $M_D(B)$ have to be contained within $\lbrace T(B)\,|\, T\in Q^E_D(n) \rbrace$.
\end{proof}
To discuss continuity of the map $M_D$ we first need a (relative) topology on the power set $\mathbb C^{n\times n}$. For this we, as before, consider the Hausdorff metric $\Delta$ on the set of all non-empty
compact subsets $\mathcal P_c(X)\subset\mathcal P(X)$ of a metric space $(X,d)$\label{symb_P_c_X}
(cf.~Appendix \ref{app_hausdorff}).
\begin{proposition}\label{prop_md_cont}
Let $d\in\mathbb R_{++}^n$. Then the map $M_D:\mathcal P_c(\mathbb C^{n\times n})\to \mathcal P_c(\mathbb C^{n\times n})$ is well-defined and non-expansive, that is,
$$
\Delta( M_D(P_1),M_D(P_2))\leq \Delta(P_1,P_2)
$$
for all $P_1,P_2\in\mathcal P_c(\mathbb C^{n\times n})$ when equipping $\mathbb C^{n\times n}$ with the trace norm. In particular $M_D$ is continuous.
\end{proposition}
\begin{proof}
Well-definedness is due to Thm.~\ref{lemma_R_d_closed} (iii). Now $M_D$ is non-expansive as
\begin{align*}
\max_{A_1\in M_D(P_1)}\min_{A_2\in M_D(P_2)}&\|A_1-A_2\|_1=\max_{\substack{T\in Q_D(n)\\B_1\in P_1}}\min_{\substack{S\in Q_D(n)\\B_2\in P_1}}\|T(B_1)-S(B_2)\|_1\\
&\leq \max_{\substack{T\in Q_D(n)\\B_1\in P_1}}\min_{B_2\in P_1}\|T(B_1)-T(B_2)\|_1\\
&\leq \max_{\substack{T\in Q_D(n)\\B_1\in P_1}}\min_{B_2\in P_1}\|T\|\|B_1-B_2\|_1=\max_{B_1\in P_1}\min_{B_2\in P_1}\|B_1-B_2\|_1
\end{align*}
where in the second-to-last step we used Prop.~\ref{thm_q_norm_1}.
\end{proof}

Finally, one finds the somewhat peculiar property that applying $M_D$ as well as $M_\mathbbm{1}$ (that is, classical matrix majorization) alternately to some initial state then, in the closure, one ends up with all states:
\begin{proposition}
Let $d\in\mathbb R_{++}^n$ such that $d$ and $\unitvector$ are linearly independent, i.e.~$d\neq c\unitvector$ for all $c\in\mathbb R_{++}$. Then for arbitrary $\rho\in\mathbb D(\mathbb C^n)$ 
$$
\lim_{m\to\infty}(M_{\mathbbm{1}}\circ M_D)^m(\rho)= \mathbb D(\mathbb C^n)
$$
with respect to the Hausdorff metric.
\end{proposition}
\begin{proof}
The case $n=1$ is obvious so consider $n>1$. Also w.l.o.g.~we may assume that $\unitvector^Td=1$ so $D\in\mathbb D(\mathbb C^n)$, else we can rescale the problem accordingly. First be aware that applying the Hausdorff metric is allowed due to the following facts:
\begin{itemize}
\item $\mathbb D(\mathbb C^n)$ is compact (Coro.~\ref{coro_states_comp})
\item $M_D$ for any $D>0$ maps non-empty compact sets to non-empty compact sets (Thm.~\ref{lemma_R_d_closed}) so 
$(M_{\mathbbm{1}}\circ M_D)^m(\rho)\subseteq \mathbb D(\mathbb C^n)$ itself is compact for all $m\in\mathbb N_0$.
\end{itemize}
Because $M$ is a closure operator---so in particular it is extensive---for any compact set $P\subseteq \mathbb D(\mathbb C^n)$ the sequence 
$\big((M_{\mathbbm{1}}\circ M_D)^m(P)\big)_{m\in\mathbb N}\subseteq \mathbb D(\mathbb C^n)$ 
is increasing with respect to $\subseteq$. Therefore \cite{Baronti86} implies that the sequence converges with respect to the Hausdorff metric $\Delta$ with compact limit set $\overline{\bigcup_{m=1}^\infty (M_{\mathbbm{1}}\circ M_D)^m(P)}\subseteq\mathbb D(\mathbb C^n)$.

The idea will be the following: First we show by explicit construction that starting from $D=\operatorname{diag}(d)\in\mathbb D(\mathbb C^n)$ we can approximately reach $ |e_1\rangle\langle e_1|$. We then may use extensiveness as well as continuity of $M$ w.r.t.~$\Delta$ to get
\begin{align*}
\mathbb D(\mathbb C^n)\supseteq \lim_{m\to\infty}(M_{\mathbbm{1}}\circ M_D)^m(\rho)&=\lim_{m\to\infty}\big(M_{\mathbbm{1}}\circ M_D\circ(M_{\mathbbm{1}}\circ M_D)^{m-2}\circ M_{\mathbbm{1}}\circ M_D\big)(\rho)\\
&= (M_{\mathbbm{1}}\circ M_D)\big(\lim_{\tilde m\to\infty} (M_{\mathbbm{1}}\circ M_D)^{\tilde m}( M_{\mathbbm{1}}\circ M_D)(\rho)\big)\\
&\supseteq (M_{\mathbbm{1}}\circ M_D)\big(\lim_{\tilde m\to\infty} (M_{\mathbbm{1}}\circ M_D)^{\tilde m}(D)\big)\\
&\supseteq (M_{\mathbbm{1}}\circ M_D)( |e_1\rangle\langle e_1|)\supseteq M_{\mathbbm{1}} ( |e_1\rangle\langle e_1|)=\mathbb D(\mathbb C^n)
\end{align*}
because $D$ is minimal in $\mathbb D(\mathbb C^n)$ w.r.t.~$\prec_D$ and 
every pure state is maximal in $\mathbb D(\mathbb C^n)$ w.r.t.~$\prec_{\mathbbm{1}}$ (that is, $\prec$) by Thm.~\ref{d_matrix_minmax}. Also in the second-to-last row we made use of Lemma \ref{lemma_5} (a). This would conclude the proof.

Carrying out this idea, by assumption we find $j\in\lbrace 1,\ldots,n-1\rbrace$ such that $d_j\neq d_{j+1}$ (w.l.o.g.~$d_j> d_{j+1}$, the other case is shown analogously). Define the $d$-stochastic matrix
\begin{align*}
T:=\begin{pmatrix} \mathbbm{1}_{j-1}&0&0&0\\0&1-\frac{d_{j+1}}{d_j}&1&0\\0&\frac{d_{j+1}}{d_j}&0&0\\0&0&0&\mathbbm{1}_{n-j-1} \end{pmatrix}\in\mathbb R^{n\times n}
\end{align*}
and let $\sigma_r=\sum_{i=1}^n |e_{i+1}\rangle\langle e_i|$ (with $e_{n+1}:=e_1$) be the cyclic right shift which in particular is doubly stochastic. Starting from any $x\in\mathbb R_{++}^n$ with $\unitvector^Tx=1$ one computes
\begin{equation}\label{eq:x_1_rec}
x^{(1)}:=\sigma_r^{n-j+1}T(\sigma_rT)^{n-2}\sigma_r^j\mathbbm{1} x=\begin{pmatrix} 1-\frac{d_{j+1}}{d_j}(1-x_1)\\\frac{d_{j+1}}{d_j}x_2\\\cdots\\\frac{d_{j+1}}{d_j}x_n \end{pmatrix}=\Big(1-\frac{d_{j+1}}{d_j}\Big)e_1+\frac{d_{j+1}}{d_j}x
\end{equation}
where $x^{(1)}\in\mathbb R_{++}^n$ and, using Coro.~\ref{coro_d_maj_diag}, $\operatorname{diag}{x^{(1)}}\in (M_{\mathbbm{1}}\circ M_D)^n(\operatorname{diag} x)$. Applying this step successively $\alpha\in\mathbb N$ times results in
$$
x^{(\alpha+1)}:=\big(\sigma_r^{n-j+1}T(\sigma_rT)^{n-2}\sigma_r^j\mathbbm{1}\big) x^{(\alpha)}=\Big(1-\Big(\frac{d_{j+1}}{d_j}\Big)^{\alpha+1}\Big)e_1+\Big(\frac{d_{j+1}}{d_j}\Big)^{\alpha+1}x
$$
as is evident by induction invoking \eqref{eq:x_1_rec}. Due to $\operatorname{diag}{x^{(\alpha)}}\in (M_{\mathbbm{1}}\circ M_D)^{n\alpha}(\operatorname{diag} x)$ for all $\alpha$ we found a sequence in $\big((M_{\mathbbm{1}}\circ M_D)^{n\alpha}(\operatorname{diag} x)\big)_{\alpha\in\mathbb N}$ which converges to $ |e_1\rangle\langle e_1|$:
\begin{align*}
\big\| |e_1\rangle\langle e_1|-\operatorname{diag}{x^{(\alpha)}}\big\|_1&=\Big\| |e_1\rangle\langle e_1|-\Big(1-\Big(\frac{d_{j+1}}{d_j}\Big)^{\alpha}\Big) |e_1\rangle\langle e_1|-\Big(\frac{d_{j+1}}{d_j}\Big)^{\alpha}\operatorname{diag} x\Big\|_1\\
&= \Big(\frac{d_{j+1}}{d_j}\Big)^{\alpha}\| |e_1\rangle\langle e_1|-\operatorname{diag} x\|_1\overset{\alpha\to\infty}\longrightarrow 0
\end{align*}
Thus by the limit point characterization of Hausdorff convergence (cf.~Lemma \ref{lem:Hausdorff})
\begin{equation*}
|e_1\rangle\langle e_1|\in \lim_{\alpha\to\infty} (M_{\mathbbm{1}}\circ M_D)^{n\alpha}(D)=\lim_{\tilde m\to\infty} (M_{\mathbbm{1}}\circ M_D)^{\tilde m}(D)
\end{equation*}
which as argued above concludes the proof.
\end{proof}
\noindent This result is non-trivial in the following sense: If the initial $\rho$ is of full rank then by Coro.~\ref{thm_full_rank} $(M_1\circ M_D)^m(\rho)$ for arbitrary $m$ can never equal all of $\mathbb D(\mathbb C^n)$, but has to be a proper subset.

\section{$C$-Numerical Range in Infinite Dimensions}\label{sec:c_num_range}

The $C$-numerical range has significant impact on quantum control and quantum information theory 
since the expression $\operatorname{tr}(\rho A)$ can be interpreted as the expectation
value of an observable $A$ with respect to the state $\rho$, that is, as the expectation value of a measurement $A$
taken on a quantum system in state $\rho$. 
While in standard quantum mechanics $A$ is self-adjoint and $\rho$ is a (trace-class) density operator, 
there are in fact important applications where $A$ or $\rho$ (or both) are allowed to be non-self-adjoint. Maximizing the absolute value \cite{NEUM-37} or the real part of
$\operatorname{tr}(\rho U^* AU)$ over the unitary orbit of $A$ relates to different optimization problems in the Euclidean geometry of the $C$-numerical range
\cite{SDHG08,SGDH08}.

In the finite-dimensional case, where $A$ and $C$ are assumed to be complex $n \times n$ matrices,
the $C$-numerical range of $A$ is defined by
\begin{align}\label{c_num_range_findim}
W_C(A)=\lbrace\operatorname{tr}(CU^* AU)\,|\,U\in\mathbb C^{n\times n}\text{ unitary}\rbrace\,.
\end{align}
Originally, it was introduced in \cite{Goldberg77} as a generalization of the $c$-numerical
range \cite{Westwick75} and the classical numerical range \cite{Hausdorff19,Toeplitz18}. Important
properties of the $C$-numerical range are convexity if $C$ is normal with collinear eigenvalues
\cite{Westwick75,Poon80}, and star-shapedness with respect to 
$(\operatorname{tr}(C)\operatorname{tr}(A)/n)$ for arbitrary complex $C$, cf.~\cite{TSING-96}. 
For a comprehensive survey, we refer to \cite{Li94}.
Now given a complex Hilbert space $\mathcal H$ define
the $C$-numerical range\index{C-numerical range@$C$-numerical range} $W_C (T)$ of a linear operator $T$ on $\mathcal H$ as follows:
\begin{definition}\label{defi_1}
For any $C\in\mathcal B^1(\mathcal H),T\in\mathcal B(\mathcal H)$
\begin{align*}
W_C (T):=\lbrace \operatorname{tr}(CU^* TU)\,|\,U\in\mathcal B(\mathcal H)\text{ unitary}\rbrace\,.
\end{align*}
This definition pertains to the case $C\in\mathcal B^p(\mathcal H)$, $T\in\mathcal B^q(\mathcal H)$ with $p,q\in [1,\infty]$ conjugate, that is, $\frac1p+\frac1q=1$. 
\end{definition}
Clearly, this is a generalization of the finite-dimensional case. Here we take advantage of the fact that
the set of all trace-class operators is a two-sided ideal in the $C^*$-algebra $\mathcal B(\mathcal H)$.
In this setting, however, symmetry in $C$ and $T$ is lost unless $C,T\in\mathcal B^2(\mathcal H)$.

The goal is to carry over star-shapedness 
or convexity of $W_C(A)$ to the infinite-dimensional setting.~Interim results on this subject were achieved by Westwick \cite{Westwick75} and Hughes \cite{Hughes90} for the $c$-numerical range and by Jones \cite{Jones92} for the
$C$-numerical range. Jones, however, pursued a different approach in \cite{Jones92}.
For $C\in\mathbb C^{k\times k}$ and $T\in\mathcal B(\mathcal H)$ he introduced the set
\begin{align}\label{eq:jones_1}
\Big\lbrace\sum\nolimits_{i,j=1}^kc_{ij}\langle f_j,Tf_i\rangle\,\Big|\,\lbrace f_1,\ldots,f_k\rbrace\text{ is orthonormal system in }\mathcal H\Big\rbrace
\end{align}
as the $C$-numerical range of $T$, where $\mathcal H$ can be any infinite-dimensional complex Hilbert
space, and proved that its closure is star-shaped. In doing so,
the essential numerical range $W_e(T)$,
or more precisely, the set $\operatorname{tr}(C)W_e(T)$ turned out to be an appropriate replacement of the finite
dimensional star-center $(\operatorname{tr}(C)\operatorname{tr}(A)/n)$. The definition and basic properties of $W_e(T)$ are given in \cite{Bonsall73}, refer also to Prop.~\ref{prop_w_e}.

Throughout this section we need some formalism to associate matrices with bounded operators on a separable Hilbert space $\mathcal H$ 
and vice versa. In doing so, let $(e_n)_{n\in\mathbb N} $ be some orthonormal basis of $\mathcal H$ and let
$(\hat e_i)_{i=1}^n$ be the standard basis of $\mathbb C^n$. For any $n\in\mathbb N$ we define 
\begin{align}\label{Gamma}
\Gamma_n:\mathbb C^n\to \mathcal H,\qquad \hat{e_i}\mapsto \Gamma_n(\hat e_i):=e_i
\end{align}
and its linear extension to all of $\mathbb C^n$. Now let 
\begin{align*}
E_n:\mathbb C^{n\times n}\to\mathcal B(\mathcal H),\qquad A\mapsto E_n(A):=\Gamma_n A\Gamma_n^*
\end{align*}
be the embedding of $\mathbb C^{n\times n}$\label{symb_embedding}
into $B(\mathcal H)$ relative to the basis $(e_n)_{n\in\mathbb N} $
and let
\begin{align}\label{cut_out_operator}
[\;\cdot\;]_n:\mathcal B(\mathcal H)\to\mathbb C^{n\times n},\qquad A\mapsto [A]_n:=\Gamma_n^* A\Gamma_n
\end{align}
be the operator which ``cuts out'' the upper $n\times n$ block of (the matrix representation of) $A$ 
with respect to $(e_n)_{n\in\mathbb N} $. 

\begin{remark}\label{rem_jones_connect}
Obviously, $W_{E_n(C)}(T)$ co{\"i}ncides with \eqref{eq:jones_1} for all $C\in\mathbb C^{n\times n}$
and $T\in\mathcal B(\mathcal H)$, where $E_n$ is the embedding operator with respect to any orthonormal basis of $\mathcal H$. Thus Definition \ref{defi_1} actually generalizes Jones' approach \cite{Jones92} who, in our words, considered only finite-rank operators $C\in\mathcal F(\mathcal H)$.
\end{remark}

The following lemma which will be needed later is a trivial consequence of the standard trace 
identity for operators acting on the \emph{same} Hilbert space.

\begin{lemma}\label{embedding_trace_preserv}
Let $n\in\mathbb N$, $A\in\mathbb C^{n\times n}$, $B\in\mathcal B(\mathcal H)$, and any orthonormal bases $(e_n)_{n\in\mathbb N}$, $(g_n)_{n\in\mathbb N}$ of $\mathcal H$ be given. Then
\begin{align*}
\operatorname{tr}\big((\Gamma^g_n)^* B\Gamma_n^eA\big)=\operatorname{tr}\big(B\Gamma_n^eA(\Gamma_n^g)^*\big)
\end{align*}
where $\Gamma_n^e$ ($\Gamma_n^g$) is the above embedding $\Gamma_n$ with respect to $(e_n)_{n\in\mathbb N}$ ($(g_n)_{n\in\mathbb N}$).
\end{lemma}

\begin{proof}
Consider the operators 
\begin{align*}
\begin{pmatrix}
B & 0 \\ 0 & 0
\end{pmatrix},
\qquad
\begin{pmatrix}
0 & \Gamma_n^e \\ (\Gamma_n^g)^* & 0
\end{pmatrix},
\quad\text{and}\quad
\begin{pmatrix}
0 & 0 \\ 0 & A
\end{pmatrix}
\end{align*}
acting on $\mathcal H \times \mathbb C^n$ and use the standard cyclicity result of the trace.
\end{proof}

\subsection{The Bounded Case}\label{ch_bounded12938}
For this section---which is based on our article \cite{DvE18}---let $\mathcal H$ be an infinite-dimensional, separable, and complex Hilbert space. 
Our strategy is to transfer the well-known properties of the finite-dimensional $[C]_n$-numerical range of 
$[T]_n$ to $W_C(T)$ via the convergence results from Lemma \ref{lemma_5}. Let $B\in\mathcal B(\mathcal H)$ and let $(e_n)_{n\in\mathbb N}$ be an orthonormal basis of $\mathcal H$. 
For any $k\in\mathbb N$ we define the $k$-th block approximation of $B$ with respect to $(e_n)_{n\in\mathbb N}$ to be
\begin{align}\label{eq:defi_block}
B_k:=\Pi_kB\Pi_k\,,\quad\text{where}\quad \Pi_k := \sum\nolimits_{j=1}^k|e_j\rangle\langle e_j|=\Gamma_k\Gamma_k^*
\end{align}
is the orthogonal projection onto $\operatorname{span}\lbrace e_1,\ldots, e_k\rbrace$. Thus one has
$
B_k = \sum_{i,j=1}^k\langle e_i, Be_j\rangle|e_i\rangle\langle e_j|
$.

\begin{lemma}\label{strong_tr_conv}
Let $(S_n)_{n\in\mathbb N}$ be a sequence in $\mathcal B(\mathcal H)$ which converges to 
$S\in\mathcal B(\mathcal H)$ in the strong operator topology. Then for all $C\in\mathcal B^1(\mathcal H)$ and $T\in\mathcal B(\mathcal H)$
one has
\begin{align*}
\lim_{n\to\infty}\operatorname{tr}(CS_n^* TS_n)=\operatorname{tr}(CS^* TS)\,.
\end{align*}
Furthermore, 
\begin{itemize}
\item the sequence of linear functionals $\big(\operatorname{tr}(CS_n^*(\cdot)S_n)\big)_{n\in\mathbb N}$ 
converges uniformly to $\operatorname{tr}(CS^* (\cdot)S)$ on bounded subsets of $\mathcal B(\mathcal H)$.
\item the sequence of linear functionals $\big(\operatorname{tr}((\cdot)S_n^* TS_n)\big)_{n\in\mathbb N}$
converges uniformly to $\operatorname{tr}((\cdot)S^* TS)$ on compact subsets of $\mathcal B^1(\mathcal H)$.
\end{itemize}
If $T$ additionally is compact, then $\big(\operatorname{tr}((\cdot)S_n^* TS_n)\big)_{n\in\mathbb N}$
converges uniformly to $\operatorname{tr}((\cdot)S^* TS)$ on (trace norm-) bounded subsets of $\mathcal B^1(\mathcal H)$.
\end{lemma}

\begin{proof}
This is a simple consequence Lemma \ref{lemma_schatten_prop_pq} \& \ref{lemma_schatten_p_approx} as
\begin{align*}
|\operatorname{tr}(CS^* TS) -\operatorname{tr}(CS_n^* TS_n)| = 
\big|\operatorname{tr}\big((SCS^*-S_nCS_n^*)T\big)\big|\leq\Vert T\Vert \|SCS^*-S_nCS_n^*\|_1 \to 0
\end{align*}
for $n\to\infty$. The remaining assertions of the lemma are evident.
\end{proof}

\begin{remark}
Note that for arbitrary bounded operators $T$, $\operatorname{tr}((\cdot)S_n^* TS_n)$ does not necessarily converge uniformly to
$\operatorname{tr}((\cdot)S^* TS)$ on (trace norm-) bounded subsets of $\mathcal B^1(\mathcal H)$. 
A counter-example is given in Appendix \ref{app_c_num_range} (Ex. \ref{ex_1}).
\end{remark}

\begin{lemma}\label{U-approximation}
Let $U \in\mathcal B(\mathcal H)$ be unitary and consider orthonormal bases $(e_n)_{n \in \mathbb N}$, $(g_n)_{n \in \mathbb N}$ of $\mathcal H$. Then there exists a
sequence $(\hat{U}_n)_{n \in \mathbb N}$ in $\mathcal B(\mathcal H)$ which satisfies the following:
\begin{itemize}
\item[(i)]
$(\hat{U}_n)_{n \in \mathbb N}$ converges strongly to $U$.\medskip
\item[(ii)]
$\Pi^g_{2n}\hat{U}_n\Pi^e_{2n}=\hat U_n$ for all $n \in \mathbb N$.\medskip
\item[(iii)]
$(\Gamma_{2n}^g)^* \hat{U}_n \Gamma_{2n}^e\in\mathbb C^{2n\times 2n}$ is unitary for all $n \in \mathbb N$.
\end{itemize}
Here, $\Gamma_k^e$, $\Pi_{k}^e$ and $\Gamma_k^g$, $\Pi_k^g$ are the maps given by \eqref{Gamma} and \eqref{eq:defi_block} with respect to $(e_n)_{n\in\mathbb N}$ and $(g_n)_{n\in\mathbb N}$, respectively.
\end{lemma}
\noindent As the proof of Lemma \ref{U-approximation} is rather technical we refer to Appendix \ref{app_c_num_range}.

\begin{lemma}\label{lemma_3}
Let $C\in\mathcal B^1(\mathcal H)$ and $T\in\mathcal B(\mathcal H)$, and let $(e_n)_{n\in\mathbb N}$, $(g_n)_{n\in\mathbb N}$ be
arbitrary orthonormal bases of $\mathcal H$. Furthermore, $[\,\cdot\,]_k^e$ and $[\,\cdot\,]_k^g$ are the maps given by \eqref{cut_out_operator} with respect to $(e_n)_{n\in\mathbb N}$ and $(g_n)_{n\in\mathbb N}$, respectively. Then for all $\varepsilon>0$ and $w\in \overline{W_C(T)}$, there exists $N\in\mathbb N$ such that the distance $d(w,W_{[C]^e_{n}}([T]^g_{n}))<\varepsilon$ for all $n\geq N$.
\end{lemma}
\begin{proof}
Let $\varepsilon>0$ as well as $w \in \overline{W_C(T)}$ be given. Then there exists unitary $U\in\mathcal B(\mathcal H)$ such that
$|w-\operatorname{tr}(CU^* TU)|<\varepsilon/2$. By Lemma \ref{U-approximation}, we can find a sequence $(\hat{U}_n)_{n \in \mathbb N}$ which converges strongly to $U$. Lemma \ref{strong_tr_conv}
then yields $N\in\mathbb N$ such that
\begin{align*}
|\operatorname{tr}(CU^* TU)-\operatorname{tr}(C\hat U_n^* T\hat U_n)|<\frac{\varepsilon}{2}
\end{align*}
for all $n\geq N$. Using Lemma \ref{embedding_trace_preserv} and \ref{U-approximation}, one gets
\begin{align*}
\operatorname{tr}(C\hat U_n^* T\hat U_n)&= \operatorname{tr}\big( C(\Pi^g_{2n}\hat U_n\Pi^e_{2n})^* T(\Pi^g_{2n}\hat U_n\Pi^e_{2n})\big)\\
&=\operatorname{tr}\big([C]^e_{2n}((\Gamma_{2n}^g)^* \hat{U}_n \Gamma_{2n}^e)^* [T]^g_{2n}(\Gamma_{2n}^g)^* \hat{U}_n \Gamma_{2n}^e\big)\in W_{[C]^e_{n}}([T]^g_{n})\,.
\end{align*}
Thus $|w-\operatorname{tr}(C\hat U_n^* T\hat U_n)|<\varepsilon$ for all $n\geq N$, which concludes the proof as the $C$-numerical 
range of any pair of matrices is compact \cite[(2.5)]{Li94}. 
\end{proof}

\noindent Note that in the above proof, $N$ depends usually on $\varepsilon$ as well as the chosen point
$w \in \overline{W_C(T)}$.

\begin{theorem}\label{lemma_2}
Let $C\in\mathcal B^1(\mathcal H)$ and $T\in\mathcal B(\mathcal H)$, and let $(e_n)_{n\in\mathbb N}$, $(g_n)_{n\in\mathbb N}$ be
arbitrary orthonormal bases of $\mathcal H$. Furthermore, $[\,\cdot\,]_k^e$ and $[\,\cdot\,]_k^g$ for all $k\in\mathbb N$ are the maps \eqref{cut_out_operator} with respect to $(e_n)_{n\in\mathbb N}$ and $(g_n)_{n\in\mathbb N}$, respectively. Given a sequence $(C_n)_{n\in\mathbb N}\subseteq\mathcal B^1(\mathcal H)$ with $\lim_{n\to\infty}\|C_n-C\|_1=0$ one finds
\begin{align*}
\lim_{n\to\infty}W_{[C]^e_{2n}}([T]^g_{2n})=\overline{W_C(T)}=\lim_{n\to\infty}\overline{W_{C_n}(T)}\,,
\end{align*}
where $W_{[C]^e_{2n}}([T]^g_{2n})$ denotes the ordinary $[C]^e_{2n}$-numerical range of $[T]^g_{2n}$ as defined in (\ref{c_num_range_findim}). If $T$ is compact and there exists a sequence $(T_n)_{n\in\mathbb N}\subseteq\mathcal K(\mathcal H)$ with $\lim_{n\to\infty}\|T_n-T\|=0$, then
\begin{align}\label{eq:lemma_2_2}
\lim_{n\to\infty}\overline{W_{C_n}(T_n)}=\overline{W_C(T)}\,.
\end{align}
\end{theorem}

\begin{proof}
W.l.o.g.~let $C_n,T_n\neq 0$ for some $n\in\mathbb N$---else all the involved sets would be trivial---so we may introduce the positive but (as seen via the reverse triangle inequality) finite numbers
$$
\kappa := \sup\{\|C\|_1,\|C_1\|_1,\|C_2\|_1,\ldots\} 
\quad\text{ and }\quad \tau :=\sup\{\|T\|,\|T_1\|,\|T_2\|,\ldots\}\,.
$$
As we want to check convergence with respect to the Hausdorff metric, we have to make sure that all occurring sets are non-empty and compact. The non-empty sets $\overline{W_{C_n}(T)}$, $\overline{W_{C_n}(T_n)}$ are bounded due to
$|\operatorname{tr}(C_nU^* T_nU)| \leq \|C_n\|_1 \Vert T_n\Vert \leq \kappa\tau$ and thus all of them are compact. Again, the $C$-numerical 
range of any pair of matrices is also compact \cite[(2.5)]{Li94}.

First, we prove the equality
$
\lim_{n\to\infty}W_{[C]^e_{2n}}([T]^g_{2n})=\overline{W_C(T)}
$ for which we w.l.o.g.~assume $C,T\neq 0$.
In view of Lemma \ref{lemma_11}, we have to consider two cases:

Let $\varepsilon>0$. Then due to compactness, there exist finitely many $w_1,\ldots,w_L\in \overline{W_C(T)}$ such that
\begin{align*}
\bigcup_{k=1}^L B_{\varepsilon/2}(w_k)\supset \overline{W_C(T)}\,.
\end{align*}
By Lemma \ref{lemma_3}, each of these $w_k$ admits $N_k\in\mathbb N$ such that $d(w_k,W_{[C]^e_{n}}([T]^g_{n}))<\varepsilon/2$ for all $n\geq N_k$. Define $N':=\max\lbrace N_1,\ldots,N_L\rbrace$. Now for any $w\in \overline{W_C(T)}$, there exists $k\in\lbrace 1,\ldots,L\rbrace$ such that $|w-w_k|<\varepsilon/2$ and thus
\begin{align*}
d(w,W_{[C]^e_{n}}([T]^g_{n}))\leq |w-w_k|+d(w_k,W_{[C]^e_{n}}([T]^g_{n}))<\varepsilon
\end{align*}
for all $n\geq N'$.

On the other hand, for $G_{2n} := \sum_{k=2n+1}^\infty |g_k\rangle\langle e_k|\in\mathcal B(\mathcal H)$ it is easy to see that $(G_{2n})_{n\in\mathbb N}$ converges strongly to the zero operator. By Lemma \ref{lemma_schatten_p_approx} we obtain $N''\in\mathbb N$
such that
\begin{align*}
\max\lbrace\|CG_{2n}\|_1,\|G_{2n}C\|_1\rbrace<\frac{\varepsilon}{3\|T\|}
\end{align*}
for all $n\geq N''$. Now let $v_n\in W_{[C]^e_{2n}}([T]^g_{2n})$, i.e.~there exists unitary $U_n\in\mathbb C^{2n\times 2n}$
such that $v_n = \operatorname{tr}([C]_{2n}^e U_n^* [T]_{2n}^g U_n )$. Again Lemma \ref{embedding_trace_preserv} yields
$v_n = \operatorname{tr}\big(C( \Gamma_{2n}^gU_n(\Gamma_{2n}^e)^*)^* T \Gamma_{2n}^gU_n(\Gamma_{2n}^e)^*\big)$. Next, we define the operator
\begin{align*}
\tilde U_n := \Gamma_{2n}^gU_n(\Gamma_{2n}^e)^* + G_{2n} \in\mathcal B(\mathcal H)
\end{align*}
with $G_{2n} $ given as above. It is readily verified that $\tilde U_n$ is unitary (cf.~Coro.~\ref{coro_unitary_approx_block_unitary}) and, therefore, we conclude
$\tilde{v}_n := \operatorname{tr}(C\tilde U^*_n T\tilde U_n)\in W_C(T)$.
Via Lemma \ref{lemma_schatten_prop_pq} we finally obtain
\begin{align*}
|v_n-\tilde{v}_n|
&=|\operatorname{tr}\big(CG_{2n}T\Gamma_{2n}^gU_n(\Gamma_{2n}^e)^*\big)+\operatorname{tr}\big(C(\Gamma_{2n}^gU_n(\Gamma_{2n}^e)^*)^* TG_{2n}\big)
+\operatorname{tr}(CG_{2n}TG_{2n})|\\
&\leq \big(\|CG_{2n}\|_1+\|G_{2n}C\|_1+\|CG_{2n}\|_1\big)\Vert T\Vert<\varepsilon
\end{align*}
which yields $d(v_n,\overline{W_C(T)})<\varepsilon$ for all $n\geq N''$. Thus, choosing $N:=\max\lbrace N',N''\rbrace$, Lemma \ref{lemma_11} implies $\Delta(W_{[C]^e_n}([T]^g_n),\overline{W_C(T)})<\varepsilon$ for all $n\geq N$.

Next, we tackle the equality
$
 \lim_{n\to\infty}\overline{W_{C_n}(T)}=\overline{W_C(T)}\,.
$
Let $\varepsilon>0$ be given, and w.l.o.g.~$T\neq 0$. By assumption there exists $\hat N\in\mathbb N$ such that
\begin{equation}\label{eq:C_n_approx}
\|C-C_n\|_1<\frac{\varepsilon}{2\Vert T\Vert}
\end{equation}
for all $n\geq\hat N$. For $w\in \overline{W_C(T)}$, there again exists unitary
$U\in\mathcal B(\mathcal H)$ such that $w' := \operatorname{tr}(CU^* TU) \in W_C(T)$ satisfies 
$|w-w'|<\varepsilon/2$. Thus, for $w_n:=\operatorname{tr}(C_nU^* TU)\in W_{C_n}(T)$ one has
\begin{align*}
|w-w_n|\leq |w-w'|+|w'-w_n|<
\frac{\varepsilon}{2}+\|C-C_n\|_1\Vert U^* TU\Vert < \varepsilon
\end{align*}
for all $n\geq N$.

On the other hand, let $v_n\in \overline{W_{C_n}(T)}$, i.e.~there exists unitary 
$U_n \in \mathcal B(\mathcal H)$ such that $v'_n := \operatorname{tr}(C_n U_n^* T U_n)$
satisfies $|v_n - v_n'| < \varepsilon/2$. Moreover, for $\tilde{v}_n := \operatorname{tr}(C U_n^* T U_n) \in W_{C}(T)$,
we obtain
\begin{align*}
|v_n-\tilde{v}_n| \leq |v_n-v'_n| + |v'_n-\tilde{v}_n|
< \frac{\varepsilon}{2} + \|C-C_n\|_1\Vert U_n^* T U_n\Vert < \varepsilon
\end{align*}
for all $n\geq N$. Again, Lemma \ref{lemma_11} implies $\lim_{n\to\infty}\overline{W_{C_n}(T)} = \overline{W_C(T)}$. 

Finally, let $T$ additionally be compact and let $\varepsilon>0$ be given. By assumption there exists $\tilde N\in\mathbb N$ such that
$
\|T-T_n\|<\frac{\varepsilon}{4\kappa}
$
for all $n\geq\hat N$ (and adjust the constant in \eqref{eq:C_n_approx} to $\frac{\varepsilon}{4\|T\|}$). As
\begin{align*}
|\operatorname{tr}(CU^* TU) &- \operatorname{tr}(C_nU^* T_nU)| \leq |\operatorname{tr}(CU^* TU)-\operatorname{tr}(C_nU^* TU)|\\
&+ |\operatorname{tr}(C_nU^* TU)-\operatorname{tr}(C_nU^* T_nU)|\,,
\end{align*}
one can choose $N:=\max\lbrace\hat N,\tilde N\rbrace$ to obtain $\Delta(\overline{W_{C_n}(T_n)},\overline{W_C(T)})<\varepsilon$ for all $n\geq N$ as above. 
\end{proof}
Reformulating (part of) the last theorem we showed that $W:\mathcal B^1(\mathcal H)\times \mathcal K(\mathcal H)\to (\mathcal P_c(\mathbb C),\Delta)$, $(C,T)\mapsto\overline{W_C(T)}$ is continuous where the domain of $W$ is again a Banach space under the norm $\|(C,T)\|=\|C\|_1+\|T\|$. 

\begin{remark}
In general, (\ref{eq:lemma_2_2}) does not hold for arbitrary bounded operators $T$ since---even if the limit exists---one has only the inclusion 
$\overline{W_C(T)} \subseteq \lim_{n\to\infty}\overline{W_{C_n}(T_n)}$ as the above
proof shows. A simple example 
which demonstrates this failing is given by Example \ref{ex_2} in Appendix \ref{app_c_num_range}. 
\end{remark}

With this we are prepared to state and prove our first main result of this section.

\begin{theorem}\label{theorem_1a}
Let $C\in\mathcal B^1(\mathcal H)$ and $T\in\mathcal B(\mathcal H)$ be given. If $C$ is normal
with collinear eigenvalues or if $T$ is essentially self-adjoint, then $\overline{W_C(T)}$ is convex.
\end{theorem}

Recall, that a set in the complex plane is said to be \emph{collinear}\index{collinear} if all of its elements lie on a common line.
Moreover, like in the matrix case \cite{Marcus79} an operator $T \in \mathcal B(\mathcal H)$ is called \emph{essentially self-adjoint}\index{operator!essentially self-adjoint (bounded)} if there exist $\theta\in\mathbb R$ and $\xi\in\mathbb C$ such 
that $e^{-i\theta} (T-\xi\mathbbm{1}_{\mathcal H})$ is self-adjoint.

\begin{proof}
First, assume that $C$ is normal with collinear eigenvalues so, as $C$ is compact (because it is trace class) Thm.~\ref{thm_compact_normal_unit_diag} states that
there exists an orthonormal basis $(e_n)_{n\in\mathbb N}$ of $\mathcal H$ such that\footnote{Note 
that $(\lambda_n(C))_{n\in\mathbb N}$ is the \textit{modified} eigenvalue sequence of $C$ as described
at the beginning of Section \ref{sect_C_spectrum}.}
$C=\sum_{n=1}^\infty\lambda_n(C)|e_n\rangle\langle e_n|$. By assumption, the eigenvalues
$\lambda_n(C)$ are collinear and $\lambda_n(C)\to 0$ as $n\to\infty$ since $C$ is compact. This implies the 
existence of $\theta\in\mathbb R$ such that $e^{i\theta}\lambda_n(C)\in\mathbb R$ for all $n\in\mathbb N$
and thus $e^{i\theta}C$ is self-adjoint (Lemma \ref{lemma_spectrum}).
By Thm.~\ref{lemma_2}
\begin{align*}
\overline{W_{C}(T)}=\overline{W_{e^{i\theta}C}(e^{-i\theta}T)}=\lim_{n\to\infty}W_{[e^{i\theta}C]_{2n}}([e^{-i\theta}T]_{2n})
\end{align*}
where $[\,\cdot\,]_{2n}$ for all $n\in\mathbb N$ are the maps \eqref{cut_out_operator} with respect to $(e_n)_{n\in\mathbb N}$. Evidently, $[B]_n^*=[B^*]_n$ for all $B\in\mathcal B(\mathcal H)$ 
and all $n\in\mathbb N$. Therefore, $[e^{i\theta}C]_{2n}$ is Hermitian and thus 
$W_{[e^{i\theta}C]_{2n}}([e^{-i\theta}T]_{2n})$ is convex for all $n\in\mathbb N$, cf.~\cite{Poon80}, meaning Lemma \ref{lemma_5} (iii) yields the desired result. The case of $T$ being essentially self-adjoint
can be handled completely along the same line as then
\begin{align*}
W_C(T)= e^{i\theta} W_C(H)+\xi\operatorname{tr}(C)
\end{align*}
where $H:=e^{-i\theta} (T-\xi\mathbbm{1}_{\mathcal H})$ is self-adjoint by definition.
\end{proof}
For now it is an open question whether $W_C(T)$ remains convex after the closure is waived. Special cases where one can answer in the affirmative include $C,T$ both being self-adjoint (cf.~also End of Ch.~\ref{sec:vonn_tr_ineq}), $C$ being a normal finite-rank operator with collinear eigenvalues (\cite[Thm.~3]{Hughes90} \& Rem.~\ref{rem_jones_connect}), and, very recently, $T$ being diagonalizable and $C\geq 0$ having either trivial or infinite-dimensional kernel \cite[Coro.~7.3]{Loreaux20}.
\begin{remark}
Unlike in finite dimensions---where $W_C(T)$ can be further located via the $C$-spectrum of $T$---it 
is intricate to obtain a similar result for infinite dimensions because there does not exist a 
meaningful counterpart of the $C$-spectrum for arbitrary bounded operators. However, if $T$ is compact
one can in fact define the $C$-spectrum of $T$ and generalize well-known properties of the matrix case, 
see Section \ref{sect_C_spectrum}.
\end{remark}

Before proceeding with the star-shapedness of $\overline{W_C(T)}$, we briefly recall the
definition\footnote{Some authors prefer a different definition which, however, is equivalent
to the stated one, cf.~\cite[Thm.~34.9]{Bonsall73}.} of 
the \emph{essential numerical range}\index{essential numerical range} $W_e(T)$ of an operator $T \in \mathcal B(\mathcal H)$,\label{symb_W_e_T}
which can be given as follows:
\begin{align*}
W_e(T) := \Big\lbrace \lim_{n \to \infty} \langle f_n,Tf_n\rangle\,\Big|\, (f_n)_{n \in \mathbb N}\;\text{ is ONS in }\mathcal H\Big\rbrace \subset \mathbb C
\end{align*}
It is well known that $W_e(T)$ is a non-empty, convex, and compact
subset of $\mathbb C$ \cite[Thm.~34.2]{Bonsall73}.

\begin{proposition}\label{prop_w_e}
Let $T\in\mathcal B(\mathcal H)$ and $\mu\in\mathbb C$ be given. The following are equivalent.
\begin{itemize}
\item[(i)] $\mu$ belongs to the essential numerical range $W_e(T)$, that is, there exists
an orthonormal system $(f_n)_{n \in \mathbb N}$ in $\mathcal H$ such that $\lim_{n \to \infty} \langle f_n,Tf_n\rangle = \mu$.\medskip
\item[(ii)] There exists an orthonormal system $(f_n)_{n\in\mathbb N}$ in $\mathcal H$ such that
\begin{align}\label{eq:theorem_app_0_1_1}
\lim_{n\to\infty}\frac{1}{n}\sum\nolimits_{j=1}^n\langle f_j,Tf_j\rangle=\mu\,.
\end{align}
\item[(iii)] There exists an orthonormal basis $(e_n)_{n\in\mathbb N}$ of $\mathcal H$ such that
\begin{align}\label{eq:theorem_app_0_1_2}
\lim_{n\to\infty}\frac{1}{n}\sum\nolimits_{j=1}^n\langle e_j,Te_j\rangle=\mu\,.
\end{align}
\end{itemize}
\end{proposition}

\begin{proof}
(i) $\Rightarrow$ (ii): It is well known that the limit of a convergent sequence and the limit of its Ces\`aro mean are equal.
(ii) $\Rightarrow$ (i): Consider any orthonormal system $(f_n)_{n\in\mathbb N}$ which satisfies 
(\ref{eq:theorem_app_0_1_1}). We will show
\begin{align}\label{eq:HP}
\mu \in \overline{\operatorname{conv}\big\lbrace \operatorname{HP} \big((\langle f_n,Tf_n\rangle)_{n\in\mathbb N}\big)\big\rbrace} =: E\,,
\end{align}
where $\operatorname{HP}(\cdot)$ denotes the set of all accumulation points of the respective sequence.
Once \eqref{eq:HP} is guaranteed we can conclude $\mu \in W_e(T)$ because the convexity and compactness
of $W_e(T)$ readily implies $E \subseteq W_e(T)$. Let us assume $\mu\notin E$. Since $E$ is obviously convex and compact, there exists a $\mathbb C$-linear 
functional $\varphi:\mathbb C\to\mathbb C$ with
\begin{align*}
\operatorname{Re}(\varphi(\mu)) < \min_{\lambda \in E}\operatorname{Re}(\varphi(\lambda))\,,
\end{align*}
cf.~\cite[Thm.~3.4]{Rudin91}. Taking into account that the sequence $(\langle f_n,Tf_n\rangle)_{n\in\mathbb N}$ is bounded (because $T$ is bounded),
a straightforward application of the Bolzano-Weierstra{\ss} theorem\index{theorem!Bolzano-Weierstra{\ss}} shows that there exist only finitely 
many indices $n_1 < n_2 < \ldots < n_k \in \mathbb N$ such that 
\begin{align*}
\operatorname{Re}\big(\varphi(\langle f_{n_j},Tf_{n_j}\rangle)\big) \leq 
\frac{1}{2}\Big(\min_{\lambda \in E}\operatorname{Re}(\varphi(\lambda))+\operatorname{Re}(\varphi(\mu))\Big) =: \kappa
\end{align*}
for all $j \in\lbrace 1,\ldots,k\rbrace$. This yields the following contradiction:
\begin{align*}
\operatorname{Re}(\varphi(\mu))&=\operatorname{Re}\Big(\varphi\Big( \lim_{n\to\infty}\frac{1}{n}\sum\nolimits_{j=1}^n\langle f_j,Tf_j\rangle \Big)\Big)\\
& = \operatorname{Re}\Big(\varphi\Big( \lim_{n\to\infty}\frac{1}{n}\sum\nolimits_{j=1}^{n_k}\langle f_j,Tf_j\rangle \Big)\Big) + \operatorname{Re}\Big(\varphi\Big( \lim_{n\to\infty}\frac{1}{n}\sum\nolimits_{j=n_k+1}^n\langle f_j,Tf_j\rangle \Big)\Big)\\
&=\lim_{n\to\infty}\frac{1}{n}\sum\nolimits_{j=n_k+1}^n\operatorname{Re}\big(\varphi(\langle f_j,Tf_j\rangle)\big)\geq 
\lim_{n\to\infty}\frac{\kappa (n - n_k)}{n} > \operatorname{Re}(\varphi(\mu))
\end{align*}
Hence $\mu\in E$.

(iii) $\Rightarrow$ (ii): $\checkmark$
(ii) $\Rightarrow$ (iii): Let $(f_n)_{n\in\mathbb N}$ be an orthonormal system in $\mathcal H$ such that
\eqref{eq:theorem_app_0_1_1} holds which we then extend to an orthonormal basis of $\mathcal H$. If,
in this procedure, we have to add only finitely many vectors (or none) we are obviously done. 
Therefore, we assume in the remaining part of the proof that we have to add countably infinitely
many vectors $(g_n)_{n\in\mathbb N}$. This allows us to define a new orthonormal basis $(e_n)_{n\in\mathbb N}$
by sorting $(g_n)_{n\in\mathbb N}$ into $(f_n)_{n\in\mathbb N}$ as follows: 
For $n = 2^k$ with $k\in\mathbb N$ choose $e_{n}=g_k$, while the gaps in between are filled up with 
the vectors of $(f_n)_{n\in\mathbb N}$, i.e.
\begin{align*}
(e_n)_{n\in\mathbb N}=(f_1,g_1,f_2,g_2,f_3,f_4,f_5,g_3,f_6, \ldots)\,.
\end{align*}
In doing so, for $2^k \leq n < 2^{k+1}$ we obtain the following identity
\begin{align*}
\frac1n\sum\nolimits_{j=1}^n\langle e_j,Te_j\rangle=\Big(1-\frac{k}{n}\Big)\bigg(\frac{1}{n-k}\sum\nolimits_{j=1}^{n-k}\langle f_j,Tf_j\rangle\bigg)+ \frac{1}{n}\sum\nolimits_{j=1}^k\langle g_j,Tg_j\rangle\,.
\end{align*}
Obviously, $ \frac{k}{n}\to 0$ as $k \to \infty$ so 
\begin{align*}
\lim_{k\to\infty}\Big|\frac{1}{n}\sum\nolimits_{j=1}^k\langle g_j,Tg_j\rangle\Big|
\leq\lim_{k\to\infty} \frac{k}{n}\Vert T\Vert=0
\end{align*}
and we conclude
\begin{align*}
\lim_{n \to \infty}\frac1n\sum\nolimits_{j=1}^n\langle e_j,Te_j\rangle = 
\lim_{n \to \infty}\frac{1}{n-k}\sum\nolimits_{j=1}^{n-k}\langle f_j,Tf_j\rangle = \mu
\end{align*}
as this is just a subsequence of \eqref{eq:theorem_app_0_1_1}.
\end{proof}

After these preliminaries, our second main result of this section reads as follows.

\begin{theorem}\label{theorem_1}
Let $C\in\mathcal B^1(\mathcal H)$ and $T\in\mathcal B(\mathcal H)$ be given. Then $\overline{W_C(T)}$
is star-shaped with respect to $\operatorname{tr}(C)W_e(T)$, that is, all $z\in\operatorname{tr}(C)W_e(T)$ are star-centers of $\overline{W_C(T)}$.
\end{theorem}

\begin{proof}
Let any $\mu\in W_e(T)$. By Prop.~\ref{prop_w_e} there exists an orthonormal basis $(e_n)_{n\in\mathbb N}$ of $\mathcal H$ such that (\ref{eq:theorem_app_0_1_2}) holds. Moreover, note that
\begin{align*}
\langle\hat e_j,[T]_{2n}\hat e_j\rangle=\langle\Gamma_{2n}\hat e_j,T\,\Gamma_{2n}\hat e_j\rangle=\langle e_j,Te_j\rangle
\end{align*}
for all $n\in\mathbb N$ and $j \in \lbrace 1,\ldots,2n\rbrace$, where $[\,\cdot\,]_n$ is the map given by \eqref{cut_out_operator} with respect to $(e_n)_{n\in\mathbb N}$. Hence
\begin{align*}
\lim_{n\to\infty} \frac{\operatorname{tr}([T]_{2n})}{2n}=\lim_{n\to\infty}\frac{1}{2n}\sum\nolimits_{j=1}^{2n}\langle e_j,Te_j\rangle=\mu\,.
\end{align*}
Additionally, by Lemma \ref{lemma_schatten_p_approx} \& \ref{embedding_trace_preserv} we find
\begin{align*}
\lim_{n\to\infty}|\operatorname{tr}(C)-\operatorname{tr}([C]_{2n})|=\lim_{n\to\infty} |\operatorname{tr}(C-C\Pi_{2n})|\leq \lim_{n\to\infty} \|C-C\Pi_{2n}\|_1=0.
\end{align*}
This shows $\operatorname{tr}([C]_{2n})\operatorname{tr}([T]_{2n})/(2n) \to \operatorname{tr}(C)\mu$ for
$n\to\infty$. On the other hand, $W_{[C]_{2n}}([T]_{2n})$ is star-shaped with respect to 
$\operatorname{tr}([C]_{2n})\operatorname{tr}([T]_{2n})/(2n)$ for all $n\in\mathbb N$,
cf.~\cite[Thm.~4]{TSING-96}. This means that the sequence of star-centers
converges to $\operatorname{tr}(C)\mu$, thus Lemma \ref{lemma_5} (iv) \& Thm.~\ref{lemma_2} imply that $\overline{W_{C}(T)}$ is star-shaped with respect to 
$\operatorname{tr}(C)\mu$. As $\mu\in W_e(T)$ was chosen arbitrarily, the proof is complete.
\end{proof}

\begin{remark}
In finite dimensions, Tsing \cite{Tsing81} showed that for normal $C\in\mathbb C^{n\times n}$ and arbitrary $A\in\mathbb C^{n\times n}$, $W_C(A)$ is star-shaped with respect to $(\operatorname{tr}(C)\operatorname{tr}(A))/n$. Nine years later Hughes \cite{Hughes90} proved, in our words, that $\overline{W_C(T)}$ is star-shaped with respect to $\operatorname{tr}(C)W_e(T)$ for all normal $C\in\mathcal F(\mathcal H)$ and all $T\in\mathcal B(\mathcal H)$. This was generalized to arbitrary $C\in\mathcal F(\mathcal H)$ by Jones \cite{Jones92} and in finite dimensions to arbitrary $C\in\mathbb C^{n\times n}$ by Cheung and Tsing \cite{TSING-96}. 

However, none of the authors provided a satisfying link between the star-center in finite dimensions
and the set of star-centers in infinite dimensions. The above proof as well as characterization (iii) of Prop.~\ref{prop_w_e}, which is new to our knowledge, now 
clearly suggest that the set $\operatorname{tr}(C)W_e(T)$ is a natural replacement of
$(\operatorname{tr}(C)\operatorname{tr}(A))/n$ in infinite dimensions.
\end{remark}

\subsection{The General Schatten Case}\label{sect_C_spectrum}
As before, $\mathcal H$ throughout this section is an infinite-dimensional separable complex Hilbert space. The following results can be found in our papers \cite{DvE18,DvE18_Schatten} and \cite[Appendix A]{vE_dirr_vonNeumann}.

The $C$-spectrum is a powerful tool in order to gain further knowledge about the $C$-numerical range,
which was first introduced for matrices in \cite{Marcus79}. We want to transfer this concept
and some of the known results to infinite dimensions.
In order to define the $C$-spectrum, we first have to fix the term \emph{eigenvalue sequence} of a \index{eigenvalue sequence!modified}
compact operator $T \in \mathcal K(\mathcal H)$ from Prop.~\ref{prop_compact_spectrum}. \label{symb_modified_ev_seq}
%
\begin{itemize}
\item 
If the image of $T$ is infinite-dimensional and the kernel of $T$ finite-dimensional
then put $\operatorname{dim}(\operatorname{ker}T)$ zeros at the beginning of the eigenvalue
sequence of $T$.
\item 
If the image and the kernel of $T$ are infinite-dimensional, mix infinitely many zeros into the
eigenvalue sequence\footnote{Since in Definition \ref{defi_3} arbitrary permutations
will be applied to the modified eigenvalue sequence, we need not specify this mixing 
procedure further, cf. also Lemma \ref{Lemma_5b}.} of $T$.
\item
If the image of $T$ is finite-dimensional, leave the eigenvalue sequence of $T$ unchanged. 
\end{itemize}

\begin{definition}[$C$-spectrum]\label{defi_3}\index{C-spectrum@$C$-spectrum}
Let $p,q\in[1,\infty]$ be conjugate. Then, for $C\in\mathcal B^p(\mathcal H)$ with modified
eigenvalue sequence $(\lambda_n(C))_{n\in\mathbb N}$ and $T\in\mathcal B^q(\mathcal H)$ with modified
eigenvalue sequence $(\lambda_n(T))_{n\in\mathbb N}$, the $C$-\emph{spectrum of} $T$ is defined via
\begin{align*}
P_C(T):=\Big\lbrace \sum\nolimits_{n=1}^\infty \lambda_n(C)\lambda_{{\pi}(n)}(T) \,\Big|\, 
{\pi}:\mathbb N \to\mathbb N \text{ is any permutation}\Big\rbrace.
\end{align*}
\end{definition}

\noindent H\"older's inequality (Lemma \ref{lemma_hoelders_ineq}) and the estimate
$\sum\nolimits_{n=1}^\infty |\lambda_n(C)|^p \leq \sum\nolimits_{n=1}^\infty s_n(C)^p$ \cite[Prop.~16.31]{MeiseVogt97en} yield
$$
\sum\nolimits_{n=1}^\infty |\lambda_n(C)\lambda_{{\pi}(n)}(T)|
\leq \Big(\sum\nolimits_{n=1}^\infty s_n(C)^p \Big)^{1/p}\Big(\sum\nolimits_{n=1}^\infty s_n(T)^q\Big)^{1/q}
=\|C\|_p\|T\|_q\,,
$$
showing that the elements of $P_C(T)$ are well-defined and bounded by $\|C\|_p\|T\|_q$.

A survey regarding the $C$-spectrum of a matrix can be found in \cite[Ch. 6]{Li94}. Recall that compact normal operators have a spectral decomposition of the form
$
T = \sum\nolimits_{n=1}^\infty \lambda_{n}(T) |f_n\rangle\langle f_n|
$
where $(f_n)_{n \in \mathbb N}$ is an orthonormal basis of $\mathcal H$ and $(\lambda_n(T))_{n \in \mathbb N}$
denotes the modified eigenvalue sequence of $T$ as defined above (cf.~Thm.~\ref{thm_compact_normal_unit_diag}).
If an operator is normal but not compact, we saw that there still is a spectral decomposition but, in general,
the above (finite or infinite) sum has to be replaced by a spectral integral \cite[Thm.~5.21]{Schmuedgen12} which makes the definition
of its $C$-spectrum quite delicate. Therefore, we will restrict our considerations	 to the compact case.\medskip

The following result is fundamental for it describes that the \textit{closure of the} $C$-spectrum is not affected by switching from the original to the modified eigenvalue sequence:

\begin{lemma}\label{Lemma_5b}
Let arbitrary sequences $a\in\ell^1(\mathbb N),b\in c_0(\mathbb N)$, or $a\in c_0(\mathbb N),b\in \ell^1(\mathbb N)$, or $a\in\ell^p(\mathbb N),b\in \ell^q(\mathbb N)$ with $p,q\in(1,\infty)$ conjugate be given. Moreover, let $(a'_n)_{n\in\mathbb N}$, $(b'_n)_{n\in\mathbb N}$ be sequences which differ from $(a_n)_{n\in\mathbb N}$, $(b_n)_{n\in\mathbb N}$ only by a finite or infinite
number of zeros; more precisely, for each $\alpha\neq 0$ one has
\begin{align}\label{eq:Lemma_5b_1}
|\{k \in \mathbb N \,|\, a_k = \alpha\}| = |\{k \in \mathbb N \,|\, a'_k = \alpha\}|
\end{align}
and similarly for $(b_n)_{n\in\mathbb N}$ and $(b'_n)_{n\in\mathbb N}$. Then the closures of the following
two sets co{\"i}ncide:
\begin{align*}
A&:=\Big\lbrace \sum\nolimits_{n=1}^\infty a_nb_{{\pi}(n)} \,\Big|\, {\pi}:\mathbb N \to\mathbb N \text{ is permutation}\Big\rbrace\\
and \quad\quad\quad& \\
A'&:=\Big\lbrace \sum\nolimits_{n=1}^\infty a'_nb'_{{\pi}(n)} \,\Big|\, {\pi}:\mathbb N \to\mathbb N \text{ is permutation}\Big\rbrace
\end{align*}
\end{lemma}

\noindent
For a proof of Lemma \ref{Lemma_5b} we refer to Appendix \ref{app_c_num_range}.
\begin{lemma}\label{lemma_6b}
Let $C\in\mathcal B^p(\mathcal H)$ and $T\in\mathcal B^q(\mathcal H)$ be both normal with $p,q\in[1,\infty]$ conjugate. Then for all $\varepsilon>0$ and $w\in \overline{P_C(T)}$ there exists $N\in\mathbb N$ such that the distance $d(w,P_{[C]^e_n}([T]^g_n))<\varepsilon$ for all $n\geq N$. Here, $[\,\cdot\,]_n^e$ and $[\,\cdot\,]_n^g$ are the maps given by (\ref{cut_out_operator}) with 
respect to the orthonormal bases $(e_n)_{n\in\mathbb N}$ and $(g_n)_{n\in\mathbb N}$ which diagonalize $C$ and
$T$, respectively.
\end{lemma}
\begin{proof}
We prove the case $p,q\in(1,\infty)$. The boundary cases $(p,q)=(1,\infty),(\infty,1)$ are shown analogously. Let $\varepsilon>0$ and $w\in\overline{P_C(T)}$ be given. There exists a permutation 
${\pi}: \mathbb N \to \mathbb N$ with
\begin{align*}
\Big| w-\sum\nolimits_{j=1}^\infty \lambda_j(C) \lambda_{{\pi}(j)}(T) \Big|<\frac{\varepsilon}{2}\,.
\end{align*}
Moreover there exists $N' \in\mathbb N$
such that
\begin{align*}
\sum\nolimits_{j=N'+1}^\infty |\lambda_j(C)|^p < \frac{\varepsilon^p}{4 \|T\|_q^p}\,.
\end{align*}
Here we used the fact that the non-vanishing singular values of a compact normal operator co{\"i}ncide
with the absolute values of its non-zero eigenvalues. Next, we define
\begin{align*}
N := \max_{1 \leq j \leq N'} {\pi}(j)\,.
\end{align*}
Note $N \geq N'$. Hence we can choose a permutation ${\pi}':\mathbb N \to \mathbb N$ such that
${\pi}'$ restricted to $\{1,\dots,N'\}$ co{\"i}ncides with ${\pi}$, and ${\pi}'(j) := j$ for $j > N$.
Then $w_n := \sum\nolimits_{j=1}^n \lambda_j(C)\lambda_{{\pi}'(j)}(T)$ belongs to $P_{[C]^e_n}([T]^g_n)$ for all $n \geq N$ as $\lbrace{\pi}'(1),\ldots,{\pi}'(n)\rbrace=\lbrace1,\ldots,n\rbrace$
and we get
\begin{align*}
| w-w_n| & \leq \Big| w- \sum\nolimits_{j=1}^\infty \lambda_j(C) \lambda_{{\pi}(j)}(T) \Big| + 
\Big| \sum\nolimits_{j=1}^\infty \lambda_j(C) \lambda_{{\pi}(j)}(T) - w_n\Big|\\
&< \frac{\varepsilon}{2} + \sum\nolimits_{j=N'+1}^\infty |\lambda_j(C)| \, |\lambda_{{\pi}(j)}(T)|
+ \sum\nolimits_{j=N'+1}^n |\lambda_j(C)| \, |\lambda_{{\pi}'(j)}(T)|\\
&< \frac{\varepsilon}{2} + 2\Big(\sum\nolimits_{j=N'+1}^\infty |\lambda_j(C)|^p\Big)^{1/p}\Big(\sum\nolimits_{j=N'+1}^\infty |\lambda_{j}(T)|^q\Big)^{1/q}
<\frac{\varepsilon}{2}+2\cdot\frac{\varepsilon}{4}=\varepsilon\,.\qedhere
\end{align*}
\end{proof}

\noindent Note that in the above proof, $N$ depends on $\varepsilon$ as well as $w$.

\begin{proposition}\label{lemma_6}
Let $C\in\mathcal B^p(\mathcal H)$, $T\in\mathcal B^q(\mathcal H)$ with $p,q\in [1,\infty]$ conjugate be given. Furthermore, let $(e_n)_{n\in\mathbb N}$ and $(g_n)_{n\in\mathbb N}$ be arbitrary orthonormal bases of $\mathcal H$. Then
\begin{align*}
\lim_{n\to\infty}W_{[C]^e_{2n}}([T]^g_{2n})=\overline{W_C(T)}
\end{align*}
where $[\,\cdot\,]_k^e$ and $[\,\cdot\,]_k^g$ are the maps given by \eqref{cut_out_operator} with respect to $(e_n)_{n\in\mathbb N}$ and $(g_n)_{n\in\mathbb N}$, respectively. Moreover, if $C$ are $T$ both are normal then
\begin{align*}
\lim_{n\to\infty}P_{[C]^e_n}([T]^g_n)= \overline{P_C(T)}\,.
\end{align*}
where $(e_n)_{n\in\mathbb N}$ and $(g_n)_{n\in\mathbb N}$ are the orthonormal bases of $\mathcal H$ which diagonalize $C$ and
$T$, respectively.
\end{proposition}
\begin{proof}
The first statement for $p=1,q=\infty$ (or vice versa) was shown in Thm.~\ref{lemma_2} and can be adjusted 
to $p,q\in(1,\infty)$ by minimal modifications.

Now for the second statement. Again, in order to apply the Hausdorff metric we have to check that all sets occurring in Prop.~\ref{lemma_6} are non-empty and compact. But this
is obviously the case as all $P_{[C]^e_n}([T]^g_n)$ are non-empty and finite, and $\overline{P_C(T)}$ is non-empty,
closed, and bounded by $\|C\|_p\|T\|_q$.

Let $(\lambda_j(C))_{j\in\mathbb N}$ and $(\lambda_{j}(T))_{j\in\mathbb N}$ denote the modified eigenvalue sequences of 
$C$ and $T$, respectively. Obviously, for arbitrary $n\in\mathbb N$, the eigenvalues of $[C]^e_n$ and $[T]^g_n$
are given by $\lbrace \lambda_1(C),\ldots,\lambda_n(C)\rbrace$ and $\lbrace \lambda_{1}(T),\ldots,\lambda_{n}(T)\rbrace$.
W.l.o.g.~$T \neq 0$. Let $\varepsilon>0$. Due to compactness, there exist finitely many $w_1,\ldots,w_L\in\overline{P_C(T)}$ such that
\begin{align*}
\bigcup_{k=1}^L B_{\varepsilon/2}(w_k)\supset\overline{P_C(T)}\,.
\end{align*}
By Lemma
\ref{lemma_6b}, each of these $w_k$ admits $N_k\in\mathbb N$ such that $d(w_k,P_{[C]^e_n}([T]^g_n))<\varepsilon/2$ for all $n\geq N_k$. Define $N':=\max\lbrace N_1,\ldots,N_L\rbrace$. Now for any $w\in \overline{P_C(T)}$, there exists $k\in\lbrace 1,\ldots,L\rbrace$ such that $|w-w_k|<\varepsilon/2$ and thus
\begin{align*}
d(w,P_{[C]^e_{n}}([T]^g_{n}))\leq |w-w_k|+d(w_k,P_{[C]^e_{n}}([T]^g_{n}))<\varepsilon
\end{align*}
for all $n\geq N'$.

Conversely, as in the previous proof there exists $N''$ such that 
such that
\begin{align*}
\sum\nolimits_{j=N''+1}^\infty |\lambda_j(C)|^p < \frac{\varepsilon^p}{\|T\|_q^p}\,.
\end{align*}
Let $v_n \in P_{[C]^e_n}([T]^g_n)$ so there exists a permutation
${\pi}_n \in S_n$ such that 
$
v_n=\sum\nolimits_{j=1}^n \lambda_j(C)\lambda_{{\pi}_n(j)}(T)
$. 
Obviously, we can extend ${\pi}_n$ to a permutation $\tilde{\pi}_n:\mathbb N\to\mathbb N$ via
\begin{align*}
\tilde{\pi}_n(j):=\begin{cases}{\pi}_n(j)&1\leq j\leq n\,,\\ j&j>n\,.\end{cases}
\end{align*}
Then for $\tilde v_n := \sum\nolimits_{j=1}^\infty\lambda_j(C)\lambda_{\tilde{\pi}_n(j)}(T) \in P_C(T) \subseteq\overline{P_C(T)}$ one by H\"older's inequality finds
\begin{align*}
|v_n-\tilde v_n| &= \Big|\sum\nolimits_{j=1}^n \lambda_j(C)\lambda_{{\pi}_n(j)}(T) - \sum\nolimits_{j=1}^\infty\lambda_j(C)\lambda_{\tilde{\pi}_n(j)}(T) \Big|\\
& = \Big|\sum_{j=n+1}^\infty\lambda_j(C)\lambda_{j}(T) \Big| \leq \sum_{j=N+1}^\infty|\lambda_j(C)||\lambda_{j}(T)| \leq 
\Vert T\Vert_q \Big(\sum_{j=N+1}^\infty|\lambda_j(C)|^p\Big)^{1/p} <\varepsilon
\end{align*}
which yields $d(v_n,\overline{P_C(T)})<\varepsilon$ for all $n\geq N''$. Thus, choosing $N:=\max\lbrace N',N''\rbrace$, Lemma \ref{lemma_11} lets us conclude $\Delta(P_{[C]^e_n}([T]^g_n),\overline{P_C(T)})<\varepsilon$ for all $n\geq N$.
\end{proof}
After obtaining a convergence result for the $C$-spectrum we are ready to connect it to the $C$-numerical range. Recall that for matrices $A,C\in\mathbb C^{n\times n}$ one has $P_C(A)\subseteq W_C(A)$ if only $A$ or $C$ is normal \cite[Eq.(4)]{Marcus79}---this can be easily seen via Schur's triangularization theorem \cite[Thm.~2.3.1]{HJ1}---and that $
W_C(A)=\operatorname{conv}(P_C(A))$
whenever $A$ and $C$ are both normal and the eigenvalues of $C$ form a collinear set in the complex plane. A generalization of this result to infinite dimensions reads as follows:

\begin{theorem}\label{theorem_3}
Let $C\in\mathcal B^p(\mathcal H)$ and $T\in\mathcal B^q(\mathcal H)$ with $p,q\in [1,\infty]$ 
conjugate be given. Then the following statements hold.
\begin{itemize}
\item[(i)] $\overline{W_C(T)}$ is star-shaped with respect to the origin.
\item[(ii)] If either $C$ or $T$ is normal with collinear eigenvalues, then $\overline{W_C(T)}$ is convex.
\item[(iii)] If $C$ and $T$ both are normal, then $P_C(T)\subseteq W_C(T)\subseteq\operatorname{conv}(\overline{P_C(T)})$. If, in addition, the eigenvalues of $C$ or $T$ are collinear then $\overline{W_C(T)}=\operatorname{conv}(\overline{P_C(T)})$.
\end{itemize}
\end{theorem}
First we need two auxiliary
results to characterize the star-center of $\overline{W_C(T)}$ in the Schatten case.

\begin{lemma}\label{lemma_2b}
Let $T\in\mathcal K(\mathcal H)$ and $(e_k)_{k\in\mathbb N}$ be any orthonormal system in $\mathcal H$. Then
\begin{itemize}
\item[(i)]
$
\displaystyle
\sum\nolimits_{k=1}^n|\langle e_k,Te_k\rangle|
\leq\sum\nolimits_{k=1}^n s_k(T)
$
for all $n\in\mathbb N$ and
\item[(ii)]
$\lim_{k \to \infty} \langle e_k,Te_k\rangle = 0\,.$
\end{itemize}
\end{lemma}

\begin{proof}
(i) Consider a Schmidt decomposition $\sum\nolimits_{m=1}^\infty s_m(T)|g_m\rangle\langle f_m|$ of $T$ so
\begin{align*}
\sum\nolimits_{k=1}^n|\langle e_k,Te_k\rangle|\leq \sum\nolimits_{m=1}^\infty s_m(T)\Big(\sum\nolimits_{k=1}^n |\langle e_k,f_m\rangle\langle g_m,e_k\rangle|\Big)\,.
\end{align*}
Defining $\xi_m:= \sum\nolimits_{k=1}^n |\langle e_k,f_m\rangle\langle g_m,e_k\rangle| $ for all $m\in\mathbb N$, using Cauchy-Schwarz and Bessel's inequality one finds
\begin{align*}
\xi_m\leq \Big(\sum\nolimits_{k=1}^n |\langle e_k,f_m\rangle|^2\Big)^{1/2}\Big(\sum\nolimits_{k=1}^n |\langle g_m,e_k\rangle|^2\Big)^{1/2}\leq 1
\end{align*}
for all $m\in\mathbb N$. On the other hand, said inequalities also imply
\begin{align*}
\sum\nolimits_{m=1}^\infty \xi_m&\leq \sum\nolimits_{k=1}^n \Big(\sum\nolimits_{m=1}^\infty |\langle e_k,f_m\rangle|^2\Big)^{1/2}\Big(\sum\nolimits_{m=1}^\infty |\langle g_m,e_k\rangle|^2\Big)^{1/2}\leq \sum\nolimits_{k=1}^n \|e_k \|^2=n\,.
\end{align*}
Hence, because $(s_m(T))_{m \in \mathbb N}$ is decreasing by construction, an upper bound of 
$\sum\nolimits_{m=1}^\infty s_m(T)\xi_m$ is obtained by choosing $\xi_1=\ldots=\xi_n=1$
and $\xi_j=0$ whenever $j>n$. This shows the 
desired inequality. A proof of (ii) can be found, e.g., in \cite[Lemma 16.17]{MeiseVogt97en}.
\end{proof}

\begin{lemma}\label{lemma_0_conv}
Let $C\in\mathcal B^p(\mathcal H)$ with $p\in( 1,\infty]$, and let $q\in[1,\infty)$ be given such that $p,q$ are conjugate. Also let $(e_n)_{n\in\mathbb N}$ be any orthonormal system in
$\mathcal H$. Then
\begin{align*}
\lim_{n\to\infty}\frac{1}{n^{1/q}}\sum\nolimits_{k=1}^n\langle e_k,Ce_k\rangle=0\,.
\end{align*}
\end{lemma}
\begin{proof}
First, let $p=\infty$, so $q=1$. As $C$ is compact, by Lemma \ref{lemma_2b} one has
$\lim_{k\to\infty}\langle e_k,Ce_k\rangle=0$, hence the sequence of arithmetic means converges
to zero as well. Next, let $p\in(1,\infty)$ and $\varepsilon>0$. Moreover, we assume w.l.o.g.~$C\neq 0$
so $s_1(C) = \Vert C\Vert\neq 0$. As $C\in \mathcal B^p(\mathcal H)$, one can choose $N_1\in\mathbb N$
such that $\sum\nolimits_{k=N_1+1}^\infty s_k(C)^p<\frac{\varepsilon^p}{2^p}$, as well as $N_2\in\mathbb N$ such 
that $\frac{1}{n^{1/q}}<\frac{\varepsilon}{2\sum\nolimits_{k=1}^{N_1}s_k(C)}$ for all $n\geq N_2$. Then for all
$n\geq N:=\max\lbrace N_1+1,N_2\rbrace$, Lemma \ref{lemma_2b}
and H\"older's inequality (Lemma \ref{lemma_hoelders_ineq}) yield the estimate
\begin{align*}
\Big|\frac{1}{n^{1/q}}\sum\nolimits_{k=1}^n\langle e_k,Ce_k\rangle\Big|&\leq \frac{1}{n^{1/q}}\sum\nolimits_{k=1}^{N_1}s_k(C)+\frac{1}{n^{1/q}}\sum\nolimits_{k=N_1+1}^{n}s_k(C)\\
&\leq \frac{1}{n^{1/q}}\sum\nolimits_{k=1}^{N_1}s_k(C)+\Big(\sum\nolimits_{k=N_1+1}^{n} s_k(C)^p \Big)^{1/p}\Big(\sum\nolimits_{k=N_1+1}^{n} \frac{1}{n} \Big)^{1/q}\\
&<\frac{\varepsilon}{2}+\Big(\sum\nolimits_{k=N_1+1}^{\infty} s_k(C)^p \Big)^{1/p}\Big( \frac{n-N_1}{n} \Big)^{1/q}<\varepsilon\,.\qedhere
\end{align*}
\end{proof}

Now we are ready to state the proof of this section's main theorem:

\begin{proof}[Proof of Thm.~\ref{theorem_3}]
(i): For arbitrary orthonormal bases $(e_n)_{n\in\mathbb N}$, $(g_n)_{n\in\mathbb N}$ of $\mathcal H$ as well as any $n\in\mathbb N$, it is readily verified that
\begin{align*}
\frac{\operatorname{tr}([C]^e_{2n})\operatorname{tr}([T]^g_{2n})}{2n} &=\frac{\operatorname{tr}([C]^e_{2n})}{(2n)^{1/q}}\frac{\operatorname{tr}([T]^g_{2n})}{(2n)^{1/p}}\\
&=\Big( \frac{1}{(2n)^{1/q}}\sum\nolimits_{j=1}^{2n} \langle e_j,Ce_j\rangle \Big)\Big( \frac{1}{(2n)^{1/p}}\sum\nolimits_{j=1}^{2n} \langle g_j,Tg_j\rangle \Big)\,.
\end{align*}
Both factors converge and, by Lemma \ref{lemma_0_conv}, at least one of them goes to $0$ as $n\to\infty$. 
Moreover, $W_{[C]^e_{2n}}([T]^g_{2n})$ is star-shaped with respect to 
$(\operatorname{tr}([C]^e_{2n})\operatorname{tr}([T]^g_{2n})/(2n)$ for all $n\in\mathbb N$,
cf.~\cite[Thm.~4]{TSING-96}. Because Hausdorff convergence preserves star-shapedness (Lemma \ref{lemma_5} (iv)), Prop.~\ref{lemma_6}
implies that $\overline{W_{C}(T)}$ is star-shaped with respect to $0 \in \mathbb{C}$.

For what follows let $(e_n)_{n\in\mathbb N},(g_n)_{n\in\mathbb N}$ be the orthonormal bases of $\mathcal H$ which diagonalize $C$ and $T$, respectively.

(ii): W.l.o.g.~let $C$ be normal with collinear eigenvalues. Since $C$ in particular is compact (i.e.~its eigenvalue sequence is a null sequence) there exists
$\phi\in[0,2\pi)$ such that $e^{i\phi}C$ is self-adjoint (Lemma \ref{lemma_spectrum}). Using Prop.~\ref{lemma_6} we obtain
\begin{align*}
\overline{W_C(T)}=\overline{W_{e^{i\phi}C}(e^{-i\phi}T)}=\lim_{n\to\infty} W_{[e^{i\phi}C]_{2n}^e}([e^{-i\phi}T]_{2n}^e)\,.
\end{align*}
Moreover, as $[e^{i\phi}C]_{2n}^e\in\mathbb C^{2n\times 2n}$ is Hermitian for all $n\in\mathbb N$ we conclude that $W_{[e^{i\phi}C]_{2n}^e}([e^{-i\phi}T]_{2n}^e)$ is convex, cf.~\cite{Poon80}. The fact that Hausdorff convergence preserves convexity \cite[Lemma 2.5 (iii)]{DvE18} then yields the desired result.

(iii): Let ${\pi}:\mathbb N\to\mathbb N$ be any permutation and define
the operator
\begin{align*}
U_{\pi}:=\sum\nolimits_{n=1}^\infty| g_{{\pi}(n)}\rangle\langle e_n| \in \mathcal B(\mathcal H)\,.
\end{align*}
Obviously, $U_{\pi}$ is unitary by Lemma \ref{lemma_unitary_ONB} and yields the following equality:
\begin{align*}
\operatorname{tr}(CU_{\pi}^* TU_{\pi}) = \sum_{n=1}^\infty \langle e_n, U_{\pi}^* TU_{\pi} Ce_n\rangle
= \sum_{n=1}^\infty\lambda_n(C)\langle g_{{\pi}(n)},T g_{{\pi}(n)} \rangle
=\sum_{n=1}^\infty\lambda_n(C)\lambda_{{\pi}(n)}(T)
\end{align*}
The fact that ${\pi}$ was chosen arbitrarily shows the first inclusion.
For the second inclusion, we note that by assumption $[C]^e_n$ and $[T]^g_n$ are diagonal and thus normal for all $n\in\mathbb N$. Hence \cite[Coro.~2.4]{Sunder82} tells us
\begin{align}\label{eq:wc_pc_incl}
W_{[C]_{2n}^e}([T]_{2n}^g)\subseteq \operatorname{conv}(P_{[C]_{2n}^e}([T]_{2n}^g))
\end{align}
for all $n\in\mathbb N$. Using that Hausdorff convergence preserves inclusions \cite[Lemma 2.5 (i)]{DvE18}, \eqref{eq:wc_pc_incl} together with Prop.~\ref{lemma_6} yields
\begin{align*}
W_C(T)\subseteq \overline{W_C(T)}=\lim_{n\to\infty} W_{[C]_{2n}^e}([T]_{2n}^g)\subseteq \lim_{n\to\infty} \operatorname{conv}( P_{[C]_{2n}^e}([T]_{2n}^g) )= \operatorname{conv}(\overline{P_C(T)})\,.
\end{align*}
Finally, applying the closure and the convex hull to the inclusions $P_C(T) \subseteq W_C(T)$
yields $\operatorname{conv}(\overline{P_C(T)})\subseteq\operatorname{conv}(\overline{W_C(T)})=\overline{W_C(T)}$, where the last equality is due to (ii), and thus 
$\overline{W_C(T)} = \operatorname{conv}(\overline{P_C(T)})$.
\end{proof}

\subsection{Von Neumann-Type Trace Inequalities}\label{sec:vonn_tr_ineq}

In the mid thirties of the last century, von Neumann \cite[Thm.~1]{NEUM-37} derived the 
following beautiful and widely used trace inequality for complex $n \times n$ matrices:

Let $A,B\in\mathbb C^{n\times n}$ with respective singular values $s_1(A)\geq s_2(A)\geq\ldots\geq s_n(A)$ and 
$s_1(B)\geq s_2(B)\geq\ldots\geq s_n(B)$ be given. Then
\begin{equation}\label{eq:von_Neumann}
\max_{U,V \in \;\mathcal U(\mathbb C^n)}|\operatorname{tr}(AUBV)|=\sum\nolimits_{j=1}^ns_j(A)s_j(B)\,,
\end{equation}
where $\mathcal U(\mathbb C^n)\subseteq\mathbb C^{n\times n}$ as usual denotes the unitary group.
This can be reinterpreted as a characterization of the image of the unitary 
double-coset $\{AUBV \,|\,U,V\in\mathbb C^{n\times n}\text{ unitary}\}$ under the trace-functional:
\begin{corollary}
Given $A,B\in\mathbb C^{n\times n}$ one finds
\begin{equation}\label{eq:von_Neumann-2}
\{\operatorname{tr}(AUBV)\,|\, U,V \in \mathcal U(\mathbb C^n)\} = \overline{B_r}(0)
\end{equation}
with $r := \sum_{j=1}^ns_j(A)s_j(B)$ where $\overline{B_r}(0)=\{z\in\mathbb C\,,\,|z|\leq r\}$ as usual.
\end{corollary}
\begin{proof}
First note that $0$ is contained in the set in question: Using the singular value decompositions $A=\sum_{j=1}^n s_j(A)|f_j\rangle\langle e_j|$, $B=\sum_{j=1}^n s_j(B)|h_j\rangle\langle g_j|$ of $A,B$ for orthonormal bases $(e_j)_{j=1}^n,(f_j)_{j=1}^n,(g_j)_{j=1}^n,(h_j)_{j=1}^n$ of $\mathbb C^n$, let us define the unitary matrices characterized by $V(f_j):=g_j$ and $U(h_j):=e_{j+1}$ for all $j=1,\ldots,n$, where $e_{n+1}:=e_1$. By a straightforward computation $\operatorname{tr}(AUBV)=\sum_{j=1}^n s_j(A)s_j(B)\langle e_j,Uh_j\rangle=0$. 

Now the result follows from \eqref{eq:von_Neumann} together with the elementary observations that the left-hand
side of \eqref{eq:von_Neumann-2} is path-connected (image of the path-connected set $\mathcal U(\mathbb C^n)\times\mathcal U(\mathbb C^n)$ under the continuous map $(U,V)\mapsto\operatorname{tr}(AUBV)$) and circular (simply replace $U$ by $e^{ i \varphi}U$).
\end{proof}
Another 
well-known consequence of \eqref{eq:von_Neumann}, a von Neumann inequality\index{von Neumann inequality} for Hermitian matrices 
\cite[Ch.~9.H.1]{MarshallOlkin}, reads as follows: Let $A,B\in\mathbb C^{n\times n}$ Hermitian with respective eigenvalues $(\lambda_j(A))_{j=1}^n$ and 
$(\lambda_j(B))_{j=1}^n$ be given. Then
\begin{equation}\label{eq:von_Neumann-Hermitian}
\sum\nolimits_{j=1}^n \lambda_j^\downarrow(A)\lambda_j^\uparrow(B) 
\leq \operatorname{tr}(AB)\leq 
\sum\nolimits_{j=1}^n \lambda_j^\downarrow(A)\lambda_j^\downarrow(B)\,,
\end{equation}
where the superindeces $\downarrow$ and $\uparrow$, as usual, denote the respective decreasing and increasing\label{symb_downarrow_uparrow}
sorting of the eigenvalue vectors.

\noindent

The area of applications of von Neumann's inequalities and, more generally, singular value 
decompositions is enormous. It ranges from operator theory \cite{KyFan51,Schatten50} and 
numerics \cite{GolubLoan} to more applied fields like control theory \cite{HM94},
neural networks \cite{Oja89}, as well as quantum dynamics and quantum control \cite{Science98,OLE-95}.
An overview can be found in \cite{MarshallOlkin, Mirsky75}. Now the goal of this short section---which was published as 
\cite{vE_dirr_vonNeumann}---is to generalize these inequalities to Schatten-class operators on infinite-dimensional Hilbert
spaces. 
Indeed let $\mathcal H,\mathcal G$ in the following denote arbitrary complex Hilbert spaces (unless specified otherwise).

\begin{definition}\label{defi_1_SC}
Let $p,q\in [1,\infty]$ be conjugate. Then, following \eqref{eq:von_Neumann-2}, for $C\in\mathcal B^p(\mathcal H,\mathcal G)$ and $T\in\mathcal B^q(\mathcal G,\mathcal H)$ define
\begin{align*}
S_C(T):= \{\operatorname{tr}(CUTV)\,|\,U\in\mathcal U(\mathcal H)\,, V\in\mathcal U(\mathcal G)\}\,.
\end{align*}
\end{definition}
\noindent Obviously this set is more general than the $C$-numerical range of $T$ as it invokes the unitary 
equivalence orbit $UTV$ of $T$ instead of the unitary similarity orbit $U^* T U$.\medskip

Recalling continuity of the map $W:\mathcal B^1(\mathcal H)\times \mathcal K(\mathcal H)\to (\mathcal P_c(\mathbb C),\Delta)$, $(C,T)\mapsto\overline{W_C(T)}$ (even more generally on the domain $\mathcal B^p(\mathcal H)\times\mathcal B^q(\mathcal H)$) from Thm.~\ref{lemma_2} we can obtain an analogous result for the more general map $(C,T)\mapsto\overline{S_C(T)}$:
\begin{proposition}\label{prop_1z}
Let $C\in\mathcal B^p(\mathcal H,\mathcal G)$, $T\in\mathcal B^q(\mathcal G,\mathcal H)$ with 
$p,q\in [1,\infty]$ conjugate be given and let $(C_n)_{n\in\mathbb N}$ and $(T_n)_{n\in\mathbb N}$ be sequences in $\mathcal B^p(\mathcal H,\mathcal G)$ and $\mathcal B^q(\mathcal G,\mathcal H)$, respectively, such that
$
\lim_{n\to\infty}\|C-C_n\|_p=\lim_{n\to\infty}\|T-T_n\|_q=0\,.
$
Then
\begin{equation*}
\lim_{n\to\infty}\overline{S_{C_n}(T_n)}=\overline{S_C(T)}\,.
\end{equation*}
\end{proposition}
\begin{proof}
W.l.o.g.~let $C_n,T_n\neq 0$ for some $n\in\mathbb N$---else all the involved sets would be trivial---so we may introduce the positive but (as seen via the reverse triangle inequality) finite numbers
$$
\kappa := \sup\{\|C\|_p,\|C_1\|_p,\|C_2\|_p,\ldots\} 
\quad\text{ and }\quad \tau :=\sup\{\|T\|_q,\|T_1\|_q,\|T_2\|_q,\ldots\}\,.
$$
Let $\varepsilon>0$. By assumption there exists $N\in\mathbb N$ such that
$$
\|C-C_n\|_p<\frac{\varepsilon}{4\tau} \qquad\text{ as well as }\qquad \|T-T_n\|_q<\frac{\varepsilon}{4\kappa}
$$
for all $n\geq N$. 
The goal will be to satisfy the assumptions of Lemma \ref{lemma_11} in order to show
$\Delta(\overline{S_C(T)},\overline{S_{C_n}(T_n)})<\varepsilon$ for all $n\geq N$. 

Let $w\in\overline{S_C(T)}$. Then one finds $U\in\mathcal U(\mathcal H)$, $V\in\mathcal U(\mathcal G)$
such that $w':=\operatorname{tr}(CUTV)$ satisfies $|w-w'|<\frac{\varepsilon}{2}$. Thus
for $w_n:=\operatorname{tr}(C_nUT_nV)$ by Lemma \ref{lemma_schatten_prop_pq}
\begin{align*}
|w-w_n|&\leq|w-w'|-|w'-w_n|\\
&<\frac{\varepsilon}{2}+|\operatorname{tr}( (C-C_n)UTV )|+|\operatorname{tr}(VC_nU(T-T_n))|\\
&\leq\frac{\varepsilon}{2}+\|C-C_n\|_p\|U\|\|T\|_q\|V\|+\|V\|\|C_n\|_p\|U\|\|T-T_n\|_q\\
&\leq \frac{\varepsilon}{2}+\|C-C_n\|_p\,\tau+\kappa\,\|T-T_n\|_q<\varepsilon
\end{align*}
for all $n \geq N$. 
Similarly for $v_n\in\overline{S_{C_n}(T_n)}$ 
one finds $U_n\in\mathcal U(\mathcal H)$, $V_n\in\mathcal U(\mathcal G)$ such that 
$v_n':=\operatorname{tr}(C_nU_nT_nV_n)$ satisfies $|v_n-v_n'|<\frac{\varepsilon}{2}$. Thus for 
$\tilde v_n:=\operatorname{tr}(CU_nTV_n)$ we obtain
\begin{align*}
|v_n-\tilde v_n|&\leq|v_n-v_n'|-|v_n'-\tilde v_n|\\
&<\frac{\varepsilon}{2}+|\operatorname{tr}( (C_n-C)U_nT_nV_n )|+|\operatorname{tr}(V_nCU_n(T_n-T))|\\
&\leq \frac{\varepsilon}{2}+\|C-C_n\|_p\,\tau+\kappa\,\|T-T_n\|_q<\varepsilon\,.\qedhere
\end{align*}
\end{proof}
%
%
%
%
%

Considering the inequalities \eqref{eq:von_Neumann} and \eqref{eq:von_Neumann-Hermitian} from the introduction, it arguably is 
easier to generalize the former, i.e.~to generalize von Neumann's ``original'' trace inequality to Schatten-class 
operators. To start with we first investigate
the finite-rank case.

\begin{lemma}\label{lemma_S_C_finite_rank}
Let $C\in\mathcal F(\mathcal H,\mathcal G)$, $T\in\mathcal F(\mathcal G,\mathcal H)$, and $k:=\max\{\operatorname{dim}(\operatorname{im}(C)),\operatorname{dim}(\operatorname{im}(T))\}$. Then $S_C(T)=\overline{B_r}(0)$ where $r:=\sum_{j=1}^ks_j(C)s_j(T)$, $k<\infty$.
\end{lemma}
\begin{proof}
Defining $k$ as above, Prop.~\ref{prop_compact_SVD} yields orthonormal systems $(e_j)_{j=1}^k$, $(h_j)_{j=1}^k$ in $\mathcal H$ and $(f_j)_{j=1}^k$, $(g_j)_{j=1}^k$ in $\mathcal G$ such that
$$
C=\sum\nolimits_{j=1}^k s_j(C)|f_j\rangle\langle e_j|\quad\text{ and }\quad T=\sum\nolimits_{j=1}^k s_j(T)|h_j\rangle\langle g_j|\,.
$$
Note that forcing both sums to have same summation range means that, potentially,
some of the singular values have to be complemented by zeros, which is not of further
importance.

``$\subseteq$'': Let any $U\in\mathcal U(\mathcal H)$, $V\in\mathcal U(\mathcal G)$ be given. Then
\begin{align*}
\operatorname{tr}(CUTV)&=\operatorname{tr}\Big( \sum\nolimits_{i,j=1}^{k}s_i(T)s_j(C)\langle e_j,Uh_i\rangle|f_j\rangle\langle V^*g_i|\Big)\\
&=\hphantom{\operatorname{tr}\Big(} \sum\nolimits_{i,j=1}^{k}s_i(T)s_j(C)\langle e_j,Uh_i\rangle\langle g_i,V f_j\rangle
\end{align*}
by direct computation. Now consider the subspaces
\begin{align*}
Z_1&:=\operatorname{span}\{e_1,\ldots,e_k,Uh_1,\ldots,Uh_k\}\subset \mathcal H\\
Z_2&:=\operatorname{span}\{f_1,\ldots,f_k,V^* g_1,\ldots,V^* g_k\}\subset\mathcal G\,.
\end{align*}
So there exist orthonormal bases of the form $$e_1,\ldots,e_k, e_{k+1},\ldots, e_N\quad\text{ and }\quad f_1,\ldots,f_k, f_{k+1},\ldots, f_{N'}$$ of $Z_1$ and $Z_2$ for some $N,N'\geq k$, respectively. W.l.o.g.\footnote{This can be done for example by sufficiently expanding the ``smaller'' orthonormal systems in $\mathcal H$ or $\mathcal G$ and possibly passing to new subspaces $Z_1'\supset Z_1$ or $Z_2'\supset Z_2$ which is always doable because we are in infinite dimensions. The particular choice of $Z_1'$ and $Z_2'$ is irrelevant because we only need the 
orthonormal systems which represent $C$ and $T$ to be contained within these finite-dimensional subspaces.} we can assume $N=N'$ and define
$$
a_j:=(\langle e_l,U h_j\rangle)_{l=1}^N\in\mathbb C^N\quad\text{ and }\quad b_j:=(\langle f_l,V^* g_j\rangle)_{l=1}^N\in\mathbb C^N
$$
for $j=1,\ldots,k$. This yields $N\times N$ matrices
\begin{align*}
C'=\operatorname{diag} (s_1(C),\ldots, s_k(C),0,\ldots,0)\quad\text{ and }\quad T'=\sum\nolimits_{j=1}^k s_j(T) |a_j\rangle\langle b_j|
\end{align*}
which satisfy $\operatorname{tr}(C'T')=\sum\nolimits_{i,j=1}^{k}s_i(T)s_j(C)\langle e_j,Uh_i\rangle\langle g_i,V f_j\rangle$. By construction $(a_j)_{j=1}^k,(b_j)_{j=1}^k$ are orthonormal systems in $\mathbb C^N$ so $s_j(T')=s_j(T)$ for all $j=1,\ldots,N$. Thus von Neumann's original result \eqref{eq:von_Neumann} yields
$$
|\operatorname{tr}(CUTV)|=|\operatorname{tr}(C'T')|\leq\sum\nolimits_{j=1}^N s_j(C')s_j(T')=\sum\nolimits_{j=1}^k s_j(C) s_j(T)\,.
$$

``$\supseteq$'': We first consider unitary operators $U_T\in\mathcal B(\mathcal H)$, $V_T\in\mathcal B(\mathcal G)$ 
such that $U_Th_j=e_j$ and $V_Tf_j=g_j$ for all $j=1,\ldots,k$. This is always possible by completing the 
respective orthonormal systems $(e_j)_{j=1}^k$, $\ldots$ to orthonormal bases $(e_j)_{j\in J}$, $\ldots$ 
which can then be transformed into each other via some unitary (Lemma \ref{lemma_unitary_ONB}). This allows us to construct $\tilde T:=U_TTV_T=\sum\nolimits_{j=1}^k s_j(T)|e_j\rangle\langle f_j|$ such that
$$
\operatorname{tr}(C\tilde U\tilde T \tilde V)=\sum\nolimits_{i,j=1}^N s_j(C)s_i(T)\langle e_j,\tilde Ue_i\rangle\langle f_i,\tilde Vf_j\rangle
$$
for any $\tilde U\in\mathcal U(\mathcal H)$, $\tilde V\in\mathcal U(\mathcal G)$. Of course $S_C(T)=S_C(\tilde T)$ and the latter satisfies
%
%
%
\begin{itemize}
\item $r\in S_C(\tilde T)$: Choose $\tilde U=\mathbbm{1}_{\mathcal H}$, $\tilde V=\mathbbm{1}_{\mathcal G}$.
\item $0\in S_C(\tilde T)$: Choose $\tilde V=\mathbbm{1}_{\mathcal G}$, and $\tilde U$ as cyclic shift on the first $k$ basis elements:
$$
\tilde U:\mathcal H\to\mathcal H\,,\qquad e_j\mapsto \begin{cases} e_{j+1}&j=1,\ldots,k-1\\ e_{1}&j=k\\ e_j&j\in J\setminus\{1,\ldots,k\} \end{cases}
$$
\end{itemize}
Now because the unitary group $\mathcal U(\mathcal G)$ on any Hilbert space $\mathcal G$ is path-connected
(Thm.~\ref{theorem_unitary_group_properties} (iv)) and because the mapping 
$f:\mathcal B(\mathcal H)\times\mathcal B(\mathcal G)\to \mathbb C$, $(U,V)\mapsto \operatorname{tr}(CU\tilde TV)$ 
is continuous, 
the image $f(\mathcal U(\mathcal H)\times\mathcal U(\mathcal G))$ has to 
be path-connected as well. In particular, $0$ and $r$ are path-connected within $S_C(T)$, i.e.~for every 
$s\in[0,r]$ there exists $\phi(s)\in[0,2\pi)$ such that $se^{i\phi(s)}\in S_C(\tilde T)=S_C(T)$. 

Finally, we can use the fact that $S_C(T)$ is circular---which follows easily by replacing 
$U$ by $e^{i\varphi} U\in\mathcal U(\mathcal H)$ with $\varphi\in [0,2\pi]$---to conclude $S_C(T)\supseteq \overline{B_r}(0)$
and thus $S_C(T) = \overline{B_r}(0)$.
\end{proof}

\begin{theorem}
Let $C\in\mathcal B^p(\mathcal H,\mathcal G)$, $T\in\mathcal B^q(\mathcal G,\mathcal H)$ with $p,q\in[1,\infty]$ conjugate. Then 
\begin{equation}\label{eq:von-Neumann-infinite-dim}
\sup_{U \in \;\mathcal U(\mathcal G) , V \in \;\mathcal U(\mathcal H)}|\operatorname{tr}(CUTV)|
= \sum\nolimits_{j=1}^\infty s_j(C)s_j(T)\,.
\end{equation}
In particular, one has $\overline{S_C(T)}=\overline{B_r}(0)$ with $r:=\sum_{j=1}^\infty s_j(C)s_j(T)$.
\end{theorem}

\begin{proof} 
Using the Schmidt decomposition $C=\sum_{j=1}^\infty s_j(C)|f_j\rangle\langle e_j|$, 
$T=\sum_{j=1}^\infty s_j(T)|h_j\rangle\langle g_j|$ for some orthonormal systems $(e_j)_{j\in\mathbb N}$,
$(h_j)_{j\in\mathbb N}$ in $\mathcal H$ and $(f_j)_{j\in\mathbb N}$, $(g_j)_{j\in\mathbb N}$ in $\mathcal G$ define the finite rank approximations $C_n:=\sum_{j=1}^n s_j(C)|f_j\rangle\langle e_j|$
and $T_n:=\sum_{j=1}^n s_j(T)|h_j\rangle\langle g_j|$. 
To pass to the original operators $C,T$, 
we use Prop.~\ref{prop_Schatten_p_properties} (i) to see
\begin{equation*}
\lim_{n\to\infty}\|C_n-C\|_p=0\qquad \text{and} \qquad\lim_{n\to\infty}\|T_n-T\|_q=0\,.
\end{equation*}
%
%
Because of this we may apply Prop.~\ref{prop_1z} and Lemma \ref{lemma_S_C_finite_rank} to obtain
$$
\overline{S_C(T)}=\lim_{n\to\infty}\overline{S_{C_n}(T_n)}=\lim_{n\to\infty} \overline{B_{r_n}}(0)
$$
with $r_n:=\sum_{j=1}^n s_j(C)s_j(T)$. Using the obvious fact $\Delta(\overline{B_r}(0),\overline{B_{r_n}}(0))=|r-r_n|$ for all $n\in\mathbb N$ 
one readily verifies
$
\overline{S_C(T)}=\lim_{n\to\infty} \overline{B_{r_n}}(0)=\overline{B_r}(0)
$ with $r=\sum_{j=1}^\infty s_j(C)s_j(T)$.
\end{proof}

\begin{remark}
To see that the supremum in \eqref{eq:von-Neumann-infinite-dim} is not necessarily a maximum, consider $\mathcal H=\ell_2(\mathbb N)$ with standard basis $(e_j)_{j\in\mathbb N}$. Now the positive definite trace-class operator $C=\sum_{j=1}^\infty \frac{1}{2^j}|e_j\rangle\langle e_j|$ as well as the compact operator $T=\sum_{k=1}^\infty \frac{1}{2^k}|e_{k+1}\rangle\langle e_{k+1}|$ satisfy
$$
\operatorname{tr}(CUTV)= \sum\nolimits_{j=1}^\infty \frac{1}{2^j}\langle e_j, UTV e_j\rangle=\sum\nolimits_{j,k=1}^\infty \frac{1}{2^j} \frac{1}{2^k}\langle e_j,Ue_{k+1}\rangle\langle e_{k+1},Ve_j\rangle
$$
for any $U,V\in\mathcal U(\mathcal H)$. We know that $\sup_{U,V\in\mathcal U(\mathcal H)}|\operatorname{tr}(CUTV)|=\sum_{j=1}^\infty(\frac{1}{2^j})^2$ but if this were a maximum, then by the above calculation $\langle e_j,Ue_{k+1}\rangle=\langle e_{k+1},Ve_j\rangle=\delta_{jk}$ for all $j,k\in\mathbb N$. The only operators which satisfy these conditions are 
the left- and the right-shift, respectively, both of which are not unitary; a contradiction.
\end{remark}

Finally, we are prepared to extend inequality \eqref{eq:von_Neumann-Hermitian} to Schatten-class operators 
on separable Hilbert spaces.

\begin{theorem}\label{thm_vN_sep_HS}
Let $\mathcal H$ be an infinite-dimensional, separable, and complex Hilbert space, $C\in\mathcal B^p(\mathcal H)$, $T\in\mathcal B^q(\mathcal H)$ both be self-adjoint with $p,q\in[1,\infty]$ conjugate, and 
let the positive semi-definite operators $C^+,T^+$ and $C^-,T^-$ denote the positive and negative part of $C,T$, 
respectively (i.e.~$C=C^+-C^-$, $T=T^+-T^-$). Then
\begin{equation}\label{eq:sup_WCT}
\sup_{U\in\mathcal U(\mathcal H)}\operatorname{tr}(CU^* TU)
= \sum\nolimits_{j=1}^\infty \big(\lambda_j^\downarrow (C^+)\lambda_j^\downarrow (T^+) 
+ \lambda_j^\downarrow (C^-)\lambda_j^\downarrow (T^-)\big)
\end{equation}
as well as
\begin{equation}\label{eq:inf_WCT}
\inf_{U\in\mathcal U(\mathcal H)}\operatorname{tr}(CU^* TU)
=- \sum\nolimits_{j=1}^\infty \big(\lambda_j^\downarrow (C^+)\lambda_j^\downarrow (T^-)
+ \lambda_j^\downarrow (C^-)\lambda_j^\downarrow (T^+)\big)\,.
\end{equation}
 In particular, one has
$$
-\sum_{j=1}^\infty \big(\lambda_j^\downarrow (C^+)\lambda_j^\downarrow (T^-)+ \lambda_j^\downarrow (C^-)\lambda_j^\downarrow (T^+)\big)
\leq \operatorname{tr}(CT)\leq 
\sum_{j=1}^\infty \big(\lambda_j^\downarrow (C^+)\lambda_j^\downarrow (T^+)+\lambda_j^\downarrow (C^-)\lambda_j^\downarrow (T^-)\big)\,.
$$
\end{theorem}

\begin{proof}
Let $C\in\mathcal B^p(\mathcal H)$, $T\in\mathcal B^q(\mathcal H)$ both be self-adjoint with $p,q$ conjugate and first 
assume that $T$ has at most $k\in\mathbb N$ non-zero eigenvalues. Then the following is straightforward to show:
\begin{align*}
\max \operatorname{conv}(\overline{P_C(T)}) &=\hphantom{-} 
\sum\nolimits_{j=1}^k\lambda_j^\downarrow(C^+) \lambda_{j}^\downarrow(T^+) 
+ \sum\nolimits_{j=1}^k \lambda_j^\downarrow(C^-) \lambda_{j}^\downarrow(T^-)\\
\min \operatorname{conv}(\overline{P_C(T)}) &=- 
\sum\nolimits_{j=1}^k\lambda_j^\downarrow(C^+) \lambda_{j}^\downarrow(T^-) 
- \sum\nolimits_{j=1}^k \lambda_j^\downarrow(C^-) \lambda_{j}^\downarrow(T^+)
\end{align*}
Note that in this case the (modified) eigenvalue sequence of $T$ contains infinitely
many zeros. Now let us address the general case. Choose any orthonormal eigenbasis $(e_n)_{n\in\mathbb N}$ of $T$ 
with corresponding modified eigenvalue sequence. Moreover, let 
$\Pi_k=\sum\nolimits_{j=1}^k|e_j\rangle\langle e_j|$ the projection onto the span
of the first $k$ eigenvectors of $T$. Then $\Pi_kT\Pi_k$ has at most $k$ non-zero eigenvalues and our 
preliminary considerations combined with Theorems \ref{lemma_2} \& \ref{theorem_3} (iii) as well as Lemma \ref{lemma_lim_max} readily imply
\begin{align*}
\sup_{U\in\mathcal U(\mathcal H)}&\operatorname{tr}(C U^* TU)
= \max \overline{W_C(T)} = \max \lim_{k\to\infty} \overline{W_C(\Pi_kT\Pi_k)}\\
& =\lim_{k\to\infty}\max \overline{W_C(\Pi_kT\Pi_k)} 
= \lim_{k\to\infty}\max \operatorname{conv}(\overline{P_C(\Pi_kT\Pi_k)})\\
& =\lim_{k\to\infty}\Big(\sum\nolimits_{j=1}^k \lambda_j^\downarrow(C^+) \lambda_{j}^\downarrow(\Pi_kT^+\Pi_k)
+ \sum\nolimits_{j=1}^k \lambda_j^\downarrow(C^-) \lambda_{j}^\downarrow(\Pi_kT^-\Pi_k)\Big)
\end{align*}
where we used the identity $(\Pi_kT\Pi_k)^\pm = \Pi_kT^\pm\Pi_k$. Now, the last step is to show 
that $(\sum\nolimits_{j=1}^k \lambda_j^\downarrow(C^+)\lambda_j^\downarrow(\Pi_kT^+\Pi_k))_{k\in\mathbb N}$ converges to
$\sum\nolimits_{j=1}^\infty\lambda_j^\downarrow(C^+)\lambda_j^\downarrow(T^+)$. Let $\varepsilon>0$
(and w.l.o.g.~$T\neq 0$). As $(\lambda_j^\downarrow(C^+))_{j\in\mathbb N}$ is a sequence in $\ell^p_+(\mathbb N)$ 
we find $N\in\mathbb N$ with
$$
\Big(\sum\nolimits_{j=N+1}^\infty\big(\lambda_j^\downarrow(C^+)\big)^p\Big)^{1/p}<\frac{\varepsilon}{2\|T\|_q}
$$
where for $p=\infty$, the left-hand side becomes $\sup_{n>N} \lambda_n^\downarrow(C^+)=\lambda_{N+1}^\downarrow(C^+)\,$. 

Either way, associated to this $N$ one can choose $K\geq N$ such that the first $N$ largest eigenvalues of $T^+$ are
listed in $(\lambda_j^\downarrow(\Pi_KT^+\Pi_K))_{j\in\mathbb N}$ and thus $\lambda_j^\downarrow(T^+)=\lambda_j^\downarrow(\Pi_KT^+\Pi_K)$ for all $j=1,\ldots,N$. Putting things together and using H\"older's inequality (Lemma \ref{lemma_hoelders_ineq}) we get
\begin{align*}
\Big| \sum\nolimits_{j=1}^K \lambda_j^\downarrow(C^+) &\lambda_{j}^\downarrow(\Pi_KT^+\Pi_K)-\sum\nolimits_{j=1}^\infty\lambda_j^\downarrow(C^+)\lambda_j^\downarrow(T^+)\Big|\\
&=\Big| \sum\nolimits_{j=N+1}^K \lambda_j^\downarrow(C^+)\lambda_{j}^\downarrow(\Pi_KT^+\Pi_K)-\sum\nolimits_{j=N+1}^\infty\lambda_j^\downarrow(C^+)\lambda_{j}^\downarrow(T^+)\Big|\\
&\leq2\|T^+\|_q \Big(\sum\nolimits_{j=N+1}^\infty\big(\lambda_j^\downarrow(C^+)\big)^p\Big)^{1/p}<2\|T\|_q\frac{\varepsilon}{2\|T\|_q}=\varepsilon\,.
\end{align*} 
The case of $C^-,T^-$ as well as the infimum-estimate are shown analogously which concludes the proof.
\end{proof}

Therefore if $C,T$ are self-adjoint (i.e.~$W_C(T)\subseteq\mathbb R$), a path-connectedness argument similar to the proof of Lemma \ref{lemma_S_C_finite_rank} shows $(a,b)\subseteq W_C(T)\subseteq [a,b]$ with $a$ ($\leq 0$) given by \eqref{eq:inf_WCT} and $b$ ($\geq 0$) given by \eqref{eq:sup_WCT}. In particular, $\overline{W_C(T)}=[a,b]$.

\section{Majorization on Trace-Class Operators}\label{maj:trace_class}

Generalizing majorization to infinite dimensions is somewhat delicate. Following \cite{gohberg66}, one may 
define majorization first on the space of all real null sequences $c_0(\mathbb N)$, and then on the space of all 
absolutely summable sequences $\ell^1(\mathbb N)$. Because we 
need a concept of majorization on density 
operators, for our purposes it suffices to introduce majorization solely on 
the summable sequences of non-negative numbers~$\ell^1_+(\mathbb N)$\label{symb_ell_1_plus},
which is rather intuitive.

Recall from Ch.~\ref{sec:maj_d_vec} that for two vectors $x,y\in\mathbb R^n$, one says $x$ is majorized by $y$ (written $x\prec y$) if $\sum_{j=1}^k x_j^\downarrow\leq\sum_{j=1}^k y_j^\downarrow$ for all $k=1,\ldots,n-1$ and $\sum_{j=1}^n x_j=\sum_{j=1}^n y_j$. By definition $x\prec y$ 
depends only on the entries of $x$ and $y$ but not on their
initial arrangement, so $\prec$ is permutation invariant. 
Now for sequences $x\in\ell^1_+(\mathbb N)$, this re-arrangement procedure works just the same way,
and all the non-zero entries of $x$ are again contained within the rearranged sequence $x^\downarrow$. However, be aware that $x$ and $x^\downarrow$ may differ in the number of their zero entries. 

\begin{definition}\label{defi_maj}
Consider $x,y\in\ell^1_+(\mathbb N)$ and $\rho,\omega\in\mathbb D(\mathcal H)$.
\begin{itemize}
\item[(i)] We say that $x$ is majorized by $y$, denoted by $x\prec y$, if $\sum_{j=1}^k x_j^\downarrow \leq\sum_{j=1}^k y_j^\downarrow$ 
holds for all $k\in\mathbb N$, and if $\sum_{j=1}^\infty x_j=\sum_{j=1}^\infty y_j$.\index{majorization!on sequences}
\item[(ii)] $\omega$ majorizes $\rho$, denoted by $\rho\prec\omega$, 
if $\lambda^\downarrow(\rho)\prec\lambda^\downarrow(\omega)$ where $\lambda^\downarrow(\cdot)\in\ell^1_+(\mathbb N)$ denotes the (non-modified) \index{majorization!on density operators}
eigenvalue sequence 
%
of the respective state.
\end{itemize}
\end{definition}
\begin{remark}\label{rem_maj}
In Definition \ref{defi_maj} (ii) it does not matter whether one considers the usual (non-modified) or the modified 
eigenvalue sequence (for the purpose of this remark denoted by $\lambda^\downarrow$ and $\lambda_m$, respectively). 
More precisely, these sequences by construction share the same non-zero entries so $\lambda^\downarrow={\lambda^\downarrow_m},$.
\end{remark}

As in finite dimensions, majorization in infinite dimensions has a number of
different characterizations, the following two being particularly advantageous
for our purposes. For this we have to introduce the notion of a \textit{bi-stochastic quantum map}, that is, a Heisenberg quantum channel which also is trace-preserving ($T(\mathcal B^1(\mathcal H))\subseteq\mathcal B^1(\mathcal G)$ with $\operatorname{tr}(T(A))=\operatorname{tr}(A)$ for all $A\in\mathcal B^1(\mathcal H)$) and its restriction to the trace class is\index{bi-stochastic quantum map}
a Schr\"odinger quantum channel. Define\label{symb_bi_stoch_qc}
\begin{align*}
\mathbb S (\mathcal H,\mathcal G):=
\lbrace T:\mathcal B(\mathcal H)\to\mathcal B(\mathcal G)\, |\, T\text{ is a bi-stochastic quantum map}\rbrace
\end{align*}
and $\mathbb S (\mathcal H):= \mathbb S (\mathcal H,\mathcal H)$.

\begin{lemma}[\cite{Li13}, Thm.~3.3]\label{lemma_li}
Let $\mathcal H$ be a separable Hilbert space. For $\rho,\omega\in\mathbb D(\mathcal H)$ the following 
are equivalent:
\begin{itemize}
\item[(i)] $\rho\prec\omega$.
\item[(ii)] There exists a bi-stochastic quantum map $T\in \mathbb S(\mathcal H)$ such that $T(\omega)=\rho$. 
\end{itemize}
\end{lemma}
\begin{proposition}\label{prop_schur_horn_gohberg}
Let $(e_n)_{n\in\mathbb N}$ be some 
orthonormal basis of a separable Hilbert space $\mathcal H$ and let $x,y\in\ell^1_+(\mathbb N)$ be non-increasing sequences. Then the following statements are equivalent:
\begin{itemize}
\item[(i)] $x\prec y$
\item[(ii)] There exists $A\in\mathcal B^1(\mathcal H)$ self-adjoint with diagonal entries $(x_n)_{n\in\mathbb N}$ 
and eigenvalues $(y_n)_{n\in\mathbb N}$.
\item[(iii)] There exists unitary $U\in\mathcal B(\mathcal H)$ such that $U\operatorname{diag}(y)U^* $ has 
diagonal entries $(x_n)_{n\in\mathbb N}$.
\end{itemize}
Here, ``diagonal'' always refers to the orthonormal basis $(e_n)_{n\in\mathbb N}$, meaning the isometric map $\operatorname{diag}:\ell^1_+(\mathbb N)\to \mathcal B^1(\mathcal H)$ is given by $x\mapsto \sum_{n=1}^\infty x_n|e_n\rangle\langle e_n|$.
\end{proposition}
\begin{proof}
``(i) $\Rightarrow$ (ii)'': Assume $x\prec y$. By \cite[Prop.~IV]{gohberg66} one finds 
orthonormal bases $(\phi_n)_{n\in\mathbb N}$ and $(\psi_n)_{n\in\mathbb N}$ of $\mathcal H$ such that 
$H=\sum_{n=1}^\infty y_n|\psi_n\rangle\langle \psi_n|$ satisfies $\langle \phi_n,H\phi_n\rangle=x_n$ 
for all $n\in\mathbb N$. Now there exists unique $U\in\mathcal U(\mathcal H)$ 
which transforms $(\phi_n)_{n\in\mathbb N}$ into $(e_n)_{n\in\mathbb N}$ (Lemma \ref{lemma_unitary_ONB} (iii)) so $A:=UHU^* \in\mathcal B^1(\mathcal H)$ does the job. ``(ii) $\Rightarrow$ (i)'': 
Follows from \cite{Fan49}. ``(ii) $\Leftrightarrow$ (iii)'': Obvious.
\end{proof}

We conclude with a classical result on sub-majorization.

\begin{lemma}[\cite{MarshallOlkin}, 3.H.3.b]\label{lemma_wprec}
Let $x,y\in\mathbb R^n$ such that $\sum_{j=1}^k x_j^\downarrow\leq\sum_{j=1}^k y_j^\downarrow$ for all $k=1,\ldots,n$. 
Then for arbitrary $c_1\geq c_2\geq\ldots\geq c_n\geq 0$ one has
$
\sum\nolimits_{j=1}^n c_jx_j^\downarrow \leq \sum\nolimits_{j=1}^n c_jy_j^\downarrow \,.
$
\end{lemma}

To simplify notation, we use the following abbreviation.

\begin{definition}\label{def_KC}
Let $C\in\mathcal B^1(\mathcal H)$, $T\in\mathcal B(\mathcal H)$ both be self-adjoint. We define
$
K_C(T):=\sup_{U \in\mathcal U(\mathcal H)}\operatorname{tr}(C U^* TU)\in\mathbb R
$
or, equivalently, $K_C(T) :=\sup W_C(T) = \max\overline{W_C(T)}$.
\end{definition}

\noindent
Note that if $C$ and $T$ are positive semi-definite, then $K_C(T)$ turns into 
the \mbox{$C$-numerical} radius $r_C(T)$ of $T$. Now this definition gives rise to the following
result, the finite-dimensional analogue of which can be found in \cite[Thm.~7.4]{Ando89}.
\begin{proposition}\label{thm_1_maj_inf}
For $\rho,\omega\in\mathbb D(\mathcal H)$ the following statements are equivalent.
\begin{itemize}
\item[(i)] $\rho \prec \omega$
\item[(ii)] $K_\rho(T)\leq K_\omega(T)$ for all self-adjoint $T\in\mathcal K(\mathcal H)$.
\item[(iii)] $K_\rho(T)\leq K_\omega(T)$ for all positive semi-definite $T\in\mathcal K(\mathcal H)$.
\end{itemize}
\end{proposition}
\begin{proof}

``(i) $\Rightarrow$ (ii)'': Keeping in mind that $\rho,\omega\geq 0$, Thm.~\ref{thm_vN_sep_HS} yields
\begin{align*}
K_\rho(T) = \max \overline{W_\rho(T)} = \sum\nolimits_{j=1}^\infty \lambda_j^\downarrow(\rho) \lambda_{j}^\downarrow(T^+)
\end{align*}
and similarly for $K_\omega(T)$. 
Moreover, $\rho\prec\omega$ by Lemma \ref{lemma_wprec} yields
\begin{equation*}
\sum\nolimits_{j=1}^n \lambda_j^\downarrow(\rho) \lambda_{j}^\downarrow(T^+)
\leq \sum\nolimits_{j=1}^n \lambda_j^\downarrow(\omega) \lambda_{j}^\downarrow(T^+)
\end{equation*}
for all $n \in \mathbb N$ and thus it follows $K_\rho(T)\leq K_\omega(T)$
for all self-adjoint $T\in\mathcal K(\mathcal H)$.

``(ii) $\Rightarrow$ (iii)'': Trivial. 
``(iii) $\Rightarrow$ (i)'': Let $k\in\mathbb N$ and let $(e_n)_{n\in\mathbb N}$ be any orthonormal basis of $\mathcal H$. Consider the (finite-rank) projection
$\Pi_k=\sum\nolimits_{j=1}^k|e_j\rangle\langle e_j|$. As $\Pi_k$ is compact with 
eigenvalues $1$ (of multiplicity $k$) and $0$ (of infinite multiplicity) one finds $\Pi_k\geq 0$ so Thm.~\ref{thm_vN_sep_HS}
yields
$
K_\rho(\Pi_k)=\sum\nolimits_{j=1}^k\lambda_j^\downarrow(\rho) 
$
and
$
K_\omega(\Pi_k)=\sum\nolimits_{j=1}^k\lambda_j^\downarrow(\omega)\,.
$
Now by assumption one has
\begin{align*}
\sum\nolimits_{j=1}^k\lambda_j^\downarrow(\rho)=K_\rho(\Pi_k)\leq K_\omega(\Pi_k)=\sum\nolimits_{j=1}^k\lambda_j^\downarrow(\omega)
\end{align*}
for all $k\in\mathbb N$ which shows $\rho\prec\omega$ and thus concludes this proof.
\end{proof}
\noindent While (ii) for now can be weakened to (iii), i.e.~from self-adjoint to positive semi-definite, this would probably be indispensable if $\rho,\omega$ were allowed to be arbitrary self-adjoint trace class operators.\smallskip

This is all we need in order to prove closedness of the set of all states majorized by some initial state. While it might seem like mathematical frills this really is the key for generalizing from bounded to unbounded drift Hamiltonians in Ch.~\ref{sec_reach_inf_dim}.
\begin{theorem}\label{lemma_maj_closed}
For all $\rho_0\in\mathbb D(\mathcal H)$ the set 
$\lbrace \rho\in\mathbb D(\mathcal H)\,|\,\rho\prec\rho_0\rbrace$ 
is closed w.r.t.~$\|\cdot\|_1$.
\end{theorem}
\begin{proof}
Given $\omega\in \overline{\lbrace \rho\in\mathbb D(\mathcal H)\,|\,\rho\prec\rho_0\rbrace}$ 
there exists a sequence $(\rho_n)_{n\in\mathbb N}$ in 
$\lbrace \rho\in\mathbb D(\mathcal H)\,|\,\rho\prec\rho_0\rbrace\subseteq\mathbb D(\mathcal H)$ 
such that $\|\omega-\rho_n\|_1\to 0$ as $n\to\infty$. Note that $\omega\in\mathbb D(\mathcal H)$ by Lemma \ref{lemma_pure_extreme_general}. Now let $T\in\mathcal K(\mathcal H)$ be 
arbitrary but self-adjoint. Then Lemma \ref{lemma_lim_max} and Lemma \ref{lemma_3} imply
\begin{align*}
K_\omega(T)=\max\overline{W_\omega(T)}&=\max\lim_{n\to\infty}\overline{W_{\rho_n}(T)}\\
&=\lim_{n\to\infty}\max\overline{W_{\rho_n}(T)}=\lim_{n\to\infty}K_{\rho_n}(T)\,.
\end{align*}
On the other hand, due to Prop.~\ref{thm_1_maj_inf} and $\rho_n\prec\omega$ for all $n\in\mathbb N$, one has
$$
K_\omega(T)=\lim_{n\to\infty}K_{\rho_n}(T)\leq K_{\rho_0}(T)
$$
which again by Prop.~\ref{thm_1_maj_inf} (as $T$ was chosen arbitrarily) implies $\omega\prec\rho_0$.
\end{proof}
\noindent The proof idea used here is fundamentally different from the proof of Thm.~\ref{lemma_R_d_closed}; while in finite dimensions we used compactness of $Q(n)$ to show closedness of, e.g., $\lbrace \rho\in\mathbb D(\mathbb C^n)\,|\,\rho\prec\rho_0\rbrace$ recall that $Q_S(\mathcal H)$ is not compact anymore once $\mathcal H$ is of infinite dimension (Prop.~\ref{thm_monoid}) so we had to resort to the characterization via compact self-adjoint operators.

\chapter{Reachable Sets for Controlled Markovian Quantum Systems}\label{sec:results}

While closed quantum control systems are relatively simple to analyze using the associated Lie algebra, studying open quantum control systems leads to the more intricate notion of Lie semigroups and Lie wedges. Even worse, although---for closed systems---having access to all unitary channels is equivalent to controllability on the unitary orbit of all initial states (Lemma \ref{lemma_unit_state_cont_equiv}) one cannot hope for such a connection between states and the group lift for open systems.

Adding to the described gap between the state and the lifted problem, specifying reachable sets for dissipative systems is rather challenging and in higher-dimensional cases almost impossible. An example where one can find an upper bound is the case of unital dynamics \cite{Yuan10}: If $\Gamma\in\mathcal B(\mathcal B(\mathcal H))$ of \textsc{gksl}-form satisfies $\Gamma(\mathbbm{1}_{\mathcal H})=0$ (and given self-adjoint $H_0,H_1,\ldots,H_m\in\mathcal B(\mathcal H)$) then the generated semigroup $(e^{-it(\operatorname{ad}_{H_0}+\Gamma+\sum_{j=1}^mu_j\operatorname{ad}_{H_j})})_{t\geq 0}$ is bi-stochastic for all $u\in\mathbb R^m$. Thus 
$
\overline{\mathfrak{reach}(\rho_0)}\subseteq \{\rho\in\mathbb D(\mathcal H)\,|\,\rho\prec\rho_0\}
$
for arbitrary initial states $\rho_0\in\mathbb D(\mathcal H)$; this is a direct consequence of Lemma \ref{thm_ando_7_1} \& Thm.~\ref{lemma_R_d_closed} (if $\operatorname{dim}(\mathcal H)<\infty$) or Lemma \ref{lemma_li} \& Thm.~\ref{lemma_maj_closed} (for complex, separable $\mathcal H$). However, this characterization becomes increasingly inaccurate the larger the system in which case one has to resort to the mentioned Lie-semigroup tools \cite{ODS11}.

As an intermediate scenario one may consider \textit{switchable} noise\index{switchable noise}, meaning the dissipative term $\Gamma$ is really of the form $\gamma(t)\Gamma$ with the additional (bang-bang) control $\gamma(t)\in\{0,1\}$\label{symb_gamma_t}.
This scenario, while allowing for rigorous mathematical results, is also of physical interest as there are instances of unitarily controllable systems in which the noise can be switched in such a
bang-bang manner. An important experimental incarnation are superconducting qubits coupled to an open transmission line~\cite{Mart14}. 
Formulated as a bilinear control system this scenario reads as follows:
\begin{equation}\label{eq:q_control_switchable}
\dot\rho(t)=-i\Big[H_0+\sum\nolimits_{j=1}^m u_j(t) H_j,\rho(t)\Big]-\gamma(t)\Gamma(\rho(t))
\end{equation}
with $\rho(0)=\rho_0\in\mathbb D(\mathcal H)$ where
\begin{equation}\label{eq:Gamma_def}
\begin{split}
\Gamma:\mathcal B^1(\mathcal H)&\to\mathcal B^1(\mathcal H)\\
\rho&\mapsto\sum_{j\in I}\Big(\frac12 (V_j^*V_j\rho +\rho V_j^*V_j)-V_j\rho V_j^*\Big)
\end{split}
\end{equation}
and the $\{V_j\}_{j\in I}\subset\mathcal B(\mathcal H)$ are chosen such that $\sum_{j\in I}V_j^*V_j$ converges weakly to a bounded operator (cf.~Ch.~\ref{ch:bilinear_control} \& Thm.~\ref{thm_gksl}). In this case---assuming piecewise constant controls as usual---$\mathfrak{reach}(\rho_0)=S_\Omega\rho_0$ where $S_\Omega$ is the system semigroup generated by
\begin{align*}
&\{e^{-i\tau(\operatorname{ad}_{H_0}+\sum_{j=1}^mu_ji\operatorname{ad}_{H_j}+\gamma\Gamma)}\,|\,\tau\geq 0,u\in\Omega,\gamma\in\{0,1\}\}\\
=&\{e^{-i\tau(\operatorname{ad}_{H_0}+\sum_{j=1}^mu_j\operatorname{ad}_{H_j})}\,|\,\tau\geq 0,u\in\Omega\}\cup\{e^{-\tau(i\operatorname{ad}_{H_0}+\Gamma+\sum_{j=1}^mu_ji\operatorname{ad}_{H_j})}\,|\,\tau\geq 0,u\in\Omega\}\,.
\end{align*}
Therefore within this model one has access to unitary channels (e.g., controllability of the closed system $\gamma(t)\equiv 0$ pertains to the state problem) as well as dissipative dynamics, thus making obtaining analytical results feasible. 

Quantum control systems with switchable noise were first studied by Bergholm et al.~\cite{BSH16} who in the case of qubits (i.e.~$\mathcal H=\mathbb C^{2^n}$, $n\in\mathbb N$) obtained the following remarkable results:
\begin{itemize}
\item Consider $H_0,H_1,\ldots,H_m\in\mathbb C^{2^n\times 2^n}$ Hermitian and\footnote{
Here $\mathbbm{1}_2^{n-1}$ is short for the $(n-1)$-fold tensor product of $\mathbbm{1}_2$.
}
$V=\mathbbm{1}_2^{n-1}\otimes |e_1\rangle\langle e_2|$. Then, assuming $\langle iH_0, iH_j\,|\,j=1,\dots,m\rangle_\textsf{Lie}=\mathfrak u(2^n)$ (or $=\mathfrak{su}(2^n)$), the reachable set of
$$
\dot\rho(t)=-i\Big[H_0+\sum\nolimits_{j=1}^m u_j(t) H_j,\rho(t)\Big]-\gamma(t)\Big(\frac12 (V^*V\rho(t) +\rho(t) V^*V)-V\rho (t)V^*\Big)
$$
satisfies $\overline{\mathfrak{reach}(\rho_0)}=\mathbb D(\mathbb C^{2^n})$ for all $\rho_0\in \mathbb D(\mathbb C^{2^n})$. 
\item Consider $H_0,H_1,\ldots,H_m\in\mathbb C^{2^n\times 2^n}$ Hermitian and $V=\mathbbm{1}_2^{n-1}\otimes V'$ with $V'\in\mathbb C^{2\times 2}$ normal. Then, assuming $\langle iH_0, iH_j\,|\,j=1,\dots,m\rangle_\textsf{Lie}=\mathfrak u(2^n)$ (or $=\mathfrak{su}(2^n)$), the reachable set of
$$
\dot\rho(t)=-i\Big[H_0+\sum\nolimits_{j=1}^m u_j(t) H_j,\rho(t)\Big]-\gamma(t)\Big(\frac12 (V^*V\rho(t) +\rho (t)V^*V)-V\rho (t)V^*\Big)
$$
satisfies $\overline{\mathfrak{reach}(\rho_0)}= \{\rho\in\mathbb D(\mathbb C^{2^n})\,|\,\rho\prec\rho_0\}$ for all $\rho_0\in \mathbb D(\mathbb C^{2^n})$. 
\end{itemize}
Based on we will tackle the following questions in this chapter:
\begin{itemize}
\item Do these results hold not only for qubit but arbitrary $d$-level (``qudit'') systems, that is, $\mathcal H=\mathbb C^{d^n}$? If so, this would be non-trivial as the proof for $d=2$ is specifically geared to the qubit structure and breaks down in the general case.
\item Can the first result be generalized to general dissipation modelling coupling the system to a bath? This question is motivated by the fact that the semigroup induced by a single Lindblad-$V$ of the form $|e_1\rangle\langle e_2|$ satisfies $\lim_{t\to\infty}e^{-t\Gamma}(\rho_0)=|e_1\rangle\langle e_1|$ meaning it can be interpreted as a temperature zero bath\index{temperature zero bath} (cf.~\cite[Appendix B]{BSH16} and the next section).
\item Can one extend these results to infinite-dimensional systems (that is, general separable, complex Hilbert spaces $\mathcal H$)? After all quantum mechanics is an infinite-dimensional theory (cf.~Ch.~\ref{sec_unbounded_op}) and one ``only'' arrives at finite-dimensional systems after tracing out sufficiently many degrees of freedom such as position or momentum\footnote{
This is \textit{not} to say that quantum mechanics in finite dimensions is uninteresting or even
useless; indeed having, e.g., finitely many energy levels is due to quantization and thus inherently quantum. However, it is also a fact that finite-dimensional systems are approximations of something originally infinite-dimensional, e.g., by ignoring spatial degrees of freedom of a spin, and not every quantum system can be treated this way.\label{footnote_inf_dim_qc}
}.
\end{itemize}

\section{Finite Dimensions}\label{sec_reach_fin_dim}

This section will be primarily concerned with control systems based on bath couplings and is mainly contained in \cite{CDC19}.

Bath couplings\index{bath coupling} and asymptotic steady states\index{steady state} of (Markovian) open quantum systems were first investigated by Spohn \cite{Spohn76,Spohn77}, Frigerio \cite{Frigerio77,Frigerio78}, and Verri \cite{FrigerioVerri82} in the late 70s and were later refined by, e.g., Fagnola \cite{Fagnola98,Fagnola99,Fagnola2008}. Given a generator $L$ of a strongly continuous \textsc{qds} $(e^{tL})_{t\geq 0}$ this semigroup is said to be \textit{relaxing}\index{relaxing semigroup} if there exists $\rho_\infty\in \mathbb D(\mathcal H)$ such that
\begin{equation*}
\lim_{t\to\infty} e^{tL}(\rho) = \rho_\infty
\end{equation*}
for all $ \rho \in \mathbb D(\mathcal H)$.
Motivated by the spin-$j$ representation\index{spin-j representation@spin-$j$ representation}\footnote{
Recall that starting from the Pauli matrices
$$
\sigma_1= \begin{pmatrix} 0&1\\1&0 \end{pmatrix}\quad \sigma_2=\begin{pmatrix} 0&-i\\i&0 \end{pmatrix}\quad \sigma_3=\begin{pmatrix} 1&0\\0&-1\end{pmatrix}
$$
one finds $\mathfrak{su}(2)=\operatorname{span}_{\mathbb R}\{\frac{i}{2}\sigma_1,\frac{i}{2}\sigma_2,\frac{i}{2}\sigma_3\}$. Using the ladder operators
$$
\sigma_+ = \sum_{k=1}^{n-1}\sqrt{k(n-k)} |e_k\rangle\langle e_{k+1}|
\quad\text{and}\quad
\sigma_- = \sum_{k=1}^{n-1}\sqrt{k(n-k)} |e_{k+1}\rangle\langle e_k|
$$
as well as $\sigma_z=\sum_{j=1}^n (n+1-2j)|e_j\rangle\langle e_j|$ for arbitrary $n\in\mathbb N\setminus\{1\}$ one readily verifies that $\{\sigma_x,\sigma_y,\sigma_z\}:=\{\sigma_++\sigma_-,i\sigma_--i\sigma_+,\sigma_z\}$ satisfy the same commutation relations as the Pauli matrices, thus leading to a representation of $\mathfrak{su}(2)$. This is called the spin-$j$ representation where the ladder operators give rise to $n=2j+1$ levels for half-integer (fermionic)
and integer (bosonic) spin quantum numbers $j\in\{\tfrac{1}{2},1,\tfrac{3}{2},2,\dots\;\}$.\label{footnote_spin_j}
}
of $\mathfrak{su}(2)$, nilpotent matrices of the form $\sum\nolimits_{j=1}^{n-1}a_j |e_{j}\rangle\langle e_{j+1}|,\sum\nolimits_{j=1}^{n-1}b_j |e_{j+1}\rangle\langle e_{j}|$ will be the Lindblad $V$'s which describe bath couplings of our qudit system. We will see later that the resulting dynamics are relaxing and that the following assumption is valid:\medskip

\noindent
\textbf{Assumption IN:}\label{not_ass_in}
Given the generator of a strongly continuous \textsc{qds} $L\in\mathcal L(\mathbb C^{n\times n})$, the set of diagonal density matrices\footnote{
Recall from Ch.~\ref{sec:maj_d_vec} that the standard simplex $\Delta^{n-1}$ is the set of all probability vectors in $\mathbb R^n$.
}\label{symb_D_n_diag}
$$
\mathrm D(n):=\{ \operatorname{diag}(x)\in\mathbb R^{n\times n}\,|\, x\in\Delta^{n-1}\}
$$
is invariant under the semiflow $(e^{tL})_{t\geq 0}$. In this case $(e^{tB_0(L)})_{t\geq 0}$, generated by the corresponding matrix 
representation $B_0(L)\in\mathbb R^{n\times n}$ of the action of $L$ on the diagonal, leaves $\Delta^{n-1}$ invariant.\medskip

This assumption will be crucial as it allows us---due to unitary controllability---to study the reduced control problem on the diagonal states $\mathrm D(n)$ instead of the original problem on $\mathbb D(\mathbb C^n)$. 

\subsection{Dynamics of Coupling to Thermal Baths}\label{ch:coupling_bath}

As hinted at before, our first results yield a rich class of physically relevant models which are relaxing and satisfy 
the invariance condition IN. 

\begin{lemma}\label{lemma_semigroup_relaxing}
Let $n\in\mathbb N$ be arbitrary and consider
\begin{equation*}
N_+:=\sum\nolimits_{j=1}^{n-1}a_j |e_{j}\rangle\langle e_{j+1}| \quad\text{and}\quad N_-:=\sum\nolimits_{j=1}^{n-1}b_j |e_{j+1}\rangle\langle e_{j}|
\end{equation*}
with arbitrary $a_1,\ldots,a_{n-1},b_1,\ldots,b_{n-1}\in\mathbb R$ and $(e_j)_{j=1}^n$ being the standard basis
of $\mathbb C^n$. Then the operator $\Gamma_N$ induced by $V_1:=N_+$ and $V_2:=N_-$ via 
\eqref{eq:Gamma_def} satisfies the following:
\begin{itemize}
\item[(i)] $\Gamma_N$ fulfills \textrm{IN}. Moreover, its matrix representation 
on diagonal matrices (with respect to the standard identification $x \to \operatorname{diag}(x)$) is given by
\end{itemize}
\begin{equation}\label{eq:action_gamma_Delta}
B_0 = \sum_{j=1}^{n-1} a_j^2 |e_{j+1}-e_j\rangle\langle e_{j+1}| + b_j^2 |e_j-e_{j+1}\rangle\langle e_j|= \begin{pmatrix} b_1^2&-a_1^2&&\\-b_1^2&a_1^2+b_2^2&-a_2^2&\\&-b_2^2&a_2^2+b_3^2&-a_3^2\\&&-b_3^2&\ddots \end{pmatrix}\in\mathbb R^{n\times n}\,.
\end{equation}
\begin{itemize}
\item[(ii)] If $a_1,\ldots,a_{n-1},b_1,\ldots,b_{n-1} \neq 0$ then $B_0$ is relaxing on $\Delta^{n-1}$, i.e.~there exists 
unique $x_\infty\in\Delta^{n-1}$, $x_\infty>0$ such that $\lim_{t\to\infty}e^{-tB_0}x=x_\infty$ for all $x\in\Delta^{n-1}$.
\end{itemize}
\end{lemma}
\begin{proof}
Let $j,k\in\lbrace 1,\ldots,n\rbrace$ and $Y\in\mathbb C^{n\times n}$. A straightforward computation yields
\begin{align*}
\big(\Gamma_N(Y)\big)_{jk} = \frac12(a_{j-1}^2+a_{k-1}^2+b_j^2+b_k^2)Y_{jk} -a_ja_kY_{(j+1)(k+1)}-b_{j-1}b_{k-1}Y_{(j-1)(k-1)}\,.
\end{align*}
This readily implies (i).
Statement (ii) can be shown via the Perron-Frobenius theorem\index{theorem!Perron-Frobenius} as follows: Let $t>0$ be arbitrary. By 
\eqref{eq:action_gamma_Delta} there exists $c\in\mathbb R_+$ such that all entries
of $ct\mathbbm{1}_n-tB_0$ are non-negative (denoted by $ct\mathbbm{1}_n-tB_0 \geq 0)$. This is still true if we take any 
power of $ct\mathbbm{1}_n-tB_0$ and due to
$a_j,b_j \neq 0$, evidently, $(ct\mathbbm{1}_n-tB_0)^{n-1}>0$ (positive entries) so
$$
0<e^{ct\mathbbm{1}_n-tB_0}=e^{ct\mathbbm{1}_n}e^{-tB_0}=e^{ct}e^{-tB_0}
$$
and thus $e^{-tB_0}>0$. Furthermore, $e^{-tB_0}$ has spectral radius 
one; this follows from \cite[Thm.~8.1.22]{HJ1} due to $\unitvector^TB_0=0 $ which 
implies $\unitvector^T e^{-tB_0}=\unitvector^T $, i.e.~$e^{-tB_0}$ leaves $\Delta^{n-1}$ invariant.
Moreover, one can show \cite[Thm.~8.2.11]{HJ1} that $0$ is a simple eigenvalue and every
other eigenvalue of $-tB_0$ has strictly negative real part. Using the Jordan canonical form of $-tB_0$ 
this readily implies convergence of $e^{-t B_0}$ to a matrix of rank one as $t\to\infty$. 
By an argument similar to the one given in Lemma \ref{lemma_row_conv} there exists $x_\infty\in\Delta^{n-1}$, $x_\infty>0$ such that $ e^{-tB_0}\to |x_\infty\rangle\langle \unitvector|$ as $t\to\infty$, cf.~\cite[Thm.~8.2.11]{HJ1}.
\end{proof}
\noindent Equivalent to (ii) of the previous lemma is the statement that $\Gamma_N$ is relaxing on $\mathrm D(n)$. 
 In fact, one can show (by means of \cite{Fagnola2015}) that $\Gamma_N$ is actually 
relaxing on all of $\mathbb D(\mathbb C^n)$.

\begin{remark}\label{rem_gamma_B0}
Because $-\Gamma_N$ is the generator of a strongly continuous \textsc{qds} (so in particular positive and trace-preserving) for which assumption IN holds, the matrix $B_0$ from \eqref{eq:action_gamma_Delta} for all $a_j,b_j$ gives rise to a semigroup $(e^{-tB_0})_{t\geq 0}$ of column-stochastic matrices. This is due to the one-to-one correspondence of the action of $\Gamma_N$ on diagonal states and the action of $B_0$ on vectors. This is also where the additional minus sign comes from (compared to Assumption IN) as $B_0(L)=B_0(-\Gamma)=-B_0(\Gamma)$.
\end{remark}

A special case not covered by the previous lemma, namely $b_1=\ldots=b_{n-1}=0$ (while $a_1,\ldots,a_{n-1}\neq 0$), still leads to a relaxing semigroup:

\begin{lemma}\label{lemma_row_conv}
Let $n\in\mathbb N$ and $c_1,\ldots,c_{n-1}>0$. Then for
\begin{equation}\label{eq:matrix_A}
A:=\begin{pmatrix}
0&-c_1&0&\ldots&0 \\
0&c_1&-c_2&\ddots&\vdots \\
\vdots&\ddots&c_2&\ddots&0 \\
\vdots & &\ddots &\ddots&-c_{n-1} \\
0&\ldots&\ldots&0&c_{n-1}
\end{pmatrix}\in\mathbb R^{n\times n}\,,
\end{equation}
one has $\lim_{t\to\infty}e^{-tA} = |e_1\rangle\langle \unitvector|$, so the resulting matrix has ones in the first
row and all other entries are zero.
\end{lemma}

\begin{proof}
Obviously the above statement is related to (but is not a special case of) Lemma \ref{lemma_semigroup_relaxing}.
Consider the following block-decomposition
\begin{equation*}
A = \begin{pmatrix} 0& A_{12}\\ 0&A_{22}\end{pmatrix} \quad\text{with}\quad 
A_{22}\in\mathbb R^{(n-1) \times (n-1)}
\end{equation*}
and note that $t \mapsto \Phi(t) := e^{-tA}$ satisfies the \textsc{ode} $\dot{\Phi}(t) = -A \Phi(t)$
with $\Phi(0) = \mathbbm{1}_n$. Now decomposing $\Phi(t)$ in the same way as $A$ and taking into account
that $\Phi(t)$ satisfies the above \textsc{ode} readily yields the following representation
\begin{equation*}
\Phi(t) = \begin{pmatrix} \Phi_{11}(t) & \Phi_{12}(t)\\0 & \Phi_{22}(t) \end{pmatrix} 
\end{equation*}
with $\Phi_{22}(t) = e^{-tA_{22}}$ and $\Phi_{11}(t) = 1$. Finally, via the variation of parameters formula we obtain
\begin{align*}
\Phi_{12}(t) = -\int_0^t A_{12} e^{-(t-s)A_{22}}\,\textrm{d}s = -A_{12} \big[A^{-1}_{22}e^{-(t-s)A_{22}}\big]_{s=0}^{s=t} = - A_{12}A^{-1}_{22} + A_{12}A^{-1}_{22}e^{-tA_{22}}\,.
\end{align*}
As $-A_{22}$ is obviously a Hurwitz matrix we conclude 
\begin{equation*}
\lim_{t\to\infty}e^{-tA} = \lim_{t\to\infty} \Phi(t) = 
\begin{pmatrix} 1 & -A_{12} A^{-1}_{22} \\0 & 0\end{pmatrix} 
\end{equation*}
and the identity $\unitvector^T A = 0$ implies the desired result.
\end{proof}

\begin{corollary}\label{coro_temp_zero}
Let $a_1,\ldots,a_{n-1}\in\mathbb R\setminus\{0\}$ be given. Then $\Gamma_0$ induced by $V=\sum\nolimits_{j=1}^{n-1}a_j |e_{j}\rangle\langle e_{j+1}|$ via 
\eqref{eq:Gamma_def} satisfies the following:
\begin{itemize}
\item[(i)] Its matrix representation $B_0$ 
on diagonal matrices (with respect to the standard identification $x \to \operatorname{diag}(x)$) is relaxing on $\Delta^{n-1}$ with steady state $e_1$.
\item[(ii)] One has $\lim_{t\to\infty}e^{-t\Gamma_0}=\operatorname{tr}(\cdot)|e_1\rangle\langle e_1|$, that is, $e^{-t\Gamma_0}(\rho_0)$ converges to $|e_1\rangle\langle e_1|$ for all $\rho_0\in\mathbb D(\mathbb C^n)$ as $t\to\infty$.
\end{itemize}
\end{corollary}
\begin{proof}
(i): Lemma \ref{lemma_semigroup_relaxing} (i) \& \ref{lemma_row_conv}. (ii): By (i) the diagonal of an arbitrary initial state $\rho_0\in\mathbb D(\mathbb C^n)$ converges to $e_1$ and because the set of quantum states is closed the limit has to be positive semi-definite, hence the semigroup relaxes $\rho_0$ into $|e_1\rangle\langle e_1|$. Using $\mathbb C^{n\times n}=\operatorname{span}_{\mathbb C}(\mathbb D(\mathbb C^n))$ we find $e^{-t\Gamma_0}\to \operatorname{tr}(\cdot)|e_1\rangle\langle e_1|$ as $t\to\infty$ in s.o.t., but in finite dimensions this is the same as norm convergence (Prop.~\ref{prop_strong_weak_op_top} (iv)). 
\end{proof}


To connect these relaxation properties to quantum systems \& bath couplings let
\begin{equation*}
\sigma_+ := \sum\nolimits_{j=1}^{n-1}\sqrt{j(n-j)} |e_j\rangle\langle e_{j+1}|
\quad\text{and}\quad
\sigma_- := (\sigma_+)^T
\end{equation*}
be the ladder operators in spin-$j$ representation\index{spin-j representation@spin-$j$ representation} (cf.~footnote \ref{footnote_spin_j}). After slightly modifying these operators they generate a relaxing semigroup according to a given steady state:

\begin{proposition}\label{thm_bath}
Let $n\in\mathbb N$ and $d\in\Delta^{n-1}$, $d>0$. Moreover, consider
\begin{equation*}
\sigma_+^d:=\sum\nolimits_{j=1}^{n-1}\sqrt{j(n-j)}\cos(\theta_j) |e_{j}\rangle\langle e_{j+1}|
\end{equation*}
and
\begin{equation*}
\sigma_-^d:=\sum\nolimits_{j=1}^{n-1}\sqrt{j(n-j)}\sin(\theta_j) |e_{j+1}\rangle\langle e_{j}|\,,
\end{equation*}
where
\begin{equation}\label{eq:thermal_angle}
\theta_j:=\arccos\Big(\big({1+\frac{d_{j+1}}{d_j}}\big)^{-\frac{1}{2}} \Big)\in\Big(0,\frac{\pi}{2}\Big)
\end{equation}
for $j= 1,\ldots,n-1$. Then $\Gamma_{d}$ induced by $V_1:=\sigma_+^d$ and $V_2:=\sigma_-^d$ via 
\eqref{eq:Gamma_def} satisfies \textrm{IN} and the generated semigroup $(e^{-tB_0(\Gamma_d)})_{t\geq 0}$ is relaxing on $\Delta^{n-1}$ into $d$.
\end{proposition}

\begin{proof}
Apply Lemma \ref{lemma_semigroup_relaxing} with
$$
a_j:=\sqrt{\frac{j(n-j)d_j}{d_j+d_{j+1}}} \quad\text{and}\quad b_j:=\sqrt{\frac{j(n-j)d_{j+1}}{d_j+d_{j+1}}}
$$
for all $j= 1,\ldots,n-1$. One readily verifies
$d\in\operatorname{ker}(B_0)$ which implies that $d$ is the 
unique attractive fixed point of $(e^{-tB_0})_{t\geq 0}$. By means of \eqref{eq:action_gamma_Delta}
one immediately gets $B_0d=0$. 
\end{proof}

Recall how the temperature $T > 0$ given as a macroscopic parameter of a bath relates to the 
equilibrium state $\rho_\textsf{Gibbs}$ (cf.~Rem.~\ref{rem_gibbs}) of an $n$-level quantum system with Hamiltonian $H_0$ once the system is `\/opened\/' by coupling it to the bath and letting it equilibrate.
The equilibration itself can be described as a Markovian relaxation process following the
\textsc{gksl}-equation with $V_1:=\sigma_+^d$ and $V_2:=\sigma_-^d$ from Prop.~\ref{thm_bath}.
To this end, $\sigma_+^d$ and $\sigma_-^d$ are designed to guarantee that $\rho_\textsf{Gibbs}=\operatorname{diag}(d)$ is
the unique fixed point\footnote{
Of course, if $H_0$ and thus $\rho_\textsf{Gibbs}$ are not diagonal in the standard basis one has to
adjust the construction of $\sigma_+^d$ and $\sigma_-^d$ by replacing $e_j$ by the corresponding
eigenvector to $H_0$. More precisely if $(g_j)_{j=1}^n$ is an orthonormal basis of $\mathbb C^n$ such that $H_0=\sum_{j=1}^n E_j|g_j\rangle\langle g_j|$ with $E_1\leq \ldots\leq E_n$ then $\sigma_+^d=\sum\nolimits_{j=1}^{n-1}\sqrt{j(n-j)}\cos(\theta_j) |g_{j}\rangle\langle g_{j+1}|$ and $\sigma_-^d:=\sum\nolimits_{j=1}^{n-1}\sqrt{j(n-j)}\sin(\theta_j) |g_{j+1}\rangle\langle g_{j}|$. 
However $B_0$ is the matrix representation of $\Gamma$ \textit{with respect to this basis} $(g_j)_{j=1}^n$ so \eqref{eq:action_gamma_Delta} is valid either way. Thus we may assume, also due to unitary controllability, that w.l.o.g.~$H_0=\operatorname{diag}(E_1,\ldots,E_n)$ with $E_1\leq\ldots\leq E_n$. 
\label{footnote_H_0_gen_diag}}
of the equilibration. 
Roughly speaking, $\sigma_+^d,\sigma_-^d$ can be interpreted to model the transition rates between
neighbouring energy levels. For this to work without ``physically'' forbidden jumps we have to require
that the energy levels (eigenvalues) $E_1,\ldots,E_n$ of $H_0$ and thus the resulting Gibbs vector\index{Gibbs vector} $d=(\frac{e^{-E_j/T}}{\operatorname{tr}(e^{-H_0/T})})_{j=1}^n\in\Delta^{n-1}$ are ordered: 
W.l.o.g.~we assume $E_1\leq\ldots\leq E_n$ to be increasing and therefore $d_1\geq d_2\geq\ldots\geq d_n$
to be decreasing.

In the sequel, we want to analyse how $(e^{-t\Gamma_d})_{t\geq 0}$ (cf.~Prop.~\ref{thm_bath}) behaves for
different choices of the Gibbs vector, that is, for different $H_0$ and at different temperatures $T$. 
The following scenarios are of special interest:\medskip

\noindent
\textbf{Equidistant energy levels:}\index{equidistant energy levels} If the neighbouring ratios $\frac{d_{k+1}}{d_{k}}$ 
are constant for all $k$ (which obviously corresponds to equidistant energy levels $E_k$) so $\theta_k=\text{const}=:\theta$ in \eqref{eq:thermal_angle}, then the generators $\sigma_+^d,\sigma_-^d$ become $\cos(\theta)\sigma_+$, 
$\sin(\theta)\sigma_-$.\medskip

\noindent
\textbf{High-temperature limit:}\index{high-temperature limit} The case $d = \unitvector/n$ (obtained via taking the limit $T\to\infty$ of $d=d(T)$) yields $\cos(\theta_k)=\sin(\theta_k)=\frac{1}{\sqrt{2}}$ for all $k= 1,\ldots,n-1$ so the generators $\sigma_+^d,\sigma_-^d$ 
become $\sigma_+,\sigma_-$ (up to a global factor).\medskip 

\noindent 
\textbf{Low-temperature limit:}\index{low-temperature limit} If the entries of $d$ are sorted and distinct, i.e.~$d_1>d_2>\ldots>d_n$, then $d$ becomes $e_1$ when taking the limit $T\to 0^+$ of $d=d(T)$---hence $\sigma_-^d\to 0$ and $\sigma_+^d\to \sigma_+$ so it is enough to consider only one generator. With this in mind the result of Coro.~\ref{coro_temp_zero} can be interpreted as relaxation properties of a temperature zero bath\index{temperature zero bath}.\medskip

\begin{remark}[Application to physics]\label{rem_control_born_markov}
The results on reachable sets we will derive are mathematically rigorous and, \textit{in principle}, independent of any physical application.\index{approximation!Born-Markov}\index{approximation!rotating wave}
Of course when implementing said results in an experiment one has to justify that the underlying dynamics follow a Markovian evolution. 
The basic assumptions usually made here are
\begin{itemize}
\item the Born approximation (also weak-coupling limit\index{weak-coupling limit}), which assumes that the influence of the system in question on the reservoir is small.
\item the Markov approximation, which requires that the time scale over which the state dissipates is large compared to the reservoir correlation time.
\item the rotating wave approximation, which holds if the time scale of the system's dynamics are small compared to relaxation of the system.
\end{itemize}
In addition, for control problems one requires that the control time scale is also notably larger than the system's dynamics.
For short, one requires little influence of the system on the environment and an appropriate separation of time scales. For a more detailed discussion of these approximations we refer to \cite[Ch.~3.3]{BreuPetr02} and \cite[Appendix C]{BSH16}.
\end{remark}

\subsection{Specification of the Toy Model}\label{sec:toy_model}

Assume, here and henceforth, that the control system \eqref{eq:q_control_switchable} for some $H_0\in\mathbb C^{n\times n}$ and some $\Gamma$ of form \eqref{eq:Gamma_def} satisfies assumption IN. Considering \eqref{eq:q_control_switchable} solely on $ D(n)$ then leads to the following core problem---dubbed \/`\textit{toy model}\/'\index{toy model} henceforth---%
on the standard simplex: 
Its controls shall amount to permutation matrices acting
instantaneously on the entries of $x(t)$ (precisely the unitary channels which are compatible with $D(n)$ when $\gamma(t)=0$) and a continuous-time one-parameter semigroup $(e^{-tB_0})_{t\in\mathbb R^+}$
of stochastic maps (because $D(n)\in\operatorname{ker}(\operatorname{ad}_{H_0})$ so if $u(t)=0$ there is only the dissipation induced by $\Gamma$).
%
More precisely, these stipulations
suggest the following hybrid/impulsive toy model $\Lambda$ on $\Delta^{n-1} \subset \mathbb R^n$, 
cf.~\cite{book_impulsive89,Leela1991,book_HybridSytems96}: 
\begin{equation}\label{eq:control-simplex_evolution}
\begin{split}
&\dot{x}(t) = -B_0 x(t)\,,\quad x(t_k) = x_k\,, \quad t \in [t_k,t_{k+1})\,,\\
& x_0 \in \Delta^{n-1}\,, \quad x_{k+1} = \pi_k e^{-(t_{k+1}-t_k)B_0}x_k\,, \quad k\geq 0\,,
\end{split}
\end{equation}
where the upper line describes the continuous-time evolution and the lower line the discrete-time part.
The switching sequence $0 =: t_0 \leq t_1 \leq t_2 \leq \dots$ and the permutation matrices $\pi_k$ are regarded
as controls for \eqref{eq:control-simplex_evolution}. For simplicity, we assume that the switching points do not 
accumulate on finite intervals. The reachable sets of $\Lambda$\label{symb_reach_lambda}
\begin{equation*}
\mathfrak{reach}_\Lambda(x_0) := \{x(t) \,|\,
\text{$x(\cdot)$ is a solution of \eqref{eq:control-simplex_evolution}, $t \geq 0$}\}
\end{equation*}
allow for the characterisation
$
\mathfrak{reach}_\Lambda(x_0) = {\mathcal S}_\Lambda x_0\,,
$
where ${\mathcal S}_\Lambda \subset \mathbb R^{n\times n}$ is the contraction semigroup generated 
by $(e^{-tB_0})_{t\geq 0}$ and the set of all permutation matrices $\pi$. Indeed if $x_0\in\Delta^{n-1}$ then $\mathfrak{reach}_\Lambda(x_0) \subseteq\Delta^{n-1}$ due to column-stochasticity of $(e^{-tB_0})_{t\in\mathbb R^+}$.

Now if the unitary part of \eqref{eq:q_control_switchable} is controllable, that is, $\langle iH_0,iH_j\,|\,j=1,\dots, m\rangle_\textsf{Lie} = \mathfrak{u}(n)$ or $=\mathfrak{su}(n)$ then\footnote{
Note that system \eqref{eq:q_control_switchable} with $\langle iH_0,iH_j\,|\,j=1,\dots, m\rangle_\textsf{Lie} = \mathfrak{u}(n)$ or $=\mathfrak{su}(n)$ is mathematically equivalent to full Hamiltonian controllability, i.e.~$\langle iH_j\,|\,j=1,\dots, m\rangle_\textsf{Lie} = \mathfrak{u}(n)$ or $=\mathfrak{su}(n)$ together with $\gamma\equiv 1$ \cite{DHKS08}. However, we will sweep this scenario under the rug because for many experiments it is hopelessly idealising.
}
for all $x_0\in\Delta^{n-1}$ and all $V\in\mathbb C^{n\times n}$ unitary
\begin{equation}\label{eq:toy_state_connect}
\{U\operatorname{diag}(x) U^*\,|\,U\in\mathbb C^{n\times n}\text{ unitary},x\in \mathfrak{reach}_\Lambda(x_0)\}\subseteq\mathfrak{reach}(V\operatorname{diag}(x_0)V^*)\,.
\end{equation}
This is obvious from the very construction of the toy model. 

Other authors used quite similar ideas to investigate reachable sets of quantum-dynamical control systems
\cite{Khaneja01b,Yuan10,rooney2018}. In particular, in \cite{rooney2018} the authors restrict themselves 
to a subsimplex of the standard simplex (which results from a Weyl-chamber type of construction) in order 
to eliminate ambiguities which result from different orderings of the eigenvalues of a density matrix. 
Moreover, their setting is more general as they avoid the invariance condition IN. However, the resulting 
conditions are hard to verify for higher-dimensional systems.\medskip


%

For $n\in\mathbb N$, consider the $n$-level toy model $\Lambda_0$ 
with controls by permutations as above and an infinitesimal generator $B_0$ which results from a
dissipative coupling to a bath of temperature $T=0$
(i.e.~$\Gamma =\Gamma_0$ from Coro.~\ref{coro_temp_zero} with $a_j=\sqrt{j(n-j)}$; then $V=\sigma_+$ as suggested by the low-temperature limit at the end of Ch.~\ref{ch:coupling_bath}). Our first result reads as follows:

\begin{theorem}\label{thm_1}
Let $n\in\mathbb N$ be arbitrary and consider $\Gamma_0$ induced by a
single generator $\sigma_+$ via \eqref{eq:Gamma_def}. Then for the toy model $\Lambda_0$
with $B_0(\Gamma_0)$, the closure of the reachable set of any initial state $x_0 \in \Delta^{n-1}$ exhausts the whole
standard simplex, i.e.
\begin{equation*}
\overline{\mathfrak{reach}_{\Lambda_0}(x_0)}=\Delta^{n-1}\,.
\end{equation*}
\end{theorem}

The idea will be to first cool the system, that is, to steer from $x_0$ to $e_1$ via dissipation and then connect the ground state $e_1$ to every other $x\in\Delta^{n-1}$. For this second step we will find a path backwards---meaning from $x$ to $e_1$---which brings on the following auxiliary result:

\begin{lemma}\label{lemma_for_thm_temp_0}
Let $n\in\mathbb N$ be arbitrary and let $A\in\mathbb R^{n\times n}$ be given by \eqref{eq:matrix_A} for 
$c_1,\ldots,c_{n-1}>0$. 
Then for any $x\in\Delta^{n-1}$ there exist $t_1,\ldots,t_{n-1}\geq 0$ and permutation matrices
$\pi_1,\ldots,\pi_{n-1}\in\mathbb R^{n\times n}$ such that
$$
\big(e^{-t_{n-1}A}\pi_{n-1}\ldots e^{-t_1A}\pi_1\big)e_1 = x\,.
$$
\end{lemma}
\begin{proof}
Note that $\unitvector^TA=0$ guarantees invariance of the hyperplane $\unitvector^T x =1$ under the flow
$(e^{-tA})_{t\in\mathbb R}$. Moreover, due to the upper triangular structure of $A$, lower-dimensional faces of
$\Delta^{n-1}$ of the form
\begin{align*}
\Delta^{n-1}_{m-1} := \Delta^{m-1}\times \{0_{n-m}\} =\lbrace (y,0_{n-m}) \,|\,y\in\Delta^{m-1}\rbrace 
\end{align*}
are left invariant, too. Now, for $x \neq e_1$ one can consider the backward evolution of $x\in\Delta^{n-1}$
and check that, eventually, the trajectory hits a face of $\Delta^{n-1}$ which can be rotated into 
$\Delta^{n-1}_{n-2}\simeq\Delta^{n-2}$ via some permutation $\pi_{n-1}$: If $x_n=0$ then this is trivial (choose $\pi_{n-1}=\mathbbm{1}_n$) so w.l.o.g.~$x_n>0$. In this case $(e^{tA}x)_n=(e^{tA})_{nn}x_n=e^{tc_{n-1}}x_n\to\infty$ as $t\to\infty$ because $c_{n-1},x_n>0$. But as stated before $\unitvector^T(e^{tA} x) =1$ at all times to there exists $\tau>0$ and $j=1,\ldots,n-1$ such that $(e^{\tau A}x)_j<0$. Using continuity one thus finds $t_{n-1}\geq 0$ such that $e^{t_{n-1}A}x\in\Delta^{n-1}$ and $(e^{t_{n-1}A}x)_j=0$ (by means of the intermediate value theorem).

Applying this procedure inductively 
$n-1$ times---which is possible due to the upper triangular structure of $A$---concludes the proof.
\end{proof}

\begin{proof}[Proof of Thm.~\ref{thm_1}]
By Lemma \ref{lemma_semigroup_relaxing}
\begin{equation}\label{eq:6b}
B_0(\Gamma_0)=\sum\nolimits_{j=1}^{n-1} j(n-j) |e_{j+1}-e_j\rangle\langle e_{j+1}|
\end{equation}
so we may apply Lemma \ref{lemma_row_conv} and \ref{lemma_for_thm_temp_0} to $B_0$. Trivially $\overline{\mathfrak{reach}_\Lambda(x_0)} \subseteq\overline{\Delta^{n-1}}=\Delta^{n-1}$ so we only have to show the converse.

Lemma \ref{lemma_row_conv} in particular shows that for arbitrary $y\in\mathbb R^n$
one has $
e^{-t B_0}(y)\to(\unitvector^T y )e_1$ as $t\to\infty$ so given $\varepsilon>0$ we find $\tau\geq 0$ such that
\begin{equation}\label{eq:relax_approx_1}
\|(\unitvector^T y)e_1-e^{-\tau B_0}(y)\|_1<\varepsilon\,.
\end{equation}

Now let $x_0,x\in\Delta^{n-1}$. As stated before, the remaining proof consists of the following steps:
\begin{equation*}
x_0 \overset{\text{Step }1}\longrightarrow e_1 \overset{\text{Step }2}\longrightarrow x\,.
\end{equation*}
For arbitrary $\varepsilon>0$ let $\tau\geq 0$ be a time such that \eqref{eq:relax_approx_1} holds (for $y=x_0$). By Lemma \ref{lemma_for_thm_temp_0} one finds times $t_1,\ldots,t_{n-1}\geq 0$ and permutation matrices
$\pi_1,\ldots,\pi_{n-1}\in\mathbb R^{n\times n}$ such that
$
(e^{-t_{n-1}B_0}\pi_{n-1}\ldots e^{-t_1B_0}\pi_1)e_1 = x$; with this we define
$$
x_F:=e^{-t_{n-1}B_0}\pi_{n-1}\ldots e^{-t_1B_0}\pi_1 e^{-\tau B_0}x_0\,.
$$
Obviously $ x_F\in \mathfrak{reach}_{\Lambda_0}( x_0)$ so we only have to show $\| x- x_F\|_1<\varepsilon$. But this is due to the following simple computation:
\begin{align*}
\|x-x_F\|_1&\leq\big\|\big(e^{-t_{n-1}B_0}\pi_{n-1}\ldots e^{-t_1B_0}\pi_1\big)e_1-\big(e^{-t_{n-1}B_0}\pi_{n-1}\ldots e^{-t_1B_0}\pi_1 \big)e^{-\tau B_0}x_0\big\|_1\\
&\leq \prod\nolimits_{j=1}^{n-1}\underbrace{\|e^{-t_jB_0}\|_{\textrm{op}}\|\pi_j\|_{\textrm{op}}}_{=1}\|e_1-e^{-\tau B_0}x_0\|<\varepsilon\,.
\end{align*}
In the last step we used \eqref{eq:doubly_stoch_trace_norm} due to column-stochasticity of $e^{-tB_0}$ for all $t\geq 0$, as well as of all permutation matrices (cf.~also Rem.~\ref{rem_gamma_B0}). 
Because $\varepsilon>0$ was chosen arbitrarily this shows $x\in \overline{\mathfrak{reach}_{\Lambda_0}( x_0)}$ which concludes the proof. 
\end{proof}

\begin{remark}\label{rem_reach_exact}
Be aware that Step 2 in the proof of Thm.~\ref{thm_1} is ``exact'' in the sense that starting from the ground
state $e_1$ (in the model $\Lambda_0$), one can reach every element of $\Delta^{n-1}$ in finite time and, moreover, in at most $n-1$ switches (permutations). In other words Lemma \ref{lemma_for_thm_temp_0} shows $\mathfrak{reach}_{\Lambda_0}(e_1)=\Delta^{n-1}$.
\end{remark}

With this in mind let us consider \textit{local} noise of temperature zero and a finite number of qudits, i.e.~a ``chain'' 
of $n$-level systems (of length $m$) and only one (say, the last) qudit is coupled to the bath. Mathematically this amounts to one Lindblad generator of the form $ \mathbbm{1}\otimes \sigma_+$ in \eqref{eq:Gamma_def}.

\begin{theorem}\label{thm_2}
Let $m,n\in\mathbb N$ be arbitrary and let $\Gamma_{0,\textsf{loc}}$ 
be solely generated by $ \mathbbm{1}_{n^{m-1}}\otimes\sigma_+$ via \eqref{eq:Gamma_def}.
Then for the corresponding toy model $\Lambda_{0,\textsf{loc}}$ with $B_0(\Gamma_{0,\textsf{loc}})$,
the closure of the reachable set of any initial state $ x_0\in\Delta^{n^m-1}$ exhausts the whole standard simplex, i.e.
$$
\overline{\mathfrak{reach}_{\Lambda_{0,\textsf{loc}}}( x_0)}=\Delta^{n^m-1}\,.
$$

\end{theorem}

For the proof of this theorem the following auxiliary result is of importance; it will tell us that combining such exact models in a block-diagonal way does not limit the reachable set:

\begin{lemma}\label{lemma_reach_oplus}
Let $k\in\mathbb N$, $\alpha_1,\ldots,\alpha_k\in\mathbb N\setminus\{1\}$, and generators of one-parameter semigroups of column-stochastic matrices $Y_j\in\mathbb R^{\alpha_j\times\alpha_j}$
for $j=1,\ldots,k$ be given. Consider the toy models $\Lambda_1,\ldots,\Lambda_k$ obtained 
by replacing $B_0$ with $Y_1,\ldots,Y_k$, respectively, and 
assume the following:
\begin{itemize}
\item[(i)] Starting from the ground state of the individual systems, every other state can be reached (in finite time).
More precisely, $\mathfrak{reach}_{\Lambda_j}(e_1)=\Delta^{\alpha_j-1}$ for all $j$.
\item[(ii)] $Y_je_1=0$ for all $j=1,\ldots,k$.
\end{itemize}
Then the toy model $\Lambda_{\textsf{diag}}$ with
$$B_0 :=\operatorname{diag}(Y_1,Y_2,\ldots,Y_k)
\in\mathbb R^{(\alpha_1+\ldots+\alpha_k)\times (\alpha_1+\ldots+\alpha_k)}$$
admits
\begin{equation*}
\mathfrak{reach}_{\Lambda_\textsf{diag}}(e_1)=\Delta^{\alpha_1+\ldots+\alpha_k-1}\,.
\end{equation*}
\end{lemma}
\begin{proof}
First $k=1$ is obvious so let us assume $k=2$. Note that starting from $e_{1}$ one can reach every state 
of the form $(r e_{1},(1-r)e_{1})\in\mathbb R^{\alpha_1}\times\mathbb R^{\alpha_2}=\mathbb R^{\alpha_1+\alpha_2}$ with $r\in [0,1]$. This is easily
achieved via (i) and appropriate permutations.
%
%
Secondly, consider an arbitrary target $ x\in\Delta^{\alpha_1+\alpha_2-1}$ which of course can be 
decomposed into $ x= (x_1,x_2)$ with $ x_j\in\mathbb R_+^{\alpha_j}$.
Again by (i) we know that there exist switching sequences and permutations such that
the dissipation operator $Y_j$ interlaced with these permutations drives 
$(\unitvector^T x_j)e_{1}$ to $ x_j$ in time $t_j\in\mathbb R_+$ for $j=1,2$. 
Assume w.l.o.g.~$t_1\geq t_2$. 

Then starting from
$((\unitvector^T x_1)e_{1}, (\unitvector^T x_2)e_{1})$ 
the control scheme goes as follows: 
Run on $(\unitvector^T x_1)e_{1}$ the switching sequence which steers to 
$ x_1$ in time $t_1$. Stay in $(\unitvector^T x_2)e_{1}$ till $(t_1-t_2)$ which is 
possible by (ii),  and then---for the remaining time---run in parallel on the second system the (shifted) switching sequence which steers to 
$ x_2$. Thus at time $t_1$ we reach $ (x_1, x_2)= x$.

Now for arbitrary $k>2$---assuming we already proved the statement for $k-1$---we can decompose $Y_1\oplus\ldots\oplus Y_m=\tilde Y_1\oplus \tilde Y_2$
 where $\tilde Y_1:=Y_1\oplus\ldots\oplus Y_{m-1}$, $\tilde Y_2:= Y_m$. Note that $\tilde Y_1,\tilde Y_2$ 
 satisfy (i) and (ii) due to the initial assumptions on the individual matrices as well as the induction hypothesis so our argument for $k=2$ concludes the proof.
\end{proof}

\begin{proof}[Proof of Thm.~\ref{thm_2}]
The case $n=1$ is covered by Thm.~\ref{thm_1} so we may assume $n>1$. 

Let $\varepsilon>0$ and $ x_0,x \in\Delta^{n^m-1}$. We have to find $ x_F\in \mathfrak{reach}_{\Lambda_{0,\textsf{loc}}}
( x_0)$ such that $\| x- x_F\|_1<\varepsilon$. The proof, similar to that of Thm.~\ref{thm_1}, 
consists of the following steps:
\begin{equation}\label{eq:steps_idea_2}
 x_0\overset{\text{Step }1}\longrightarrow e_{1} \overset{\text{Step }2}
\longrightarrow x\,.
\end{equation}


For applying Lemma \ref{lemma_reach_oplus} in Step 2 check that
$Y_j=B_0(\Gamma_0)$ from \eqref{eq:6b} for $j =1, \dots, n^{m-1}$
satisfies conditions (i) and (ii), which obviously hold due to Thm.~\ref{thm_1}, Remark 
\ref{rem_reach_exact}, and Eq.~\eqref{eq:6b}\footnote{Here we use $\mathbbm{1}_k\otimes\sigma_+ = \operatorname{diag}(\sigma_+,\ldots,\sigma_+)$ which for any $X\in\mathrm D(kn)$ (when decomposed into $X=\operatorname{diag}(X_1,\ldots,X_k)$) implies $\Gamma_{\mathbbm{1}_k\otimes\sigma_+}(X)=\operatorname{diag}(\Gamma_0(X_1),\ldots,\Gamma_0(X_k))$ and thus $B_0(\Gamma_{\mathbbm{1}_k\otimes\sigma_+})= \operatorname{diag}(B_0(\Gamma_0),\ldots,B_0(\Gamma_0))$ as is readily verified.}.
%
%
%
Thus we know $\mathfrak{reach}_{\Lambda_{0,\textsf{loc}}}(e_1)=\Delta^{n^m-1}$ and in 
particular $ x\in \mathfrak{reach}_{\Lambda_{0,\textsf{loc}}}(e_1)$. 

For the first step in \eqref{eq:steps_idea_2}, we may decompose $ x_0$ into $( x_{1},\ldots,x_{n^{m-1}})$ with $ x_j\in\mathbb R_+^{n}$. Then
\begin{align*}
\lim_{t\to\infty}e^{-tB_0}x_0=\lim_{t\to\infty}\big(e^{-tB_0(\Gamma_0)} x_{1},\ldots,e^{-tB_0(\Gamma_0)} x_{n^{m-1}}\big)=\big( (\unitvector^T x_{1})e_1, \ldots, (\unitvector^T x_{n^{m-1}})e_1\big)
\end{align*}
by Lemma \ref{lemma_row_conv} so applying an appropriate permutation yields
$$
\big( \unitvector^T x_{1},\ldots,\unitvector^T x_{n^{m-1}},0_{n^m-n^{m-1}}\big)\in\Delta^{n^m-1}\,.
$$
Repeating this scheme $m$ times in total leaves us with
$$
\Big( \sum\nolimits_{j=1}^{n^{m-1}}\unitvector^T x_{j},0_{n^m-1}\Big)=e_1
$$
because $ x_0\in\Delta^{n^m-1}$ so $\unitvector^T x_0=\sum_{j=1}^{n^{m-1}}\unitvector^T x_{j}=1$. 
Clearly, above limits (for $t\to\infty$) have to be approximated, that is, for every $y\in\mathbb R^{n}$ we find 
$\tau \geq 0$ such that
$$
\|(\unitvector^T y)e_1-e^{-\tau B_0(\Gamma_0)}(y)\|_1<\tfrac{\varepsilon}{m\cdot n^{m-1}}\,.
$$
Just like in the proof of Thm.~\ref{thm_1} one then finds $ x_F\in \mathfrak{reach}_{\Lambda_{0,\textsf{loc}}}( x_0)$ with 
$$
\|e_1- x_F\|_1<m\cdot\big( n^{m-1}\cdot\tfrac{\varepsilon}{m\cdot n^{m-1}}\big)=\varepsilon
$$ 
as each of the $m$ relaxation steps has precision $\frac{\varepsilon}{m}$ (and the second step does not alter the precision due to $\|\cdot\|_\textrm{op}=1$).
\end{proof}

This is all we need to analyze the corresponding quantum control problem by means of \eqref{eq:toy_state_connect}:
\begin{corollary}\label{cor:toy_quantum}
Let $d,n\in\mathbb N$, $H_0,H_1,\ldots,H_m\in\mathbb C^{d^n\times d^n}$ Hermitian, and $V=\mathbbm{1}_d^{n-1}\otimes \sigma_+$ be given. Then, assuming $\langle iH_0, iH_j\,|\,j=1,\dots,m\rangle_\textsf{Lie}=\mathfrak u(d^n)$ (or $=\mathfrak{su}(d^n)$), the reachable set of
$$
\dot\rho(t)=-i\Big[H_0+\sum\nolimits_{j=1}^m u_j(t) H_j,\rho(t)\Big]-\gamma(t)\Big(\frac12 (V^*V\rho(t) +\rho(t) V^*V)-V\rho (t)V^*\Big)
$$
satisfies $\overline{\mathfrak{reach}(\rho_0)}=\mathbb D(\mathbb C^{d^n})$ for all $\rho_0\in \mathbb D(\mathbb C^{d^n})$.
\end{corollary}
\begin{proof}
Using unitary controllability for $\gamma(t)=0$ (Coro.~\ref{coro_contr_su} \& \ref{coro_contr_u}), starting from any 
$\rho_0\in\mathbb D(\mathbb C^n)$ we can steer the system into $\rho_0\to U\rho_0U^*$ such that it is diagonal in an eigenbasis of $H_0$ (with $E_1\leq\ldots\leq E_n$). This in particular 
means $[H_0,U\rho_0U^*]=0$ so we are in the diagonal case (cf.~footnote \ref{footnote_H_0_gen_diag}) 
with effectively no coherent drift, but only dissipation and coherent controls, 
i.e.~in the realm of the toy model via the obvious one-to-one correspondence $\mathrm D(n) \leftrightarrow
\Delta^{n-1}$. Here one can (approximately) reach every other diagonal state (Thm.~\ref{thm_2}) which 
by finally rotating back gives the desired result for the quantum control system.
\end{proof}
This result covers a single qudit being fully coupled to the bath ($n=1$) as well as multiple qudits where only the last qudit is coupled (local noise, $n>1$), thus generalizing \cite[Thm.~1]{BSH16} from qubits ($d=2$) to arbitrary qudits ($d\in\mathbb N$). Note that because Lemma \ref{lemma_for_thm_temp_0} and \ref{lemma_reach_oplus} were proven constructively, the underlying control scheme of Coro.~\ref{cor:toy_quantum} is constructive as well, which adds to the strength of this result from an application point of view.
\begin{remark}\label{rem_control_sp_lamb}
\begin{itemize}
\item[(i)] Leaving out the closure in Coro.~\ref{cor:toy_quantum} would make the result impossible: If the initial state $\rho$ is positive definite (i.e.~all eigenvalues are $>0$) then
$$
\mathfrak{reach}(\rho_0)\subseteq\{\rho\in\mathbb D(\mathbb C^n)\,|\,\rho>0\}\subsetneq\mathbb D(\mathbb C^n)
$$
meaning such Markovian control systems can never be controllable on $\mathbb D(\mathbb C^n)$ and approximate controllability is the best result one can obtain.
 
This is due to Rem.~\ref{rem_markov} because the system semigroup $S_\Omega$ only contains finite products Markovian channels so $S_\Omega\subseteq\textsc{sp}$.
Thus we found an alternative proof for the ``no-go result'' regarding controllability of open quantum systems \cite[Thm.~3.10]{DiHeGAMM08}.\index{open quantum system!never controllable}
\item[(ii)] When coupling a quantum system to the environment the Hamiltonian $H_0$ of the 
closed system may change. This is known as \textit{Lamb shift}\index{Lamb shift} \cite[Ch.~3.3.1]{BreuPetr02} 
meaning our control problem strictly speaking is of the form
$$
\dot\rho(t)=-i\Big[H_0+\sum\nolimits_{j=1}^m u_j(t) H_j,\rho(t)\Big]-\gamma(t)(\Gamma+(\operatorname{ad}_{H_\textrm{LS}}-\operatorname{ad}_{H_0}))
$$
so the system semigroup $S_\Omega$ is generated by
$$
\{e^{-i\tau(\operatorname{ad}_{H_0}+\sum_{j=1}^mu_j\operatorname{ad}_{H_j})}\,|\,\tau\geq 0,u\in\Omega\}\cup\{e^{-\tau(i\operatorname{ad}_{H_\textrm{LS}}+\Gamma+\sum_{j=1}^mu_ji\operatorname{ad}_{H_j})}\,|\,\tau\geq 0,u\in\Omega\}\,,
$$
cf.~also \cite[(3.140) ff.]{BreuPetr02}. However one finds $[H_0,H_\textrm{LS}]=0$ so $H_0$ and $H_\textrm{LS}$ are diagonal in the same basis \cite[Thm.~4.5.15]{HJ1}; thus our proof of Coro.~\ref{cor:toy_quantum} goes through unchanged: We can still apply every unitary channel and once we are in the eigenbasis of $H_0$ we have $[H_\textrm{LS},U\rho U^*]=0$, as well, meaning we still have access to pure dissipation $-\Gamma$. 
\item[(iii)] If the Lamb shift ever poses a problem then one can always use Trotter's formula\index{Trotter's formula} \cite[Thm.~VIII.29]{ReedSimonI} to obtain
$$
e^{-t\Gamma}(\rho)=\lim_{n\to\infty}\big( e^{t\operatorname{ad}_{H_\textrm{LS}}/n}e^{-t(i\operatorname{ad}_{H_\textrm{LS}}+\Gamma)/n})^n(\rho)\,.
$$
for all $\rho\in\mathbb D(\mathbb C^n)$. Obviously $e^{t\operatorname{ad}_{H_\textrm{LS}}/n}$ is a unitary channel for all $t,n$ so unitary controllability guarantees access to it, hence $e^{-t\Gamma}\in\overline{S_\Omega}$. While this is mathematically fine---after all one always has to resort to approximate controllability for open systems by (i) of this remark---from a physics perspective this should be avoided whenever possible. Indeed Trotterization usually leads to a considerably worse performance of numerics and experiments, let it be precision (fidelity), time of control sequences, etc. 
\end{itemize}
\end{remark}




Now for all temperatures $T\in[0,\infty]$, in the qubit case (for unitary controllability and switchable bath coupling $\Gamma_d$) the
closure of the reachable set for any initial state $ \rho_0\in\mathbb D(\mathbb C^2)$ equals
$$
\lbrace \rho\in\mathbb D(\mathbb C^2)\,|\, \rho\prec \rho_\textsf{Gibbs} \vee \rho\prec \rho_0\rbrace
$$
as can be seen easily, cf.~\cite{RBR12}. One might hope that this extends to general $n$-level systems with $n>2$
at finite temperatures. However, this is not true even if the above is taken as an upper bound for the reachable
set, as the following example shows.

\begin{example}
Let
$$
d=\frac1{e^{0.64}+1+e^{-0.64}}\begin{pmatrix} e^{0.64}\\1\\e^{-0.64} \end{pmatrix}\approx\begin{pmatrix}0.5539\\0.2921\\0.1540
\end{pmatrix}\in\Delta^2\,.
$$ 
Then for $ \rho_0 = \operatorname{diag}(0.55,0.4,0.05) \in\mathbb D(\mathbb C^3)$ and the semigroup 
$(e^{-t\Gamma_d})_{t\geq 0}$ (cf.~Prop.~\ref{thm_bath}) one gets for $t=1$
$$
e^{-\Gamma_d}( \rho_0)=\operatorname{diag}\Big(e^{-B_0(\Gamma_d)}\begin{pmatrix}0.55\\0.4\\0.05 \end{pmatrix}\Big)\approx\operatorname{diag}\begin{pmatrix}0.5783\\0.3098\\0.1119\end{pmatrix}\,.
$$
Evidently, $e^{-\Gamma_d}( \rho_0)\not\prec \operatorname{diag}(d) = \rho_\textsf{Gibbs}$ and 
$e^{-\Gamma_d}( \rho_0)\not\prec \rho_0$.
\end{example}

To obtain some analytic results for $0<T<\infty$ we restrict ourselves to the case of equidistant energy
levels, which is the best one can hope for, cf.~Ex.~\ref{ex_equid_ev}. Thus $d$ is of the form
\begin{equation}\label{eq:equidist_d_vec}
d=\frac{1-\alpha}{1-\alpha^n}\begin{pmatrix} 1\\\alpha\\\vdots\\\alpha^{n-1} \end{pmatrix}
\end{equation}
for some $\alpha\in(0,1)$. This includes the so-called diagonal spin case:\index{diagonal spin case}

\begin{theorem}\label{thm_3}
Let $n\in\mathbb N$ and $d\in\Delta^{n-1}$, $d>0$ such that $\frac{d_{j+1}}{d_j}$
is constant for $j=1,\ldots,n-1$. Also let $\Gamma_d$ be induced by $\sigma_-^d,\sigma_+^d$ (cf.~Prop.~\ref{thm_bath}). 
Then the reachable set for the toy model $\Lambda_d$ with $B_0(\Gamma_d)$
for all $x_0\in\Delta^{n-1}$ satisfies
\begin{equation*}
\overline{\mathfrak{reach}_{\Lambda_d}(x_0)}\subseteq \lbrace x\in\Delta^{n-1}\,|\, x\prec z \rbrace\,.
\end{equation*}
Here $z\in\Delta^{n-1}$ is any vector such that $x_0\prec z$, and that $d$ and $\frac{z}{d}$ are similarly ordered. Moreover, such a vector $z$ always exists, and if $x_0>0$ then one can choose $z>0$.
\end{theorem}

\noindent
Note that $d$ is the unique fixed point of $(e^{-tB_0(\Gamma_d)})_{t\geq 0}$. We proved this result in \cite{CDC19} for the special case $x_0=d$ (or more generally $x_0\prec d$) in which case $z$ can be chosen to be $d$. The idea of the proof stays the same; the difficulty was to show existence and crucial properties of the extremal point $z$, which we did in Ch.~\ref{sec_d_maj_poly}.
\begin{lemma}\label{lemma_perm_partition}
Let $n,k\in\mathbb N$ with $k\leq n$ and let $\pi$ be any permutation on $\lbrace1,\ldots,n\rbrace$. 
Then there exist unique non-empty subsets $\square_1,\ldots,\square_q\subseteq\pi(\lbrace1,\ldots,k\rbrace)$
(henceforth called ``blocks'') 
with the following properties.
\begin{itemize}
\item[(i)] The blocks $\square_1,\ldots,\square_q$ yield a disjoint partition of 
$\pi(\lbrace1,\ldots,k\rbrace)$, i.e.~$\square_{j}\cap\square_l=\emptyset$ for $j \neq l$
and $\bigcup_{j=1}^q \square_j=\pi(\lbrace1,\ldots,k\rbrace)$.
\item[(ii)] The blocks are the ``connected components'' of $\pi(\lbrace1,\ldots,k\rbrace)$. 
More precisely, for each $j\in\lbrace 1,\ldots,q\rbrace$ there exist $b_j^-,b_j^+\in\lbrace 1,\ldots,n \rbrace$ 
such that
$$
\square_j= \lbrace b_j^-, b_j^- +1, \ldots, b_j^+-1, b_j^+\rbrace
$$
and $b_j^--1, b_j^++1 \notin\pi(\lbrace1,\ldots,k\rbrace)$ so the nearest neighbours of the blocks 
are not in $\pi(\lbrace1,\ldots,k\rbrace)$.
\end{itemize}
\end{lemma}

Instead of proving the above lemma, let us quickly illustrate what is going on here by considering an example. 
With this the proof will be obvious.
\begin{example}
Let $\pi$ be the permutation (in cycle notation) 
$
\pi = (1,6,2,3,4)(5)
$ 
on $\lbrace 1,\ldots,6\rbrace$. First, consider $k=3$ so $\pi(\lbrace 1,2,3\rbrace)=\lbrace 3,4,6\rbrace$. 
The connected block-components of this set are $\square_1=\lbrace 3,4\rbrace$, $\square_2=\lbrace6\rbrace$ 
as they satisfy 
$$
\square_1\cap\square_2 =\emptyset \quad\text{and}\quad \square_1\cup\square_2=\pi(\lbrace 1,2,3\rbrace)
$$ 
and neither of their neighbouring numbers (i.e.~$2,5,7$) are contained within 
$\pi(\lbrace 1,2,3\rbrace)$. To finish off this example, for $k=5$ one gets $\pi(\lbrace 1,2,3,4,5\rbrace)
=\lbrace 1,3,4,5,6\rbrace$. Here, the blocks obviously are 
$\square_1=\lbrace 1\rbrace$, $\square_2=\lbrace3,4,5,6\rbrace$.
\end{example}

\begin{proof}[Proof of Thm.~\ref{thm_3}]
If suffices to prove the inclusion in question without the closure (because the r.h.s.~is closed). 
Using \eqref{eq:action_gamma_Delta} and \eqref{eq:equidist_d_vec} for $B_0=B_0(\Gamma_d)$ gives
\begin{equation}\label{eq:75}
B_0 = \begin{pmatrix} c_1\alpha&-c_1&&&&\\
-c_1\alpha&c_1+c_2\alpha&-c_2&&\\
&-c_2\alpha&c_2+c_3\alpha&-c_3&\\
&&\ddots&\ddots&\ddots\\
 \end{pmatrix}
\end{equation}
with $\alpha=\frac{d_{j+1}}{d_j}\in(0,1)$ and $c_j:=j(n-j)/(1+\alpha)\geq 0$. In order to show that $\mathfrak{reach}_{\Lambda_d}(x_0)$ for some $x_0\in\Delta^{n-1}$ is upper bounded 
by $\lbrace x\in\Delta^{n-1}\,|\, x\prec z\rbrace$ one has to show that the latter
\begin{itemize}
\item[(i)] contains the initial state.
\item[(ii)] is invariant under permutation channels.
\item[(iii)] is invariant under the semigroup $(e^{-tB_0})_{t\geq 0}$.
\end{itemize}
Here (ii) is evident and (i) holds by assumption so we only have to show (iii). As $\exp(-tB_0)$ is linear and the 
set $\lbrace x\in\Delta^{n-1}\,|\, x\prec z\rbrace$ is a convex 
polytope (Coro.~\ref{thm_convex_poly}) it suffices to prove that the semigroup acts contractively on its extreme points 
$\underline{\pi} z$, where $\pi$ is any permutation.
Thus we have to show that for every such $\pi$ there exists $t_0 > 0$ such that
\begin{equation}\label{eq:76}
\exp(-tB_0)\underline{\pi} z\prec z \quad \text{for all $t\in [0,t_0)$}\,.
\end{equation}
Again the fact that $\lbrace x\in\Delta^{n-1}\,|\, x\prec z\rbrace$
is a compact, convex polytope implies that \eqref{eq:76} can be replaced by the tangential condition
\begin{equation}\label{eq:77}
\forall_{\pi\in S_n}\, \exists_{\mu>0}\quad( \mathbbm{1}_n-\mu B_0) \underline{\pi} z\prec z\,.
\end{equation}
By assumption
\begin{equation}\label{eq:77z}
z_j>\alpha z_j\geq z_{j+1}
\end{equation}
for all $j=1,\ldots,n-1$. Therefore---using $(\underline{\pi} z)_{\pi^{-1}(j)}=z_j=z_j^\downarrow$ (footnote \ref{footnote_permutation_matrix})---one finds $\mu>0$ such that\footnote{
Given $y\in\mathbb R^n$ such that $y_{\pi(i)}>\ldots> y_{\pi(n)}$ for some $\pi\in S_n$ as well as arbitrary $z\in\mathbb R^n$, define
$$
\mu_j:=\begin{cases} 1&z_{\pi(j)}\leq z_{\pi(j+1)}\\ \frac{1}{2}\frac{y_{\pi(j)}-y_{\pi(j+1)}}{z_{\pi(j)}-z_{\pi(j+1)}}&\text{else} \end{cases}
$$
for all $j=1,\ldots,n-1$ as well as $\mu:=\min_{j=1,\ldots,n-1}\mu_j>0$. One readily computes $y_{\pi(j)}-\mu z_{\pi(j)}> y_{\pi(j+1)}-\mu z_{\pi(j+1)}$ so $(y-\mu z)^\downarrow_j=y_{\pi(j)}-\mu z_{\pi(j)}$ for all $j=1,\ldots,n-1$ meaning $\mu>0$ can always be chosen small enough such that the ``order of $y$'' is preserved.
}
$(( \mathbbm{1}_n-\mu B_0) \underline{\pi} z)^\downarrow_j=z_j-\mu (B_0\underline{\pi} z)_{\pi^{-1}(j)}$ for all $j$. Thus the partial sum condition for \eqref{eq:77} reads
$$
\sum\nolimits_{j=1}^k z_j=\sum\nolimits_{j=1}^k z_j^\downarrow\geq \sum\nolimits_{j=1}^k (( \mathbbm{1}_n-\mu B_0) \underline{\pi} z)^\downarrow_j=\Big(\sum\nolimits_{j=1}^k z_j\Big)-\mu \sum\nolimits_{j=1}^k(B_0\underline{\pi} z)_{\pi^{-1}(j)}
$$
for all $k=1,\ldots,n-1$ as well as $\unitvector^T z=\unitvector^T ((\mathbbm{1}_n-\mu B_0) \underline{\pi}z)$ (but the latter is evident as $\unitvector^T B_0=0$, cf.~\eqref{eq:action_gamma_Delta}). Thus if can show $ \sum\nolimits_{j=1}^k(B_0\underline{\pi}z)_{\pi^{-1}(j)}\geq 0$ for all $k=1,\ldots,n-1$ then \eqref{eq:77} holds and we are done.

Indeed let $k\in\lbrace 1,\ldots,n\rbrace$ be arbitrary and consider the ``connected components''
$\square_1,\ldots,\square_q$ of the set 
$\pi^{-1}(\lbrace 1,\ldots,k\rbrace)$, cf.~Lemma 
\ref{lemma_perm_partition}. Then 
$$
\sum\nolimits_{j=1}^{k} (B_0\underline{\pi}z)_{\pi^{-1}(j)} = 
\sum\nolimits_{j=1}^q \sum\nolimits_{a\in\square_j} (B_0\underline{\pi}z)_{a} \,.
$$
Thus it would suffice to show that every $\square_j$-sum individually yields something non-negative. 
%
Using \eqref{eq:75} and the properties of the $\square_j$
\begin{align}
\sum_{a\in\square_j} (B_0\underline{\pi}z)_{a}&=\sum\nolimits_{k=b_j^-}^{b_j^+}(B_0\underline{\pi}z)_{k}=\sum\nolimits_{k=b_j^-}^{b_j^+}c_{k-1}\big( (\underline{\pi}z)_{k}-\alpha (\underline{\pi}z)_{k-1})+c_{k}\big(\alpha(\underline{\pi}z)_{k}- (\underline{\pi}z)_{k+1})\big)\notag\\
&=c_{b_j^--1}\big((\underline{\pi}z)_{b_j^-}-\alpha (\underline{\pi}z)_{b_j^--1}) + c_{b_j^+}\big(\alpha (\underline{\pi}z)_{b_j^+}-(\underline{\pi}z)_{b_j^++1})\big)\,. \label{eq:79}
\end{align}
We know that $b_j^-,b_j^+\in\pi^{-1}(\lbrace 1,\ldots,k\rbrace)\not\ni b_j^--1,b_j^++1$ so \eqref{eq:77z} shows
$$
(\underline{\pi}z)_{b_j^-}\geq (\underline{\pi}z)_{\pi^{-1}(k)}=z_k>z_{k+1}=(\underline{\pi}z)_{\pi^{-1}(k+1)}\geq (\underline{\pi}z)_{b_j^--1}\geq \alpha (\underline{\pi}z)_{b_j^--1}
$$
as well as
$$
\alpha (\underline{\pi}z)_{b_j^+}\geq \alpha z_k\geq z_{k+1}\geq \alpha (\underline{\pi}z)_{b_j^++1}\,.
$$
Because 
$c_j\geq 0$ for all $j$, the summands involved in \eqref{eq:79} are non-negative.
Finally, existence of a vector $z$ with the desired properties was shown in Thm.~\ref{theorem_max_corner_maj}. This concludes the proof.
\end{proof}
%
\noindent One can show that for all $x_0\in\Delta^{n-1}$, $d>0$ the set of possible ``upper bounds'' from Thm.~\ref{thm_3}
\begin{equation}\label{eq:outwards_maj}
\{y\in\Delta^{n-1}\,|\,x_0\prec y\ \wedge\ d\text{ and }\tfrac{y}{d}\text{ are similarly ordered}\}
\end{equation}
forms a convex polytope. In particular, this allows one to find an ``optimal'' upper bound, e.g., by considering the vector which attains the (well-defined) quantity $\min_{y\in\eqref{eq:outwards_maj}}\|y-\frac{\unitvector^T}{n}\|_1$. Of course one can, in principle, minimize over any continuous function $f:\Delta^{n-1}\to\mathbb R$.\medskip

One may wonder whether it is necessary to restrict oneself to Hamiltonians with equidistant eigenvalues.
The following example gives a positive answer.
\begin{example}\label{ex_equid_ev}
Let $$d=\frac1{ 1+e^{-1/4}+e^{-17/4}}{\begin{pmatrix} 1\\e^{-1/4}\\e^{-17/4} \end{pmatrix}}\approx{\begin{pmatrix}0.5577\\0.4343\\0.0080
\end{pmatrix}}\in\Delta^2$$ 
so the semigroup $( e^{-tB_0})_{t\geq 0}$ (cf.~Prop.~\ref{thm_bath}) acts like
$$
e^{-tB_0}{\begin{pmatrix} 0.0080\\0.5577\\0.4343\end{pmatrix} }\approx{\begin{pmatrix}0.0683\\0.5730\\0.3587\end{pmatrix}}\quad\text{ for }t=1/10\,.
$$
Therefore majorization is violated (the largest eigenvalue grows) and the set $\lbrace x\in
\Delta^{n-1}\,|\, x\prec d\rbrace$ is not left invariant by $(e^{-tB_0})_{t\geq 0}$, 
although $d$ satisfies the ``physical'' ordering condition.	
\end{example}

Anyway Thm.~\ref{thm_3} is a promising first step towards an upper bound for the reachable set of the quantum control problem
$$
\dot\rho(t)=-i\Big[H_0+\sum\nolimits_{j=1}^m u_j(t) H_j,\rho(t)\Big]-\gamma(t)\Gamma_d(\rho(t))
$$
with $\Gamma_d$ from Prop.~\ref{thm_bath}. However carrying over this result is not trivial as only \textit{lower} bounds pertain from the toy model to the general control problem, cf.~\eqref{eq:toy_state_connect}; more on this in Ch.~\ref{ch:concl_outlook}.

%
%


\section{Infinite Dimensions}\label{sec_reach_inf_dim}

Next let us tackle reachability of unital systems, that is, for normal Lindblad $V$'s; after all, these systems are exceptional as they allow for an upper bound $\overline{\mathfrak{reach}(\rho_0)}\subseteq \{\rho\in\mathbb D(\mathcal H)\,|\,\rho\prec\rho_0\}$ via majorization (cf.~start of this chapter).

However we also learned that this upper bound becomes increasingly inaccurate the larger the system size, unless the dissipation term $-\Gamma$ becomes an additional control by being switchable by $\gamma(t)$. Indeed, it turns out that for switchable noise and a single Lindblad-$V$ this upper bound is (almost) always saturated:

\begin{proposition}\label{prop_reach_normal}
Consider $H_0,H_1,\ldots,H_m\in\mathbb C^{n\times n}$ Hermitian and $V\in\mathbb C^{n\times n}$ normal. Then, assuming $\langle iH_0, iH_j\,|\,j=1,\dots,m\rangle_\textsf{Lie}=\mathfrak u(n)$ (or $=\mathfrak{su}(n)$) and $V\neq\lambda\mathbbm{1}$ for all $\lambda\in\mathbb C$ the reachable set of
$$
\dot\rho(t)=-i\Big[H_0+\sum\nolimits_{j=1}^m u_j(t) H_j,\rho(t)\Big]-\gamma(t)\Big(\frac12 (V^*V\rho(t) +\rho (t)V^*V)-V\rho (t)V^*\Big)
$$
satisfies $\overline{\mathfrak{reach}(\rho_0)}= \{\rho\in\mathbb D(\mathbb C^{n})\,|\,\rho\prec\rho_0\}$ for all $\rho_0\in \mathbb D(\mathbb C^{n})$. 
\end{proposition}
\begin{proof}[Proof idea]
The upper bound is obvious from our previous considerations so let $\rho\in\mathbb D(\mathbb C^{n})$ with $\rho\prec\rho_0$ be given. Because $V$ is normal we can write it as $V=\sum_{j=1}^n v_j|f_j\rangle\langle f_j|$ with eigenvalues $v_j\in\mathbb C$ and orthonormal basis $\{f_j\}_{j=1}^n\subset\mathbb C^n$. Then 
\begin{align*}
\langle f_j,\Gamma_V(A) f_k\rangle =\big(\tfrac{1}{2}{|v_j-v_k|^2}-i\Im(v_j\overline{v_k})\big) \langle f_j,Af_k\rangle
\end{align*}
for all $A\in\mathbb C^{n\times n}$ and all $j,k=1,\ldots,n$, implying
$$
\langle f_j,e^{-t\Gamma_V}(A) f_k\rangle =e^{-\frac{t}{2}(|v_j-v_k|^2)}e^{it\Im(v_j\overline{v_k})} \langle f_j,Af_k\rangle
$$
for all $t\in\mathbb R$.
The control scheme now heavily relies on the Schur-Horn theorem\index{theorem!Schur-Horn} \cite{Schur23,Horn54} which (due to $\rho\prec\rho_0$) guarantees the existence of $U\in\mathbb C^{n\times n}$ such that the eigenvalues of $\rho$ appear on the diagonal of $U\rho_0 U^*$ (w.r.t.~$(f_j)_{j=1}^n$). If the eigenvalues of $V$ were pairwise different $\lim_{t\to\infty}e^{-t\Gamma_V}(A)=\sum_{j=1}^n \langle f_j,Af_j\rangle|f_j\rangle\langle f_j|$ meaning the state undergoes full decoherence (w.r.t.~the eigenbasis of $V$) and an idealized control scheme would go as follows:
$$
\rho_0\longrightarrow U\rho_0U^*\overset{\text{pure noise}}\longrightarrow \sum\nolimits_{j=1}^n \lambda_j(\rho)|f_j\rangle\langle f_j|\overset{V\text{ unitary}}\longrightarrow \rho\,.
$$
While the first and last step are unitary channels and thus exact in finite dimensions, the second step is in need of three approximations:
\begin{itemize}
\item The reachable set only covers finite times so given $\varepsilon$ one has to choose $t\geq 0$ big enough such that $e^{-t\Gamma_V}(U\rho_0U^*)$ is sufficiently close to $\sum_{j=1}^n \lambda_j(\rho)|f_j\rangle\langle f_j|$.
\item We do not have access to pure noise meaning we have to approximate it, e.g., by means of Trotter's formula (cf.~Rem.~\ref{rem_control_sp_lamb} (iii)).
\item If the eigenvalues of $V$ do not differ pairwise then even ideal dissipation does not reduce the state to its diagonal. However, $V\neq\lambda\mathbbm{1}$ for all $\lambda\in\mathbb C$ guarantees that at least two eigenvalues of $V$ differ from each other so by appropriate permutation channels one can make all off-diagonal elements of $U\rho_0 U^*$ arbitrarily small.
\end{itemize}
In total, this is enough to show $\rho\in \overline{\mathfrak{reach}(\rho_0)}$. 
\end{proof}
\noindent We intentionally waived the details here as they will appear in the proof of the generalized result, and conveying the idea of the control scheme is more important for now.\medskip

At this point there are two results which seem reasonable to try to generalize to infinite dimensions: On one hand approximate controllability for a switchable temperature zero bath (Coro.~\ref{cor:toy_quantum}) and on the other hand approximate controllability on the set of majorized states for a switchable normal (but non-trivial) Lindblad-$V$ (Prop.~\ref{prop_reach_normal}). 

Generalizing the temperature zero result bath means replacing $\sigma_+$ by an unbounded operator so we would first have to check well-posedness of the corresponding control system. This is the reason we will aim for transferring Prop.~\ref{prop_reach_normal} to \textit{infinite-dimensional 
systems on separable complex} Hilbert spaces $\mathcal H$. Motivated by Ch.~\ref{ch:bilinear_control} we allow for arbitrary bounded control Hamiltonians $H_j\in\mathcal B(\mathcal H)$, while the drift $H_0$ may be any unbounded self-adjoint operator. The following is based on our article \cite{OSID19}:

\begin{theorem}\label{thm_normal_V}
Given the Markovian control system $\Sigma_V$ 
\begin{equation*}
\dot\rho(t) = - i \Big[H_0 + \sum_{j=1}^mu_j(t)H_j,\rho\Big] - 
\gamma(t) \big(\tfrac{1}{2}(V^* V \rho(t) + \rho(t) V^* V) - V\rho(t) V^* \big)\,,\;\text{where}
\end{equation*}
\begin{itemize}
\item[(i)] the drift $H_0$ is self-adjoint and the controls $H_1,\ldots,H_m$ are self-adjoint and bounded,
\item[(ii)] the Hamiltonian part $\dot U(t)=-i(H_0+\sum\nolimits_{j=1}^mu_j(t)H_j) U(t)$ with $U(0)=\mathbbm{1}_{\mathcal H}$ is strongly approximately controllable 
on $\mathcal U(\mathcal H)$ in the sense of Def.~\ref{def_inf_dim_unit_contr},
\item[(iii)] the noise term $V\in\mathcal K(\mathcal H)\setminus\lbrace 0\rbrace$ is compact, normal, and switchable by $\gamma(t) \in\{0,1\}$.
\end{itemize}
Then the $\|\cdot\|_1$-closure of the reachable set of any initial state $\rho_0\in\mathbb D(\mathcal H)$ under 
the system $\Sigma_V$\label{symb_reach_sigma}
exhausts all states majorized by the initial state $\rho_0$ 
\begin{align*}
\overline{\mathfrak{reach}_{\Sigma_V}(\rho_0)}=\lbrace \rho\in\mathbb D(\mathcal H)\,|\,\rho\prec\rho_0\rbrace\,.
\end{align*}
\end{theorem}
%
Because the spectral behaviour of $V$ was crucial to the proof in finite dimensions it is natural to choose $V$ compact (cf.~Ch.~\ref{compact_wot_trace_class}). Indeed this guarantees the same beautiful eigenspace structure of the corresponding noise, as follows by direct computation:
\begin{lemma}\label{lemma_normal_generator}
Let $V\in\mathcal K(\mathcal H)$ be normal, $(f_j)_{j\in\mathbb N}$ its orthonormal
eigenbasis, and $(v_j)_{j\in\mathbb N}$ its modified eigenvalue sequence, hence 
$ V =\sum_{j=1}^\infty v_j|f_j\rangle\langle f_j|$ (cf.~Thm.~\ref{thm_compact_normal_unit_diag} \& Ch.~\ref{sect_C_spectrum}). 
Then for all $B\in\mathcal B(\mathcal H)$, the noise operator $\Gamma_V$ given by Eq.~\eqref{eq:Gamma_def}
acts like
\begin{equation}\label{eq:action_noise}
\begin{split}
\langle f_j,\Gamma_V(B) f_k\rangle 
& =\big(\tfrac{1}{2}{|v_j-v_k|^2}-i\Im(v_j\overline{v_k})\big) \langle f_j,Bf_k\rangle
\end{split}
\end{equation}
for all $j,k\in\mathbb N$. In particular, each rank-$1$ operator of the form 
$|f_j\rangle\langle f_k|$ is an eigenvector of $\Gamma_V$ to the eigenvalue 
$\tfrac{1}{2}{|v_j-v_k|^2}-i\Im(v_j\overline{v_k})$ and the kernel of
$\Gamma_V$ contains $ \operatorname{span}\{|f_j\rangle\langle f_j| \,|\, j \in\mathbb N\}$.
Moreover, it follows
\begin{equation*}
\exp(-t\Gamma_V)(|f_j\rangle\langle f_k|)
= \exp\big(-\tfrac{t}2 |v_j-v_k|^2\big)\exp(it\Im(v_j\overline{v_k}))|f_j\rangle\langle f_k|
\end{equation*}
for all $t\in\mathbb R$ and $j,k\in\mathbb N$. 
\end{lemma} 


Now the proof we have in mind---inspired by Prop.~\ref{prop_reach_normal}---roughly goes as follows: As for ``$\subseteq$'', normality of $V$ guarantees $(i\operatorname{ad}_{H}(t) + \gamma(t)\Gamma_V)(\mathbbm{1})=0$. Therefore the corresponding semigroup is bi-stochastic, thus obeying majorization (Lemma \ref{lemma_li}). As for ``$\supseteq$'', because $V\in\mathcal K(\mathcal H)$ is normal we can diagonalize it 
with orthonormal eigenbasis $(f_j)_{j \in \mathbb N}$.
Now let $\varepsilon>0$ and $\rho\in\mathbb D(\mathcal H)$ with $\rho\prec\rho_0$ be given. We have 
to find $\rho_F\in \mathfrak{reach}_{\Sigma_V}(\rho_0)$ such that $\|\rho-\rho_F\|_1<\varepsilon$. 
By assumption there exist $x,y\in\ell^1_+(\mathbb N)$, $x,y\neq 0$ as well as $W_1,W_2\in\mathcal U(\mathcal H)$ such that
$
\rho=W_1\operatorname{diag}(x)W_1^*$, $ \rho_0=W_2\operatorname{diag}(y)W_2^*
$ with $x\prec y$ (here, $\operatorname{diag}$ refers to the above eigenbasis of $V$). Applying Prop.~\ref{prop_schur_horn_gohberg} to $x,y$ gives us unitary $U\in\mathcal B(\mathcal H)$ such that $U\operatorname{diag}(y)U^*$ has diagonal entries $(x_n)_{n\in\mathbb N}$. Now all we have to do is execute the three steps shown here:
\begin{equation}\label{eq:steps_idea3}
\begin{split}
\rho_0=W_2\operatorname{diag}(y)W_2^*\overset{\text{Step }1}\longrightarrow U \operatorname{diag}(y)U^* \overset{\text{Step }2}\longrightarrow \operatorname{diag}(x) \overset{\text{Step }3}\longrightarrow W_1\operatorname{diag}(x)W_1^*=\rho\,.
\end{split}
\end{equation}
Step 1 and 3 merely apply a unitary channel; assuming strong operator controllability, 
we may use unitary channels with arbitrary precision (in the strong operator topology, cf.~Lemma 
\ref{lemma_state_approx}).
Step 2 
again is about getting rid of all off-diagonal elements of $U \operatorname{diag}(y)U^*$ 
by applying pure noise $\exp(-t\Gamma_V)$ in the limit $t\to\infty$ (cf.~Lemma 
\ref{lemma_normal_generator}). As before there are a few delicate issues:
\begin{itemize}
\item 
We have no access to pure noise, as in our setting we cannot switch off $H_0$. 
Yet by a Trotter-type 
argument---now adjusted to an unbounded drift $H_0$---we can approximate the desired noise with arbitrary precision in a weaker topology. On top of that the ideal pure noise ($t\to\infty$) has to be approximated with $t\geq 0$ sufficiently large.
\item 
If the eigenvalues of $V$ are not pairwise different, then there are some ``matrix'' elements left 
untouched by the noise as a consequence of \eqref{eq:action_noise}. So one may need permutation channels (which in particular are unitary) to rearrange those elements into ``spots'' where the noise affects them. 
\item 
As in Step 1 and 3 we have to approximate these permutation channels. Here we use the 
approximation property of the Schatten classes (cf.~Lemma~\ref{lemma_schatten_p_approx}), that is, we invoke 
decoherence on a sufficiently large but finite ``block'' of the density operator so we only need
finitely many permutations.
\end{itemize}
While these issues were to be expected from the finite-dimensional proof we will face an additional problem exclusive to infinite dimensions:
\begin{itemize}
\item
Applying Prop.~\ref{prop_schur_horn_gohberg} requires that $\rho_0$, $\rho$ are unitarily diagonalized so that the 
original and the modified eigenvalue sequences of these states co{\"i}ncide (which either means the states 
are finite-rank or have trivial kernel)---else the zeros that have to be added for the modified eigenvalue sequence 
prevent this. In the latter case we can proceed to states $\rho'$, $\rho_0'$ which satisfy the assumptions of 
Prop.~\ref{prop_schur_horn_gohberg} and which are close (in trace norm) to the original states.
\end{itemize}
Altogether this should be enough to perform the scheme suggested in Eq.~\eqref{eq:steps_idea3} with arbitrary 
precision, so $\rho$ $\prec\rho_0$ is in the $\|\cdot\|_1$-closure of the reachable set.\medskip

Before working out the proof idea in detail we need some further tools. 
First is the Trotter product formula for contraction 
semigroups on Banach spaces:\index{Trotter's formula}

\begin{lemma}[\cite{ReedSimonII}, Thm.~X.51]\label{lemma_trotter_banach}
Let ${A}_1$ and ${A}_2$ be generators of contraction semigroups on a Banach space $X$, i.e.~strongly continuous semigroups of operator norm less or
equal one for all $t\geq 0$. Suppose that the closure $\overline{({A}_1+{A}_2)}$ of $({A}_1+{A}_2)$
generates a contraction semigroup on $X$. Then for all $\rho\in X$
and all (fixed) $t \geq 0$
$$
\lim_{n\to\infty}\Vert (e^{t{A}_1/n}e^{t{A}_2/n})^n(\rho)-e^{t\overline{({A}_1+{A}_2)}} (\rho)\Vert_1=0\,.
$$
\end{lemma}
\noindent Of course if ${A}_1+{A}_2$ already generates a contraction semigroup we can waive the closure as such generators are always closed \cite[Ch.~II, Thm.~1.4]{EngelNagel00}.\medskip

The next auxiliary result is readily verified via a simple induction argument:
\begin{lemma}\label{lem:telescope}
Let $m\in\mathbb N$ and let $A_1,\ldots,A_m$, $B_1,\ldots,B_m: D \to D$ be arbitrary maps acting on some common domain $D$. Then
$$
\prod_{j=1}^m \,A_j - \prod_{j=1}^m\, B_j 
= \sum_{j=1}^m\Big( \prod_{k=1}^{j-1} A_k \circ (A_j-B_j) \circ \prod_{k=j+1}^m B_j \Big)\,.
$$
Here and henceforth, the order of the ``product'' $\prod_{j=1}^m \,A_j$ shall be fixed by
$A_1 \circ \cdots \circ A_m$.
\end{lemma}
Thus as a special case of Lemma \ref{lemma_trotter_banach} we obtain:

\begin{corollary}\label{lemma_trotter_approx}
Let $V\in\mathcal B(\mathcal H)$, $H_0$ self-adjoint on $\mathcal H$, and $\rho\in\mathbb D(\mathcal H)$ be arbitrary. Moreover let $R\subseteq \mathcal U(\mathcal H)$ with $\overline{R}^{\,\mathrm{s}}=\mathcal U(\mathcal H)$,
where the closure is taken in $\mathcal U(\mathcal H)$. Then for all $t\geq 0$ and all $\varepsilon>0$ there exist $m\in\mathbb N$ and $U_1,\ldots,U_m\in R$ such that
\begin{equation*}
\Big\| \exp(-t\Gamma_V)(\rho)-\prod_{j=1}^m \Big(\operatorname{Ad}_{U_j}\circ\exp\Big( \frac{-it\operatorname{ad}_{H_0}-t\Gamma_V}{m} \Big) \Big)(\rho) \Big\|_1<\varepsilon
\end{equation*}
\end{corollary}
\begin{proof}
Recall that $-\Gamma_V$, $i\operatorname{ad}_{H_0}$, and $-i\operatorname{ad}_{H_0}-\Gamma_V$ are generators of strongly continuous quantum-dynamical semigroups (Prop.~\ref{prop_bounded_pert}) so in particular they are generators of contraction semigroups (Prop.~\ref{thm_q_norm_1}). Thus by Lemma \ref{lemma_trotter_banach}, given $\varepsilon>0$ there exists $m\in\mathbb N$ with\footnote{
While $i\operatorname{ad}_{H_0}+(-i\operatorname{ad}_{H_0}-\Gamma_V)$ equals $-\Gamma_V$ \textit{on the dense domain} $D(\operatorname{ad}_{H_0})$, boundedness of $\Gamma_V$ guarantees $\overline{i\operatorname{ad}_{H_0}+(-i\operatorname{ad}_{H_0}-\Gamma_V)}=\overline{ -\Gamma_V|_{D(\operatorname{ad}_{H_0})} }=-\Gamma_V$ so we may apply Lemma \ref{lemma_trotter_banach} as below.
}
\begin{equation*}
\Big\Vert \exp(-t\Gamma_V)(\rho)-\Big( \underbrace{\exp\Big(\frac{it\operatorname{ad}_{H_0}}{m}\Big)}_{=:F}\circ\underbrace{\exp\Big(\frac{-it\operatorname{ad}_{H_0}-t\Gamma_V}{m}\Big)}_{=:G}\Big)^m(\rho)\Big\Vert_1<\frac{\varepsilon}{2}\,.
\end{equation*}
%
For convenience define $\rho_j:=(G\circ (F\circ G)^{m-j})(\rho)$ for $j=1,\ldots,m$. Then, 
Lemma \ref{lemma_state_approx} yields $U_j\in R\subseteq\mathcal U(\mathcal H)$ with
$
\| F(\rho_j)-U_j\rho_jU_j^*\|_1 < \frac{\varepsilon}{2m}
$.
Finally, Prop.~\ref{thm_q_norm_1} and Lemma \ref{lem:telescope} imply
 \begin{align*}
\Big\| \exp(-t\Gamma_V)&(\rho)-\prod_{j=1}^m (\operatorname{Ad}_{U_j}{}\!\circ G ) (\rho)\Big\|_1\\
& \leq \Big\| \exp(-t\Gamma_V)(\rho)-(F\circ G)^m(\rho) \Big\|_1
+ \Big\| (F \circ G)^m (\rho)-\prod_{j=1}^m (\operatorname{Ad}_{U_j}{}\!\circ G ) (\rho)\Big\|_1\\
& <\frac{\varepsilon}{2} + \sum_{j=1}^{m} \Big\| \prod_{k=1}^{j-1}(\operatorname{Ad}_{U_k}{}\!\circ G) \circ (\operatorname{Ad}_{U_j}-F) \circ \underbrace{G \circ (F \circ G)^{m-j} (\rho)}_{=\rho_j}\Big\|_1\\
&\leq \frac{\varepsilon}{2} + \sum_{j=1}^{m} \underbrace{\Big( \prod_{k=1}^{j-1} \| \operatorname{Ad}_{U_k} {} \|_\textrm{op}\, \| G\|_\textrm{op} \Big)}_{=1} \big\|F(\rho_j)-U_j\rho_jU_j^*\big\|_1
<\frac{\varepsilon}{2}+m\cdot\frac{\varepsilon}{2m}=\varepsilon\,. \qedhere
\end{align*}
\end{proof}

Now we are finally prepared to generalize Prop.~\ref{prop_reach_normal} to infinite dimensions.

\begin{proof}[Proof of Thm.~\ref{thm_normal_V}]
First note that this control problem is well-defined with unique (mild) solutions as guaranteed by Prop.~\ref{prop_control_eq} with $\mathfrak{reach}(\rho_0)=S_\Omega\rho_0$ 
(under assumption PK) where $S_\Omega$ is the system semigroup generated by
\begin{align*}
\{e^{-i\tau(\operatorname{ad}_{H_0}+\sum_{j=1}^mu_j\operatorname{ad}_{H_j})}\,|\,\tau\geq 0,u\in\mathbb R^m\}\cup\{e^{-\tau(i\operatorname{ad}_{H_0}+\Gamma_V+\sum_{j=1}^mu_ji\operatorname{ad}_{H_j})}\,|\,\tau\geq 0,u\in\mathbb R^m\}\,.
\end{align*}
Moreover, all of these maps are in $Q_S(\mathcal H)$ by Prop.~\ref{prop_bounded_pert} so $S_\Omega\subseteq \overline{S_\Omega}^{\,\tops}\subseteq Q_S(\mathcal H)$ (Prop.~\ref{thm_monoid})
Thus assumption (ii) by the state approximation lemma guarantees
$$
 \{U(\cdot)U^*\,|\,U\in\mathcal U(\mathcal H)\} \subseteq\overline{S_\Omega}^{\,\tops}\subseteq Q_S(\mathcal H)
$$
and, in particular, we are dealing exclusively with contraction semigroups (Prop.~\ref{thm_q_norm_1}).\medskip

``$\subseteq$'': 
As $V\in\mathcal B(\mathcal H)$ is assumed to be normal, one has $\Gamma_V\in\mathcal B(\mathcal B(\mathcal H))$ and $\Gamma_V(\mathbbm{1})=0$. Thus the 
corresponding one-parameter semigroup is in $\mathbb S(\mathcal H)$, i.e.~it consists of bi-stochastic 
quantum maps. To see that $e^{-it\operatorname{ad}_{H}-t\Gamma_V}(\rho)\prec\rho$ for all $\rho\in\mathbb D(\mathcal H)$, all $H$ self-adjoint, and all $t\geq 0$ we note
\begin{itemize}
\item $e^{-it\operatorname{ad}_{H}}(\rho)\prec\rho$ as unitary channels (Lemma \ref{lemma_ad_H_unbounded}) do not change the eigenvalues. Thus, majorization 
cannot increase if the noise $\Gamma_V$ is switched off.
\item $e^{-t\Gamma_V}(\rho)\prec\rho$ by Lemma \ref{lemma_li}.
\end{itemize}
Due to this and the fact that $\prec$ is a preorder (so in particular transitive) one finds
\begin{equation*}
(e^{-it\operatorname{ad}_{H}/n}e^{-t\Gamma_V/n})^n(\rho)\in \lbrace \omega\in\mathbb D(\mathcal H)\,|\,\omega\prec\rho\rbrace
\end{equation*}
for all $n\in
\mathbb N_0$. As stated before $(e^{-it\operatorname{ad}_{H}})_{t\geq 0}$, $(e^{-t\Gamma_V})_{t\geq 0}$, and $(e^{-it\operatorname{ad}_{H}-t\Gamma_V})_{t\in
\mathbb R_+}$ all are contraction semigroups. Therefore Lemma \ref{lemma_trotter_banach} yields
$$
\lim_{n\to\infty}\Vert (e^{-it\operatorname{ad}_{H}/n}e^{-t\Gamma_V/n})^n(\rho)-e^{-it\operatorname{ad}_{H}-t\Gamma_V} (\rho)\Vert_1=0
$$
for all $\rho\in\mathcal B^1(\mathcal H)$, which shows
$$
e^{-it\operatorname{ad}_{H}-t\Gamma_V} (\rho)=\lim_{n\to\infty}(e^{-it\operatorname{ad}_{H}/n}e^{-t\Gamma_V/n})^n(\rho)\in\overline{ \lbrace \omega\in\mathbb D(\mathcal H)\,|\,\omega\prec\rho\rbrace}= \lbrace \omega\in\mathbb D(\mathcal H)\,|\,\omega\prec\rho\rbrace
$$
for all $t\geq 0$, $\rho\in\mathbb D(\mathcal H)$ as claimed. In the last step we used that the set of majorized states is 
trace-norm closed (Thm.~\ref{lemma_maj_closed}).\medskip
%

``$\supseteq$'': As $V\in\mathcal K(\mathcal H)$ is normal, by Thm.~\ref{thm_compact_normal_unit_diag} there exists an 
orthonormal basis $(f_j)_{j\in\mathbb N}$ of $\mathcal H$ such that $V=\sum_{j=1}^\infty v_j|f_j\rangle\langle f_j|$ 
with modified eigenvalue sequence $(v_j)_{j\in\mathbb N}$. Whenever we use the term ``diagonal'' 
or ``diag'' in the following it always refers to $(f_j)_{j\in\mathbb N}$.

Let $\varepsilon>0$ and $\rho\in\mathbb D(\mathcal H)$ with $\rho\prec\rho_0$ be given. We now have to find 
$\rho_F\in \mathfrak{reach}_{\Sigma_V}(\rho_0)$ such that $\|\rho-\rho_F\|_1<\varepsilon$. As seen before there exist 
$x,y\in\ell^1_+(\mathbb N)$, $x,y\neq 0$ as well as unitary $W_1,W_2\in\mathcal B(\mathcal H)$ such that
\begin{equation}\label{eq:W_1-W_2}
\rho=W_1\operatorname{diag}(x)W_1^*\quad\text{and}\quad \rho_0=W_2\operatorname{diag}(y)W_2^*
\end{equation}
with $x\prec y$ (so $x$ and $y$ denote the modified eigenvalue sequence of $\rho$
and $\rho_0$, respectively).

First assume that the original and the modified eigenvalue sequence of $\rho$ as well as $\rho_0$ co{\"i}ncide, i.e.~$x=x^\downarrow$, $y=y^\downarrow$ from the start (necessary to apply Prop.~\ref{prop_schur_horn_gohberg}).
The subsequent steps of the proof were sketched in \eqref{eq:steps_idea3}, 
where \textbf{Step 1 \& 3} are the mere application of a suitable unitary channel whereas 
\textbf{Step 2} is about (approximately) getting rid of enough ``off-diagonal'' elements 
$\langle f_j,UW_2^*\rho_0 W_2U^* f_k\rangle$ of 
$UW_2^*\rho_0 W_2U^* = U\operatorname{diag}(y)U^*$.

\textbf{Step 1:} By assumption \& by Lemma \ref{lemma_state_approx} we find $\tilde U\in\mathcal U(\mathcal H)$ such that $\tilde U\rho_0\tilde U^*\in \mathfrak{reach}_{\Sigma_V}(\rho_0)$ with
$$
\|U\operatorname{diag}(y)U^*-\tilde U\rho_0\tilde U^*\|_1=\|UW_2^* \rho_0 W_2U^*-\tilde U\rho_0\tilde U^*\|_1<\varepsilon / 3\,.
$$

\textbf{Step 2:} By Lemma \ref{lemma_normal_generator} the pure noise generator $\Gamma_V$ acts like
\begin{align}
|\langle f_j,\exp(-t\Gamma_V)&(B) f_k\rangle|=\Big|\exp\Big(-\frac{t|v_j-v_k|^2}2\Big)\exp(it\Im(v_j\overline{v_k}))\langle f_j,Bf_k\rangle\Big|\nonumber\\
&=\exp\Big(-\frac{t|v_j-v_k|^2}2\Big)|\langle f_j,Bf_k\rangle|\leq |\langle f_j,Bf_k\rangle|\label{eq_approx_3a2}
\end{align}
on arbitrary $B\in\mathcal B(\mathcal H)$ for all $j,k\in\mathbb N$ and $t\in\mathbb R_0^+$. Evidently,
\begin{equation*}
\lim_{t\to\infty}\langle f_j,\exp(-t\Gamma_V)(B) f_k\rangle=\begin{cases} 0 & \text{if }v_j\neq v_k\,, \\ 
\langle f_j,Bf_k\rangle & \text{else}\,. \end{cases}
\end{equation*}
If we assume $v_j\neq v_k$ for all $j\neq k$, all the off-diagonal terms of $B$ vanish in the limit $t\to\infty$ 
and one is left with $\sum_{j=1}^\infty \langle f_j,Bf_j\rangle |f_j\rangle\langle f_j|=:P(B)$. Note that this 
projection map has Kraus operators $(|f_j\rangle\langle f_j|)_{j=1}^\infty$ so $P\in \mathbb S(\mathcal H)$. 
Since we want to approximate a density operator in trace norm, we only have to care about 
a sufficiently large upper left block of the matrix representation $(\langle f_j,Bf_k\rangle)_{j,k \in \mathbb N}$ 
as the rest is ``already small'' in the trace norm. More formally, by Lemma \ref{lemma_schatten_p_approx} \& \ref{lemma_approx_strong_top} there exists 
$N_1\in\mathbb N$ such that 
\begin{equation}\label{eq_approx_4a}
\|\tilde U\rho_0\tilde U^*-\Pi_n \tilde U\rho_0\tilde U^*\Pi_n\|_1<\varepsilon / 24
\end{equation}
for all $n\geq N_1$, where $\Pi_n := \sum_{j=1}^n|f_j\rangle\langle f_j|$ for all 
$n\in\mathbb N$.

Of course, there is no reason for the eigenvalues of $V$ to be pairwise different. Therefore we have to make sure 
that the upper left block is large enough such that it corresponds to at least two different 
eigenvalues of $V$; then we have access to partial decoherence, which we may spread anywhere needed via 
permutation channels.

Due to $V\neq 0$ and $v_j\to 0$ as $j\to\infty$ (compactness of $V$), there exists $M\in\mathbb N$ such that 
$v_1\neq v_M$. On the other hand \eqref{eq_approx_4a} still holds if we define $N:=\max\lbrace N_1,M\rbrace$. 
Then, by construction and \eqref{eq_approx_3a2}, we know that $\langle f_1,Bf_M\rangle$ (and 
$\langle f_M,Bf_1\rangle$) tend to zero when pure noise is applied.


Thus we find $\alpha\in\mathbb N_0$, $\alpha\leq N(N-1)/2$ (number of matrix elements above the diagonal), 
permutation operators $\sigma_1,\ldots,\sigma_\alpha\in\mathcal U(\mathcal H)$, and relaxation times $s_1,\ldots,s_\alpha\in\mathbb R_0^+$ such that
\begin{itemize}
\item the permutations only operate non-trivially on the $N\times N$-block, i.e.~for all $l=1,\ldots,\alpha$ 
and $k>N$ one has $\sigma_lf_k=f_k$.
\item
for every matrix element $|f_j\rangle\langle f_k|$ with $j,k=1,\ldots,N$, $j\neq k$ there
exists a permutation $\sigma_l$ with $1\leq l \leq \alpha$ such that $|f_j\rangle\langle f_k|$ sits 
in the ``relaxation'' spot (i.e.~$|f_M\rangle \langle f_1|$ or $|f_1\rangle\langle f_M|$).
More precisely, 
\begin{equation}\label{eq:perm_relax}
\big\|\operatorname{Ad}_{\sigma_l^*} \circ \exp(-s_l\Gamma_V) \circ \operatorname{Ad}_{\sigma_l}
\big(|f_j\rangle\langle f_k|\big)\big\|_1\leq \frac{\varepsilon}{12N^2} \,.
\end{equation}
\item after having successively applied all operations from \eqref{eq:perm_relax}, every matrix element $|f_j\rangle\langle f_k|$ is in its 
original spot because all $|f_j\rangle\langle f_k|$ are eigenvectors of $\exp(-s_l\Gamma_V)$.
\end{itemize}
Now, using linearity of the involved maps, the estimate in question reads
\begin{align*}
\| P( \tilde U\rho_0 &\tilde U^*)-\prod_{m=1}^{\alpha}\big( \operatorname{Ad}_{\sigma_m^*}{}\!\circ \exp(-s_m\Gamma_V) \circ \operatorname{Ad}_{\sigma_m} \big)( \tilde U\rho_0 \tilde U^*)\|_1<\\
& \Big( \|P\|_\textrm{op} + \prod_{m=1}^\alpha \| \operatorname{Ad}_{\sigma_m} \|^2_\textrm{op}\, \| \exp(-s_m\Gamma_V) \|_\textrm{op} \Big) \| \tilde U\rho_0 \tilde U^*-\Pi_N \tilde U\rho_0 \tilde U^*\Pi_N\|_1 \\
+ &\Big\| \sum_{j,k=1}^{N} \langle f_j,\tilde U\rho_0 \tilde U^* f_k\rangle\Big( P-\prod_{m=1}^{\alpha} \operatorname{Ad}_{\sigma_m^*}{}\!\circ \exp(-s_m\Gamma_V) \circ \operatorname{Ad}_{\sigma_m} \Big)(|f_j\rangle\langle f_k|)\Big\|_1\,.
\end{align*}
The first summand is smaller than $2\cdot \frac{\varepsilon}{24}=\frac{\varepsilon}{12}$ by Prop.~\ref{thm_q_norm_1} and \eqref{eq_approx_4a}. For the second one notice that
\begin{align*}
\Big( P-\prod_{m=1}^{\alpha} \operatorname{Ad}_{\sigma_m^*}{}\!\circ \exp (-s_m\Gamma_V) \circ \operatorname{Ad}_{\sigma_m} \Big)(|f_j\rangle\langle f_j|) = 0
\end{align*}
for all $j\in\mathbb N$. 
Now $P(|f_j\rangle\langle f_k|)=0$ whenever $j\neq k$ and, moreover, 
\begin{align*}
\Big\|\prod_{m=1}^{\alpha} \big(\operatorname{Ad}_{\sigma_m^*}{}\!\circ \exp(-s_m\Gamma_V) \circ \operatorname{Ad}_{\sigma_m} \big)(|f_j\rangle\langle f_k|)\Big\|_1 \leq\frac{\varepsilon}{12N^2}\,.
\end{align*}
by 
\eqref{eq_approx_3a2} and \eqref{eq:perm_relax}. 
Putting 
together gives the estimate
\begin{align*}
\Big\| P&(\tilde U\rho_0 \tilde U^*)- \prod_{m=1}^{\alpha}\big( \operatorname{Ad}_{\sigma_m^*}{}\!\circ \exp(-s_m\Gamma_V) \circ \operatorname{Ad}_{\sigma_m} \big)( \tilde U\rho_0 \tilde U^*)\Big\|_1\\
< & \frac{\varepsilon}{12}+ \sum_{\substack{j,k=1\\ j\neq k}}^{N}\underbrace{|\langle f_j,\tilde U\rho_0 \tilde U^* f_k\rangle|}_{\leq 1}\Big\|\prod_{m=1}^{\alpha} \big(\operatorname{Ad}_{\sigma_m^*}{}\!\circ \exp(-s_m\Gamma_V) \circ \operatorname{Ad}_{\sigma_m} \big)(|f_j\rangle\langle f_k|)\Big\|_1\\
< & \frac{\varepsilon}{12} + \sum\limits_{j,k=1,\, j\neq k}^n \frac{\varepsilon}{12N^2}\leq \frac{\varepsilon}{6}\,.
\end{align*}
This leaves us with two problems:
\begin{itemize}
\item[1.] We have to approximate all permutation channels.
\item[2.] We do not have access to pure noise $(\exp(-t\Gamma_V))_{t\in\mathbb R_0^+}$ within the given control problem.
\end{itemize}
For solving the first problem we exploit that we can strongly approximate every unitary 
channel. First, to simplify the upcoming computations, let us assume w.l.o.g.~that $\sigma_\alpha$
is the identity and let us introduce the notation $\pi_l := \sigma_l \circ \sigma_{l-1}^*$ for 
$l\in\lbrace 2,\ldots,\alpha\rbrace$ and $\pi_1 := \sigma_1$. Moreover, define
$$
\omega_l:=\Big( \exp(-s_l\Gamma_V) \circ \prod_{m=l+1}^\alpha \big(\operatorname{Ad}_{\pi_m^*}\circ \exp(-s_m\Gamma_V) \big)\Big)(\tilde U\rho_0\tilde U^*)\in\mathbb D(\mathcal H)
$$
for every $l\in\lbrace1,\ldots,\alpha\rbrace$ as then by Lemma \ref{lemma_state_approx}
we find $\tilde\pi_l\in\mathcal U(\mathcal H)$ which we have access to within the system semigroup such that
\begin{equation*}
\|\tilde\pi_l^* \omega_l \tilde\pi_l - \pi_l^* \omega_l \pi_l\|_1 < \frac{\varepsilon}{12\alpha}\,.
\end{equation*}
Then a telescope argument (cf.~Lemma \ref{lem:telescope})
yields the estimate
\begin{align*}
\Big\|\Big(&\prod_{m=1}^{\alpha}\big( \operatorname{Ad}_{\tilde\pi_m^*}{}\!\circ \exp(-s_m\Gamma_V) \big)-\prod_{m=1}^{\alpha}\big( \operatorname{Ad}_{\pi_m^*}{}\!\circ \exp(-s_m\Gamma_V) \big)\Big)(\tilde U\rho_0\tilde U^*) \Big\|_1\\
&\leq \sum_{m=1}^\alpha \Big\| \Big(\prod_{l=1}^{m-1}( \operatorname{Ad}_{\tilde\pi_l^*}{}\!\circ \exp(-s_l\Gamma_V) )\circ ( \operatorname{Ad}_{\tilde\pi_m^*}- \operatorname{Ad}_{\pi_m^*})\Big)(\omega_m )\Big\|_1\\
&\leq\sum_{m=1}^\alpha\Big( \prod_{l=1}^{m-1}\| \operatorname{Ad}_{\tilde\pi_l^*}{}\|_\textrm{op}\, \|\exp(-s_l\Gamma_V)\|_\textrm{op}\Big) \|\tilde\pi_m^* \omega_m \tilde\pi_m-\pi_m^* \omega_m \pi_m\|_1
<\frac{\varepsilon}{12}
\end{align*}
where in the last step we once again used Prop.~\ref{thm_q_norm_1}.

For the second problem we luckily may approximate the pure noise as precisely as needed using
Coro.~\ref{lemma_trotter_approx}. For every $l=1,\ldots,\alpha$ define 
$$
\rho_l:=\prod_{m=l+1}^\alpha (\operatorname{Ad}_{\tilde\pi_m^*}{}\!\circ \exp(-s_m\Gamma_V))(\tilde U\rho_0\tilde U^*)\in\mathbb D(\mathcal H)\,.
$$ 
Then by Coro.~\ref{lemma_trotter_approx} there exists a \textsc{cptp} map $F_l$ which we have access to within the system semigroup
such that $\|\exp(-s_l\Gamma_V)(\rho_l) - F_l(\rho_l)\|_1<\frac{\varepsilon}{12\alpha}$. Just as before
\begin{align*}
\Big\|\Big( \prod_{m=1}^{\alpha}\big( \operatorname{Ad}_{\tilde\pi_m^*}{}\!\circ \exp(-s_m\Gamma_V) \big)-\prod_{m=1}^{\alpha}\big( \operatorname{Ad}_{\tilde\pi_m^*}{}\!\circ \,F_m \big)\Big)(\tilde U\rho_0\tilde U^*) \Big\|_1<\frac{\varepsilon}{12}\,.
\end{align*}

\textbf{Step 3:} The current state $\tilde\rho:=\prod_{m=1}^{\alpha}\big( \operatorname{Ad}_{\tilde\pi_m^*}{}\!\circ \,F_m \big)(\tilde U\rho_0\tilde U^*)$ of the system is ``close to $\operatorname{diag}(x)$'' in 
the trace distance as we saw before. Now we want to apply the unitary channel generated by $W_1$ so
again by Lemma \ref{lemma_state_approx} one finds unitary $\tilde W\in\mathcal B(\mathcal H)$
such that 
$
\|W_1\tilde\rho W_1^*-\tilde W\tilde\rho\tilde W^*\|_1<\frac{\varepsilon}{3}\,.
$
Then one has
$\rho_F = \operatorname{Ad}_{\tilde W}\circ\prod_{m=1}^{\alpha}\big( \operatorname{Ad}_{\tilde\pi_m^*}{}\!\circ \,F_m\big)(\tilde U\rho_0\tilde U^*)\in \mathfrak{reach}_{\Sigma_V}(\rho_0)$ and by \eqref{eq:W_1-W_2}
\begin{align*}
\|\rho-\rho_F\|_1 & \leq \|W_1P(UW_2^*\rho_0 W_2U^*) W_1^*-W_1P(\tilde U\rho_0\tilde U^*) W_1^*\|_1\\
&\quad+\|W_1P(\tilde U\rho_0\tilde U^*) W_1^*-W_1\tilde\rho W_1^*\|_1+\|W_1\tilde\rho W_1^*-\rho_F\|_1\,.
\end{align*}
Using Prop.~\ref{thm_q_norm_1} we 
ultimately obtain
\begin{align*}
\|\rho-&\rho_F\|_1\leq \|\operatorname{Ad}_{W_1}\|_\textrm{op} \|P\|_\textrm{op} \|UW_2^*\rho_0 W_2U^*-\tilde U\rho_0\tilde U^*\|_1\\
&\quad+\|\operatorname{Ad}_{W_1}\|_\textrm{op}\Big\|P(\tilde U\rho_0\tilde U^*) -\prod_{m=1}^{\alpha}\big( \operatorname{Ad}_{\pi_m^*}{}\!\circ \exp(-s_m\Gamma_V) \big)(\tilde U\rho_0\tilde U^*)\,\Big\|_1\\
&\quad+ \|\operatorname{Ad}_{W_1}\|_\textrm{op} \Big\|\big(\prod_{m=1}^{\alpha}\big( \operatorname{Ad}_{\pi_m^*}{}\!\circ \exp(-s_m\Gamma_V) \big)-\\
&\hspace*{126pt}-\prod_{m=1}^{\alpha}\big( \operatorname{Ad}_{\tilde\pi_m^*}{}\!\circ \exp(-s_m\Gamma_V) \big)\big)(\tilde U\rho_0\tilde U^*) \Big\|_1\\
&\quad+\|\operatorname{Ad}_{W_1}\|_\textrm{op}\Big\|\prod_{m=1}^{\alpha}\big( \operatorname{Ad}_{\tilde\pi_m^*}{}\!\circ \exp(-s_m\Gamma_V) \big)(\tilde U\rho_0\tilde U^*) -\tilde\rho\,\Big\|_1\\
&\quad+\|W_1\tilde\rho W_1^*-\rho_F\|_1<\frac{\varepsilon}{3}+\frac{\varepsilon}{6}+\frac{\varepsilon}{12}+\frac{\varepsilon}{12}+\frac{\varepsilon}{3}=\varepsilon\,.
\end{align*}
Finally, what happens if we cannot apply Prop.~\ref{prop_schur_horn_gohberg} directly, that is, if the original and the modified eigenvalue sequence of $\rho=W_1\operatorname{diag}(x)W_1^*$ or $\rho_0=W_2\operatorname{diag}(y)W_2^*$ do not co{\"i}ncide? Given $\varepsilon>0$, first of all we find $N\in\mathbb N$ such that
\begin{equation}\label{eq:ineq_x_y}
\sum\nolimits_{j=N+1}^\infty x_j^\downarrow <\frac{\varepsilon}{12}\qquad \sum\nolimits_{j=N+1}^\infty y_j^\downarrow <\frac{\varepsilon}{12}\,.
\end{equation}
There clearly exist unitaries $ X, Y\in\mathcal U(\mathcal H)$ such that
$
 X\rho X^*=\operatorname{diag}(x_1^\downarrow,\ldots,x_N^\downarrow,*,*,\ldots)$ and \mbox{$ Y\rho_0 Y^*=\operatorname{diag}(y_1^\downarrow,\ldots,y_N^\downarrow,*,*,\ldots)
$}
where the diagonal entries differ from the original ones only by a permutation on a finite block. As the tail of these new diagonals is ``already small'' we may change these elements within the realm of approximation. Given $\sum_{j=1}^N x_j^\downarrow\leq \sum_{j=1}^N y_j^\downarrow$ (because $\rho\prec\rho_0$) where this inequality may or may not be strict, we want to fill up $ X\rho X^*$ with small entries such that the traces match. Define $\varphi:=\sum_{j=1}^N(y_j^\downarrow-x_j^\downarrow)$ where $0\leq\varphi <\frac{\varepsilon}{12}$ due to \eqref{eq:ineq_x_y} and $\rho\geq 0$, as well as $m:=\lceil \varphi/x_k^\downarrow\rceil\in\mathbb N$. Here $k\in\lbrace 1,\ldots,N\rbrace$ is chosen such that $x_k^\downarrow$ is the smallest non-zero entry of $(x_1^\downarrow,\ldots,x_N^\downarrow)$. The new (eigenvalue) sequences then are
$
\hat x:=(x_1^\downarrow,\ldots,x_k^\downarrow,\tfrac{\varphi}{m},\ldots,\tfrac{\varphi}{m},0,0,\ldots)$ (where $\varphi/m$ occurs $m$ times) and $\hat y:=(y_1^\downarrow,\ldots,y_N^\downarrow,0,0,\ldots)\,.
$
These sequences satisfy $\hat x^\downarrow=\hat x$, $\hat y^\downarrow=\hat y$, and $\hat x\prec\hat y$ (for this note that if $k<N$ then majorization forces $\sum_{j=1}^k x_j^\downarrow=\sum_{j=1}^k y_j^\downarrow=1$ and thus $\varphi=0$) so we could apply Prop.~\ref{prop_schur_horn_gohberg} to them. Now to $$
\omega:=\frac{\operatorname{diag}(\hat x)}{\sum_{j=1}^N y_j^\downarrow}\quad\text{ and }\quad\omega_0:=\frac{\operatorname{diag}(\hat y)}{\sum_{j=1}^N y_j^\downarrow}	\,,
$$
which are both in $\mathbb D(\mathcal H)$, we can apply the original scheme which yields a \textsc{cptp} map $f$ on $\mathcal H$ such that $f(\omega_0)=\omega_F\in\mathfrak{reach}_{ \Sigma_V}(\omega_0)$ and $\|\omega-\omega_F\|_1<\frac{\varepsilon}{6}$. Of course linearity implies $\|\operatorname{diag}(\hat x)-f(\operatorname{diag}(\hat y)) \|_1<\frac{\varepsilon}{6}$. The final scheme goes as follows:
$$
\rho_0\overset{ Y}\longrightarrow Y\rho_0 Y^*\approx\operatorname{diag}(\hat y)\overset{f}\longrightarrow \operatorname{diag}(\hat x)\approx X\rho X^*\overset{ X^*}\longrightarrow \rho
$$
More precisely, by Lemma \ref{lemma_state_approx} we find unitaries $\tilde X,\tilde Y\in\mathcal B(\mathcal H)$ such that 
$$
\| Y\rho_0 Y^*-\tilde Y\rho_0\tilde Y^*\|_1<\tfrac{\varepsilon}{4},\quad \|\tilde X^* (f\circ\operatorname{Ad}_{\tilde Y})(\rho_0)\tilde X- X^* (f\circ\operatorname{Ad}_{\tilde Y})(\rho_0) X\|_1<\tfrac{\varepsilon}{4}
$$
and $\rho_F:=(\operatorname{Ad}_{\tilde X^*}\circ f\circ \operatorname{Ad}_{\tilde Y})(\rho_0) \in\mathfrak{reach}_{ \Sigma_V}(\rho_0)$. Putting things together,
\begin{align*}
\|\rho-\rho_F\|_1&\leq \|\rho- X^*\operatorname{diag}(\hat x) X\|_1+\| X^*\operatorname{diag}(\hat x) X- X^* f(\operatorname{diag}(\hat y)) X \|_1\\
&\hphantom{\leq \|\rho- X^*\operatorname{diag}(\hat x) X\|_1}\ +\| X^* f(\operatorname{diag}(\hat y)) X -\rho_F \|_1\\
&<\frac{\varepsilon}{6}+\|\operatorname{Ad}_{ X^*}\|_{\textrm{op}}\cdot\frac{\varepsilon}{6}+\|(\operatorname{Ad}_{ X^*}\circ f)(\operatorname{diag}(\hat y)) -\rho_F \|_1\\	
&\leq\frac{\varepsilon}{3}+\|\operatorname{Ad}_{ X^*}\|_{\textrm{op}}\|f \|_{\textrm{op}}\| \operatorname{diag}(\hat y)- Y\rho_0 Y^* \|_1\\
&\hphantom{\leq\frac{\varepsilon}{4}}\ +\|\operatorname{Ad}_{ X^*}\|_{\textrm{op}}\|f \|_{\textrm{op}} \| Y\rho_0 Y^*-\tilde Y\rho_0\tilde Y^*\|_1\\
&\hphantom{\leq\frac{\varepsilon}{4}}\ +\| (\operatorname{Ad}_{ X^*}\circ f\circ \operatorname{Ad}_{ \tilde Y})(\rho_0) -\rho_F \|_1<\varepsilon
\end{align*}
so $\rho\in\overline{\mathfrak{reach}_{ \Sigma_V}(\rho_0)}$, which concludes the proof.
\end{proof}

\chapter{Conclusion and Outlook}\label{ch:concl_outlook}

For the first time we have derived sufficient conditions under which a quantum-dynamical 
system can actually (approximately) reach all quantum states majorized by the respective initial 
state in an infinite-dimensional quantum system following a 
controlled Markovian master equation. 
To this end, we have extended the standard unital \textsc{gksl} master equation to an infinite-dimensional 
bilinear control system with unbounded drift, the unitary part of which has to be (strongly approximately) controllable
and the dissipative part (generated by a single normal compact noise term $V$) 
has to be bang-bang switchable. For this is was important to generalize the $C$-numerical range and some of its properties to the trace class and general Schatten classes and, more fundamentally, to generalize the notion of (approximate) unitary controllability to infinite dimensions. The latter is reasonable from a control theoretic perspective as---even in finite dimensions---it is equivalent to unitary controllability on the quantum states, which in turn is \textit{strictly weaker} than controllability on state vectors. Also such an approach allows for using powerful tools from operator and general Lie group and topological group theory.

We also showed that coupling a finite-dimensional system, either locally or globally, to a bath of temperature zero\index{temperature zero bath} in a switchable manner suffices to (approximately) generate \textit{every} state from every initial state. Indeed this is the best result one can obtain as exact controllability of the state problem is impossible as long as one is restricted to Markovian dynamics.
All of this takes recent results on qubit systems~\cite{BSH16,OSID17} to more general finite, as well as infinite dimensions.

Moreover for the problem of coupling an $n$-level system to a bath of finite temperature we made significant progress on finding a non-trivial upper bound of the corresponding reachable set. The key for doing so was to fully characterize the $d$-majorization polytope and all of its (now easily computable) extreme points, and to show that for all $y\in\mathbb R_+^n$ there exists a unique extreme point $z=z(y,d)$ which classically majorizes every point from said polytope. This result enabled us to find an upper bound for the reachable set of the toy model (i.e.~for the simpler control problem on the standard simplex), that is, $\mathfrak{reach}_{\Lambda_d}(x_0)\subseteq \{x\in\Delta^{n-1}\,|\,x\prec z\}$ for all $x_0\in\Delta^{n-1}$.\medskip

%
This also makes for the first and most natural point of an outlook section: Does this upper bound pertain from the toy model to the general problem, that is, does one have
\begin{equation}\label{eq:reach_outlook_d}
\mathfrak{reach}_{\Sigma_d}(\rho_0)\subseteq\{\rho\in\mathbb D(\mathbb C^n)\,|\,\rho\prec\operatorname{diag}(z)\}
\end{equation}
where $z$ would of course depend on the initial state $\rho_0\in\mathbb D(\mathbb C^n)$ and be non-trivial (i.e.~$z$ is not pure, and $z>0$ whenever $\rho_0>0$)? We saw that only lower
bounds transfer directly from the classical (i.e.~toy model) to the quantum control problem \eqref{eq:toy_state_connect}. Of course this does not rule out the validity of \eqref{eq:reach_outlook_d}, but if this were true then a more creative proof is needed.
Also one cannot directly copy the toy model-proof as it relied heavily on the convex polytope structure in the vector case. This does not hold for general $D$-majorization: in the matrix case the number of extreme points is infinite\footnote{
Similarly while the set of $d$-stochastic matrices forms a convex polytope, the set of channels with a common fixed point $D$ has infinitely many extreme points. The argument, just as below, relies on the concatenation with suitable unitary channels from left and right; after all the bijective quantum channels the inverse of which is a channel again are precisely the unitary ones (Prop.~\ref{ch_3_Theorem_14}).
}; a straightforward argument shows that if $X$ is extremal in $ M_D(A)$ then so is $U^* XU$ for all unitaries $U$ which satisfy $[U,D]=0$. A way of bridging this gap could be to define the unitary equivalence relation
$$
\sim\;:=\{(X,UXU^*)\,|\,X\in M_D(A), U\in\mathbb C^{n\times n}\text{ unitary with }[U,D]=0\}
$$
for arbitrary $d\in\mathbb R_{++}^n$ and $A\in\mathbb C^{n\times n}$, and then look at the equivalence class of an extreme point $X$ of $M_D(A)$ under $\sim$. Whether one ends up with finitely many extreme points after factoring out the unitary equivalents---and whether this would even be of use for the corresponding control problem---for now remains open.\medskip

Unsurprisingly, this is just one of many natural follow-up questions:
\begin{itemize}
\item Does our temperature zero result hold as well in infinite dimensions? An answer to this would go via the unavoidable problem of proving that the control system is still well-defined if $V$ is unbounded and, more fundamentally, that even the uncontrolled dynamics are well-defined in this case (cf.~also Rem.~\ref{rem_unbounded_lindblad_genV}).
%
%
\item For \textit{applying} the results to broader classes of physical systems, 
the current setup restricts us to system Hamiltonians $H_0$ 
with discrete spectrum such as
bound systems where particles are trapped within an unbounded potential (e.g., harmonic oscillators). 
To look at more interesting setups where processes like ionization,
tunneling, and evaporation play a role, we have to use operators with
continuous spectrum. However, such more general results for strong approximate controllability for the unitary propagators are still amiss, and in this area even coherent control is
not understood well enough (if at all). 
\item Which of the assumptions of our infinite-dimensional control result are necessary and which can be relaxed? While obtaining such a result was remarkable in the first place, the Lindblad-$V$ being compact is rather restrictive from the application point of view (unless, of course, one considers only finite-dimensional systems). More precisely: can $V$ chosen to be bounded and normal, or can one add time-dependence of $V$, or does the result still hold for more than one Lindblad-$V$ (assuming the environment is \textit{globally} switchable and the $V$'s do not commute)? Note that the latter is unclear even in finite dimensions.
\item While we took care of the Lamb shift\index{Lamb shift} in finite dimensions (cf.~Rem.~\ref{rem_control_sp_lamb}) is this still the case for our infinite-dimensional result? That is, even if the system's Hamiltonian $H_0$ and the Lamb shift Hamiltonian $H_\textrm{LS}$ still share an eigenbasis can one guarantee boundedness of $H_\textrm{LS}-H_0$? If not then, as in the first bullet point of this list, one first needs controllability and well-posedness results for unbounded control operators.
\item Is the control problem induced by switchable coupling to a temperature zero bath accessible? From the proof it would not be surprising if $\{\rho\in\mathbb D(\mathbb C^n)\,|\,\rho>0\}$ were always a subset of the reachable set for all initial states, but this is only a conjecture for now. Unfortunately this is not answered by the general result that approximate controllability together with accessibility implies exact controllability (cf.~Lemma \ref{rem_approx_contr_acc}) as this only holds for control problems on closed (Lie) groups. 
\item The temperature zero result proves that every finite-dimensional state transfer can be done (approximately) in a Markovian way. Of course non-Markovian dynamics\index{quantum dynamics!non-Markovian} \textit{may} allow for exact reachability---as those do not have to obey strict positivity---or may make for better numerics by allowing for faster control sequences (given a fixed precision), compared to the Markovian setting.
\end{itemize}
Thus we see that---although our results are rich, non-trivial, and make for connections to numerous different fields of mathematics and physics---this was but a first step to explore quantum control problems involving switchable (Markovian) noise and, more generally, quantum control problems on $\mathbb D(\mathcal H)$ for infinite-dimensional Hilbert spaces $\mathcal H$.

\appendix{}


\renewcommand{\thechapter}{\Alph{chapter}}
\chapter{Appendix}\label{appendix}
\section{The Functional Analysis Funfair}\label{sec_refresh_func_ana}

All norms on finite-dimensional vector spaces are equivalent which means that for questions of convergence, closure, and the like, it does not matter which norm we consider. Indeed most of the concepts we will introduce in this section (such as completeness, weak topologies, reflexivity, etc.) are without consequences in finite dimensions.

However for infinite-dimensional spaces we cannot find any way around fundamental concepts of topology. As an example we will need a notion of convergence on bounded linear operators which is weaker than the usual operator norm to solve some problems arising in infinite-dimensional quantum dynamics (such as approximation of observables or missing continuity of quantum-dynamical groups). 

\subsection{Topological Basics}
To start with a fundamental question: What does it mean for a subset of some general set to be open or closed---notions very familiar from everyday spaces such as $\mathbb R$ or $\mathbb R^n$? This brings us to the definition of a topology and well-studied related concepts, cf.~\cite[Ch.~3 ff.]{Willard70} as well as \cite{MeiseVogt97en,Rudin91,Munkres00}, which we shall go through as quickly as possible but at the same time as slowly as necessary.

\begin{definition}\label{defi_topology}
Let $X$ be a non-empty set and $\tau$ a system of subsets of $X$ (i.e.~$\tau\subseteq \mathcal P(X)$ with the latter being the power set of $X$). Then $\tau$ is a topology\index{topology} on $X$ if
\begin{enumerate}
\item $\emptyset\in\tau$ and $X\in\tau$
\item $\bigcup_{i\in I}U_i\in\tau$ for any family $(U_i)_{i\in I}\subset\tau$
\item $\bigcap_{i\in I}U_i\in\tau$ for any finite family $(U_i)_{i\in I}\subset\tau$ (that is, $|I|<\infty$)
\end{enumerate}
In this case $(X,\tau)$ is a topological space\index{space!topological} and the elements of $\tau$ are called the open\index{open (in topology)} sets of $X$. A subset $A\subseteq X$ is closed\index{closed} if its complement is open, i.e.~$X\setminus A\in\tau$. 
\end{definition}

Most topologies one deals with in practice have the somewhat intuitive property of ``separating points'' which is formalized as follows, see \cite[Def.~4.1 \& Def.~13.5]{Willard70}.

\begin{definition}
Let $(X,\tau)$ be a topological space.
\begin{itemize}
\item[(i)] Given $x\in X$ and $A\subseteq X$ then $A$ is called a neighborhood\index{neighborhood} of $x$ if there exists $U\in\tau$ such that $x\in U\subseteq A$. If $A$ additionally is open itself then it is called an open neighborhood\index{neighborhood!open} of $x$.
\item[(ii)] A topological space $(X,\tau)$ is called a Hausdorff space\index{space!Hausdorff} if for every $x_1,x_2\in X$ with $x_1\neq x_2$ there exist neighborhoods $A_1$ of $x_1$ and $A_2$ of $x_2$ such that $A_1\cap A_2=\emptyset$. 
\end{itemize}
\end{definition}
\noindent Be aware---to avoid possible confusion---that some authors \textit{define} a neighborhood to be open and some authors include the Hausdorff condition into their definition of a topology.\medskip

With this out of the way we can specify what interior, closure, boundary, and compactness means in topological spaces, cf.~\cite[Ch.~3 \& Ch.~17]{Willard70}.

\begin{definition}
Let $(X,\tau)$ be a topological space and let $A\subseteq X$ be given. 
\begin{itemize}
\item[(i)] Define
$$
\operatorname{int}(A):=\bigcup_{\substack{U\text{ open}\\U\subseteq A}} U\quad \quad \quad \overline{A}:=\bigcap_{\substack{U\text{ closed}\\A\subseteq U}} U\quad \quad \quad \partial A:=\overline{A}\setminus \operatorname{int}(A)\,.
$$
One calls $\operatorname{int}(A)$\label{symb_interior}
the interior\index{interior}, $\overline{A}$\label{symb_closure}
the closure\index{closure}, and $\partial A$\label{symb_boundary}
the boundary\index{boundary} of $A$. 
\item[(ii)] $A$ is said to be compact\index{compact} if every open cover of $A$ has a finite subcover, i.e.~for every family $(U_i)_{i\in I}\subseteq\tau$ with $A\subseteq\bigcup_{i\in I}U_i$ there exists $J\subseteq I$ finite such that $A\subseteq\bigcup_{j\in J}U_j$. 
\item[(iii)] $A$ is relatively compact\index{relatively compact} if $\overline{A}$ is compact.
\end{itemize}
\end{definition}
\noindent It is straightforward to see that $\operatorname{int}(A)\subseteq A\subseteq\overline{A}$ and that $\partial A=\overline{A}\cap\overline{X\setminus A}$. Moreover $\operatorname{int}(A)$ is open, $\overline{A}$ is closed, $\operatorname{int}(A)=A$ if and only if $A$ is open, and $\overline{A}=A$ if and only if $A$ is closed. \medskip

The closure will be an essential notion when studying reachable sets of controlled dynamical systems later on. Roughly speaking it introduces the possibility of ``reaching something approximately'' or ``reaching something in infinite time''
. We will see the reasoning behind this when talking about metric spaces in Section \ref{subsec:metric_normed_inner_product}.\medskip

For the following definition see \cite[Ch.~7 \& Ch.~11]{Willard70}.

\begin{definition}[Continuity \& Nets]
Let topological spaces $X,Y$, some $x_0\in X$, and a function $f:X\to Y$ be given.
\begin{itemize}
\item[(i)] $f$ is continuous\index{continuous} in $x_0$ for every neighborhood $V$ of $f(x_0)$ there exists an neighborhood $U$ of $x_0$ such that $f(U)\subset V$. 
\item[(ii)] $f$ is continuous if it is continuous in every $x\in X$. Equivalently, $f$ is continuous if for every open set $V\subset Y$ the pre-image $f^{-1}(V)=\{x\in X\,|\,f(x)\in V\}$ is an open subset of $X$. 
\item[(iii)] A net\index{net} on $X$ is a function $f$ from a directed set\footnote{A relation $\succeq$ on a set $I$ is called a partial order\index{partial order} if $i\succeq j$, $j\succeq i$ implies $i=j$, and $i\succeq j$, $j\succeq k$ implies $i\succeq k$, and finally $i\succeq i$ for all $i,j,k\in I$. Then a set $I$ with partial order $\succeq$ is called a directed set\index{directed set} if for all $i,j\in I$ there exists $k\in I$ such that $k\succeq i$ and $k\succeq j$.} $I$ into $X$. Such a net $f$ will also be denoted by $(x_i)_{i\in I}$ when identifying $f(i)=x_i$ for all $i\in I$. 
\item[(iv)] A net $(x_i)_{i\in I}$ on $X$ converges to $x\in X$ if for every neighborhood $N$ of $x$ there exists $i_N\in I$ such that $x_i\in N$ for all $i\succeq i_N$. We will occasionally write $x_i\overset{\tau}\to x$ or $x_i\to x$ in $\tau$ if the latter is the topology on $X$. 
\end{itemize}
\end{definition}
\noindent The notion of a net generalizes usual sequences which is why nets sometimes are called ``generalized sequences''\index{sequence!generalized}. Indeed a net $(x_i)_{i\in I=\mathbb N}$ in a topological space $X$ is called a sequence\index{sequence}. \medskip

Having introduced all those concepts we want to quickly explore some connections between them. Taking the intuition from $\mathbb R^n$ with the standard topology the generalized sequences (nets) from above should characterize continuity, closedness, and compactness. This is indeed the case, cf.~\cite[Thms.~11.5--11.8, 13.7, 17.4]{Willard70}.
\begin{lemma}\label{lemma_topol_connect}
Let topological spaces $(X,\tau_X), (Y,\tau_Y)$, a subset $A\subseteq X$, and $f:X\to Y$ be given.
\begin{itemize}
\item[(i)] $f$ is continuous\index{continuous} in $x_0\in X$ if and only if for every net $(x_i)_{i\in I}$ in $X$ which converges to $x$, $(f(x_i))_{i\in I}$ converges to $f(x)$. 
\item[(ii)] A point $x\in X$ belongs to $\overline{A}$ if and only if there exists a net on $A$ (i.e.~a net on $X$ with elements in $A$) which converges to $x$. Thus $A$ is closed if and only if for every net in $A$, all of its limits also belong to $A$.
\item[(iii)] $A$ is compact\index{compact} if and only if every net on $A$ has a convergent subnet.
\item[(iv)] $\tau_X$ is Hausdorff if and only if no net in $X$ converges to more than one point.
\end{itemize}
\end{lemma}

To wrap this section up we recall the notion of separability of a topological space.

\begin{definition}\label{defi_top_separable}
Let $(X,\tau)$ be a topological space. 
\begin{itemize}
\item[(i)] A subset $A\subseteq X$ is called dense\index{dense} in $X$ if $\overline{A}=X$.
\item[(ii)] If there exists a countable dense subset, i.e.~if there exists a subset $\{x_n\,|\,n\in\mathbb N\}\subseteq X$ such that $\overline{\{x_n\,|\,n\in\mathbb N\}}=X$, then $X$ is called separable\index{separable}.
\end{itemize}
\end{definition}

A standard argument to show non-separability of a topological space goes as follows.

\begin{lemma}\label{lemma_non_sep}
Let $(X,\tau)$ be a topological space. If there exists an uncountable family of pairwise disjoint non-empty open sets, i.e.~a family $\{U_i\}_{i\in I}\subseteq\tau\setminus\{\emptyset\}$ with $I$ uncountable and $U_i\cap U_j=\emptyset$ for all $i,j\in I$ with $i\neq j$, then $(X,\tau)$ is not separable.
\end{lemma}
\begin{proof}
Let $\{U_i\}_{i\in I}\subseteq\tau\setminus\{\emptyset\}$ be an uncountable family of pairwise disjoint open sets and assume to the contrary that $(X,\tau)$ were separable. Thus one finds $\{x_n\}_{n\in\mathbb N}\subseteq X$ such that $\overline{\{x_n\}_{n\in\mathbb N}}=X$ which is equivalent to $U\cap \{x_n\}_{n\in\mathbb N}\neq\emptyset$ for all $U\in\tau$ non-empty, cf.~\cite[Thm.~17.5]{Munkres00}. In particular---by applying this to $U_i$---we find $n_i\in\mathbb N$ such that $x_{n_i}\in U_i$. This yields a map $f:I\to\mathbb N$, $i\mapsto n_i$ which is injective: If $f(i)=f(j)=:n$ then $x_n=U_i\cap U_j$ but by assumption the latter is empty unless $i=j$. The fact that we found an injective map $f:I\to\mathbb N$ shows that $I$ is countable \cite[Thm.~7.1]{Munkres00}, a contradiction. Hence $(X,\tau)$ cannot be separable.
\end{proof}

\subsection{Generating and Comparing Topologies}
Similar to the idea behind bases in finite-dimensional vector spaces one can simplify some of the fundamental concepts presented above by introducing bases of a topology. This will pay off even more when introducing the concept of a metric and the induced topology later on. The following is based on Willard \cite[Ch.~5]{Willard70}.
\begin{definition}\label{defi_basis_topo}
Let $X$ be a non-empty set. A collection $\mathcal B$ of subsets of $X$---the elements of which are called ``basis elements''---is a basis (for a topology)\index{topology!basis for a}\index{basis (topology)|see{topology, basis for a}} if the following statements hold.
\begin{itemize}
\item[(i)] $\bigcup_{B\in\mathcal B}B=X$
\item[(ii)] If $x\in B_1\cap B_2$ for some $B_1,B_2\in\mathcal B$ then there exists $B_3\in\mathcal B$ such that $x\in B_3\subseteq B_1\cap B_2$. 
\end{itemize}
In this case the topology $\tau_{\mathcal B}$ generated by $\mathcal B$ is defined to be the collection of arbitrary unions of elements of $\mathcal B$.
\end{definition}

Actually it suffices to specify a basis at each point of the set to obtain a basis of the whole topology.

\begin{definition}\label{defi_neighb_basis_topo}
Let $(X,\tau)$ be a topological space and let $x\in X$. A neighborhood basis\index{neighborhood basis} $\mathcal B_x$ at $x$ is a collection of neighborhoods of $x$ with the following property: For every neighborhood $U$ of $x$ there exists $B\in\mathcal B_x$ such that $x\in B\subseteq U$. 
\end{definition}
The connection between those concepts reads as follows, cf.~\cite[Thm.~5.4]{Willard70}.
\begin{lemma}\label{lemma_neighb_basis_topo}
Let $(X,\tau)$ be a topological space. If $\mathcal B$ is a collection of open sets in $X$ then $\mathcal B$ is a basis of $\tau$ if and only if for each $x\in X$ the collection $\mathcal B_x=\{B\in\mathcal B\,|\,x\in B\}$ is a neighborhood basis at $x$.
\end{lemma}
\noindent Thus the concepts of specifying a basis of the whole topology or specifying a neighborhood basis at each point are equivalent and both fully characterize the underlying (or generated) topology.\medskip

Even if one does not have access to property (ii) in Definition \ref{defi_basis_topo} one can still specify a generated topology.
\begin{definition}\label{defi_subbasis_topo}
Let $X$ be a non-empty set. A collection $\mathcal S$ of subsets of $X$ is called a subbasis (for a topology)\index{topology!subbasis for a}\index{subbasis|see{topology, subbasis for a}} if $\bigcup_{U\in\mathcal S} U=X$. The topology $\tau_{\mathcal S}$ generated by $\mathcal S$ is defined to be the collection of arbitrary unions of finite intersections of elements of $\mathcal S$.
\end{definition}
\noindent It is easy to see that $\tau_{\mathcal B},\tau_{\mathcal S}$ are indeed topologies on $X$ in the sense of Definition \ref{defi_topology} and that every topology on $X$ forms a basis and a subbasis of itself. \begin{remark}\label{rem_nbh_basis_top}
The following statements are immediate: 
\begin{itemize}
\item[(i)] If $\mathcal B$ is a basis on $X$ then $U\subseteq X$ is in $\tau_{\mathcal B}$ if and only if for every $x\in U$ there exists $B\in\mathcal B$ such that $x\in B\subseteq U$.
\item[(ii)] If $\mathcal S$ is a subbasis of $X$ then $\mathcal B_{\mathcal S}:=\{\bigcap_{j=1}^nS_j\,|\, n\in\mathbb N, S_1,\ldots,S_n\in \mathcal S\}$ is a basis of $\tops$. 
\end{itemize} 
\end{remark}

Another way of (indirectly) specifying a topology on a set $X$ is via a continuity requirement of a family of functions with domain equal to $X$, cf.~\cite[Def.~8.9]{Willard70}.

\begin{definition}[Initial Topology]\label{def_initial_top}\index{topology!initial}
Let $X$ be a set and $\mathcal F:=(f_j)_{j\in J}$ be a family of functions $f_j:X\to Y_j$ where $(Y_j,\tau_j)$ is a topological space for every $j\in J$. The topology $\sigma(X,\mathcal F)$ generated by all those functions is the weakest topology on $X$ such that $f_j:(X,\sigma(X,\mathcal F))\to Y_j$ is continuous for all $j\in J$. More precisely the collection $\{f_j^{-1}(U)\,|\,j\in J, U\in\tau_j\}$ forms a subbasis of $\sigma(X,\mathcal F)$.
\end{definition}

The initial topology will be useful later on when specifying topologies on spaces of bounded linear operators (beyond the norm topology) to see that those can be defined either via basic neighborhoods or via requiring continuity of a family of seminorms. 

\begin{lemma}\label{lemma_initial_top}
Let $X$ be a set and $\mathcal F:=(f_j)_{j\in J}$ be a family of functions as above. A net $(x_i)_{i\in I}$ converges to $x$ with respect to $\sigma(X,\mathcal F)$ if and only if $f_j(x_i)\to f_j(x)$ for all $j\in J$ (in the respective space $Y_j$). 
\end{lemma}
\begin{proof}
``$\Rightarrow$'': Simple consequence of Lemma \ref{lemma_topol_connect} because all $f$ are continuous.

``$\Leftarrow$'': Let $N$ be a neighborhood of $x$ so using the subbasis property of $\sigma(X,\mathcal F)$ we find $n\in\mathbb N$, $j_1,\ldots,j_n\in J$, and open sets $U_1\in\tau_1,\ldots,U_n\in\tau_n$ such that $x\in\bigcap_{k=1}^n f_{j_k}^{-1}(U_k)\subseteq N$. By assumption we know that $f_{j_k}(x_i)\to f_{j_k}(x)$ for all $j=1,\ldots,n$ so because $f_{j_k}(x)\in U_k$ and the latter is open we find $i_k\in I$ such that $f_{j_k}(x_i)\in U_k$ for all $i\succeq i_k$. By the directed set property---because we just obtained \textit{finitely many} indices---we also find $i_0\in I$ with $i_0\succeq i_k$ for all $k=1,\ldots,n$. Thus $x_i\in f_{j_k}^{-1}(U_k)$ for all $k=1,\ldots,n$ and $i\succeq i_0$ which shows $x_i\to x$ in $\sigma(X,\mathcal F)$ because
\begin{equation}
x_i\in\bigcap\nolimits_{k=1}^n f_{j_k}^{-1} (U_k)\subseteq N\qquad\text{ for all }i\succeq i_0\,.\tag*{\qedhere}
\end{equation}
\end{proof}

To conclude this section we want to see that the concept of a (topological) basis is a nice way to simplify handling continuity and convergence.
\begin{lemma}\label{lemma_top_properties_basis}
Let topological spaces $(X,\tau_X), (Y,\tau_Y)$ with respective basis $\mathcal B_X,\mathcal B_Y$ and with respective neighborhood basis $\tilde{\mathcal B}_x$ at every $x\in X$, $\tilde{\mathcal B}_y$ at every $y\in Y$ be given. The following statements hold.
\begin{itemize}
\item[(i)] A map $f:X\to Y$ is continuous\index{continuous} if and only if the pre-image of every basis element is open, that is, $f^{-1}(B)\in\tau_X$ for all $B\in\mathcal B_Y$.
\item[(ii)] A map $f:X\to Y$ is continuous in $x$ if and only if $f^{-1}(B)\in\tau_X$ for all $B\in\tilde{\mathcal B}_{f(x)}$. 
\item[(iii)] For a net $(x_i)_{i\in I}$ in $X$ and $x\in X$ the following statements are equivalent.
\begin{itemize}
\item[(a)] $x_i\to x$ in $\tau_X$.
\item[(b)] For all $B\in\mathcal B_X$ with $x\in B$ there exists $i_0\in I$ such that $x_i\in B$ for all $i\succeq i_0$.
\item[(c)] For all $B\in\tilde{\mathcal B_x}$ there exists $i_0\in I$ such that $x_i\in B$ for all $i\succeq i_0$.
\end{itemize}
\item[(iv)] Given a subset $A\subseteq X$ and $x\in X$ one has $x\in\overline{A}$ if and only if $B\cap A\neq\emptyset$ for all $B\in\tilde{\mathcal B}_x$.
\end{itemize}
\end{lemma}
\begin{proof}
(i): ``$\Rightarrow$'': Obvious. ``$\Leftarrow$'': Let $V\in\tau_Y$ so there exist $(B_j)_{j\in J}\subset\mathcal B_Y$ such that $V=\bigcup_{j\in J} B_j$ by definition of a basis. Then
$$
f^{-1}(V)=f^{-1}\Big(\bigcup\nolimits_{j\in J} B_j\Big)=\bigcup\nolimits_{j\in J} f^{-1}(B_j)
$$
and because each $f^{-1}(B_j)$ is open by assumption, $f^{-1}(V)$ is open as a union of open sets. Hence $f$ is continuous.\smallskip

(ii): ``$\Rightarrow$'': Obvious. ``$\Leftarrow$'': Let $V$ be a neighborhood of $f(x)$ so there exists $B\in\tilde{\mathcal B}_{f(x)}$ such that $f(x)\in B\subseteq V$. Now $f^{-1}(B)$ is open by assumption, hence it is a neighborhood of $x$ which additionally satisfies $f(f^{-1}(B))\subseteq B\subseteq V$. This shows continuity of $f$ in $x$.\smallskip

(iii): ``(a) $\Rightarrow$ (b)'', ``(a) $\Rightarrow$ (c)'': Obvious. ``(b) $\Leftarrow$ (a)'': For every neighborhood $N$ of $x$ we find $U\in\tau_X$ such that $x\in U\subseteq N$. Then, as stated in Remark \ref{rem_nbh_basis_top}, there exists $B\in\mathcal B_X$ such that $x\in B\subseteq U \subseteq N$. But on each basis element we already know about convergence by assumption. One argues analogously for ``(c) $\Leftarrow$ (a)''.\smallskip

(iv): \cite[Thm.~4.7 (c)]{Willard70}.
\end{proof}
\noindent This result is essential when recovering the well-known $\varepsilon$--criterion of sequence convergence as well as the $\varepsilon$--$\delta$--criterion for continuity of maps in the case of metric spaces.\medskip


Now the idea of comparing different topologies on a common set is a rather intuitive: For a non-empty set $X$ and topologies $\tau_1,\tau_2$ on $X$ we will say that $\tau_1$ is \textit{weaker}\index{topology!weaker} than $\tau_2$ or, equivalently, $\tau_2$ is \textit{stronger}\index{topology!stronger} than $\tau_1$, if $\tau_1\subseteq\tau_2$. 

Given our topological knowledge up until here such a comparison should be decidable via the basis of the topologies and, moreover, influence what nets converge in the compared topology.

\begin{proposition}\label{prop_compare_topo}
Let $X$ be a set and $\tau_1,\tau_2$ be topologies on $X$ with basis $\mathcal B_1,\mathcal B_2$ and neighborhood basis $\mathcal B_{1,x},\mathcal B_{2,x}$ at each $x\in X$, respectively. The following statements are equivalent.
\begin{itemize}
\item[(i)] $\tau_1\subseteq\tau_2$.
\item[(ii)] $\mathbbm{1}_X:(X,\tau_2)\to(X,\tau_1)$ is continuous.
\item[(iii)] For every net $(x_i)_{i\in I}$ in $X$ which converges to $x$ with respect to $\tau_2$ one has $x_i\overset{\tau_1}\to x$.
\item[(iv)] For every $x\in X$ and every $U_1\in \mathcal B_1$ containing $x$ there exists $U_2\in \mathcal B_2$ with $x\in U_2\subseteq U_1$.
\item[(v)] For every $x\in X$ and every $U_1\in \mathcal B_{1,x}$ there exists $U_2\in \mathcal B_{2,x}$ with $U_2\subseteq U_1$.
\end{itemize}
\end{proposition}
\begin{proof}
``(i) $\Leftrightarrow$ (ii)'': Definition of continuity. ``(ii) $\Leftrightarrow$ (iii)'': Lemma \ref{lemma_topol_connect} (i). ``(i) $\Leftrightarrow$ (iv)'': \cite[Thm.~4.8]{Willard70}. ``(v) $\Rightarrow$ (iv)'': Let $U_1\in\mathcal B_1$ with $x\in U_1$ be given. By definition of a neighborhood basis there exists $U\in\mathcal B_{1,x}$ with $x\in U\subseteq U_1$ so by (v) we find $V\in\mathcal B_{2,x}$ with $ V\subseteq U\subseteq U_1$. But because $V$ is a neighborhood of $x$ with respect to $\tau_2$ we (by definition of a neighborhood and) by Remark \ref{rem_nbh_basis_top} obtain $U_2\in\mathcal B_2$ with $x\in U_2\subseteq V\subseteq U\subseteq U_1$ as claimed. ``(iv) $\Rightarrow$ (v)'': Can be shown analogously.
\end{proof}

This limitation of convergence of nets has a converse: \textit{Unique} limit points in the weaker topology carry over to the stronger topology, if the net converges in the latter as we will see now.

\begin{corollary}
Let a non-empty set $X$, an element $x\in X$, topologies $\tau_1,\tau_2$ on $X$ with $\tau_1\subseteq\tau_2$, and a net $(x_i)_{i\in I}$ on $X$ be given. If $(x_i)_{i\in I}$ converges in $\tau_1$ with unique limit point $x$, then this net either does not converge in $\tau_2$ or it converges with unique limit point $x$.
\end{corollary}
\begin{proof}
For $(x_i)_{i\in I}$---when considering $\tau_2$ as topology on $X$---there are precisely three possible scenarios: \textit{Either} the net does not converge \textit{or} the net converges and $x$ is the unique limit point \textit{or} the net converges and there exists a limit point $y\neq x$\,. Thus to prove the statement we only have to ensure that the third scenario cannot occur. But because $\tau_1\subseteq\tau_2$ applying Prop.~\ref{prop_compare_topo} (iii) yields that $x_i\not\to y$ in $\tau_1$ implies $x_i\not\to y$ in $\tau_2$ and we are done.
\end{proof}
\noindent A more general formulation of this would be that if $A\subseteq X$ denotes the set of limit points of a net $(x_i)_{i\in I}$ (with respect to $\tau_1$), then its limit points (with respect to $\tau_2$) are included in $A$.\medskip

If two topologies are comparable then one can relate further topological concepts:

\begin{lemma}\label{lemma_separable_topology_comp}
Let $(X,\tau_1)$ be a topological space and $\tau_2$ another topology on $X$ such that $\tau_1\subseteq\tau_2$. The following statements hold.
\begin{itemize}
\item[(i)] If $A\subseteq X$ is dense in $(X,\tau_2)$, then it is dense in $(X,\tau_1)$.
\item[(ii)] If $(X,\tau_2)$ is separable, then $(X,\tau_1)$ is separable.
\item[(iii)] If $A\subseteq X$ is compact in $\tau_2$, then it is compact in $\tau_1$.
\end{itemize}
\end{lemma}
\begin{proof}
(i): Let $A\subseteq X$ be dense in $(X,\tau_2)$, i.e.~$\overline{A}^{\,\tau_2}=X$. It is straightforward to see that every $U\subseteq X$ which is closed in $\tau_1$ is closed in $\tau_2$---this implies
$$
X\supseteq \overline{A}^{\,\tau_1}= \bigcap_{\substack{U\text{ closed in }\tau_1\\A\subseteq U}} U\supseteq \bigcap_{\substack{U\text{ closed in }\tau_2\\A\subseteq U}} U=\overline{A}^{\,\tau_2}=X\,,
$$
because intersecting over more sets (when going from $\tau_1$ to $\tau_2$) means the set can only become smaller. Hence $\overline{A}^{\,\tau_1}=X$ so $A$ is dense in $(X,\tau_1)$. 

(ii): If $(X,\tau_2)$ is separable then there exists $\{x_n\}_{n\in\mathbb N}\subseteq X$ which is dense in $(X,\tau_2)$. Thus $\{x_n\}_{n\in\mathbb N}$ is dense in $(X,\tau_1)$ by (i) so the latter is separable as well.

(iii): Let $(U_i)_{i\in I}\subseteq\tau_1$ be an open cover of $A\subseteq X$. Now $\tau_1\subseteq\tau_2$ implies that $(U_i)_{i\in I}\subseteq\tau_2$ is an open cover of $A$ which---because $A$ is compact in $\tau_2$---has a finite subcover, i.e.~there exists $J\subseteq I$ finite such that $A\subseteq\bigcup_{j\in J}U_j$. But $(U_j)_{j\in J}$ is still in $\tau_1$ so we found a finite subcover of $A$ with respect to $\tau_1$, as well.
\end{proof}

\subsection{Product \& Subspace Topology}\label{section_product_subsp_top}

Luckily we can take the easy route through this section because we will only need a topology on a \textit{finite} product of topological spaces. This concept will enable us to talk about, e.g., continuity of the addition $+:X\times X\to X$ on a vector space $X$ endowed with some topology $\tau$. We refer to Munkres \cite[Ch.~15]{Munkres00}, in particular for all results stated without proof.

\begin{definition}
Let $(X,\tau_X),(Y,\tau_Y)$ be topological spaces. The product topology\index{topology!product} on the Cartesian product $X\times Y=\{(x,y)\,|\,x\in X,y\in Y\}$ is the topology generated by the basis $\mathcal B=\{U\times V\,|\,U\in\tau_X,V\in\tau_Y\}$.
\end{definition}
\noindent Checking that $\mathcal B$ is a basis in the sense of Definition \ref{defi_basis_topo} is straightforward. More interesting are the other (equivalent) ways to generate the product topology.

\begin{lemma}\label{lemma_basis_prod_top}
Let $(X,\tau_X),(Y,\tau_Y)$ be topological spaces and let $\tau$ denote the product topology on $X\times Y$.
\begin{itemize}
\item[(i)] If $\mathcal B_X,\mathcal B_Y$ are a basis of $\tau_X,\tau_Y$, respectively, then $\mathcal B:=\{B\times C\,|\,B\in\mathcal B_X,C\in\mathcal B_Y\}$ forms a basis of $\tau$.
\item[(ii)] Let $\pi_1:X\times Y\to X$, $\pi_2:X\times Y\to Y$ be the projection onto $X$, $Y$, respectively, i.e.~$\pi_1(x,y)=x$, $\pi_2(x,y)=y$ for all $x,y\in Y$. Then $\tau=\sigma(X,\{\pi_1,\pi_2\})$. 
\end{itemize}
\end{lemma}

Now given a topological space one expects that the topology of the whole space somehow transfers onto any non-empty subset---this is another important special case of the initial topology \cite[Ch.~6]{Willard70}.

\begin{definition}\label{defi_subspace_top}
Let $(X,\tau)$ be a topological space and $A\subseteq X$ non-empty be given. Then $\tau_A:=\sigma(A,\iota_A)$ is called the subspace topology\index{topology!subspace} on $A$ (induced by $X$) where $\iota_A:A\to (X,\tau)$, $x\mapsto x$ is the canonical embedding\index{canonical embedding}\label{symb_can_emb_2}.
Moreover such subset $A$ is said to be separable\index{separable} if $(A,\tau_A)$ is separable in the usual sense.
\end{definition}
One readily verifies that $\tau_A= \{A\cap U\,|\,U\in\tau\}$ because $\iota_A^{-1}(U)=A\cap U$ for all $U\subseteq X$. This has the following immediate consequence.
\begin{lemma}\label{lemma_isolated_sub}
Let a topological space $(X,\tau)$ and a non-empty subset $A\subseteq X$ be given. Then $x\in A$ is isolated in $A$ (with respect to the subspace topology)\,\footnote{A point $x\in X$ is called isolated\index{isolated point} if $\{x\}\in\tau$ \cite[p.~176]{Munkres00}.}
if and only if there exists $U\in\tau$ such that $U\cap A=\{x\}$. 
\end{lemma}
Moreover the simple characterization of the subspace topology allows for an interesting characterization of separability.

\begin{lemma}\label{lemma_separable_countable_union}
Let $(X,\tau)$ be a topological space. The following statements hold.
\begin{itemize}
\item[(i)] Let $A\subseteq X$. Then $A$ is separable (in the subspace topology) if and only if there exists a countable subset $A_0$ of $A$ such that $\overline{A_0}^{\,\tau}\supseteq A$.
\item[(ii)] Let a family $(X_n)_{n\in\mathbb N}$ of subsets $X_n\subseteq X$ be given. If $X_n$ is separable for every $n\in\mathbb N$ then $\bigcup_{n\in\mathbb N}X_n$ is separable. In other words the countable union of separable sets is separable (provided some common overarching topological space).
\end{itemize}
\end{lemma}
\begin{proof}
(i): By definition $(A,\tau_A)$---where $\tau_A$ denotes the subspace topology---is separable if and only if there exists a countable subset $A_0$ of $A$ such that $\overline{A_0}^{\,\tau_A}=A$. But $\overline{A_0}^{\,\tau_A}=\overline{A_0}^{\,\tau}\cap A$ as shown in \cite[Thm.~17.4]{Munkres00} and $\overline{A_0}^{\,\tau}\cap A=A$ is obviously equivalent to $\overline{A_0}^{\,\tau}\supseteq A$.

(ii): For every $n\in\mathbb N$ there exists $X_{n,0}\subseteq X_n$ countable such that $\overline{X_{n,0}}\supseteq X_n$ (by (i); here $\overline{(\cdot)}$ denotes the closure with respect to $\tau$). Then $\bigcup_{n\in\mathbb N} X_{n,0}$ is a countable union of countable sets, hence a countable subset of $\bigcup_{n\in\mathbb N}X_n$ \cite[Thm.~7.5]{Munkres00}. Moreover
$
\overline{\bigcup_{n\in\mathbb N} X_{n,0}}\supseteq \bigcup_{n\in\mathbb N} \overline{X_{n,0}}\supseteq \bigcup_{n\in\mathbb N} X_n
$ 
which together with (i) shows that $\bigcup_{n\in\mathbb N}X_n$ is separable.
\end{proof}

\subsection{Metric Spaces and Metrizability}\label{subsec:metric_normed_inner_product}

All concepts presented above are as general as possible, thus also capturing the topological nuances of more well-structured spaces such as the in the quantum theory well-known Hilbert spaces. We kick things off by recalling the definition of a metric.

\begin{definition}\label{def_metric}
Let $X$ be a non-empty set. A metric\index{metric} on $X$ is a function $d:X\times X\to [0,\infty)$ with the following properties:
\begin{itemize}
\item[(i)] $d(x,y)=d(y,x)$ for all $x,y\in X$ (symmetry).
\item[(ii)] $d(x,z)\leq d(x,y)+d(y,z)$ for all $x,y,z\in X$ (triangle equality).\index{triangle inequality}
\item[(iii)] $d(x,y)=0$ if and only if $x=y$ (definiteness).
\end{itemize}
In this case $(X,d)$ is called a metric space\index{space!metric}. Moreover $B_r(x):=\{y\in X\,|\,d(x,y)<r\}$\label{symb_metric_ball}
 is the open ball of radius $r$ around $x$ and $S_r(x):=\{y\in X\,|\,d(x,y)=r\}$\label{symb_metric_sphere} is the sphere with radius $r$ around $x$ (for all $r>0$, $x\in X$).
\end{definition}
The requirement of $d$ mapping into $[0,\infty)$ is actually redundant as the defining properties of a metric force $0= \frac12 d(x,x)\leq \frac12 d(x,y)+\frac12 d(y,x)=d(x,y) $ for all $x,y\in X$.

\begin{remark}[Reverse triangle inequality]\index{reverse triangle inequality}
Let $X$ be a non-empty set and $f:X\times X\to\mathbb R$ be a function which is symmetric and satisfies the triangle equality. Then the reverse triangle inequality holds:
\begin{equation}\label{eq:reverse_triangle}
|f(x,z)-f(y,z)|\leq f(x,y)\qquad\text{ for all }x,y,z\in X
\end{equation}
as is seen easily. In particular \eqref{eq:reverse_triangle} is true for every metric.
\end{remark}

Now every metric space becomes a topological space as follows.

\begin{lemma}\label{lemma_metric_topo}
Let $(X,d)$ be a metric space. Then $\mathcal B_d:=\{B_r(x)\,|\,x\in X,r>0\}$ is a basis and the topology $\tau_d$ generated by this basis is called the topology induced by the metric. Moreover a neighborhood basis of $\tau_d$ at any $x\in X$ is given by $\{B_r(x)\,|\,r>0\}$, and $\tau_d$ is Hausdorff.
\end{lemma}
\begin{proof}
For all $x\in X$ we have $x\in\mathcal B_r(x)$ for any $r>0$. Now given $x_1,x_2,x\in X$ and $r_1,r_2>0$ such that $x\in B_{r_1}(x_1)\cap B_{r_2}(x_2)$, if we can find $r>0$ such that $B_r(x)\subset B_{r_1}(x_1)\cap B_{r_2}(x_2)$ then $\mathcal B_d$ is a basis by Definition \ref{defi_basis_topo}---and this would also show the neighborhood basis claim (by choosing $r_1=r_2$, $x_1=x_2$). 

Indeed choose $r:=\min\{r_1-d(x,x_1),r_2-d(x,x_2)\}$ which is obviously positive. Then for all $z\in B_r(x)$ we see that
\begin{align*}
d(z,x_i)\leq d(z,x)+d(x,x_i)<r+d(x,x_i)\leq r_i-d(x,x_i)+d(x,x_i)=r_i
\end{align*}
for $i=1,2$. Hence $z\in B_{r_1}(x_1)\cap B_{r_2}(x_2)$ which as $z$ was chosen arbitrarily shows $B_r(x)\subset B_{r_1}(x_1)\cap B_{r_2}(x_2)$.

To see that $\tau_d$ is Hausdorff let $x_1,x_2\in X$ with $x_1\neq x_2$ be given. Then $\varepsilon:=d(x_1,x_2)>0$ lets us consider the basis elements $B_{\varepsilon/2}(x_1), B_{\varepsilon/2}(x_2)$. Suppose there exists $z\in B_{\varepsilon/2}(x_1)\cap B_{\varepsilon/2}(x_2)$ so
$$
\varepsilon= d(x_1,x_2)\leq d(x_1,z)+d(z,x_2)<\frac{\varepsilon}{2}+\frac{\varepsilon}{2}=\varepsilon\,,
$$
a contradiction. Thus $B_{\varepsilon/2}(x_1)\cap B_{\varepsilon/2}(x_2)=\emptyset$ which concludes the proof.
\end{proof}
\noindent Albeit straightforward we quickly sketched a proof here because it captures the idea of some more involved constructions in Section \ref{sec_top_B_X_Y}.
\begin{definition}
A topological space $(X,\tau)$ is said to be metrizable\index{space!metrizable} if there exists a metric $d$ on $X$ such that $\tau=\tau_d$.
\end{definition}
\begin{remark}\label{rem_metrizable_topo}
As soon as we are in a metric (or a metrizable) space some topological notions simplify considerably.
\begin{itemize}
\item[(i)] Lemma \ref{lemma_topol_connect} (i), (ii) \& (iii) remain true if one replaces ``net'' by ``sequence'' \cite[Lemma 21.2, Thm.~21.3 \& 28.2]{Munkres00}.
This is connected to---although not \textit{fully} explained by---the fact that metric spaces are first countable meaning that at every point $x\in X$ there is a countable neighborhood basis at $x$. Following Lemma \ref{lemma_metric_topo} this can be done by choosing $\{B_{1/n}(x)\,|\,n\in\mathbb N\}$. 
\item[(ii)] Unsurprisingly by Lemma \ref{lemma_top_properties_basis}, convergence of a sequence $(x_n)_{n\in\mathbb N}$ in $X$ (to some $x\in X$) reduces to the usual $\varepsilon$--criterion: For all $\varepsilon>0$ there exists $N\in\mathbb N$ such that $d(x,x_n)<\varepsilon$ for all $n\geq N$.
\item[(iii)] Also continuity of a map $f$ between metric spaces $X,Y$ at some point $x\in X$ is equivalent to the well-known $\varepsilon$--$\delta$--criterion \cite[Thm.~21.1]{Munkres00}\footnote{
Recall that continuity in metric spaces reduces to: for all $\varepsilon>0$ there exists $\delta>0$ such that $f(B_\delta(x))\subseteq B_\varepsilon(f(x))$, i.e.~for every $\varepsilon$-ball around $f(x)$ one finds a $\delta$-ball around $x$ which is fully mapped into the former. 
}. This is a direct consequence of Lemma \ref{lemma_top_properties_basis} (ii) together with Remark \ref{rem_nbh_basis_top}.
\item[(iv)] Finally given $A\subseteq X$ and $x\in X$ one has $x\in\overline{A}$ if and only if $B_\varepsilon(x)\cap A\neq\emptyset$ for all $\varepsilon>0$. This is a direct consequence of Lemma \ref{lemma_top_properties_basis} \& \ref{lemma_metric_topo}.
\end{itemize}
\end{remark}

It turns out that there is a nice connection between compactness and separability for metric spaces:

\begin{lemma}\label{lemma_comp_met_sep}
Every compact metric space is separable.
\end{lemma}
\begin{proof}
Let $(X,d)$ be a metric space which is compact. For every $n\in\mathbb N$, $\{B_{1/n}(x)\,|\,x\in X\}$ is an open cover of $X$ so by compactness there exists a finite subcover, that is, one finds $i_n\in\mathbb N$ as well as $x_1^n,\ldots,x_{i_n}^n\in X$ such that $X\subseteq\bigcup_{j=1}^{i_n} B_{1/n}(x_j^n)$. Now the set
$\bigcup_{n\in\mathbb N}\{x_1^n,\ldots,x_{i_n}^n\}$
is obviously countable and, by construction, dense in $X$ which shows separability of the latter.
\end{proof}

Another intuitive requirement of metric spaces is the idea that if the elements of a sequence become arbitrarily close eventually, then there has to exist some limit the sequence converges to. This is captured by the following.

\begin{definition}
Let $(X,d)$ be a metric space. 
\begin{itemize}
\item[(i)] A sequence $(x_n)_{n\in\mathbb N}$ in $X$ is called a Cauchy sequence\index{sequence!Cauchy} if for every $\varepsilon>0$ there exists $N\in\mathbb N$ such that $d(x_m,x_n)<\varepsilon$ for all $m,n\geq N$. 
\item[(ii)] The metric space $X$ is said to be complete\index{space!complete} if every Cauchy sequence in $X$ is convergent.
\end{itemize}
\end{definition}
\noindent Of course every convergent sequence in a metric space is a Cauchy sequence but the converse need not be true: Take $X=(0,2)\subset\mathbb R$ with the standard metric $d(x,y)=|x-y|$, then $(\frac{1}{n})_{n\in\mathbb N}$ is a Cauchy sequence in $X$ but its limit point would be $0$ which is not in $X$, hence $(X,d)$ is not complete.

\begin{lemma}\label{lemma_image_closed_isometry}
Let $X,Y$ be metric spaces such that $X$ is complete. Moreover let an isometry $f:X\to Y$ (i.e.~$d(f(x),f(y))=d(x,y)$ for all $x,y\in X$) be given. If $Z\subset X$ is closed then $f(Z)$ is closed.
\end{lemma}
\begin{proof}
Consider a sequence $(z_n)_{n\in\mathbb N}\subset Z$ as well as $y\in Y$ such that $f(z_n)\to y$ for some $y\in Y$. If we can show that $y\in f(Z)$ then by Remark \ref{rem_metrizable_topo} $\overline{f(Z)}=f(Z)$, i.e.~$f(Z)$ is closed as claimed.

Because $(f(z_n))_{n\in\mathbb N}$ converges it is a Cauchy sequence, hence $(z_n)_{n\in\mathbb N}$ is a Cauchy sequence in $X$ due to $f$ being an isometry: $d(z_m,z_n)=d(f(z_m),f(z_n))$. But $X$ is assumed to be complete so one finds $x\in X$ with $\lim_{n\to\infty}d(z_n,x)=0$. Moreover $Z$ being closed implies $x\in Z$. Then $d(f(z_n),f(x))=d(z_n,x)\to 0$ so $f(z_n)\to f(x)$. But this shows $y= f(x)\in f(Z)$ because every metric space is Hausdorff (Lemma \ref{lemma_metric_topo}) and limits in Hausdorff spaces are unique (Lemma \ref{lemma_topol_connect} (iv)). 
\end{proof}

\subsection{Topological Vector Spaces}

Up until now we dealt with topological spaces, that is, arbitrary non-empty sets endowed with some topology. Of course the spaces one deals with in everyday mathematics---as well as quantum physics---are much more structured. Thus we pass over to vector spaces over some complete field\footnote{Here completeness of a field $\mathbb F$ refers to completeness with respect to some given metric $d$ on $\mathbb F$.} which for now we want to be either $\mathbb F=\mathbb R$ or $\mathbb F=\mathbb C$\label{symb_base_field} with the standard topology (induced by the standard metric $d(x,y)=|x-y|$). 

For the topic of general topological vector spaces we orient ourselves towards Meise \& Vogt \cite[Ch.~22]{MeiseVogt97en}.

\begin{definition}
A topological vector space\index{space!topological vector} $X$ is a vector space over $\mathbb F$ equipped with a topology for which addition $+:X\times X\to X$ and scalar multiplication $\cdot:\mathbb F\times X\to X$ are continuous (with respect to the product topology on $X\times X$, $\mathbb F\times X$). 
\end{definition}
\noindent Thus for a topological vector space the translation $T_y:X\to X$, $x\mapsto x+y$ is continuous so the neighborhoods of each $x\in X$ are of the form $x+V$ with $V$ being a neighborhood of $0\in X$. In a way requiring continuity of $+$ and $\cdot$ allows us to shift topological properties to the origin ``in spirit of the linear nature'' of $X$.\medskip

Of course every vector space equipped with a metric is a topological space by Lemma \ref{lemma_metric_topo}. However not every such space is a topological \textit{vector} space as the following classic example shows.

\begin{example}\label{example_metric_not_top_vec}
Equip $\mathbb R^2$ with the following metric (sometimes called the \textit{Paris metric})
$$
d_{\mathrm p}(x,y):=\begin{cases} \|x-y\|&\text{if there exist }( \lambda,\mu )\in\mathbb R^2\setminus\{(0,0)\}\text{ with }\lambda x=\mu y\\\|x\|+\|y\|&\text{else} \end{cases}
$$
where $\|\cdot\|$ denotes the usual euclidean norm on the vector space $\mathbb R^2$. Verifying that $d_{\mathrm p}$ is a metric on $\mathbb R^2$ is straightforward. To see that $(\mathbb R^2,d_{\mathrm p})$ is not a topological vector space we will construct a sequence $(x_n,y_n)_{n\in\mathbb N}$ in $\mathbb R^2\times\mathbb R^2$ which converges to $(x,y)$ but $x_n+y_n\not\to x+y$ as $n\to\infty$---this would imply that $+:\mathbb R^2\times\mathbb R^2\to\mathbb R^2$ is not continuous by Lemma \ref{lemma_topol_connect} (i) as desired.

Indeed for all $n\in\mathbb N$ one computes
$$
d_{\mathrm p}\Big(\begin{pmatrix} 0\\1-\frac{1}{n} \end{pmatrix},\begin{pmatrix} 0\\1 \end{pmatrix}\Big)=\Big\|\begin{pmatrix} 0\\1-\frac{1}{n} \end{pmatrix}-\begin{pmatrix} 0\\1 \end{pmatrix}\Big\|=\frac{1}{n}\overset{n\to\infty}\to 0
$$
so by Lemma \ref{lemma_basis_prod_top} (ii) (\& Def.~\ref{def_initial_top}) we find that $\big({\scriptsize\begin{pmatrix} 0\\1-\frac{1}{n} \end{pmatrix},\begin{pmatrix} 1\\0 \end{pmatrix}}\big)_{n\in\mathbb N}$ converges to $\big({\scriptsize\begin{pmatrix} 0\\1\end{pmatrix},\begin{pmatrix} 1\\0 \end{pmatrix}}\big)$ in the product topology on $\mathbb R^2\times\mathbb R^2$ induced by $d_{\mathrm p}$. However 
\begin{align*}
d_{\mathrm p}\Big(\begin{pmatrix} 0\\1-\frac{1}{n} \end{pmatrix}+\begin{pmatrix} 1\\0 \end{pmatrix},\begin{pmatrix} 0\\1 \end{pmatrix}+\begin{pmatrix} 1\\0 \end{pmatrix}\Big)&=d_{\mathrm p}\Big(\begin{pmatrix} 1\\1-\frac{1}{n} \end{pmatrix},\begin{pmatrix} 1\\1 \end{pmatrix}\Big)\\
&=\Big\|\begin{pmatrix} 1\\1-\frac{1}{n} \end{pmatrix}\Big\|+\Big\|\begin{pmatrix} 1\\1 \end{pmatrix}\Big\|\geq\sqrt{2}
\end{align*}
for all $n\in\mathbb N$ so the sum of those two sequences does not converge to the sum of their two limits. Thus $(\mathbb R^2,d_{\mathrm p})$ is a vector space with topology induced by a metric but is not a topological vector space.
\end{example}

Being in a topological vector space already simplifies the notion of separability as one has access to the linear span (in a continuous manner).

\begin{lemma}\label{lemma_separable_linear_span}
Let $(X,\tau)$ be a topological $\mathbb F$-vector space. The following are equivalent.
\begin{itemize}
\item[(i)] $(X,\tau)$ is separable.\index{separable}
\item[(ii)] There exists a countable subset $\{x_n\}_{n\in\mathbb N}$ on $X$ such that its linear span\footnote{Recall that the linear span of a subset $A$ of some $\mathbb F$-vector space $V$ is $\{\sum_{j=1}^m \lambda_jv_j\,|\,m\in\mathbb N,v_j\in A,\lambda_j\in\mathbb F\}$.\label{footnote_linear_span}} is dense.
\end{itemize}
\end{lemma}
\begin{proof}
``(i) $\Rightarrow$ (ii)'': Let $\{x_n\}_{n\in\mathbb N}$ be dense in $(X,\tau)$. Then $X=\overline{\{x_n\}_{n\in\mathbb N}}\subseteq\overline{\operatorname{span}(\{x_n\}_{n\in\mathbb N})}\subseteq X$ so (ii) holds. ``(ii) $\Rightarrow$ (i)'': Let $\{x_n\}_{n\in\mathbb N}\subseteq X$ be given such that its linear span is dense in $(X,\tau)$. Consider any $x\in\operatorname{span}(\{x_n\}_{n\in\mathbb N})$ so there exist $m\in\mathbb N$, $\lambda_1,\ldots,\lambda_m\in\mathbb F$ and $n_1,\ldots,n_m\in\mathbb N$ such that $x=\sum_{j=1}^m\lambda_jx_{n_j}$. Because $\mathbb Q$ is dense in $\mathbb F$ (as the latter is assumed to be $\mathbb R$ or $\mathbb C$) for every $j\in\{1,\ldots,m\}$ one finds $(\lambda_j^{(k)})_{k\in\mathbb N}\subseteq\mathbb Q$ such that $\lim_{k\to\infty}|\lambda_j^{(k)}-\lambda_j|= 0$. Then $\lambda_j^{(k)}x_{n_j}\overset{k\to\infty}\to\lambda_jx_{n_j}$ in $(X,\tau)$ because scalar multiplication on $X$ is continuous and $\sum_{j=1}^m\lambda_j^{(k)}x_{n_j}\overset{k\to\infty}\to\sum_{j=1}^m\lambda_jx_{n_j}$ in $(X,\tau)$ because addition on $X$ is continuous (and this sum is finite). Hence the countable\footnote{One can write $\operatorname{span}_\mathbb Q(\{x_n\}_{n\in\mathbb N})=\bigcup_{m\in\mathbb N}\{\sum_{j=1}^m \lambda_jx_{n_j}\,|\,\lambda_j\in\mathbb Q,n_1,\ldots,n_m\in\mathbb N\}$ so the latter is the image of $\mathbb Q^m\times\mathbb N^m$ under the function $(\lambda_1,\ldots,\lambda_m,n_1,\ldots,n_m)\mapsto\sum_{j=1}^m\lambda_jx_{n_j}$. Because $\mathbb Q^m\times\mathbb N^m$ is countable as a finite Cartesian product of countable sets \cite[Thm.~7.6]{Munkres00} so is its image under the above function. Finally the $\mathbb Q$-span of $\{x_n\}_{n\in\mathbb N}$ is a countable union of countable sets hence countable itself \cite[Thm.~7.5]{Munkres00}.
}
set $\operatorname{span}_\mathbb Q(\{x_n\}_{n\in\mathbb N})\subseteq X$ is dense in $\operatorname{span}(\{x_n\}_{n\in\mathbb N})$ and thus it is dense in $(X,\tau)$. 
\end{proof}

\subsection{Normed \& Locally Convex Spaces}

Example \ref{example_metric_not_top_vec} works as it does because the metric $d_{\mathrm p}$ is not translation invariant, that is, one does \textit{not} have $d_{\mathrm p}(x+a,y+a)=d_{\mathrm p}(x,y)$ for all $x,y,a$. Yet there is an important class of metrics which are translation invariant, namely those which are induced by a norm or at least by a family of seminorms.

\begin{definition}
Let $X$ be a vector space over $\mathbb F$. A map $\|\cdot\|:X\to[0,\infty)$ is called a seminorm\index{seminorm} on $X$ if
\begin{itemize}
\item[(i)] $\|\lambda x\|=|\lambda|\|x\|$ for all $\lambda\in\mathbb F$, $x\in X$.
\item[(ii)] $\|x+y\|\leq\|x\|+\|y\|$ for all $x,y\in X$.
\end{itemize}
A seminorm $\|\cdot\|$ is a norm\index{norm} on $X$ if, additionally, $\|x\|=0$ holds only if $x=0$. In this case $(X,\|\cdot\|)$ is called a normed space.\index{space!normed}
\end{definition}
\noindent A normed space $(X,\|\cdot\|)$ is a metric space under the metric $d(x,y):=\|x-y\|$ induced by the norm. This turns $X$ into a topological space by Lemma \ref{lemma_metric_topo}. It even turns $X$ into a \textit{topological} vector space with uniformly continuous norm, cf.~\cite[Prop.~5.1]{MeiseVogt97en}, thus enabling the following definition:
\begin{definition}
A normed space $(X,\|\cdot\|)$ which is complete with respect to the metric induced by the norm is called a Banach space.\index{space!Banach}
\end{definition}
An important example of normed spaces are the $p$-summable sequence spaces (see \cite[Ch.~2]{Bollobas99}) which will be indispensable when introducing certain classes of operators later on. 
\begin{example}\label{ex_ell_p_space}
For $p\in [1,\infty)$ the space $\ell^p(\mathbb N)$\index{space!lp@$\ell^p(\mathbb N)$}
consists of all sequences $(x_1,x_2,x_3,\ldots)$ with values in $\mathbb F$ (more precisely all maps $f:\mathbb N\to\mathbb F$ when identifying $f(n)=x_n$) such that
$
\sum\nolimits_{n=1}^\infty |x_n|^p<\infty
$. This is obviously an $\mathbb F$-vector space and turns into a normed space under
$$
\|x\|_p:=\Big(\sum\nolimits_{n=1}^\infty|x_n|^p\Big)^{1/p}\qquad\text{ for all }x\in\ell^p(\mathbb N)\,.
$$
Similarly the space $\ell^\infty(\mathbb N)$\index{space!linf@$\ell^\infty(\mathbb N)$} is defined to consist of all bounded sequences (more precisely all $f:\mathbb N\to\mathbb F$ such that $\sup_{n\in\mathbb N}|f(n)|<\infty$) and becomes a normed space under
$$
\|x\|_\infty:=\sup_{n\in\mathbb N}|x_n|\qquad\text{ for all }x\in\ell^\infty(\mathbb N)\,.
$$
Further important sequence spaces are\index{space!c0@$c_0(\mathbb N)$}\index{space!c00@$c_{00}(\mathbb N)$}
\begin{align*}
c_0(\mathbb N)&:=\{(x_n)_{n\in\mathbb N}\,|\,\lim_{n\to\infty}x_n=0\}\\
\text{as well as }\qquad c_{00}(\mathbb N)&:=\{(x_n)_{n\in\mathbb N}\,|\,\exists_{N\in\mathbb N}\,\forall_{n\geq N}\ x_n=0\}\,.
\end{align*}
Now $c_{00}(\mathbb N)$ is dense in $(\ell^p(\mathbb N),\|\cdot\|_p)$ for all $p\in[1,\infty)$ as well as $(c_0(\mathbb N),\|\cdot\|_\infty)$. Hence these are separable Banach spaces whereas $(\ell^\infty(\mathbb N),\|\cdot\|_\infty)$ is a non-separable Banach space\footnote{
The standard arguments go as follows: Consider the sequences $\{e^{(n)}\}_{n\in\mathbb N}$ (where $e^{(n)}$ has a $1$ in the $n$-th place and is $0$ otherwise). Because this set is countable and $\operatorname{span}\{e^{(n)}\}_{n\in\mathbb N}=c_{00}(\mathbb N)$ is dense in $(\ell^p(\mathbb N),\|\cdot\|_p)$ as well as $(c_0(\mathbb N),\|\cdot\|_\infty)$, these are separable by Lemma \ref{lemma_separable_linear_span}.

For non-separability of $\ell^\infty(\mathbb N)$ on the other hand one can explicitly construct an uncountable set of disjoint open balls (so Lemma \ref{lemma_non_sep} implies non-separability): Given $M\subseteq\mathbb N$ define $x^M$ via $x^M(n)=1$ if $n\in M$ and $0$ otherwise. Then $M,M'\in\mathcal P(\mathbb N)$, $M\neq M'$ implies $\|x^M-x^{M'}\|_\infty=1$ so the family $\{B_{1/2}(x^M)\}_{M\in\mathcal P(\mathbb N)}$ does the job (because $\mathcal P(\mathbb N)$ is uncountable).\label{footnote_sequence_space_separable}}.
\end{example}

As soon as we get to Hilbert spaces---which allow for expanding every element in terms of a (possibly uncountable) basis---we need a concept of convergence for summation over arbitrary unordered sets. For more on this topic, which we only briefly touch upon here, we refer to Ringrose \cite[Ch.~1.2]{Ringrose71}.
Given an arbitrary non-empty set $I$ define $\mathcal P_\text{fin}(I):=\{J\subseteq I\,|\,J\text{ finite}\}$ which together with the usual set inclusion $\subseteq$ is a directed set. Thus the following definition is reasonable.

\begin{definition}\label{defi_unordered_summation}
Let a normed space $X$, a non-empty set $I$, and $(x_i)_{i\in I}\subseteq X$ be given. Then $(x_i)_{i\in I}$ is said to be summable\index{summable} to $x\in X$ if the net $(\sum_{i\in J}x_i)_{J\in \mathcal P_\text{fin}(I)}$ converges to $x$ in norm. In this case one writes $\sum_{i\in I}x_i=x$.
\end{definition}

Unsurprisingly if $I$ is finite or countably infinite then this concept co{\"i}ncides with usual summation. Given our knowledge of nets it is easy to characterize summability as follows.

\begin{lemma}\label{lemma_unordered_summation}
Let $X$ be a normed space, $I$ be a non-empty set, and $(x_i)_{i\in I}\subseteq X$ be given. The following statements hold.
\begin{itemize}
\item[(i)] $(x_i)_{i\in I}$ is summable to $x\in X$ if and only if for all $\varepsilon>0$ there exists $J_0\subseteq I$ finite such that $\|x-\sum_{i\in J}x_i\|<\varepsilon$ for all $J$ finite with $J_0\subseteq J\subseteq I$.
\end{itemize}
Let $(x_i)_{i\in I}$ be summable to $x\in X$.
\begin{itemize}
\item[(ii)] Given any $\varepsilon>0$ one finds $J_\varepsilon\subseteq I$ finite such that $\|\sum_{i\in J} x_i\|<\varepsilon$ for all $J\subseteq I\setminus J_\varepsilon$ finite.
\item[(iii)] Given any $\varepsilon>0$ the set $\{i\in I\,|\,\|x_i\|\geq\varepsilon\}$ is finite and $\{i\in I\,|\,x_i\neq 0\}$ is at most countable.
\item[(iv)] If $I$ is infinite and $f:\mathbb N\to I$ an injective mapping then $\lim_{n\to\infty}\|x_{f(n)}\|=0$.
\end{itemize}
\end{lemma}
\begin{proof}
(i): Direct consequence of Lemma \ref{lemma_top_properties_basis} (iii) because a neighborhood basis at $x$ is given by $\{B_\varepsilon(x)\,|\,\varepsilon>0\}$.
(ii): This is shown in \cite[Lemma 1.2.2]{Ringrose71}. Note that in their proof completeness of $X$ is not used so this holds for all normed spaces. 
(iii): Following \cite[Coro.~1.2.3]{Ringrose71} given $\varepsilon>0$ one finds $J_\varepsilon\subseteq I$ finite such that $\|x_i\|<\varepsilon$ for all $i\in I\setminus J_\varepsilon$. Thus $|\{i\in I\,|\,\|x_i\|\geq\varepsilon\}|\leq |J_\varepsilon|<\infty$. The second statement then is obvious because $\{i\in I\,|\,x_i\neq 0\}=\bigcup_{n\in\mathbb N}\{i\in I\,|\,\|x_i\|\geq\frac{1}{n}\}$, i.e.~the former is a countable union of finite sets, hence countable \cite[Thm.~7.5]{Munkres00}.
(iv): Let $\varepsilon>0$ be given so by (iii) the set $\{n\in\mathbb N\,|\,\|x_{f(n)}\|\geq\varepsilon\}\subseteq\{i\in I\,|\,\|x_i\|\geq\varepsilon\}$ is finite. Defining $N:=\max \{n\in\mathbb N\,|\,\|x_{f(n)}\|\geq\varepsilon\}$ we get $\|x_{f(n)}\|<\varepsilon$ for all $n\geq N$ (here we use injectivity of $f$). But this is precisely the definition of $\lim_{n\to\infty}\|x_{f(n)}\|=0$.
\end{proof}

Having introduced normed spaces, even if one does not have access to a norm but ``only'' seminorms on some vector space one can still induce some interesting topological structure. This is the theory of locally convex spaces for which we orient ourselves towards Conway \cite[Ch.~IV.1]{Conway90}.
More precisely given a vector space $X$ over some field $\mathbb F$ and an arbitrary family of seminorms $(p_i)_{i\in I}$ on $X$, these induce a topology $\tau_p$ via the subbasis
\begin{equation}\label{eq:seminorms_subbasis}
\{N(x_0,i,\varepsilon)\,|\,x_0\in X, i\in I,\varepsilon>0\}\quad\text{ where }\quad N(x_0,i,\varepsilon):=\{x\in X\,|\,p_i(x-x_0)<\varepsilon\}\,.
\end{equation}
Given this definition one intuitively expects a connection to the initial topology on $X$ somehow generated by this family of seminorms. 
\begin{lemma}\label{lemma_locally_convex}
Let $X$ be a vector space and $(p_i)_{i\in I}$ be a family of seminorms on $X$. Then $\tau_p=\sigma(X,\{x\mapsto p_i(x- x_0)\}_{i\in I,x_0\in X})$ and $(X,\tau_p)$ is a topological vector space.
\end{lemma}
\begin{proof}
``$\subseteq$'': Because all the maps $x\mapsto f_{i,x_0}(x):=p_i(x- x_0)$ are continuous in $\tau:=\sigma(X,\{x\mapsto p_i(x- x_0)\}_{i\in I,x_0\in X})$ we know that $f_{i,x_0}^{-1}(B_\varepsilon(0))\in\tau_1$ for all $x_0\in X, i\in I,\varepsilon>0$. But
$$
f_{i,x_0}^{-1}(B_\varepsilon(0))=\{x\in X\,|\ |f_{i,x_0}(x)|<\varepsilon\}=\{x\in X\,|\,p_i(x-x_0)<\varepsilon\}\,.
$$
Thus the subbasis of $\tau_p$ \eqref{eq:seminorms_subbasis} is contained in $\tau$ so because the latter is a topology itself this shows $\tau_p\subseteq\tau$.

``$\supseteq$'': 
Let arbitrary $z\in f_{i,x_0}^{-1}(B_\varepsilon(\alpha))$ be given, i.e.~$| p_i(z-x_0) -\alpha|<\varepsilon$. If we can find $\tilde\varepsilon>0$ such that
\begin{equation}\label{eq:seminorm_subset_basis}
z\in \underbrace{\{x\in X\,|\,p_i(x-z)<\tilde\varepsilon\}}_{=N(z,i,\tilde\varepsilon)\in\tau_p} \subseteq f_{i,x_0}^{-1}(B_\varepsilon(\alpha))
\end{equation}
then $f_{i,x_0}^{-1}(B_\varepsilon(\alpha))\in\tau_p$ which would show that all $f_{i,x_0}:(X,\tau_p)\to\mathbb R$ are continuous as desired. Indeed define $\tilde\varepsilon:=\varepsilon-| p_i(z-x_0) -\alpha|>0$. If $x\in X$ satisfies $p_i(x-z)<\tilde\varepsilon$ then\footnote{Given a seminorm $p$ on a vector space $X$ and $x,y,z\in X$, $\alpha\in\mathbb R$ using the reverse triangle inequality (which is allowed because $p$ is symmetric and satisfies the triangle inequality) one gets
\begin{align*}
|p(x-y)-\alpha|-|p(z-y)-\alpha|&\leq\big|\ |p(x-y)-\alpha|-|p(z-y)-\alpha|\ \big|\\
&\leq |p(x-y)-p(z-y)|\leq p(x-y-z+y)=p(x-z)
\end{align*} 
which shows $|p(x-y)-\alpha|\leq p(x-z)+|p(z-y)-\alpha|$.}
$$
|p_i(x-x_0)-\alpha| \leq p_i(x-z)+|p_i(z-x_0)-\alpha|<\tilde\varepsilon+|p_i(z-x_0)-\alpha|=\varepsilon
$$
so \eqref{eq:seminorm_subset_basis} holds.\smallskip

For the second statement recall that given a net $(x_j)_{j\in J}$ and $x$ in $X$ one has $x_i\to x$ in $\tau=\tau_p$ if and only if $p_i(x_j-x_0)\to p_i(x-x_0)$ for all $i\in I$, $x_0\in X$ by Lemma \ref{lemma_initial_top}. In particular one has $p_i(x_j-x)\to 0$ for all $i\in I$ (choose $x_0=x$).

First off given some net $(x_j,y_j)_{j\in J}$ in $X\times X$ which converges to $(x,y)\in X\times X$ in the product topology one finds $x_j\to x$, $y_j\to y$ in $\tau_p$ by Lemma \ref{lemma_basis_prod_top} \& \ref{lemma_initial_top}. Hence by the (reverse) triangle inequality
\begin{align*}
|p_i(x_j+y_j-x_0)-p_i(x+y-x_0)|\leq p_i(x_j+y_j-x-y)\leq p_i(x_j-x)+p_i(y_j-y)\to 0\,.
\end{align*}
This shows $x_j+y_j\to x+y$ in $\tau_p$, meaning addition on $X$ is continuous. For continuity of scalar multiplication one proceeds likewise by means of the identity
\begin{equation}
\lambda_jx_j-\lambda x=(\lambda_j-\lambda)x+\lambda(x_j-x)+(\lambda_j-\lambda)(x_j-x)\,. \tag*{\qedhere}
\end{equation}
\end{proof}
Thus the following definition is reasonable.
\begin{definition}\label{defi_locally_convex}
A locally convex space\index{space!locally convex} is a topological vector space $(X,\tau)$ whose topology is induced by a family of seminorms $(p_i)_{i\in I}$ which is fundamental (i.e.~if $p_i(x)=0$ for all $i\in I$ then $x=0$) so $\tau=\sigma(X,\{x\mapsto p_i(x- x_0)\}_{i\in I,x_0\in X})$. 
\end{definition}
\begin{remark}\label{remark_normed_locally_convex}
Unsurprisingly the norm topology $\tau$ of a normed space $(X,\|\cdot\|)$ co{\"i}ncides with the topology induced by the fundamental family of seminorms $\{\|\cdot\|\}$ so every normed space is a locally convex Hausdorff space. This follows directly from $\tau=\sigma(X,\{x\mapsto\|x-x_0\|\}_{x_0\in X})$ (``$\supseteq$'' is a direct consequence of the reverse triangle inequality as above and ``$\subseteq$'' holds because continuity of all $f_{x_0}(x):=\|x-x_0\|$ implies $B_r(x_0)=f_{x_0}^{-1}(B_r(0))\in \sigma(X,\{x\mapsto\|x-x_0\|\}_{x_0\in X})$).
\end{remark}

The term ``locally convex'' is motivated by the convexity of the defining neighborhoods \eqref{eq:seminorms_subbasis}. Imposing that the family of seminorms is fundamental is of course not necessary but ensures that the resulting topology is Hausdorff. With this we---in absence of a norm---get a reasonable generalization of Banach spaces.

\begin{definition}[\cite{Rudin91}, Def.~1.8]
A locally convex space $(X,\tau)$ is called a Fr{\'e}chet space\index{space!Fr{\'e}chet} if its topology is metrizable and if $X$ is complete with respect to this metric.
\end{definition}

Part of this criterion can be decided by the family of seminorms itself, refer to \cite[Prop.~2.1]{Conway90}.

\begin{lemma}
Let $(X,\tau)$ be a locally convex space. Then there exists a metric $d$ on $X$ such that $\tau=\tau_d$ if and only if $\tau$ is determined by a countable family of seminorms. Indeed if $\{p_j\}_{j\in\mathbb N}$ is a fundamental family of seminorms which induce $\tau$ then the metric 
$$
d(x,y):=\sum\nolimits_{n=1}^\infty \frac{1}{2^n}\frac{p_n(x-y)}{1+p_n(x-y)}\qquad\text{ for all }x,y\in X
$$
satisfies $\tau=\tau_d$.
\end{lemma}
\noindent This common trick of constructing a metric from a countable number of seminorms will also come in handy later.

\subsection{Inner Product Spaces}\label{sec_inner_product}

Vector spaces with an inner product and thus an associated notion of bra- $\langle\Psi|$ and ket-vectors $|\Psi\rangle$ are central objects in quantum physics. This is also justified mathematically as ``Hilbert spaces are perfect generalizations of euclidean spaces'' whereas infinite-dimensional Banach spaces may lack a ``notion of perpendicular vectors and [...] good notion of a basis'' \cite[p.~79]{Pedersen89}. For this section we follow Meise \& Vogt \cite[Ch.~11 \& 12]{MeiseVogt97en}

\begin{definition}
Let $X$ be a vector space over $\mathbb F$. An inner product\index{inner product} on $X$ is a mapping $\langle\cdot,\cdot\rangle:X\times X\to\mathbb F$ such that
\begin{itemize}
\item[(i)] $\langle x,\lambda y+\mu z\rangle=\lambda \langle x,y\rangle+\mu\langle x,z\rangle$ for all $\lambda,\mu\in\mathbb F$, $x,y,z\in X$.
\item[(ii)] $\langle x,y\rangle=\overline{\langle y,x\rangle}$ for all $x,y\in\mathbb F$. Here $\overline{(\cdot)}$ stands for the complex conjugate (which can of course be waived if $\mathbb F=\mathbb R$). 
\item[(iii)] $\langle x,x\rangle\geq 0$ for all $x\in X$ with $\langle x,x\rangle=0$ if and only if $x=0$.
\end{itemize}
In this case $(X,\langle \cdot,\cdot\rangle)$ is called an inner product space\index{space!inner product} (or sometimes a pre-Hilbert space).\index{space!pre-Hilbert|see{space, inner product}}
\end{definition}
\noindent Now ``a mathematical physicist is a mathematician believing that a sesquilinear form is conjugate linear in the first variable and linear in the second'' \cite[p.~80]{Pedersen89}---in other words most of the mathematics literature defines an inner product to be linear in the \textit{first} argument which is something one should take note of.\medskip

An inner product space $(X,\langle\cdot,\cdot\rangle)$ is a normed space under the norm (induced by the inner product) $\|x\|:=\sqrt{\langle x,x\rangle}$. More importantly under this norm the Cauchy-Schwarz inequality\index{Cauchy-Schwarz inequality} holds:
\begin{equation*}
|\langle x,y\rangle|\leq \|x\|\cdot\|y\|\qquad\text{ for all }x,y\in X\,,
\end{equation*}
with equality if and only if $x$ and $y$ are linearly dependent. 

\begin{lemma}[Pythagorean theorem]\label{lemma_pyth_thm}\index{theorem!Pythagorean}
Let $(X,\langle \cdot,\cdot\rangle)$ be an inner product space and $\{g_i\}_{i\in I}\subset X$ be a set of pairwise orthogonal vectors, that is, for all $i,j\in I$, $i\neq j$ one has $\langle g_i,g_j\rangle=0$. Then $\sum\nolimits_{i\in I} g_i$ converges if and only if $\sum\nolimits_{i\in I} \|g_i\|^2$ converges in which case
$$
\Big\|\sum\nolimits_{i\in I} g_i\Big\|^2=\sum\nolimits_{i\in I} \|g_i\|^2\,.
$$
\end{lemma}
\begin{proof}
We only prove the case $|I|<\infty$ as this showcases the whole idea of the proof; the details are carried out, e.g., in \cite[Lemma 1.6.1]{Ringrose71}. Because the inner product is sesquilinear we get
\begin{align*}
\Big\|\sum\nolimits_{j=1}^n g_j\Big\|^2=\Big\langle \sum\nolimits_{j=1}^n g_j,\sum\nolimits_{k=1}^n g_k\Big\rangle=\sum\nolimits_{j,k=1}^n \underbrace{\langle g_j,g_k\rangle}_{=\langle g_j,g_j\rangle\delta_{jk}}=\sum\nolimits_{j=1}^n\|g_j\|^2\,.\tag*{\qedhere}
\end{align*}
\end{proof}

Interestingly enough one has the following characterization \cite[Lemma 11.2.~ff.]{MeiseVogt97en}.
\begin{lemma}
Let $(X,\|\cdot\|)$ be a normed space over $\mathbb F$. The norm on $X$ is induced by an inner product (via $\|x\|=\sqrt{\langle x,x\rangle}$) if and only if the parallelogram law holds, i.e.~
$$
\|x+y\|^2+\|x-y\|^2=2(\|x\|^2+\|y\|^2)\qquad\text{ for all }x,y\in X\,.
$$
\end{lemma}

Because every inner product space is a normed space it is also a topological vector space with (translation-invariant) metric. Thus we have access to a notion of completeness.
\begin{definition}
An inner product space $(X,\langle \cdot,\cdot\rangle)$ which is complete with respect to the metric induced by the norm $\|x\|:=\sqrt{\langle x,x\rangle}$ is called a Hilbert space.\index{space!Hilbert}
\end{definition}

We made it all the way from general topological spaces to Hilbert spaces. For the latter, assuming the axiom of choice, one can guarantee the existence of an orthonormal basis which induces a number of interesting characterizations and expansions. We will summarize all these results \cite[Ch.~12]{MeiseVogt97en} in the following proposition.

\begin{proposition}\label{prop_hilbert_space_basis}
Let $(X,\langle \cdot,\cdot\rangle)$ be an inner product space and $(e_i)_{i\in I}$ an orthonormal system\footnote{Given a set $I\neq\emptyset$ and a family $(e_i)_{i\in I}$ in $X$ this family is called an \textit{orthonormal system}\index{orthonormal system} if $\langle e_i,e_j\rangle=\delta_{ij}$, for all $i,j\in I$, i.e.~$\langle e_i,e_j\rangle=1$ if $i=j$ and $\langle e_i,e_j\rangle=0$ if $i\neq j$.} in $X$.
\begin{itemize}
\item[(i)] For all $x\in X$ Bessel's inequality holds:\index{Bessel's inequality}
$
\sum\nolimits_{i\in I}|\langle e_i,x\rangle|^2\leq\|x\|^2
$
\item[(ii)] The following statements are equivalent.
\begin{itemize}
 \item[(a)] $\overline{\operatorname{span}\{e_i\,|\,i\in I\}}=X$
 \item[(b)] For every $x\in X$ the Fourier expansion\index{Fourier expansion} 
 $
 x=\sum_{i\in I}\langle e_i,x\rangle e_i
 $
 holds in the sense of Def.~\ref{defi_unordered_summation} ff.
 \item[(c)] For every $x\in X$ Parseval's equation\index{Parseval's equation} holds:
 $
 \|x\|^2=\sum\nolimits_{i\in I}|\langle e_i,x\rangle|^2
 $
\end{itemize}
If any of these equivalent conditions hold then $(e_i)_{i\in I}$ is called orthonormal basis\index{orthonormal basis} of $X$.
\end{itemize}
Now let $X$ be complete.
\begin{itemize}
\item[(iii)] Every orthonormal system in $X$ can be extended to an orthonormal basis of $X$. In particular every non-trivial Hilbert space has an orthonormal basis.
\item[(iv)] Let $X$ be infinite-dimensional (i.e.~for all finite subsets $\{x_1,\ldots,x_n\}\subset X$ one has $\operatorname{span}\{x_1,\ldots,x_n\}\neq X$). Then the following are equivalent.
\begin{itemize}
\item[(a)] $X$ is separable.\index{separable!Hilbert space}
\item[(b)] $X$ has a countable orthonormal basis, i.e.~an orthonormal basis of the form $(e_i)_{i\in\mathbb N}$.
\item[(c)] Every orthonormal system in $X$ is countable.
\end{itemize}
\end{itemize}
\end{proposition}

Arguably, the most elementary Hilbert space of infinite dimension is to be found in the zoo of sequence spaces:

\begin{example}\label{ex_ell2_Hilbert}\index{space!l2@$\ell^2(\mathbb N)$}
The space of square-summable sequences $\ell^2(\mathbb N)$ from Ex.~\ref{ex_ell_p_space} turns into a pre-Hilbert space via the inner product
\begin{align*}
\langle\cdot,\cdot\rangle:\ell^2(\mathbb N)\times \ell^2(\mathbb N)&\to\mathbb C\\
(x,y)&\mapsto \sum\nolimits_{n=1}^\infty \overline{x_n}y_n\,.
\end{align*}
The induced norm co{\"i}ncides with the $2$-norm on $\ell^2(\mathbb N)$ so the latter complete, hence a Hilbert space. Moreover an orthonormal basis of $\ell^2(\mathbb N)$ is given by $(e^{(n)})_{n\in\mathbb N}$ consisting of the standard basis vectors $e^{(n)}=(\delta_{jn})_{j=1}^\infty=(0,\ldots,0,1,0,\ldots)$. Notably this orthonormal basis is countable which by Prop.~\ref{prop_hilbert_space_basis} (iv) implies that $\ell^2(\mathbb N)$ is a separable Hilbert space.
\end{example}

\begin{remark}\label{rem_ell2_sep_HS}
Actually every separable, infinite-dimensional Hilbert space $\mathcal H$ is isometrically isomorphic to $\ell^2(\mathbb N)$ (\cite[Coro.~12.9]{MeiseVogt97en} \& Def.~\ref{def_isom_isom}). Later on we will formulate this as follows: To every such $\mathcal H$ there exists a unitary transformation $U:\mathcal H\to\ell^2(\mathbb N)$ (refer to Rem.~\ref{rem_op_between_diff_spaces}).

Although all separable infinite-dimensional Hilbert spaces are ``structurally equivalent'' there surely are useful Hilbert spaces besides $\ell^2(\mathbb N)$. Examples of such spaces include, but are not limited to, the square-integrable functions $L^2(\mathbb R,\mathbb C)$ (Ex.~\ref{ex_quantum_harm_osc}, footnote \ref{footnote_qhosc}) or general square-integrable functions defined on some subset of $\mathbb R^n$, such as $L^2([0,1])$.
\end{remark}

While one has access to the Cauchy-Schwarz inequality on the Hilbert space $\ell^2(\mathbb N)$ this is just a special case of the famous H\"older inequality\index{Hoelders inequality@H{\"o}lder's inequality} which we state here for the sake of completeness:\index{space!lp@$\ell^p(\mathbb N)$}
\begin{lemma}[\cite{Ringrose71}, Lemma 1.3.2]\label{lemma_hoelders_ineq}
Let $p,q\in[1,\infty]$ be conjugate, i.e.~$\frac{1}{p}+\frac{1}{q}=1$. Then for all $x\in\ell^p(\mathbb N)$, $y\in\ell^q(\mathbb N)$ one has $(x_jy_j)_{j\in\mathbb N}\in\ell^1(\mathbb N)$ with
$$
\sum\nolimits_{j=1}^\infty |x_jy_j|\leq \|x\|_p\|y\|_q\,.
$$
Here $\|\cdot\|_p,\|\cdot\|_q$ are the respective norms from Ex.~\ref{ex_ell_p_space}.
\end{lemma}

\section{Spectral Measures and Spectral Integrals}\label{sec:spectral_integrals}
Generalizing $A=\sum_{\lambda\in\sigma(A)}\lambda |g_\lambda\rangle\langle g_\lambda|$ to something like $T=\int_{-\infty}^\infty t\,dE(t)$, of course, requires making sense of the latter. This leads us to spectral measures and spectral integrals for which we will strongly orient ourselves towards Schm\"udgen \cite[Ch.~4]{Schmuedgen12}.

\begin{definition}
Let $\mathcal H$ be a Hilbert space. Then $\{E(\lambda)\}_{\lambda\in\mathbb R}\subset\mathcal B(\mathcal H)$ is a resolution of the identity\index{resolution of the identity} if all of the following statements hold.
\begin{itemize}
\item[$\bullet$] For all $\lambda\in\mathbb R$, $E(\lambda)$ is an orthogonal projection.
\item[$\bullet$] If $\lambda_1\leq\lambda_2$ then $E(\lambda_2)-E(\lambda_1)$ is positive semi-definite.
\item[$\bullet$] For all $\lambda_0\in\mathbb R$ one has strong right continuity, that is, $\lim_{\lambda\to\lambda_0^+}E(\lambda)x=E(\lambda_0)(x)$ for all $x\in\mathcal H$.
\item[$\bullet$] For all $x\in\mathcal H$, $\lim_{\lambda\to-\infty}E(\lambda)x=0$ and $\lim_{\lambda\to+\infty}E(\lambda)x=x$.
\end{itemize}
\end{definition}
\begin{example}\label{ex_spectr_separable}
Let $\mathcal H$ be a separable Hilbert space with orthonormal basis $(e_n)_{n\in\mathbb N}$ and $(\lambda_n)_{n\in\mathbb N}$ be any real-valued sequence. Then the map
\begin{equation}\label{eq:spectr_separable}
E:\mathbb R\to\mathcal B(\mathcal H)\qquad \lambda\mapsto \sum_{\{n\in\mathbb N\,|\,\lambda_n\leq\lambda\}}| e_n\rangle\langle e_n|
\end{equation}
is a resolution of the identity. The sum converges in the strong operator topology so the codomain of $E$, strictly speaking, is $(\mathcal B(\mathcal H),\tops)$.
\end{example}
First we are concerned with compact intervals $J:=[a,b]$ where $a,b\in\mathbb R$, $a<b$, as well as partitions\footnote{
A partition of an interval $[a,b]$ in this context is a finite set $\{\lambda_0,\ldots,\lambda_n\}=:Z\subset\mathbb R$ such that
$$
a-1<\lambda_0<a<\lambda_1<\ldots<\lambda_n=b\,.
$$ 
Its norm is defined to be $|Z|:=\max_{k=1,\ldots,n}|\lambda_k-\lambda_{k-1}|$. Given a continuous function $f:[a,b]\to\mathbb C$, a resolution of the identity $\{E(\lambda)\}_{\lambda\in\mathbb R}$, and $z_k\in[\lambda_{k-1},\lambda_k]$ the corresponding Riemann sum is given by
$$
S(f,Z):=\sum\nolimits_{k=1}^n f(z_k)\big( E(\lambda_k)-E(\lambda_{k-1})\big)\,.
$$\label{footnote_partition}
}
of such intervals which allow us to define an operator-valued Stieltjes integral via approximating Riemann sums.
\begin{lemma}
Let non-empty $J=[a,b]\subset\mathbb R$, a continuous function $f:J\to\mathbb C$, and a resolution of the identity $\{E(\lambda)\}_{\lambda\in\mathbb R}$ be given. Then there exists a bounded operator $\int_J f\,dE$ on $\mathcal H$ which is uniquely defined by the following property:

For all $\varepsilon>0$ there exists $\delta>0$ such that for all partitions $Z$ of $J$ with $|Z|<\delta$ one has $$
\Big\|\int_J f\,dE-S(f,Z)\Big\|\leq\varepsilon\,.
$$
Moreover for every $x\in\mathcal H$
\begin{align*}
\Big\langle x,\Big(\int_J f\,dE\Big)x\Big\rangle=\int_J f(\lambda)\,d\langle x,E(\lambda)x\rangle\quad \text{ and }\quad\Big\|\Big(\int_J f\,dE\Big)x\Big\|^2=\int_J |f(\lambda)|^2\,d\langle x,E(\lambda)x\rangle\,.
\end{align*}
\end{lemma}

Now let us extend this from compact intervals $J$ to the whole real number line:

\begin{lemma}
Let a continuous function $f:\mathbb R\to\mathbb C$ as well as a resolution of the identity $\{E(\lambda)\}_{\lambda\in\mathbb R}$ be given. Defining
$$
D:=\Big\{x\in\mathcal H\,\Big|\,\int_\mathbb R|f(\lambda)|^2d\langle x,E(\lambda)x\rangle<\infty\Big\}
$$
there exists a linear operator $\int_\mathbb Rf\,dE:D\to\mathcal H$ such that 
\begin{align*}
\Big(\int_\mathbb Rf\,dE\Big)x&=\lim_{a\to-\infty}\lim_{b\to\infty}\Big(\int_{[a,b]}f\,dE\Big)x\\
\text{and}\qquad \Big\langle x,\Big(\int_{\mathbb R} f\,dE\Big)x\Big\rangle&=\int_{\mathbb R} f(\lambda)\,d\langle x,E(\lambda)x\rangle
\end{align*}
for all $x\in D$.
\end{lemma}

Usually more general integrals are defined using measure theory and things are no different here:
\begin{definition}
Let $X$ be a non-empty set, $\mathcal A$ some $\sigma$-algebra\index{sigma-algebra@$\sigma$-algebra}\footnote{
Given a non-empty set $X$, a collection of subsets $\mathcal A\subseteq\mathcal P(X)$ is called a $\sigma$-algebra if
\begin{itemize}
\item[$\bullet$] $X\in\mathcal A$
\item[$\bullet$] for all $S\in\mathcal A$ one has $X\setminus S\in\mathcal A$.
\item[$\bullet$] for any sequence $\{S_n\}_{n\in\mathbb N}$ in $\mathcal A$ one has $\bigcup_{n\in\mathbb N}S_n\in \mathcal A$.
\end{itemize}
If, additionally, $X$ is a topological space (with topology $\tau$) then the Borel-$\sigma$-algebra $\mathbb B(X,\tau)$ (for short: $\mathbb B(X)$) is the smallest $\sigma$-algebra which contains $\tau$.\label{footnote_sigma_algebra_borel}
}
on $X$, and $\mathcal H$ a Hilbert space. A spectral measure\index{spectral measure} is a map $E:\mathcal A\to\mathcal B(\mathcal H)$ which satisfies the following properties.
\begin{itemize}
\item[$\bullet$] For all $S\in\mathcal A$, $E(S)$ is an orthogonal projection.
\item[$\bullet$] $E(X)=\mathbbm{1}_{\mathcal H}$.
\item[$\bullet$] For any sequence $(S_n)_{n\in\mathbb N}$ of pairwise disjoint sets from $\mathcal A$ one has
$$
E\Big(\bigcup\nolimits_{n=1}^\infty S_n\Big)=\sum\nolimits_{n=1}^\infty E(S_n)
$$
where the sum converges in the strong operator topology.
\end{itemize}
\end{definition}

This concept directly relates to resolutions of the identity from the start of this section:

\begin{lemma}
Let $X$ be a non-empty set, $\mathcal A$ some $\sigma$-algebra on $X$, and $\mathcal H$ a Hilbert space. The following statements hold.
\begin{itemize}
\item[(i)] A map $E:\mathcal A\to\mathcal B(\mathcal H)$ is a spectral measure if and only if $E(X)=\mathbbm{1}_{\mathcal H}$ and the map $E_x:\mathcal A\to[0,\infty)$, $E_x(S):=\langle x,E(S)x\rangle$ is a measure\footnote{
Given a non-empty set $X$ and a $\sigma$-algebra $\mathcal A$ on $X$, a measure is a map $\mu:\mathcal A\to[0,\infty]$ which is $\sigma$-additive, i.e.~for every sequence $(M_n)_{n\in\mathbb N}$ of disjoint sets in $\mathcal A$ one has $\mu(\bigcup_{n=1}^\infty M_n)=\sum_{n=1}^\infty \mu(M_n)$. Also a complex measure is a map $\mu:\mathcal A\to\mathbb C$ which is $\sigma$-additive.
}
for all $x\in\mathcal H$. 
\item[(ii)] If $E:\mathbb B(\mathbb R)\to\mathcal B(\mathcal H)$ is a spectral measure (where $\mathbb B(\mathbb R)$ is the Borel-$\sigma$-algebra on $\mathbb R$ with the standard topology, cf.~footnote \ref{footnote_sigma_algebra_borel}) then $\{E((-\infty,\lambda])\}_{\lambda\in\mathbb R}$ is a resolution of the identity. 
\item[(iii)] Conversely, if $\{E(\lambda)\}_{\lambda\in\mathbb R}$ is a resolution of the identity then there exists a unique spectral measure $E$ on $\mathbb B(\mathbb R)$ such that $E(\lambda)=E((-\infty,\lambda])$ holds for all $\lambda\in\mathbb R$.
\end{itemize}
\end{lemma}

This suffices to introduce spectral integrals as we aimed for in this section. Henceforth let $X$ be a non-empty set, $\mathcal A$ a $\sigma$-algebra on $X$, and $E$ a spectral measure on $(X,\mathcal A)$. Our goal now is to investigate spectral integrals $I(f)=\int_X f(t)\,dE(t):\mathcal H\to\mathcal H$ of $\mathcal A$-measurable functions\index{measurable}\footnote{
A function $f:(X,\mathcal A)\to\mathbb C$ is called measurable if
$
f^{-1}(S)=\{x\in X\,|\,f(x)\in S\}\in\mathcal A$ for all $S\in\mathbb B(\mathbb C)$.
}
$f:X\to\mathbb C\cup\{\infty\}$ which are finite almost everywhere (with respect to $E$).

Now if one considers the Banach space $\mathcal B(X,\mathcal A)$ of all bounded $\mathcal A$-measurable functions on $X$ equipped with the norm $\|f\|_X=\sup_{t\in X}|f(t)|$ one obtains the following results:

\begin{proposition}\label{prop_properties_func_calc}
Given $f,g\in\mathcal B(X,\mathcal A)$, $\alpha,\beta\in\mathbb C$, and $x,y\in\mathcal H$ the following hold:
\begin{itemize}
\item[(i)] $\|I(f)\|\leq\|f\|_X$ so in particular $I(f)\in\mathcal B(\mathcal H)$. Moreover $I(f)$ is normal.
\item[(ii)] $I(\overline{f})=I(f)^*$, $I(\alpha f+\beta g)=\alpha I(f)+\beta I(g)$ and $I(fg)=I(f)I(g)$.
\item[(iii)] $\langle x,I(f)y\rangle=\int_X f(t)\,d\langle x,E(t)y\rangle$ and $\|I(f)x\|^2=\int_X|f(t)|^2\,d\langle x,E(t)x\rangle$.
\item[(iv)] $f\equiv g$ almost everywhere if and only if $I(f)=I(g)$.
\item[(v)] $f\neq 0$ almost everywhere if and only if $I(f)$ is invertible. In this case $I(f)^{-1}=I(f^{-1})$.
\item[(vi)] $\sigma(I(f))=\{\lambda\in\mathbb C\,|\,\forall_{\varepsilon>0}\;E(\{t\in X\,|\,|f(t)-\lambda|<\varepsilon\})\neq 0\}$
\end{itemize}
\end{proposition}
One can also makes sense of this integral if $f$ is an unbounded measurable function but we will omit this for it is beyond our needs. 
\begin{example}[Example \ref{ex_spectr_separable} continued]\label{ex_sep_spectr_integral}
Given a measurable, almost everywhere-finite function $f$ the spectral integral corresponding to \eqref{eq:spectr_separable} is given by $I(f)x=\sum_{n=1}^\infty f(\lambda_n)\langle e_n,x\rangle e_n$ for all $x$ from the domain
$$
D(I(f))=\Big\{x\in\mathcal H\,\Big|\,\sum\nolimits_{n=1}^\infty |f(\lambda_n)|^2|\langle e_n,x\rangle|^2<\infty\Big\}\,.
$$
\end{example}

\section{Tensor Products of Hilbert Spaces}\label{app_tensor_hilbert}

We will only give a short introduction to this topic, summarizing main concept and results. For this we orient ourselves towards Kadison \& Ringrose \cite[Ch.~2.6, p.~125 ff.]{Kadison83}. Given two vector spaces $V,W$ there exists a vector space $V\odot W$ as well as a bilinear map $\eta:V\times W\to V\odot W$, $(x,y)\mapsto x\otimes y$ with the following property: For every vector space $Z$ and every bilinear map $\xi:V\times W\to Z$ there exists a unique linear map $\zeta:V\odot W\to Z$ such that $\xi=\zeta\circ\eta$. The pair $(V\odot W,\eta)$, which is unique up to isomorphism, is called the algebraic tensor product\index{tensor product!algebraic} of $V,W$, and given bases $(e_i)_{i\in I}$, $(f_j)_{j\in J}$ of $V,W$, respectively, $(e_i\otimes f_j)_{i\in I,j\in J}$ is a basis of $V\odot W$.

Now if $V,W$ are Hilbert spaces (at least one of which of infinite dimension) this concept has a serious drawback: The algebraic tensor product $V\odot W$---while it can be turned into an inner product space with a unique inner product satisfying $\langle x_1\otimes y_1,x_2\otimes y_2\rangle=\langle x_1,x_2\rangle\langle y_1,y_2\rangle$---this space is not complete, meaning not a Hilbert space itself. There are two ways to repair this:

\begin{itemize}
\item To put it simply, the map $\eta$ lacks some form of continuity to carry over completeness of the initial spaces. For this one can adjust $\eta$ to be a weak Hilbert-Schmidt mapping \cite[Def.~2.6.3]{Kadison83}.
\item Every metric space $(X,d)$ can be completed, that is, there exists a complete metric space $(\tilde X,\tilde d)$ and an injective isometry $\iota: X\to\tilde X$; in this case $\tilde X$ is unique up to isometry and thus is called the completion of $X$ \cite[Thm.~43.7]{Munkres00}. Then given Hilbert spaces $\mathcal H_1,\mathcal H_2$ one considers the completion of their algebraic tensor product $\mathcal H_1\odot\mathcal H_2$, which then yields a Hilbert space.
\end{itemize}

Luckily both approaches are equivalent \cite[Rem.~2.6.7]{Kadison83} and given Hilbert spaces $\mathcal H_1,\mathcal H_2$ the Hilbert tensor product\index{tensor product!Hilbert} $\mathcal H_1\otimes \mathcal H_2$ has the following properties \cite[Thm.~2.6.4 ff.]{Kadison83}:
\begin{itemize}
\item The Hilbert tensor product $\mathcal H_1\otimes \mathcal H_2$ is unique up to isomorphism and is complete with respect to the inner product which satisfies $\langle x_1\otimes y_1,x_2\otimes y_2\rangle=\langle x_1,x_2\rangle\langle y_1,y_2\rangle$.
\item Given orthonormal bases $(e_i)_{i\in I}$, $(f_j)_{j\in J}$ of $\mathcal H_1,\mathcal H_2$ respectively, $(e_i\otimes f_j)_{i\in I,j\in J}$ is an orthonormal basis of $\mathcal H_1\otimes \mathcal H_2$.
\item If $\mathcal H_1, \mathcal H_2$ are both separable then so is $\mathcal H_1\otimes \mathcal H_2$. This follows from the previous point together with Prop.~\ref{prop_hilbert_space_basis} (iv). 
\item For all $n\in\mathbb N$ one has $\mathcal H\otimes\mathbb C^n\simeq\mathcal H^n=\mathcal H\times\ldots\times\mathcal H$ \cite[Rem.~2.6.8]{Kadison83}. In particular $\mathcal B(\mathcal H\otimes\mathbb C^n)\simeq \mathcal B(\mathcal H)\otimes\mathbb C^{n\times n}$ can be identified with the $n\times n$-matrices with entries in $\mathcal B(\mathcal H)$ \cite[p.~147 ff.]{Kadison83} and similarly for the trace class.
\item Given Hilbert spaces $\mathcal H_1,\mathcal H_2,\mathcal G_1,\mathcal G_2$ and operators $B_1\in\mathcal B(\mathcal H_1,\mathcal G_1)$, $B_2\in\mathcal B(\mathcal H_2,\mathcal G_2)$ there exists unique $B_1\otimes B_2\in\mathcal B(\mathcal H_1\otimes\mathcal H_2,\mathcal G_1\otimes\mathcal G_2)$ which satisfies $(B_1\otimes B_2)(x_1\otimes x_2)=B_1x_1\otimes B_2x_2$ for all $x_1\in\mathcal H_1$, $x_2\in\mathcal H_2$ \cite[Prop.~2.6.12]{Kadison83}.
\item The tensor product of bounded operators is itself bilinear and satisfies $(B_1\otimes B_2)(B_3\otimes B_4)=B_1B_3\otimes B_2B_4$, $(B_1\otimes B_2)^*=B_1^*\otimes B_2^*$, and $(B_1\otimes B_2)^{-1}=B_1^{-1}\otimes B_2^{-1}$. Moreover $\|B_1\otimes B_2\|=\|B_1\|\|B_2\|$ and for trace class operators $\operatorname{tr}(A_1\otimes A_2)=\operatorname{tr}(A_1)\operatorname{tr}(A_2)$ as well as $\|A_1\otimes A_2\|_1=\|A_1\|_1\|A_2\|_1$.
\item The tensor product of positive semi-definite operators is positive semi-definite again. This follows from $\sqrt{B_1\otimes B_2}=\sqrt{B_1}\otimes\sqrt{B_2}$ for $B_1,B_2\geq 0$.
\end{itemize}
This holds analogously for the Hilbert tensor product of finitely many Hilbert spaces $\mathcal H_1,\ldots,\mathcal H_n$. We conclude this section with the following small result:

\begin{lemma}\label{lemma_V_tensor}
Let $\mathcal H$, $\mathcal G$ be Hilbert spaces and let $\psi\in\mathcal G$ with $\langle \psi,\psi\rangle=1$ be given. Then there exists an isometry $V\in\mathcal B(\mathcal H,\mathcal H\otimes \mathcal G)$ such that for all $B\in\mathcal B(\mathcal H)$
\begin{equation*}
VBV^*=B\otimes |\psi\rangle\langle\psi|\,.
\end{equation*}
\end{lemma}
\begin{proof}
Define $V\in\mathcal B(\mathcal H,\mathcal H\otimes \mathcal G)$ via $Vx:=x\otimes\psi$ for all $x\in\mathcal H$; then its adjoint satisfies $V^*(x\otimes y)=\langle \psi,y\rangle x$ and $V^*V=\mathbbm{1}_{\mathcal H}$, that is, $V$ is an isometry as is readily verified. One computes
\begin{align*}
(VBV^*)(x\otimes y)=\langle \psi,y\rangle (VB)x=\langle \psi,y\rangle (Bx\otimes\psi)=(Bx)\otimes(\langle\psi,y\rangle\psi) =(B\otimes |\psi\rangle\langle\psi|)(x\otimes y)\,.
\end{align*}
Because all involved operators are linear and bounded (hence continuous) this extends from the pure tensors to the whole space $\mathcal H\otimes \mathcal G=\overline{ \operatorname{span}\{ x\otimes y \,|\, x\in\mathcal H,y\in\mathcal G \} }$.
\end{proof}

\section{The Hausdorff Metric}\label{app_hausdorff}

In order to study continuity of majorization polytopes, and to transfer the known results about convexity and star-shapedness of the $C$-numerical
range of matrices to trace-class operators, we need some basic facts about set convergence\index{set convergence}. 
We will use the Hausdorff metric\index{Hausdorff metric} on compact subsets and the associated notion
of convergence, see, e.g.~\cite{Nadler78} and \cite[p.~260 ff.]{Munkres00}. 

Let $(X,d)$ be a metric space. The distance between $z \in X$ and any non-empty compact subset $A \subseteq X$ given by
\begin{align}\label{eq.Hausdorff-1}
d(z,A) := \min_{w \in A} d(z,w) 
\end{align}
is well-defined (compactness ensures that the minimum is attained). 
Based on \eqref{eq.Hausdorff-1}, the \emph{Hausdorff metric} $\Delta$ on the set of all non-empty
compact subsets of $(X,d)$ is given by
\begin{align*}
\Delta(A,B) := \max\Big\lbrace \max_{z \in A}d(z,B),\max_{z \in B}d(z,A) \Big\rbrace\,.
\end{align*}
For the sake of completeness let us give a quick proof that $\Delta$ is indeed a metric. Well-definedness and finiteness of $\Delta$ as well as symmetry are evident. For definiteness: given $A,B\subset X$ non-empty and compact, $\max_{a\in A}d(a,B)=0$ readily implies $a\in B$ for all $a\in A$ meaning $A\subseteq B$. Similarly one finds $B\subseteq A$ so $A=B$. Also $\Delta(A,A)=0$ is trivial.

For the triangle inequality note that $d(a,B)\leq d(a,c)+d(c,B)$ for all $a,c\in X$, $B\subseteq X$ non-empty and compact. Now choose $c=c_0$ to be the point which satisfies $d(a,c_0)=d(a,C)$, i.e.~the point which attains the minimum in $d(a,C)$. Then
$$
d(a,B)\leq d(a,c_0)+d(c_0,B)= d(a,C)+d(c_0,B)\leq\Delta(A,C)+\Delta(C,B)\,.
$$
Taking the maximum over all $a\in A$, interchanging the roles of $A,B$, and using symmetry yields $\Delta(A,B)\leq\Delta(A,C)+\Delta(C,B)$ which concludes the proof.\medskip

The following characterization of the Hausdorff metric will be essential throughout this thesis.

\begin{lemma}\label{lemma_11}
Let $A,B \subseteq X$ be two non-empty compact sets in a metric space $(X,d)$, and let $\varepsilon > 0$ be given.
Then $\Delta(A,B) \leq \varepsilon$ if and only if for all $z \in A$, there exists $w \in B$ 
with $d(z,w) \leq \varepsilon$ and vice versa.
\end{lemma}
\begin{proof}
By definition, $\Delta(A,B) \leq \varepsilon$ is equivalent to $\max_{z \in A}d(z,B)\leq\varepsilon$ and $\max_{z \in B}d(z,A) \leq\varepsilon$. This in turn means
\begin{align}\label{eq:Hausdorff19}
\max_{z \in A}\min_{w \in B} d(z,w)\leq\varepsilon\qquad\text{and}\qquad\max_{z \in B}\min_{w \in A} d(z,w) \leq\varepsilon\,.
\end{align}
Evidently, \eqref{eq:Hausdorff19} holds if and only if for all $z \in A$, there exists $w \in B$ 
with $d(z,w) \leq \varepsilon$ and vice versa.
\end{proof}

With this metric at hand, one can introduce the notion of convergence of a sequence 
$(A_n)_{n\in\mathbb N}$ of non-empty compact subsets. Alternatively, one can introduce the notion of Kuratowski convergence as
follows:
Consider a sequence $(A_n)_{n\in\mathbb N}$ of non-empty compact subsets of $(X,d)$ and define
\begin{itemize}
\item 
$\liminf_{n\to\infty} A_n$ as the set of all $z \in X$ such that for all $\varepsilon > 0$ 
one has $B_\varepsilon(z)\cap A_n\neq\emptyset$ for all but finitely many indices.
\item
$\limsup_{n\to\infty} A_n$ as the set of all $z \in X$ such that for all $\varepsilon>0$
one has $B_\varepsilon(z)\cap A_n\neq\emptyset$ for infinitely many indices.
\end{itemize}

If $\liminf_{n\to\infty} A_n = \limsup_{n\to\infty} A_n =:A$ one says that $(A_n)_{n\in\mathbb N}$ converges 
to $A$ and writes
\begin{align*}
\lim_{n\to\infty}A_n=A\,.
\end{align*}

\noindent The following Lemma shows that both approaches are essentially equivalent, cf.~\cite[Thm. 0.7]{Nadler78}.

\begin{lemma}\label{lem:Hausdorff}
Let $(A_n)_{n\in\mathbb N}$ be a bounded sequence of non-empty compact subsets of $(X,d)$.
\begin{itemize}
\item[(i)]
If $(A_n)_{n\in\mathbb N}$ converges to $A$ with respect to the Hausdorff metric, then
$\liminf_{n\to\infty} A_n = \limsup_{n\to\infty} A_n = A$.
\item[(ii)]
If $\liminf_{n\to\infty} A_n = \limsup_{n\to\infty} A_n =:A$, then $A$ is non-empty and compact, and 
$(A_n)_{n\in\mathbb N}$ converges to $A$ with respect to the Hausdorff metric.
\end{itemize}
\end{lemma}
The Hausdorff metric has the following nice properties:

\begin{lemma}\label{lemma_5}
Let $(A_n)_{n\in\mathbb N}$ and $(B_n)_{n\in\mathbb N}$ be bounded sequences of non-empty compact subsets
of a metric space $(X,d)$ such that $\lim_{n\to\infty}A_n = A$, $\lim_{n\to\infty}B_n = B$. Then the following statements hold.
\begin{itemize}
\item[(i)] 
If $A_n\subseteq B_n$ for all $n\in\mathbb N$, then $A \subseteq B$.
\end{itemize}
Now let $X$ be a finite-dimensional normed space.
\begin{itemize}
\item[(ii)] 
The sequence $(\operatorname{conv}(A_n))_{n\in\mathbb N}$ of compact subsets converges to $\operatorname{conv}(A)$, i.e.
\begin{align*}
\lim_{n\to\infty}\operatorname{conv}(A_n) = \operatorname{conv}(A)\,.
\end{align*}
\item[(iii)] 
If $A_n$ is convex for all $n\in\mathbb N$, then $A$ is convex.
\item[(iv)] Let $(z_n)_{n\in\mathbb N}$ be sequence of complex numbers with $\lim_{n\to\infty}z_n = z$ and assume $X=\mathbb C$. If $A_n$ is star-shaped with respect to $z_n$ for all $n\in\mathbb N$, then $A$ is star-shaped w.r.t.~$z$.
\end{itemize}
\end{lemma}

Note that for bounded sequences $(A_n)_{n\in\mathbb N}$ of non-empty
compact subsets of a metric space which converges to $A$ with respect to the Hausdorff metric
one has the following characterization of the limit set (Lemma \ref{lem:Hausdorff}):
\begin{align*}
x \in A \;\Longleftrightarrow\;\text{there exists a sequence $(a_n)_{n\in\mathbb N}$ with 
$a_n \in A_n$ and $a_n \to x$ for $n \to \infty$\,.}
\end{align*}

\begin{proof}[Proof of Lemma \ref{lemma_5}]
(i): Let $x\in A$ be given. Then there exists a sequence $(a_n)_{n\in\mathbb N}$ with 
$a_n \in A_n$ and $a_n \to x$ for $n \to \infty$. By assumption, we have $A_n \subset B_n$
and thus $a_n \in B_n$. Hence, by the above characterization of the limit set we obtain $x \in B$.

(ii): We prove this for $X=\mathbb C$, the general case is done analogously. Let $\varepsilon>0$ be given. By assumption there exists $N \in\mathbb N$ such 
that for all $n \geq N$, $\Delta(A_n,A)<\varepsilon$. By Lemma \ref{lemma_11}, the latter 
is equivalent to the assertion that for all $a\in A$ there exists $a_n \in A_n$ satisfying
$|a-a_n|<\varepsilon$ and for all $a'_n \in A_n$ there exists $a' \in A$ with $|a'-a'_n|<\varepsilon$. 
First, let $x \in \operatorname{conv}(A)$ be arbitrary. By Carath{\'e}odory's theorem\index{theorem!Carath{\'e}odory's},
$x \in \operatorname{conv}(A)$ can be written as
\begin{align*}
x = r a + s b + t c 
\end{align*}
with $a,b,c\in A$, $r,s,t \geq 0$, and $r + s + t = 1$. 
Then for all $n\geq N$ we can choose $a_n,b_n,c_n\in A_n$ with distance less than 
$\varepsilon$ to $a,b,c$, respectively. This yields for
\begin{align*}
x_n := r a_n + s b_n + t c_n \in \operatorname{conv}(A_n)
\end{align*}
the estimate
\begin{align*}
|x-x_n|\leq r|a-a_n| + s|b-b_n| + t|c-c_n| < \varepsilon\,.
\end{align*}
Similarly, for every $x'_n\in\operatorname{conv}(A_n)$ one can choose $x'\in\operatorname{conv}(A)$ 
with $|x'-x'_n|<\varepsilon$ for all $n\geq N$. This proves (ii) according to Lemma \ref{lemma_11}.

(iii): If $A_n$ is convex, one has $A_n = \operatorname{conv}(A_n)$ for all $n\in\mathbb N$ so by (ii) we immediately obtain
\begin{align*}
A=\lim_{n\to\infty}A_n=\lim_{n\to\infty}\operatorname{conv}(A_n)=\operatorname{conv}(A)\,.
\end{align*}
Hence, $A$ is convex.

(iv): We have to show $t z + (1-t)a \in A$ for all $a \in A$ and $t \in[0,1]$. To this end, let 
$a \in A$ and choose $a_n \in A_n$ such that $a_n \to a$ for $n \to \infty$. Since $A_n$ is
star-shaped with respect to $z_n$ one has $t z_n + (1-t)a_n \in A_n$ for all $n \in \mathbb N$.
Moreover, $t z_n + (1-t)a_n$ obviously converges to $t z + (1-t)a$ and therefore by the 
above characterization of the limit set we conclude $t z + (1-t)a \in A$.
\end{proof}

With this one finds that for sequences $(A_n)_{n\in\mathbb N}$ 
of non-empty compact subsets of $\mathbb R$ the maximum- as well as the minimum-operation are continuous in the following sense:

\begin{lemma}\label{lemma_lim_max}
Let $(A_n)_{n\in\mathbb N}$ be a bounded sequence of non-empty, compact subsets of $\mathbb R$ 
which converges to $A\subset\mathbb R$. Then the sequences of real numbers $(\max A_n)_{n\in\mathbb N}$
and $(\min A_n)_{n\in\mathbb N}$ are convergent with 
\begin{align*}
\lim_{n\to\infty}(\max A_n) 
=\max A \quad\text{and}\quad \lim_{n\to\infty}(\min A_n)=\min A\,.
\end{align*}
\end{lemma}
\begin{proof}
Let $\varepsilon>0$. By assumption, there exists $N\in\mathbb N$ such that $\Delta(A_n,A)<\varepsilon$ 
for all $n\geq N$. Hence by Lemma \ref{lemma_11} one finds $a_n\in A_n$ with 
$|\max A - a_n| < \varepsilon$ and thus
$
\max A < a_n + \varepsilon < \max A_n + \varepsilon\,.
$
Similarly, there exists $a \in A$ such that $|\max A_n - a| < \varepsilon$, so
$
\max A_n < a + \varepsilon < \max A + \varepsilon\,.
$
Combining both estimates, we get $|\max A - \max A_n| < \varepsilon$. The case of the minimum 
is shown analogously.
\end{proof}

\section{Proofs That are Obvious to the Gentle Reader}
This section is dedicated to proofs which are too long or too technical to appear in the main text so we outsourced them and present them now.

\subsection{Proposition \ref{prop_strong_weak_op_top}}\label{sec:app_proof_op_topologies}
Unless specified otherwise we will prove all the statements about $\tops$ because the proofs for $\topw$ are analogous.\smallskip

(i),(a): Let $T\in\mathcal B(X,Y)$ and $U$ be a neighborhood of $T$ with respect to $\tops$. By Remark \ref{rem_nbh_basis_top} there exist $T_1\in\mathcal B(X,Y)$, $A\subset X$ finite, and $\varepsilon_1>0$ with $T\in N(T_1,A,\varepsilon_1)\subseteq U$. Just like in the proof of Lemma \ref{lemma_topsw_basis} one sees that $T\in N(T,A,\varepsilon)\subseteq N(T_1,A,\varepsilon_1)$ when defining $\varepsilon:=\varepsilon_1-\max_{x\in A}\|Tx-T_1x\|>0$. In particular we have $T\in N(T,A,\varepsilon)\subseteq U$, that is, we found an element of the neighborhood basis at $T$ which is contained in $U$.\smallskip

(ii): By (i),(a) and Lemma \ref{lemma_top_properties_basis} (iii) one has $T_i\to T$ in $\tops$ if and only if for all $A\subset X$ finite, $\varepsilon>0$ there exists $i_0\in I$ such that $T_i\in N(T,A,\varepsilon)$ for all $i\succeq i_0$. 
``$\Rightarrow$'': Let $x\in X$ and assume $T_i\to T$ in $\tops$. Choosing $A:=\{x\}$ shows $\|T_ix-Tx\|<\varepsilon$ for all $i\succeq i_0$ so $T_ix\to Tx$ as claimed.
``$\Leftarrow$'': Let $A=\{x_1,\ldots,x_n\}\subset X$ and $\varepsilon>0$ be given. For all $j=1,\ldots,n$ by assumption there exists $i_j\in I$ such that $\|Tx_j-T_ix_j\|<\varepsilon$ for all $i\succeq i_j$. Now $I$ is a directed set so inductively we find $i_0\in I$ with $i_0\succeq i_j$ for all $j=1,\ldots,n$. Evidently $\|T_ix-Tx\|<\varepsilon$ given $i\succeq i_0$ so in other words $T_i\in N(T,A,\varepsilon)$ which by the above characterization proves $T_i\to T$ in $\tops$.\smallskip

(iii): Let $(T_i)_{i\in I}$ be a net in $\mathcal B(X,Y)$ such that $T_i\to T_1$ and $T_i\to T_2$ in $\tops$ for some $T_1,T_2\in\mathcal B(X,Y)$. If we can show that $T_1=T_2$ then by Lemma \ref{lemma_topol_connect} (iv) this proves that $\tops$ Hausdorff. Using (ii) we know $T_ix\to T_1x$, $T_ix\to T_2x$ for all $x\in X$. But the topology induced by a metric is Hausdorff (Lemma \ref{lemma_metric_topo}) so $T_1x=T_2x$ for all $x\in X$ which shows $T_1=T_2$. (For $\topw$ make use of Lemma \ref{lemma_point_separation_dual_space}.)\smallskip

(iv),(a): Given $T\in\mathcal B(X,Y)$, $A\subset X$ and $B\subset Y^*$ both finite, and $\varepsilon>0$ arbitrary we for all $x\in X$, $y\in Y^*$ have
$$
|y((T-T_i)x)|\leq \|y\| \|(T-T_i)(x)\|\leq\|y\|\|T-T_i\|\|x\|
$$
by Lemma \ref{lemma_normed_space_properties} (i). This shows
$$
 B_{\varepsilon/\alpha}(T) \subseteq N(T,A,\varepsilon) \subseteq N(T,A,B,\beta\varepsilon)
$$
where $\alpha:=\max\{\max_{x\in A}\|x\|,1\}$ and $\beta:=\max\{\max_{y\in B}\|y\|,1\}$. Thus $\topw\subseteq\tops\subseteq\topn$ by (i) together with Prop.~\ref{prop_compare_topo} (v).\smallskip

(iv),(b): ``$\Leftarrow$'': Let $\operatorname{dim}(X)<\infty$ so there exists a basis $B:=\{x_1,\ldots,x_n\}$ of $(X,\|\cdot\|_X)$ for some $n\in\mathbb N$. Thus given $x\in X$ we find unique coefficients $c_1,\ldots,c_n\in\mathbb F$ such that $x=\sum_{j=1}^n c_jx_j$ which enables defining a norm $\|x\|_B:=\sum_{j=1}^n |c_j|$ on $X$ as is readily verified. But on finite-dimensional normed spaces all norms are equivalent \cite[Lemma 5.14]{MeiseVogt97en} so there exists $C\geq 1$ such that $\frac{1}{C}\|x\|_B\leq\|x\|_X\leq C\|x\|_B$ for all $x\in X$.
Thus given $\varepsilon>0$ and a net $(T_i)_{i\in I}$ in $\mathcal B(X,Y)$ which converges to $T\in\mathcal B(X,Y)$ in $\tops$, there exists $i_0\in I$ such that for all $i\succeq i_0$
$$
T_i\in N\Big(T,B,\frac{\varepsilon}{C}\Big)\qquad\text{ that is, }\qquad \|T_ix_j-Tx_j\|_X\leq\frac{\varepsilon}{C}\text{ for all }j=1,\ldots,n\,.
$$
Thus for all $x\in X$ and all $i\succeq i_0$
\begin{align*}
\|T_ix-Tx\|_X&=\Big\|\sum\nolimits_{j=1}^n c_j(T_ix_j-Tx_j)\Big\|_X\\
&\leq \sum\nolimits_{j=1}^n |c_j|\|T_ix_j-Tx_j\|_X<\frac{\varepsilon}{C}\sum\nolimits_{j=1}^n |c_j|=\frac{\varepsilon\|x\|_B}{C}\leq\varepsilon\|x\|_X\,.
\end{align*}
But this shows $\|T_i-T\|=\sup_{x\in X,\|x\|=1}\|T_ix-Tx\|<\varepsilon$ for all $i\succeq i_0$ so $T_i\to T$ in $\topn$ by Lemma \ref{lemma_top_properties_basis} (iii),(c) \& Lemma \ref{lemma_metric_topo}. Thus every net which converges in $\tops$ converges in $\topn$, i.e.~$\topn\subseteq\tops$ (Prop.~\ref{prop_compare_topo} (iii)). Together with (iv),(a) we get the claimed equality of the two topologies.

``$\Rightarrow$'': Let $\operatorname{dim}(X)=\infty$. If we can show $N(0,A,\varepsilon)\not\subset B_1(0)$ for all $A\subset X$ finite and all $\varepsilon>0$, then $\topn\not\subset\tops$ by Prop.~\ref{prop_compare_topo} (v) (together with (i)). Indeed given such a subset $A$ we can find\footnote{The existence of such $x_0$ is guaranteed due to $\operatorname{dim}(X)=\infty$ and moreover, $\overline{\operatorname{span}(A)}=\operatorname{span}(A)$ as the latter is a finite-dimensional subspace of a normed space \cite[Thm.~1.21]{Rudin91}.} $x_0\in X\setminus\overline{\operatorname{span}(A)}=X\setminus\operatorname{span}(A)$ with $\|x_0\|=1$. Thus by Lemma \ref{lemma_hahn_banach_coro} one finds $f\in X^*$ such that $f(x_0)=1$ but $f(x)=0$ for all $x\in \operatorname{span}(A)$. By choosing some $y\in Y$, $\|y\|=1$---which is possible because $Y$ is non-trivial---we can turn this functional into a bounded operator via
$$
T:X\to Y\qquad x\mapsto f(x)\cdot y
$$
because $\|T\|=\sup_{x\in X,\|x\|=1}|f(x)|=\|f\|<\infty$. This operator satisfies
$$
\|T\|=\sup_{x\in X,\|x\|=1}\|Tx\|\geq \|Tx_0\|=|f(x_0)|=1
$$
so $T\not\in B_1(0)$ but $\|Tx\|=|f(x)|=0<\varepsilon$ for all $x\in A$ by construction of $f$, i.e.~$T\in N(0,A,\varepsilon)$. Thus we found $T\in N(0,A,\varepsilon)\setminus B_1(0)$ which shows $N(0,A,\varepsilon)\not\subset B_1(0)$.\smallskip

(iv),(c): ``$\Leftarrow$'': Note that $\operatorname{dim}(Y^*)=\operatorname{dim}(Y)<\infty$ by Lemma \ref{lemma_hahn_banach_coro} so one finds a basis $\{y_1,\ldots,y_n\}$ of $Y^*$. As in (iv),(b) this yields an equivalent norm on $Y^*$ which by a similar argument leads to $T_i\to T $ in $\topw$ $\Rightarrow$ $T_i\to T$ in $\tops$. The only difference here is that the norm is calculated via the dual space, i.e.~$
\|T_ix-Tx\|=\sup_{y\in Y^*,\|y\|\leq 1}|y(T_ix-Tx)|$, cf.~\cite[Thm.~4.3]{Rudin91}. ``$\Rightarrow$'': Let any $x_0\in X$, $\|x_0\|=1$ be given. The idea again will be to show that $N(0,A,B,\varepsilon)\not\subset N(0,\{x_0\},1)$ for all $A\subset X$, $B\subset Y^*$ both finite and all $\varepsilon>0$ by constructing an operator $T\in N(0,A,B,\varepsilon)\setminus N(0,\{x_0\},1)$. Indeed given any $B=\{f_1,\ldots,f_n\}\subset Y^*$ we can find $y_0\in \bigcap_{j=1}^n \operatorname{ker}(f_j)\subset Y$, $\|y_0\|=1$ by Lemma \ref{lemma_ker_functionals}. On the other hand there exists $f\in Y^*$ with $f(x_0)=\|x_0\|=1$ and $\|f\|\leq 1$ \cite[Prop.~6.10]{MeiseVogt97en} so define $T:X\to Y$ via $x\mapsto f(x)\cdot y_0$. Evidently $T\in\mathcal B(X,Y)$ with $\|Tx_0\|=\|y_0\|=1$ so $T\not\in N(0,\{x_0\},1)$ but
$$
|f_j(Tx)|=|f(x)|\underbrace{|f_j(y_0)|}_{=0}=0<\varepsilon
$$
for all $j=1,\ldots,n$ so $T\in N(0,A,B,\varepsilon)$ as desired.\smallskip

(iv),(d): Obtained by combining (iv),(b) with (iv),(c).\smallskip

(v),(a): Following Lemma \ref{lemma_locally_convex} and Definition \ref{def_initial_top} we have to show that
$$
\tops=\underbrace{\sigma(\mathcal B(X,Y), \{T\mapsto Tx\}_{x\in X} )}_{=:\tau_1}=\underbrace{\sigma(\mathcal B(X,Y), \{T\mapsto \|Tx-Sx\|\}_{x\in X,S\in\mathcal B(X,Y)} )}_{=:\tau_2}\,.
$$
The easiest path is the one via Prop.~\ref{prop_compare_topo} (iii) so let a net $(T_i)_{i\in I}$ in $\mathcal B(X,Y)$ as well as $T\in\mathcal B(X,Y)$ be given. 

``$\tops=\tau_1$'': Assume $T_i\to T$ in $\tops$ so $T_ix\to Tx$ for all $x\in X$ (in $(X,\|\cdot\|)$) by (ii). But this characterizes $T_i\overset{\tau_1}\to T$ by Lemma \ref{lemma_initial_top}.

``$\tops\supseteq\tau_2$'': Assume $T_i\to T$ in $\tops$ so $\|T_ix-Tx\|\to 0$ for all $x\in X$ by (ii). Then 
$$
\big|\,\|T_ix-Sx\|-\|Tx-Sx\|\,\big|\leq \|T_ix-Tx\|\to 0
$$
by the reverse triangle inequality so $\|T_ix-Sx\|\to \|Tx-Sx\|$ for all $x\in X$, $S\in\mathcal B(X,Y)$. But this means $T_i\to T$ in $\tau_2$ by Lemma \ref{lemma_initial_top}.

``$\tau_2\supseteq\tops$'': Let $T_i\to T$ in $\tau_2$ so again by Lemma \ref{lemma_initial_top} $\|T_ix-Sx\|\to \|Tx-Sx\|$ for all $x\in X$, $S\in\mathcal B(X,Y)$. Choose $S=T$ so $\|T_ix-Tx\|\to 0$ (i.e.~$T_ix\to Tx$) for all $x\in X$ which shows $T_i\to T$ in $\tops$ by (ii).

Finally $\{T\mapsto \|Tx\|\}_{x\in X}$ is a fundamental family of seminorms because $\|Tx\|=0$ for all $x\in X$ implies $Tx=0$ for all $x\in X$ so $T=0$. Now the final statement is precisely Def.~\ref{defi_locally_convex}.

\subsection{The Lowering Operator is Closed}\label{appendix_lowering_op_proof}
Consider the lowering operator\index{operator!lowering}
$$
a:D(a)\to\ell^2(\mathbb N)\qquad x=(x_1,x_2,x_3,\ldots)\mapsto ( x_2,\sqrt{2}x_3,\sqrt{3}x_4 ,\ldots)
$$
with $D(a)=\{\sum_{n=1}^\infty n|x_{n+1}|^2<\infty\,|\,x\in\ell^2(\mathbb N)\}$ from Rem.~\ref{rem_I_plus_a_a_star}. We want to show that this operator is closed. 
\begin{proof}
First note that $a$ is surjective: Given $y\in\ell^2(\mathbb N)$ define $\tilde y:=(0,y_1,\frac{y_2}{\sqrt{2}},\frac{y_3}{\sqrt{3}},\ldots)$. Obviously $\tilde y\in\ell^2(\mathbb N)$ and $a\tilde y=y$ with
$$
\sum\nolimits_{n=1}^\infty n|\tilde y_{n+1}|^2=\sum\nolimits_{n=1}^\infty|y_n|^2=\|y\|^2<\infty\quad\Rightarrow\quad \tilde y\in D(a)\,.
$$
With this in mind consider a sequence $(x^{(n)})_{n\in\mathbb N}$ in $D(a)$ such that $x^{(n)}\to x$ and $ax^{(n)}\to y$ for some $x,y\in\ell^2(\mathbb N)$. If we can show that $x\in D(a)$ and $ax=y$ then the graph of $a$ is closed as desired. By surjectivity we find $\tilde y\in D(a)$ with $y=a\tilde y$. Consider the usual left shift $L\in\mathcal B(\ell^2(\mathbb N))$ given by $L(x_1,x_2,\ldots):=(x_2,x_3,\ldots)$. Then
$$
\|Lx^{(n)}-L\tilde y\|^2=\sum\nolimits_{k=1}^\infty |x^{(n)}_{k+1}-\tilde y_{k+1}|^2\leq \sum\nolimits_{k=1}^\infty k|x^{(n)}_{k+1}-\tilde y_{k+1}|^2=\|ax^{(n)}-a\tilde y\|^2\overset{n\to\infty}\to 0
$$
by assumption. On the other hand $\|Lx^{(n)}-Lx\|\leq\|L\|\|x^{(n)}-x\|\to 0$ as $n\to\infty$ so as the limit in normed spaces is unique (Lemma \ref{lemma_metric_topo} \& \ref{lemma_topol_connect} (iv)) we find $Lx=L\tilde y$. Hence there exists $c\in\mathbb C$ such that $x=(c,\tilde y_2,\tilde y_3,\ldots)$. With this it is obvious that $x\in D(a)$ (because $\tilde y$ is) as well as $ax=a\tilde y=y$.
\end{proof}
\subsection{Proposition \ref{prop_kraus_op_general}---the General Case}\label{app_kraus_gen}

Due to the proofs of Davies or Kraus we already know that Prop.~\ref{prop_kraus_op_general} holds for $\mathcal H=\mathcal G$ and we only have to extend the result to different Hilbert spaces. The idea we will follow to show (i) $\Rightarrow$ (ii)---after all the proof of the converse needs no adjustment---is rather simple: We will transform $S:\mathcal B(\mathcal G)\to\mathcal B(\mathcal H)$ into a map with domain and codomain $\mathcal B(\mathcal G\otimes\mathcal H)$ so we can apply our knowledge of the special case, and then we transfer the obtained form to the original $S$. 

Let $\psi\in\mathcal H$, $\phi\in\mathcal G$ with $\langle\psi,\psi\rangle=\langle\phi,\phi\rangle=1$ be given. Because $\iota_{|\psi\rangle\langle\psi|}:\mathcal B^1(\mathcal G)\to\mathcal B^1(\mathcal G\otimes\mathcal H)$, $A\mapsto A\otimes|\psi\rangle\langle\psi|$ is well-defined, linear, and positive (the tensor product ``carries over'' positive semi-definiteness) we may consider its dual map $\operatorname{tr}_{|\psi\rangle\langle\psi|}:=(\iota_{|\psi\rangle\langle\psi|})^*$ in the sense of Coro.~\ref{coro_pos_dual_equal} which then satisfies $\operatorname{tr}_{|\psi\rangle\langle\psi|}(B_1\otimes B_2)=\langle\psi,B_2\psi\rangle B_1$. This lets us define
\begin{align*}
\tilde S:\mathcal B(\mathcal G\otimes\mathcal H)&\to\mathcal B(\mathcal G\otimes\mathcal H)\\
\tilde B&\mapsto |\phi\rangle\langle\phi|\otimes S(\operatorname{tr}_{|\psi\rangle\langle\psi|}(\tilde B))\,.
\end{align*}
Writing $\tilde S=\iota_{|\phi\rangle\langle\phi|}\circ S\circ \operatorname{tr}_{|\psi\rangle\langle\psi|}$ we see that $\tilde S$ is completely positive and ultraweakly continuous as a composition completely positive, ultraweakly continuous maps\footnote{
By Lemma \ref{lemma_V_tensor} the map $\iota$ which extends by a pure state can be written as $B\mapsto VBV^*$ for some bounded $V$ so its (pre-)dual by \eqref{eq:duality_trace_map} is given by $\tilde B\mapsto V^*\tilde BV$. While ultraweak continuity is obvious this is also a prime example of a positive and thus a completely positive map due to $VB_1V^*\otimes B_2=(V\otimes\mathbbm{1})(B_1\otimes B_2)(V\otimes \mathbbm{1})^*$.
}.
Therefore one finds Kraus operators $(\tilde K_i)_{i\in I}\subset\mathcal B(\mathcal G\otimes\mathcal H)$ such that $\tilde S(\tilde B)=\sum_{i\in I}\tilde K_i^*\tilde B\tilde K_i$ for all $\tilde B\in\mathcal B(\mathcal G\otimes\mathcal H)$ with the sum converging strongly. Thus the only thing left to do is to go back from $\tilde S$ to $S$.
Indeed for all $B\in\mathcal B(\mathcal G)$ we compute
\begin{align*}
S(B)&=(\operatorname{tr}_{|\phi\rangle\langle\phi|}\circ \tilde S\circ \iota_{|\psi\rangle\langle\psi|})(B)=\operatorname{tr}_{|\phi\rangle\langle\phi|}(\tilde S(B\otimes|\psi\rangle\langle\psi|))\\
&=\operatorname{tr}_{|\phi\rangle\langle\phi|}\Big(\sum\nolimits_{i\in I}\tilde K_i^*(B\otimes|\psi\rangle\langle\psi|)\tilde K_i\Big)\\
&=\sum\nolimits_{i\in I}V_\phi^*\tilde K_i^*V_\psi BV_\psi^*\tilde K_iV_\phi=\sum\nolimits_{i\in I}(V_\psi^*\tilde K_iV_\phi)^* BV_\psi^*\tilde K_iV_\phi
\end{align*} 
so \eqref{eq:kraus_form} holds for $K_i:=V_\psi^*\tilde K_iV_\phi\in\mathcal B(\mathcal G,\mathcal H)$ where $V_\psi,V_\phi$ are the maps from Lemma \ref{lemma_V_tensor} with respect to $\psi,\phi$. For the last statement, if $\mathcal H,\mathcal G$ are both separable then so is $\mathcal H\otimes\mathcal G$ which means $I$ can be chosen to be countable. This concludes the proof.

\subsection{Theorem \ref{theorem_unitary_group_properties}}\label{subsec:unitary_group}
(i): Let $U\in\mathcal U(\mathcal H)$. Because $\|Ux\|^2=\langle Ux,Ux\rangle=\langle x,U^*Ux\rangle=\langle x,x\rangle=\|x\|^2$ for all $x\in\mathcal H$ one finds $\|U\|=\sup_{x\in\mathcal H,\|x\|=1}\|Ux\|=\sup_{x\in\mathcal H,\|x\|=1}\|x\|=1$ as claimed.\smallskip

(ii): Let arbitrary $U_1,U_2\in\mathcal U(\mathcal H)$ be given. Now the (well-defined) composition of bijective maps is bijective with $(U_1U_2)^{-1}=U_2^{-1}U_1^{-1}=U_2^*U_1^*=(U_1U_2)^*$ as well as $(U_1^{-1})^*=(U_1^*)^*=U_1=(U_1^{-1})^{-1}$ (Prop.~\ref{prop_adjoint_op_hilbert}). Also $\|U_1^{-1}\|=\|U_1^*\|=\|U_1\|<\infty$ so $U_1U_2,U_1^{-1}\in \mathcal U(\mathcal H)$ which shows that the latter is indeed a group.\smallskip

(iii): To show closedness (because we are in a metric space, see Remark \ref{rem_metrizable_topo} (i)) we have to show that if a every sequence $(U_n)_{n\in\mathbb N}\subseteq\mathcal U(\mathcal H)$ converges to some $T\in\mathcal B(\mathcal H)$ in norm then $T\in\mathcal U(\mathcal H)$, i.e.~$T^*T=TT^*=\mathbbm{1}_{\mathcal H}$. Using that ${}^*$ is a conjugate-linear isometry (Prop.~\ref{prop_adjoint_op_hilbert} (iii)) we get
\begin{align*}
\|\mathbbm{1}_{\mathcal H}-T^*T\|&=\|U_n^*U_n-T^*T\|\leq \|U_n^*U_n-T^*U_n\|+\|T^*U_n-T^*T\|\\
&\leq \underbrace{\|U_n^*-T^*\|}_{\|U_n-T\|}\underbrace{\|U_n\|}_{=1\text{ by (i)}}+\|T^*\|\|U_n-T\|=(1+\|T\|)\|U_n-T\|\overset{n\to\infty}\to 0
\end{align*}
so $T^*T=\mathbbm{1}_{\mathcal H}$. Analogously one obtains $TT^*=\mathbbm{1}_{\mathcal H}$ so $T\in\mathcal U(\mathcal H)$ as desired.\smallskip

(iv): For every $U\in\mathcal U(\mathcal H)$ there exists $Q\in\mathcal B(\mathcal H)$ self-adjoint such that $U=\exp(iQ)$ (cf.~\cite[Proof of Thm.~12.37]{Rudin91}, as usual $\exp(iQ)=\sum_{n=0}^\infty\frac{(iQ)^n}{n!}$ converges in norm because $Q$ is bounded). 
Then $t\mapsto T(t):=\exp(itQ)$ is a continuous mapping of $[0,1]$ into $(\mathcal U(\mathcal H),\topn)$ with 
$T(0)=\mathbbm{1}_{\mathcal H}$ and $T(1)=U $ (continuity remains when replacing the norm by the weaker topologies $\topw,\tops$). Thus every unitary operator is connected to the 
identity in a continuous manner which implies path-connectedness of $\mathcal U(\mathcal H$).

(v): Let $(U_i)_{i\in I}$ be a net in $\mathcal U(\mathcal H)$ which converges to $T\in\mathcal B(\mathcal H)$ in $\tops$. Then $U_ix\to Tx$ for all $x\in\mathcal H$ which implies $\|U_ix\|\to \|Tx\|$ by the reverse triangle inequality. But
$$
\|U_ix\|=\sqrt{\langle U_ix,U_ix\rangle}=\sqrt{\langle x,U_i^*U_ix\rangle}=\sqrt{\langle x,x\rangle}=\|x\|
$$
for all $x\in\mathcal H$, $i\in I$ because every $U_i$ is unitary so $\|Tx\|=\|x\|$ for all $x\in\mathcal H$, i.e.~$T\in\mathcal B(\mathcal H)$ is an isometry. Now if $(U_i)_{i\in I}$ is assumed to converge in $\topw$ then for all $x,y\in\mathcal H$ one has $|\langle y,Tx\rangle|\leq\sup_{i\in I}|\langle y,U_ix\rangle|\leq\|x\|\|y\|\sup_{i\in I}\|U_i\|=\|x\|\|y\|$ by (i). Thus \cite[Thm.~4.3]{Rudin91} together with Lemma \ref{lemma_riesz_frechet} lets us estimate the norm of $T$ as follows:
$$
\|T\|=\sup_{x,y\in\mathcal H,\|x\|=\|y\|=1}|\langle y,Tx\rangle|\leq \sup_{x,y\in\mathcal H,\|x\|=\|y\|=1}\|x\|\|y\|=1\,.
$$

(vi): If $\operatorname{dim}(\mathcal H)<\infty$ then $\topw=\tops=\topn$ (on $\mathcal B(\mathcal H)$) so (iii) implies that in this case $(\mathcal U(\mathcal H),\tops)$, $(\mathcal U(\mathcal H),\topw)$ are closed.
Now let $\mathcal H$ be infinite-dimensional. All we have to show that the unitary group is not closed in $\tops$ (then it cannot be closed in $\topw$ either because $\topw\subseteq\tops$). The following argument is from \cite[Ch.~II, Rem.~4.10]{Takesaki79}: Let $(e_i)_{i\in I}$ be an orthonormal basis of $\mathcal H$ so by assumption we can extract a countable orthonormal system $(e_{i_n})_{n\in\mathbb N}$ from this (i.e.~one finds an injective mapping $n\mapsto i_n$ from $\mathbb N$ into $I$). Then for every $n\in\mathbb N$ there exists unique $U_n\in\mathcal U(\mathcal H)$ which maps $e_{i_j}\mapsto e_{i_{j+1}}$ for all $j=1,\ldots,n-1$, $e_{i_n}\mapsto e_{i_1}$ and maps $e_{i_{n+1}},e_{i_{n+2}},\ldots$ as well as $(e_i)_{I\setminus \{i_n\,|\,n\in\mathbb N\}}$ to itself (Lemma \ref{lemma_unitary_ONB} (iii)). Now let $T\in\mathcal B(\mathcal H)$ be the unique operator which maps $e_{i_n}\mapsto e_{i_{n+1}}$ for all $n\in\mathbb N$ and maps $(e_i)_{I\setminus \{i_n\,|\,n\in\mathbb N\}}$ to itself (Lemma \ref{lemma_unitary_ONB} (iii)). We claim that $U_n\to T$ in $\tops$ but $T$ is not unitary (because $T$ is not surjective: $e_{i_1}\not\in \operatorname{im}(T)$) which would prove that the unitary group is not closed in $\tops$. Indeed by the Fourier expansion
\begin{align*}
\|U_nx-Tx\|&=\Big\|\sum\nolimits_{i\in I} \langle e_i,x\rangle (U_ne_i-Te_i)\Big\|=\Big\|\sum\nolimits_{j=1}^\infty \langle e_{i_j},x\rangle (U_ne_{i_j}-Te_{i_j})\Big\|\\
&\leq \Big\| \langle e_{i_n},x\rangle (e_{i_1}-e_{i_n})\|+\Big\| \sum\nolimits_{j=n+1}^\infty \langle e_{i_j},x\rangle (e_{i_j}-Te_{i_j})\Big\|\\
&\leq 2|\langle e_{i_n},x\rangle|+(1+\|T\|)\Big\| \sum\nolimits_{j=n+1}^\infty \langle e_{i_j},x\rangle e_{i_j}\Big\|\,.
\end{align*}
The first summand converges to $0$ as $n\to\infty$ by Lemma \ref{lemma_unordered_summation} (iv) and the second summand converges to $0$ by (convergence of) the Fourier expansion.\smallskip

(vii): We will only show that $(\mathcal U(\mathcal H),\tops)$ is a topological group. The norm case is done analogously and the weak case follows from the fact that on $\mathcal U(\mathcal H)$ (the subspace topology induced by) $\tops$ co{\"i}ncides with $\topw$ \cite[Coro.~9.4]{LNM1552}. Indeed let $(U_i,V_i)_{i\in I}$ be a net in $\mathcal U(\mathcal H)\times\mathcal U(\mathcal H)$ which converges to $(U,V)\in\mathcal U(\mathcal H)\times\mathcal U(\mathcal H)$ in the product topology induced by $\tops$. Then $U_i\to U$, $V_i\to V$ in $\tops$ (cf.~Chapter \ref{section_product_subsp_top}) so for all $x\in\mathcal H$ we get
\begin{align*}
\|U_iV_ix-UVx\|&\leq \|U_iV_ix-U_iVx\|+\|U_iVx-UVx\|\\
&\leq\underbrace{\|U_i\|}_{=1\text{ by (i)}}\underbrace{\|V_ix-Vx\|}_{\to 0}+\underbrace{\|U_i(Vx)-U(Vx)\|}_{\to 0}\to 0
\end{align*}
which shows $U_iV_i\to UV$ in $\tops$ (so multiplication on $\mathcal U(\mathcal H)$ is $\tops$-continuous). Similarly because every unitary in particular is an isometry one sees
\begin{align*}
\|U_i^{-1}x-U^{-1}x\|=\|U_i(U_i^{-1}x-U^{-1}x)\|=\|U(U^{-1}x)-U_i(U^{-1}x)\|\to 0
\end{align*}
so taking the inverse is also continuous in $\tops$ which implies the claim.\smallskip

(viii): If $\operatorname{dim}(\mathcal H)<\infty$ then norm-separability of $\mathcal B(\mathcal H)$ transfers onto $\mathcal U(\mathcal H)$ as every subspace of a separable metric space is separable	 \cite[Ex.~16G]{Willard70}. If $\operatorname{dim}(\mathcal H)=\infty$ then as usual one finds an orthonormal basis $(e_i)_{i\in I}$ with countable orthonormal subset $(e_{i_n})_{n\in\mathbb N}$. For any bijective map $f:\mathbb N\to\mathbb N$ let $U_f\in\mathcal U(\mathcal H)$ denote the unique unitary operator which maps $e_{i_n}$ to $e_{i_{f(n)}}$ for all $n\in\mathbb N$ (and leaves $(e_i)_{i\in I\setminus\{i_n\,|\,n\in\mathbb N\}}$ invariant), refer to Lemma \ref{lemma_unitary_ONB} (iii). Then, similarly to Step 2 in the proof of Prop.~\ref{prop_bounded_op_norm_sep}, $(B_{1/2}(U_f))_{f:\mathbb N\to\mathbb N\text{ bijective}}$ is an uncountable set\footnote{To see that the set of all bijective maps $f:\mathbb N\to\mathbb N$ is uncountable keep in mind that $\{0,1\}^{\mathbb N}$ is uncountable (cf.~footnote \ref{footnote_0_1_N}) \cite[Thm.~2.14]{Rudin76}: For every $z\in\{0,1\}^{\mathbb N}$ define $f_z:\mathbb N\to\mathbb N$ via
$$
f_z(2j-1)=\begin{cases} 2j&\text{ if }z_j=1\\2j-1&\text{ else} \end{cases}\qquad f_z(2j)=\begin{cases} 2j-1&\text{ if }z_j=1\\2j&\text{ else} \end{cases}
$$
for all $j\in\mathbb N$. One readily verifies that $f_z$ is bijective and that the map $z\mapsto f_z$ is injective so uncountability of $\{0,1\}^{\mathbb N}$ transfers as claimed.} of disjoint open balls, showing non-separability of $\mathcal U(\mathcal H)$ in norm in this case (Lemma \ref{lemma_non_sep}).\smallskip

(ix): If $\operatorname{dim}(\mathcal H)<\infty$ then separability of the unitary group in $\tops$ (and $\topw$) follows from (viii) together with Lemma \ref{lemma_separable_topology_comp} (ii) (because $\topw\subseteq\tops\subseteq\topn$). Thus we may assume $\mathcal H$ to be (complex) infinite-dimensional and separable\footnote{This case might seem like a trivial corollary of Coro.~\ref{coro_bounded_op_strong_sep}. However subsets of general separable spaces need not be separable \cite[Thm.~16.4]{Willard70}. Such a conclusion could for example be drawn if the subset in question was open or if the topology on the whole space was metrizable \cite[Ex.~16G]{Willard70}. However $\mathcal U(\mathcal H)$ is not open (it is easy to see that the complement is not closed in $\tops$ or $\topw$) and $\tops$ is not metrizable in infinite dimensions, cf.~footnote \ref{footnote_tops_not_metrizable}.

Another proof strategy would be to use that every open ball in $\mathcal B(\mathcal H)$ (e.g., $B_2(0)$) is metrizable in $\tops$ if $\mathcal H$ is separable (Prop.~\ref{prop_op_top_metrizable}) so $\mathcal U(\mathcal H)$ by (i) is a subset of the separable metrizable space $(B_2(0),\tops)$ hence $\tops$-separable itself \cite[Ex.~16G]{Willard70}. However we want to pursue a constructive approach using the previous results on approximations of unitary operators, which hopefully turns out to be more illuminating to you, the reader.} which lets us find a (countable) orthonormal basis $(e_n)_{n\in\mathbb N}$ of $\mathcal H$ (Prop.~\ref{prop_hilbert_space_basis}). Our strategy will be to show that the set of all unitary ``block approximations'' from Coro.~\ref{coro_unitary_approx_block_unitary} is strongly separable and dense in $(\mathcal U(\mathcal H),\tops)$. Indeed by said corollary we know
\begin{equation}\label{eq:unitary_approx_union}
\bigcup_{n\in\mathbb N} J_n'(\mathcal U(\mathbb C^n))\subseteq\mathcal U(\mathcal H)\quad\text{ as well as }\quad\overline{\bigcup_{n\in\mathbb N} J_n'(\mathcal U(\mathbb C^n))}^{\,\tops}\supseteq\mathcal U(\mathcal H)
\end{equation}
because for every unitary operator there exists a sequence in $\bigcup_{n\in\mathbb N} J_n'(\mathcal U(\mathbb C^n))$ which $\tops$-converges to said operator. Now the proof can be subdivided into two steps.\smallskip

\textit{Step 1:} $J_n'(\mathcal U(\mathbb C^n))$ is $\tops$-separable for all $n\in\mathbb N$.\smallskip

\noindent By (vii) $\mathcal U(\mathbb C^n)$ is norm-separable for all $n\in\mathbb N$ so let $(U_k)_{k\in\mathbb N}$ denote a countable norm-dense subset of $\mathcal U(\mathbb C^n)$. We claim that $(J_n'(U_k))_{k\in\mathbb N}$ is norm-dense in $J_n'(\mathcal U(\mathbb C^n))$. Given $A\in J_n'(\mathcal U(\mathbb C^n))$, i.e.~$A=J_n'(U)$ for some $U\in \mathcal U(\mathbb C^n)$, there exists a subsequence $(U_{k_j})_{j\in\mathbb N}$ which converges to $U$ as $j\to\infty$. Then 
\begin{align*}
\|J_n'(U_{k_j})-A\|&=\|J_n'(U_{k_j})-J_n'(U)\|=\|J_n(U_{k_j})-J_n(U)\|\\
&=\|J_n(U_{k_j}-U)\|\leq\underbrace{\|J_n\|}_{\leq 1} \|U_{k_j}-U\|\to 0\quad\text{ as }j\to\infty
\end{align*}
because $J_n$ is linear and contractive (Lemma \ref{lemma_unitary_approx_matrices}) which shows that $\overline{(J_n'(U_{k_j}))_{j\in\mathbb N}}^{\,\topn}\supseteq J_n'(\mathcal U(\mathbb C^n))$. Therefore $J_n'(\mathcal U(\mathbb C^n))$ is norm-separable (Lemma \ref{lemma_separable_countable_union}) and thus this set is separable in every weaker topology such as, e.g., $\tops$ (Lemma \ref{lemma_separable_topology_comp}).\smallskip

\textit{Step 2:} $(\mathcal U(\mathcal H),\tops)$ is separable.\smallskip

\noindent Step 1 together with Lemma \ref{lemma_separable_countable_union} shows that $\bigcup_{n\in\mathbb N} J_n'(\mathcal U(\mathbb C^n))$ is $\tops$-separable so one finds a countable subset $(U_n)_{n\in\mathbb N}$ of $\bigcup_{n\in\mathbb N} J_n'(\mathcal U(\mathbb C^n))$ such that $\overline{\{U_n\,|\,n\in\mathbb N\}}^{\,\tops}\supseteq \bigcup_{n\in\mathbb N} J_n'(\mathcal U(\mathbb C^n))$. But then $(U_n)_{n\in\mathbb N}$ is a countable subset of $\mathcal U(\mathcal H)$ (by \eqref{eq:unitary_approx_union}) which satisfies
\begin{align*}
\overline{\{U_n\,|\,n\in\mathbb N\}}^{\,\tops}=\overline{\overline{\{U_n\,|\,n\in\mathbb N\}}}^{\,\tops}\supseteq \overline{\bigcup_{n\in\mathbb N} J_n'(\mathcal U(\mathbb C^n))}^{\,\tops}\overset{\eqref{eq:unitary_approx_union}}\supseteq\mathcal U(\mathcal H)\,.
\end{align*}
This by Lemma \ref{lemma_separable_countable_union} proves that $(\mathcal U(\mathcal H),\tops)$ is separable.\smallskip

(x): Simple consequence of Prop.~\ref{prop_op_top_metrizable} together with (i) \& Lemma \ref{lemma_riesz_frechet}.\smallskip

(xi): If $\mathcal H$ is of finite dimension then closedness (by (iii)) and boundedness (by (i)) implies norm compactness by the Heine-Borel theorem\index{theorem!Heine-Borel}. Also the three topologies in question co{\"i}ncide in finite dimensions (Prop.~\ref{prop_op_top_metrizable}) so one finds compactness in either of them. As for the converse: if $\operatorname{dim}(\mathcal H)=\infty$ then $\mathcal U(\mathcal H)$ is not closed in $\topw$ by (vi). Therefore $\mathcal U(\mathcal H)$ is not compact in $\topw$ (\cite[Thm.~26.3]{Munkres00} because $\topw$ is Hausdorff) meaning it cannot be compact in any stronger topology, either (Lemma \ref{lemma_separable_topology_comp} (iii)).\smallskip

(xii): Consider arbitrary $U\in\mathcal U(\mathcal H)$. Then $U^n\in\mathcal U(\mathcal H)$ as the latter forms a group (as shown in (ii)) so $\|U^n\|=1$ for all $n\in\mathbb N$ by (i). Thus $U$ is a power-bounded operator on a reflexive space (Coro.~\ref{coro_hilbert_reflexive}) so $U$ is mean ergodic by Prop.~\ref{prop_mean_ergodic} (ii).
\subsection{Theorem \ref{thm_Ebsigma_extreme}}\label{app_thm_minmax_matrix}
First we need the following technical lemma:

\begin{lemma}\label{lemma_minmax_matrix}
Let $A\in\{0,1\}^{m\times n}$ be a matrix such that
\begin{itemize}
\item[$\bullet$] $\operatorname{rank}(A)=n$.
\item[$\bullet$] $\unitvector^T$ is a row of $A$.
\item[$\bullet$] For any two rows $a_1^T$, $a_2^T$ of $A$ their minimum $\min\{a_1^T,a_2^T\}$ and maximum $\max\{a_1^T,a_2^T\}$ are rows of $A$ as well.
\end{itemize}
Then the following statements hold.
\begin{itemize}
\item[(i)] There exists a row $a^T$ of $A$ such that $\unitvector^Ta=n-1$.
\item[(ii)] For every row $a^T$ of $A$ with $\unitvector^Ta>1$ one finds a row $\tilde a^T$ of $A$ such that $\unitvector^T\tilde a=\unitvector^Ta-1$ and $\tilde a\leq a$.
\item[(iii)] There exist rows $a_1^T,\ldots,a_{n-1}^T$ of $A$ and a permutation ${\tau}\in S_n$ such that
\begin{equation}\label{eq:aj_rows_perm}
{\footnotesize\begin{pmatrix}
a_1^T\\\vdots\\a_{n-1}^T\\\unitvector^T
\end{pmatrix}}=
{\footnotesize\begin{pmatrix}
1&0&\cdots&0\\
\vdots&\ddots&\ddots&\vdots\\
\vdots&&\ddots&0\\
1&\cdots&\cdots&1
\end{pmatrix}}\underline{{\tau}}\,.
\end{equation}
\end{itemize}
\end{lemma}
\begin{proof}
(i): For all $j=1,\ldots,n$ define
$
S_j:=\{a^T\,|\,a^T\text{ is a row of }A\text{ and }e_j^Ta=0\}\setminus\{0\}
$
as the collection of all (non-zero) rows of $A$ the $j$-th entry of which vanishes\footnote{
It may happen that $A$ contains (at most, due to rank condition) one column of ones so (at most) one of the $S_j$ might be empty, but one can still guarantee the existence of some $j$ such that $S_j\neq\emptyset$ (because $n\geq 2$, the case $n=1$ is trivial). 
}. Defining $a_j^T:=\max S_j$ this is a row of $A$ (due to the maximum property) with $\unitvector^Ta_j\leq n-1$. It is obvious that a row $a^T$ of $A$ is in $S_j$ if and only if $a^T\leq a_j^T$; hence $a_j^T=a_k^T$ for any two $j,k$ implies $S_j=S_k$. Now there exists $k\in\{1,\ldots,n\}$ such that
\begin{equation}\label{eq:ak_aj_rel}
\unitvector^Ta_k=\max_{j=1,\ldots,n}\unitvector^Ta_j\quad(\geq 1)\,.
\end{equation}
If $\unitvector^Ta_k=n-1$ then we are done. Otherwise $\unitvector^Ta_k<n-1$ so one finds an index $i\in\{1,\ldots,n\}\setminus\{k\}$ such that $e_i^Ta_k=0$. Therefore $a_k\in S_i$ so $a_k^T\leq a_i^T$ but $\unitvector^Ta_k\geq \unitvector^Ta_i$ by \eqref{eq:ak_aj_rel}; this shows $a_k^T=a_i^T$ and thus $S_k=S_i$. We claim that now \textit{all} rows $a^T$ of $A$ satisfy $(e_i^T+e_k^T)a\in\{0,2\}$; then the linear span of all rows of $A$ has this property as well so it cannot contain $e_k^T$, contradicting $\operatorname{rank}(A)=n$. 

Indeed let $a^T$ be any row of $A$. If $a^T\in S_k=S_i$ then $e_i^Ta=e_k^Ta=0=e_i^Ta+e_k^Ta$. If $a^T\not\in S_k=S_i$ then $e_i^Ta=e_k^Ta=1$ so $e_i^Ta+e_k^Ta=2$.

(ii): We prove this via induction. The case $n=1$ is trivial. Now for $n\to n+1$ let $A\in\{0,1\}^{m\times(n+1)}$ with the above properties be given. Be aware of the following argument:
For ${\tau}_j={\footnotesize\begin{pmatrix} 1&2&\cdots&j-1&j&j+1&\cdots&n+1\\2&3&\cdots&j&1&j+1&\cdots&n+1 \end{pmatrix}}$ and all $j=1,\ldots,n+1$ the matrix
\begin{align*}
A_j:=A\underline{\tau_j}{\footnotesize\begin{pmatrix} 0&\cdots&0\\1&&0\\&\ddots&\\0&&1 \end{pmatrix}}\in\mathbb R^{m\times n}
\end{align*}
is the original matrix $A$ but without the $j$-th column. It is easy to see that $\operatorname{rank}(A_j)=n$ (follows from, e.g., \cite[Thm.~0.4.5.(c)]{HJ1}), $\unitvector^T\in A_j$ and the min-max condition for the rows of $A_j$ holds for all $j=1,\ldots,n+1$. Hence we may apply the induction hypothesis to any of these matrices $A_j$.

Now consider any row $a^T$ of $A$ with $\unitvector^Ta>1$. There are two cases which, once verified, conclude the proof of (ii).
\begin{itemize}
\item[Case 1:] $a^T=\unitvector^T$. By (i) we find $j\in\{1,\ldots,n+1\}$ such that $\unitvector^T-e_j^T$ is a row of $A$.
\item[Case 2:] $a^T\neq \unitvector^T$ so there exists $j\in\{1,\ldots,n+1\}$ such that $e_j^Ta=a_j=0$. Consider $A_j\in\mathbb R^{m\times n}$ and the truncated row $b^T$ corresponding to $a^T$. By induction hypothesis ($\unitvector^Tb=\unitvector^Ta>1$) we find $\tilde b^T\in A_j$ such that $\unitvector^T\tilde b=\unitvector^Tb-1$ and $\tilde b\leq b$. Now there exists a row $\hat a$ in $A$ which becomes $\tilde b$ when removing the $j$-th entry. Defining $\tilde a^T:=\min\{\hat a^T,a^T\}$ we know that this is a row of $A$ (min-max-property of $A$) and $\tilde a\leq a$ as well as $\unitvector^T\tilde a=\unitvector^T\tilde b=\unitvector^Tb-1=\unitvector^Ta-1$.
\end{itemize}

(iii): By assumption $\unitvector^T\in A$ so using (ii) $A$ contains some $a_{n-1}^T$ of row sum $n-1$, which in turn yields $a_{n-2}^T\in A$ of row sum $n-2$ with $a_{n-2}^T\leq a_{n-1}^T$ and so forth. Eventually one ends up with rows $a_1^T,\ldots,a_{n-1}^T$ of $A$ which satisfy $\unitvector^Ta_j=j$ for all $j=1,\ldots,n-1$ as well as $a_1^T\leq\ldots\leq a_{n-1}^T$; this readily implies the existence of a permutation ${\tau}\in S_n$ such that \eqref{eq:aj_rows_perm} holds.
\end{proof}
\begin{remark}\label{rem_extend_lemma_minmax}
With an analogous argument one can show that every such matrix $A$ contains a standard basis vector as a row and that for every row with $\unitvector^Ta<n$ one finds $\tilde a\in A$ with $\unitvector^T\tilde a=\unitvector^Ta+1$ and $\tilde a\geq a$. This suffices to prove that \textit{every} row of $A$ can be completed to a matrix of the form \eqref{eq:aj_rows_perm}.
\end{remark}

\begin{proof}[Proof of Thm.~\ref{thm_Ebsigma_extreme}]
Let $p\in \{x\in\mathbb R^n\,|\,Mx\leq b\}$ be extreme so by Lemma \ref{lemma_extreme_points_h_descr} there exists a submatrix $M'\in\mathbb R^{n\times n}$ of $M$ of full rank, one row of $M'$ being equal to $\unitvector^T$, such that $M'p=\mathfrak b(M')$. By Minkowski's theorem\index{theorem!Minkowski} \cite[Thm.~5.10]{Brondsted83} if we can show that $p=E_b({\pi})$ for some ${\pi}\in S_n$ then this theorem is proven.

Indeed consider any two rows $m_1^T,m_2^T$ of $M'$. Then the vectors $m_\textsf{min}^T:=\min\{m_1^T,m_2^T\}$, $m_\textsf{max}^T:=\max\{m_1^T,m_2^T\}$ satisfy
\begin{itemize}
\item $m_\textsf{min}^T+m_\textsf{max}^T=m_1^T+m_2^T\,$.
\item $m_\textsf{min}^T\leq m_\textsf{max}^T$, so one finds $\pi\in S_n$ such that $m_\textsf{min}^T$, $ m_\textsf{max}^T$ are rows of
$$
{\footnotesize\begin{pmatrix}
1&0&\cdots&0\\
\vdots&\ddots&\ddots&\vdots\\
\vdots&&\ddots&0\\
1&\cdots&\cdots&1
\end{pmatrix}}\underline{\pi}=:M_\pi\,.
$$
If $m_\textsf{min}^T=0$ then one can trivially find $\pi\in S_n$ such that $ m_\textsf{max}^T$ is a row of $M_\pi$.
\end{itemize}
This has two immediate consequences:
\begin{align*}
\mathfrak b(m_1^T)+\mathfrak b(m_2^T)&\overset{M'p=\mathfrak b(M')}=(m_1^T+m_2^T)p\\
&\overset{\hphantom{M'p=\mathfrak b(M')}}=(m_\textsf{min}^T+ m_\textsf{max}^T)p\overset{Mp\leq b}\leq \mathfrak b( m_\textsf{min}^T) +\mathfrak b(m_\textsf{max}^T)\,.
\end{align*}
as well as
\begin{align*}
\mathfrak b( m_\textsf{min}^T) +\mathfrak b(m_\textsf{max}^T)&\overset{M_\pi E_b(\pi)=b_\pi}=(m_\textsf{min}^T+ m_\textsf{max}^T)E_b(\pi)\\
&\overset{\hphantom{M_\pi E_b(\pi)=b_\pi}}=(m_1^T+m_2^T)E_b(\pi)\overset{ME_b(\pi)\leq b}\leq \mathfrak b(m_1^T)+\mathfrak b(m_2^T)
\end{align*}
(because $E_b(\pi)\in \{x\in\mathbb R^n\,|\,Mx\leq b\}$ by assumption). Combining these two we get
$$
\mathfrak b(m_1^T)+\mathfrak b(m_2^T)=(m_\textsf{min}^T+ m_\textsf{max}^T)p\leq \mathfrak b( m_\textsf{min}^T) +\mathfrak b(m_\textsf{max}^T)=\mathfrak b(m_1^T)+\mathfrak b(m_2^T)\,.
$$
i.e.~$(m_\textsf{min}^T+ m_\textsf{max}^T)p= \mathfrak b( m_\textsf{min}^T) +\mathfrak b(m_\textsf{max}^T)$. But we know $m_\textsf{min}^Tp\leq \mathfrak b( m_\textsf{min}^T)$, $ m_\textsf{max}^Tp\leq \mathfrak b(m_\textsf{max}^T)$ (due to $Mp\leq b$) so this implies
$$
m_\textsf{min}^Tp= \mathfrak b( m_\textsf{min}^T)\qquad\text{ and }\qquad m_\textsf{max}^Tp= \mathfrak b(m_\textsf{max}^T)\,.
$$
This is the key to finishing this proof as $p$ does not only satisfy $M'p=\mathfrak b(M')$ but
$$
\begin{pmatrix}
M'\\m_\textsf{min}^T\\m_\textsf{max}^T
\end{pmatrix}p=\mathfrak b
\begin{pmatrix}
M'\\m_\textsf{min}^T\\m_\textsf{max}^T
\end{pmatrix}
$$
meaning that taking \textit{any} two rows of $M'$ we may extend the matrix by their entrywise minimum and maximum and $p$ still satisfies $M'p=\mathfrak b(M')$ (now for the possibly enlarged $M'$). Of course if $m_\textsf{min}^T$ (or $m_\textsf{max}^T$) was already a row of $M'$ then we need not add it to $M'$.

Repeating this enlargement process over and over will terminate eventually: $M'$ can only grow but the set of possible rows is upper bounded by $\{0,1\}^n$, i.e.~by something finite. The final matrix $M'$ then is of full rank, contains $\unitvector^T$, and, most importantly, for any two rows $m_1^T$, $m_2^T$ of $M'$, $m_{\textsf{min}}^T,m_{\textsf{max}}^T$ are rows of $M'$ as well. Therefore by Lemma \ref{lemma_minmax_matrix} (iii) one finds a permutation ${\pi}\in S_n$ such that $p=E_b({\pi})$ (because $M'p=\mathfrak b(M')$) which concludes the proof.
\end{proof}

\section{Miscellaneous}\label{sec:ex}
\subsection{Appendix to Section \ref{sec:maj_d_vec}}

\begin{lemma}\label{lemma_extreme_points}
Let $d\in\mathbb R_{++}^3$ with $d_1>d_2>d_3$. 
\begin{itemize}
\item[(i)] If $d_1\geq d_2+d_3$, then the 10 extreme points $s_d(3)$ are given by
\begin{align*}
\mathbbm{1}_3\quad\begin{pmatrix} 1&0&0\\0&1-\frac{d_3}{d_2}&1\\0&\frac{d_3}{d_2}&0 \end{pmatrix}\quad  &\begin{pmatrix} 1-\frac{d_3}{d_1}&0&1\\0&1&0\\\frac{d_3}{d_1}&0&0 \end{pmatrix}\quad\begin{pmatrix} 1-\frac{d_2}{d_1}&1&0\\\frac{d_2-d_3}{d_1}&0&1\\\frac{d_3}{d_1}&0&0 \end{pmatrix}\\
\begin{pmatrix} 1-\frac{d_3}{d_1}&\frac{d_3}{d_2}&0\\0&1-\frac{d_3}{d_2}&1\\\frac{d_3}{d_1}&0&0 \end{pmatrix}\quad&\begin{pmatrix} 1-\frac{d_2}{d_1}&1&0\\\frac{d_2}{d_1}&0&0\\0&0&1 \end{pmatrix}\quad\begin{pmatrix} 1-\frac{d_3}{d_1}&0&1\\\frac{d_3}{d_1}&1-\frac{d_3}{d_2}&0\\0&\frac{d_3}{d_2}&0 \end{pmatrix}\\
\begin{pmatrix} 1-\frac{d_2-d_3}{d_1}&1-\frac{d_3}{d_2}&0\\\frac{d_2-d_3}{d_1}&0&1\\0&\frac{d_3}{d_2}&0 \end{pmatrix}\quad&\begin{pmatrix} 1-\frac{d_2}{d_1}&1-\frac{d_3}{d_2}&1\\\frac{d_2}{d_1}&0&0\\0&\frac{d_3}{d_2}&0 \end{pmatrix}\quad\begin{pmatrix} 1-\frac{d_2+d_3}{d_1}&1&1\\\frac{d_2}{d_1}&0&0\\\frac{d_3}{d_1}&0&0 \end{pmatrix}
\end{align*}
\item[(ii)] If $d_1< d_2+d_3$, then the 13 extreme points $s_d(3)$ are given by
\begin{align*}
\mathbbm{1}_3\quad\begin{pmatrix} 1&0&0\\0&1-\frac{d_3}{d_2}&1\\0&\frac{d_3}{d_2}&0 \end{pmatrix}\quad  &\begin{pmatrix} 1-\frac{d_3}{d_1}&0&1\\0&1&0\\\frac{d_3}{d_1}&0&0 \end{pmatrix}\quad\begin{pmatrix} 1-\frac{d_2}{d_1}&1&0\\\frac{d_2-d_3}{d_1}&0&1\\\frac{d_3}{d_1}&0&0 \end{pmatrix}\\
\begin{pmatrix} 1-\frac{d_3}{d_1}&\frac{d_3}{d_2}&0\\0&1-\frac{d_3}{d_2}&1\\\frac{d_3}{d_1}&0&0 \end{pmatrix}\quad&\begin{pmatrix} 1-\frac{d_2}{d_1}&1&0\\\frac{d_2}{d_1}&0&0\\0&0&1 \end{pmatrix}\quad\begin{pmatrix} 1-\frac{d_3}{d_1}&0&1\\\frac{d_3}{d_1}&1-\frac{d_3}{d_2}&0\\0&\frac{d_3}{d_2}&0 \end{pmatrix}\\
\begin{pmatrix} 1-\frac{d_2-d_3}{d_1}&1-\frac{d_3}{d_2}&0\\\frac{d_2-d_3}{d_1}&0&1\\0&\frac{d_3}{d_2}&0 \end{pmatrix}\quad&\begin{pmatrix} 1-\frac{d_2}{d_1}&1-\frac{d_3}{d_2}&1\\\frac{d_2}{d_1}&0&0\\0&\frac{d_3}{d_2}&0 \end{pmatrix}\quad\begin{pmatrix} 0&1&\frac{d_1-d_2}{d_3}\\\frac{d_2}{d_1}&0&0\\1-\frac{d_2}{d_1}&0&1-\frac{d_1-d_2}{d_3} \end{pmatrix}\\
\begin{pmatrix} 0&\frac{d_1-d_3}{d_2}&1\\1-\frac{d_3}{d_1}&1-\frac{d_1-d_3}{d_2}&0\\\frac{d_3}{d_1}&0&0 \end{pmatrix}\quad&\begin{pmatrix} 0&\frac{d_1-d_3}{d_2}&1\\ \frac{d_2}{d_1} & 0 & 0 \\ 1-\frac{d_2}{d_1} &1-\frac{d_1-d_3}{d_2}&0 \end{pmatrix}\quad\begin{pmatrix} 0&1& \frac{d_1-d_2}{d_3} \\ 1-\frac{d_3}{d_1} &0&  1-\frac{d_1-d_2}{d_3}\\ \frac{d_3}{d_1} &0&0 \end{pmatrix}
\end{align*}
\end{itemize}
\end{lemma}
\begin{proof}
The respective number of extreme points is stated in \cite[Remark 4.5]{Joe90} or, more recently, \cite[Ch.~IV]{Mazurek18}. Then one only has to verify that the above matrices (under the given assumptions) are in fact extremal in $s_d(3)$.
\end{proof}
Once we allow components of $d$ to coincide, the above extreme points simplify slightly (as already observed in \cite[Remark 4.5]{Joe90}). Within the setting of (i) if $d_2=d_3$ then one is left with 7 extreme points. For (ii) if either $d_1=d_2$ or $d_2=d_3$ then one has 10 extreme points and if $d_1=d_2=d_3$ then there are $6$ extreme points---namely the $3\times 3$ permutation matrices---which recovers Birkhoff's theorem, cf.~\cite[Thm.~2.A.2]{MarshallOlkin}.\medskip

We follow up with a rather general result on the comparison of convex or concave functions:
\begin{lemma}\label{lemma_convex_cpl_compare}
Let $a,b\in\mathbb R$ with $a<b$, and functions $f,g:[a,b]\to\mathbb R$ be given. Assume that $g$ is continuous piecewise linear, and let $a_1<\ldots<a_n$ denote the points in $(a,b)$ where $g$ changes slope. Defining $a_0:=a$, $a_{n+1}:=b$, the following statements hold:
\begin{itemize}
\item[(i)] If $f$ is concave, then $f(c)\geq g(c)$ for all $c\in[a,b]$ holds if and only if $f(a_i)\geq g(a_i)$ for all $i=0,\ldots,n+1$.\smallskip
\item[(ii)] If $f$ is convex, then $f(c)\leq g(c)$ for all $c\in[a,b]$ if and only if $f(a_i)\leq g(a_i)$ for all $i=0,\ldots,n+1$.
\end{itemize}
\end{lemma}
\begin{proof}
The only non-trivial statement is ``(i), $\Leftarrow$''; its converse is obvious and (ii) directly follows from (i) by changing $f,g$ to $-f,-g$.

``(i), $\Leftarrow$'': Assume to the contrary that there exists $c\in[a,b]$ where $f(c)<g(c)$. By assumption $c\not\in\{a,a_1,\ldots,a_n,b\}$ so one finds a unique index $i=0,\ldots,n$ such that $a_i<c<a_{i+1}$. Using concavity of $f$ and setting $\lambda:=\frac{a_{i+1}-c}{a_{i+1}-a_i}\in(0,1)$ we get
\begin{align*}
g(c)>f(c)=f(\lambda a_i+(1-\lambda)a_{i+1})&\geq \lambda f(a_i)+(1-\lambda)f(a_{i+1})\\
&\geq \lambda g(a_i)+(1-\lambda)g(a_{i+1})\\
&=g(\lambda a_i+(1-\lambda)a_{i+1})=g(c)\,.
\end{align*}
In the second-to-last step we used that $g$ is affine linear on $[a_i,a_{i+1}]$. Thus we arrived at the contradiction $g(c)>g(c)$ which concludes the proof.
\end{proof}

Next let us collect further auxiliary lemmata. For this section 
\begin{itemize}
\item $\operatorname{ext}(S)$ denotes the set of extreme points of an arbitrary convex compact set $S\subset\mathbb R^n$.
\item $P_A(b)$ will be short for $\{x\in\mathbb R^n\,|\,Ax\leq b\}$.
\item Given matrices $A,A_0$ then $A_0\subset A$ means ``Every row of $A_0$ is also a row of $A$''.
\item $\textsf{GL}(n,\mathbb R)$ as usual denotes the invertible real $n\times n$ matrices.
\end{itemize}
Also keep in mind that (cf.~\cite[Thm.~8.4 ff.]{Schrijver86})
\begin{equation}\label{eq:poly_corner_submat}
\operatorname{ext}(P_A(b))= \{A_0^{-1}\mathfrak b(A_0) \,|\,A_0\subset A, A_0\in\textsf{GL}(n,\mathbb R)\} \cap P_A(b)
\end{equation}
with $\mathfrak b(\cdot)$ the map from Def.~\ref{def_mathfrak_b}.

\begin{lemma}\label{lemma_cont_path_extreme}
Let $m,n\in\mathbb N$, $A\in\mathbb R^{m\times n}$ with $P_A(0)=\{0\}$, $b\in\mathbb R^m$, and $j\in\{1,\ldots,m\}$ be given. Moreover let $a_j^T$ denote the $j$-th row of $A$ and $b_j=\mathfrak b(a_j^T)$ the corresponding entry of $b$. The following statements hold.
\begin{itemize}
\item[(i)] Let $x\in \operatorname{ext}(P_A(b))$ such that $a_j^Tx=b_j$ and assume $P_A(b-\xi e_j)\neq\emptyset$ for some $\xi>0$. Then there exists $y\in\operatorname{ext}(P_A(b))$ with $a_j^Ty<b_j$ such that
\begin{equation}\label{eq:x_t_convex}
x(t):=\frac{t}{b_j-a_j^Ty}y+\Big(1-\frac{t}{b_j-a_j^Ty}\Big)x\in\operatorname{ext}(P_A(b-te_j))
\end{equation}
for all $t\in[0,s]$ where $s:=b_j-a_j^Ty>0$.
\item[(ii)] Let $\lambda<0$ be given such that $P_A(b+\lambda e_j)\neq\emptyset$. For every $y\in \operatorname{ext}(P_A(b))$ there exist $k\in\mathbb N$
, $A_1,\ldots,A_k$ ($\in \textsf{GL}(n,\mathbb R)$, $\subset A$), and $0=t_0<t_1<\ldots<t_k=1$ such that
\begin{equation}\label{eq:cont_path_gamma}
\gamma:[0,1]\to\mathbb R^n\qquad \gamma(t):=\begin{cases}
 A_1^{-1}\mathfrak b_t(A_1) &t\in[0,t_1]\\
 A_2^{-1}\mathfrak b_t(A_2) &t\in[t_1,t_2]\\
 \vdots&\vdots\\
 A_{k}^{-1}\mathfrak b_t(A_{k}) &t\in[t_{k-1},1]
\end{cases}
\end{equation}
is well-defined (i.e.~$A_\ell^{-1}\mathfrak b_{t_\ell}(A_\ell)=A_{\ell+1}^{-1}\mathfrak b_{t_\ell}(A_{\ell+1})$ for all $\ell=1,\ldots,k-1$) and satisfies $\gamma(0)=y$ as well as $\gamma(t)\in P_A(b+\lambda te_j)$ for all $t\in[0,1]$. Here $\mathfrak b_t$ is the usual map $\mathfrak b$ from Def.~\ref{def_mathfrak_b} but with respect to the $t$-dependent inequality $Ax\leq b+t\lambda e_j$.
\end{itemize}
\end{lemma}
\begin{proof}
(i): Consider any $x\in \operatorname{ext}(P_A(b))$ with $a_j^Tx=b_j$. Because $P_A(b-\xi e_j)\neq\emptyset$ for some $\xi>0$ there certainly exist $z\in P_A(b)$ with $a_j^Tz<b_j$ so, because $x$ is extremal, one can find an edge $e$ of $P_A(b)$ such that $x\in e$ but $e\not\subset\{z\in\mathbb R^n\,|\,a_j^Tz=b_j\}$. Because $P_A(b)$ is bounded (due to $P_A(0)=\{0\}$, cf.~\cite[Ch.~8.2]{Schrijver86}) $e$ is finite so in particular it is the convex hull of $x$ and some $y\in\operatorname{ext}(P_A(b))\setminus\{x\}$. This $y$ satisfies $a_j^Ty<b_j$ (else $e$ would be in $\{z\in\mathbb R^n\,|\,a_j^Tz=b_j\}$) and, by convexity, $x(t)$ from \eqref{eq:x_t_convex} is in $ P_A(b)$ for all $t\in[0,s]$. Moreover $a_j^Tx(t)=b_j-t$ as is readily verified so one even has $x(t)\in P_A(b-te_j)$ for all $t\in[0,s]$. 

Now $e$, just like every edge of a convex polytope, is characterized by $n-1$ linearly independent rows $\tilde a^T_1,\ldots,\tilde a^T_{n-1}$ from $A$ in the sense that \cite[Ch.~8.7]{Schrijver86}
$$
e=\{x\in P_A(b)\,|\,\tilde a^T_ix=b_i\text{ for all }i=1,\ldots,n-1\}\supseteq \{x(t)\,|\,t\in[0,s]\}\,.
$$
Because $x,y\in e$ and $a_j^Tx=b_j$ but $a_j^Ty<b_j$ the matrix $A_0$ consisting of $\tilde a^T_1,\ldots,\tilde a^T_{n-1},a_j^T$ is of full rank and $A_0x(t)=\mathfrak b_t(A_0)$. This---again by \eqref{eq:poly_corner_submat}---shows $x(t)\in\operatorname{ext}(P_A(b-te_j))$ for all $t\in[0,s]$ as claimed.

(ii): Let $y\in\operatorname{ext}(P_A(b))$ so by \eqref{eq:poly_corner_submat} there exists $A_0\subset A, A_0\in\textsf{GL}(n,\mathbb R)$ such that $y=A_0^{-1}\mathfrak b(A_0)$. If $a_j^Ty<b_j$ then $A_0^{-1}\mathfrak b(A_0)\in P_A(b+t\lambda e_j)$ for all $t\in[0,\frac{a_j^Ty-b_j}{\lambda}]$ and thus $A_0^{-1}\mathfrak b(A_0)$ is extremal for all those $t$ because determined by a submatrix of $A$ which does not contain $a_j^T$. Moreover for $t=\frac{a_j^Ty-b_j}{\lambda}$ one has $a_j^T(A_0^{-1}\mathfrak b_t(A_0))=a_j^Ty=b_j+t\lambda$. 

Thus we may assume w.l.o.g.~that $a_j^Ty=b_j$ from the start. As shown in the proof of (i) one finds $y_1\in\operatorname{ext}(P_A(b))$ with $a_j^Ty_1<b_j$ as well as submatrix $A_1\in\textsf{GL}(n,\mathbb R)$ of $A$ such that
$$
\frac{\lambda t}{a_j^Ty_1-b_j}y_1+\Big(1-\frac{\lambda t}{a_j^Ty_1-b_j}\Big)y=A_1^{-1}\mathfrak b_t(A_1)\in P_A(b+t\lambda e_j)
$$
for all $t\in[0,t_1]$ where $t_1:=\frac{a_j^Ty_1-b_j}{\lambda}>0$. If $t_1\geq 1$ then we are done. Otherwise repeat the process, i.e.~find another corner $y_2$ of $P_A(b)$ such that $a_j^Ty_2<b_j+t_1\lambda$ and a corresponding invertible submatrix $A_2$ of $A$ such that $\operatorname{conv}\{y_1,y_2\}=\{A_2^{-1}\mathfrak b_t(A_2)\,|\,t\in [t_1,t_2]\}$ with $t_2:=\frac{a_j^Ty_2-b_j}{\lambda}>0$ (this is always possible by (i) because $P_A(b+\lambda e_j)\neq\emptyset$). Note that this process has to be repeated at most finitely many times because $|\operatorname{ext}(P_A(b))|<\infty$ and no element of $\operatorname{ext}(P_A(b))$ can be re-used as $a_j^Ty_i=b_j+t_i\lambda>b_j+t_{i+1}\lambda$. This concludes the proof.
%
%
%
\end{proof}

\begin{lemma}\label{lemma_cont_in_b}
Let $m,n\in\mathbb N$, $A\in\mathbb R^{m\times n}$, and $b,b'\in\mathbb R^m$ be given such that $P_A(0)=\{0\}$ and $P_A(b),P_A(b')\neq\emptyset$. Then
$$
\Delta(P_A(b),P_A(b'))\leq C\|b-b'\|_1
$$
where the constant
\begin{equation}\label{eq:C_A_Lipschitz}
C=\max_{\{A_0\in \textsf{GL}(n,\mathbb R)\,|\,A_0\subset A\}}\|A_0^{-1}\|_{1\to 1}\,.
\end{equation}
is independent of $b$. Here $\|\cdot\|_{1\to 1}$ is the operator norm on $(\mathbb R^n,\|\cdot\|_1)$, i.e.~the column sum norm of the respective matrix.
\end{lemma}
\begin{proof}
Let us subdivide the proof into the following four steps.

\noindent\textit{Step 1:} Let $\lambda<0$, $j\in\{1,\ldots,m\}$ such that $P_A( b+\lambda e_j)\neq\emptyset$. For all $y\in\operatorname{ext}(P_A(b))$ there exists $z\in\operatorname{ext}(P_A( b+\lambda e_j))$ such that $\|y-z\|_1\leq C|\lambda|$.

By Lemma \ref{lemma_cont_path_extreme} one finds a path $\gamma:[0,1]\to\mathbb R^n$ of form \eqref{eq:cont_path_gamma} such that $\gamma(0)=y$ and $\gamma(t)\in\operatorname{ext}(P_A(b+\lambda te_j))$ for all $t\in[0,1]$. Thus $z:=\gamma(1)\in \operatorname{ext}(P_A(b+\lambda e_j))$ satisfies
\begin{align*}
\|y-z\|_1&=\|A_1^{-1}\mathfrak b_0(A_1)-A_k^{-1} \mathfrak b_1(A_k)\|_1\\
&\leq\sum\nolimits_{\ell=1}^k \|A_\ell^{-1}\mathfrak b_{t_{\ell-1}}(A_\ell)-A_\ell^{-1}\mathfrak b_{t_\ell}(A_\ell)\|_1\\
&\leq \sum\nolimits_{\ell=1}^k \underbrace{\|A_\ell^{-1}\|_{1\to 1}}_{\leq C}\|\mathfrak b_{t_{\ell-1}}(A_\ell)-\mathfrak b_{t_\ell}(A_\ell)\|_1\\
&\leq C\sum\nolimits_{\ell=1}^k\|(b+\lambda t_{\ell-1}e_j)-(b+\lambda t_{\ell}e_j)\|_1=C|\lambda|\underbrace{\sum\nolimits_{\ell=1}^k (t_{\ell}-t_{\ell-1})}_{=t_k-t_0=1}=C|\lambda|\,.
\end{align*}

\noindent\textit{Step 2:} Given $\lambda\in\mathbb R$, $j\in\{1,\ldots,m\}$ such that $P_A(b),P_A( b+\lambda e_j)\neq\emptyset$ one has
$
\Delta(P_A(b),P_A( b+\lambda e_j))\leq C|\lambda|
$.

W.l.o.g.\footnote{
The case $\lambda=0$ is trivial. If $\lambda>0$ define $\tilde\lambda:=-\lambda<0$ and $\tilde b:=b+\lambda e_j$ as then $\Delta(P_A(b),P_A(b+\lambda e_j))=\Delta(P_A(\tilde b+\tilde\lambda e_j),P_A(\tilde b))=\Delta(P_A(\tilde b),P_A(\tilde b+\tilde\lambda e_j))$.
}~$\lambda<0$ so $b+\lambda e_j\leq b$ which implies $P_A(b+\lambda e_j)\subseteq P_A(b)$ (cf.~Remark \ref{rem:halfspace_actions}). Thus given $z\in P_A(b)$ it suffices to find $y\in P_A(b+\lambda e_j)$ such that $\|z-y\|_1\leq C|\lambda |$ (by definition of the Hausdorff metric). Indeed for such $z$ one finds $\lambda_1,\ldots,\lambda_{n+1}\in [0,1]$ as well as $z_1,\ldots,z_{n+1}\in\operatorname{ext}(P_A(b))$ such that $z=\sum_{j=1}^{n+1}\lambda_jz_j$. As shown in Step 1 for every $z_j$ one finds $y_j\in \operatorname{ext}(P_A(b+\lambda e_j))$ with $\|z_j-y_j\|_1\leq C|\lambda|$ which for $y:=\sum_{j=1}^{n+1} \lambda_jy_j\in P_A(b+\lambda e_j)$ shows
$$
\|z-y\|_1\leq\sum\nolimits_{j=1}^{n+1} \lambda_j\|z_j-y_j\|_1\leq C|\lambda|\sum\nolimits_{j=1}^{n+1} \lambda_j=C|\lambda|\,.
$$
Here we used Carath\'{e}odory's theorem\index{theorem!Carath{\'e}odory's} as well as convexity and compactness of $P_A(b),P_A(b+\lambda e_j)$ (boundedness comes from $P_A(0)=\{0\}$, cf.~\cite[Ch.~8.2]{Schrijver86}).

\noindent\textit{Step 3:} If $b\leq b'$ then $\Delta(P_A(b),P_A(b'))\leq C\cdot\|b-b'\|_1$.

We recursively define vectors $b^{(0)},\ldots,b^{(m)}\in\mathbb R^m$ as follows:
\begin{align*}
b^{(0)}&:=b\\
b^{(j)}&:=b^{(j-1)}+(b_j'-b_j)e_j\qquad\text{ for all }j=1,\ldots,m
\end{align*}
so $b^{(m)}=b'$ and $b^{(0)}\leq b^{(1)}\leq\ldots\leq b^{(m)}$ which implies
$$
\emptyset\neq P_A(b)= P_A(b^{( 0 )})\subseteq P_A(b^{( 1 )})\subseteq\ldots\subseteq P_A(b^{( m)})=P_A(b')\,.
$$
Now Step 2 together with the triangle inequality yields
\begin{align*}
\Delta(P_A(b),P_A(b'))&\leq\sum\nolimits_{j=1}^{m}\Delta(P_A(b^{(j-1)}),P_A(b^{(j)}))\\
&= \sum\nolimits_{j=1}^{m}\Delta(P_A(b^{(j-1)}),P_A(b^{(j-1)}+(b_j'-b_j)e_j))\\
&\leq \sum\nolimits_{j=1}^{m} C|b_j'-b_j|=C\|b-b'\|_1\,.
\end{align*}

\noindent\textit{Step 4:} 
Given $b,\tilde b\in\mathbb R^m$ define $b':=\max\{b,\tilde b\}$ so $\emptyset\neq \{P_A(b),P_A(\tilde b)\}\subseteq P_A(b')$ and again by the triangle inequality
\begin{align*}
\Delta(P_A(b),P_A(\tilde b))&\leq\Delta(P_A(b),P_A(b'))+\Delta(P_A(b'),P_A(\tilde b))\leq C(\|b-b'\|_1+\|b'-\tilde b\|_1)\\
&=C\Big(\sum_{\{j\,|\,b_j<\tilde b_j\}}|b_j-\tilde b_j|+\sum_{\{j\,|\,b_j>\tilde b_j\}}|\tilde b_j-b_j|\Big)=C\|b-\tilde b\|_1\,.\qedhere
\end{align*}
\end{proof}
The Lipschitz-type constant $C$ from \eqref{eq:C_A_Lipschitz} might be related (at least in terms of idea) to the condition number of a matrix. 
\begin{remark}
For the matrix $M$ from \eqref{eq:M_maj} one gets $C=2$ in two dimensions and $C=3$ in three dimensions. Thus it seems reasonable to conjecture $C=C(M,n)=n$ for all $n\in\mathbb N$---however this is way beyond the scope or interest of this thesis and we will not pursue this question further here.
\end{remark}

\begin{lemma}\label{lemma_maj_sum_recovery}
Let $y\in\mathbb R^n$, $d\in\mathbb R_{++}^n$ be arbitrary and consider ${\pi}\in S_n$ which satisfies\footnote{Obviously such a permutation ${\pi}$ always exists as it is just the decreasing ordering of the vector $\frac{y}{d}:=(\frac{y_i}{d_i})_{i=1}^n$.}
\begin{equation}\label{eq:sigma_ordering_yd}
\frac{y_{{\pi}(1)}}{d_{{\pi}(1)}}\geq \frac{y_{{\pi}(2)}}{d_{{\pi}(2)}}\geq\ldots\geq\frac{y_{{\pi}(n)}}{d_{{\pi}(n)}}\,.
\end{equation}
Then the following statements hold.
\begin{itemize}
\item[(i)] For all $c\in\mathbb R$, $k=1,\ldots,n$. 
$$
\unitvector^T\Big( y-\frac{y_{{\pi}(k)}}{d_{{\pi}(k)}} d\Big)_++\frac{y_{{\pi}(k)}}{d_{{\pi}(k)}} c=\frac{y_{{\pi}(1)}}{d_{{\pi}(1)}} c-\sum_{i=1}^{k-1}\Big(\frac{y_{{\pi}(i)}}{d_{{\pi}(i)}}-\frac{y_{{\pi}(i+1)}}{d_{{\pi}(i+1)}}\Big)\Big( c-\sum_{j=1}^i d_{{\pi}(j)} \Big)\,.
$$
\item[(ii)] Let $c\in(0,\unitvector^Td]$ be arbitrary. Then there exists unique $k\in\{1,\ldots,n\}$ such that $c-\sum_{i=1}^{k-1}d_{{\pi}(i)}> 0$ but $c-\sum_{i=1}^k d_{{\pi}(i)}\leq 0$. This $k$ satisfies
$$
\min_{i=1,\ldots,n}\Big(\unitvector^T\Big( y-\frac{y_{i}}{d_{i}} d\Big)_++\frac{y_{i}}{d_{i}} c\Big)=\Big(\sum\nolimits_{i=1}^{k-1}y_{{\pi}(i)}\Big)+\frac{y_{{\pi}(k)}}{d_{{\pi}(k)}} \Big(c-\sum\nolimits_{i=1}^{k-1} d_{{\pi}(i)}\Big) \,.
$$
\item[(iii)] If $d=\unitvector$ then for all $k=1,\ldots,n$ 
$$
\min_{i=1,\ldots,n} \unitvector^T(y-y_i\unitvector)_++ky_i=\sum\nolimits_{i=1}^k y_i^\downarrow\,.
$$
\end{itemize}
\end{lemma}
\begin{proof}
(i): This identity comes from
$$
\frac{y_{{\pi}(1)}}{d_{{\pi}(1)}} c-\sum_{i=1}^{k-1}\Big(\frac{y_{{\pi}(i)}}{d_{{\pi}(i)}}-\frac{y_{{\pi}(i+1)}}{d_{{\pi}(i+1)}}\Big)c=\frac{y_{{\pi}(1)}}{d_{{\pi}(1)}} c-\frac{y_{{\pi}(1)}}{d_{{\pi}(1)}} c+\frac{y_{{\pi}(k)}}{d_{{\pi}(k)}} c=\frac{y_{{\pi}(k)}}{d_{{\pi}(k)}} c
$$
as well as
\begin{align*}
\unitvector^T\Big( y-\frac{y_{{\pi}(k)}}{d_{{\pi}(k)}} d\Big)_+&=\sum_{j=1}^{k-1}\Big(\frac{y_{{\pi}(j)}}{d_{{\pi}(j)}}-\frac{y_{{\pi}(k)}}{d_{{\pi}(k)}}\Big)d_{{\pi}(j)}\\
&=\sum_{j=1}^{k-1}\sum_{i=j}^{k-1}\Big(\frac{y_{{\pi}(i)}}{d_{{\pi}(i)}}-\frac{y_{{\pi}(i+1)}}{d_{{\pi}(i+1)}}\Big)d_{{\pi}(j)}=\sum_{i=1}^{k-1}\sum_{j=1}^{i}\Big(\frac{y_{{\pi}(i)}}{d_{{\pi}(i)}}-\frac{y_{{\pi}(i+1)}}{d_{{\pi}(i+1)}}\Big)d_{{\pi}(j)}
\end{align*}
where in the last step we just changed the way how to enumerate the index set $\{(i,j)\,|\,1\leq j\leq i\leq k-1\}$. 

(ii): Using (i)
\begin{align*}
\min_{i=1,\ldots,n}\unitvector^T\Big( y-\frac{y_{i}}{d_{i}} d\Big)_++&\frac{y_{i}}{d_{i}} c=\min_{\ell=1,\ldots,n}\unitvector^T\Big( y-\frac{y_{{\pi}(\ell)}}{d_{{\pi}(\ell)}} d\Big)_++\frac{y_{{\pi}(\ell)}}{d_{{\pi}(\ell)}} c\\
&=\min_{\ell=1,\ldots,n}\frac{y_{{\pi}(1)}}{d_{{\pi}(1)}} c-\sum_{i=1}^{\ell-1}\Big(\frac{y_{{\pi}(i)}}{d_{{\pi}(i)}}-\frac{y_{{\pi}(i+1)}}{d_{{\pi}(i+1)}}\Big)\Big( c-\sum_{j=1}^i d_{{\pi}(j)} \Big)\\
&=\frac{y_{{\pi}(1)}}{d_{{\pi}(1)}} c-\max_{\ell=1,\ldots,n}\sum_{i=1}^{\ell-1}\Big(\frac{y_{{\pi}(i)}}{d_{{\pi}(i)}}-\frac{y_{{\pi}(i+1)}}{d_{{\pi}(i+1)}}\Big)\Big( c-\sum_{j=1}^i d_{{\pi}(j)} \Big)\,.
\end{align*}
There are two important things to notice here: The expression $\frac{y_{{\pi}(i)}}{d_{{\pi}(i)}}-\frac{y_{{\pi}(i+1)}}{d_{{\pi}(i+1)}}$ is always non-negative by \eqref{eq:sigma_ordering_yd} and, moreover, the map 
$$
f:\{0,\ldots,n\}\to \mathbb R\qquad i\mapsto c-\sum\nolimits_{j=1}^i d_{{\pi}(j)}
$$
satisfies $f(0)=c>0$, $f(n)=c-\unitvector^Td\leq 0$ and is strictly monotonically decreasing. Thus the index $k$ described above exists, is unique, and we get
$$
\max_{\ell=1,\ldots,n}\sum_{i=1}^{\ell-1}\Big(\frac{y_{{\pi}(i)}}{d_{{\pi}(i)}}-\frac{y_{{\pi}(i+1)}}{d_{{\pi}(i+1)}}\Big)\Big( c-\sum_{j=1}^i d_{{\pi}(j)} \Big)=\sum_{i=1}^{k-1}\Big(\frac{y_{{\pi}(i)}}{d_{{\pi}(i)}}-\frac{y_{{\pi}(i+1)}}{d_{{\pi}(i+1)}}\Big)\Big( c-\sum_{j=1}^i d_{{\pi}(j)} \Big)
$$
which shows
\begin{align*}
\min_{i=1,\ldots,n}\unitvector^T\Big( y-\frac{y_{i}}{d_{i}} d\Big)_++\frac{y_{i}}{d_{i}} c
&=\frac{y_{{\pi}(1)}}{d_{{\pi}(1)}} c-\sum\nolimits_{i=1}^{k-1}\Big(\frac{y_{{\pi}(i)}}{d_{{\pi}(i)}}-\frac{y_{{\pi}(i+1)}}{d_{{\pi}(i+1)}}\Big)\Big( c-\sum\nolimits_{j=1}^i d_{{\pi}(j)} \Big)\\
&\overset{\text{(i)}}=\unitvector^T\Big( y-\frac{y_{{\pi}(k)}}{d_{{\pi}(k)}} d\Big)_++\frac{y_{{\pi}(k)}}{d_{{\pi}(k)}} c\\
&= \sum\nolimits_{i=1}^{k-1}d_{{\pi}(i)} \Big( \frac{y_{{\pi}(i)}}{d_{{\pi}(i)}}-\frac{y_{{\pi}(k)}}{d_{{\pi}(k)}} \Big)+\frac{y_{{\pi}(k)}}{d_{{\pi}(k)}} c \\
&= \Big(\sum\nolimits_{i=1}^{k-1}y_{{\pi}(i)}\Big)+\frac{y_{{\pi}(k)}}{d_{{\pi}(k)}} \Big(c-\sum\nolimits_{i=1}^{k-1} d_{{\pi}(i)}\Big)\,.
\end{align*}

(iii): Direct consequence of (ii).
\end{proof}
\begin{lemma}\label{lemma_d_ordering_maj}
Let $d\in\mathbb R_{++}^n$, $k=1,\ldots,n-1$, pairwise different $\alpha_1,\ldots,\alpha_k\in\{1,\ldots,n\}$, and $\tau\in S_n$ be given. Then
$$
v:=\begin{pmatrix} \sum_{i=1}^{\alpha_1-1}d_{\tau(i)}\\\vdots\\\sum_{i=1}^{\alpha_k-1}d_{\tau(i)}\\\sum_{i=1}^kd_{\tau(\alpha_i)} \end{pmatrix}\prec\begin{pmatrix} \sum_{i=1}^{\alpha_1}d_{\tau(i)}\\\vdots\\\sum_{i=1}^{\alpha_k}d_{\tau(i)}\\0 \end{pmatrix}=:w\in\mathbb R^{k+1}\,.
$$
\end{lemma}
\begin{proof}
W.l.o.g.~$\alpha_1>\ldots>\alpha_k$; reordering the $\alpha_i$ amounts to reordering $v,w$ but classical majorization is permutation invariant. We know $v\prec w$ is equivalent to the partial sum conditions $\sum_{i=1}^\ell v_i^\downarrow\leq\sum_{i=1}^\ell w_i^\downarrow$ for all $\ell=1,\ldots,k$ together with $\unitvector^Tv=\unitvector^Tw$ (the latter is readily verified). Because the $\alpha_i$ are ordered one finds unique $\xi\in\{1,\ldots,k+1\}$ such that
$$
\sum\nolimits_{i=1}^{\alpha_\xi-1}d_{\tau(i)}< \sum\nolimits_{i=1}^kd_{\tau(\alpha_i)}\leq \sum\nolimits_{i=1}^{\alpha_{\xi-1}-1}d_{\tau(i)}
$$
(where $\alpha_0:=n+1$ and $\alpha_{k+1}:=0$). Thus $v\prec w$ is equivalent to
$$
v^\downarrow= \begin{pmatrix} \sum_{i=1}^{\alpha_1-1}d_{\tau(i)}\\\vdots\\ \sum_{i=1}^{\alpha_{\xi-1}-1}d_{\tau(i)} \\ \sum_{i=1}^kd_{\tau(\alpha_i)} \\ \sum_{i=1}^{\alpha_\xi-1}d_{\tau(i)}\\\vdots \\\sum_{i=1}^{\alpha_k-1}d_{\tau(i)} \\\sum_{i=1}^{\alpha_{k-1}-1}d_{\tau(i)} \end{pmatrix} 
\prec
\begin{pmatrix} \sum_{i=1}^{\alpha_1}d_{\tau(i)}\\\vdots\\ \sum_{i=1}^{\alpha_{\xi-1}}d_{\tau(i)}\\ \sum_{i=1}^{\alpha_\xi}d_{\tau(i)} \\ \sum_{i=1}^{\alpha_{\xi+1}}d_{\tau(i)}\\\vdots\\ \sum_{i=1}^{\alpha_k}d_{\tau(i)}\\0 \end{pmatrix} =w^\downarrow
$$
The first $\xi-1$ partial sum conditions are evident (because $v^\downarrow_j\leq w^\downarrow_j$ for all $j=1,\ldots,\xi-1$ individually). Consider any $\ell\in\{\xi,\ldots,k\}$. Then
\begin{align*}
\sum_{j=1}^\ell w_j^\downarrow-\sum_{j=1}^\ell v_j^\downarrow&=\sum_{j=1}^\ell \Big(\sum_{i=1}^{\alpha_j}d_{\tau(i)}\Big)-\sum_{j=1}^{\ell-1}\Big(\sum_{i=1}^{\alpha_j-1}d_{\tau(i)}\Big)-\sum_{i=1}^kd_{\tau(\alpha_i)}\\
&= \sum_{i=1}^{\alpha_\ell}d_{\tau(i)}+\sum_{j=1}^{\ell-1}d_{\tau(\alpha_j)}-\sum_{i=1}^kd_{\tau(\alpha_i)}= \sum_{i=1}^{\alpha_\ell}d_{\tau(i)}-\sum_{i=\ell}^kd_{\tau(\alpha_i)}\geq0\,.
\end{align*}
In the last step we used that the entries of $d$ are non-negative and, more importantly, that $\{\alpha_{k},\alpha_{k-1},\ldots,\alpha_{\ell+1},\alpha_{\ell}\}\subseteq\{1,2,\ldots,\alpha_{\ell}-1,\alpha_{\ell}\}$ due to the ordering of the $\alpha_i$
\end{proof}

\begin{lemma}\label{lemma_minmax}
Let $m,n\in\mathbb N$ and $x\in\mathbb R^n$, $y\in\mathbb R_+^n$, $z\in\mathbb R^m$ be given. Then
$$
\max_{k=1,\ldots,m}\min_{i=1,\ldots,n}(x_i+y_iz_k)=\min_{i=1,\ldots,n}\big(x_i+y_i\big(\max_{k=1,\ldots,m} z_k\big)\big)\,.
$$
\end{lemma}
\begin{proof}
We subdivide the proof into the following three steps.
\begin{itemize}
\item[$\bullet$] $\boxed{\max_{k=1,\ldots,m}\min_{i=1,\ldots,n}(x_i+y_iz_k)\leq\min_{i=1,\ldots,n}\max_{k=1,\ldots,m}(x_i+y_iz_k) }$

This is the usual max-min inequality: 
for all $l=1,\ldots,n$ and $j=1,\ldots,m$
$$
\min_{i=1,\ldots,n}(x_i+y_iz_j)\leq x_l+y_lz_j\leq \max_{k=1,\ldots,m}(x_l+y_lz_k)\,.
$$
The inequality $\min_{i=1,\ldots,n}(x_i+y_iz_j)\leq \max_{k=1,\ldots,m}(x_l+y_lz_k)$ is preserved by taking the maximum over $j$ (only the lower bound depends on $j$) and, afterwards, taking the minimum over $l$ (only the upper bound depends on $l$).

\item[$\bullet$] $\boxed{\min_{i=1,\ldots,n}\max_{k=1,\ldots,m}(x_i+y_iz_k)=\min_{i=1,\ldots,n}\big(x_i+y_i\big(\max_{k=1,\ldots,m} z_k\big)\big)}$

Because $y$ is non-negative one for arbitrary but fix $i=1,\ldots,n$ finds
$$
\max_{k=1,\ldots,m}(x_i+y_iz_k)= x_i+\max_{k=1,\ldots,m}y_iz_k =x_i+y_i\big(\max_{k=1,\ldots,m} z_k\big)\,.
$$
Thus this remains true after taking the minimum over $i$ on both sides.

\item[$\bullet$] $\boxed{\min_{i=1,\ldots,n}\big(x_i+y_i\big(\max_{k=1,\ldots,m} z_k\big)\big)\leq \max_{k=1,\ldots,m}\min_{i=1,\ldots,n}(x_i+y_iz_k) }$

For all $l=1,\ldots,m$, obviously,
$$
\min_{i=1,\ldots,n}\big(x_i+y_iz_l)\leq \max_{k=1,\ldots,m}\min_{i=1,\ldots,n}(x_i+y_iz_k) \,.
$$
In particular this holds for the index $l$ which satisfies $z_l=\max_{k=1,\ldots,m} z_k$.\qedhere
\end{itemize}
\end{proof}

Finally here are some examples relevant to the $d$-majorization polytope.
\begin{example}\label{ex_proof_ext_point_fail}
Let $n=4$ so
$$
M={\footnotesize\begin{pmatrix} 1&0&0&0\\
0&1&0&0\\
0&0&1&0\\
0&0&0&1\\
1&1&0&0\\
1&0&1&0\\
1&0&0&1\\
0&1&1&0\\
0&1&0&1\\
0&0&1&1\\
1&1&1&0\\
1&1&0&1\\
1&0&1&1\\
0&1&1&1\\
1&1&1&1\\
-1&-1&-1&-1 \end{pmatrix}}
\qquad\text{ and choose }\qquad
b={\footnotesize\begin{pmatrix} 0\\
0\\
0\\
0\\
0\\
-1/2\\
-1/4\\
0\\
0\\
0\\
-1/2\\
-1/2\\
-5/8\\
0\\
-1\\
1 \end{pmatrix}}\,.
$$
By Definition \ref{def_Ebsigma} and Lemma \ref{lemma_E_b_sigma_properties}
\begin{align*}
\{E_b({\pi})\,|\,{\pi}\in S_4\}=\Big\{& {\footnotesize
\begin{pmatrix} 0 \\0 \\-1/2 \\-1/2 \end{pmatrix},
\begin{pmatrix} 0 \\ -3/8 \\ -1/2 \\ -1/8 \end{pmatrix},
\begin{pmatrix} 0 \\ -1/4 \\ -1/2 \\ -1/4 \end{pmatrix}, 
\begin{pmatrix} 0 \\ -3/8 \\ -3/8 \\ -1/4 \end{pmatrix}, 
\begin{pmatrix} -1/2 \\ 0 \\ 0 \\ -1/2 \end{pmatrix}, 
\begin{pmatrix} -1 \\ 0 \\ 0 \\ 0 \end{pmatrix}, }\\
&{\footnotesize
\begin{pmatrix} -1/2 \\ 0 \\ -1/2 \\ 0 \end{pmatrix}, 
\begin{pmatrix} -1/2 \\ -3/8 \\ 0 \\ -1/8 \end{pmatrix}, 
\begin{pmatrix} -5/8 \\ -3/8 \\ 0 \\ 0 \end{pmatrix}, 
\begin{pmatrix} -1/4 \\ -1/4 \\ -1/2 \\ 0 \end{pmatrix}, 
\begin{pmatrix} -1/4 \\ -3/8 \\ -3/8 \\ 0 \end{pmatrix}
}\Big\}\,.
\end{align*}
The second and the fourth vector from this list are the solutions to
\begin{equation*}
{\footnotesize\begin{pmatrix} 1&0&0&0\\1&0&1&0\\1&0&1&1\\1&1&1&1 \end{pmatrix}}p={\footnotesize\begin{pmatrix} 0\\-1/2\\-5/8\\-1 \end{pmatrix}}\quad\text{ and }\quad
{\footnotesize\begin{pmatrix} 1&0&0&0\\1&0&0&1\\1&0&1&1\\1&1&1&1 \end{pmatrix}}p={\footnotesize\begin{pmatrix}0\\-1/4\\-5/8\\-1 \end{pmatrix}}\,,
\end{equation*}
respectively, and are not in $\{x\in\mathbb R^4\,|\,Mx\leq b\}$ but every other point of $\{E_b({\pi})\,|\,{\pi}\in S_4\}$ is in. On the other hand one readily verifies that $p=-\frac{1}{8}(1,3,3,1)^T$ satisfies $Mp\leq b$ and solves
\begin{equation*}
{\footnotesize\begin{pmatrix} 1&0&1&0\\1&0&0&1\\1&0&1&1\\1&1&1&1 \end{pmatrix}}p={\footnotesize\begin{pmatrix} -1/2\\-1/4\\-5/8\\-1 \end{pmatrix}}
\end{equation*}
so it is extreme in $\{x\in\mathbb R^4\,|\,Mx\leq b\}$ by Lemma \ref{lemma_extreme_points_h_descr}, but $p\not\in\{E_b({\pi})\,|\,{\pi}\in S_4\}$. Thus there exist extreme points of $\{x\in\mathbb R^4\,|\,Mx\leq b\}$ not of the form $E_b({\pi})$.
\end{example}

\begin{example}\label{example_2}
Let $d=(4,2,1)^T$, $y=(4,-2,2)^T$. To compute $M_d(y)$ we first need the vector $b\in\mathbb R^8$ from the corresponding halfspace description. Now $\{\frac{y_i}{d_i}\,|\,i=1,2,3\}=\{1,-1,2\}$ so by Thm.~\ref{thm_maj_halfspace} $M_d(y)=\{x\in\mathbb R^3\,|\,Mx\leq b\}$ with
$$
M={\footnotesize\begin{pmatrix} 1&0&0\\
0&1&0\\
0&0&1\\
1&1&0\\
1&0&1\\
0&1&1\\
1&1&1\\
-1&-1&-1 \end{pmatrix}}\in\mathbb R^{8\times 3}
\qquad
b=\min\Big\{{\footnotesize
\begin{pmatrix} 1+4\\1+2\\1+1\\1+6\\1+5\\1+3\\4\\-4 \end{pmatrix} ,
\begin{pmatrix} 11-4\\11-2\\11-1\\11-6\\11-5\\11-3\\4\\-4 \end{pmatrix} ,
\begin{pmatrix} 0+8\\0+4\\0+2\\0+12\\0+10\\0+6\\4\\-4 \end{pmatrix} 
}\Big\}={\footnotesize\begin{pmatrix}
5\\3\\2\\5\\6\\4\\4\\-4
\end{pmatrix}}\,.
$$
Using Thm.~\ref{thm_Eb_sigma} one can easily generate the extreme points of $M_d(y)$:
$$
M_d(y)=\operatorname{conv}\Big\{
\begin{pmatrix} 5\\0\\-1 \end{pmatrix},
\begin{pmatrix} 5\\-2\\1 \end{pmatrix},
\begin{pmatrix} 2\\3\\-1 \end{pmatrix},
\begin{pmatrix} 0\\3\\1 \end{pmatrix},
\begin{pmatrix} 4\\-2\\2 \end{pmatrix},
\begin{pmatrix} 0\\2\\2 \end{pmatrix}
\Big\}\,.
$$
One can verify this using the corresponding extreme points of $s_d(3)$ from Lemma \ref{lemma_extreme_points}; after all if $z$ is an extreme point of $M_d(y)$ then there exists an extreme point $A$ of $s_d(n)$ such that $z=Ay$\,\footnote{
A short proof for the sake of completeness: Assume that $z\neq Ay$ for all extreme points $A$ of $s_d(n)$ but $z\in M_d(y)$. Then one finds $A'\in s_d(n)$ such that $z=A'y$. But $A'$ can in turn be written as $A'=\sum_i\lambda_i A_i$ with $A_i$ being the extreme points of $s_d(n)$ and $\lambda_i\in[0,1)$, $\sum_i\lambda_i=1$. Hence
$
z=\sum_i\lambda_i (A_iy)$ but none of the $A_iy$ are equal to $z$ so the latter cannot be extremal in $M_d(y)$.
}.
\end{example}

\begin{figure}[!htb]
\centering
\includegraphics[width=0.49\textwidth]{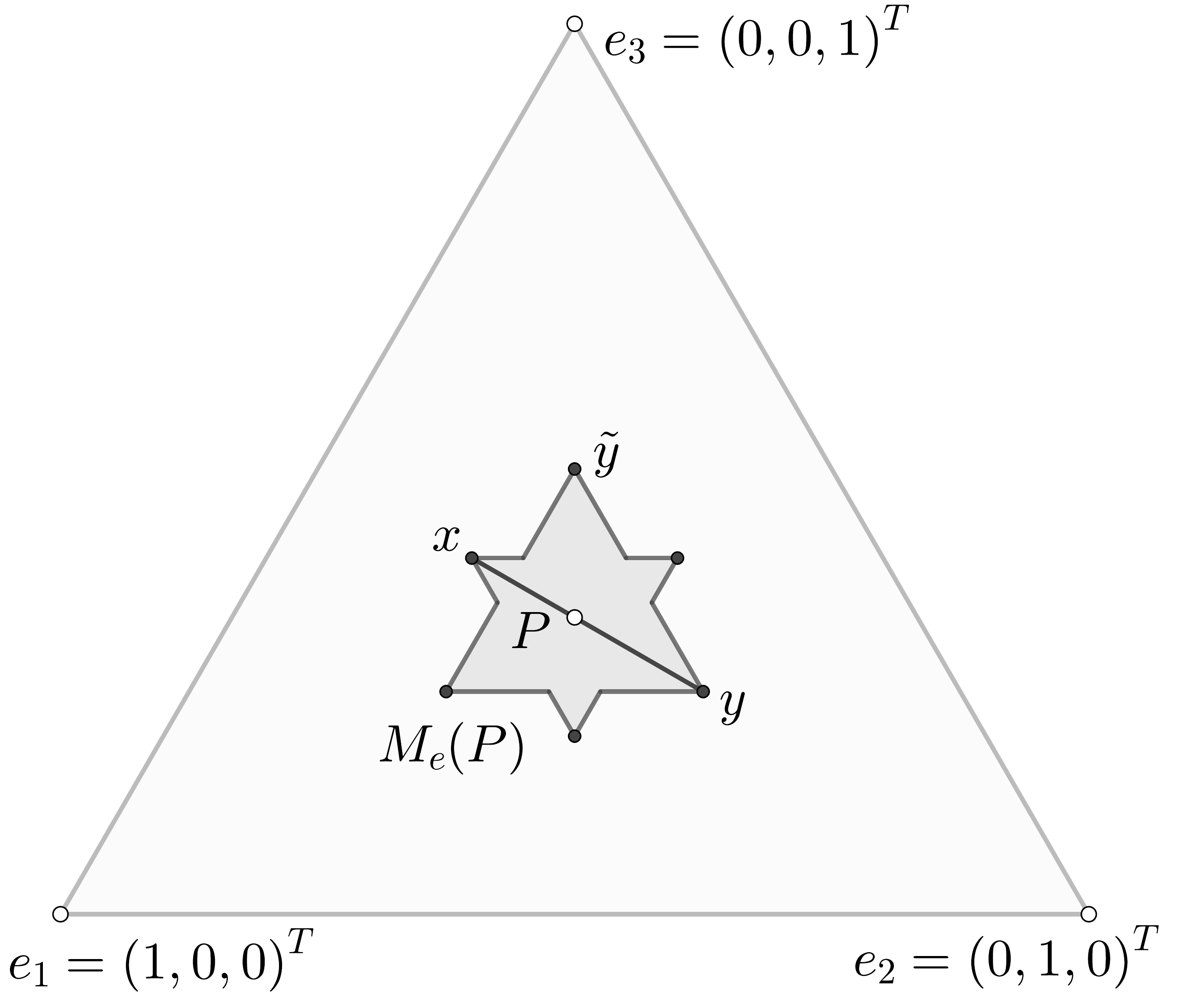}
\includegraphics[width=0.49\textwidth]{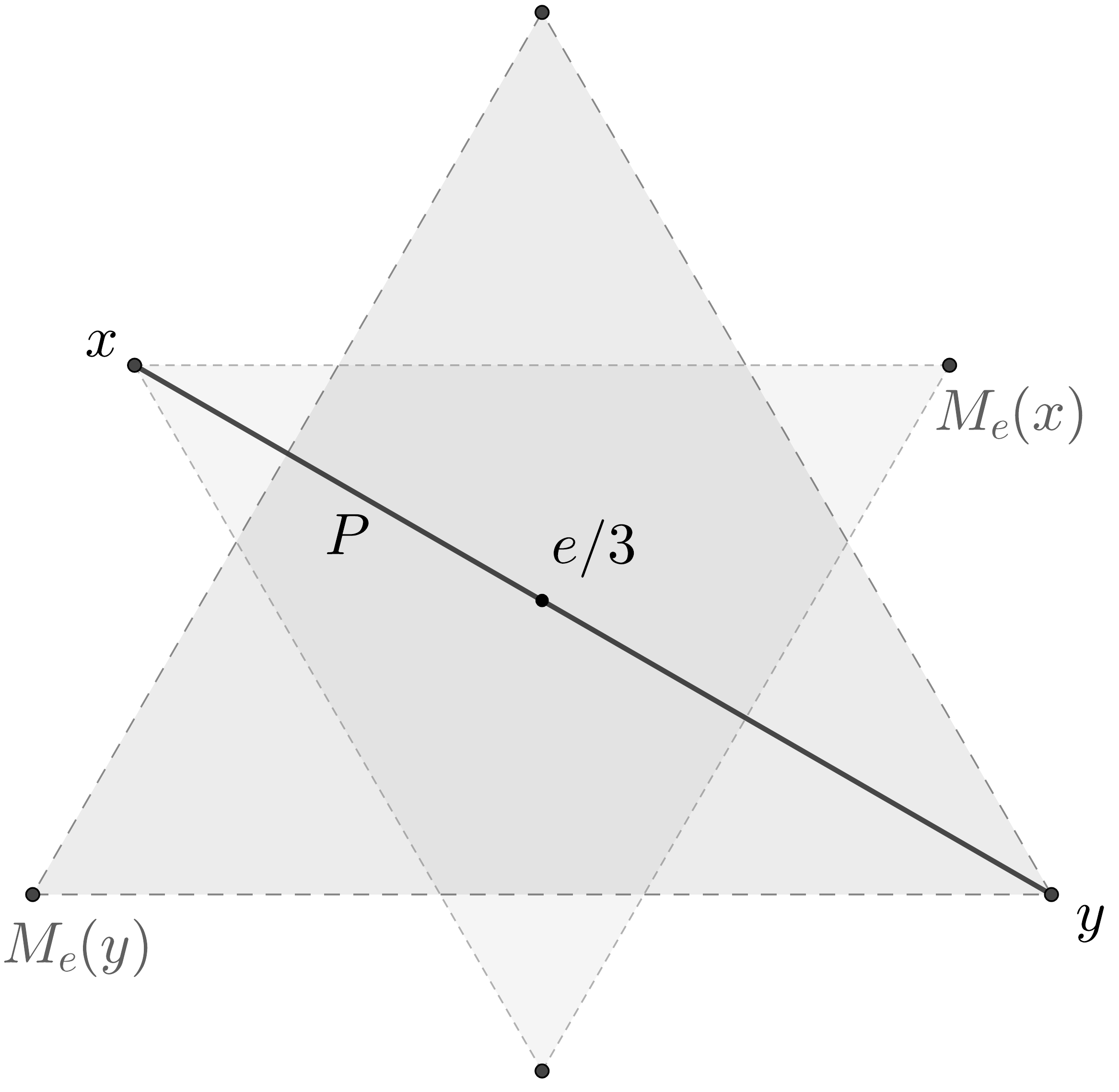}
\caption{
Visualization of Example \ref{example_1} on the 3-dimensional standard simplex. 
The image on the right zooms in on $M_{\unitvector}(P)$ and shows the decomposition into $M_{\unitvector}(x)$ 
and $M_{\unitvector}(y)$. In particular, one sees that for all $z\in P$ one has either $z\prec x$ ($\Leftrightarrow z\in M_{\unitvector}(x)$) 
or $z\prec y$ ($\Leftrightarrow z\in M_{\unitvector}(y)$) which implies \eqref{eq:maj_decomp}.}\label{Abb1}
\end{figure}

\begin{example}[Convexity counterexample]\label{example_1}
Let $n=3$ and $d=\unitvector$ (so $\prec_d$ becomes $\prec$). Consider the probability vectors
$$
x=\frac15\begin{pmatrix} 2\\1\\2 \end{pmatrix}=\begin{pmatrix} 0.4\\0.2\\0.4 \end{pmatrix}\qquad\text{ and }\qquad y=\frac14\begin{pmatrix} 1\\2\\1 \end{pmatrix}=\begin{pmatrix} 0.25\\0.5\\0.25 \end{pmatrix}
$$
and their joining line segment $P:=\operatorname{conv}\lbrace x,y\rbrace$. 
Be aware that $P$ as well as $M_{\unitvector}(P)=\bigcup_{z\in P}\lbrace v\in\mathbb R_+^n\,|\,v\prec z\rbrace$ are subsets of $\Delta^2$
. 
One readily verifies
\begin{equation}\label{eq:maj_decomp}
M_{\unitvector}(P)=\lbrace v\in\mathbb R_+^n\,|\, v\prec x\,\vee\,v\prec y\rbrace=M_{\unitvector}(x)\cup M_{\unitvector}(y)\,,
\end{equation}
refer also to Figure \ref{Abb1}. Now although $x, \tilde y:=(0.25,0.25,0.5)\in M_{\unitvector}(P)$ one has
$$
\frac12 x+\frac12\tilde y=\frac1{40}\begin{pmatrix} 13\\9\\18 \end{pmatrix} =\begin{pmatrix} 0.325\\0.225\\0.45 \end{pmatrix} \overset{\eqref{eq:maj_decomp}}{\not\in} M_{\unitvector}(P)
$$
as neither $x$ nor $y$ majorizes it so $M_{\unitvector}(P)$ is not convex, although $P$ is.
\end{example}


\begin{example}\label{ex_pos_y_necess}
Let $y=(1,1,-1)^T$, $d=(1,2,3)^T$. As in Example \ref{example_2} one easily sees
\begin{align*}
M_d(y)&=\Big\{x\in\mathbb R^3\,\Big|\, Mx\leq {\footnotesize\begin{pmatrix}
1\\3/2\\2\\2\\5/3\\4/3\\1\\-1
\end{pmatrix}}\Big\}=\operatorname{conv}\big\{{\footnotesize
\begin{pmatrix} 1\\1\\-1 \end{pmatrix},
\begin{pmatrix} 1\\-2/3\\2/3 \end{pmatrix},
\begin{pmatrix} 1/2\\3/2\\-1 \end{pmatrix},
\begin{pmatrix} -1/3\\3/2\\-1/6 \end{pmatrix},
\begin{pmatrix} -1/3\\-2/3\\2 \end{pmatrix}}
\big\}
\end{align*}
so the only possible candidate for the point $z$ from Thm.~\ref{theorem_max_corner_maj} is the vector $z=(-1/3,2/3,2)$ (because $z_1^\downarrow=2> x_1^\downarrow$ for all other extreme points $x$). However, $ y\not\prec z$ as
$$
y_1^\downarrow+y_2^\downarrow=1+1=2 > \frac{5}{3} = z_1^\downarrow+z_2^\downarrow\,.
$$
\end{example}

\begin{figure}[!htb]
\centering
(i)\includegraphics[width=0.45\textwidth]{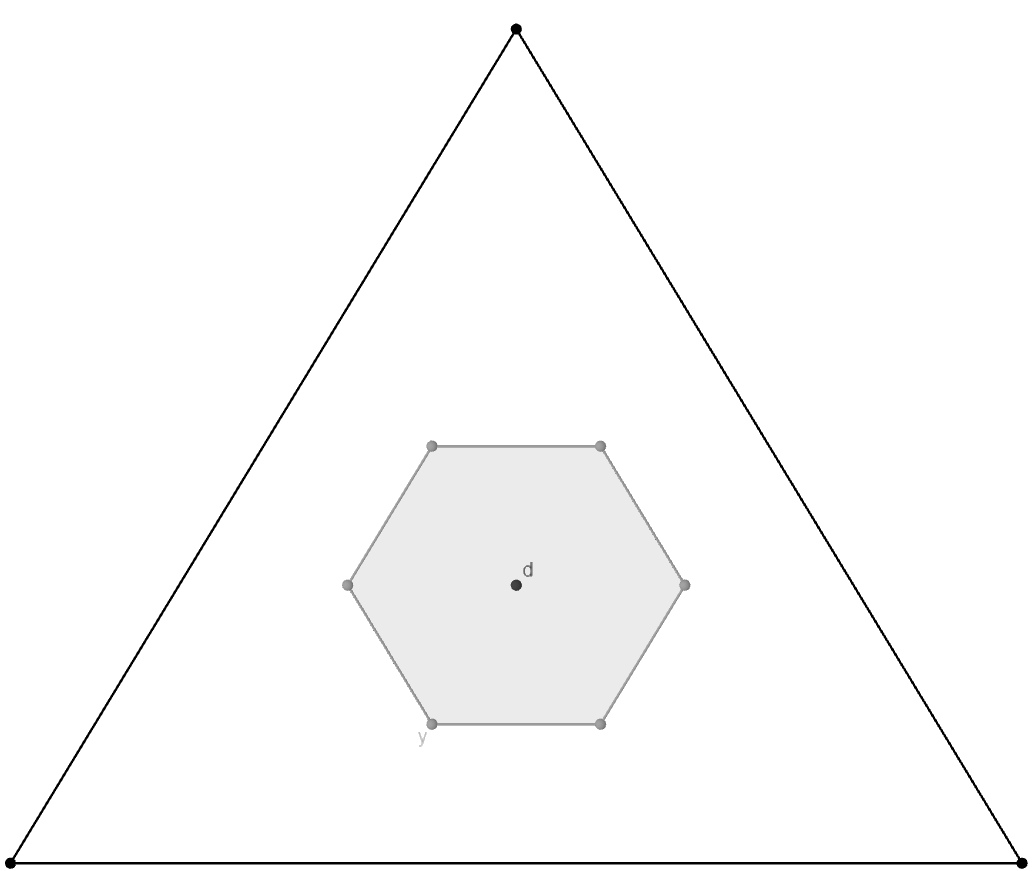}
(ii)\includegraphics[width=0.45\textwidth]{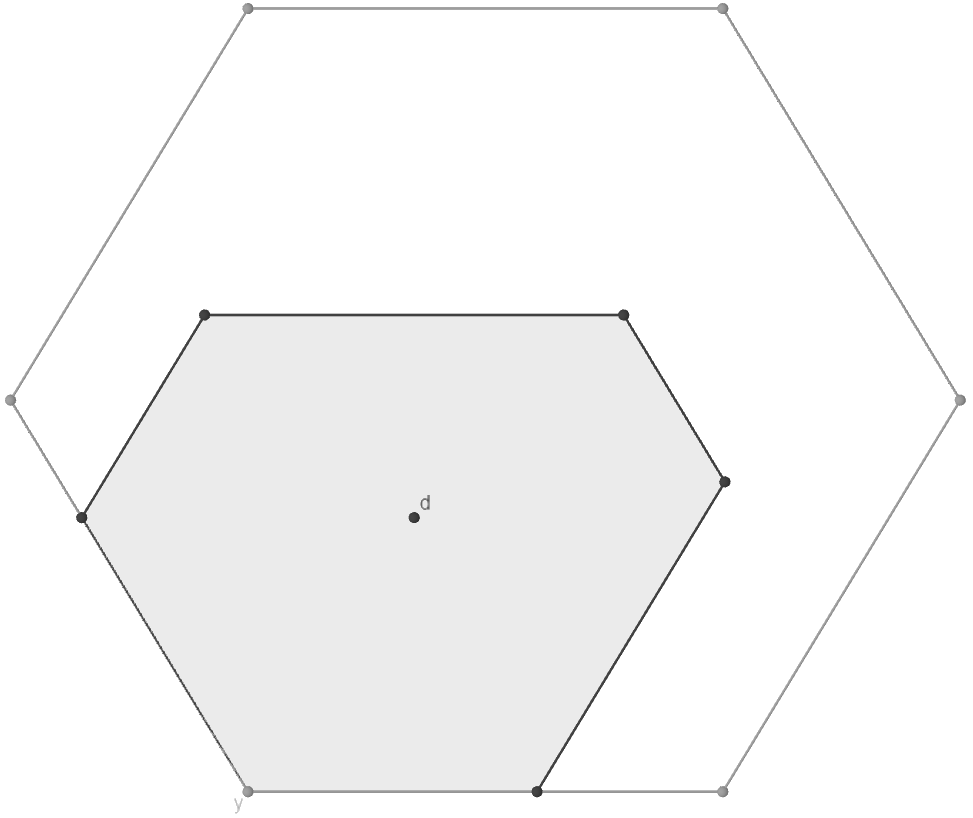}
(iii)\includegraphics[width=0.45\textwidth]{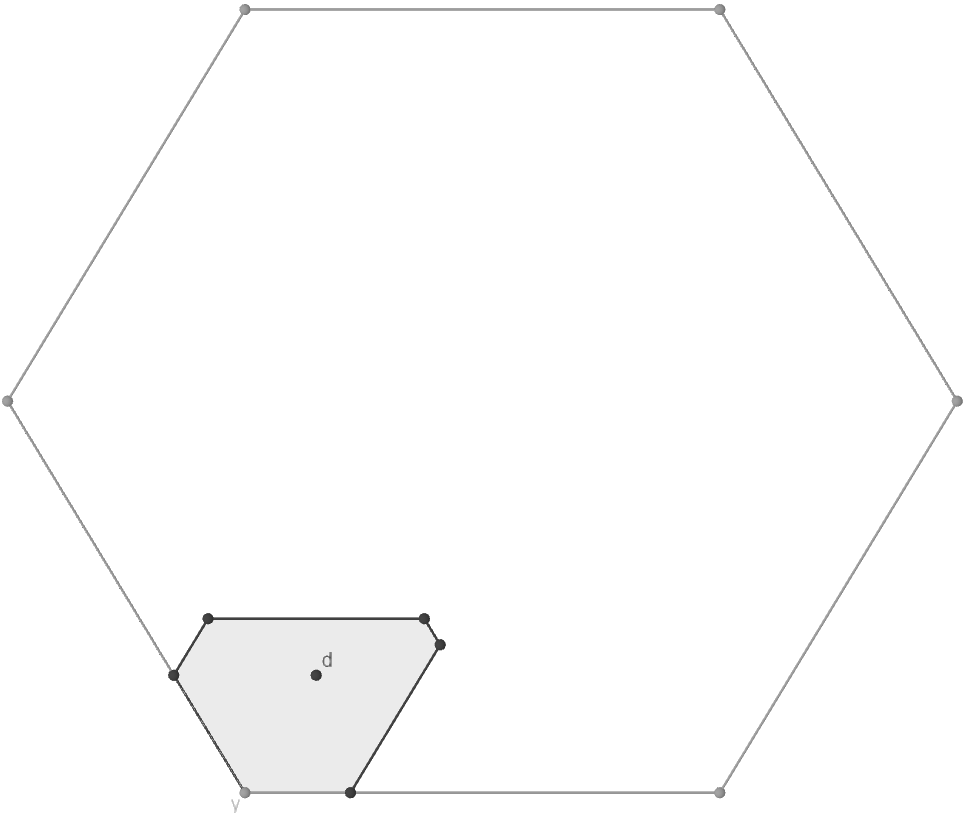}
(iv)\includegraphics[width=0.45\textwidth]{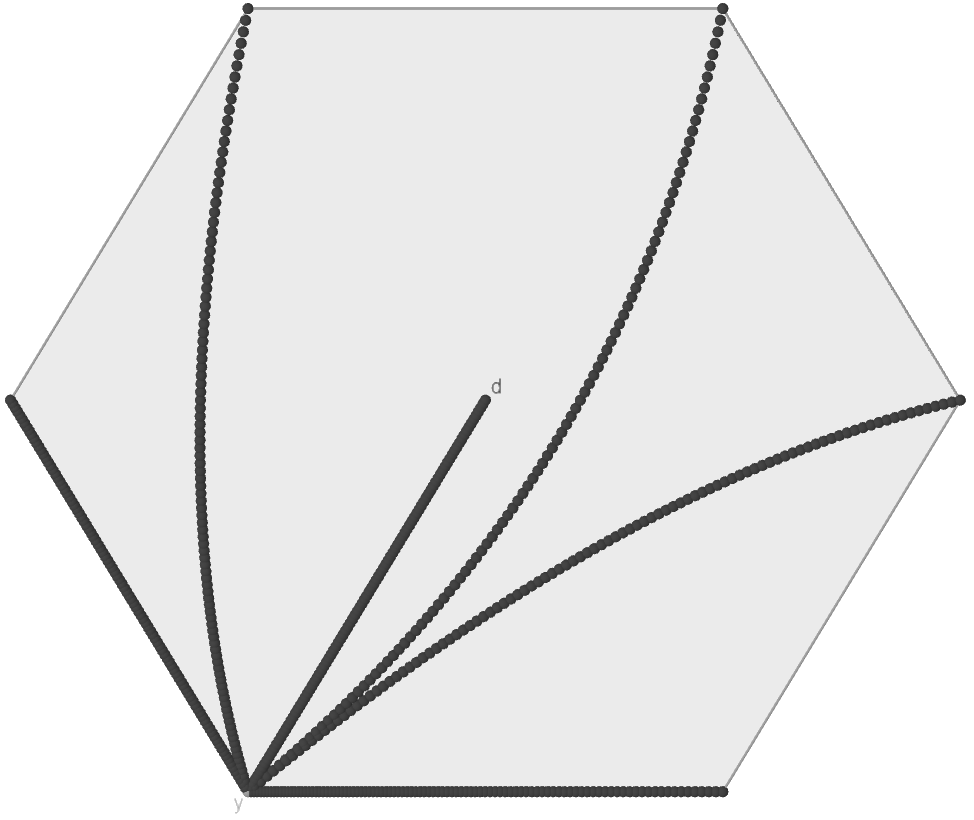}
\caption{
Visualization of Example \ref{ex_wandering_d_vector}. 
(i): Shows $M_{d(0)}(y)=M_\unitvector(y)=\operatorname{conv}\{\underline{{\pi}}y\,|\,{\pi}\in S_3\}$ inside (a multiple of) the 3-dimensional standard simplex. (ii): Zooms in on the classical majorization polytope $M_\unitvector(y)$. The shaded area is $M_{d(\lambda)}(y)$ for $\lambda=0.3$. (iii): Shows $M_{d(\lambda)}(y)$ for $\lambda=0.7$. (iv): The graph of the map $\lambda\mapsto\operatorname{ext}(M_{d(\lambda)}(y))$.}\label{Abb2}
\end{figure}

\begin{example}\label{example_not_cont_A}
To see discontinuity of the map
$$
P(b):D(P)\to \mathcal P_c(\mathbb R^n)\qquad A\mapsto P_A(b)=\{x\in\mathbb R^n\,|\,Ax\leq b\}
$$
for arbitrary but fix $b\in\mathbb R^m$ and domain\footnote{This choice of domain ensures that the codomain of $P$ is $P_c(\mathbb R^n)$, i.e.~that all $P_A(b)$ are non-empty and bounded (hence compact), cf.~\cite[Ch.~8.2]{Schrijver86}.} $D(P)$ consisting of all $A\in\mathbb R^{m\times n}$ such that $P_A(0)=\{0\}$ and $P_A(b)\neq\emptyset$, consider the following: Let
$$
A={\footnotesize\begin{pmatrix}
1&0\\-1&0\\0&1\\0&-1
\end{pmatrix}}\ ,\ A_t={\footnotesize\begin{pmatrix}
1&0\\-1&0\\\sin(t)&\cos(t)\\0&-1
\end{pmatrix}}\qquad\text{ and }\qquad b={\footnotesize\begin{pmatrix}
1\\1\\0\\0
\end{pmatrix}}
$$
for all $t\in(0,1]$. It is readily verified that
\begin{align*}
P_{A}(b)&=\operatorname{conv}\big\{{\footnotesize \begin{pmatrix} -1\\0 \end{pmatrix},\begin{pmatrix} 1\\0 \end{pmatrix} }\big\}\hspace*{64pt}\text{ as well as}\\
P_{A_t}(b)&=\operatorname{conv}\big\{{\footnotesize \begin{pmatrix} 0\\0 \end{pmatrix},\begin{pmatrix} -1\\0 \end{pmatrix},\begin{pmatrix} -1\\\tan(t) \end{pmatrix} }\big\}\qquad\text{ for all }t>0
\end{align*}
and $(A_t)_{t\geq 0}\subset D(P)$. Thus by definition of the Hausdorff metric
$$
\Delta(P_{A_t}(b),P_A(b))\geq \max_{z\in P_A(b)}\min_{w\in P_{A_t}(b)}\|z-w\|_1\geq \min_{w\in P_{A_t}(b)}\big\| {\footnotesize \begin{pmatrix} 1\\0 \end{pmatrix}}-w\big\|_1=1
$$
for all $t>0$ but, obviously, $\lim_{t\to 0^+}\|A_t-A\|=0$ so $P(b)$ cannot be continuous. 
\end{example}
\begin{example}\label{ex_discont_IR_plus}
Let $y=(1,1,1)^T$, $\lambda\in[0,\frac12]$, and $d(\lambda)=(1,\lambda,\lambda^2)$. For all $\lambda\in(0,\frac12]$ one readily verifies (cf.~also Example \ref{example_2})
\begin{align*}
M_{d(\lambda)}(y)&=\Big\{x\in\mathbb R^3\,\Big|\, Mx\leq {\footnotesize\begin{pmatrix}
3-\lambda-\lambda^2 \\ 2-\lambda \\ 1 \\ 3-\lambda^2 \\ 3-\lambda \\ 2 \\ 3 \\ -3
\end{pmatrix}}\Big\}\\
&=\operatorname{conv}\Big\{{\footnotesize
\begin{pmatrix} 3-\lambda-\lambda^2\\\lambda\\\lambda^2 \end{pmatrix},
\begin{pmatrix} 1+\lambda-\lambda^2\\2-\lambda\\\lambda^2 \end{pmatrix},
\begin{pmatrix} 1\\2-\lambda\\\lambda\end{pmatrix},
\begin{pmatrix} 2-\lambda\\\lambda\\1 \end{pmatrix},
\begin{pmatrix} 1\\1\\1 \end{pmatrix}
}
\Big\}\,.
\end{align*}
as well as\footnote{
For $\lambda=0$, i.e.~$d=d(0)=(1,0,0)$ it is easy so see that every $d$-stochastic matrix is of the form
$$
A=\begin{pmatrix} 1&&\\0&v&w\\0&& \end{pmatrix}\quad\text{ with arbitrary }v,w\in\Delta^2\,.
$$
Thus $M_{d(0)}(y)=\{Ay\,|\,A\in s_{d(0)}(3)\}=\{(1+v_1+w_1,v_2+w_2,v_3+w_3)^T\,|\,v,w\in\Delta^2\}$ which has extreme points $(3,0,0)^T, (1,2,0)^T, (1,0,2)^T$.
}
\begin{align*}
M_{d(\lambda)}(y)\overset{\lambda\to 0^+}\to\operatorname{conv}\Big\{{\footnotesize
\begin{pmatrix} 3\\0\\0 \end{pmatrix},
\begin{pmatrix} 1\\2\\0 \end{pmatrix},
\begin{pmatrix} 2\\0\\1 \end{pmatrix},
\begin{pmatrix} 1\\1\\1 \end{pmatrix}
}
\Big\}\neq \operatorname{conv}\Big\{{\footnotesize
\begin{pmatrix} 3\\0\\0 \end{pmatrix},
\begin{pmatrix} 1\\2\\0 \end{pmatrix},
\begin{pmatrix} 1\\0\\2 \end{pmatrix}
}
\Big\}=M_{d(0)}(y)\,.
\end{align*}
\end{example}
\begin{example}\label{ex_wandering_d_vector}
Let $y=(3,2,1)^T$, $\lambda\in[0,1]$, and $d(\lambda)=(2+\lambda,2,2-\lambda)$ so 
$$
M_{d(0)}(y)=M_\unitvector(y)=\operatorname{conv}\{\underline{{\pi}}y\,|\,{\pi}\in S_3\}\qquad\text{ and }\qquad M_{d(1)}(y)=M_y(y)=\{y\}
$$
(cf.~also Example \ref{example_2}). Thus the parameter $\lambda\in[0,1]$ describes the deformation of a classical majorization polytope into a singleton. Indeed one readily computes (cf.~Fig.~\ref{Abb2})
\begin{align*}
M_{d(\lambda)}(y)=\Big\{x\in\mathbb R^3\,\Big|\, Mx\leq {\footnotesize\begin{pmatrix}
3\\\frac{6}{2+\lambda}\\\frac{6-3\lambda}{2+\lambda}\\5\\5-\lambda\\5-2\lambda\\6\\-6
\end{pmatrix}}\Big\}=\operatorname{conv}\Big\{{\footnotesize
\begin{pmatrix} 3\\2\\1 \end{pmatrix},
\begin{pmatrix} 3\\1+\lambda\\2-\lambda \end{pmatrix},
\frac{1}{2+\lambda}\begin{pmatrix} 4+5\lambda\\6\\2+\lambda \end{pmatrix},}&\\
{\footnotesize\frac{1}{2+\lambda}\begin{pmatrix} 2\lambda^2+5\lambda+2\\6\\-2\lambda^2+\lambda+4 \end{pmatrix},
\frac{1}{2+\lambda}\begin{pmatrix} -\lambda^2+6\lambda+4\\\lambda^2+3\lambda+2\\6-3\lambda \end{pmatrix},
\frac{1}{2+\lambda}\begin{pmatrix} 2\lambda^2+5\lambda+2\\-2\lambda^2+4\lambda+4\\6-3\lambda \end{pmatrix}}&
\Big\}\,.
\end{align*}
\end{example}

\subsection{Appendix to Section \ref{sec:maj_d_mat}}

\begin{example}\label{example_1z}
The linear map
\begin{align*}
T:\mathbb C^{2\times 2}&\to\mathbb C^{2\times 2}\\
 \begin{pmatrix} a_{11}&a_{12}\\a_{21}&a_{22}\end{pmatrix}&\mapsto\begin{pmatrix} a_{11}+\frac12a_{22}&0\\0&\frac12a_{22}\end{pmatrix}
\end{align*}
is obviously \textsc{cptp} and strictly positive ($T(\mathbbm{1})>0$) but the only fixed points of $T$ are of the form $\footnotesize\begin{pmatrix}x&0\\0&0 \end{pmatrix}$, that is, not of full rank.
\end{example}
\begin{example}\label{example_2z}
Consider the channel
\begin{align*}
T:\mathbb C^{3\times 3}&\to \mathbb C^{3\times 3}\\
\begin{pmatrix} a_{11}&a_{12}&a_{13}\\a_{21}&a_{22}&a_{23}\\a_{31}&a_{32}&a_{33} \end{pmatrix}&\mapsto \begin{pmatrix}a_{22}+a_{33}&0&0\\0& \frac12a_{11}&0\\0&0& \frac12a_{11} \end{pmatrix}\,.
\end{align*}
In particular this map is strictly positive by Prop.~\ref{prop_positivity} (iii) as $\operatorname{diag}(2,1,1)$ is a fixed point. However
\begin{align*}
1=\operatorname{rank}(|e_1\rangle\langle e_1|)&<\operatorname{rank}(T(|e_1\rangle\langle e_1|))=2\\
2=\operatorname{rank}(|e_2\rangle\langle e_2|+|e_3\rangle\langle e_3|)&>\operatorname{rank}(T(|e_2\rangle\langle e_2|+|e_3\rangle\langle e_3|))=1\,.
\end{align*}
\end{example}

\begin{example}\label{example_3new}
For any $m\in\mathbb N$ define $T_m:\mathbb C^{2\times 2}\to \mathbb C^{2\times 2}$ via
$$
T_m\begin{pmatrix} a_{11}&a_{12}\\a_{21}&a_{22}\end{pmatrix}=\begin{pmatrix} (1+\frac1m)a_{11}-\frac1ma_{22}&a_{12}\\a_{21}&(1+\frac1m)a_{22}-\frac1ma_{11} \end{pmatrix}\,.
$$
Obviously every $T_m$ is trace-preserving but not positive as
$$
T_m\begin{pmatrix} 1&0\\0&0 \end{pmatrix}=\begin{pmatrix} 1+\frac1m&0\\0&-\frac1m \end{pmatrix}\,.
$$
However $\lim_{m\to\infty}T_m=\mathbbm{1}_{2\times 2}$ so for every $\varepsilon>0$ there exists $T\in B_\varepsilon(\mathbbm{1}_{2\times 2})$ which is not strictly positive, although the identity itself is strictly positive. 
This example can easily be generalized to arbitrary sizes of domain and codomain. 
\end{example}

\begin{example}\label{example_5}
\begin{itemize}
\item[(i)] The Choi matrix of the linear map
\begin{align*}
T:\mathbb C^{3\times 3}&\to \mathbb C^{3\times 3}\\
\begin{pmatrix} a_{11}&a_{12}&a_{13}\\a_{21}&a_{22}&a_{23}\\a_{31}&a_{32}&a_{33} \end{pmatrix}&\mapsto \begin{pmatrix} a_{11}&\frac{i}{\sqrt2}(a_{12}+a_{13})&0\\-\frac{i}{\sqrt2}(a_{21}+a_{31})&a_{22}+a_{33}&0\\0&0&0 \end{pmatrix}
\end{align*}
has simple eigenvalues $2,1$ and the $7$-fold eigenvalue $0$ so $T$ is \textsc{cptp}, not \textsc{sp}, and not a trace projection, that is, not of the form $A\mapsto\operatorname{tr}(A)\rho$ for any state $\rho$.
\item[(ii)] Via $\operatorname{tr}(T(A)B)=\operatorname{tr}(AT^*(B))$ for all $A,B\in\mathbb C^{3\times 3}$ the dual of $T$ from (i) is given by
\begin{align*}
T^*:\mathbb C^{3\times 3}&\to \mathbb C^{3\times 3}\\
\begin{pmatrix} b_{11}&b_{12}&b_{13}\\b_{21}&b_{22}&b_{23}\\b_{31}&b_{32}&b_{33} \end{pmatrix}&\mapsto \begin{pmatrix} b_{11} & -\frac{i}{\sqrt{2}}b_{12} & -\frac{i}{\sqrt{2}}b_{12} \\ \frac{i}{\sqrt{2}}b_{21} & b_{22} & 0 \\ \frac{i}{\sqrt{2}}b_{21} & 0 & b_{22} \end{pmatrix}\,.
\end{align*}
Note that the action of $T^*$ is determined by a subalgebra of the domain because 
$$
T^*(B)=T^*\begin{pmatrix} b_{11}&b_{12}&0\\b_{21}&b_{22}&0\\0&0&0 \end{pmatrix}\,.
$$
\end{itemize}
\end{example}
\begin{example}\label{example_5c}
Consider the unitary matrix $\sigma={\footnotesize\begin{pmatrix} 0&1\\1&0 \end{pmatrix}}$ and the induced channel $T:\mathbb C^{2\times 2}\to \mathbb C^{2\times 2}$, $\rho\mapsto \sigma\rho\sigma$. Then
$$
\|T-\mathbbm{1}\|\geq \big\|T\big(|e_1\rangle\langle e_1|\big)-|e_1\rangle\langle e_1|\,\big\|_1=\big\|\,|e_2\rangle\langle e_2|-|e_1\rangle\langle e_1|\,\big\|_1=2
$$
but as a unitary channel, $T$ preserves the identity and thus is strictly positive. 
\end{example}
\subsection{Appendix to Section \ref{sec:c_num_range}}\label{app_c_num_range}

To prove Lemma \ref{U-approximation} we need the following auxiliary result.

\begin{lemma}\label{lem:unitary-dilation}
Let $U\in\mathbb C^{n\times n}$ with $\Vert U\Vert_\textrm{op}\leq 1$. Then one finds matrices
$Q,R,S\in\mathbb C^{n\times n}$ such that
\begin{align*}
V:=\begin{pmatrix} U&Q\\R&S \end{pmatrix}\in\mathbb C^{2n\times 2n}
\end{align*}
is unitary.
\end{lemma}

\begin{proof}
Obviously, $\Vert U\Vert\leq 1$ implies $\mathbbm{1}_n-U U^*\geq 0$, where $\mathbbm{1}_n$ denotes 
the $n\times n$ identity matrix. Hence 
$Q:=\sqrt{\mathbbm{1}_n-UU^* }$ is well-defined. Now the upper $n$ rows of $V$ form an orthonormal system in 
$\mathbb C^{2n}$ as
\begin{align*}
\begin{pmatrix} U&Q \end{pmatrix}\begin{pmatrix} U^*\\Q^* \end{pmatrix}
= UU^* +QQ^*=\mathbbm{1}_n.
\end{align*}
Completing this orthonormal system to an orthonormal basis of $\mathbb C^{2n}$ gives $R,S$ such that,
in total, $V$ is unitary.
\end{proof}

\begin{proof}[Proof of Lemma \ref{U-approximation}]
Let $U\in\mathcal B(\mathcal H)$ be unitary and consider arbitrary orthonormal bases 
$(e_n)_{n\in\mathbb N}$, $(g_n)_{n\in\mathbb N}$ of $\mathcal H$. For all $n\in\mathbb N$ one has $\Vert (\Gamma_n^g)^* U\Gamma_n^e\Vert\leq 1$ so Lemma
\ref{lem:unitary-dilation} yields $Q_n,R_n,S_n\in\mathbb C^{n\times n}$ such that
\begin{align*}
V_n:=\begin{pmatrix} (\Gamma_n^g)^* U\Gamma_n^e&Q_n\\R_n&S_n \end{pmatrix}\in\mathbb C^{2n\times 2n}
\end{align*}
is unitary. Define $\hat U_n:=\Gamma_{2n}^gV_n(\Gamma_{2n}^e)^*\in\mathcal B(\mathcal H)$. Then, obviously, (ii) and (iii) 
of Lemma \ref{U-approximation} hold. To show that $(\hat U_n)_{n\in\mathbb N}$ converges strongly to
$U$ we first observe $\|\hat U_n x - Ux\| \leq \|\hat U_n x - \Pi^g_{n}U\Pi^e_{n}x\| + \|\Pi^g_{n}U\Pi^e_{n}x - Ux\|$
and 
\begin{align*}
\|\Pi^g_{n}U\Pi^e_{n}x - Ux\| \leq\Vert \Pi^g_{n}U\Pi^e_{n}x-\Pi^g_nUx\Vert+ \Vert\Pi^g_nUx-Ux\Vert\leq \Vert \Pi^e_{n}x-x\Vert+ \Vert\Pi^g_nUx-Ux\Vert \,.
\end{align*}
Hence, $(\Pi^g_{n}U\Pi^e_{n})_{n\in\mathbb N}$ converges strongly to $U$ by Lemma  \ref{lemma_approx_strong_top}, meaning it suffices to show that
\begin{align*}
Z_n:=\hat U_n-\Pi^g_{n}U\Pi^e_{n}=\Gamma_{2n}^g\begin{pmatrix} 0&Q_n\\R_n&S_n \end{pmatrix}(\Gamma_{2n}^e)^*
\end{align*}
strongly converges to 0. Let $x \in\mathcal H\setminus\lbrace 0\rbrace$ and $\varepsilon>0$ be given.
By Lemma \ref{lemma_approx_strong_top} one can choose $N\in\mathbb N$ such that
\begin{align}
\begin{split}
\Vert x\Vert^2-\Vert\Pi_n^gUx\Vert^2 &= \Vert Ux\Vert^2-\Vert\Pi_n^gUx\Vert^2 = \Vert\Pi_n^gUx-Ux\Vert^2<\frac{\varepsilon^2}{8}\\
\text{and} \quad \Vert\Pi_n^e x-x\Vert&<\min\Big\lbrace\frac{\varepsilon^2}{16\Vert x\Vert},\frac{\varepsilon}{2\sqrt{2}}\Big\rbrace\end{split}\label{pi_ineqs}
\end{align}
for all $n\geq N$. Now let $\Lambda_n^e:\mathbb C^n\to\mathcal H$ be the unique linear operator
given by $\hat e_j\mapsto e_{j+n}$ for $j\in\lbrace1,\ldots,n\rbrace$. So basically $(\Lambda_n^e)^*$ ``cuts out'' the components $x_{n+1},\ldots,x_{2n}$ of $x\in\mathcal H$ with respect to $(e_n)_{n\in\mathbb N}$. Next, we decompose $x$ as follows
\begin{align*}
x=\Pi_n^ex+(\Pi^e_{2n}-\Pi^e_n)x+(\mathbbm{1}_{\mathcal H}-\Pi_{2n}^e)x\,.
\end{align*}
Then $\Pi_n^ex \in \mathcal H$ and $x_n:=(\Gamma_n^e)^* x\in\mathbb C^n$ are essentially the same vectors, as those differ only by the isometric embedding $\Gamma_n^e$. The same holds for $(\Pi^e_{2n}-\Pi^e_n)x \in \mathcal H$ and $y_n:=(\Lambda_n^e)^* x\in\mathbb C^n$. 
Taking into account that $\Gamma_{2n}^g$ is an isometry, we obtain
\begin{align*}
\Vert x\Vert^2\geq\Vert\hat U_nx\Vert^2=\Vert(\Gamma_n^g)^* U\Gamma_n^ex_n+Q_ny_n\Vert^2+\Vert R_nx_n+S_ny_n\Vert^2
\end{align*}
and thus
\begin{align}\label{ineqs-2}
\Vert R_nx_n+S_ny_n\Vert^2&\leq\Vert x\Vert^2-\Vert(\Gamma_n^g)^* U\Gamma_n^ex_n+Q_ny_n\Vert^2\nonumber\\
&= \Vert x\Vert^2-\Vert(\Gamma_n^g)^* Ux - ((\Gamma_n^g)^* Ux-(\Gamma_n^g)^* U\Gamma_n^ex_n-Q_ny_n)\Vert^2\nonumber\\
&\leq \Vert x\Vert^2 - \big|\Vert(\Gamma_n^g)^* Ux \Vert - \Vert (\Gamma_n^g)^* Ux-(\Gamma_n^g)^* U\Gamma_n^ex_n-Q_ny_n\Vert \big|^2 \,,
\end{align}
where the last estimate follows from the reverse triangle inequality. Then, using again that $\Gamma_n^g$ is an isometry satisfying $\Gamma_n^g(\Gamma_n^g)^*=\Pi_n^g$ and further $\Vert Q_n\Vert\leq 1$ by construction, we from 
\eqref{pi_ineqs} and \eqref{ineqs-2} deduce the estimate
\begin{align*}
\Vert R_nx_n+S_ny_n\Vert^2&\leq\Vert x\Vert^2-\Vert\Pi_n^g Ux\Vert^2+2\Vert \Pi_n^g Ux \Vert\Vert\Pi_n^gUx-\Pi_n^gU\Pi_n^ex-\Gamma_n^gQ_ny_n\Vert\\
&<\frac{\varepsilon^2}{8}+2\Vert\Pi_n^g Ux\Vert \big(\Vert\Pi_n^gU\Vert\Vert x-\Pi_n^ex\Vert+\Vert Q_n\Vert\Vert\Pi_{2n}^ex-\Pi_n^ex\Vert\big)\\
&\leq\frac{\varepsilon^2}{8}+2\Vert x\Vert\big(\Vert x-\Pi_n^ex\Vert+\Vert \Pi_{2n}^ex-x\Vert+\Vert x-\Pi_n^ex\Vert\big)
< \frac{\varepsilon^2}{2}
\end{align*}
for all $n\geq N$. Finally,
\begin{align*}
\Vert Z_nx\Vert^2&=\Vert Q_ny_n\Vert^2+\Vert R_nx_n+S_ny_n\Vert^2<\Vert Q_n\Vert^2\Vert\Pi_{2n}^ex-\Pi_n^ex\Vert^2+\frac{\varepsilon^2}{2}\\
&\leq (\Vert \Pi_{2n}^ex-x\Vert+\Vert x-\Pi_n^ex\Vert)^2+\frac{\varepsilon^2}{2}<\Big( 2\frac{\varepsilon}{2\sqrt{2}} \Big)^2+\frac{\varepsilon^2}{2}=\varepsilon^2
\end{align*}
for all $n\geq N$. This proves part (i) and, in total, Lemma \ref{U-approximation}.
\end{proof}

\begin{proof}[Proof of Lemma \ref{Lemma_5b}]
Consider the following intermediate sets:
\begin{align*}
A&:=\Big\lbrace \sum\nolimits_{n=1}^\infty a_nb_{{\pi}(n)} \,\Big|\, {\pi}:\mathbb N \to\mathbb N \text{ is permutation}\Big\rbrace\,,\\
A_1&:=\Big\lbrace \sum\nolimits_{n=1}^\infty a_nb'_{{\pi}(n)} \,\Big|\, {\pi}:\mathbb N \to\mathbb N \text{ is permutation}\Big\rbrace\,,\\
A_2&:=\Big\lbrace \sum\nolimits_{n=1}^\infty a'_nb_{{\pi}(n)} \,\Big|\, {\pi}:\mathbb N \to\mathbb N \text{ is permutation}\Big\rbrace\,,\\
A'&:=\Big\lbrace \sum\nolimits_{n=1}^\infty a'_nb'_{{\pi}(n)} \,\Big|\, {\pi}:\mathbb N \to\mathbb N \text{ is permutation}\Big\rbrace\,.
\end{align*}
We will proceed as follows: First we will show that the closure of $A$ and $A_1$ co{\"i}ncides, then 
that of $A$ and $A_2$, and finally that of $A_2$ and $A'$. In the following let $p,q\in (1,\infty)$ as the case $\ell^1, c_0$ (and $c_0,\ell^1$) are proven analogously, cf.~\cite[Proof of Lemma 3.6]{DvE18}.

Assume w.l.o.g.~that $\|a\|_p,\|b\|_q\neq 0$. 
Now for every $\varepsilon>0$ there exists $N\in\mathbb N$ such that
\begin{align*}
\sum_{j=N+1}^\infty |a_j|^p < \frac{\varepsilon}{4\|b\|_q} \quad\text{and}\quad 
\sum_{j=N+1}^\infty |b_j'|^q < \frac{\varepsilon}{4\|a\|_p} \,.
\end{align*}

To prove $\overline A=\overline{A}_1$ let $\varepsilon>0$ and $x\in\overline A$ be given. Hence there exists a 
permutation ${\pi}: \mathbb N \to \mathbb N$ such that $x':=\sum_{n=1}^\infty a_nb_{{\pi}(n)}$ satisfies 
$|x-x'| < \varepsilon/4$. Now by \eqref{eq:Lemma_5b_1} one can construct a permutation 
$\hat{\pi}: \mathbb N \to \mathbb N$ which for all $k\in\lbrace1,\ldots,N\rbrace$ satisfies the following:
\begin{itemize}
\item If $b_{{\pi}(k)}\neq 0$, then $b_{{\pi}(k)}=b'_{\hat{\pi}(k)}$.
\item If $b_{{\pi}(k)}= 0$, then $\hat{\pi}(k)> N$.
\end{itemize}
Then for $y:=\sum_{n=1}^\infty a_n b'_{\hat{\pi}(n)} \in A_1$, using H\"older's inequality (Lemma \ref{lemma_hoelders_ineq}) one finds
\begin{align*}
|x-y| & < \frac{\varepsilon}{4}+\Big|\sum_{n=1}^N a_n(b_{{\pi}(n)}-b'_{\hat{\pi}(n)})\Big|
+ \sum_{n=N+1}^\infty |a_n|\big(|b_{{\pi}(n)}|+|b'_{\hat{\pi}(n)}|\big)\\
& \leq\frac{\varepsilon}{4} +\Big(\sum_{n=1}^N |a_n|^p\Big)\Big(\sum_{n=1}^N|b_{\pi(n)}-b'_{\hat\pi(n)}|^q\Big)+2\|b\|_q \sum_{n=N+1}^\infty |a_n|^p\\
&< \frac{3\varepsilon}{4}+\|a\|_p\sum_{\{1\leq n\leq N\,|\, b_{\pi(n)}=0 \}}|b'_{\hat\pi(n)}|^q\leq \frac{3\varepsilon}{4}+\|a\|_p\sum_{n=N+1}^\infty |b'_n|^q< \varepsilon\,.
\end{align*}
This shows the inclusion $\overline A \subset \overline{A}_1$. Obviously, the role of
$(b_n)_{n \in \mathbb N}$ and $(b_n')_{n \in \mathbb N}$ is interchangeable and thus the converse is shown the same way.

Next, we prove $\overline A = \overline{A}_2$. As by assumption all sums converge absolutely, 
rearranging them via permutations does not change their value and thus
\begin{align*}
A=\Big\lbrace \sum\nolimits_{n=1}^\infty a_{{\pi}(n)}b_{n} \,\Big|\, {\pi}:\mathbb N \to\mathbb N \text{ is permutation}\Big\rbrace
\end{align*}
and analogously for $A_2$. But now this follows from the previous step because $p,q\in (1,\infty)$ were chosen arbitrarily (and thus are interchangeable).

Finally, $\overline{A}=\overline{A}_1$ implies $\overline{A}_2 = \overline{A'}$ by choosing
$(a'_n)_{n\in\mathbb N}=(a_n)_{n\in\mathbb N}$, so $\overline{A}=\overline{A}_2=\overline{A'}$.
\end{proof}

\noindent
Note that Lemma \ref{Lemma_5b} becomes false if one replaces $c_0(\mathbb N)$ by $\ell^\infty(\mathbb N)$: For this consider $(a_n)_{n\in\mathbb N} = (a_n')_{n\in\mathbb N} :=( \frac12,\frac14,\frac18,\ldots)$,
$(b_n)_{n\in\mathbb N}:=(1,1,1,\ldots)$ as well as $(b_n')_{n\in\mathbb N}:=(0,1,1,1,\ldots)$. One readily verifies $A = \lbrace 1\rbrace$ and
$A'=\lbrace 1-\frac{1}{2^n}\,|\,n\in\mathbb N\rbrace$, hence $\overline{A} \subsetneq \overline{A'}$.

\begin{example}\label{ex_1}
Consider the set
$E:=\lbrace C\in\mathcal B^1(\mathcal H)\,|\,\|C\|_1\leq 1\rbrace\subset \mathcal B^1(\mathcal H)$
and define $C_n=|e_{n+1}\rangle\langle e_{n+1}|$, where $(e_n)_{n\in\mathbb N}$ is some orthonormal 
basis of $\mathcal H$. Obviously, $C_n \in E$ as $\|C_n\|_1=1$. Moreover, let $\Pi_n$ be the 
corresponding orthogonal projections as in \eqref{eq:defi_block} and set $T:=\mathbbm{1}_{\mathcal H}$
and $S_n=\Pi_n$ for all $n\in\mathbb N$. Then, by Lemma \ref{lemma_approx_strong_top}, the projections
$\Pi_n$ converge strongly to $\mathbbm{1}_{\mathcal H}$ but
\begin{align*}
\sup_{C\in E}|\operatorname{tr}(CS_n^* TS_n-C)|=\sup_{C\in E}|\operatorname{tr}(C\Pi_n-C)|
\geq |\operatorname{tr}(C_n\Pi_n - C_n)|=1
\end{align*}
as $C_n\Pi_n = 0$. Hence, $\lim_{n\to\infty}\sup_{C\in E}|\operatorname{tr}(CS_n^* TS_n-CS^* TS)| \geq 1$,
i.e. $\operatorname{tr}\big((\cdot)S_n^* TS_n\big)_{n \in \mathbb N}$ does not converge uniformly
to $\operatorname{tr}\big((\cdot)S^* TS\big)$ on $E$.
\end{example}

\begin{example}\label{ex_2}
Let $(e_n)_{n\in\mathbb N}$ of $\mathcal H$ be an orthonormal basis of $\mathcal H$ 
and choose $C := |e_1\rangle\langle e_1|$ and $T := \mathbbm{1}$. Then for
the corresponding block approximations one has $C_n = C$ and $T_n = \Pi_n$ for all $n \in \mathbb N$,
where $\Pi_n$ denotes the orthogonal projection onto $\operatorname{span}\{e_1, \dots, e_n\}$.
Therefore
\begin{align*}
W_{C_n}(T_n) = \{\operatorname{tr}(C_nU^* T_nU)\,|\, U \in \mathcal B(\mathcal H)\;\text{unitary}\}
= \{\langle x,\Pi_nx \rangle\,|\, \Vert x \Vert = 1\} = [0,1]
\end{align*}
and thus $1 = \overline{W_C(T)} \subsetneq \lim_{n\to\infty}\overline{W_{C_n}(T_n)} = [0,1]$.
\end{example}

\microtypesetup{protrusion=false}
\microtypesetup{protrusion=true}

\chapter*{Notation}\pagestyle{plain}
\addcontentsline{toc}{chapter}{Notation}
\renewcommand{\arraystretch}{1.3}
\section*{List of Abbreviations}
\begin{center}
\begin{tabular}{p{3cm}p{2cm}p{9.3cm}}
\textbf{Abb.}&\textbf{Reference}& \textbf{Description}\\\hline
Assumption IN & p.~\pageref{not_ass_in}& invariance condition for diagonal states \\
Assumption PK & p.~\pageref{not_ass_pk} & piecewise constant control amplitudes \\
 \textsc{cp} & p.~\pageref{not_cp}& completely positive \\
 \textsc{cptp} & p.~\pageref{not_cptp}& completely positive, trace preserving \\
 \textsc{csp} & p.~\pageref{not_csp} & completely strictly positive \\
\textsc{gksl}& p.~\pageref{not_gksl} & Gorini, Kossakowski, Sudarshan, Lindblad \\
\textsc{ode} &--- & ordinary differential equation \\
 \textsc{p} & p.~\pageref{not_p}& positive \\
 \textsc{ptp} & p.~\pageref{not_ptp} & positive, trace preserving \\
 \textsc{qds} & p.~\pageref{not_qds} & quantum-dynamical semigroup \\
s.o.t.& p.~\pageref{def_strong_weak_op_top} &strong operator topology \\
 \textsc{sp} & p.~\pageref{not_sp}& strictly positive \\
 w.o.t. & p.~\pageref{def_strong_weak_op_top}& weak operator topology 
\end{tabular}
\end{center}

\section*{List of Symbols}
In the following list $(\cdot)$ denotes one or sometimes multiple arguments of the expression in question. For example $\mathcal B(\cdot)$ can appear in the main text as $\mathcal B(\mathcal H)$ or $\mathcal B(X,Y)$ or similar.
\begin{center}
\begin{longtable}{p{3cm}p{1.9cm}p{9.4cm}}
\textbf{Symbol}&\textbf{Reference}&\textbf{Description}\\\hline
 $|\,\cdot\,\rangle\langle\,\cdot\,|$ & p.~\pageref{symb_bra_ket} & bra-ket notation, $|y\rangle\langle x|$ denotes the linear map $:\mathcal H\to\mathcal H$, $z\mapsto \langle x,z\rangle y$ \\
 $\|\cdot\|_\textrm{op}$& p.~\pageref{symb_op_norm} & operator norm \\
$\|\cdot\|_p$ & p.~\pageref{def_schatten_class} & Schatten-$p$ norm, $p\in[1,\infty]$. \textbf{Exception:} If the argument is a vector or sequence then this denotes the vector-$p$- or $\ell^p$-norm (p.~\pageref{ex_ell_p_space}) \\
 $(\cdot)_+$ & p.~\pageref{lemma_char_d_vec} & --- \\
\textbf{Symbol}&\textbf{Reference}&\textbf{Description}\\\hline
 $(\cdot)^+$, $(\cdot)^-$ & p.~\pageref{thm_vN_sep_HS} & positive and negative part of a hermitian matrix or a compact self-adjoint operator \\
 $(\cdot)'$ & p.~\pageref{def_dual_op} & dual operator \\
 $(\cdot)^*$ & --- & \textit{dual space} if argument is normed or topological vector space (p.~\pageref{def_dual_space_tvs}), \textit{adjoint operator} if argument is Hilbert space operator (p.~\pageref{prop_adjoint_op_hilbert}), \textit{dual channel} if argument is Schr\"odinger quantum channel (p.~\pageref{eq:duality_trace_map}) \\
 $(\cdot)^\downarrow$ &p.~\pageref{symb_downarrow_vec}& components of (real) vector or (non-negative) sequence in decreasing order \\
 $\langle\,\cdot\,\rangle_\textsf{Lie}$ & p.~\pageref{symb_Lie_closure} & smallest linear subspace of a given Lie algebra which contains the argument together with all iterated Lie brackets \\
 $\simeq$ & p.~\pageref{symb_iso_iso} & isometrically isomorphic \\
$\geq 0$ & p.~\pageref{symb_positive_semi_def} & positive semi-definite operator \\
$>0$ & p.~\pageref{symb_positive_def} & positive definite operator \\
$\sqrt{(\cdot)}$ & p.~\pageref{symb_sqrt} & square root of a positive semi-definite operator \\
 $|\cdot|$ & p.~\pageref{symb_abs} & absolute value of an operator \\
 $\overline{(\cdot)}$ & p.~\pageref{symb_closure} & closure in a topological space. \textbf{Exception:} smallest closed extension if the argument is an operator (p.~\pageref{symb_closure_op}) \\
 $\overline{(\cdot)}^{\,\tops}$ & --- & closure in the strong operator topology $\tops$ \\
 $\overline{(\cdot)}^{\,\topw}$ & --- & closure in the weak operator topology $\topw$ \\
 $\overline{(\cdot)}^{\,\mathrm{u}}$ & p.~\pageref{symb_subspace_top_cl} & closure in the subspace topology induced by the operator norm topology \\
 $\overline{(\cdot)}^{\,\mathrm{s}}$ & p.~\pageref{symb_subspace_top_cl} & closure in the subspace topology induced by the strong operator topology \\
 $\underline{(\cdot)}$ & p.~\pageref{footnote_permutation_matrix} & permutation matrix $\underline{\pi}$ corresponding to a permutation $\pi$ (footnote \ref{footnote_permutation_matrix}) \\
 $\otimes$ & p.~\pageref{app_tensor_hilbert} ff. & tensor product of Hilbert spaces or Hilbert space operators. If the arguments are matrices then this is the Kronecker product (p.~\pageref{footnote_vec}, footnote \ref{footnote_vec}) \\
 $\oplus$ & --- & direct sum. In abuse of notation $A\oplus B={\footnotesize\begin{pmatrix}
A&0\\0&B \end{pmatrix}}$ for square matrices or operators $A,B$ \\
$[\cdot,\cdot]$ & p.~\pageref{footnote_lie_algebra} & Lie bracket (footnote \ref{footnote_lie_algebra}). Usually $[A,B]=AB-BA$ \\
 $[\,\cdot\,]_n^{(\cdot)}$ & p.~\pageref{cut_out_operator} & cut-out operator \\&&\\
\textbf{Symbol}&\textbf{Reference}&\textbf{Description}\\\hline
 $\prec$ & p.~\pageref{symb_maj_1}, \pageref{symb_maj_2}, \pageref{defi_maj} & classical majorization \\
 $\prec_d$ & p.~\pageref{defi_d_stochastic_matrix} & $d$-majorization (on vectors) \\
 $\prec_D$ & p.~\pageref{defi_matrix_D_maj} & $D$-majorization (on matrices) \\
 $\mathbbm{1}$ & --- & identity operator $\mathbbm{1}_X:X\to X$, $x\mapsto x$. also $\mathbbm{1}_n:=\mathbbm{1}_{\mathbb C^n}$ is the $n\times n$ identity operator (identity matrix)\\
 $\operatorname{Ad}$ & --- & $\operatorname{Ad}_A(B)=ABA^{-1}$ 
 (adjoint representation) \\
 $\operatorname{ad}$ & p.~\pageref{symb_ad_rep} & $\operatorname{ad}_A(B)=[A,B]$ (adjoint representation of a Lie algebra) \\
 $B_r(\cdot)$ & p.~\pageref{symb_metric_ball} & open ball of radius $r$ in a metric space \\
 $\mathcal B(\cdot)$ & p.~\pageref{def_bounded_op} & collection of all bounded linear operators \\
 $\mathcal B$ & p.~\pageref{defi_basis_topo} & basis for a topology (no argument) \\
$\mathcal B^p(\cdot)$ & p.~\pageref{def_schatten_class} & Schatten-$p$ class, $p\in[1,\infty]$ \\
$\mathbb B(\cdot)$ & p.~\pageref{footnote_sigma_algebra_borel}, footnote \ref{footnote_sigma_algebra_borel} & Borel-$\sigma$ algebra \\
$B_0$, $B_0(\cdot)$ & p.~\pageref{eq:action_gamma_Delta}, \pageref{eq:control-simplex_evolution} & dissipative action in toy model, action of matrix operator on diagonal \\
$\mathfrak{b}(\cdot)$ & p.~\pageref{def_mathfrak_b} & --- \\
 $C(\cdot)$ & p.~\pageref{lemma_choi_matrix} & Choi matrix \\
 $c_0(\cdot)$, $c_{00}(\cdot)$ & p.~\pageref{ex_ell_p_space} & sequence spaces \\
 $\operatorname{conv}$ & --- & convex hull \\
 $\Gamma$ & p.~\pageref{symb_gamma_qds} & dissipative part of \textsc{qds}-generator \\
$\Gamma_n$, $\Gamma_{(\cdot)}^{(\cdot)}$ & p.~\pageref{Gamma} & embedding of $\mathbb C^n$ into infinite-dimensional space \\
 $\gamma(t)$ & p.~\pageref{symb_gamma_t} & control function for the dissipative part of a \textsc{qds}-generator \\
 $D(\cdot)$ & --- & domain of a map. \textbf{Exception:} diagonal density matrices if the argument is a number (e.g., $D(n)$, p.~\pageref{symb_D_n_diag}) \\
 $\mathbb D(\cdot)$ & p.~\pageref{symb_q_states} & collection of all quantum states / density operators \\
$\Delta^{(\cdot)}$ & p.~\pageref{symb_simplex} & standard simplex \\
$\Delta$ & p.~\pageref{app_hausdorff} ff. & Hausdorff metric \\
$d$ & p.~\pageref{def_metric} & metric. \textbf{Exception:} $d(z,A)$ denotes the distance of a point $z\in X$ to a subset $A$ of a metric space $(X,d)$ (p.~\pageref{eq.Hausdorff-1}) \\
 $\delta_{jk}$ & --- & Kronecker delta, is $1$ if $j=k$ and $0$ else \\
 $\partial(\cdot)$ & p.~\pageref{symb_boundary} & boundary in a topological space \\&&\\
\textbf{Symbol}&\textbf{Reference}&\textbf{Description}\\\hline
$\operatorname{diag}(\cdot)$ & --- & maps a vector or sequence to the diagonal of an operator. more precisely $:\mathcal H\to\mathcal L(\mathcal H)$, $x\mapsto \sum_j\langle e_j,x\rangle|e_j\rangle\langle e_j|$ for an orthonormal basis $(e_j)_j$ of a separable Hilbert space $\mathcal H$ \\
 $E_b(\pi)$ & p.~\pageref{def_Ebsigma} & extreme points of certain convex polytopes \\
 $E_n(\cdot)$ & p.~\pageref{symb_embedding} & embedding operator \\
 $\unitvector$ & p.~\pageref{symb_unit_vector} & column-vector of all ones \\
 $\mathcal F(\cdot)$ & p.~\pageref{symb_finite_rank} & collection of all finite-rank operators \\
 $\mathbb F$ & p.~\pageref{symb_base_field} & base field of vector space, usually $\mathbb R$ or $\mathbb C$ \\
$f(\cdot)$ & p.~\pageref{eq:f_b_vec} & in Ch.~\ref{sec_d_maj_poly}: function which characterizes $\prec_d$-polytope \\
$\operatorname{fix}(\cdot)$ & p.~\pageref{symb_fixed_points} & set of fixed points of an operator \\
 $\operatorname{GL}(\cdot)$ & p.~\pageref{symb_GL} & general linear group \\
$\mathrm{gr}(\cdot)$ & p.~\pageref{symb_graph} & graph of a function \\
$\mathcal H,\mathcal G$ & --- & common symbols for Hilbert spaces \\
 $H_0,H_j$ & --- & usually Hamiltonians, i.e.~self-adjoint Hilbert space operators \\
$\iota$ & p.~\pageref{symb_can_emb_1}, \pageref{symb_can_emb_2} & canonical embedding \\
$\Im$ & --- & imaginary part of a complex number \\
$\operatorname{im}(\cdot)$ & p.~\pageref{symb_image} & image of a function \\
$\operatorname{int}(\cdot)$ & p.~\pageref{symb_interior} & interior in a topological space \\
$\mathcal K(\cdot)$ & p.~\pageref{def_compact_op} & collection of all compact operators \\
 $K_C(T)$ & p.~\pageref{def_KC} & --- \\
 $\operatorname{ker}(\cdot)$ & --- & kernel / null space of a map \\
 $L$ & p.~\pageref{eq:gksl_gen} & usually: generator of a \textsc{qds} \\
 $\mathcal L(\cdot)$ & p.~\pageref{def_linear_op} & collection of all linear operators \\
 $\ell^p(\cdot)$, $\ell^\infty(\cdot)$ & p.~\pageref{ex_ell_p_space} & sequence spaces \\
$\ell^1_+(\cdot)$ & p.~\pageref{symb_ell_1_plus} & collection of all summable sequences with non-negative entries \\
 $L^p(\cdot)$, $L^\infty(\cdot)$ & p.~\pageref{eq:control_eq} & collection of all Lebesgue measurable functions on the respective domain which satisfy $\int |f(t)|^p\,dt<\infty$ (for $p\in[1,\infty)$) or are essentially bounded ($p=\infty$, p.~\pageref{symb_L_infty}) \\
 $\mathcal{LI}$ & p.~\pageref{symb_LI} & locally integrable control functions \\
$\lambda_n(\cdot)$ & p.~\pageref{prop_compact_spectrum} & eigenvalue sequence, possibly modified (p.~\pageref{symb_modified_ev_seq}) \\
 $\lambda_n^\downarrow(\cdot)$ & p.~\pageref{symb_downarrow_uparrow} & decreasing arrangement of (real) eigenvalue sequence \\
\textbf{Symbol}&\textbf{Reference}&\textbf{Description}\\\hline
 $\lambda_n^\uparrow(\cdot)$ & p.~\pageref{symb_downarrow_uparrow} & increasing arrangement of (real) eigenvalue sequence \\
 $M_d(\cdot)$ & p.~\pageref{sec_d_maj_poly} & set of elements $d$-majorized by the argument \\
 $M_D(\cdot)$ & p.~\pageref{eq:def_MD} & set of elements $D$-majorized by the argument \\
 $N(\cdot)$ & p.~\pageref{lemma_topsw_basis}, \pageref{eq:seminorms_subbasis} & neighbourhood with respect to a seminorm \\
$\Pi_k$ & p.~\pageref{symb_proj_k_el_1}, \pageref{eq:defi_block} & $\Pi_k=\sum_{j=1}^k|e_j\rangle\langle e_j|$ \\
 $P_T$ & p.~\pageref{def_mean_erg} & mean ergodic projection \\
$P_C(T)$ & p.~\pageref{defi_3} & $C$-spectrum of a compact operator \\
$\mathcal P(\cdot)$ & --- & power set; collection of all subsets of the argument \\
 $\mathcal P_c(\cdot)$ & p.~\pageref{symb_P_c_X} & collection of non-empty compact subsets of a metric space \\
$\mathcal{PC}$ & p.~\pageref{symb_PC} & piecewise continuous control functions \\
 $\mathcal{PK}$ & p.~\pageref{symb_PK} & piecewise constant control functions \\
 $\mathfrak{pos}(\cdot)$ & p.~\pageref{symb_positive_semi_def} & collection of all positive semi-definite operators \\
 $Q_H(\cdot)$ & p.~\pageref{symb_heisenberg_ch} & collection of all Heisenberg quantum channels \\
 $Q_S(\cdot)$ & p.~\pageref{symb_schrodinger_ch} & collection of all Schr\"odinger quantum channels \\
 $Q(\cdot)$ & p.~\pageref{symb_Q_n} & finite-dimensional special case: $Q(n,k)=Q_S(\mathbb C^n,\mathbb C^k)$ \\
 $Q_{(\cdot)}(\cdot)$ & p.~\pageref{symb_Q_X_n} & all channels with a common fixed point \\
 $Q_{(\cdot)}^E(\cdot)$ & p.~\pageref{symb_Q_X_E_n} & all extreme points of $Q_{(\cdot)}(\cdot)$ \\
 $\rho$ & --- & usual notation for a state (i.e.~an element of $\mathbb D(\cdot)$) \\
$\rho_\textsf{Gibbs}$ & p.~\pageref{symb_gibbs_state} & Gibbs state \\
 $\mathbb R_+$ & p.~\pageref{symb_R_plus} & $[0,\infty)$ \\
$\mathbb R_{++}$ & p.~\pageref{symb_R_plus} & $(0,\infty)$ \\
 $\mathbbm{r}(\cdot)$ & p.~\pageref{symb_resolvent} & resolvent of an operator \\
$\mathfrak{reach}_{[0,T]}$, $\mathfrak{reach}$ & p.~\pageref{def_contr_bounded} & reachable set of some control system \\
$\mathfrak{reach}_{(\cdot)}$ & p.~\pageref{symb_reach_lambda}, \pageref{symb_reach_sigma} & reachable set of a particular control problem \\
 $\sigma(\cdot)$ & p.~\pageref{symb_spectrum} & spectrum of an operator \\
 $\sigma_{\mathrm{c}}(\cdot)$ & p.~\pageref{symb_spectrum} & continuous spectrum of an operator \\
$\sigma_{\mathrm{p}}(\cdot)$ & p.~\pageref{symb_spectrum} & point spectrum of an operator \\
$\sigma_{\mathrm{r}}(\cdot)$ & p.~\pageref{symb_spectrum} & residual spectrum of an operator \\
 $\sigma(X,(\cdot))$ & p.~\pageref{def_initial_top} & initial topology \\
 $\sigma(X,X^*)$ & p.~\pageref{symb_weak_top} & weak topology (special case of initial topology) \\
 $\sigma(X^*,X)$ & p.~\pageref{symb_weak_star_top} & weak* topology (special case of initial topology) \\
\textbf{Symbol}&\textbf{Reference}&\textbf{Description}\\\hline
 $\sigma_+$, $\sigma_-$ & p.~\pageref{footnote_spin_j} & ladder operators in spin-$j$ representation (footnote \ref{footnote_spin_j}) \\
 $\sigma_+^d$, $\sigma_-^d$ & p.~\pageref{thm_bath} & weighted ladder operators \\
$s_j(\cdot)$ & p.~\pageref{eq:compact_SVD} & singular value of a compact operator \\
$s_d(n)$ & p.~\pageref{symb_d_stoc_mat} & collection of all $d$-stochastic matrices \\
 $\mathcal S$ & p.~\pageref{defi_subbasis_topo} & subbasis for a topology (no argument) \\
 $S_C(T)$ & p.~\pageref{defi_1_SC} & --- \\
$S_r(\cdot)$ & p.~\pageref{symb_metric_sphere} & sphere of radius $r$ in a metric space \\
$S_\Omega(\cdot)$ & p.~\pageref{def_contr_bounded}, \pageref{eq:system_semigroup_unbounded} & system semigroup \\
 $S_n$ & p.~\pageref{symb_sym_group} & symmetric group (collection of permutations of order $n$) \\
 $\mathbb S(\cdot)$ & p.~\pageref{symb_bi_stoch_qc} & collection of all bi-stochastic quantum maps \\
 $\operatorname{span}(\cdot)$ & p.~\pageref{footnote_linear_span} & linear span (footnote \ref{footnote_linear_span}) \\
$SU(\mathcal H)$ & p.~\pageref{symb_SU} & special unitary group \\
 $\mathfrak{su}(\mathcal H)$ & p.~\pageref{symb_su} & special unitary algebra \\
$\tau$ & p.~\pageref{defi_topology} & general symbol for topology. \textbf{Exception:} In Ch.~\ref{ch_maj_cnr} $\tau$ is sometimes used for permutations \\
 $\topn$ & p.~\pageref{prop_strong_weak_op_top} & operator norm topology \\
 $\tops$ & p.~\pageref{def_strong_weak_op_top} & strong operator topology \\
 $\topuw$ & p.~\pageref{def_uw_topology} & ultraweak operator topology \\
 $\topw$ ($\topw^*$) & p.~\pageref{def_strong_weak_op_top} (\pageref{defi_weak_star_top}) & weak (weak*) operator topology \\
$\operatorname{tr}(\cdot)$ & p.~\pageref{def_trace} & trace functional \\
$\operatorname{tr}_{\mathcal K}(\cdot)$ & p.~\pageref{eq:partial_trace_1} & partial trace with respect to a space \\
$\operatorname{tr}_{\omega}(\cdot)$ & p.~\pageref{eq:partial_trace_2} & partial trace with respect to a state \\
 $\mathcal U(\cdot)$ & p.~\pageref{symb_unitary} & collection of all unitary operators \\
$\mathfrak u(\cdot)$ & p.~\pageref{symb_un_alg} & unitary algebra \\
$u(\cdot)$, $u_j(\cdot)$ & p.~\pageref{eq:bilinear_1} & usually control functions \\
 $V_j$ & p.~\pageref{symb_lindblad_V} & Lindblad-$V$, generators of the dissipative part $\Gamma$ of $L$ \\
 $\operatorname{vec}$ & p.~\pageref{footnote_vec} & vectorization (footnote \ref{footnote_vec}) \\ 
$W_C(T)$ & p.~\pageref{defi_1} & $C$-numerical range of an operator \\
 $W_e(T)$ & p.~\pageref{symb_W_e_T} & essential numerical range of an operator \\
 $X,Y$ & --- & common symbols for normed or Banach spaces 
\end{longtable}
\end{center}
\newpage
\addcontentsline{toc}{chapter}{Index}
\printindex

\newpage
\printbibliography

\end{document}